\documentclass{aa}  

\usepackage{float}
\usepackage{graphicx}
\usepackage{txfonts}
\usepackage{longtable}
\usepackage{caption}
\usepackage{hyperref}
\begin{document}

   \title{DCO$^+$ and DCN 1-0 survey toward a sample of Planck cold clumps}

   \author{Fu Mo\inst{1}
          \and
          Junzhi Wang\inst{1}
          \and
          Shu Liu\inst{2}
          \and
          Yan Duan\inst{3}\fnmsep\inst{2}
          \and
          Huanxue Feng\inst{4}
          \and
          Yuqiang Li\inst{5}\fnmsep\inst{6}
          \and
          Zhe Lu\inst{1}\fnmsep\inst{7}
          \and
          Rui Luo\inst{1}
          \and
          Chao Ou\inst{1}
          \and
          Yani Xu\inst{1}
          \and
          Zhuoying Yan\inst{1}
          }

   \institute{Guangxi Key Laboratory for Relativistic Astrophysics, School of Physical Science and Technology, Guangxi University, Nanning 530004,  People’s Republic of China\\
   \email{junzhiwang@gxu.edu.cn}
         \and
         National Astronomical Observatories, Chinese Academy of Sciences, Beijing 100101, People’s Republic of China
         \and
         Space Engineering University, Beijing 101416, People’s Republic of China
         \and
         School of Physics and Astronomy, Sun Yat-sen University, Zhuhai, 519082, People’s Republic of China
         \and
         Shanghai Astronomical Observatory, Chinese Academy of Sciences, No. 80 Nandan Road, Shanghai, 200030, People’s Republic of China
         \and
         School of Astronomy and Space Sciences, University of Chinese Academy of Sciences, No. 19A Yuquan Road, Beijing 100049, People’s Republic of China
         \and
         Department of Electrical and Electronic Engineering, Guilin University of Technology at Nanning, Nanning 530001, People’s Republic of China
             }

   \date{Received xx; accepted xx}

 
  \abstract
   {Deuterated molecules can be used to study the physical conditions and the astro-chemical evolution of molecular clouds. }
   {large-sample surveys for deuterated molecules are needed  to understand the enhancement of deuterated molecules from diffuse molecular gas to cold cores. }
   {A single-pointing survey toward the 559 Planck cold clumps of the Early Cold Core Catalogue (ECC) has been conducted using the Arizona Radio Observatory 12-meter telescope, focusing on the $J$=1-0 transitions of DCO$^+$ and DCN. The surve included observations of 309 cores for DCO$^+$ and DCN 1-0 simultaneously, followed by 71 of these cores where DCO$^+$ 1-0 was detected for H$^{13}$CO$^+$ and H$^{13}$CN 1-0 simultaneously, aiming to determine the deuterated fraction ($D_{\rm frac}$). Additionally, 250 cores were observed for  DCO$^+$, DCN, H$^{13}$CO$^+$ and H$^{13}$CN 1-0 simultaneously. }
   {Among the 309 sources, DCO$^+$ and DCN 1-0 were detected in 79 and 11 sources, with a detection rates of 25.6\% and 3.6\% respectively. In the 250 sources observed for all four species, DCO$^+$, DCN, H$^{13}$CO$^+$ and H$^{13}$CN 1-0 were detected in 58, 9, 57 and 13 sources, with a detection rate of 23.2\%, 3.6\%, 22.8\% and 5.2\% respectively. The $D_{\rm frac}$(HCO$^+$) values in 112 sources range from 0.89\% to 7.4\% with a median value of 3.1\%, while $D_{\rm frac}$(HCN) values in 11 sources range from 1.5\% to 5.5\% with a median value of 2.3\%. The line widths of DCO$^+$ and H$^{13}$CO$^+$ 1-0  detections are mostly within 1 km s$^{-1}$. }
   { The similarity in $D_{\rm frac}$ values between HCO$^+$ and HCN indicates that the higher detection rate of DCO$^+$ 1-0 compared with DCN 1-0 is due to the lower critical density of DCO$^+$ 1-0. We suggest that the enhancement of DCO$^+$ and DCN likely begins in the early diffuse stage of the molecular cloud, rather than during the cold core formation stage.
}


   \maketitle
%
\section{Introduction} \label{sec:intro}

Deuterated molecules are useful tools for studying the physical conditions and astro-chemical evolution of molecular clouds   \citep{2002A&A...381.1026R,2006A&A...448L...5G}.  Deuterium fractionation,  i.e., the abundance ratios of deuterated molecules to their hydrogenated counterparts are much enhanced over the cosmic D/H elemental abundance ratio $\sim$1.5 $\times$ 10$^{-5}$  \citep{2000A&A...361..388R}. Due to the lower zero-point energy of deuterated molecules compared to their non-deuterated counterparts which ensures that  deuterium is preferentially bonded into molecules compared to hydrogen, deuterated molecules can be synthesized effectively in the cold gas phase during the early stage of the evolution of molecular clouds \citep{2000A&A...361..388R}. The process of deuterium fractionation is sensitive to various physical conditions such as temperature, density, CO depletion \citep{1989ApJ...340..906M,{2020ApJ...901..145F}}, ionization  \citep{2002P&SS...50.1133C}, etc., with low temperatures and high densities having particularly significant effects  \citep{2011A&A...529L...7F, 2011A&A...530A.118P}. Therefore, deuterated molecules are considered as excellent tracers for cold (T $\sim$10 K) and dense (n $\ge$10$^4$ cm$^{-3}$) regions within molecular clouds \citep{2007A&A...471..849R}.

After the detection of DCN in the Orion Nebula molecular cloud  \citep{1973ApJ...179L..57J}, many other deuterated species have been identified in molecular clouds over the past decades. These include
DCO$^+$  \citep{1976ApJ...209L..83H, 2000A&A...356.1039T, 2002P&SS...50.1133C}, 
DNC  \citep{1977ApJ...216L.111S, 2009A&A...507..347V, 2024ApJS..270...35Y},
N$_2$D$^+$ \citep{1977ApJ...218L..61S, 2005ApJ...619..379C, 2006A&A...460..709F}, 
deuterated NH$_3$ \citep{1978ApJ...219L..43T, 2005A&A...438..585R, 2007A&A...467..207P,2022MNRAS.512.4934L}, and 
deuterated H$_2$CO \citep{1975ApJ...201..102W, 2002A&A...381.1026R, 2007A&A...471..849R}. 
Among these deuterated molecules, DCO$^+$ and DCN are particularly valuable for studying the physical conditions and astro-chemical evolution of molecular clouds. This is due to their relatively high abundances, simple rotational spectra, accessible rotational transitions \citep{2024ApJS..270...35Y}, and their synthetic pathways, which are closely  linked to the deuterium fractionation process  \citep{1989ApJ...340..906M}. 
Observational studies  \citep{1973ApJ...181L.129W, 2002A&A...381.1026R, 2020ApJ...901..145F} and theoretical models  \citep{1982A&A...111...76H, 1989ApJ...340..906M}
 of deuterium chemistry suggest that DCO$^+$ and DCN are important for understanding the evolution of molecular cloud.

The deuterated fraction ($D_{\rm frac}$) is defined as the abundance ratio of a deuterated molecule to its hydrogenated counterpart. This ratio has been studied for many deuterated species \citep{2001ApJ...547..814H, 2001ApJS..136..579T, 2002P&SS...50.1133C, 2002A&A...381.1026R}, including DCO$^+$ and DCN.
DCO$^+$ was first detected in the molecular clouds NGC 2264 and DR (OH) with H$^{13}$CO$^+$/DCO$^+$ abundance ratios of 0.54 and 1.18 \citep{1976ApJ...209L..83H}. 
The DCO$^+$/HCO$^+$ ratio was found to be $\sim$0.18 in the dark cloud L134N and 0.02 in TMC1-N  \citep{2000A&A...356.1039T}. 
 For massive starless clump candidates, the DCO$^+$/HCO$^+$ and DCN/HCN ratios were estimated to be 0.011-0.040 and 0.004-0.045, respectively \citep{2024ApJS..270...35Y}. 
 The DCN/HCN ratio increases from 0.001 in the hot core gas close to the infrared source IRc2 to values of 0.01-0.06 in the OMC-1 ridge region \citep{1992A&A...256..595S}.  
The abundance of deuterated molecules is considered crucial for understanding both the physical conditions and deuterium chemistry in the cold gas phase of molecular clouds.

However,  large-sample surveys of DCO$^+$ and DCN lines toward cold dense cores are still lacking in the literature. This gap limits our understanding of the enhancement of deuterated molecules in the phase from diffuse molecular gas to cold cores,  including the measurements of $D_{\rm frac}$ for these molecules. The Planck cold clumps from the Early Cold Core Catalogue (ECC), which were selected from \cite{2012ApJ...756...76W}, lack feedback from star formation, represent some of the most quiescent regions, and are ideal for such large-sample surveys. These Planck cold clumps have a typical mass of $\sim$5 $M_{\odot}$, a size of $\sim$0.5 pc \citep{2016A&A...594A..28P}, and an excitation temperature of $\sim$10 K \citep{2012ApJ...756...76W}.

In this paper, we present a single-pointing survey of DCO$^+$ and DCN 1-0 toward the 559 Planck cold clumps observed to date. The observational details are described in Section \ref{sec:observations}, the results are presented in Section \ref{sec:results}, and the discussion and summary are provided in Section \ref{sec:discussion} and Section \ref{sec:summary} respectively.

\section{Observations} \label{sec:observations}

In total, 559 sources selected from the 674 Planck cold clumps with CO line detections \citep{2012ApJ...756...76W} were  observed using the Arizona Radio Observatory (ARO) 12-meter telescope in 2020 and 2023. Among the 559 observed sources, the masses of 83 sources can be found from \cite{2016A&A...594A..28P}, with mass values ranging from 0.29 to 1.85 $\times$ 10$^4$ $M_{\odot}$. There are 75 sources with masses below 100 $M_{\odot}$, 6 sources with masses between 100 and 1000 $M_{\odot}$, and only 2 sources with masses exceeding 1000 $M_{\odot}$. The kinematic distances of these 559 sources range from 0.10 to 21.58 kpc, while the excitation temperatures derived from CO 1-0 ($T_{\rm ex}$(CO) hereafter) range from 3.9 to 27.1 K \citep{2012ApJ...756...76W}. The beam size at 72 GHz is $\sim$87 $^{\prime \prime}$. The focus was checked at the beginning of each observing block. Pointing was checked every two hours on a nearby quasar or planet. The main beam brightness temperature ($T_{\rm mb}$) is calculated from  $T_{\rm A}^{\ast}$ = $T_{\rm mb}\times$ $\eta$, where $T_{\rm A}^{\ast}$ is the antenna temperature and $\eta$ is the average beam efficiency of the 4 mm receiver, with a value of 0.92 $\pm$ 0.06.

In early 2020, 258 sources were observed in DCO$^+$ and DCN 1-0 using the 4 mm receiver (66-90 GHz) with dual polarizations and ARO Wideband Spectrometer (AROWS) with mode 3, which provides 78.125 kHz channel spacing ($\sim$0.32 km s$^{-1}$ at 72GHz) and 500 MHz bandwidth, covering DCO$^+$ 1-0 at 72.039312 GHz and DCN 1-0 at 72.414927 GHz simultaneously. Due to the limited  velocity resolution of $\sim$0.32 km s$^{-1}$ with AROWS mode 3, accurate measurements of full width at half maximum (FWHM) for sources with narrow line widths were challenging. To address this, high spectral resolution supplementary observations were conducted for 32 sources with narrow DCO$^+$ 1-0 lines using AROWS mode 5, which provides 19.531 kHz channel spacing ($\sim$0.081 km s$^{-1}$ at 72 GHz) and 125 MHz bandwidth. To derive $D_{\rm frac}$, 41 sources with DCO$^+$ 1-0 detections were observed for H$^{13}$CO$^+$ 1-0 at 86.754288 GHz and H$^{13}$CN 1-0 at 86.3401764 GHz simultaneously, using AROWS mode 3 with a velocity resolution $\sim$0.27 km s$^{-1}$ at 86 GHz. Standard position switching mode was used with 3 minutes on and 3 minutes off for each source. The typical system temperatures ($T_{\rm sys}$) were 150 K at 72 GHz and 120 K at 86 GHz, resulting in a root-mean-square (rms) noise level of $\sim$50 mK at 78.125 kHz channel spacing for the final spectrum after averaging both polarizations. 

In 2023, 51 sources were observed for DCO$^+$ and DCN 1-0, with 30 sources additionally observed for H$^{13}$CO$^+$  and H$^{13}$CN 1-0, using the same setup as in 2020.

In 2023,  250 sources were observed using the 4 mm receiver and the newly updated AROWS multi-window mode 13, which provides up to eight spectral windows of the same size with 19.53 kHz channel spacing ($\sim$0.081 km s$^{-1}$ at 72 GHz) and 40 MHz bandwidth. This setup covers DCO$^+$ and DCN 1-0 in the lower sideband (LSB) and H$^{13}$CO$^+$ and H$^{13}$CN 1-0 in the upper sideband (USB) simultaneously. The standard position switching mode was used with 30 seconds on  and 30 seconds off, repeating 6 times for each source. The typical $T_{\rm sys}$ values are 150 K at 72 GHz and 120 K at 86 GHz, resulting in an rms noise level of $\sim$100 mK at 19.53 kHz channel spacing  for the final spectrum after averaging both polarizations. In observations using AROWS mode 13 in 2023, due to a software bug  in the multi-window configuration code during shared risking period for the wrong sign of Doppler correction for the second sideband, even though all the parameters obtained for the lines in the first sideband, which is the LSB including  DCO$^+$ and DCN 1-0, the  H$^{13}$CO$^+$ and the $V_{\rm LSR}$ measurements  of H$^{13}$CN 1-0 of  in USB exhibit offsets of 1 to 7 km s$^{-1}$, while all the other parameters are right. Therefore, the $V_{\rm LSR}$ values of H$^{13}$CO$^+$ and H$^{13}$CN 1-0 are not provided in Section \ref{sec:results} and not used for scientific discussions.

Data reduction was performed using the CLASS package in GILDAS\footnote{http://www.iram.fr/IRAMFR/GILDAS} software.  After checking each scan,  bad scans (fewer than 3\%) were discarded. A first-order baseline was applied to all spectral lines, and a single-component Gaussian fitting was applied to obtain parameters, including velocity-integrated intensity ($\int T_{\rm mb}{\rm d}v$ denoted as $W$ hereafter), $V_{\rm LSR}$ and FWHM for each source.

\section{Results} \label{sec:results}
\begin{figure}
\centering
\includegraphics[width=0.4\textwidth]{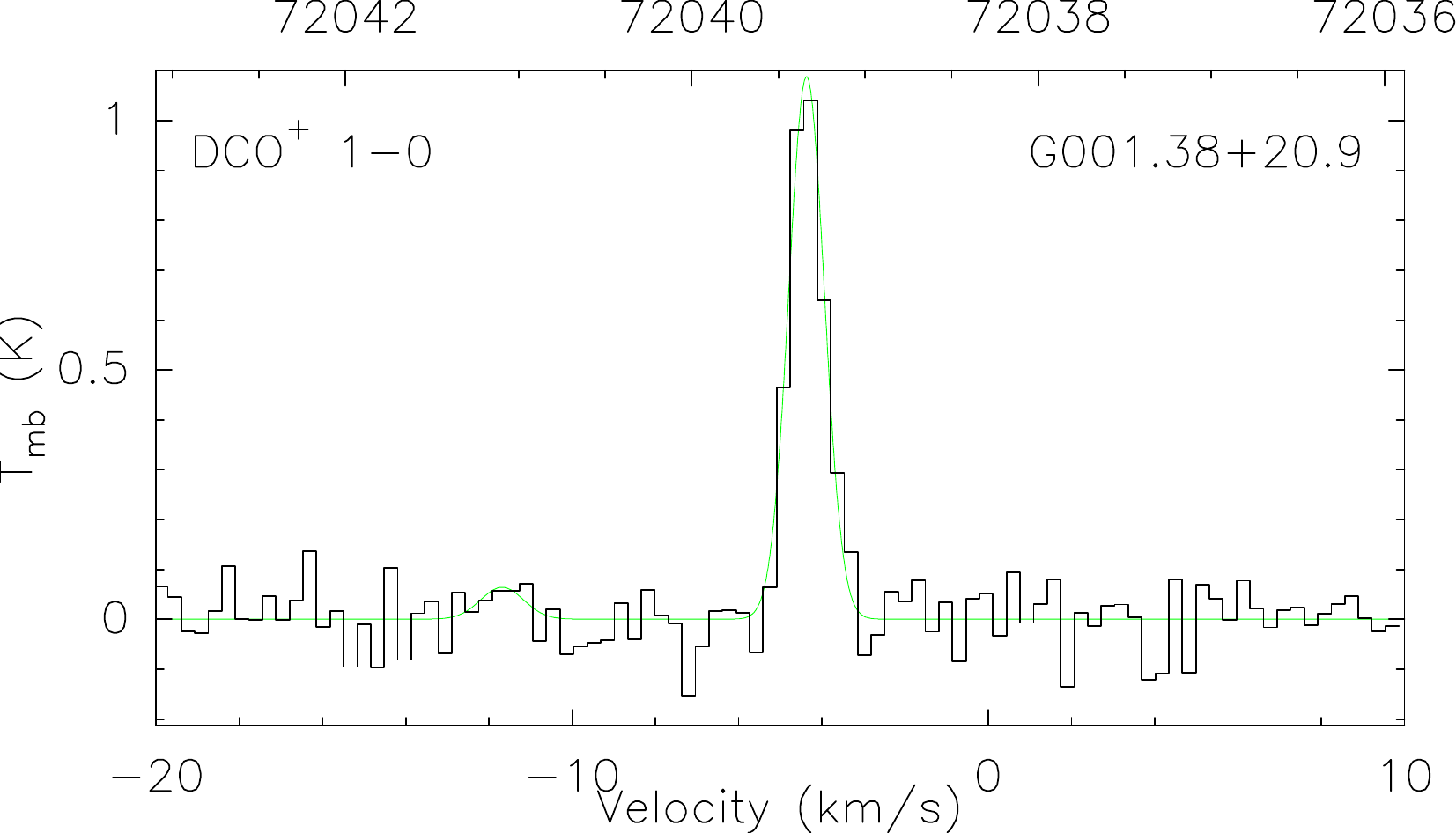}
\includegraphics[width=0.4\textwidth]{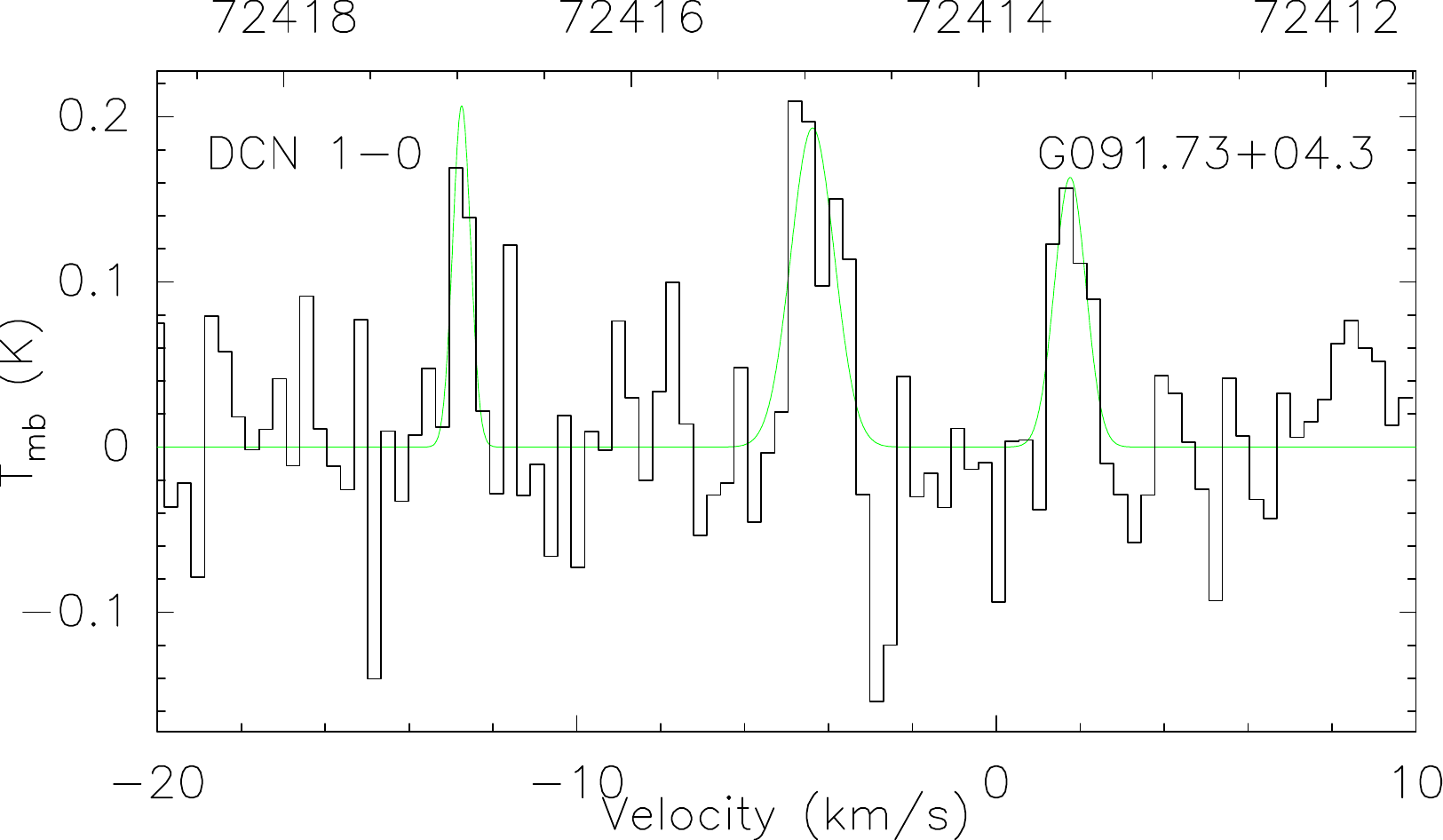}

\caption{Spectral lines of DCO$^+$ and DCN 1-0 (black line) overlaid with Gaussian fitting results (green line) toward G091.73+04.3, observed in 2020 using the low velocity resolution mode (AROWS mode 3) with velocity resolution $\sim$0.32 km s$^{-1}$. The hyperfine components of DCN 1-0 are visible.
\label{Dlow}}
\end{figure}

\begin{figure}
\centering
\includegraphics[width=0.4\textwidth]{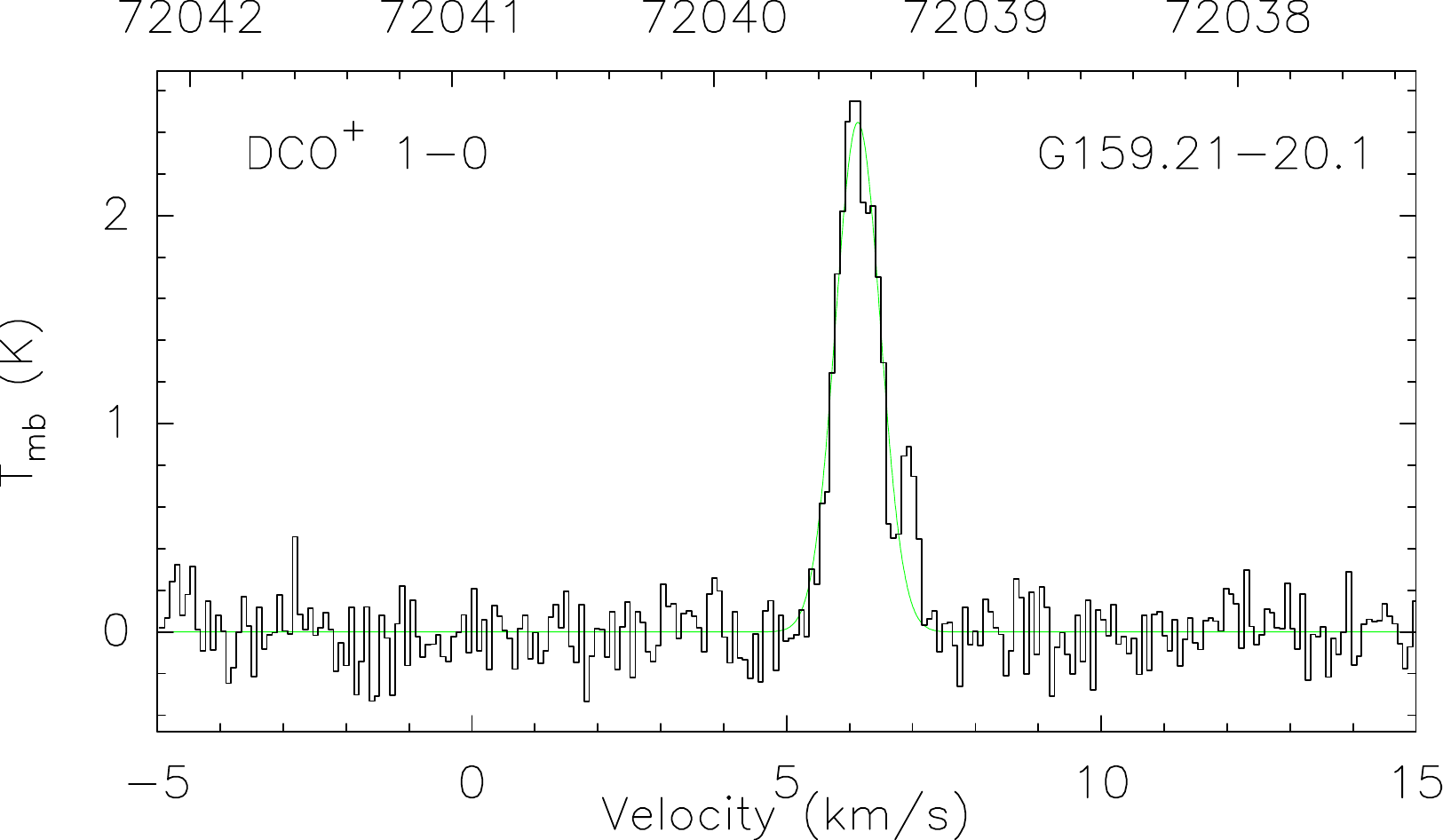}
\includegraphics[width=0.4\textwidth]{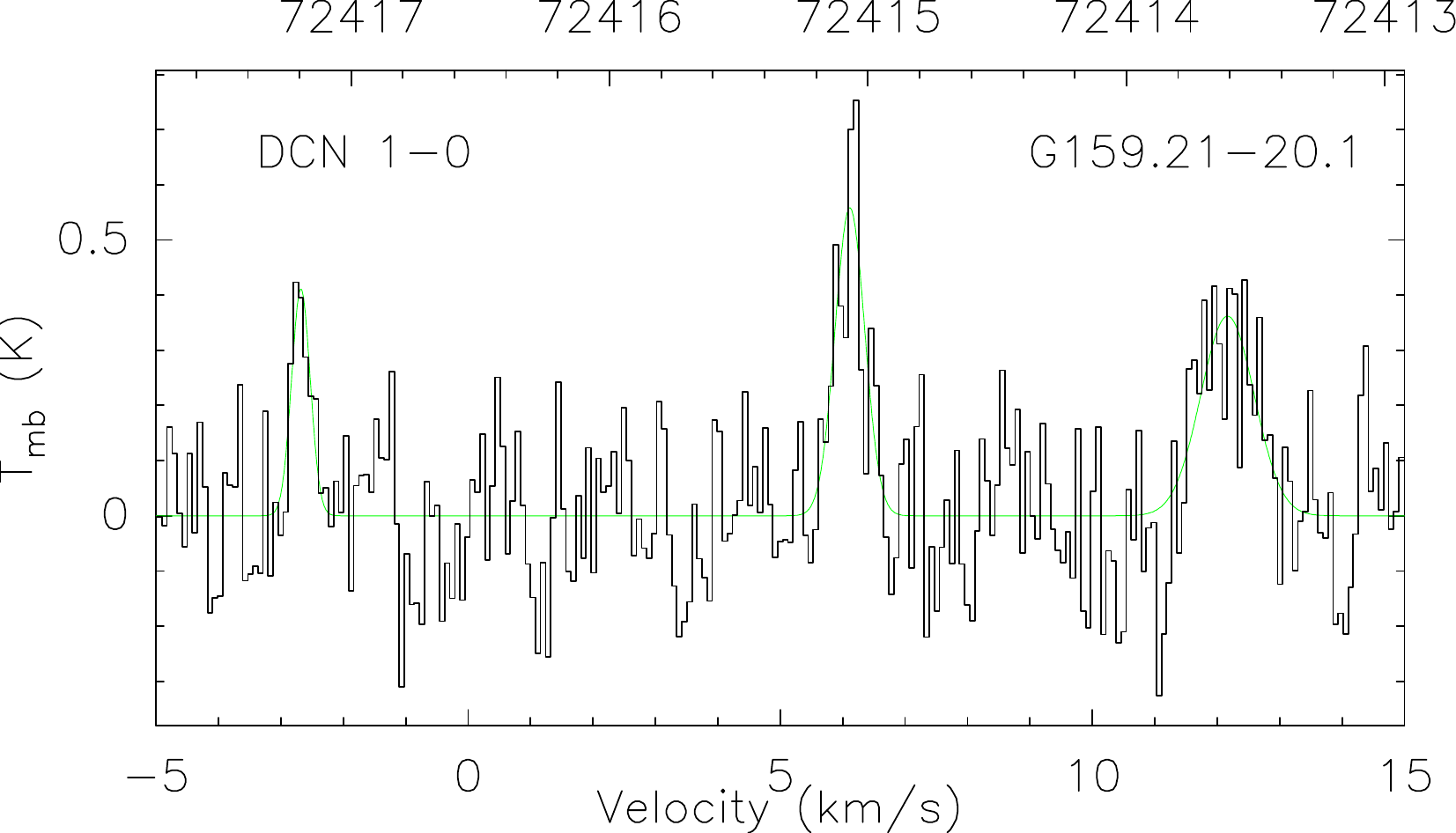}

\caption{Spectral lines of DCO$^+$ and DCN 1-0 (black line) overlaid with Gaussian fitting results (green line) toward G159.21-20.1, observed in 2023 using the high velocity resolution mode (AROWS mode 13) with velocity resolution $\sim$0.081 km s$^{-1}$. textcolor{red}{The} hyperfine components of DCN 1-0 are visible.
\label{Dhigh}}
\end{figure}

Sources considered with detections were judged by $W$ greater than 3$\sigma$, where both $W$ and $\sigma$ were obtained from the Gaussian fitting in CLASS. $W$ of the DCO$^+$ 1-0 line is denoted as $W$(DCO$^+$) hereafter,  with analogous notation applied to the DCN, H$^{13}$CO$^+$, and H$^{13}$CN 1-0 lines.
Examples of detections for the DCO$^+$ and DCN 1-0 lines toward G091.73+04.3 observed in 2020 using the low velocity resolution mode (AROWS  mode 3), are shown in Figure \ref{Dlow}. While detections toward G159.21-20.1 observed in 2023 using high resolution mode ( AROWS mode 13), are shown in Figure \ref{Dhigh}. Since each source was observed for approximately the same duration of six minutes, the noise levels of the spectra are similar, though the different rms levels are caused by different velocity resolution.  DCN and H$^{13}$CN $J$=1-0 exhibit hyperfine splitting into $F$=1-0, 2-1, and 0-0 components, with $F$=2-1 being the strongest. For the results and discussion in this paper,  $J$=1-0, $F$=2-1 lines of DCN and H$^{13}$CN are focused on, which are written as  DCN 1-0 and H$^{13}$CN 1-0 hereafter.

\subsection{Detection rates of DCO$^+$, DCN, H$^{13}$CO$^+$ and H$^{13}$CN 1-0} \label{subsec:detection}

Using the low velocity resolution mode (AROWS mode 3), DCO$^+$ and DCN 1-0 were detected in 79 and 11 out of 309 sources with detection rates of 25.6\% and 3.6\%, respectively. Using the high velocity resolution mode (AROWS mode 13), DCO$^+$, DCN, H$^{13}$CO$^+$, and H$^{13}$CN 1-0 were detected in  58, 9, 57, and 13 out of 250 sources with detection rates of 23.2\%, 3.6\%, 22.8\% and 5.2\%, respectively. The detection rates of DCO$^+$ and DCN 1-0 are similar between the two AROWS modes. Overall, the detection rates for DCO$^+$ and DCN 1-0 are  24.5\% and 3.6\%, with detections in 137 and 20 out of 559 sources, respectively. The detection rates are summarized in Table \ref{detection_rate}. The detailed detection status is presented in Table \ref{sources_with_detections}.

Among the 250 sources observed with the high velocity resolution mode (AROWS mode 13), the detection rates of DCO$^+$ and H$^{13}$CO$^+$ 1-0  are similar, and detections of these two lines occurred simultaneously in most sources. There are 46 sources where both DCO$^+$ and H$^{13}$CO$^+$ 1-0 were detected, with a median $W$(DCO$^+$) of 6.0$\sigma$ and a median $W$(H$^{13}$CO$^+$) of 7.0$\sigma$. 
While 12 sources showed detections of DCO$^+$ 1-0 only, with a median $W$(DCO$^+$) of 4.0$\sigma$. These sources include 
G168.85-15.8,
G161.85-08.6,
G163.32-08.4,
G165.69-09.1,
G178.48-06.7,
G195.00-16.9,
G177.14-01.2,
G200.34-10.9,
G185.33-02.1,
G181.84+00.3,
G199.88+00.9,
and G201.13+00.3.
There were 11 sources with only H$^{13}$CO$^+$ 1-0 detections, suggesting that deuterium enhancement may not occur, with median a $W$(H$^{13}$CO$^+$) of 6.6$\sigma$, These sources are
G159.23-34.4,
G162.64-31.6,
G168.00-15.6,
G170.13-16.0,
G170.99-15.8,
G165.16-07.5,
G195.09-16.4,
G159.34+11.2,
G191.00-04.5,
G215.00-15.1,
and G216.18-15.2.

\subsection{Derived parameters of detected DCO$^+$ and DCN 1-0}
\label{paraD}

\begin{figure}
\centering
\includegraphics[width=0.4\textwidth]{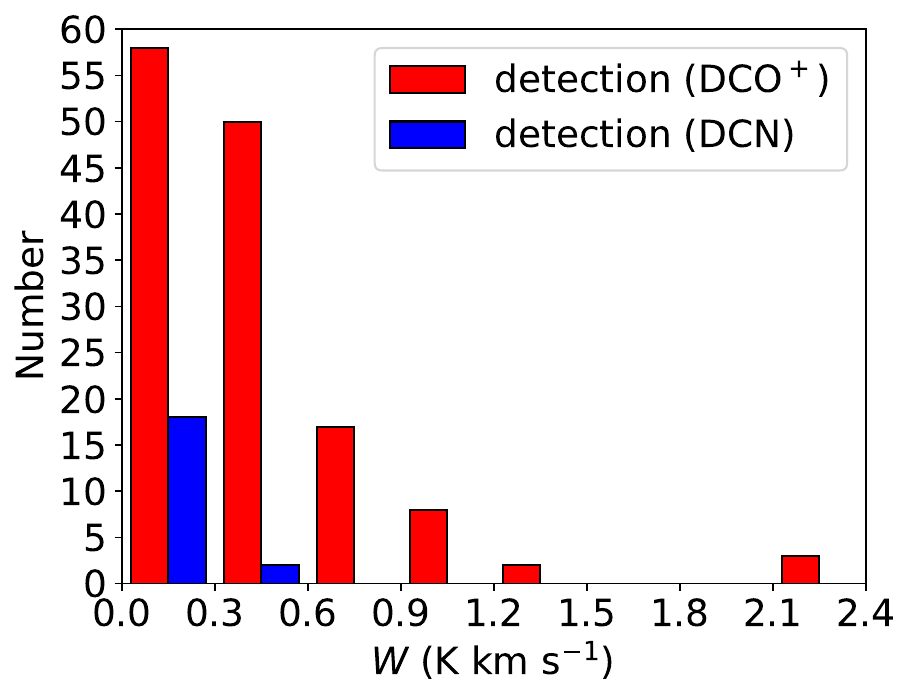}
\caption{Distribution of $W$(DCO$^+$) (red bars) and $W$(DCN) (blue bars).
\label{WD}}
\end{figure}

The parameters for sources with DCO$^+$ 1-0 detections are listed in Table \ref{dcop}, while those with DCN 1-0 detections are in Table \ref{dcn}. DCO$^+$ and DCN 1-0 were detected in 137 and 20 out of 559 sources, respectively. There were 16 sources with detections of both DCO$^+$ and DCN 1-0, 121 sources with DCO$^+$ but without DCN, 4 sources with DCN but without DCO$^+$, and 418 sources with neither detection. The distributions of $W$(DCO$^+$) and $W$(DCN) are shown in Figure \ref{WD}. $W$(DCO$^+$) ranges from 0.06 to 2.40 K km s$^{-1}$ with a median value of 0.35 K km s$^{-1}$, while the FWHM ranges from 0.2 to 2.5 km s$^{-1}$ with a median value of 0.7 km s$^{-1}$. $W$(DCN) ranges from 0.10 to 0.44 K km s$^{-1}$ with a median value of 0.15 K km s$^{-1}$, while the FWHM  ranges from 0.2 to 1.2 km s$^{-1}$ with a median value of 0.4 km s$^{-1}$.

\begin{figure}
\centering
\includegraphics[width=0.4\textwidth]{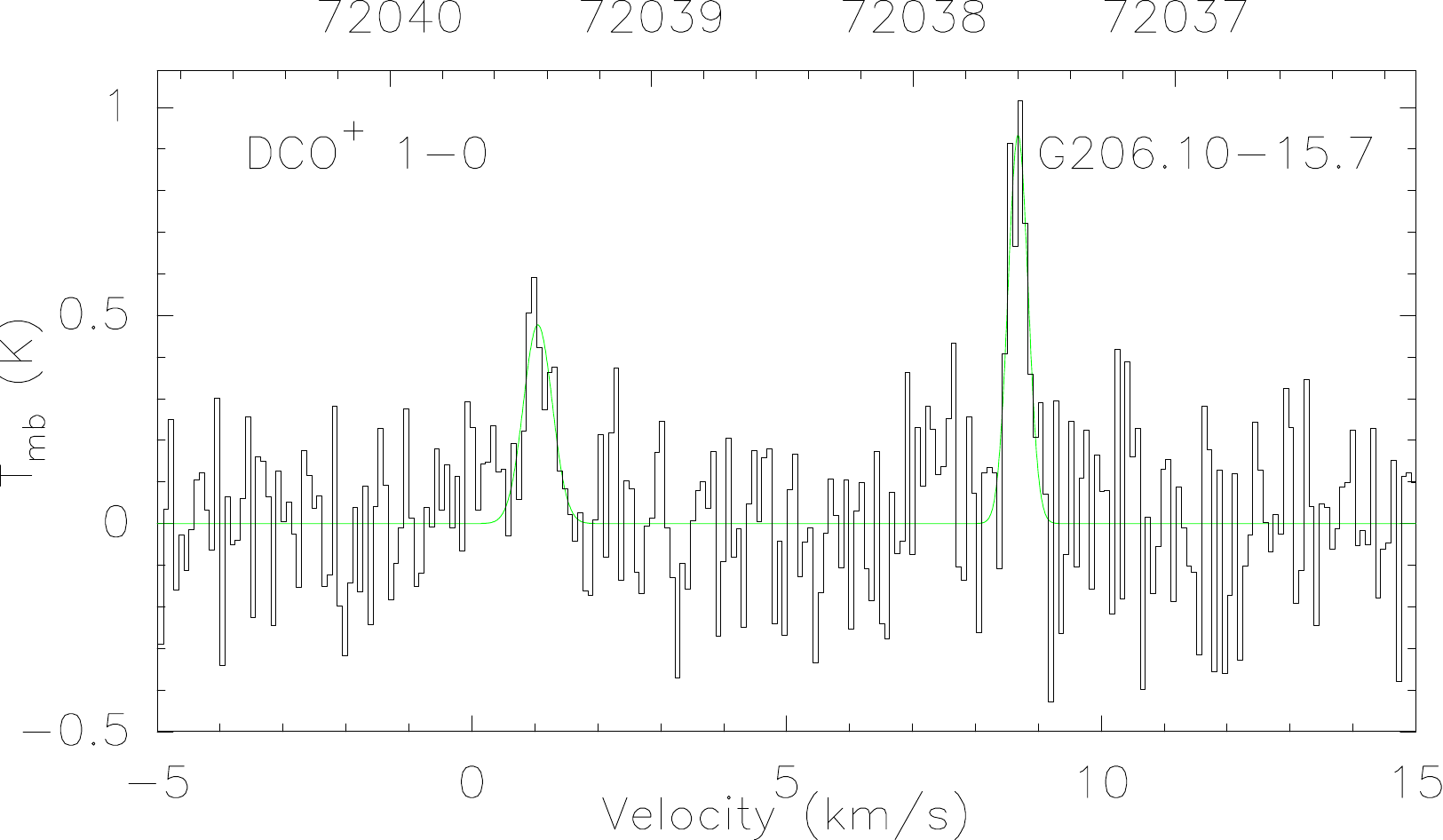}
\includegraphics[width=0.4\textwidth]{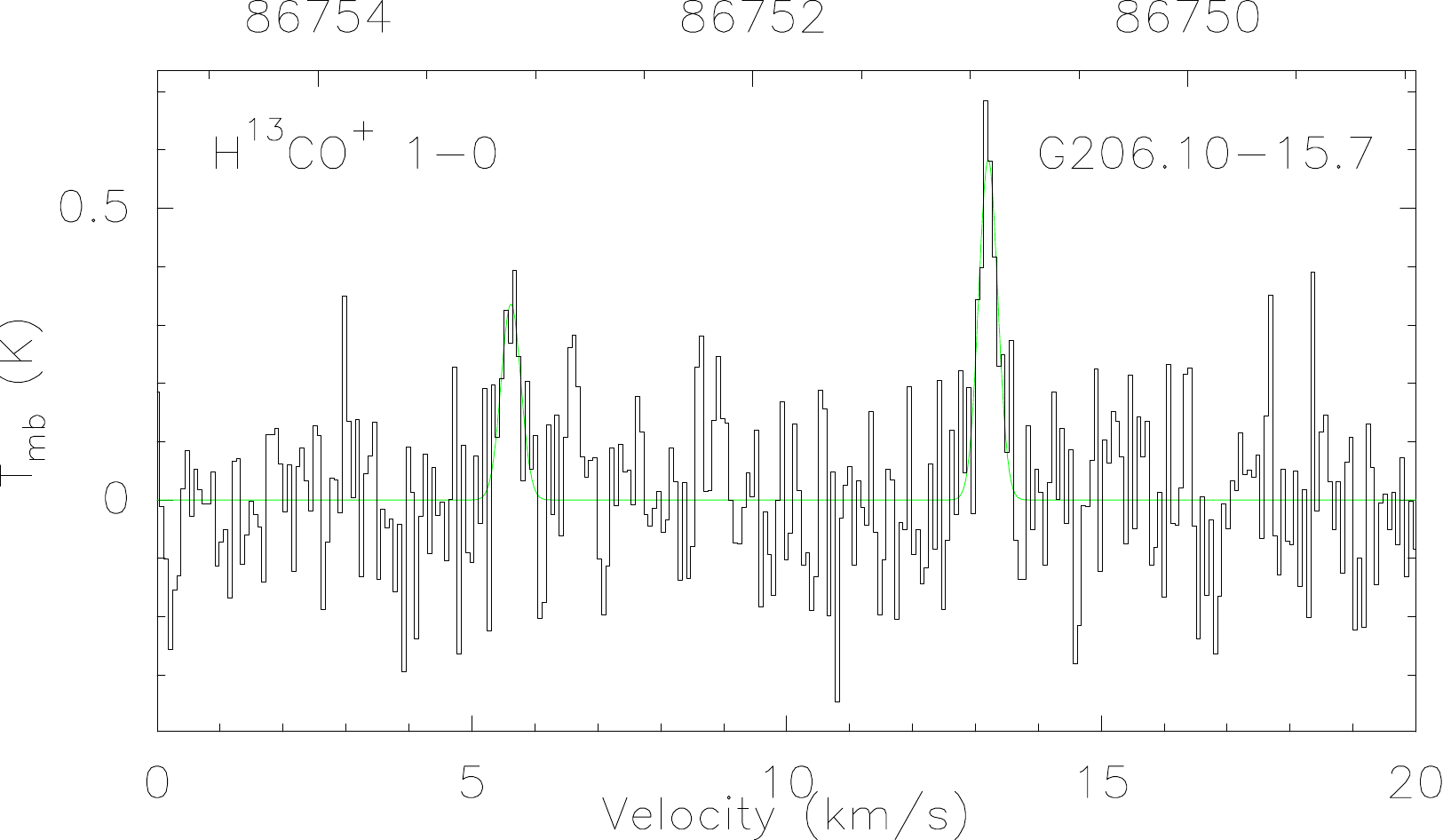}
\caption{Spectral lines of DCO$^+$ 1-0 (black line) overlaid with Gaussian fitting results (green line) toward G206.10-15.7. Double velocity components in DCO$^+$ and H$^{13}$CO$^+$ 1-0 were detected.
\label{G206}}
\end{figure}

In G209.28-19.6 and G206.10-15.7, multiple velocity components of DCO$^+$ 1-0 were found. In G209.28-19.6, $W$ was derived by directly integrating the velocity range of the line. In G206.10-15.7, two components are shown in Figure \ref{G206}, with $W$ values derived from a double-components Gaussian fitting.

\subsection{Derived parameters of detected H$^{13}$CO$^+$ and H$^{13}$CN 1-0}

A total of 71 sources, selected from 79 sources with DCO$^+$ 1-0 detections using the low velocity resolution mode (AROWS mode 3), were observed for H$^{13}$CO$^+$ and H$^{13}$CN 1-0 simultaneously in the low velocity resolution mode (AROWS mode 3). H$^{13}$CO$^+$ 1-0 was detected in 66 sources out of these 71 observed sources. Among these 66 sources with  H$^{13}$CO$^+$ 1-0, there were 20 sources with H$^{13}$CN 1-0 detections.

\begin{figure}
\centering
\includegraphics[width=0.4\textwidth]{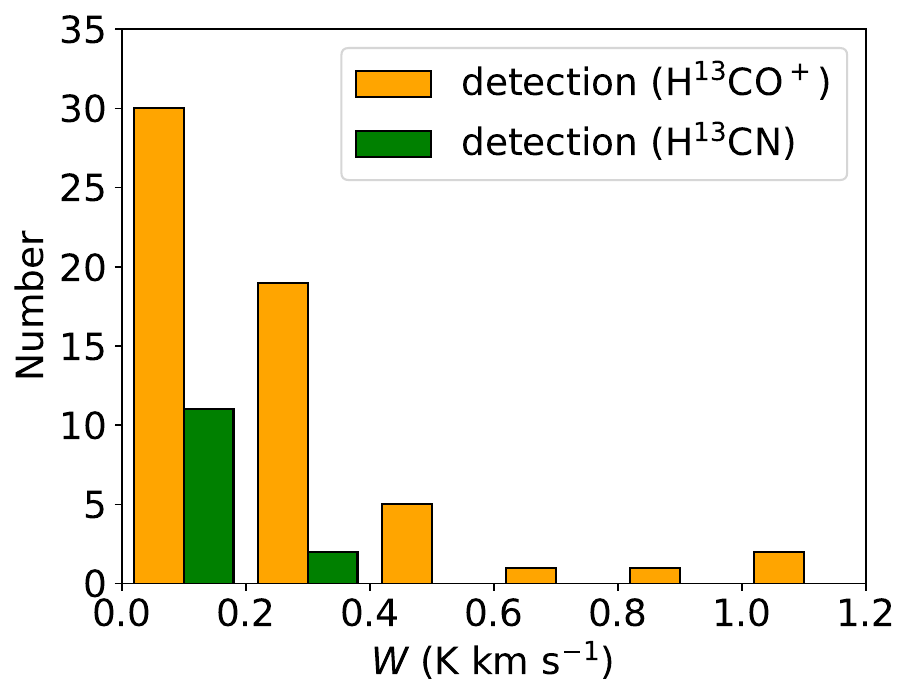}
\caption{Distribution of $W$(H$^{13}$CO$^+$) (orange bars) and $W$(H$^{13}$CN) (green bars).
\label{WH13}}
\end{figure}

H$^{13}$CO$^+$ and H$^{13}$CN 1-0 were detected using the high velocity resolution mode (AROWS mode 13) in  57 and 13 out of 250 sources respectively, with 12 sources showing detections of both H$^{13}$CO$^+$ and H$^{13}$CN 1-0. 
 The distributions of $W$(H$^{13}$CO$^+$) and $W$(H$^{13}$CN) with this mode are shown in Figure \ref{WH13}.

The parameters for H$^{13}$CO$^+$ and H$^{13}$CN 1-0 are listed in Table \ref{h13cop} and Table \ref{h13cn}, respectively. $W$(H$^{13}$CO$^+$) ranges from 0.06 to 1.11 K km s$^{-1}$ with a median value of 0.24 K km s$^{-1}$, while the FWHM ranges from 0.1 to 2.4 km s$^{-1}$ with a median value of 0.5 km s$^{-1}$. 
$W$(H$^{13}$CN) ranges from 0.05 to 0.28 km s$^{-1}$ with a median value of 0.10 K km s$^{-1}$, while the FWHM ranges from 0.2 to 1.5 km s$^{-1}$ with a median value of 0.4 km s$^{-1}$.

As mentioned in Section \ref{paraD}, G206.10-15.7 and G209.28-19.6 also exhibited multiple velocity components of H$^{13}$CO$^+$ 1-0. For G209.28-19.6, $W$(H$^{13}$CO$^+$) was derived by directly integrating the velocity range of the line.  For G206.10-15.7, parameters for H$^{13}$CO$^+$ 1-0 with two velocity components were both listed in Table \ref{h13cop}.

\subsection{Deuterated fraction of HCO$^+$ and HCN}

The $D_{\rm frac}$ values of HCO$^+$ and HCN are estimated by calculating the column density ratios of deuterated molecules to their hydrogenated counterparts, using their $^{13}$C-isotopologue counterparts. Column densities are derived from Equation (\ref{n}), assuming  local thermodynamic equilibrium (LTE) and that both deuterated molecules and their $^{13}$C-isotopologue counterparts are optically thin. 

\begin{equation}\label{n}
N_{\rm{tot}}=\frac{8 \pi k \nu^2 Q(T_{\rm{ex}})}{h c^3 g_u A _{ul} } e^{\frac{E_l + h \nu}{k T_{\rm{ex}}}} \int T_{\rm{mb}} dv \rm{(cm^{-2})}
\end{equation}
Where, $k=1.38\times 10^{-13}$ erg K$^{-1}$ is the Boltzmann constant, $\nu$ is the frequency of the transition, $h=6.624\times 10^{-27}$erg s is the Planck constant, $c=2.998 \times 10^{10}$ cm s$^{-1}$ is the speed of light, $T_{\rm{ex}}$ is the excitation temperature (assumed to be 9.375 K), $Q(T_{\rm{ex}})$ is the partition function dependent on the $T_{\rm{ex}}$, $A_{ul}$ is the Einstein emission coefficient, $g_u$ is the upper level degeneracy, $E_l$ is the energy of the ground state. $Q(T_{\rm{ex}})$, $A_{ul}$, $g_u$, and $E_l$ for the four transitions are given by the CDMS  \citep{2009A&A...507..347V}. Since all the four lines are the transitions between $J$=1 and $J$=0,  $E_l$ for each line is 0.  

According to Equation (\ref{n}), the column density ratios of deuterated molecules to their $^{13}$C-isotopologue counterparts are given by: 
 N(DCO$^+$)/N(H$^{13}$CO$^+$)$\approx$1.33$\times$$W$(DCO$^+$)/$W$(H$^{13}$CO$^+$) and
 N(DCN)/N(H$^{13}$CN)$\approx$1.30$\times$$W$(DCN)/$W$(H$^{13}$CN).  $T_{\rm{ex}}$=9.375 K is a reasonable assumption, since these sources are cold cores. Even if the gas density is below the critical densities of these lines, which might invalidate the LTE assumption, the derived column densities of DCO$^+$ and H$^{13}$CO$^+$  may have large uncertainties.
However,  the relative abundance ratio of DCO$^+$/H$^{13}$CO$^+$ estimated by there $J$=1-0 lines does not vary much, because the two molecules do have similar excitation conditions which can cancel the bias  in column density estimation. It is the same for estimating  the relative abundance ratio of DCN/H$^{13}$CN. 
 
Kinematic distances of Planck cold clumps are obtained for 741 $^{13}$CO components, with 51\% ranging from 0.5 to 1.5 kpc \citep{2012ApJ...756...76W}, indicating  that these sources are in the solar neighborhood. The corresponding physical scale of the beam size 87$^{\prime \prime}$ for source at 1 kpc  is $\sim$4 pc.  The $^{12}$C/$^{13}$C ratio  as a function of Galactocentric distance $D\rm_{GC}$, ${\rm^{12}C/^{13}C}=6.21D\rm_{GC}+18.71$, provided in  \cite{2005ApJ...634.1126M}, where  the $D\rm_{GC}$ of the sun is 8.15 kpc  \citep{2019ApJ...885..131R}. A  $^{12}$C/$^{13}$C ratio of 68, representative of the local interstellar medium, is used for these Planck cold clumps.
$D_{\rm frac}$ values are derived from  the $^{12}$C/$^{13}$C ratios and column density ratios: $D_{\rm frac}$(HCO$^+$)=1/68$\times$N(DCO$^+$)/N(H$^{13}$CO$^+$) and $D_{\rm frac}$(HCN)=1/68$\times$N(DCN)/N(H$^{13}$CN), which are listed in Table \ref{df}.
The 3$\sigma$ upper limit values are also provided, calculated as 
$3 rms \sqrt{\delta v \cdot \Delta v}$ $\rm{(K \cdot km \cdot s^{-1})}$, 
where $\delta v$ is  the channel spacing of velocity as well as $\Delta v$ is the FWHM.

\begin{figure}
\centering
\includegraphics[width=0.4\textwidth]{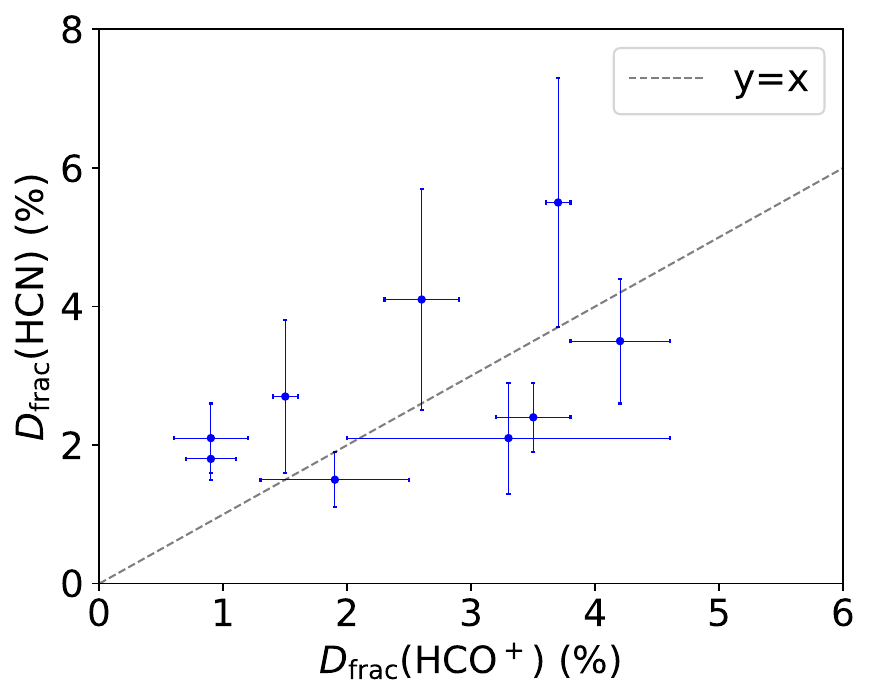}
\caption{$D_{\rm frac}$(HCO$^+$) v.s.  $D_{\rm frac}$(HCN) for 9 source with both deuterated fractions obtained.
\label{Df1vsDf2}}
\end{figure}

$D_{\rm frac}$(HCO$^+$) are derived for 113 components from 112 sources, including two velocity components in G206.10-15.7, ranging from 0.89\% to 7.4\% with a median of 3.1\%. 
$D_{\rm frac}$(HCN) are derived for 11 sources, ranging from 1.5\% to 5.5\% with a median of 2.3\%. 
Among the 9 sources with both $D_{\rm frac}$(HCO$^+$) and $D_{\rm frac}$(HCN) values, 4 show  $D_{\rm frac}$(HCO$^+$) greater than $D_{\rm frac}$(HCN), while the remaining 5 show the opposite (see Figure \ref{Df1vsDf2}).

\subsection{FWHMs of DCO$^+$ and H$^{13}$CO$^+$ 1-0}

\begin{figure}
\centering
\includegraphics[width=0.4\textwidth]{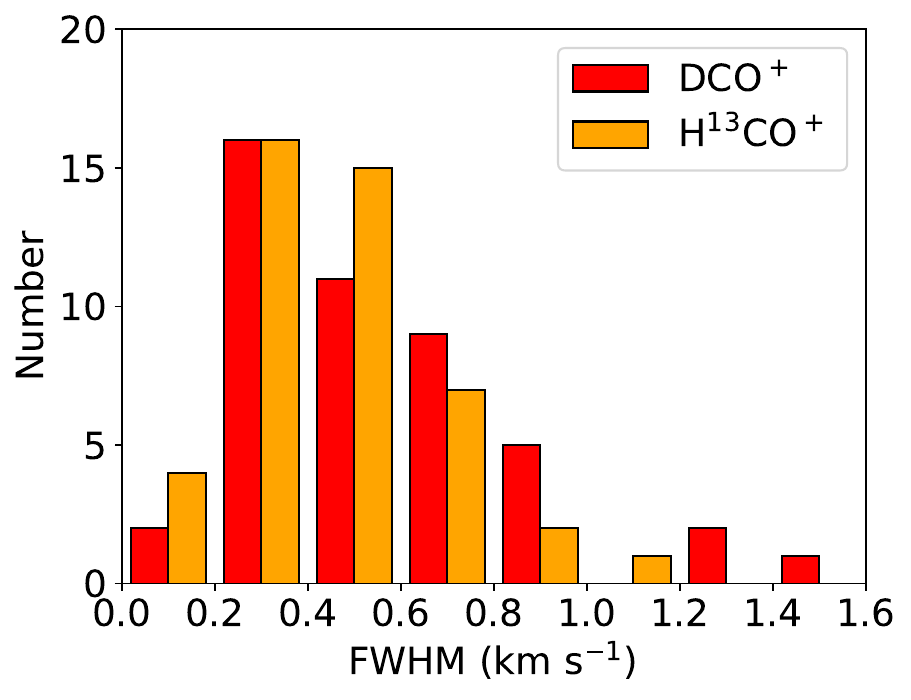}
\includegraphics[width=0.4\textwidth]{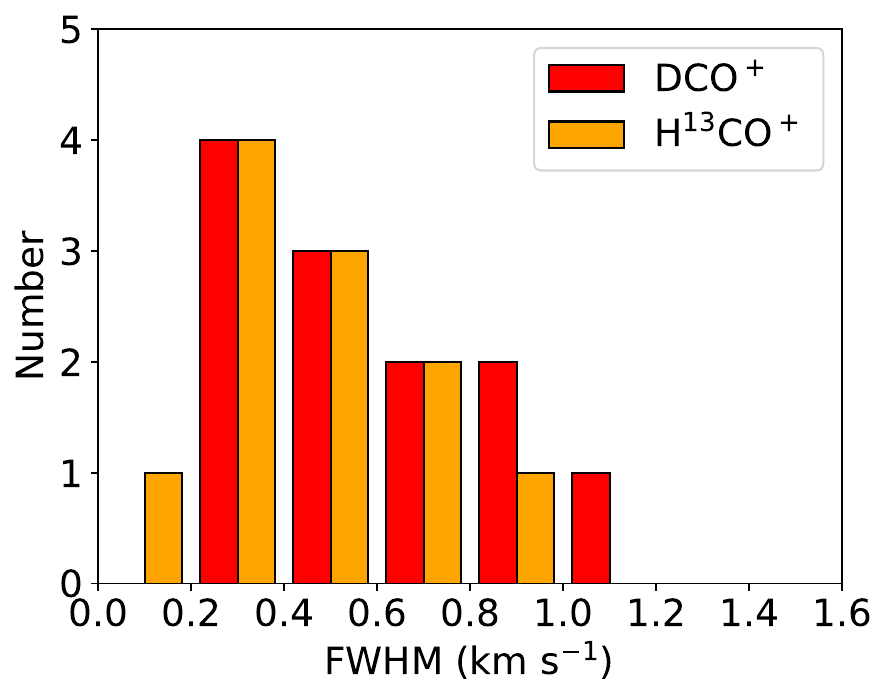}
\caption{Top:  Distributions of FWHM for DCO$^+$ (red bars) and H$^{13}$CO$^+$ (orange bars) 1-0 in 46 sources where DCO$^+$ and H$^{13}$CO$^+$ 1-0 were detected using the high velocity resolution mode (AROWS mode 13). Bottom: Distribution of FWHM of DCO$^+$ in sources with only DCO$^+$ 1-0 detections (red bars) and distribution of FWHM of H$^{13}$CO$^+$ in sources with only H$^{13}$CO$^+$ 1-0 detections (orange bars).
\label{FWHM_dh}}
\end{figure}

\begin{figure}
\centering
\includegraphics[width=0.4\textwidth]{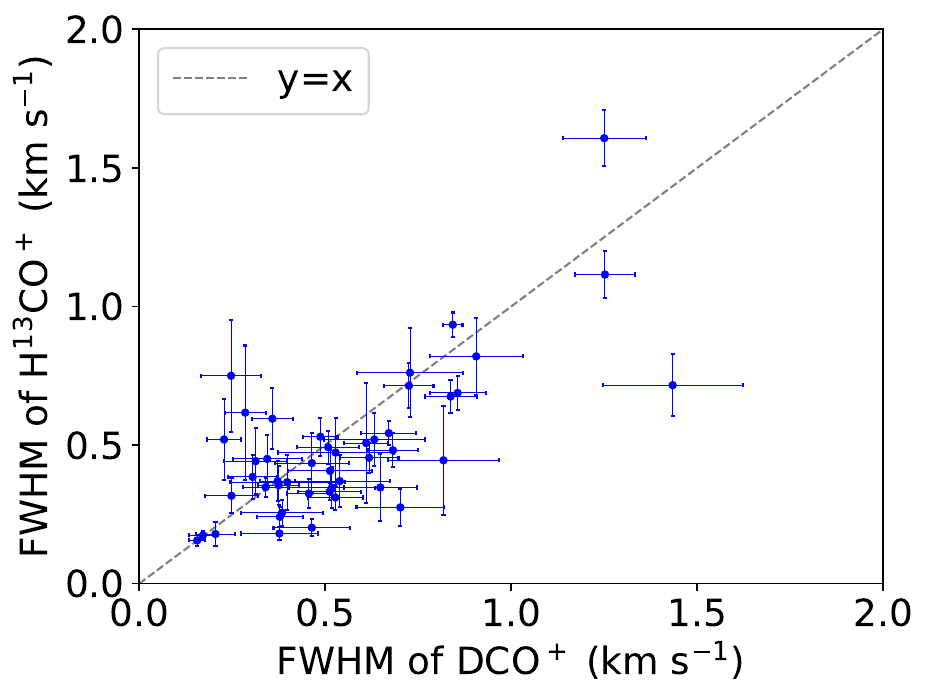}
\caption{FWHM of DCO$^+$ 1-0 v.s.  H$^{13}$CO$^+$ 1-0 of 46 sources, in which both lines were detected with the high velocity resolution mode (AROWS mode 13).
\label{FWHMofDCO+vsH13CO+}}
\end{figure}

\begin{figure}
\centering
\includegraphics[width=0.4\textwidth]{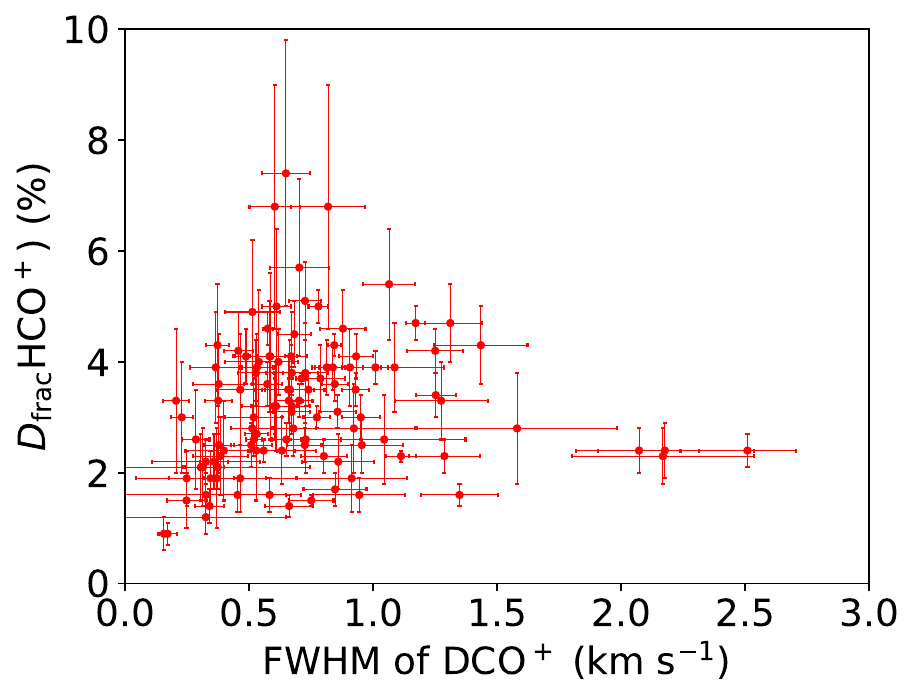}
\includegraphics[width=0.4\textwidth]{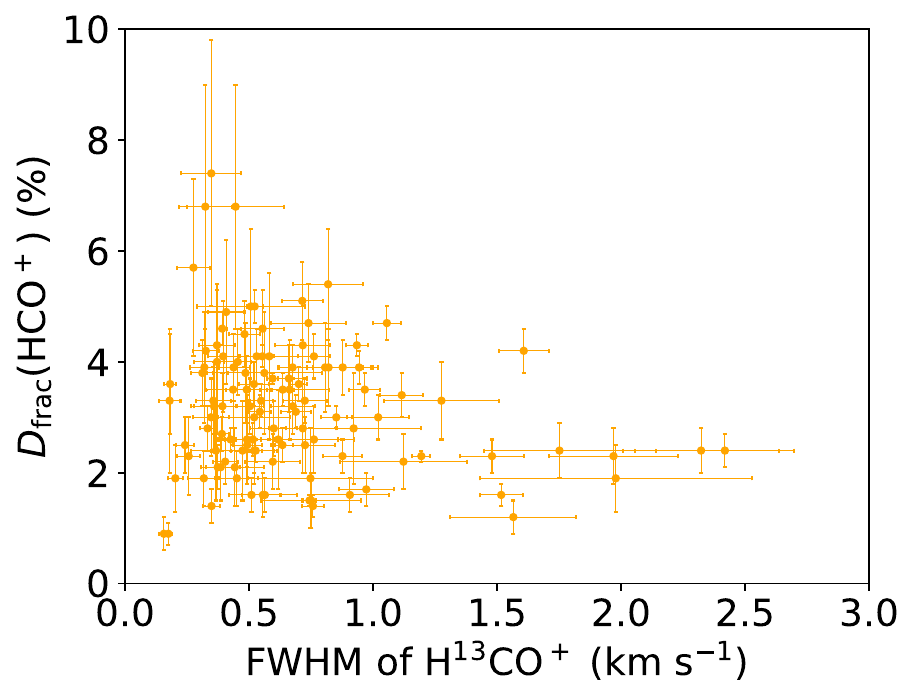}
\caption{Top: Comparison between the FWHM of DCO$^+$ and $D_{\rm frac}$(HCO$^+$). Bottom: Comparison between the FWHM of H$^{13}$CO$^+$ and $D_{\rm frac}$(HCO$^+$). Each plot contains 113 data points.
\label{FWHMvsDf1}}
\end{figure}

Among the 250 sources observed using the high velocity resolution mode (AROWS mode 13), the number distributions of the FWHM of DCO$^+$ and H$^{13}$CO$^+$ 1-0 in 46 sources with detections of both molecules are shown in the top of Figure \ref{FWHM_dh}. The FWHMs of these detections are mainly below 1 km s$^{-1}$,  with the peak at  0.3 to 0.4 km s$^{-1}$ corresponding to 11 and 12 out of 46 detections for DCO$^+$ and H$^{13}$CO$^+$ respectively. 
The number distributions of FWHM of DCO$^+$ and H$^{13}$CO$^+$ 1-0 in the 12 sources with only DCO$^+$ 1-0 detections and the 11 sources with only H$^{13}$CO$^+$ 1-0 detections are shown in the bottom of Figure \ref{FWHM_dh}. The FWHM of sources with single detections shows no significant difference compared to those with both DCO$^+$ and H$^{13}$CO$^+$ 1-0 detections. A comparison of the FWHMs between DCO$^+$ and H$^{13}$CO$^+$ is shown in Figure \ref{FWHMofDCO+vsH13CO+}, which demonstrates that the line widths are roughly similar within the error bars, with DCO$^+$ 1-0 having a slightly larger line width than H$^{13}$CO$^+$ 1-0. Figure \ref{FWHMvsDf1} displays the comparisons between FWHM of DCO$^+$ and $D_{\rm frac}$(HCO$^+$), as well as between FWHM of H$^{13}$CO$^+$ and $D_{\rm frac}$(HCO$^+$), without clear trend. The FWHM of DCO$^+$ 1-0 seem to be larger than that of H$^{13}$CO$^+$ 1-0 in sources with $D_{\rm frac}$ greater than 5\%. However, it is hard to make a solid conclusion for such differences, since signal to noise ratios of H$^{13}$CO$^+$ 1-0 are not good enough in these sources.

\section{Discussion} \label{sec:discussion}

\subsection{The properties of these Planck cold clumps with and without DCO$^+$ 1-0 detection}

\begin{figure}
\centering
\includegraphics[width=0.4\textwidth]{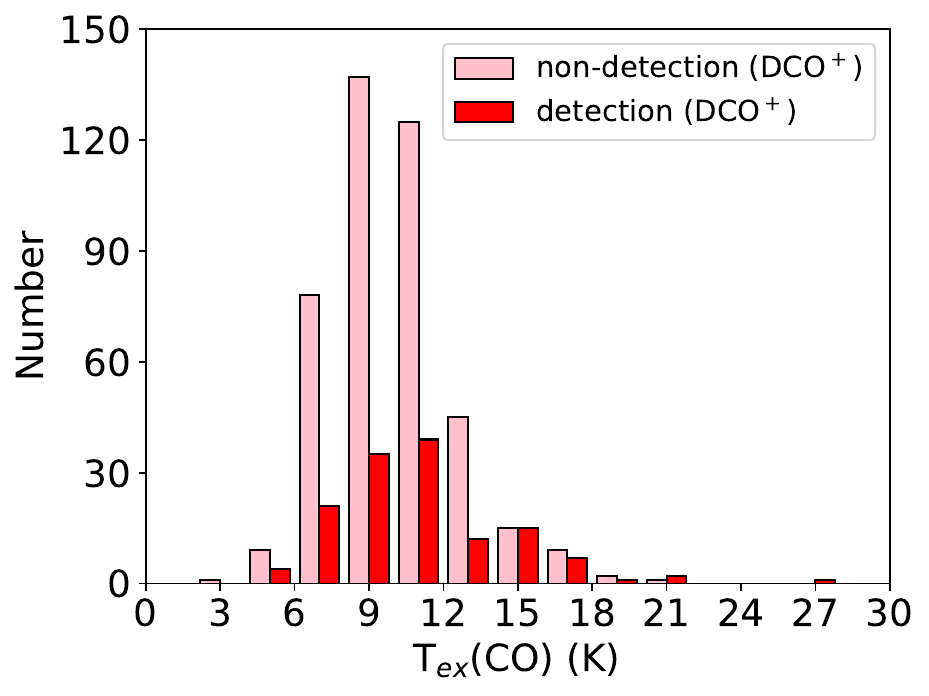}
\caption{Distribution of $T_{\rm ex}$(CO) of sources where DCO$^+$ 1-0 is detected (red bars) and not (pink bars).
\label{hist_Tex}}
\end{figure}

\begin{figure}
\centering
\includegraphics[width=0.4\textwidth]{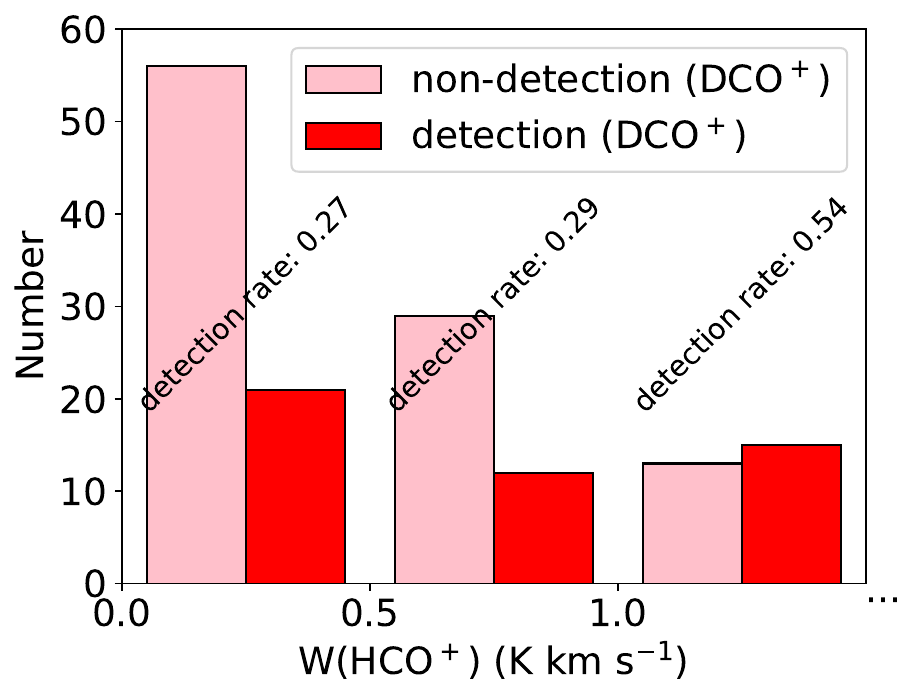}
\caption{Distribution of $W$(HCO$^+$) of sources where DCO$^+$ 1-0 is detected (red bars) and not (pink bars)
\label{hist_WHCOp}}
\end{figure}

In order to determine whether the properties of sources with DCO$^+$ 1-0 detections differ from those without detections, the distributions of $T_{\rm ex}$(CO) and $W$(HCO$^+$) for sources with and without DCO$^+$ 1-0 detections are provided, where $W$(HCO$^+$) data are from \cite{2016yCat..18200037Y}. The distributions of $T_{\rm ex}$(CO) for sources with and without DCO$^+$ 1-0 detection are similar, as shown in Figure \ref{hist_Tex}. The $T_{\rm ex}$(CO)s of these Planck cold clumps are below 30 K, mainly ranging from 7 to 12 K. The distributions of $W$(HCO$^+$) for sources with and without DCO$^+$ 1-0 detection are also similar (see Figure \ref{hist_WHCOp}). The detection rates of DCO$^+$ 1-0 is slightly higher in sources with higher $W$(HCO$^+$) compared to those with lower $W$(HCO$^+$). The similarity of distributions suggests that detection/non-detection status of DCO$^+$ 1-0 is not limited by $T_{\rm ex}$(CO) and $W$(HCO$^+$).

\subsection{Possible reason of different detection rates and similar $D_{\rm frac}$s of HCO$^+$ and HCN in Planck cold clumps}

The detection rates for DCO$^+$ and H$^{13}$CO$^+$ 1-0 are 24.5\% and 22.8\%, respectively, whereas those for DCN and H$^{13}$CN 1-0 are 3.6\% and 5.2\%. This indicates that both the deuterated and $^{13}$C-isotopologue counterparts of HCO$^+$ are detected more frequently than their HCN counterparts. The higher detection rates of DCO$^+$ compared to DCN 1-0, are consistent with the studies of 70 $\mu$m dark high-mass clumps  \citep{2022ApJ...939..102L} and massive starless clump candidates  \citep{2024ApJS..270...35Y}.

$D_{\rm frac}$(HCO$^+$) values  in 112 sources range from 0.89\% to 7.4\% with a median value of 3.1\%, while  $D_{\rm frac}$(HCN) values in 11 sources range from 1.5\% to 5.5\% with a median value of 2.3\%. These values do not show a significant difference between the two molecular pairs. The $D_{\rm frac}$ values are consistent with those observed in molecular clouds (e.g.  \citealt{1976ApJ...209L..83H}), low-mass starless cores (e.g.  \citealt{2006A&A...455..577T}), and massive starless clump candidates (e.g.  \citealt{2024ApJS..270...35Y}). Among the 9 sources where both $D_{\rm frac}$(HCO$^+$) and $D_{\rm frac}$(HCN) were estimated, 4 sources show $D_{\rm frac}$(HCO$^+$) greater than $D_{\rm frac}$(HCN), while the remaining 5 sources show the opposite (see Figure \ref{Df1vsDf2}). Therefore, the significantly lower detection rate of DCN 1-0 compared to DCO$^+$ 1-0, despite similar noise levels, cannot be attributed to differences in $D_{\rm frac}$ values.

Deuterium fractionation is generally regarded as starting from the molecules-ions reaction primarily at temperature below 30 K. At these low temperatures, the reaction, $\rm{HD} + H_3^+ \rightleftharpoons  H_2+H_2D^+ + \Delta E $ (where $\Delta$E = 232 K),  proceeds forward, enhancing the abundance of H$_2$D$^+$, which is a precursor to many deuterated molecules \citep{1989ApJ...340..906M}. 
DCO$^+$ is thought to form in the gas phase and to be abundant below $\sim$30 K, 
while DCN is thought to form through multiple pathways in the interstellar medium, with its abundance peaking at higher temperatures compared to DCO$^+$  \citep{1989ApJ...340..906M}.
Consequently, deuterated molecules are often used as chemical clocks to trace evolutionary stages during star formation \citep{2005A&A...433..535F,2010A&A...517L...6B,2014MNRAS.440..448F,2022ApJ...925..144S}. For example, in  \cite{2020ApJS..249...33K}, the detection rates of deuterated molecules suggest chemical evolutionary stages, while the $D_{\rm frac}$s should be used for such study.

 Given the similar $D_{\rm frac}$ values for HCO$^+$ and HCN in Planck cold clumps, we propose that the disparity in detection rates between DCO$^+$ and DCN 1-0 is primarily due to differences in their critical densities. This explanation was also suggested for massive starless clump candidates  \citep{2024ApJS..270...35Y}. Ignoring background emission, the optically thin critical density can be approximated by $n^{thin}_{crit}$ $\sim$$A_{jk}$/$\gamma_{jk}$, where $A_{jk}$ (s$^{-1}$) is the Einstein A and $\gamma_{jk}$ (cm$^{3}$ s$^{-1}$) is collision rate out of upper level j to level k  \citep{2015PASP..127..299S}. At 10 K, the critical densities for DCO$^+$ and DCN 1-0 are 3.2$\times$10$^{4}$ and 2.6$\times$10$^{5}$ cm$^{-3}$  \citep{2020ApJ...901..145F} , while those for H$^{13}$CO$^+$ and H$^{13}$CN 1-0 are 6.2$\times$10$^{4}$ and 5.3$\times$10$^{5}$ cm$^{-3}$  \citep{2015PASP..127..299S}. Although the $A_{jk}$ values for DCO$^+$ and DCN 1-0 are similar,  $\gamma_{jk}$ related to the collision cross section for collisions with H$_{2}$ for DCO$^+$ (a molecular ion) is larger than DCN (a neutral molecule), resulting in a critical density for DCO$^+$ that is an order of magnitude lower than DCN \citep{2015PASP..127..299S}.  It is the same for H$^{13}$CO$^+$ and H$^{13}$CN as that for DCO$^+$ and DCN.


\subsection{Origin of deuterium fractionation}

\begin{figure}
\centering
\includegraphics[width=0.4\textwidth]{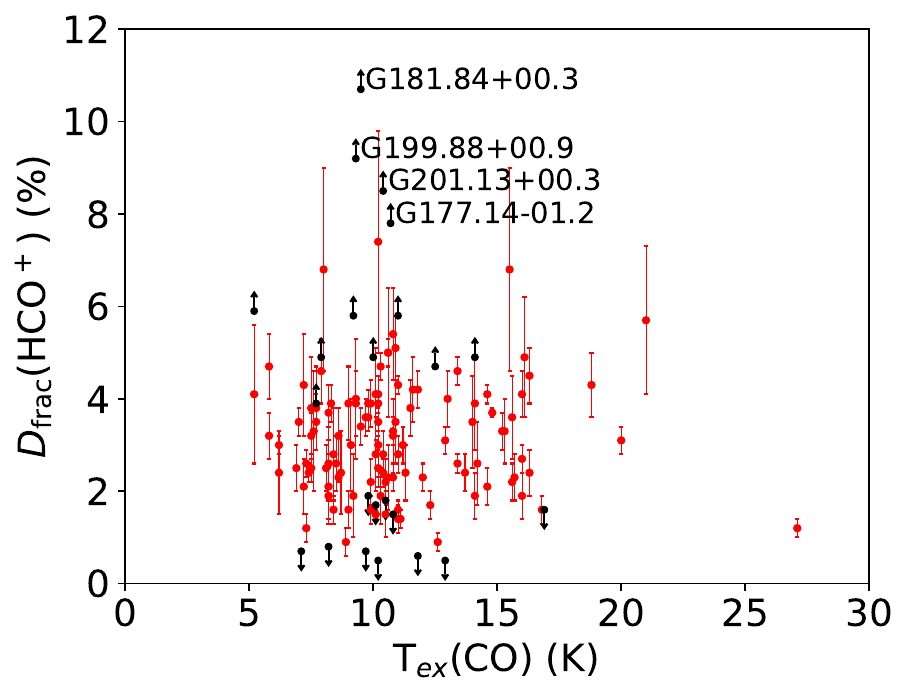}
\caption{The comparison of $T_{\rm ex}$ and $D_{\rm frac}$(HCO$^+$) (red points), including 3$\sigma$ upper and lower limit value of $D_{\rm frac}$(HCO$^+$) (firebrick points and arrows) for sources with only H$^{13}$CO$^+$ and only DCO$^+$ detections, respectively.
\label{Tex_vs_Dfrac}}
\end{figure}

The comparison between $T_{\rm{ex}}$(CO) and $D_{\rm frac}$(HCO$^+$) are found to be without any trend including upper and lower limits of $D_{\rm frac}$(HCO$^+$) (see Figure \ref{Tex_vs_Dfrac}). Most upper and lower limits of $D_{\rm frac}$(HCO$^+$) fall within expected ranges, suggesting that most sources with detections of only DCO$^+$ or H$^{13}$CO$^+$ do not exhibit significantly different properties compared to sources with both detections. 
The small range of $T_{\rm ex}$(CO)s for these sources makes it challenging to discuss a relationship between $T_{\rm ex}$(CO) and deuterium fractionation. Observations of deuterium fractionation in late stages of star-forming regions are needed for further study.
Chemical models suggest that deuterated molecules can be destroyed in hot molecular gas \citep{1989ApJ...340..906M, 1998A&A...332..695H}. Observations suggest that, in hot cores, $D_{\rm frac}$(HCN) decreases generally from ``off-core'' to ``on-core'' positions, where temperatures vary significantly \citep{1998A&A...332..695H}. 

Among the observed sources, 12 sources exhibit lower limits of $D_{\rm frac}$(HCO$^+$), with only DCO$^+$ 1-0 detections, possibly due to the lower critical density of 3.2$\times$10$^{4}$ cm$^{-3}$ for DCO$^+$ 1-0 compared to 6.2$\times$10$^{4}$ cm$^{-3}$ for H$^{13}$CO$^+$ 1-0 at 10 K \citep{2015PASP..127..299S,2020ApJ...901..145F}. Among these 12 sources, four sources exhibit particularly high $D_{\rm frac}$ values, namely G177.14-01.2, G181.84+00.3, G199.88+00.9 and G201.13+00.3  (see Figure \ref{Tex_vs_Dfrac}), which deserve further study. 

Conversely, there are 11 sources with upper limits of $D_{\rm frac}$(HCO$^+$), with only H$^{13}$CO$^+$ 1-0 detections, corresponding to low $D_{\rm frac}$ values. No significant differences in the number distributions of $T_{\rm ex}$(CO) are found between sources with only DCO$^+$ 1-0 detections, only H$^{13}$CO$^+$ 1-0 detections, and those with both detections (see Figure \ref{Tex_vs_Dfrac}). The FWHM number distributions for sources with only DCO$^+$ 1-0 detections, only H$^{13}$CO$^+$ 1-0 detections, and both detections show no significant differences, with FWHMs mainly within 1 km s$^{-1}$ (see Figure \ref{FWHM_dh}). It has been suggested that $D_{\rm frac}$ may decrease with increasing microturbulent velocity fields  \citep{2015A&A...579A..80G, 2022MNRAS.512.4934L}. However, our results do not reveal any trend between $D_{\rm frac}$ and line widths for either DCO$^+$ or H$^{13}$CO$^+$ 1-0 (see Figure \ref{FWHMvsDf1}). The limited range of FWHMs for these emission lines in Planck cold clumps makes it difficult to discuss the relationship between $D_{\rm frac}$ and FWHM.

Deuterium chemistry suggests that deuterium fractionation starts during the pre-collapse phase of molecular cloud evolution, when the gas becomes cold (T $\sim$10 K) and dense (n $\ge$ 10$^4$ cm$^{-3}$) \citep{2000A&A...361..388R,2001A&A...372..998C,2007A&A...471..849R}. This mechanism has been used to explain the high $D_{\rm frac}$ observed in cold molecular cores \citep{2001A&A...372..998C,2003A&A...403L..37C}. According to this framework, in the early stages of cold core formation ($T$ $\sim$15-20 K, $n$ $\sim$10$^4$ cm$^{-3}$), $D_{\rm frac}$ would remain too low ($\sim$10$^{-4}$-10$^{-3}$) to produce detectable DCO$^+$ 1-0, while H$^{13}$CO$^+$ 1-0 could be detected.
However, among our 250 observed sources using the high velocity resolution mode (AROWS mode 13), DCO$^+$ and H$^{13}$CO$^+$ 1-0 are detected in 58 and 57, with similar detection rates of 23.2\% and 22.8\%, respectively. More importantly, there are 46 sources with both DCO$^+$ and H$^{13}$CO$^+$ 1-0 detections. DCO$^+$ and H$^{13}$CO$^+$ 1-0 emissions are detected almost simultaneously in the same sources.
Such results indicate that DCO$^+$  can already be widespread in cold molecular clouds, rather than just beginning to become abundant. 
Thus, the phenomenon of high  $D_{\rm frac}$ observed in cold cores may not be necessarily explained by cold core deuterium chemistry.

Moreover, the $D_{\rm frac}$ values for HCO$^+$ and HCN, reflecting the enhancement of deuterium for these two species, are similar, range from 0.89\% to 7.4\% and 1.5\% to 5.5\%, respectively. Detection rates of DCN and H$^{13}$CN 1-0, as 3.6\% and 5.2\% respectively,  are significantly lower than those of DCO$^+$ and H$^{13}$CO$^+$ 1-0, as 24.5\% and 22.8\% respectively. It is possibly due to the higher critical densities of DCN and H$^{13}$CN 1-0 compared to those of DCO$^+$ and H$^{13}$CO$^+$ 1-0, respectively. Thus, perhaps because the volume density is not high enough to excite DCO$^+$ and DCN to $J$=1 energy level, DCO$^+$ and DCN lines cannot be detected in diffuse gas.
The method for testing the origin of DCO$^+$ and DCN in diffuse molecular clouds is to observe the DCO$^+$ and DCN absorption lines toward diffuse molecular clouds with background quasar emitting millimeter-wave continuum sources.
The differences in $D_{\rm frac}$ during evolution of molecular clouds and late stages of star formation caused by the synthesis and depletion of deuterated molecules in different physical conditions. Thus, comparison with large-sample surveys of deuterated molecular lines toward hot molecular cores at late stages of star formation, as well as updated astro-chemical models for these deuterated molecules, are needed to fully understand enhancement of deuterium in molecular clouds. 


\section{Summary} \label{sec:summary}

Aiming to understand the enhancement of deuterated molecules from diffuse molecular gas to cold cores, a single-pointing survey toward 559 Planck cold clumps of ECC for DCO$^+$ and DCN 1-0 have been conducted using the ARO 12-m telescope. It included observations of 309 cores for DCO$^+$ and DCN 1-0 simultaneously, followed by 71 of these cores where DCO$^+$ 1-0 was detected for H$^{13}$CO$^+$ and H$^{13}$CN 1-0 simultaneously, aiming to determine $D_{\rm frac}$. Additionally, 250 cores were observed for  DCO$^+$, DCN, H$^{13}$CO$^+$, and H$^{13}$CN 1-0 simultaneously. The observed parameters, including  $W$, $V_{\rm LSR}$ and FWHM,  were presented. The detection rates and $D_{\rm frac}$ for HCO$^+$ and HCN are also presented. Potential reasons of the different detection rates and the similar $D_{\rm frac}$ for HCO$^+$ and HCN in Planck cold clumps are discussed. The main results of this survey are as follows:
\begin{enumerate}
\item DCO$^+$ and DCN 1-0 are detected in 79 and 11 out of 309 sources with the low velocity resolution mode (AROWS mode 3), which gives the  detection rates of 25.6\% and 3.6\%, respectively. DCO$^+$, DCN, H$^{13}$CO$^+$ and H$^{13}$CN 1-0 detected in 58, 9, 57 and 13 out of 250 sources with  the high velocity resolution mode (AROWS mode 13), which gives the detection rate of 23.2\%, 3.6\%, 22.8\% and 5.2\%, respectively. Overall, DCO$^+$ and DCN 1-0 detected in 137 and 20 out of 559 sources, with detection rates of 24.5\% and 3.6\%, respectively, with about 6 minutes telescope time for each source. 
\item $D_{\rm frac}$(HCO$^+$) values in 112 sources range from 0.89\% to 7.4\% with a median value of 3.1\%, and $D_{\rm frac}$(HCN) values  in 11 sources range from 1.5\% to 5.5\% with median value of 2.3\%. 
\item The line widths (FWHMs) of DCO$^+$ and H$^{13}$CO$^+$ 1-0 detections are mainly distributed within 1 km s$^{-1}$,  with most measurements falling between 0.3 and 0.4 km s$^{-1}$. 
\item Similar $D_{\rm frac}$ values for DCO$^+$ and DCN suggest that the higher detection rate of DCO$^+$ compared to DCN is due to the lower critical density of DCO$^+$. 
\item The nearly simultaneous detections of DCO$^+$ and H$^{13}$CO$^+$ 1-0 in the same sources suggest that deuterium fractionation may start during diffuse molecular gas phase prior to cold core formation.
\end{enumerate}


\section*{Acknowledgements}
We would like to acknowledge the help of the staff of the ARO 12m  for assistance with the observations. This work is supported by National Key R$\&$D Program of China under grant 2023YFA1608204 and the National Natural Science Foundation of China grant 12173067.

\clearpage
\onecolumn
\captionsetup[table]{justification=centering}
\begin{longtable}{ccccc}
\caption{Detection rates of different  molecular lines and observing modes.}
\label{detection_rate}\\
	\hline 
	\hline              
	Mode & Number of sources observed & Line & Number of sources with detections & Detection rate\\
\hline
\endfirsthead
\caption[]{continue.}\\
	\hline 
	\hline
	Mode & Number of sources observed & Line & Number of sources with detections & Detection rate\\
	         \hline
\endhead
    
\hline
\endfoot
low resolution mode (mode 3) & 309 & DCO$^+$ 1-0 & 79 & 25.6$\%$\\
&& DCN 1-0 & 11 & 3.6$\%$\\
\hline
high resolution mode (mode 13) & 250 & DCO$^+$ 1-0 &58& 23.2$\%$\\
&&DCN 1-0&9&3.6$\%$\\
&&H$^{13}$CO$^+$ 1-0 & 57 & 22.8$\%$\\
&&H$^{13}$CN 1-0 &13&5.2$\%$\\
\hline
total&559&DCO$^+$ 1-0&137&24.5$\%$\\
&&DCN 1-0&20&3.6$\%$\\
\end{longtable}	
\begin{center}

\begin{longtable}{crrccccrrrr}

	\caption{Sources with detection of DCO$^+$, DCN, H$^{13}$CO$^+$ or H$^{13}$CN.}
	\label{sources_with_detections}\\
	\hline 
	\hline
	Source & R.A. & Dec & DCO$^+$ & DCN & H$^{13}$CO$^+$ & H$^{13}$CN & $D$ & $T_{\rm ex}$ \\
	Name   & (hh:mm:ss)  & (dd:mm:ss) & & & & & (kpc)   &  (K) \\
\hline

\endfirsthead
\caption[]{continue.}\\
	         
	\hline 
	\hline
	Source & R.A. & Dec & DCO$^+$ & DCN & H$^{13}$CO$^+$ & H$^{13}$CN & $D$ & $T_{\rm ex}$ \\
	Name   & (hh:mm:ss)  & (dd:mm:ss) & & & & & (kpc)   &  (K)  \\
	         \hline
  
    \endhead
\hline
\endfoot

        G001.38+20.9  &      16:34:38.06   &      -15:46:40.71  &Y  &Y  &Y  &Y  &     1.1   &     14.8  \\    
        G001.84+16.5  &      16:50:12.91   &      -18:04:22.37  &Y  &Y  &Y  &/  &     4.16  &     14.6  \\    
        G003.73+16.3  &      16:55:21.78   &      -16:43:35.31  &Y  &/  &Y  &/  &     2.75  &     14.6  \\    
        G003.73+18.3  &      16:48:54.69   &      -15:36:02.09  &Y  &/  &Y  &/  &     2.14  &     16.8  \\    
        G004.46+16.6  &      16:56:11.73   &      -16:00:51.08  &Y  &/  &Y  &Y  &     2.23  &     15.6  \\    
        G006.04+36.7  &      15:54:10.81   &      -02:50:56.32  &Y  &/  &Y  &Y  &     1.19  &     10.6  \\    
        G006.32+20.4  &      16:47:40.85   &      -12:22:03.36  &Y  &/  &Y  &/  &     1.38  &     13.4  \\    
        G006.41+20.5  &      16:47:28.67   &      -12:13:52.53  &Y  &/  &Y  &/  &     1.42  &     15.2  \\    
        G006.70+20.6  &      16:47:46.71   &      -11:57:20.05  &Y  &/  &Y  &Y  &     1.17  &     12.3  \\    
        G006.94+05.8  &      17:39:41.49   &      -19:58:05.07  &Y  &/  &Y  &/  &     2.37  &     10.1  \\    
        G006.98+20.7  &      16:48:12.55   &      -11:42:10.17  &Y  &/  &/  &/  &     0.95  &     15.1  \\    
        G007.14+05.9  &      17:39:47.47   &      -19:45:05.14  &Y  &/  &Y  &/  &     2.36  &     9.0   \\    
        G008.52+21.8  &      16:47:48.42   &      -09:53:09.82  &Y  &/  &Y  &/  &     0.94  &     13.4  \\    
        G008.67+22.1  &      16:47:07.71   &      -09:35:50.08  &Y  &/  &Y  &Y  &     0.84  &     14.1  \\    
        G021.20+04.9  &      18:12:01.72   &      -08:05:27.73  &Y  &/  &Y  &Y  &     0.28  &     5.8   \\    
        G021.66+03.7  &      18:17:06.23   &      -08:14:35.15  &Y  &Y  &Y  &/  &     0.62  &     9.8   \\    
        G025.48+06.1  &      18:15:46.82   &      -03:45:19.28  &Y  &/  &Y  &Y  &     0.63  &     7.4   \\    
        G026.85+06.7  &      18:16:10.50   &      -02:16:39.94  &Y  &Y  &Y  &Y  &     0.55  &     7.3   \\    
        G027.66+05.7  &      18:21:28.05   &      -02:03:24.93  &Y  &/  &Y  &/  &     0.59  &     7.5   \\    
        G028.45-06.3  &      19:06:09.21   &      -06:52:51.78  &Y  &Y  &Y  &Y  &     0.93  &     10.1  \\    
        G028.71+03.8  &      18:29:55.29   &      -01:58:10.45  &Y  &Y  &Y  &Y  &     0.55  &     7.0   \\    
        G028.87+04.2  &      18:28:56.46   &      -01:40:12.11  &Y  &/  &Y  &/  &     0.52  &     7.3   \\    
        G030.43+02.3  &      18:38:30.46   &      -01:09:02.56  &Y  &/  &Y  &Y  &     0.64  &     10.8  \\    
        G030.78+05.2  &      18:28:54.49   &      +00:28:40.02  &Y  &/  &Y  &Y  &     0.57  &     8.6   \\    
        G031.44+04.1  &      18:33:50.27   &      +00:35:00.48  &Y  &/  &Y  &Y  &     0.78  &     7.5   \\    
        G032.93+02.6  &      18:41:52.83   &      +01:13:45.39  &Y  &/  &Y  &Y  &     0.84  &     6.2   \\    
        G038.36-00.9  &      19:04:45.73   &      +04:23:50.61  &Y  &/  &Y  &Y  &     1.11  &     11.0  \\    
        G043.02+08.3  &      18:39:34.51   &      +12:44:43.47  &Y  &/  &Y  &/  &     0.15  &     10.1  \\    
        G048.40-05.8  &      19:41:10.67   &      +10:56:50.84  &Y  &/  &Y  &/  &     0.57  &     7.7   \\    
        G057.10+03.6  &      19:23:50.21   &      +23:07:48.89  &Y  &/  &Y  &/  &     0.77  &     12.0  \\    
        G058.16+03.5  &      19:26:34.58   &      +23:59:17.87  &Y  &/  &Y  &Y  &     0.71  &     8.4   \\    
        G060.75-01.2  &      19:50:13.23   &      +23:55:19.72  &Y  &/  &Y  &/  &     0.79  &     9.9   \\    
        G065.43-03.1  &      20:08:19.49   &      +26:53:55.61  &Y  &/  &/  &/  &     0.35  &     7.6   \\    
        G070.44-01.5  &      20:14:44.32   &      +31:57:59.95  &Y  &/  &Y  &/  &     1.25  &     15.7  \\    
        G074.11+00.1  &      20:17:58.44   &      +35:56:15.15  &Y  &/  &Y  &/  &     5.23  &     10.4  \\    
$\star$ G084.79-01.1  &      20:56:46.83   &      +43:42:17.88  &Y  &Y  &Y  &Y  &     2.22  &     11.8  \\    
        G089.36-00.6  &      21:12:19.32   &      +47:24:19.39  &Y  &/  &Y  &/  &     0.73  &     9.9   \\    
        G089.62+02.1  &      21:00:44.18   &      +49:30:29.76  &Y  &Y  &Y  &/  &     1.62  &     9.9   \\    
        G089.93-07.0  &      21:39:51.55   &      +43:14:08.87  &Y  &/  &Y  &/  &      &     10.8  \\    
        G091.29-38.1  &      23:08:26.74   &      +18:18:20.84  &/  &Y  &   &   &     2.82  &     8.6   \\    
        G091.73+04.3  &      20:58:55.57   &      +52:32:37.12  &Y  &Y  &Y  &/  &     2.07  &     9.8   \\    
        G092.02+03.9  &      21:02:17.26   &      +52:28:34.24  &Y  &/  &Y  &/  &     1.63  &     8.1   \\    
        G092.26+03.8  &      21:03:58.96   &      +52:34:12.00  &Y  &/  &Y  &/  &     1.6   &     5.8   \\    
        G093.20+09.5  &      20:37:09.59   &      +56:54:14.31  &Y  &/  &/  &/  &     1.46  &     6.2   \\    
        G093.62-04.4  &      21:45:59.91   &      +47:35:50.06  &Y  &/  &Y  &/  &      &     10.8  \\    
        G093.99-04.9  &      21:49:28.13   &      +47:28:37.93  &Y  &/  &Y  &/  &      &     8.2   \\    
        G094.08+09.4  &      20:41:24.99   &      +57:33:22.41  &Y  &/  &Y  &/  &     0.98  &     8.2   \\    
        G095.51+09.9  &      20:44:46.24   &      +58:59:14.74  &Y  &/  &Y  &/  &     1.37  &     9.0   \\    
        G096.94+10.2  &      20:49:47.93   &      +60:16:24.52  &Y  &/  &Y  &/  &     1.26  &     10.2  \\    
        G097.09+10.1  &      20:51:23.79   &      +60:18:36.21  &Y  &/  &Y  &/  &     1.27  &     7.7   \\    
        G097.20+09.8  &      20:53:29.55   &      +60:14:26.34  &Y  &/  &Y  &/  &     1.22  &     7.9   \\    
        G097.38+09.9  &      20:53:53.73   &      +60:25:22.60  &Y  &/  &Y  &/  &     1.17  &     8.2   \\    
        G102.34+15.9  &      20:35:50.75   &      +67:53:21.94  &Y  &/  &Y  &/  &      &     10.2  \\    
        G102.72+15.3  &      20:43:15.74   &      +67:50:41.15  &Y  &/  &Y  &Y  &     0.32  &     9.7   \\    
        G103.90+13.9  &      21:02:23.20   &      +67:54:43.25  &Y  &/  &Y  &Y  &      &     11.1  \\    
        G108.10+13.1  &      21:40:19.64   &      +70:20:25.48  &Y  &/  &Y  &/  &     1.06  &     8.6   \\    
        G108.85-00.8  &      22:58:51.53   &      +58:57:27.12  &Y  &/  &Y  &/  &     5.4   &     15.3  \\    
        G110.65+09.6  &      22:28:00.22   &      +69:01:48.14  &Y  &/  &Y  &/  &     0.84  &     11.2  \\    
        G111.66+20.2  &      20:57:22.65   &      +77:35:01.72  &Y  &/  &Y  &Y  &     1.3   &     10.2  \\    
        G113.75+14.9  &      22:24:16.19   &      +75:05:01.80  &Y  &/  &/  &/  &     0.88  &     8.6   \\    
        G114.16+14.8  &      22:30:22.86   &      +75:13:16.60  &Y  &/  &Y  &/  &     0.77  &     10.4  \\    
        G114.56+14.7  &      22:36:09.22   &      +75:21:18.22  &Y  &/  &Y  &/  &     0.79  &     9.2   \\    
        G114.67+14.4  &      22:39:35.54   &      +75:11:34.02  &Y  &Y  &Y  &/  &     0.77  &     10.3  \\    
        G115.81-03.5  &      23:56:35.69   &      +58:33:58.49  &Y  &/  &Y  &/  &     0.37  &     8.0   \\    
        G116.23+20.3  &      21:57:06.33   &      +80:52:32.55  &/  &Y  &   &   &     1.09  &     5.6   \\    
        G117.11+12.4  &      23:25:30.22   &      +74:19:06.45  &Y  &/  &Y  &/  &     0.11  &     10.8  \\    
        G120.16+03.0  &      00:24:26.01   &      +65:49:27.56  &Y  &/  &Y  &/  &     2.0   &     13.7  \\    
        G120.67+02.6  &      00:29:41.95   &      +65:26:40.02  &Y  &/  &Y  &/  &     1.82  &     16.3  \\    
        G120.98+02.6  &      00:32:38.94   &      +65:28:07.05  &Y  &/  &/  &/  &     1.75  &     12.1  \\    
        G121.35+03.3  &      00:35:48.66   &      +66:13:13.29  &Y  &/  &Y  &Y  &     0.72  &     7.5   \\    
        G126.95-01.0  &      01:25:19.48   &      +61:32:36.24  &Y  &/  &   &   &     1.24  &     9.2   \\    
        G127.66+13.9  &      02:09:18.76   &      +76:06:34.73  &Y  &/  &   &   &      &     7.2   \\    
        G127.85+14.1  &      02:13:28.41   &      +76:15:01.14  &Y  &/  &   &   &      &     8.3   \\    
        G127.88+02.6  &      01:38:39.10   &      +65:05:06.53  &Y  &/  &   &   &     1.16  &     12.6  \\    
        G128.89+13.6  &      02:26:09.65   &      +75:27:12.35  &Y  &/  &   &   &     0.13  &     9.4   \\    
        G130.36+11.2  &      02:32:13.50   &      +72:39:51.09  &Y  &/  &Y  &/  &     1.56  &     8.4   \\    
        G132.03+08.9  &      02:39:33.57   &      +69:53:21.13  &Y  &/  &Y  &/  &     1.34  &     5.2   \\    
        G133.48+09.0  &      02:54:44.49   &      +69:19:57.55  &Y  &/  &   &   &     1.68  &     12.8  \\    
        G136.31-01.7  &      02:36:07.01   &      +58:21:09.13  &Y  &/  &   &   &     0.9   &     8.1   \\    
$\star$ G142.29+07.6  &      03:58:53.93   &      +63:14:15.37  &/  &/  &/  &Y  &     1.43  &     9.1   \\    
        G144.66+00.1  &      03:36:47.67   &      +55:53:53.31  &Y  &/  &Y  &/  &     0.85  &     10.9  \\    
        G147.01+03.3  &      04:04:41.37   &      +56:56:16.82  &Y  &/  &   &   &     0.59  &     10.9  \\    
$\star$ G150.44+03.9  &      04:25:07.08   &      +54:58:32.42  &Y  &/  &Y  &/  &     0.17  &     10.2  \\    
$\star$ G151.08+04.4  &      04:30:42.88   &      +54:51:53.94  &Y  &/  &Y  &/  &     0.15  &     10.2  \\    
$\star$ G156.20+05.2  &      04:57:00.65   &      +51:31:08.92  &Y  &/  &Y  &/  &     0.43  &     11.3  \\    
$\star$ G157.54-04.0  &      04:20:48.94   &      +44:18:06.00  &Y  &Y  &Y  &Y  &     5.75  &     7.6   \\    
$\star$ G158.20-20.2  &      03:29:19.31   &      +31:33:54.87  &Y  &/  &Y  &/  &     0.78  &     16.0  \\    
$\star$ G158.22-20.1  &      03:29:47.61   &      +31:39:49.38  &Y  &/  &Y  &/  &     0.86  &     16.0  \\    
$\star$ G158.37-20.7  &      03:28:41.99   &      +31:06:59.39  &Y  &/  &Y  &/  &     0.73  &     18.8  \\    
$\star$ G159.01-08.4  &      04:09:55.13   &      +40:06:54.78  &Y  &/  &Y  &/  &     0.73  &     9.1   \\    
$\star$ G159.21-20.4  &      03:32:23.18   &      +30:51:44.84  &Y  &/  &Y  &Y  &     0.71  &     14.1  \\    
$\star$ G159.21-20.1  &      03:33:17.90   &      +31:06:48.54  &Y  &Y  &Y  &/  &     0.71  &     11.0  \\    
$\star$ G159.23-34.4  &      02:56:04.75   &      +19:26:23.35  &/  &/  &Y  &/  &     1.17  &     10.2  \\    
$\star$ G159.34+11.2  &      05:40:13.24   &      +52:17:44.74  &/  &Y  &Y  &Y  &     0.17  &     11.8  \\    
$\star$ G159.65-19.6  &      03:36:04.66   &      +31:12:02.60  &Y  &/  &Y  &/  &     0.82  &     16.1  \\    
$\star$ G160.53-09.8  &      04:10:37.94   &      +38:05:04.82  &Y  &/  &Y  &/  &     0.65  &     14.2  \\    
$\star$ G160.64-35.0  &      02:58:48.02   &      +18:20:11.95  &Y  &Y  &Y  &Y  &     1.1   &     8.9   \\    
$\star$ G161.85-35.7  &      03:00:26.79   &      +17:11:16.08  &Y  &Y  &Y  &Y  &     1.74  &     10.3  \\    
$\star$ G161.85-08.6  &      04:19:31.10   &      +38:00:39.22  &Y  &/  &/  &/  &     0.47  &     9.2   \\    
$\star$ G162.64-31.6  &      03:12:56.11   &      +20:04:33.85  &/  &Y  &Y  &Y  &     0.34  &     10.8  \\    
$\star$ G163.32-08.4  &      04:25:31.80   &      +37:08:19.15  &Y  &/  &/  &/  &     0.41  &     7.9   \\    
$\star$ G163.67-08.3  &      04:27:00.26   &      +36:56:18.24  &Y  &/  &Y  &/  &     0.44  &     7.2   \\    
$\star$ G164.75-24.1  &      03:39:34.02   &      +24:42:59.94  &Y  &Y  &Y  &Y  &      &     12.6  \\    
$\star$ G164.94-08.5  &      04:30:27.46   &      +35:51:45.39  &Y  &/  &Y  &/  &     0.24  &     8.3   \\    
$\star$ G165.16-07.5  &      04:34:54.40   &      +36:23:13.28  &/  &/  &Y  &/  &     0.25  &     8.2   \\    
$\star$ G165.69-09.1  &      04:30:51.85   &      +34:56:09.42  &Y  &/  &/  &/  &     0.24  &     5.2   \\    
$\star$ G167.23-15.3  &      04:14:30.92   &      +29:35:16.09  &Y  &/  &Y  &/  &     1.06  &     8.5   \\    
$\star$ G168.00-15.6  &      04:15:40.46   &      +28:48:00.55  &/  &/  &Y  &/  &     1.29  &     12.9  \\    
$\star$ G168.13-16.3  &      04:13:47.56   &      +28:13:22.16  &Y  &/  &Y  &/  &     1.1   &     11.6  \\    
$\star$ G168.72-15.4  &      04:18:34.58   &      +28:26:34.64  &Y  &/  &Y  &Y  &     1.29  &     12.9  \\    
$\star$ G168.85-15.8  &      04:17:49.49   &      +28:06:39.04  &Y  &/  &/  &/  &     1.18  &     10.0  \\    
$\star$ G169.32-16.1  &      04:18:07.30   &      +27:33:43.22  &Y  &/  &Y  &/  &     1.3   &     10.9  \\    
$\star$ G169.43-16.1  &      04:18:23.08   &      +27:28:18.97  &Y  &/  &Y  &/  &     1.08  &     10.5  \\    
$\star$ G169.76-16.1  &      04:19:25.47   &      +27:15:16.21  &Y  &/  &Y  &/  &     1.35  &     13.0  \\    
$\star$ G169.98-18.9  &      04:10:57.74   &      +25:09:42.12  &Y  &/  &Y  &Y  &     1.58  &     11.0  \\    
$\star$ G170.00-16.1  &      04:20:12.04   &      +27:05:52.89  &Y  &/  &Y  &/  &     1.36  &     7.2   \\    
$\star$ G170.13-16.0  &      04:20:50.61   &      +27:03:29.08  &/  &/  &Y  &/  &     1.39  &     9.7   \\    
$\star$ G170.26-16.0  &      04:21:21.47   &      +26:59:29.26  &Y  &/  &Y  &/  &     1.35  &     10.2  \\    
$\star$ G170.99-15.8  &      04:24:10.41   &      +26:37:16.68  &/  &/  &Y  &/  &     1.41  &     7.1   \\    
$\star$ G171.14-17.5  &      04:18:48.20   &      +25:18:55.88  &Y  &/  &Y  &/  &     1.66  &     10.6  \\    
$\star$ G171.34-10.6  &      04:42:45.75   &      +29:44:27.70  &Y  &/  &Y  &/  &     1.37  &     6.2   \\    
$\star$ G171.51-10.5  &      04:43:32.45   &      +29:39:25.68  &Y  &/  &Y  &Y  &     1.35  &     6.9   \\    
$\star$ G171.84-05.2  &      05:04:08.91   &      +32:44:34.59  &Y  &/  &Y  &/  &     1.55  &     11.5  \\    
$\star$ G177.14-01.2  &      05:33:52.81   &      +30:42:36.33  &Y  &/  &/  &/  &      &     10.7  \\    
$\star$ G177.86+01.0  &      05:44:35.77   &      +31:17:57.39  &Y  &/  &Y  &/  &      &     8.7   \\    
$\star$ G178.48-06.7  &      05:16:16.57   &      +26:29:58.11  &Y  &/  &/  &/  &     4.82  &     12.5  \\    
$\star$ G178.98-06.7  &      05:17:37.38   &      +26:05:53.18  &Y  &/  &Y  &/  &     5.82  &     11.0  \\    
$\star$ G181.84+00.3  &      05:51:10.62   &      +27:31:08.15  &Y  &/  &/  &/  &     8.99  &     9.5   \\    
$\star$ G185.33-02.1  &      05:49:48.29   &      +23:16:02.13  &Y  &/  &/  &/  &     0.13  &     7.7   \\    
$\star$ G191.00-04.5  &      05:53:00.77   &      +17:08:25.35  &/  &/  &Y  &/  &     0.37  &     10.5  \\    
$\star$ G192.32-11.8  &      05:29:57.12   &      +12:16:50.79  &Y  &/  &Y  &/  &     3.43  &     20.0  \\    
$\star$ G195.00-16.9  &      05:17:50.47   &      +07:24:43.08  &Y  &/  &/  &/  &     0.17  &     14.1  \\    
$\star$ G195.09-16.4  &      05:19:53.82   &      +07:37:24.15  &/  &/  &Y  &/  &     0.12  &     16.9  \\    
$\star$ G198.56-09.1  &      05:52:17.25   &      +08:23:18.98  &Y  &Y  &Y  &/  &     2.03  &     16.3  \\    
$\star$ G199.88+00.9  &      06:30:41.17   &      +12:01:22.98  &Y  &/  &/  &/  &     0.97  &     9.3   \\    
$\star$ G200.34-10.9  &      05:49:08.80   &      +05:56:35.00  &Y  &/  &/  &/  &     2.27  &     11.0  \\    
$\star$ G201.13+00.3  &      06:30:49.51   &      +10:37:40.87  &Y  &/  &/  &/  &     0.79  &     10.4  \\    
$\star$ G201.44+00.6  &      06:32:36.91   &      +10:30:37.66  &Y  &/  &Y  &/  &     0.98  &     10.6  \\    
$\star$ G202.30-08.9  &      06:00:08.88   &      +05:15:06.16  &Y  &/  &Y  &Y  &     1.36  &     15.6  \\    
$\star$ G203.20-11.2  &      05:53:46.20   &      +03:22:38.79  &Y  &/  &Y  &/  &     1.54  &     9.5   \\    
$\star$ G204.49-11.3  &      05:55:43.23   &      +02:11:43.90  &Y  &/  &Y  &/  &     0.33  &     14.0  \\    
$\star$ G206.10-15.7  &      05:43:00.61   &      -01:16:00.53  &Y  &/  &Y  &/  &     0.33  &     9.3   \\    
$\star$ G207.35-19.8  &      05:30:50.16   &      -04:12:16.22  &Y  &/  &Y  &/  &     1.58  &     21.0  \\    
$\star$ G209.28-19.6  &      05:34:52.22   &      -05:44:06.10  &Y  &/  &Y  &/  &     1.08  &     27.1  \\    
$\star$ G211.48-19.2  &      05:39:52.49   &      -07:25:49.74  &Y  &/  &Y  &/  &     0.61  &     9.3   \\    
$\star$ G212.10-19.1  &      05:41:20.21   &      -07:53:59.69  &Y  &/  &Y  &/  &     0.65  &     15.5  \\    
$\star$ G213.96-19.6  &      05:42:38.05   &      -09:42:04.55  &Y  &/  &Y  &/  &     0.51  &     10.5  \\    
$\star$ G214.43-19.9  &      05:42:26.71   &      -10:12:14.46  &Y  &/  &Y  &/  &     0.46  &     8.2   \\    
$\star$ G215.00-15.1  &      06:00:51.24   &      -08:37:54.08  &/  &/  &Y  &/  &     1.43  &     9.8   \\    
$\star$ G216.18-15.2  &      06:02:26.88   &      -09:42:47.78  &/  &/  &Y  &/  &     1.49  &     10.1  \\    
        G359.31+17.0  &      16:42:30.08   &      -19:43:23.07  &Y  &/  &Y  &/  &      &     16.0  \\    
\end{longtable}
\vspace{-1em}
\hspace{-0em}
\begin{minipage}{\linewidth}
{\bf Note.}  Column 1: source name, the sources marked with “$\star$” before source name were observed with the high velocity resolution mode (AROWS mode 13) and the rest were observed with the low velocity resolution mode (AROWS mode 3); Column 2: right ascension; Column 3: declination; Column 4$\sim$7: the emission corresponding to the table head in this source was detected (``Y''), observed but not detected(``/'') or not observed(`` ''); Column 8: kinematic distance, the data obtained from \cite{2012ApJ...756...76W}; Column 9: excitation temperature derived from CO, obtained from \cite{2012ApJ...756...76W}.
\end{minipage}
\setcounter{footnote}{0}
\begin{longtable}{crrr}
	\caption{Parameters of detected DCO$^+$ 1-0.}
	\label{dcop}\\
	\hline 
	\hline              
	Source & $W$ & $V_{\rm LSR}$ & FWHM \\
	Name      & (K·km\,s$^{-1}$)  & (km\,s$^{-1}$)   &   (km\,s$^{-1}$) \\
\hline
\endfirsthead
\caption[]{continue.}\\
	\hline 
	\hline
		Source & $W$ & $V_{\rm LSR}$ & FWHM \\
	Name      & (K·km\,s$^{-1}$)  & (km\,s$^{-1}$)   &   (km\,s$^{-1}$) \\

	         \hline
\endhead
    
\hline
\endfoot

        G001.38+20.9 &      1.44   $\pm$  0.04   &      0.8    $\pm$  0.1    &      0.7    $\pm$  0.1    \\     
        G001.84+16.5 &      1.35   $\pm$  0.04   &      5.8    $\pm$  0.1    &      0.7    $\pm$  0.1    \\     
        G003.73+16.3 &      0.22   $\pm$  0.03   &      6.0    $\pm$  0.1    &      0.4    $\pm$  0.1    \\     
        G003.73+18.3 &      0.23   $\pm$  0.04   &      4.2    $\pm$  0.1    &      0.6    $\pm$  0.1    \\     
        G004.46+16.6 &      0.16   $\pm$  0.02   &      5.6    $\pm$  0.1    &      0.3    $\pm$  0.2    \\     
        G006.04+36.7 &      1.03   $\pm$  0.05   &      2.5    $\pm$  0.1    &      0.8    $\pm$  0.1    \\     
        G006.32+20.4 &      0.77   $\pm$  0.03   &      4.2    $\pm$  0.1    &      0.6    $\pm$  0.1    \\     
        G006.41+20.5 &      0.48   $\pm$  0.03   &      4.5    $\pm$  0.1    &      0.7    $\pm$  0.1    \\     
        G006.70+20.6 &      0.34   $\pm$  0.05   &      3.7    $\pm$  0.1    &      0.8    $\pm$  0.1    \\     
        G006.94+05.8 &      0.23   $\pm$  0.05   &      10.0   $\pm$  0.1    &      0.9    $\pm$  0.2    \\     
        G006.98+20.7 &      0.14   $\pm$  0.03   &      3.0    $\pm$  0.1    &      0.4    $\pm$  0.1    \\     
        G007.14+05.9 &      0.4    $\pm$  0.06   &      10.4   $\pm$  0.1    &      1.1    $\pm$  0.2    \\     
        G008.52+21.8 &      0.48   $\pm$  0.03   &      3.8    $\pm$  0.1    &      0.5    $\pm$  0.1    \\     
        G008.67+22.1 &      0.22   $\pm$  0.04   &      3.6    $\pm$  0.1    &      0.4    $\pm$  0.1    \\     
        G021.20+04.9 &      0.39   $\pm$  0.05   &      3.5    $\pm$  0.1    &      0.6    $\pm$  0.1    \\     
        G021.66+03.7 &      1.14   $\pm$  0.06   &      6.7    $\pm$  0.1    &      0.8    $\pm$  0.1    \\     
        G025.48+06.1 &      0.57   $\pm$  0.05   &      7.8    $\pm$  0.1    &      0.6    $\pm$  0.1    \\     
        G026.85+06.7 &      0.45   $\pm$  0.04   &      7.2    $\pm$  0.1    &      0.7    $\pm$  0.1    \\     
        G027.66+05.7 &      0.33   $\pm$  0.04   &      8.3    $\pm$  0.1    &      0.7    $\pm$  0.1    \\     
        G028.45-06.3 &      0.43   $\pm$  0.04   &      12.2   $\pm$  0.1    &      0.8    $\pm$  0.1    \\     
        G028.71+03.8 &      0.96   $\pm$  0.05   &      7.6    $\pm$  0.1    &      0.9    $\pm$  0.1    \\     
        G028.87+04.2 &      0.16   $\pm$  0.03   &      6.6    $\pm$  0.1    &      0.4    $\pm$  0.1    \\     
        G030.43+02.3 &      0.37   $\pm$  0.04   &      9.0    $\pm$  0.1    &      0.8    $\pm$  0.1    \\     
        G030.78+05.2 &      1.1    $\pm$  0.05   &      8.2    $\pm$  0.1    &      1.1    $\pm$  0.1    \\     
        G031.44+04.1 &      0.42   $\pm$  0.04   &      11.1   $\pm$  0.1    &      0.7    $\pm$  0.1    \\     
        G032.93+02.6 &      0.78   $\pm$  0.04   &      11.8   $\pm$  0.1    &      0.8    $\pm$  0.1    \\     
        G038.36-00.9 &      0.67   $\pm$  0.06   &      16.8   $\pm$  0.1    &      1.3    $\pm$  0.2    \\     
        G043.02+08.3 &      0.21   $\pm$  0.03   &      4.2    $\pm$  0.1    &      0.6    $\pm$  0.1    \\     
        G048.40-05.8 &      0.25   $\pm$  0.03   &      9.5    $\pm$  0.1    &      0.7    $\pm$  0.1    \\     
        G057.10+03.6 &      0.49   $\pm$  0.05   &      11.8   $\pm$  0.1    &      1.3    $\pm$  0.1    \\     
        G058.16+03.5 &      0.21   $\pm$  0.03   &      9.9    $\pm$  0.1    &      0.5    $\pm$  0.1    \\     
        G060.75-01.2 &      0.25   $\pm$  0.04   &      10.8   $\pm$  0.1    &      0.9    $\pm$  0.2    \\     
        G065.43-03.1 &      0.2    $\pm$  0.06   &      5.8    $\pm$  0.3    &      1.7    $\pm$  0.5    \\     
        G070.44-01.5 &      0.43   $\pm$  0.07   &      10.8   $\pm$  0.2    &      2.2    $\pm$  0.4    \\     
        G074.11+00.1 &      0.74   $\pm$  0.06   &      -1.1   $\pm$  0.1    &      2.5    $\pm$  0.2    \\     
$\star$ G084.79-01.1 &      2.37   $\pm$  0.19   &      2.0    $\pm$  0.1    &      1.2    $\pm$  0.1    \\     
        G089.36-00.6 &      0.3    $\pm$  0.04   &      2.7    $\pm$  0.1    &      0.9    $\pm$  0.1    \\     
        G089.62+02.1 &      0.58   $\pm$  0.04   &      -0.3   $\pm$  0.1    &      0.8    $\pm$  0.1    \\     
        G089.93-07.0 &      0.18   $\pm$  0.03   &      12.8   $\pm$  0.1    &      0.6    $\pm$  0.1    \\     
        G091.73+04.3 &      1.17   $\pm$  0.05   &      -4.4   $\pm$  0.1    &      1.0    $\pm$  0.1    \\     
        G092.02+03.9 &      0.26   $\pm$  0.04   &      -1.9   $\pm$  0.1    &      1.0    $\pm$  0.2    \\     
        G092.26+03.8 &      0.53   $\pm$  0.04   &      -1.8   $\pm$  0.1    &      1.3    $\pm$  0.1    \\     
        G093.20+09.5 &      0.18   $\pm$  0.04   &      -1.8   $\pm$  0.1    &      0.5    $\pm$  0.3    \\     
        G093.62-04.4 &      0.53   $\pm$  0.05   &      3.7    $\pm$  0.1    &      1.1    $\pm$  0.1    \\     
        G093.99-04.9 &      0.23   $\pm$  0.05   &      4.7    $\pm$  0.1    &      0.9    $\pm$  0.2    \\     
        G094.08+09.4 &      0.36   $\pm$  0.05   &      0.1    $\pm$  0.1    &      0.8    $\pm$  0.1    \\     
        G095.51+09.9 &      0.12   $\pm$  0.02   &      -2.6   $\pm$  0.1    &      0.4    $\pm$  0.1    \\     
        G096.94+10.2 &      0.36   $\pm$  0.04   &      -2.6   $\pm$  0.1    &      0.5    $\pm$  0.1    \\     
        G097.09+10.1 &      0.31   $\pm$  0.04   &      -2.6   $\pm$  0.1    &      0.7    $\pm$  0.1    \\     
        G097.20+09.8 &      0.54   $\pm$  0.05   &      -2.2   $\pm$  0.1    &      0.9    $\pm$  0.1    \\     
        G097.38+09.9 &      0.2    $\pm$  0.05   &      -2.4   $\pm$  0.1    &      1.0    $\pm$  0.3    \\     
        G102.34+15.9 &      0.63   $\pm$  0.04   &      2.5    $\pm$  0.1    &      0.7    $\pm$  0.1    \\     
        G102.72+15.3 &      0.44   $\pm$  0.04   &      1.3    $\pm$  0.1    &      0.6    $\pm$  0.1    \\     
        G103.90+13.9 &      0.3    $\pm$  0.04   &      2.9    $\pm$  0.1    &      0.7    $\pm$  0.1    \\     
        G108.10+13.1 &      0.36   $\pm$  0.04   &      -5.5   $\pm$  0.1    &      0.6    $\pm$  0.1    \\     
        G108.85-00.8 &      0.41   $\pm$  0.05   &      -49.5  $\pm$  0.1    &      1.3    $\pm$  0.2    \\     
        G110.65+09.6 &      0.78   $\pm$  0.05   &      -4.5   $\pm$  0.1    &      0.9    $\pm$  0.1    \\     
        G111.66+20.2 &      0.91   $\pm$  0.05   &      -8.2   $\pm$  0.1    &      0.9    $\pm$  0.1    \\     
        G113.75+14.9 &      0.1    $\pm$  0.03   &      -5.0   $\pm$  0.1    &      0.4    $\pm$  0.1    \\     
        G114.16+14.8 &      0.35   $\pm$  0.04   &      -4.1   $\pm$  0.1    &      0.7    $\pm$  0.1    \\     
        G114.56+14.7 &      0.11   $\pm$  0.03   &      -4.4   $\pm$  0.1    &      0.4    $\pm$  0.3    \\     
        G114.67+14.4 &      2.12   $\pm$  0.06   &      -4.6   $\pm$  0.1    &      1.2    $\pm$  0.1    \\     
        G115.81-03.5 &      0.35   $\pm$  0.04   &      -1.0   $\pm$  0.1    &      0.6    $\pm$  0.1    \\     
        G117.11+12.4 &      0.63   $\pm$  0.04   &      4.0    $\pm$  0.1    &      0.7    $\pm$  0.1    \\     
        G120.16+03.0 &      0.67   $\pm$  0.06   &      -19.5  $\pm$  0.1    &      2.1    $\pm$  0.2    \\     
        G120.67+02.6 &      0.38   $\pm$  0.06   &      -18.0  $\pm$  0.2    &      2.2    $\pm$  0.4    \\     
        G120.98+02.6 &      0.17   $\pm$  0.05   &      -17.2  $\pm$  0.2    &      1.4    $\pm$  0.3    \\     
        G121.35+03.3 &      0.29   $\pm$  0.04   &      -5.3   $\pm$  0.1    &      0.7    $\pm$  0.1    \\     
        G126.95-01.0 &      0.21   $\pm$  0.04   &      -12.1  $\pm$  0.1    &      0.7    $\pm$  0.1    \\     
        G127.66+13.9 &      0.34   $\pm$  0.06   &      3.3    $\pm$  0.1    &      0.9    $\pm$  0.2    \\     
        G127.85+14.1 &      0.35   $\pm$  0.06   &      2.7    $\pm$  0.1    &      0.6    $\pm$  0.1    \\     
        G127.88+02.6 &      0.26   $\pm$  0.07   &      -11.1  $\pm$  0.1    &      0.9    $\pm$  0.4    \\     
        G128.89+13.6 &      0.8    $\pm$  0.04   &      3.9    $\pm$  0.1    &      0.7    $\pm$  0.1    \\     
        G130.36+11.2 &      0.23   $\pm$  0.06   &      -15.9  $\pm$  0.3    &      1.6    $\pm$  0.4    \\     
        G132.03+08.9 &      0.21   $\pm$  0.04   &      -13.3  $\pm$  0.1    &      0.6    $\pm$  0.1    \\     
        G133.48+09.0 &      0.74   $\pm$  0.13   &      -16.8  $\pm$  0.2    &      2.3    $\pm$  0.6    \\     
        G136.31-01.7 &      0.27   $\pm$  0.05   &      -8.6   $\pm$  0.1    &      1.1    $\pm$  0.2    \\     
        G144.66+00.1 &      0.32   $\pm$  0.05   &      -8.0   $\pm$  0.1    &      0.7    $\pm$  0.1    \\     
        G147.01+03.3 &      0.44   $\pm$  0.05   &      -4.7   $\pm$  0.1    &      0.6    $\pm$  0.1    \\     
$\star$ G150.44+03.9 &      0.38   $\pm$  0.05   &      3.5    $\pm$  0.1    &      0.6    $\pm$  0.1    \\     
$\star$ G151.08+04.4 &      0.29   $\pm$  0.05   &      3.1    $\pm$  0.1    &      0.5    $\pm$  0.1    \\     
$\star$ G156.20+05.2 &      0.27   $\pm$  0.05   &      5.1    $\pm$  0.1    &      0.6    $\pm$  0.1    \\     
$\star$ G157.54-04.0 &      0.12   $\pm$  0.03   &      -31.0  $\pm$  0.1    &      0.2    $\pm$  0.1    \\     
$\star$ G158.20-20.2 &      0.69   $\pm$  0.06   &      7.5    $\pm$  0.1    &      0.5    $\pm$  0.1    \\     
$\star$ G158.22-20.1 &      0.13   $\pm$  0.03   &      8.2    $\pm$  0.1    &      0.2    $\pm$  0.1    \\     
$\star$ G158.37-20.7 &      0.68   $\pm$  0.08   &      6.8    $\pm$  0.1    &      1.4    $\pm$  0.2    \\     
$\star$ G159.01-08.4 &      0.17   $\pm$  0.03   &      -4.8   $\pm$  0.1    &      0.2    $\pm$  0.1    \\     
$\star$ G159.21-20.4 &      0.18   $\pm$  0.04   &      6.5    $\pm$  0.1    &      0.3    $\pm$  0.1    \\     
$\star$ G159.21-20.1 &      2.2    $\pm$  0.05   &      6.1    $\pm$  0.1    &      0.8    $\pm$  0.1    \\     
$\star$ G159.65-19.6 &      0.3    $\pm$  0.06   &      7.4    $\pm$  0.1    &      0.5    $\pm$  0.1    \\     
$\star$ G160.53-09.8 &      0.19   $\pm$  0.03   &      -3.9   $\pm$  0.1    &      0.3    $\pm$  0.1    \\     
$\star$ G160.64-35.0 &      0.06   $\pm$  0.02   &      -4.5   $\pm$  0.1    &      0.2    $\pm$  0.1    \\     
$\star$ G161.85-35.7 &      0.19   $\pm$  0.05   &      -6.6   $\pm$  0.1    &      0.5    $\pm$  0.1    \\     
$\star$ G161.85-08.6 &      0.21   $\pm$  0.05   &      -2.7   $\pm$  0.1    &      0.5    $\pm$  0.2    \\     
$\star$ G163.32-08.4 &      0.1    $\pm$  0.03   &      -2.3   $\pm$  0.1    &      0.3    $\pm$  0.2    \\     
$\star$ G163.67-08.3 &      0.17   $\pm$  0.04   &      -2.1   $\pm$  0.1    &      0.3    $\pm$  0.1    \\     
$\star$ G164.75-24.1 &      0.1    $\pm$  0.02   &      1.1    $\pm$  0.1    &      0.2    $\pm$  0.1    \\     
$\star$ G164.94-08.5 &      1.01   $\pm$  0.07   &      -0.8   $\pm$  0.1    &      0.8    $\pm$  0.1    \\     
$\star$ G165.69-09.1 &      0.27   $\pm$  0.06   &      -0.8   $\pm$  0.1    &      0.5    $\pm$  0.1    \\     
$\star$ G167.23-15.3 &      0.29   $\pm$  0.05   &      6.5    $\pm$  0.1    &      0.7    $\pm$  0.1    \\     
$\star$ G168.13-16.3 &      0.47   $\pm$  0.05   &      6.1    $\pm$  0.1    &      0.5    $\pm$  0.1    \\     
$\star$ G168.72-15.4 &      0.73   $\pm$  0.06   &      7.0    $\pm$  0.1    &      0.7    $\pm$  0.1    \\     
$\star$ G168.85-15.8 &      0.15   $\pm$  0.04   &      5.8    $\pm$  0.1    &      0.4    $\pm$  0.1    \\     
$\star$ G169.32-16.1 &      0.76   $\pm$  0.06   &      6.7    $\pm$  0.1    &      0.7    $\pm$  0.1    \\     
$\star$ G169.43-16.1 &      0.15   $\pm$  0.04   &      5.4    $\pm$  0.1    &      0.2    $\pm$  0.1    \\     
$\star$ G169.76-16.1 &      0.54   $\pm$  0.06   &      6.5    $\pm$  0.1    &      0.6    $\pm$  0.1    \\     
$\star$ G169.98-18.9 &      0.22   $\pm$  0.04   &      7.4    $\pm$  0.1    &      0.3    $\pm$  0.1    \\     
$\star$ G170.00-16.1 &      0.31   $\pm$  0.04   &      6.5    $\pm$  0.1    &      0.4    $\pm$  0.1    \\     
$\star$ G170.26-16.0 &      0.48   $\pm$  0.06   &      6.3    $\pm$  0.1    &      0.5    $\pm$  0.1    \\     
$\star$ G171.14-17.5 &      0.21   $\pm$  0.05   &      7.2    $\pm$  0.1    &      0.4    $\pm$  0.1    \\     
$\star$ G171.34-10.6 &      0.16   $\pm$  0.05   &      5.9    $\pm$  0.1    &      0.4    $\pm$  0.2    \\     
$\star$ G171.51-10.5 &      0.23   $\pm$  0.04   &      5.6    $\pm$  0.1    &      0.4    $\pm$  0.1    \\     
$\star$ G171.84-05.2 &      0.47   $\pm$  0.05   &      6.7    $\pm$  0.1    &      0.5    $\pm$  0.1    \\     
$\star$ G177.14-01.2 &      0.32   $\pm$  0.06   &      -17.6  $\pm$  0.1    &      0.7    $\pm$  0.2    \\     
$\star$ G177.86+01.0 &      0.17   $\pm$  0.05   &      -18.7  $\pm$  0.1    &      0.5    $\pm$  0.2    \\     
$\star$ G178.48-06.7 &      0.12   $\pm$  0.03   &      7.3    $\pm$  0.1    &      0.3    $\pm$  0.1    \\     
$\star$ G178.98-06.7 &      0.45   $\pm$  0.05   &      7.2    $\pm$  0.1    &      0.5    $\pm$  0.1    \\     
$\star$ G181.84+00.3 &      0.44   $\pm$  0.06   &      2.9    $\pm$  0.1    &      0.8    $\pm$  0.1    \\     
$\star$ G185.33-02.1 &      0.16   $\pm$  0.05   &      -0.9   $\pm$  0.1    &      0.6    $\pm$  0.2    \\     
$\star$ G192.32-11.8 &      0.77   $\pm$  0.06   &      11.2   $\pm$  0.1    &      0.9    $\pm$  0.1    \\     
$\star$ G195.00-16.9 &      0.15   $\pm$  0.04   &      -2.3   $\pm$  0.1    &      0.3    $\pm$  0.1    \\     
$\star$ G198.56-09.1 &      0.44   $\pm$  0.04   &      10.5   $\pm$  0.1    &      0.7    $\pm$  0.1    \\     
$\star$ G199.88+00.9 &      0.47   $\pm$  0.06   &      5.1    $\pm$  0.1    &      0.9    $\pm$  0.1    \\     
$\star$ G200.34-10.9 &      0.12   $\pm$  0.03   &      12.8   $\pm$  0.1    &      0.2    $\pm$  0.1    \\     
$\star$ G201.13+00.3 &      0.39   $\pm$  0.07   &      4.4    $\pm$  0.1    &      1.1    $\pm$  0.2    \\     
$\star$ G201.44+00.6 &      0.49   $\pm$  0.05   &      5.8    $\pm$  0.1    &      0.6    $\pm$  0.1    \\     
$\star$ G202.30-08.9 &      0.26   $\pm$  0.05   &      11.4   $\pm$  0.1    &      0.4    $\pm$  0.1    \\     
$\star$ G203.20-11.2 &      1.06   $\pm$  0.07   &      9.6    $\pm$  0.1    &      1.3    $\pm$  0.1    \\     
$\star$ G204.49-11.3 &      0.2    $\pm$  0.04   &      1.0    $\pm$  0.1    &      0.5    $\pm$  0.1    \\     
$\star$ G206.10-15.7 &      0.27   $\pm$  0.06   &      1.0    $\pm$  0.1    &      0.5    $\pm$  0.1    \\     
\footnote{G206.10-15.7 contains 2 velocity components, and the parameters are derived from double-component Gaussian fitting.}&      0.37   $\pm$  0.05   &      8.7    $\pm$  0.1    &      0.4    $\pm$  0.1    \\     
$\star$ G207.35-19.8 &      0.38   $\pm$  0.06   &      10.6   $\pm$  0.1    &      0.7    $\pm$  0.1    \\     
$\star$ G209.28-19.6\footnote{The profile of the line in G209.28-19.6 is not Gaussian, so $V_{\rm LSR}$ and FWHM are not provided.}  &      0.62   $\pm$  0.08   &      ...       &      ...    \\     
$\star$ G211.48-19.2 &      0.62   $\pm$  0.08   &      3.3    $\pm$  0.1    &      0.9    $\pm$  0.1    \\     
$\star$ G212.10-19.1 &      0.59   $\pm$  0.09   &      3.7    $\pm$  0.1    &      0.8    $\pm$  0.1    \\     
$\star$ G213.96-19.6 &      0.27   $\pm$  0.05   &      2.7    $\pm$  0.1    &      0.4    $\pm$  0.1    \\     
$\star$ G214.43-19.9 &      0.2    $\pm$  0.05   &      2.1    $\pm$  0.1    &      0.3    $\pm$  0.1    \\     
        G359.31+17.0 &      0.47   $\pm$  0.04   &      5.4    $\pm$  0.1    &      0.5    $\pm$  0.1    \\     
\end{longtable}	
\vspace{-1em}
\hspace{-0em}
\begin{minipage}{\linewidth}
{\bf Note.}  Column 1: source name, the sources marked with “$\star$” before source name were observed with the high velocity resolution mode (AROWS mode 13) and the rest were observed with the low velocity resolution mode (AROWS mode 3); Column 2: velocity integrated intensity; Column 3: local standard of rest velocity; Column 4: full width at half maximum.
\end{minipage}

\begin{longtable}{crrr}

	\caption{Parameters of detected DCN 1-0.}
	\label{dcn}\\
	\hline 
	\hline
	Source & $W$ & $V_{\rm LSR}$ & FWHM \\
	Name      & (K·km\,s$^{-1}$)  & (km\,s$^{-1}$)   &   (km\,s$^{-1}$) \\
\hline

\endfirsthead
\caption[]{continue.}\\
	         
	\hline 
	\hline
	Source & $W$ & $V_{\rm LSR}$ & FWHM \\
	Name      & (K·km\,s$^{-1}$)  & (km\,s$^{-1}$)   &   (km\,s$^{-1}$) \\
	         \hline
  
    \endhead
\hline
\endfoot

        G001.38+20.9 &      0.23   $\pm$  0.05   &      0.6    $\pm$  0.1    &      0.4    $\pm$  0.1    \\     
        G001.84+16.5 &      0.13   $\pm$  0.03   &      5.8    $\pm$  0.1    &      0.4    $\pm$  0.1    \\     
        G021.66+03.7 &      0.21   $\pm$  0.06   &      6.8    $\pm$  0.1    &      0.8    $\pm$  0.2    \\     
        G026.85+06.7 &      0.15   $\pm$  0.04   &      7.0    $\pm$  0.1    &      0.9    $\pm$  0.3    \\     
        G028.45-06.3 &      0.1    $\pm$  0.03   &      12.2   $\pm$  0.1    &      0.4    $\pm$  0.1    \\     
        G028.71+03.8 &      0.19   $\pm$  0.03   &      7.4    $\pm$  0.1    &      0.4    $\pm$  0.1    \\     
$\star$ G084.79-01.1 &      0.44   $\pm$  0.08   &      1.8    $\pm$  0.1    &      1.1    $\pm$  0.2    \\     
        G089.62+02.1 &      0.14   $\pm$  0.04   &      -0.4   $\pm$  0.1    &      0.5    $\pm$  0.2    \\     
        G091.29-38.1 &      0.17   $\pm$  0.05   &      -3.9   $\pm$  0.1    &      1.0    $\pm$  0.4    \\     
        G091.73+04.3 &      0.25   $\pm$  0.05   &      -4.4   $\pm$  0.1    &      1.2    $\pm$  0.2    \\     
        G114.67+14.4 &      0.28   $\pm$  0.06   &      -4.5   $\pm$  0.1    &      1.1    $\pm$  0.2    \\     
        G116.23+20.3 &      0.11   $\pm$  0.02   &      -7.9   $\pm$  0.1    &      0.4    $\pm$  0.1    \\     
$\star$ G157.54-04.0 &      0.12   $\pm$  0.03   &      -31.1  $\pm$  0.1    &      0.3    $\pm$  0.1    \\     
$\star$ G159.21-20.1 &      0.34   $\pm$  0.04   &      6.1    $\pm$  0.1    &      0.6    $\pm$  0.1    \\     
$\star$ G159.34+11.2 &      0.13   $\pm$  0.03   &      2.5    $\pm$  0.1    &      0.2    $\pm$  0.1    \\     
$\star$ G160.64-35.0 &      0.14   $\pm$  0.03   &      -4.4   $\pm$  0.1    &      0.2    $\pm$  0.1    \\     
$\star$ G161.85-35.7 &      0.13   $\pm$  0.03   &      -6.6   $\pm$  0.1    &      0.2    $\pm$  0.1    \\     
$\star$ G162.64-31.6 &      0.12   $\pm$  0.03   &      -0.9   $\pm$  0.1    &      0.2    $\pm$  0.1    \\     
$\star$ G164.75-24.1 &      0.2    $\pm$  0.03   &      1.1    $\pm$  0.1    &      0.2    $\pm$  0.1    \\     
$\star$ G198.56-09.1 &      0.1    $\pm$  0.02   &      10.6   $\pm$  0.1    &      0.2    $\pm$  0.1    \\     
\end{longtable}
\vspace{-1em}
\hspace{-0em}
\begin{minipage}{\linewidth}
{\bf Note.} Column 1: source name, the sources marked with “$\star$” before source name were observed with the high velocity resolution mode (AROWS mode 13) and the rest were observed with the low velocity resolution mode (AROWS mode 3); Column 2: velocity integrated intensity; Column 3: local standard of rest velocity; Column 4: full width at half maximum.
\end{minipage}

\setcounter{footnote}{0}
\begin{longtable}{crrrrrccc}

	\caption{Parameters of detected H$^{13}$CO$^+$ 1-0}
	\label{h13cop}\\
	\hline 
	\hline
	Source & $W$ & $V_{\rm LSR}$ & FWHM \\
	Name      & (K·km\,s$^{-1}$)  & (km\,s$^{-1}$)   &   (km\,s$^{-1}$) \\
\hline

\endfirsthead
\caption[]{continue.}\\
	\hline 
	\hline
	Source & $W$ & $V_{\rm LSR}$ & FWHM \\
	Name      & (K·km\,s$^{-1}$)  & (km\,s$^{-1}$)   &   (km\,s$^{-1}$) \\
	         \hline
    \endhead
\hline
\endfoot

        G001.38+20.9 &      0.76   $\pm$  0.02   &      0.6    $\pm$  0.1    &      0.6    $\pm$  0.1    \\     
        G001.84+16.5 &      0.64   $\pm$  0.02   &      5.8    $\pm$  0.1    &      0.6    $\pm$  0.1    \\     
        G003.73+16.3 &      0.2    $\pm$  0.02   &      6.1    $\pm$  0.1    &      0.4    $\pm$  0.1    \\     
        G003.73+18.3 &      0.28   $\pm$  0.02   &      4.2    $\pm$  0.1    &      0.5    $\pm$  0.1    \\     
        G004.46+16.6 &      0.14   $\pm$  0.02   &      5.5    $\pm$  0.1    &      0.4    $\pm$  0.1    \\     
        G006.04+36.7 &      0.4    $\pm$  0.02   &      2.4    $\pm$  0.1    &      0.5    $\pm$  0.1    \\     
        G006.32+20.4 &      0.33   $\pm$  0.02   &      4.2    $\pm$  0.1    &      0.4    $\pm$  0.1    \\     
        G006.41+20.5 &      0.28   $\pm$  0.03   &      4.5    $\pm$  0.1    &      0.7    $\pm$  0.1    \\     
        G006.70+20.6 &      0.39   $\pm$  0.03   &      3.6    $\pm$  0.1    &      1.0    $\pm$  0.1    \\     
        G006.94+05.8 &      0.16   $\pm$  0.03   &      9.9    $\pm$  0.1    &      0.7    $\pm$  0.1    \\     
        G007.14+05.9 &      0.2    $\pm$  0.03   &      10.3   $\pm$  0.1    &      0.8    $\pm$  0.2    \\     
        G008.52+21.8 &      0.36   $\pm$  0.02   &      3.7    $\pm$  0.1    &      0.4    $\pm$  0.1    \\     
        G008.67+22.1 &      0.11   $\pm$  0.02   &      3.3    $\pm$  0.1    &      0.3    $\pm$  0.1    \\     
        G021.20+04.9 &      0.24   $\pm$  0.02   &      3.5    $\pm$  0.1    &      0.5    $\pm$  0.1    \\     
        G021.66+03.7 &      0.61   $\pm$  0.03   &      6.6    $\pm$  0.1    &      0.7    $\pm$  0.1    \\     
        G025.48+06.1 &      0.46   $\pm$  0.02   &      7.7    $\pm$  0.1    &      0.5    $\pm$  0.1    \\     
        G026.85+06.7 &      0.34   $\pm$  0.02   &      7.1    $\pm$  0.1    &      0.5    $\pm$  0.1    \\     
        G027.66+05.7 &      0.2    $\pm$  0.02   &      8.2    $\pm$  0.1    &      0.7    $\pm$  0.1    \\     
        G028.45-06.3 &      0.57   $\pm$  0.02   &      12.1   $\pm$  0.1    &      0.7    $\pm$  0.1    \\     
        G028.71+03.8 &      0.53   $\pm$  0.03   &      7.5    $\pm$  0.1    &      1.0    $\pm$  0.1    \\     
        G028.87+04.2 &      0.25   $\pm$  0.03   &      6.9    $\pm$  0.1    &      1.6    $\pm$  0.3    \\     
        G030.43+02.3 &      0.32   $\pm$  0.03   &      9.1    $\pm$  0.1    &      0.9    $\pm$  0.1    \\     
        G030.78+05.2 &      0.93   $\pm$  0.03   &      8.2    $\pm$  0.1    &      1.2    $\pm$  0.1    \\     
        G031.44+04.1 &      0.33   $\pm$  0.02   &      11.1   $\pm$  0.1    &      0.6    $\pm$  0.1    \\     
        G032.93+02.6 &      0.51   $\pm$  0.03   &      11.8   $\pm$  0.1    &      0.9    $\pm$  0.1    \\     
        G038.36-00.9 &      0.81   $\pm$  0.04   &      16.5   $\pm$  0.1    &      1.5    $\pm$  0.1    \\     
        G043.02+08.3 &      0.1    $\pm$  0.02   &      4.1    $\pm$  0.1    &      0.4    $\pm$  0.1    \\     
        G048.40-05.8 &      0.14   $\pm$  0.02   &      9.6    $\pm$  0.1    &      0.5    $\pm$  0.1    \\     
        G057.10+03.6 &      0.41   $\pm$  0.03   &      11.8   $\pm$  0.1    &      1.5    $\pm$  0.1    \\     
        G058.16+03.5 &      0.25   $\pm$  0.02   &      9.9    $\pm$  0.1    &      0.6    $\pm$  0.1    \\     
        G060.75-01.2 &      0.3    $\pm$  0.04   &      10.8   $\pm$  0.1    &      0.9    $\pm$  0.2    \\     
        G070.44-01.5 &      0.36   $\pm$  0.04   &      10.9   $\pm$  0.1    &      2.0    $\pm$  0.3    \\     
        G074.11+00.1 &      0.6    $\pm$  0.06   &      -0.9   $\pm$  0.1    &      2.4    $\pm$  0.3    \\     
$\star$ G084.79-01.1 &      1.11   $\pm$  0.06   &      ...                  &      1.6    $\pm$  0.1    \\     
        G089.36-00.6 &      0.26   $\pm$  0.04   &      2.7    $\pm$  0.1    &      1.1    $\pm$  0.3    \\     
        G089.62+02.1 &      0.29   $\pm$  0.03   &      -0.3   $\pm$  0.1    &      0.9    $\pm$  0.1    \\     
        G089.93-07.0 &      0.11   $\pm$  0.02   &      12.8   $\pm$  0.1    &      0.4    $\pm$  0.1    \\     
        G091.73+04.3 &      0.58   $\pm$  0.04   &      -4.4   $\pm$  0.1    &      0.9    $\pm$  0.1    \\     
        G092.02+03.9 &      0.2    $\pm$  0.03   &      -2.0   $\pm$  0.1    &      0.7    $\pm$  0.1    \\     
        G092.26+03.8 &      0.22   $\pm$  0.03   &      -1.6   $\pm$  0.1    &      0.7    $\pm$  0.2    \\     
        G093.62-04.4 &      0.19   $\pm$  0.03   &      3.7    $\pm$  0.1    &      0.8    $\pm$  0.1    \\     
        G093.99-04.9 &      0.24   $\pm$  0.05   &      4.6    $\pm$  0.2    &      2.0    $\pm$  0.5    \\     
        G094.08+09.4 &      0.19   $\pm$  0.02   &      0.0    $\pm$  0.1    &      0.7    $\pm$  0.1    \\     
        G095.51+09.9 &      0.15   $\pm$  0.03   &      -2.6   $\pm$  0.1    &      0.6    $\pm$  0.1    \\     
        G096.94+10.2 &      0.18   $\pm$  0.02   &      -2.5   $\pm$  0.1    &      0.4    $\pm$  0.1    \\     
        G097.09+10.1 &      0.16   $\pm$  0.03   &      -2.6   $\pm$  0.1    &      0.5    $\pm$  0.1    \\     
        G097.20+09.8 &      0.23   $\pm$  0.03   &      -2.2   $\pm$  0.1    &      0.6    $\pm$  0.1    \\     
        G097.38+09.9 &      0.15   $\pm$  0.03   &      -2.4   $\pm$  0.1    &      0.5    $\pm$  0.1    \\     
        G102.34+15.9 &      0.35   $\pm$  0.02   &      2.6    $\pm$  0.1    &      0.6    $\pm$  0.1    \\     
        G102.72+15.3 &      0.24   $\pm$  0.02   &      1.2    $\pm$  0.1    &      0.5    $\pm$  0.1    \\     
        G103.90+13.9 &      0.41   $\pm$  0.02   &      2.8    $\pm$  0.1    &      0.8    $\pm$  0.1    \\     
        G108.10+13.1 &      0.22   $\pm$  0.03   &      -5.4   $\pm$  0.1    &      0.4    $\pm$  0.1    \\     
        G108.85-00.8 &      0.24   $\pm$  0.04   &      -49.7  $\pm$  0.1    &      1.3    $\pm$  0.2    \\     
        G110.65+09.6 &      0.5    $\pm$  0.05   &      -4.5   $\pm$  0.1    &      1.0    $\pm$  0.1    \\     
        G111.66+20.2 &      0.43   $\pm$  0.03   &      -8.1   $\pm$  0.1    &      0.8    $\pm$  0.1    \\     
        G114.16+14.8 &      0.24   $\pm$  0.03   &      -4.0   $\pm$  0.1    &      0.6    $\pm$  0.1    \\     
        G114.56+14.7 &      0.11   $\pm$  0.04   &      -4.7   $\pm$  0.1    &      0.7    $\pm$  0.3    \\     
        G114.67+14.4 &      0.88   $\pm$  0.04   &      -4.5   $\pm$  0.1    &      1.1    $\pm$  0.1    \\     
        G115.81-03.5 &      0.1    $\pm$  0.03   &      -0.9   $\pm$  0.1    &      0.3    $\pm$  0.1    \\     
        G117.11+12.4 &      0.37   $\pm$  0.03   &      4.0    $\pm$  0.1    &      0.5    $\pm$  0.1    \\     
        G120.16+03.0 &      0.54   $\pm$  0.07   &      -19.3  $\pm$  0.1    &      2.3    $\pm$  0.3    \\     
        G120.67+02.6 &      0.31   $\pm$  0.05   &      -18.2  $\pm$  0.1    &      1.8    $\pm$  0.3    \\     
        G121.35+03.3 &      0.15   $\pm$  0.04   &      -5.2   $\pm$  0.1    &      0.6    $\pm$  0.2    \\     
        G130.36+11.2 &      0.16   $\pm$  0.04   &      -15.7  $\pm$  0.1    &      0.9    $\pm$  0.3    \\     
        G132.03+08.9 &      0.1    $\pm$  0.03   &      -13.4  $\pm$  0.1    &      0.6    $\pm$  0.2    \\     
        G144.66+00.1 &      0.18   $\pm$  0.04   &      -7.7   $\pm$  0.1    &      0.7    $\pm$  0.2    \\     
$\star$ G150.44+03.9 &      0.1    $\pm$  0.03   &      ...                  &      0.3    $\pm$  0.1    \\     
$\star$ G151.08+04.4 &      0.19   $\pm$  0.03   &      ...                  &      0.3    $\pm$  0.1    \\     
$\star$ G156.20+05.2 &      0.22   $\pm$  0.03   &      ...                  &      0.5    $\pm$  0.1    \\     
$\star$ G157.54-04.0 &      0.07   $\pm$  0.02   &      ...                  &      0.2    $\pm$  0.1    \\     
$\star$ G158.20-20.2 &      0.33   $\pm$  0.03   &      ...                  &      0.5    $\pm$  0.1    \\     
$\star$ G158.22-20.1 &      0.13   $\pm$  0.02   &      ...                  &      0.3    $\pm$  0.1    \\     
$\star$ G158.37-20.7 &      0.31   $\pm$  0.04   &      ...                  &      0.7    $\pm$  0.1    \\     
$\star$ G159.01-08.4 &      0.11   $\pm$  0.03   &      ...                  &      0.5    $\pm$  0.1    \\     
$\star$ G159.21-20.4 &      0.18   $\pm$  0.03   &      ...                  &      0.5    $\pm$  0.1    \\     
$\star$ G159.21-20.1 &      1.0    $\pm$  0.04   &      ...                  &      0.9    $\pm$  0.1    \\     
$\star$ G159.23-34.4 &      0.28   $\pm$  0.03   &      ...                  &      0.5    $\pm$  0.1    \\     
$\star$ G159.34+11.2 &      0.19   $\pm$  0.02   &      ...                  &      0.2    $\pm$  0.1    \\     
$\star$ G159.65-19.6 &      0.12   $\pm$  0.02   &      ...                  &      0.4    $\pm$  0.1    \\     
$\star$ G160.53-09.8 &      0.14   $\pm$  0.04   &      ...                  &      0.6    $\pm$  0.2    \\     
$\star$ G160.64-35.0 &      0.13   $\pm$  0.02   &      ...                  &      0.2    $\pm$  0.1    \\     
$\star$ G161.85-35.7 &      0.2    $\pm$  0.03   &      ...                  &      0.2    $\pm$  0.1    \\     
$\star$ G162.64-31.6 &      0.08   $\pm$  0.02   &      ...                  &      0.2    $\pm$  0.1    \\     
$\star$ G163.67-08.3 &      0.16   $\pm$  0.03   &      ...                  &      0.4    $\pm$  0.1    \\     
$\star$ G164.75-24.1 &      0.22   $\pm$  0.02   &      ...                  &      0.2    $\pm$  0.1    \\     
$\star$ G164.94-08.5 &      0.51   $\pm$  0.04   &      ...                  &      0.7    $\pm$  0.1    \\     
$\star$ G165.16-07.5 &      0.22   $\pm$  0.03   &      ...                  &      0.6    $\pm$  0.1    \\     
$\star$ G167.23-15.3 &      0.22   $\pm$  0.04   &      ...                  &      0.8    $\pm$  0.2    \\     
$\star$ G168.00-15.6 &      0.42   $\pm$  0.03   &      ...                  &      0.5    $\pm$  0.1    \\     
$\star$ G168.13-16.3 &      0.22   $\pm$  0.03   &      ...                  &      0.3    $\pm$  0.1    \\     
$\star$ G168.72-15.4 &      0.46   $\pm$  0.03   &      ...                  &      0.5    $\pm$  0.1    \\     
$\star$ G169.32-16.1 &      0.29   $\pm$  0.03   &      ...                  &      0.7    $\pm$  0.1    \\     
$\star$ G169.43-16.1 &      0.19   $\pm$  0.04   &      ...                  &      0.8    $\pm$  0.2    \\     
$\star$ G169.76-16.1 &      0.26   $\pm$  0.03   &      ...                  &      0.5    $\pm$  0.1    \\     
$\star$ G169.98-18.9 &      0.31   $\pm$  0.03   &      ...                  &      0.3    $\pm$  0.1    \\     
$\star$ G170.00-16.1 &      0.14   $\pm$  0.03   &      ...                  &      0.4    $\pm$  0.1    \\     
$\star$ G170.13-16.0 &      0.42   $\pm$  0.04   &      ...                  &      0.8    $\pm$  0.1    \\     
$\star$ G170.26-16.0 &      0.38   $\pm$  0.04   &      ...                  &      0.5    $\pm$  0.1    \\     
$\star$ G170.99-15.8 &      0.31   $\pm$  0.04   &      ...                  &      0.5    $\pm$  0.1    \\     
$\star$ G171.14-17.5 &      0.18   $\pm$  0.03   &      ...                  &      0.3    $\pm$  0.1    \\     
$\star$ G171.34-10.6 &      0.13   $\pm$  0.03   &      ...                  &      0.4    $\pm$  0.1    \\     
$\star$ G171.51-10.5 &      0.18   $\pm$  0.02   &      ...                  &      0.2    $\pm$  0.1    \\     
$\star$ G171.84-05.2 &      0.24   $\pm$  0.03   &      ...                  &      0.3    $\pm$  0.1    \\     
$\star$ G177.86+01.0 &      0.14   $\pm$  0.03   &      ...                  &      0.5    $\pm$  0.1    \\     
$\star$ G178.98-06.7 &      0.31   $\pm$  0.02   &      ...                  &      0.3    $\pm$  0.1    \\     
$\star$ G191.00-04.5 &      0.1    $\pm$  0.02   &      ...                  &      0.3    $\pm$  0.1    \\     
$\star$ G192.32-11.8 &      0.49   $\pm$  0.04   &      ...                  &      0.7    $\pm$  0.1    \\     
$\star$ G195.09-16.4 &      0.06   $\pm$  0.02   &      ...                  &      0.1    $\pm$  0.1    \\     
$\star$ G198.56-09.1 &      0.19   $\pm$  0.02   &      ...                  &      0.5    $\pm$  0.1    \\     
$\star$ G201.44+00.6 &      0.19   $\pm$  0.05   &      ...                  &      0.5    $\pm$  0.2    \\     
$\star$ G202.30-08.9 &      0.14   $\pm$  0.02   &      ...                  &      0.2    $\pm$  0.1    \\     
$\star$ G203.20-11.2 &      0.6    $\pm$  0.05   &      ...                  &      1.1    $\pm$  0.1    \\     
$\star$ G204.49-11.3 &      0.11   $\pm$  0.02   &      ...                  &      0.4    $\pm$  0.1    \\     
$\star$ G206.10-15.7 &      0.13   $\pm$  0.03   &      ...                  &      0.4    $\pm$  0.1    \\     
\footnote{G206.10-15.7 contains 2 velocity components, and the parameters are derived from double-component Gaussian fitting.} &      0.22   $\pm$  0.03   &      ...                  &      0.4    $\pm$  0.1    \\     
$\star$ G207.35-19.8 &      0.13   $\pm$  0.03   &      ...                  &      0.3    $\pm$  0.1    \\     
$\star$ G209.28-19.6\footnote{The profile of the line in G209.28-19.6 is not Gaussian, so $V_{\rm LSR}$ and FWHM are not provided.} &      0.99   $\pm$  0.06   &      ...                  &      ...    \\     
$\star$ G211.48-19.2 &      0.31   $\pm$  0.04   &      ...                  &      0.8    $\pm$  0.1    \\     
$\star$ G212.10-19.1 &      0.17   $\pm$  0.05   &      ...                  &      0.4    $\pm$  0.2    \\     
$\star$ G213.96-19.6 &      0.24   $\pm$  0.04   &      ...                  &      0.6    $\pm$  0.1    \\     
$\star$ G214.43-19.9 &      0.19   $\pm$  0.04   &      ...                  &      0.4    $\pm$  0.1    \\     
$\star$ G215.00-15.1 &      0.08   $\pm$  0.02   &      ...                  &      0.2    $\pm$  0.1    \\     
$\star$ G216.18-15.2 &      0.15   $\pm$  0.04   &      ...                  &      0.6    $\pm$  0.2    \\     
        G359.31+17.0 &      0.34   $\pm$  0.02   &      5.4    $\pm$  0.1    &      0.4    $\pm$  0.1    \\     

\end{longtable}	
\vspace{-1em}
\hspace{-0em}
\begin{minipage}{\linewidth}
{\bf Note.}  Column 1: source name, the sources marked with “$\star$” before source name were observed with the high velocity resolution mode (AROWS mode 13) and the rest were observed with the low velocity resolution mode (AROWS mode 3); Column 2: velocity integrated intensity; Column 3: local standard of rest velocity; Column 4: full width at half maximum. In the high velocity resolution mode (AROWS mode 13) observations, due to errors in the Doppler tracking for different sidebands during the shared risking period, the $V_{\rm LSR}$ measurements of H$^{13}$CO$^+$ and H$^{13}$CN 1-0 have an uncertainty of 1 to 7 km s$^{-1}$, while the other parameters are corrected. Therefore, the $V_{\rm LSR}$ values are not provided here and not used for scientific discussions.
\end{minipage}

\begin{longtable}{crrr}

	\caption{Parameters of detected H$^{13}$CN 1-0.}
	\label{h13cn}\\
	\hline 
	\hline
	Source & $W$ & $V_{\rm LSR}$ & FWHM \\
	Name      & (K·km\,s$^{-1}$)  & (km\,s$^{-1}$)   &   (km\,s$^{-1}$) \\
\hline

\endfirsthead
\caption[]{continue.}\\
	         
	\hline 
	\hline
	Source & $W$ & $V_{\rm LSR}$ & FWHM \\
	Name      & (K·km\,s$^{-1}$)  & (km\,s$^{-1}$)   &   (km\,s$^{-1}$) \\
	         \hline
  
    \endhead
\hline
\endfoot

        G001.38+20.9 &      0.08   $\pm$  0.02   &      0.9    $\pm$  0.1    &      0.5    $\pm$  0.1    \\     
        G004.46+16.6 &      0.09   $\pm$  0.03   &      5.3    $\pm$  0.2    &      0.8    $\pm$  0.3    \\     
        G006.04+36.7 &      0.06   $\pm$  0.02   &      2.6    $\pm$  0.1    &      0.3    $\pm$  0.2    \\     
        G006.70+20.6 &      0.06   $\pm$  0.02   &      3.9    $\pm$  0.1    &      0.4    $\pm$  0.1    \\     
        G008.67+22.1 &      0.06   $\pm$  0.02   &      3.7    $\pm$  0.1    &      0.4    $\pm$  0.2    \\     
        G021.20+04.9 &      0.07   $\pm$  0.02   &      3.7    $\pm$  0.1    &      0.6    $\pm$  0.2    \\     
        G025.48+06.1 &      0.1    $\pm$  0.02   &      7.7    $\pm$  0.1    &      0.5    $\pm$  0.1    \\     
        G026.85+06.7 &      0.07   $\pm$  0.02   &      7.3    $\pm$  0.1    &      0.7    $\pm$  0.2    \\     
        G028.45-06.3 &      0.07   $\pm$  0.02   &      12.2   $\pm$  0.1    &      0.4    $\pm$  0.2    \\     
        G028.71+03.8 &      0.15   $\pm$  0.02   &      7.7    $\pm$  0.1    &      1.0    $\pm$  0.2    \\     
        G030.43+02.3 &      0.05   $\pm$  0.02   &      9.1    $\pm$  0.1    &      0.4    $\pm$  0.2    \\     
        G030.78+05.2 &      0.28   $\pm$  0.04   &      8.0    $\pm$  0.1    &      1.5    $\pm$  0.2    \\     
        G031.44+04.1 &      0.05   $\pm$  0.02   &      11.2   $\pm$  0.1    &      0.4    $\pm$  0.2    \\     
        G032.93+02.6 &      0.07   $\pm$  0.03   &      11.6   $\pm$  0.1    &      0.7    $\pm$  0.4    \\     
        G038.36-00.9 &      0.16   $\pm$  0.03   &      16.8   $\pm$  0.1    &      1.3    $\pm$  0.2    \\     
        G058.16+03.5 &      0.05   $\pm$  0.02   &      10.0   $\pm$  0.1    &      0.3    $\pm$  0.1    \\     
$\star$ G084.79-01.1 &      0.24   $\pm$  0.04   &      ...                  &      1.1    $\pm$  0.2    \\     
        G102.72+15.3 &      0.05   $\pm$  0.02   &      1.2    $\pm$  0.1    &      0.4    $\pm$  0.1    \\     
        G103.90+13.9 &      0.15   $\pm$  0.03   &      2.8    $\pm$  0.1    &      0.8    $\pm$  0.2    \\     
        G111.66+20.2 &      0.11   $\pm$  0.03   &      -8.0   $\pm$  0.1    &      0.9    $\pm$  0.3    \\     
        G121.35+03.3 &      0.07   $\pm$  0.02   &      -5.9   $\pm$  0.1    &      0.5    $\pm$  0.2    \\     
$\star$ G142.29+07.6 &      0.09   $\pm$  0.02   &      ...                  &      0.2    $\pm$  0.1    \\     
$\star$ G157.54-04.0 &      0.11   $\pm$  0.03   &      ...                  &      0.3    $\pm$  0.1    \\     
$\star$ G159.21-20.4 &      0.13   $\pm$  0.03   &      ...                  &      0.4    $\pm$  0.1    \\     
$\star$ G159.34+11.2 &      0.16   $\pm$  0.02   &      ...                  &      0.2    $\pm$  0.1    \\     
$\star$ G160.64-35.0 &      0.13   $\pm$  0.02   &      ...                  &      0.2    $\pm$  0.1    \\     
$\star$ G161.85-35.7 &      0.17   $\pm$  0.02   &      ...                  &      0.2    $\pm$  0.1    \\     
$\star$ G162.64-31.6 &      0.1    $\pm$  0.03   &      ...                  &      0.3    $\pm$  0.1    \\     
$\star$ G164.75-24.1 &      0.21   $\pm$  0.02   &      ...                  &      0.2    $\pm$  0.1    \\     
$\star$ G168.72-15.4 &      0.14   $\pm$  0.04   &      ...                  &      0.6    $\pm$  0.2    \\     
$\star$ G169.98-18.9 &      0.1    $\pm$  0.03   &      ...                  &      0.2    $\pm$  0.1    \\     
$\star$ G171.51-10.5 &      0.07   $\pm$  0.02   &      ...                  &      0.2    $\pm$  0.1    \\     
$\star$ G202.30-08.9 &      0.13   $\pm$  0.03   &      ...                  &      0.2    $\pm$  0.1    \\     
\end{longtable}
\vspace{-1em}
\hspace{-0em}
\begin{minipage}{\linewidth}
{\bf Note.}  Column 1: source name, the sources marked with “$\star$” before source name were observed with the high velocity resolution mode (AROWS mode 13) and the rest were observed with the low velocity resolution mode (AROWS mode 3); Column 2: velocity integrated intensity; Column 3: local standard of rest velocity; Column 4: full width at half maximum. The $V_{\rm LSR}$ values are not provided here, as described in the notes for Table 5.
\end{minipage}

\clearpage
\setcounter{footnote}{0}
\begin{longtable}{crrrrrr}

	\caption{$D_{\rm frac}$(HCO$^+$) and $D_{\rm frac}$(HCN).}
	\label{df}\\
	\hline 
	\hline
	Source & $W$(DCO$^+$) & $W$(DCN) & $W$(H$^{13}$CO$^+$) & $W$(H$^{13}$CN) & $D_{\rm frac}$(HCO$^+$) & $D_{\rm frac}$(HCN)\\
	Name     & (K·km\,s$^{-1}$) & (K·km\,s$^{-1}$) & (K·km\,s$^{-1}$) & (K·km\,s$^{-1}$)   & $\%$  & $\%$   \\
\hline

\endfirsthead
\caption{continue.}\\
	         
	\hline 
	\hline
	Source & $W$(DCO$^+$) & $W$(DCN) & $W$(H$^{13}$CO$^+$) & $W$(H$^{13}$CN) & $D_{\rm frac}$(HCO$^+$) & $D_{\rm frac}$(HCN)\\
	Name     & (K·km\,s$^{-1}$) & (K·km\,s$^{-1}$) & (K·km\,s$^{-1}$) & (K·km\,s$^{-1}$)   & $\%$  & $\%$   \\
	         \hline
  
    \endhead
\hline
\endfoot
G001.38+20.9   &       1.44  $\pm$  0.04   &       0.23  $\pm$  0.05   &       0.76  $\pm$  0.02   &       0.08  $\pm$  0.02   &       3.7     $\pm$ 0.1      &       5.5     $\pm$ 1.8     \\
G001.84+16.5   &       1.35  $\pm$  0.04   &       0.13  $\pm$  0.03   &       0.64  $\pm$  0.02   &  $\le$0.04                &       4.1     $\pm$ 0.2      &  $\ge$6.2                   \\
G003.73+16.3   &       0.22  $\pm$  0.03   &  $\le$0.06                &       0.2  $\pm$  0.02    &  $\le$0.04                &       2.1     $\pm$ 0.4      &                             \\
G003.73+18.3   &       0.23  $\pm$  0.04   &  $\le$0.07                &       0.28  $\pm$  0.02   &  $\le$0.04                &       1.6     $\pm$ 0.3      &                             \\
G004.46+16.6   &       0.16  $\pm$  0.02   &  $\le$0.05                &       0.14  $\pm$  0.02   &       0.09  $\pm$  0.03   &       2.2     $\pm$ 0.4      &  $\le$1.1                   \\
G006.04+36.7   &       1.03  $\pm$  0.05   &  $\le$0.11                &       0.4  $\pm$  0.02    &       0.06  $\pm$  0.02   &       5.0     $\pm$ 0.3      &  $\le$3.5                   \\
G006.32+20.4   &       0.77  $\pm$  0.03   &  $\le$0.06                &       0.33  $\pm$  0.02   &  $\le$0.04                &       4.6     $\pm$ 0.3      &                             \\
G006.41+20.5   &       0.48  $\pm$  0.03   &  $\le$0.07                &       0.28  $\pm$  0.03   &  $\le$0.05                &       3.3     $\pm$ 0.4      &                             \\
G006.70+20.6   &       0.34  $\pm$  0.05   &  $\le$0.1                 &       0.39  $\pm$  0.03   &       0.06  $\pm$  0.02   &       1.7     $\pm$ 0.3      &  $\le$3.2                   \\
G006.94+05.8   &       0.23  $\pm$  0.05   &  $\le$0.11                &       0.16  $\pm$  0.03   &  $\le$0.07                &       2.8     $\pm$ 0.8      &                             \\
G006.98+20.7   &       0.14  $\pm$  0.03   &  $\le$0.07                &                           &                           &                              &                             \\
G007.14+05.9   &       0.4  $\pm$  0.06    &  $\le$0.12                &       0.2  $\pm$  0.03    &  $\le$0.06                &       3.9     $\pm$ 0.8      &                             \\
G008.52+21.8   &       0.48  $\pm$  0.03   &  $\le$0.07                &       0.36  $\pm$  0.02   &  $\le$0.04                &       2.6     $\pm$ 0.2      &                             \\
G008.67+22.1   &       0.22  $\pm$  0.04   &  $\le$0.07                &       0.11  $\pm$  0.02   &       0.06  $\pm$  0.02   &       3.9     $\pm$ 1.0      &  $\le$2.2                   \\
G021.20+04.9   &       0.39  $\pm$  0.05   &  $\le$0.11                &       0.24  $\pm$  0.02   &       0.07  $\pm$  0.02   &       3.2     $\pm$ 0.5      &  $\le$3.0                   \\
G021.66+03.7   &       1.14  $\pm$  0.06   &       0.21  $\pm$  0.06   &       0.61  $\pm$  0.03   &  $\le$0.05                &       3.6     $\pm$ 0.3      &  $\ge$8.1                   \\
G025.48+06.1   &       0.57  $\pm$  0.05   &  $\le$0.09                &       0.46  $\pm$  0.02   &       0.1  $\pm$  0.02    &       2.4     $\pm$ 0.2      &  $\le$1.7                   \\
G026.85+06.7   &       0.45  $\pm$  0.04   &       0.15  $\pm$  0.04   &       0.34  $\pm$  0.02   &       0.07  $\pm$  0.02   &       2.6     $\pm$ 0.3      &       4.1     $\pm$ 1.6     \\
G027.66+05.7   &       0.33  $\pm$  0.04   &  $\le$0.08                &       0.2  $\pm$  0.02    &  $\le$0.04                &       3.2     $\pm$ 0.5      &                             \\
G028.45-06.3   &       0.43  $\pm$  0.04   &       0.1  $\pm$  0.03    &       0.57  $\pm$  0.02   &       0.07  $\pm$  0.02   &       1.5     $\pm$ 0.1      &       2.7     $\pm$ 1.1     \\
G028.71+03.8   &       0.96  $\pm$  0.05   &       0.19  $\pm$  0.03   &       0.53  $\pm$  0.03   &       0.15  $\pm$  0.02   &       3.5     $\pm$ 0.3      &       2.4     $\pm$ 0.5     \\
G028.87+04.2   &       0.16  $\pm$  0.03   &  $\le$0.04                &       0.25  $\pm$  0.03   &  $\le$0.08                &       1.2     $\pm$ 0.3      &                             \\
G030.43+02.3   &       0.37  $\pm$  0.04   &  $\le$0.08                &       0.32  $\pm$  0.03   &       0.05  $\pm$  0.02   &       2.3     $\pm$ 0.3      &  $\le$3.1                   \\
G030.78+05.2   &       1.1  $\pm$  0.05    &  $\le$0.11                &       0.93  $\pm$  0.03   &       0.28  $\pm$  0.04   &       2.3     $\pm$ 0.1      &  $\le$0.8                   \\
G031.44+04.1   &       0.42  $\pm$  0.04   &  $\le$0.09                &       0.33  $\pm$  0.02   &       0.05  $\pm$  0.02   &       2.5     $\pm$ 0.3      &  $\le$3.5                   \\
G032.93+02.6   &       0.78  $\pm$  0.04   &  $\le$0.09                &       0.51  $\pm$  0.03   &       0.07  $\pm$  0.03   &       3.0     $\pm$ 0.2      &  $\le$2.5                   \\
G038.36-00.9   &       0.67  $\pm$  0.06   &  $\le$0.16                &       0.81  $\pm$  0.04   &       0.16  $\pm$  0.03   &       1.6     $\pm$ 0.2      &  $\le$1.9                   \\
G043.02+08.3   &       0.21  $\pm$  0.03   &  $\le$0.06                &       0.1  $\pm$  0.02    &  $\le$0.03                &       4.1     $\pm$ 1.0      &                             \\
G048.40-05.8   &       0.25  $\pm$  0.03   &  $\le$0.06                &       0.14  $\pm$  0.02   &  $\le$0.03                &       3.5     $\pm$ 0.6      &                             \\
G057.10+03.6   &       0.49  $\pm$  0.05   &  $\le$0.09                &       0.41  $\pm$  0.03   &  $\le$0.07                &       2.3     $\pm$ 0.3      &                             \\
G058.16+03.5   &       0.21  $\pm$  0.03   &  $\le$0.06                &       0.25  $\pm$  0.02   &       0.05  $\pm$  0.02   &       1.6     $\pm$ 0.3      &  $\le$2.3                   \\
G060.75-01.2   &       0.25  $\pm$  0.04   &  $\le$0.09                &       0.3  $\pm$  0.04    &  $\le$0.06                &       1.6     $\pm$ 0.3      &                             \\
G065.43-03.1   &       0.2  $\pm$  0.06    &  $\le$0.14                &                           &                           &                              &                             \\
G070.44-01.5   &       0.43  $\pm$  0.07   &  $\le$0.12                &       0.36  $\pm$  0.04   &  $\le$0.08                &       2.3     $\pm$ 0.5      &                             \\
G074.11+00.1   &       0.74  $\pm$  0.06   &  $\le$0.16                &       0.6  $\pm$  0.06    &  $\le$0.12                &       2.4     $\pm$ 0.3      &                             \\
G084.79-01.1   &       2.37  $\pm$  0.19   &       0.44  $\pm$  0.08   &       1.11  $\pm$  0.06   &       0.24  $\pm$  0.04   &       4.2     $\pm$ 0.4      &       3.5     $\pm$ 0.9     \\
G089.36-00.6   &       0.3  $\pm$  0.04    &  $\le$0.08                &       0.26  $\pm$  0.04   &  $\le$0.07                &       2.2     $\pm$ 0.5      &                             \\
G089.62+02.1   &       0.58  $\pm$  0.04   &       0.14  $\pm$  0.04   &       0.29  $\pm$  0.03   &  $\le$0.06                &       3.9     $\pm$ 0.5      &  $\ge$4.5                   \\
G089.93-07.0   &       0.18  $\pm$  0.03   &  $\le$0.07                &       0.11  $\pm$  0.02   &  $\le$0.04                &       3.2     $\pm$ 0.8      &                             \\
G091.29-38.1   &  $\le$0.09                &       0.17  $\pm$  0.05   &                           &                           &                              &                             \\
G091.73+04.3   &       1.17  $\pm$  0.05   &       0.25  $\pm$  0.05   &       0.58  $\pm$  0.04   &  $\le$0.07                &       3.9     $\pm$ 0.3      &  $\ge$6.9                   \\
G092.02+03.9   &       0.26  $\pm$  0.04   &  $\le$0.09                &       0.2  $\pm$  0.03    &  $\le$0.06                &       2.5     $\pm$ 0.5      &                             \\
G092.26+03.8   &       0.53  $\pm$  0.04   &  $\le$0.11                &       0.22  $\pm$  0.03   &  $\le$0.07                &       4.7     $\pm$ 0.7      &                             \\
G093.20+09.5   &       0.18  $\pm$  0.04   &  $\le$0.07                &                           &                           &                              &                             \\
G093.62-04.4   &       0.53  $\pm$  0.05   &  $\le$0.09                &       0.19  $\pm$  0.03   &  $\le$0.07                &       5.4     $\pm$ 1.0      &                             \\
G093.99-04.9   &       0.23  $\pm$  0.05   &  $\le$0.1                 &       0.24  $\pm$  0.05   &  $\le$0.12                &       1.9     $\pm$ 0.6      &                             \\
G094.08+09.4   &       0.36  $\pm$  0.05   &  $\le$0.1                 &       0.19  $\pm$  0.02   &  $\le$0.05                &       3.7     $\pm$ 0.6      &                             \\
G095.51+09.9   &       0.12  $\pm$  0.02   &  $\le$0.05                &       0.15  $\pm$  0.03   &  $\le$0.05                &       1.6     $\pm$ 0.4      &                             \\
G096.94+10.2   &       0.36  $\pm$  0.04   &  $\le$0.07                &       0.18  $\pm$  0.02   &  $\le$0.05                &       3.9     $\pm$ 0.6      &                             \\
G097.09+10.1   &       0.31  $\pm$  0.04   &  $\le$0.09                &       0.16  $\pm$  0.03   &  $\le$0.05                &       3.8     $\pm$ 0.9      &                             \\
G097.20+09.8   &       0.54  $\pm$  0.05   &  $\le$0.08                &       0.23  $\pm$  0.03   &  $\le$0.06                &       4.6     $\pm$ 0.7      &                             \\
G097.38+09.9   &       0.2  $\pm$  0.05    &  $\le$0.1                 &       0.15  $\pm$  0.03   &  $\le$0.06                &       2.6     $\pm$ 0.8      &                             \\
G102.34+15.9   &       0.63  $\pm$  0.04   &  $\le$0.08                &       0.35  $\pm$  0.02   &  $\le$0.05                &       3.5     $\pm$ 0.3      &                             \\
G102.72+15.3   &       0.44  $\pm$  0.04   &  $\le$0.08                &       0.24  $\pm$  0.02   &       0.05  $\pm$  0.02   &       3.6     $\pm$ 0.4      &  $\le$3.1                   \\
G103.90+13.9   &       0.3  $\pm$  0.04    &  $\le$0.08                &       0.41  $\pm$  0.02   &       0.15  $\pm$  0.03   &       1.4     $\pm$ 0.2      &  $\le$1.0                   \\
G108.10+13.1   &       0.36  $\pm$  0.04   &  $\le$0.08                &       0.22  $\pm$  0.03   &  $\le$0.05                &       3.2     $\pm$ 0.6      &                             \\
G108.85-00.8   &       0.41  $\pm$  0.05   &  $\le$0.1                 &       0.24  $\pm$  0.04   &  $\le$0.09                &       3.3     $\pm$ 0.7      &                             \\
G110.65+09.6   &       0.78  $\pm$  0.05   &  $\le$0.1                 &       0.5  $\pm$  0.05    &  $\le$0.09                &       3.0     $\pm$ 0.4      &                             \\
G111.66+20.2   &       0.91  $\pm$  0.05   &  $\le$0.13                &       0.43  $\pm$  0.03   &       0.11  $\pm$  0.03   &       4.1     $\pm$ 0.4      &  $\le$2.3                   \\
G113.75+14.9   &       0.1  $\pm$  0.03    &  $\le$0.08                &                           &                           &                              &                             \\
G114.16+14.8   &       0.35  $\pm$  0.04   &  $\le$0.1                 &       0.24  $\pm$  0.03   &  $\le$0.05                &       2.8     $\pm$ 0.5      &                             \\
G114.56+14.7   &       0.11  $\pm$  0.03   &  $\le$0.07                &       0.11  $\pm$  0.04   &  $\le$0.08                &       1.9     $\pm$ 0.9      &                             \\
G114.67+14.4   &       2.12  $\pm$  0.06   &       0.28  $\pm$  0.06   &       0.88  $\pm$  0.04   &  $\le$0.08                &       4.7     $\pm$ 0.3      &  $\ge$6.7                   \\
G115.81-03.5   &       0.35  $\pm$  0.04   &  $\le$0.09                &       0.1  $\pm$  0.03    &  $\le$0.05                &       6.8     $\pm$ 2.2      &                             \\
G116.23+20.3   &  $\le$0.08                &       0.11  $\pm$  0.02   &                           &                           &                              &                             \\
G117.11+12.4   &       0.63  $\pm$  0.04   &  $\le$0.11                &       0.37  $\pm$  0.03   &  $\le$0.06                &       3.3     $\pm$ 0.3      &                             \\
G120.16+03.0   &       0.67  $\pm$  0.06   &  $\le$0.15                &       0.54  $\pm$  0.07   &  $\le$0.13                &       2.4     $\pm$ 0.4      &                             \\
G120.67+02.6   &       0.38  $\pm$  0.06   &  $\le$0.12                &       0.31  $\pm$  0.05   &  $\le$0.1                 &       2.4     $\pm$ 0.5      &                             \\
G120.98+02.6   &       0.17  $\pm$  0.05   &  $\le$0.1                 &                           &                           &                              &                             \\
G121.35+03.3   &       0.29  $\pm$  0.04   &  $\le$0.07                &       0.15  $\pm$  0.04   &       0.07  $\pm$  0.02   &       3.8     $\pm$ 1.1      &  $\le$1.9                   \\
G126.95-01.0   &       0.21  $\pm$  0.04   &  $\le$0.09                &                           &                           &                              &                             \\
G127.66+13.9   &       0.34  $\pm$  0.06   &  $\le$0.12                &                           &                           &                              &                             \\
G127.85+14.1   &       0.35  $\pm$  0.06   &  $\le$0.1                 &                           &                           &                              &                             \\
G127.88+02.6   &       0.26  $\pm$  0.07   &  $\le$0.12                &                           &                           &                              &                             \\
G128.89+13.6   &       0.8  $\pm$  0.04    &  $\le$0.08                &                           &                           &                              &                             \\
G130.36+11.2   &       0.23  $\pm$  0.06   &  $\le$0.16                &       0.16  $\pm$  0.04   &  $\le$0.08                &       2.8     $\pm$ 1.0      &                             \\
G132.03+08.9   &       0.21  $\pm$  0.04   &  $\le$0.08                &       0.1  $\pm$  0.03    &  $\le$0.07                &       4.1     $\pm$ 1.5      &                             \\
G133.48+09.0   &       0.74  $\pm$  0.13   &  $\le$0.21                &                           &                           &                              &                             \\
G136.31-01.7   &       0.27  $\pm$  0.05   &  $\le$0.12                &                           &                           &                              &                             \\
G142.29+07.6   &  $\le$0.06                &  $\le$0.05                &  $\le$0.04                &       0.09  $\pm$  0.02   &                              &  $\le$1.1                   \\
G144.66+00.1   &       0.32  $\pm$  0.05   &  $\le$0.1                 &       0.18  $\pm$  0.04   &  $\le$0.09                &       3.5     $\pm$ 0.9      &                             \\
G147.01+03.3   &       0.44  $\pm$  0.05   &  $\le$0.09                &                           &                           &                              &                             \\
G150.44+03.9   &       0.38  $\pm$  0.05   &  $\le$0.1                 &       0.1  $\pm$  0.03    &  $\le$0.06                &       7.4     $\pm$ 2.4      &                             \\
G151.08+04.4   &       0.29  $\pm$  0.05   &  $\le$0.09                &       0.19  $\pm$  0.03   &  $\le$0.06                &       3.0     $\pm$ 0.7      &                             \\
G156.20+05.2   &       0.27  $\pm$  0.05   &  $\le$0.1                 &       0.22  $\pm$  0.03   &  $\le$0.07                &       2.4     $\pm$ 0.6      &                             \\
G157.54-04.0   &       0.12  $\pm$  0.03   &       0.12  $\pm$  0.03   &       0.07  $\pm$  0.02   &       0.11  $\pm$  0.03   &       3.3     $\pm$ 1.3      &       2.1     $\pm$ 0.8     \\
G158.20-20.2   &       0.69  $\pm$  0.06   &  $\le$0.09                &       0.33  $\pm$  0.03   &  $\le$0.07                &       4.1     $\pm$ 0.5      &                             \\
G158.22-20.1   &       0.13  $\pm$  0.03   &  $\le$0.05                &       0.13  $\pm$  0.02   &  $\le$0.05                &       1.9     $\pm$ 0.5      &                             \\
G158.37-20.7   &       0.68  $\pm$  0.08   &  $\le$0.15                &       0.31  $\pm$  0.04   &  $\le$0.08                &       4.3     $\pm$ 0.7      &                             \\
G159.01-08.4   &       0.17  $\pm$  0.03   &  $\le$0.06                &       0.11  $\pm$  0.03   &  $\le$0.07                &       3.0     $\pm$ 1.0      &                             \\
G159.21-20.4   &       0.18  $\pm$  0.04   &  $\le$0.07                &       0.18  $\pm$  0.03   &       0.13  $\pm$  0.03   &       1.9     $\pm$ 0.5      &  $\le$1.0                   \\
G159.21-20.1   &       2.2  $\pm$  0.05    &       0.34  $\pm$  0.04   &       1.0  $\pm$  0.04    &  $\le$0.09                &       4.3     $\pm$ 0.2      &  $\ge$7.3                   \\
G159.23-34.4   &  $\le$0.07                &  $\le$0.07                &       0.28  $\pm$  0.03   &  $\le$0.06                &  $\le$0.5                    &                             \\
G159.34+11.2   &  $\le$0.06                &       0.13  $\pm$  0.03   &       0.19  $\pm$  0.02   &       0.16  $\pm$  0.02   &  $\le$0.6                    &       1.6     $\pm$ 0.4     \\
G159.65-19.6   &       0.3  $\pm$  0.06    &  $\le$0.08                &       0.12  $\pm$  0.02   &  $\le$0.06                &       4.9     $\pm$ 1.3      &                             \\
G160.53-09.8   &       0.19  $\pm$  0.03   &  $\le$0.06                &       0.14  $\pm$  0.04   &  $\le$0.06                &       2.6     $\pm$ 0.9      &                             \\
G160.64-35.0   &       0.06  $\pm$  0.02   &       0.14  $\pm$  0.03   &       0.13  $\pm$  0.02   &       0.13  $\pm$  0.02   &       0.9     $\pm$ 0.3      &       2.1     $\pm$ 0.5     \\
G161.85-35.7   &       0.19  $\pm$  0.05   &       0.13  $\pm$  0.03   &       0.2  $\pm$  0.03    &       0.17  $\pm$  0.02   &       1.9     $\pm$ 0.6      &       1.5     $\pm$ 0.4     \\
G161.85-08.6   &       0.21  $\pm$  0.05   &  $\le$0.09                &  $\le$0.07                &  $\le$0.07                &  $\ge$5.8                    &                             \\
G162.64-31.6   &  $\le$0.06                &       0.12  $\pm$  0.03   &       0.08  $\pm$  0.02   &       0.1  $\pm$  0.03    &  $\le$1.5                    &       2.3     $\pm$ 0.9     \\
G163.32-08.4   &       0.1  $\pm$  0.03    &  $\le$0.05                &  $\le$0.04                &  $\le$0.04                &  $\ge$4.9                    &                             \\
G163.67-08.3   &       0.17  $\pm$  0.04   &  $\le$0.07                &       0.16  $\pm$  0.03   &  $\le$0.07                &       2.1     $\pm$ 0.6      &                             \\
G164.75-24.1   &       0.1  $\pm$  0.02    &       0.2  $\pm$  0.03    &       0.22  $\pm$  0.02   &       0.21  $\pm$  0.02   &       0.9     $\pm$ 0.2      &       1.8     $\pm$ 0.3     \\
G164.94-08.5   &       1.01  $\pm$  0.07   &  $\le$0.12                &       0.51  $\pm$  0.04   &  $\le$0.09                &       3.9     $\pm$ 0.4      &                             \\
G165.16-07.5   &  $\le$0.09                &  $\le$0.09                &       0.22  $\pm$  0.03   &  $\le$0.09                &  $\le$0.8                    &                             \\
G165.69-09.1   &       0.27  $\pm$  0.06   &  $\le$0.09                &  $\le$0.09                &  $\le$0.09                &  $\ge$5.9                    &                             \\
G167.23-15.3   &       0.29  $\pm$  0.05   &  $\le$0.11                &       0.22  $\pm$  0.04   &  $\le$0.09                &       2.6     $\pm$ 0.6      &                             \\
G168.00-15.6   &  $\le$0.1                 &  $\le$0.09                &       0.42  $\pm$  0.03   &  $\le$0.08                &  $\le$0.5                    &                             \\
G168.13-16.3   &       0.47  $\pm$  0.05   &  $\le$0.09                &       0.22  $\pm$  0.03   &  $\le$0.06                &       4.2     $\pm$ 0.7      &                             \\
G168.72-15.4   &       0.73  $\pm$  0.06   &  $\le$0.11                &       0.46  $\pm$  0.03   &       0.14  $\pm$  0.04   &       3.1     $\pm$ 0.3      &  $\le$1.5                   \\
G168.85-15.8   &       0.15  $\pm$  0.04   &  $\le$0.08                &  $\le$0.06                &  $\le$0.06                &  $\ge$4.9                    &                             \\
G169.32-16.1   &       0.76  $\pm$  0.06   &  $\le$0.11                &       0.29  $\pm$  0.03   &  $\le$0.08                &       5.1     $\pm$ 0.7      &                             \\
G169.43-16.1   &       0.15  $\pm$  0.04   &  $\le$0.06                &       0.19  $\pm$  0.04   &  $\le$0.09                &       1.5     $\pm$ 0.5      &                             \\
G169.76-16.1   &       0.54  $\pm$  0.06   &  $\le$0.09                &       0.26  $\pm$  0.03   &  $\le$0.06                &       4.0     $\pm$ 0.6      &                             \\
G169.98-18.9   &       0.22  $\pm$  0.04   &  $\le$0.09                &       0.31  $\pm$  0.03   &       0.1  $\pm$  0.03    &       1.4     $\pm$ 0.3      &  $\le$1.7                   \\
G170.00-16.1   &       0.31  $\pm$  0.04   &  $\le$0.08                &       0.14  $\pm$  0.03   &  $\le$0.06                &       4.3     $\pm$ 1.1      &                             \\
G170.13-16.0   &  $\le$0.14                &  $\le$0.11                &       0.42  $\pm$  0.04   &  $\le$0.09                &  $\le$0.7                    &                             \\
G170.26-16.0   &       0.48  $\pm$  0.06   &  $\le$0.09                &       0.38  $\pm$  0.04   &  $\le$0.08                &       2.5     $\pm$ 0.4      &                             \\
G170.99-15.8   &  $\le$0.11                &  $\le$0.1                 &       0.31  $\pm$  0.04   &  $\le$0.08                &  $\le$0.7                    &                             \\
G171.14-17.5   &       0.21  $\pm$  0.05   &  $\le$0.09                &       0.18  $\pm$  0.03   &  $\le$0.06                &       2.3     $\pm$ 0.7      &                             \\
G171.34-10.6   &       0.16  $\pm$  0.05   &  $\le$0.08                &       0.13  $\pm$  0.03   &  $\le$0.06                &       2.4     $\pm$ 0.9      &                             \\
G171.51-10.5   &       0.23  $\pm$  0.04   &  $\le$0.08                &       0.18  $\pm$  0.02   &       0.07  $\pm$  0.02   &       2.5     $\pm$ 0.5      &  $\le$2.2                   \\
G171.84-05.2   &       0.47  $\pm$  0.05   &  $\le$0.1                 &       0.24  $\pm$  0.03   &  $\le$0.06                &       3.8     $\pm$ 0.6      &                             \\
G177.14-01.2   &       0.32  $\pm$  0.06   &  $\le$0.11                &  $\le$0.08                &  $\le$0.08                &  $\ge$7.8                    &                             \\
G177.86+01.0   &       0.17  $\pm$  0.05   &  $\le$0.09                &       0.14  $\pm$  0.03   &  $\le$0.08                &       2.4     $\pm$ 0.9      &                             \\
G178.48-06.7   &       0.12  $\pm$  0.03   &  $\le$0.06                &  $\le$0.05                &  $\le$0.05                &  $\ge$4.7                    &                             \\
G178.98-06.7   &       0.45  $\pm$  0.05   &  $\le$0.06                &       0.31  $\pm$  0.02   &  $\le$0.05                &       2.8     $\pm$ 0.4      &                             \\
G181.84+00.3   &       0.44  $\pm$  0.06   &  $\le$0.11                &  $\le$0.08                &  $\le$0.08                &  $\ge$10.7                   &                             \\
G185.33-02.1   &       0.16  $\pm$  0.05   &  $\le$0.1                 &  $\le$0.08                &  $\le$0.09                &  $\ge$3.9                    &                             \\
G191.00-04.5   &  $\le$0.09                &  $\le$0.07                &       0.1  $\pm$  0.02    &  $\le$0.07                &  $\le$1.8                    &                             \\
G192.32-11.8   &       0.77  $\pm$  0.06   &  $\le$0.13                &       0.49  $\pm$  0.04   &  $\le$0.09                &       3.1     $\pm$ 0.3      &                             \\
G195.00-16.9   &       0.15  $\pm$  0.04   &  $\le$0.08                &  $\le$0.06                &  $\le$0.06                &  $\ge$4.9                    &                             \\
G195.09-16.4   &  $\le$0.05                &  $\le$0.05                &       0.06  $\pm$  0.02   &  $\le$0.04                &  $\le$1.6                    &                             \\
G198.56-09.1   &       0.44  $\pm$  0.04   &       0.1  $\pm$  0.02    &       0.19  $\pm$  0.02   &  $\le$0.06                &       4.5     $\pm$ 0.6      &  $\ge$3.2                   \\
G199.88+00.9   &       0.47  $\pm$  0.06   &  $\le$0.13                &  $\le$0.1                 &  $\le$0.1                 &  $\ge$9.2                    &                             \\
G200.34-10.9   &       0.12  $\pm$  0.03   &  $\le$0.05                &  $\le$0.04                &  $\le$0.04                &  $\ge$5.8                    &                             \\
G201.13+00.3   &       0.39  $\pm$  0.07   &  $\le$0.13                &  $\le$0.09                &  $\le$0.09                &  $\ge$8.5                    &                             \\
G201.44+00.6   &       0.49  $\pm$  0.05   &  $\le$0.1                 &       0.19  $\pm$  0.05   &  $\le$0.06                &       5.0     $\pm$ 1.4      &                             \\
G202.30-08.9   &       0.26  $\pm$  0.05   &  $\le$0.08                &       0.14  $\pm$  0.02   &       0.13  $\pm$  0.03   &       3.6     $\pm$ 0.9      &  $\le$1.2                   \\
G203.20-11.2   &       1.06  $\pm$  0.07   &  $\le$0.14                &       0.6  $\pm$  0.05    &  $\le$0.11                &       3.4     $\pm$ 0.4      &                             \\
G204.49-11.3   &       0.2  $\pm$  0.04    &  $\le$0.06                &       0.11  $\pm$  0.02   &  $\le$0.05                &       3.5     $\pm$ 1.0      &                             \\
G206.10-15.7\footnote{G206.10-15.7 contains 2 velocity components}   &       0.27  $\pm$  0.06   &  $\le$0.08                &       0.13  $\pm$  0.03   &  $\le$0.06                &       4.0     $\pm$ 1.3      &                             \\
   &       0.37  $\pm$  0.05   &  $\le$0.1                 &       0.22  $\pm$  0.03   &  $\le$0.07                &       3.3     $\pm$ 0.6      &                             \\
G207.35-19.8   &       0.38  $\pm$  0.06   &  $\le$0.11                &       0.13  $\pm$  0.03   &  $\le$0.06                &       5.7     $\pm$ 1.6      &                             \\
G209.28-19.6   &       0.62  $\pm$  0.08   &  $\le$0.0                 &       0.99  $\pm$  0.06   &  $\le$0.0                 &       1.2     $\pm$ 0.2      &                             \\
G211.48-19.2   &       0.62  $\pm$  0.08   &  $\le$0.11                &       0.31  $\pm$  0.04   &  $\le$0.1                 &       3.9     $\pm$ 0.7      &                             \\
G212.10-19.1   &       0.59  $\pm$  0.09   &  $\le$0.14                &       0.17  $\pm$  0.05   &  $\le$0.08                &       6.8     $\pm$ 2.2      &                             \\
G213.96-19.6   &       0.27  $\pm$  0.05   &  $\le$0.09                &       0.24  $\pm$  0.04   &  $\le$0.1                 &       2.2     $\pm$ 0.5      &                             \\
G214.43-19.9   &       0.2  $\pm$  0.05    &  $\le$0.09                &       0.19  $\pm$  0.04   &  $\le$0.08                &       2.1     $\pm$ 0.7      &                             \\
G215.00-15.1   &  $\le$0.08                &  $\le$0.07                &       0.08  $\pm$  0.02   &  $\le$0.06                &  $\le$1.9                    &                             \\
G216.18-15.2   &  $\le$0.13                &  $\le$0.12                &       0.15  $\pm$  0.04   &  $\le$0.08                &  $\le$1.7                    &                             \\
G359.31+17.0   &       0.47  $\pm$  0.04   &  $\le$0.06                &       0.34  $\pm$  0.02   &  $\le$0.05                &       2.7     $\pm$ 0.3      &                             \\
\end{longtable}	
\vspace{-1em}
\hspace{-0em}
\begin{minipage}{\linewidth}
{\bf Note.}  Column 1: source name; Column 2$\sim$5: velocity integrated intensity or The 3$\sigma$ upper limit value of DCO$^{+}$, DCN, H$^{13}$CN and H$^{13}$CO$^{+}$ 1-0; Column 6$\sim$7: deuterated fraction of HCO$^+$ and HCN.
\end{minipage}
\end{center}

\clearpage
\appendix

\section{High spectral resolution supplementary observations with AROWS mode 5 }

As described in Section \ref{sec:observations}, high spectral resolution supplementary observations with AROWS mode 5 were conducted for DCO$^+$ 1-0 narrow lines toward 32 sources. The fluxes of DCO$^+$ 1-0 observed with the high velocity resolution mode (AROWS mode 5) are similar to those observed with the low velocity resolution mode (AROWS mode 3) (see Table \ref{hr_mode5}). 

The FWHMs of  DCO$^+$ 1-0 using the high velocity resolution mode (AROWS mode 5) are compared with those using the low velocity resolution mode (AROWS mode 3) in Figure \ref{FWHMMode5vsMode3}, showing similar values within the error bars. Therefore, even though the FWHMs from low spectral resolution observations may not be precise, they are still useful as a reference.
\begin{longtable}{crrrrr}
\caption{High spectral resolution DCO$^+$ 1-0  observed with the high velocity resolution mode (AROWS mode 5).}
	\label{hr_mode5}\\
	\hline 
	\hline
	Source &  $W$ & $V_{\rm LSR}$ & FWHM & $W$(DCO$^+$)~with~mode~3 & $\frac{W{\rm(DCO^+)~with~mode~5}}{W{\rm(DCO^+)~with~mode~3}}$ ratio\\

	Name     & (K·km\,s$^{-1}$)  & (km\,s$^{-1}$)   &   (km\,s$^{-1}$)     &  (K·km\,s$^{-1}$)   &     \\

\hline

\endfirsthead
\caption[]{continue.}\\
	         
	\hline 
	\hline
	Source &  $W$ & $V_{\rm LSR}$ & FWHM & $W$(DCO$^+$)~with~mode~3 & $\frac{W{\rm(DCO^+)~with~mode~5}}{W{\rm(DCO^+)~with~mode~3}}$ ratio\\

	Name     & (K·km\,s$^{-1}$)  & (km\,s$^{-1}$)   &   (km\,s$^{-1}$)     &  (K·km\,s$^{-1}$)   &     \\
	         \hline
  
    \endhead
    \hline
\endfoot
G001.38+20.9   &    1.57     $\pm$  0.04     &    0.7      $\pm$  0.1      &    0.6      $\pm$  0.1    &      1.44   $\pm$  0.04   &      1.09   $\pm$  0.04   \\     
G001.84+16.5   &    1.46     $\pm$  0.05     &    5.8      $\pm$  0.1      &    0.6      $\pm$  0.1    &      1.35   $\pm$  0.04   &      1.08   $\pm$  0.05   \\     
G003.73+16.3   &    0.25     $\pm$  0.04     &    6.0      $\pm$  0.1      &    0.4      $\pm$  0.1    &      0.22   $\pm$  0.03   &      1.14   $\pm$  0.21   \\     
G003.73+18.3   &    0.31     $\pm$  0.05     &    4.3      $\pm$  0.1      &    0.7      $\pm$  0.1    &      0.23   $\pm$  0.04   &      1.35   $\pm$  0.24   \\     
G004.46+16.6   &    0.27     $\pm$  0.04     &    5.6      $\pm$  0.1      &    0.5      $\pm$  0.1    &      0.16   $\pm$  0.02   &      1.69   $\pm$  0.19   \\     
G006.04+36.7   &    1.18     $\pm$  0.04     &    2.5      $\pm$  0.1      &    0.6      $\pm$  0.1    &      1.03   $\pm$  0.05   &      1.15   $\pm$  0.06   \\     
G006.32+20.4   &    0.77     $\pm$  0.04     &    4.2      $\pm$  0.1      &    0.5      $\pm$  0.1    &      0.77   $\pm$  0.03   &      1.0    $\pm$  0.06   \\     
G006.41+20.5   &    0.57     $\pm$  0.04     &    4.4      $\pm$  0.1      &    0.6      $\pm$  0.1    &      0.48   $\pm$  0.03   &      1.19   $\pm$  0.09   \\     
G006.70+20.6   &    0.36     $\pm$  0.04     &    3.5      $\pm$  0.1      &    0.6      $\pm$  0.1    &      0.34   $\pm$  0.05   &      1.06   $\pm$  0.18   \\     
G007.14+05.9   &    0.51     $\pm$  0.07     &    10.3     $\pm$  0.1      &    0.9      $\pm$  0.1    &      0.4    $\pm$  0.06   &      1.27   $\pm$  0.2    \\     
G008.52+21.8   &    0.52     $\pm$  0.03     &    3.8      $\pm$  0.1      &    0.4      $\pm$  0.1    &      0.48   $\pm$  0.03   &      1.08   $\pm$  0.09   \\     
G008.67+22.1   &    0.27     $\pm$  0.03     &    3.5      $\pm$  0.1      &    0.3      $\pm$  0.1    &      0.22   $\pm$  0.04   &      1.23   $\pm$  0.21   \\     
G021.20+04.9   &    0.53     $\pm$  0.04     &    3.5      $\pm$  0.1      &    0.5      $\pm$  0.1    &      0.39   $\pm$  0.05   &      1.36   $\pm$  0.15   \\     
G021.66+03.7   &    1.18     $\pm$  0.05     &    6.7      $\pm$  0.1      &    0.7      $\pm$  0.1    &      1.14   $\pm$  0.06   &      1.04   $\pm$  0.07   \\     
G025.48+06.1   &    0.55     $\pm$  0.04     &    7.7      $\pm$  0.1      &    0.6      $\pm$  0.1    &      0.57   $\pm$  0.05   &      0.96   $\pm$  0.11   \\     
G026.85+06.7   &    0.49     $\pm$  0.04     &    7.1      $\pm$  0.1      &    0.6      $\pm$  0.1    &      0.45   $\pm$  0.04   &      1.09   $\pm$  0.12   \\     
G027.66+05.7   &    0.3      $\pm$  0.04     &    8.2      $\pm$  0.1      &    0.6      $\pm$  0.1    &      0.33   $\pm$  0.04   &      0.91   $\pm$  0.18   \\     
G028.45-06.3   &    0.55     $\pm$  0.04     &    12.2     $\pm$  0.1      &    0.6      $\pm$  0.1    &      0.43   $\pm$  0.04   &      1.28   $\pm$  0.12   \\     
G028.71+03.8   &    1.16     $\pm$  0.05     &    7.6      $\pm$  0.1      &    0.9      $\pm$  0.1    &      0.96   $\pm$  0.05   &      1.21   $\pm$  0.07   \\     
G028.87+04.2   &    0.25     $\pm$  0.04     &    6.6      $\pm$  0.1      &    0.6      $\pm$  0.1    &      0.16   $\pm$  0.03   &      1.56   $\pm$  0.25   \\     
G030.43+02.3   &    0.37     $\pm$  0.05     &    9.1      $\pm$  0.1      &    0.9      $\pm$  0.1    &      0.37   $\pm$  0.04   &      1.0    $\pm$  0.17   \\     
G030.78+05.2   &    1.08     $\pm$  0.05     &    8.2      $\pm$  0.1      &    1.1      $\pm$  0.1    &      1.1    $\pm$  0.05   &      0.98   $\pm$  0.06   \\     
G031.44+04.1   &    0.41     $\pm$  0.04     &    11.0     $\pm$  0.1      &    0.7      $\pm$  0.1    &      0.42   $\pm$  0.04   &      0.98   $\pm$  0.14   \\     
G032.93+02.6   &    0.63     $\pm$  0.04     &    11.7     $\pm$  0.1      &    0.5      $\pm$  0.1    &      0.78   $\pm$  0.04   &      0.81   $\pm$  0.08   \\     
G038.36-00.9   &    0.63     $\pm$  0.05     &    16.6     $\pm$  0.1      &    1.2      $\pm$  0.1    &      0.67   $\pm$  0.06   &      0.94   $\pm$  0.12   \\     
G043.02+08.3   &    0.16     $\pm$  0.02     &    4.0      $\pm$  0.1      &    0.3      $\pm$  0.1    &      0.21   $\pm$  0.03   &      0.76   $\pm$  0.19   \\     
G048.40-05.8   &    0.27     $\pm$  0.03     &    9.6      $\pm$  0.1      &    0.5      $\pm$  0.1    &      0.25   $\pm$  0.03   &      1.08   $\pm$  0.16   \\     
G057.10+03.6   &    0.52     $\pm$  0.05     &    11.7     $\pm$  0.1      &    1.3      $\pm$  0.1    &      0.49   $\pm$  0.05   &      1.06   $\pm$  0.14   \\     
G058.16+03.5   &    0.4      $\pm$  0.04     &    9.9      $\pm$  0.1      &    0.6      $\pm$  0.1    &      0.21   $\pm$  0.03   &      1.9    $\pm$  0.17   \\     
G060.75-01.2   &    0.27     $\pm$  0.04     &    10.8     $\pm$  0.1      &    0.5      $\pm$  0.1    &      0.25   $\pm$  0.04   &      1.08   $\pm$  0.22   \\     
G065.43-03.1   &    0.13     $\pm$  0.03     &    6.2      $\pm$  0.1      &    0.3      $\pm$  0.1    &      0.2    $\pm$  0.06   &      0.65   $\pm$  0.38   \\     
G070.44-01.5   &    0.34     $\pm$  0.06     &    11.0     $\pm$  0.1      &    1.5      $\pm$  0.2    &      0.43   $\pm$  0.07   &      0.79   $\pm$  0.24   \\     
\end{longtable}
\vspace{-1em}
\hspace{-0em}
\begin{minipage}{\linewidth}
{\bf Note.}  Column 1: source name; Column 2: velocity integrated intensity of DCO$^+$ 1-0 with the high velocity resolution mode (AROWS mode 5); Column 3: local standard of rest velocity; Column 4: full width at half maximum; Column 5: velocity integrated intensity of DCO$^+$ 1-0 with the low velocity resolution mode (AROWS mode 3); Column 6: ratio of velocity integrated intensity of DCO$^+$ 1-0 with AROWS mode 5 to which with mode 3.
\end{minipage}

\setcounter{figure}{0}
\renewcommand{\thefigure}{A\arabic{figure}}
\begin{figure}[H]
\centering
\includegraphics[width=0.4\textwidth]{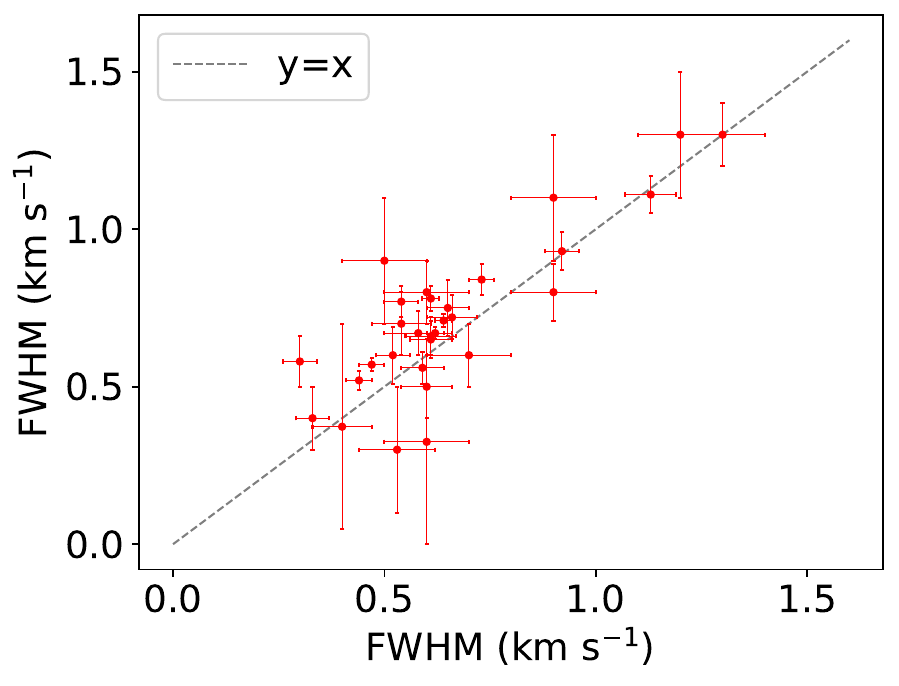}
\caption{Comparison between FWHM of DCO$^+$ with velocity resolution $\sim$0.081 km s$^{-1}$ with mode 5 and FWHM of DCO$^+$ with velocity resolution $\sim$0.32 km s$^{-1}$ with mode 3. There are 32 points in the plot. 
\label{FWHMMode5vsMode3}}
\end{figure}

\section{The relationships of the kinematic distances of sources and detection results}

To determine whether DCO$^+$ 1-0 detection is limited by distance, the distributions of kinematic distances for sources with and without DCO$^+$ 1-0 detections are shown in Figure \ref{hist_distance}. To reduce uncertainty in kinematic distances, sources located at galactic longitudes 0$^\circ \pm$15$^\circ$ and 90$^\circ \pm$15$^\circ$, as well as those with a galactic scale height greater than 1 kpc, were excluded. Among the observed sources, the kinematic distances for 378 sources that meet the coordinate requirements are from \cite{2012ApJ...756...76W}. The kinematic distances of these sources, regardless of DCO$^+$ 1-0 detection, show a similar distribution (see Figure \ref{hist_distance}). The kinematic distances of 82 sources with DCO$^+$ 1-0 detections range from 0.11 to 5.75 kpc, with a mean of 1.08 kpc and a median value of 0.81 kpc. For the 296 sources without DCO$^+$ 1-0 detections, kinematic distances range from 0.10 to 5.42 kpc, with a mean of 1.11 kpc and a median value of 1.01 kpc. The kinematic distances of sources with and without detections show similar distributions, means, and medians, which rules out the effect of beam dilution on the detection or non-detection status.

\setcounter{figure}{0}
\renewcommand{\thefigure}{B\arabic{figure}}
\begin{figure}
\centering
\includegraphics[width=0.4\textwidth]{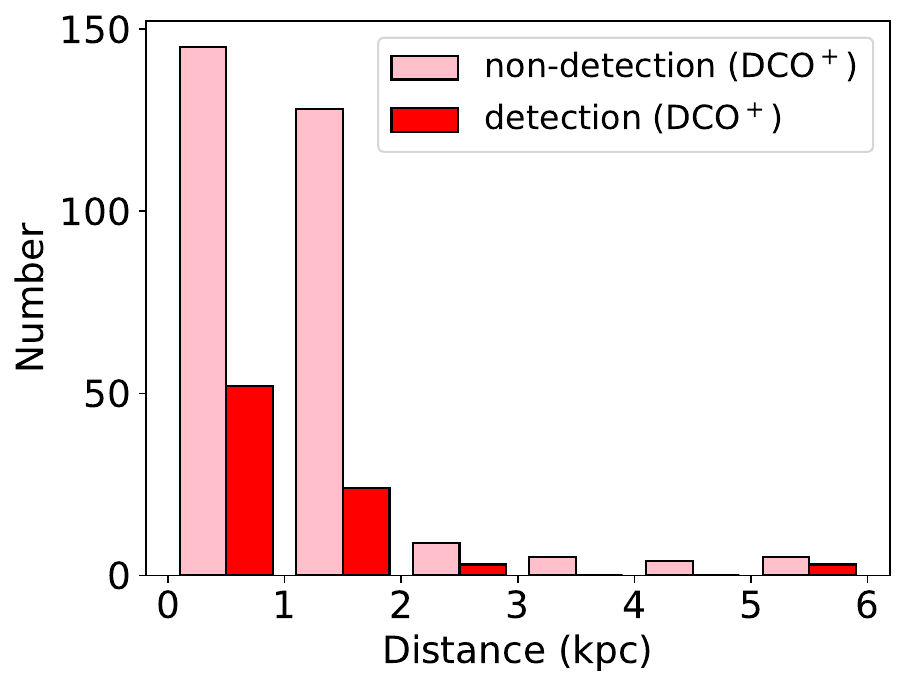}
\caption{Distribution of kinematic distance of sources where DCO$^+$ 1-0 is detected or not.
\label{hist_distance}}
\end{figure}

\section{The sources without any detection}

A total of 228 sources were observed for DCO$^+$ and DCN 1-0 using the low velocity resolution mode (AROWS mode 3), with no detections of either molecule. The details for these 228 sources, including name, R.A., and Decl., are provided in Table \ref{228sources}. Additionally, Table \ref{180sources} lists details for 180 sources observed with the high velocity resolution mode (AROWS mode 13) that showed no detection of DCO$^+$, DCN, H$^{13}$CO$^+$, or H$^{13}$CN 1-0.

\setcounter{table}{0}
\renewcommand{\thetable}{C\arabic{table}}
\begin{longtable}{crr|crr}

	\caption{228 sources without DCO$^+$ or DCN 1-0 detection with the low velocity resolution mode (AROWS mode 3).}
	\label{228sources}\\
	\hline 
	\hline
	Source & R.A. & Decl. & Source & R.A. & Decl.  \\
	Name     & (hh:mm:ss)  & (dd:mm:ss)  & Name     & (hh:mm:ss)  & (dd:mm:ss)\\
\hline

\endfirsthead
\caption[]{continue.}\\
	         
	\hline 
	\hline
	Source & R.A. & Decl. & Source & R.A. & Decl.  \\
	Name     & (hh:mm:ss)  & (dd:mm:ss)  & Name     & (hh:mm:ss)  & (dd:mm:ss)\\
	         \hline
  
    \endhead
\hline
\endfoot
G004.02+16.6 & 16:55:10.45	 & -16:21:23.27	 &
G004.15+35.7 & 15:53:29.82	 & -04:38:52.39	 \\
G004.17+36.6 & 15:50:42.66	 & -04:04:20.84	 &
G004.19+18.0 & 16:50:42.47	 & -15:22:25.13	 \\
G004.41+15.9 & 16:58:35.99	 & -16:28:36.07	 &
G004.54+36.7 & 15:51:14.27	 & -03:47:40.14	 \\
G004.81+37.0 & 15:50:52.79	 & -03:27:20.38	 &
G004.92+17.9 & 16:52:50.41	 & -14:53:40.51	 \\
G005.03+19.0 & 16:49:19.74	 & -14:09:20.80	 &
G005.29+11.0 & 17:17:19.97	 & -18:30:57.71	 \\
G005.29+14.4 & 17:05:31.09	 & -16:36:07.96	 &
G005.31+10.7 & 17:18:23.02	 & -18:39:22.30	 \\
G005.69+36.8 & 15:53:11.88	 & -03:00:56.50	 &
G005.80+19.9 & 16:48:13.56	 & -13:04:14.26	 \\
G006.08+20.2 & 16:47:44.35	 & -12:39:21.78	 &
G006.96+00.8 & 17:57:59.28	 & -22:29:20.42	 \\
G007.53+21.1 & 16:48:08.77	 & -11:03:53.56	 &
G007.80+21.1 & 16:48:43.20	 & -10:51:47.54	 \\
G008.43+36.3 & 16:00:02.55	 & -01:32:25.18	 &
G009.62+21.2 & 16:52:01.63	 & -09:21:33.98	 \\
G011.09+03.4 & 17:57:24.13	 & -17:38:42.83	 &
G011.40+36.1 & 16:06:03.58	 & +00:19:32.84	 \\
G013.86+04.5 & 17:59:03.17	 & -14:41:26.77	 &
G014.74+04.0 & 18:02:34.88	 & -14:09:59.84	 \\
G014.87+19.6 & 17:08:24.20	 & -06:07:18.26	 &
G015.16+07.2 & 17:52:06.35	 & -12:14:27.93	 \\
G016.67-02.7 & 18:31:16.07	 & -15:42:41.22	 &
G017.22-01.4 & 18:27:35.34	 & -14:37:44.09	 \\
G018.39+19.3 & 17:16:04.31	 & -03:24:13.42	 &
G018.61+08.8 & 17:53:10.33	 & -08:27:42.53	 \\
G019.86+20.4 & 17:15:20.84	 & -01:41:55.70	 &
G020.52+01.6 & 18:22:24.56	 & -10:14:01.42	 \\
G021.26+12.1 & 17:46:54.19	 & -04:35:51.36	 &
G023.33+11.8 & 17:51:52.37	 & -02:58:23.38	 \\
G023.37+08.2 & 18:04:38.79	 & -04:38:50.91	 &
G023.68+07.5 & 18:07:37.30	 & -04:41:59.16	 \\
G026.45+08.0 & 18:11:02.76	 & -02:02:51.60	 &
G026.93-20.6 & 19:56:07.08	 & -14:21:30.47	 \\
G027.09-20.7 & 19:56:26.08	 & -14:14:06.47	 &
G027.70-21.0 & 19:58:36.21	 & -13:50:25.40	 \\
G028.56-00.2 & 18:44:19.42	 & -03:59:49.18	 &
G031.26-05.3 & 19:07:35.28	 & -03:55:38.37	 \\
G032.38-15.2 & 19:45:07.40	 & -07:22:03.75	 &
G033.70-00.0 & 18:52:54.94	 & +00:40:49.67	 \\
G034.69-06.5 & 19:18:04.25	 & -01:26:07.30	 &
G035.13+11.3 & 18:14:42.06	 & +07:05:34.57	 \\
G037.51+44.5 & 16:10:54.35	 & +21:45:37.26	 &
G046.75-07.6 & 19:44:32.57	 & +08:36:47.86	 \\
G047.74-05.5 & 19:38:56.88	 & +10:30:04.67	 &
G048.25-05.7 & 19:40:32.47	 & +10:51:33.54	 \\
G048.82-03.8 & 19:34:50.76	 & +12:16:54.25	 &
G049.06-04.1 & 19:36:35.77	 & +12:19:16.42	 \\
G049.76-07.2 & 19:48:57.29	 & +11:25:36.57	 &
G052.99+03.0 & 19:17:48.77	 & +19:13:58.55	 \\
G054.03-02.3 & 19:40:07.84	 & +17:31:24.94	 &
G056.84+04.8 & 19:18:50.12	 & +23:26:16.08	 \\
G057.08+04.4 & 19:20:41.76	 & +23:29:11.33	 &
G057.17+03.4 & 19:24:54.22	 & +23:04:26.60	 \\
G057.26+04.0 & 19:22:47.03	 & +23:25:56.16	 &
G058.02+03.0 & 19:28:10.41	 & +23:38:31.75	 \\
G058.97-01.6 & 19:47:54.71	 & +22:10:14.34	 &
G061.76-10.8 & 20:27:23.05	 & +19:38:21.54	 \\
G065.30-08.4 & 20:27:28.71	 & +23:51:10.63	 &
G070.72-00.6 & 20:11:52.55	 & +32:42:33.25	 \\
G084.11+15.5 & 19:30:41.46	 & +52:11:22.10	 &
G089.03-41.2 & 23:08:43.43	 & +14:43:35.71	 \\
G089.29+04.0 & 20:50:37.89	 & +50:27:04.95	 &
G089.64-06.5 & 21:37:06.74	 & +43:22:02.80	 \\
G089.64-06.8 & 21:38:04.87	 & +43:10:20.80	 &
G089.75-02.1 & 21:20:13.68	 & +46:39:03.34	 \\
G089.93-01.9 & 21:20:01.36	 & +46:56:00.17	 &
G090.76-04.5 & 21:34:00.07	 & +45:37:39.81	 \\
G092.48-02.6 & 21:33:49.01	 & +48:11:19.11	 &
G092.79+09.1 & 20:37:43.27	 & +56:20:13.73	 \\
G092.83-11.0 & 22:05:41.29	 & +41:56:59.39	 &
G093.12-10.4 & 22:04:52.01	 & +42:37:09.24	 \\
G093.16+09.6 & 20:36:31.49	 & +56:54:48.32	 &
G093.22-04.5 & 21:44:47.75	 & +47:13:40.19	 \\
G093.31-11.6 & 22:09:47.90	 & +41:42:41.79	 &
G093.36-12.0 & 22:11:10.90	 & +41:25:42.25	 \\
G093.44-04.6 & 21:46:03.92	 & +47:18:45.57	 &
G093.51-04.3 & 21:45:00.01	 & +47:37:34.10	 \\
G093.66+04.6 & 21:05:57.29	 & +54:11:19.51	 &
G093.69+09.9 & 20:36:30.71	 & +57:33:22.77	 \\
G093.75-04.5 & 21:47:09.99	 & +47:34:03.01	 &
G093.91+10.0 & 20:37:13.90	 & +57:45:13.58	 \\
G095.11+06.5 & 21:03:02.13	 & +56:29:14.76	 &
G095.29-09.6 & 22:12:00.39	 & +44:28:46.38	 \\
G095.75+08.1 & 20:56:39.58	 & +58:03:19.36	 &
G095.97+08.1 & 20:57:43.02	 & +58:13:20.01	 \\
G096.30+10.0 & 20:48:15.70	 & +59:37:39.53	 &
G096.37+10.2 & 20:46:47.60	 & +59:51:10.72	 \\
G097.71+15.3 & 20:17:47.47	 & +63:46:21.34	 &
G097.77+08.5 & 21:04:16.83	 & +59:50:45.96	 \\
G097.80+15.2 & 20:19:05.20	 & +63:47:07.60	 &
G097.84+15.3 & 20:18:23.05	 & +63:52:53.91	 \\
G098.06+13.9 & 20:29:40.72	 & +63:20:29.47	 &
G098.10+15.8 & 20:15:36.11	 & +64:21:24.43	 \\
G098.50-03.2 & 22:05:00.08	 & +51:33:11.74	 &
G098.52+13.8 & 20:32:37.51	 & +63:39:38.00	 \\
G098.96+13.6 & 20:36:14.53	 & +63:54:15.88	 &
G099.11+13.9 & 20:34:50.84	 & +64:11:26.96	 \\
G099.97+14.7 & 20:33:04.55	 & +65:19:14.44	 &
G100.32+14.8 & 20:34:17.00	 & +65:38:45.35	 \\
G100.45+14.9 & 20:34:02.37	 & +65:48:58.56	 &
G101.14-15.2 & 22:54:59.56	 & +42:36:10.85	 \\
G101.49+16.8 & 20:23:43.33	 & +67:38:41.41	 &
G101.62-28.8 & 23:24:47.63	 & +30:25:04.98	 \\
G101.88+15.1 & 20:40:04.56	 & +67:04:28.98	 &
G102.17+07.2 & 21:37:08.54	 & +62:02:13.09	 \\
G102.19+15.2 & 20:41:10.70	 & +67:21:44.30	 &
G102.70+15.1 & 20:44:37.79	 & +67:43:34.88	 \\
G102.72-25.9 & 23:24:03.29	 & +33:26:00.51	 &
G103.18-17.0 & 23:08:44.22	 & +41:48:19.98	 \\
G103.22-15.2 & 23:04:48.15	 & +43:31:27.62	 &
G103.49+13.0 & 21:06:51.57	 & +67:01:15.21	 \\
G103.77+13.9 & 21:02:09.15	 & +67:45:51.82	 &
G104.06+12.1 & 21:17:20.80	 & +66:51:02.09	 \\
G104.78+11.4 & 21:27:31.69	 & +66:52:14.83	 &
G105.55+10.4 & 21:40:37.38	 & +66:36:13.43	 \\
G106.30+13.5 & 21:22:40.15	 & +69:23:07.62	 &
G106.89+16.7 & 20:58:22.50	 & +71:52:40.57	 \\
G107.51-09.3 & 23:13:52.32	 & +50:33:37.76	 &
G107.64-09.2 & 23:14:27.39	 & +50:40:44.41	 \\
G108.23+15.6 & 21:20:05.68	 & +72:08:51.84	 &
G110.41+11.5 & 22:13:42.51	 & +70:27:51.52	 \\
G110.52+19.1 & 21:00:37.84	 & +76:06:21.30	 &
G110.54+11.9 & 22:12:18.79	 & +70:50:58.56	 \\
G110.76+19.4 & 20:59:29.33	 & +76:26:33.85	 &
G110.76+20.1 & 20:48:51.71	 & +76:52:51.23	 \\
G110.96+19.4 & 21:00:35.97	 & +76:37:38.72	 &
G111.46+20.1 & 20:55:39.95	 & +77:25:18.28	 \\
G111.68+19.7 & 21:04:19.81	 & +77:19:35.11	 &
G111.77+13.7 & 22:10:39.25	 & +73:04:17.94	 \\
G111.77+20.2 & 20:57:36.58	 & +77:42:08.09	 &
G111.97+20.5 & 20:55:40.82	 & +78:00:13.92	 \\
G112.03+17.8 & 21:32:42.81	 & +76:21:26.12	 &
G112.25+13.8 & 22:15:37.52	 & +73:22:36.95	 \\
G112.52+08.3 & 22:52:47.61	 & +68:49:28.29	 &
G112.60+08.5 & 22:52:54.76	 & +68:59:53.89	 \\
G112.63+20.8 & 20:58:19.23	 & +78:40:03.24	 &
G113.42+16.9 & 21:59:58.99	 & +76:34:08.73	 \\
G113.62+15.0 & 22:21:37.31	 & +75:06:33.49	 &
G114.41-13.9 & 00:01:41.52	 & +48:06:29.81	 \\
G115.92+09.4 & 23:24:04.62	 & +71:08:08.73	 &
G116.01+09.1 & 23:26:27.92	 & +70:50:36.54	 \\
G116.08-02.3 & 23:56:41.78	 & +59:45:13.21	 &
G116.12+08.9 & 23:28:14.03	 & +70:45:12.43	 \\
G116.25+20.1 & 22:00:24.54	 & +80:45:10.68	 &
G116.30+12.2 & 23:15:19.75	 & +73:54:48.36	 \\
G118.14+08.8 & 23:52:17.36	 & +71:09:30.48	 &
G118.25-52.7 & 00:39:55.42	 & +10:03:41.94	 \\
G118.34+08.6 & 23:55:06.71	 & +71:02:12.60	 &
G118.36+21.7 & 22:16:54.81	 & +83:11:56.16	 \\
G121.57+23.3 & 23:37:23.24	 & +86:04:27.40	 &
G121.99+24.2 & 23:45:20.96	 & +86:59:07.74	 \\
G122.73+09.6 & 00:48:54.75	 & +72:29:01.64	 &
G123.66+24.8 & 01:57:28.86	 & +87:40:07.76	 \\
G123.74+24.7 & 02:01:24.04	 & +87:31:40.15	 &
G125.17+11.6 & 01:24:12.50	 & +74:22:20.01	 \\
G125.70+12.5 & 01:33:57.65	 & +75:09:47.66	 &
G126.62+24.5 & 04:23:16.95	 & +85:47:11.88	 \\
G126.65-71.4 & 00:56:13.82	 & -08:36:07.67	 &
G127.31-70.0 & 00:57:27.24	 & -07:16:29.84	 \\
G128.25+20.7 & 03:25:46.87	 & +82:00:25.06	 &
G128.95-00.1 & 01:43:15.17	 & +62:04:39.07	 \\
G128.95+04.4 & 01:52:32.86	 & +66:35:51.19	 &
G129.90+11.6 & 02:28:23.83	 & +73:10:13.46	 \\
G129.94+11.7 & 02:29:32.62	 & +73:15:36.66	 &
G129.96+13.7 & 02:41:53.27	 & +75:07:36.03	 \\
G130.14+11.0 & 02:28:35.31	 & +72:34:12.58	 &
G130.14+11.6 & 02:31:43.16	 & +73:08:00.74	 \\
G130.14+13.7 & 02:44:17.12	 & +75:03:12.38	 &
G130.56+11.5 & 02:35:56.34	 & +72:48:55.42	 \\
G131.35-45.7 & 01:15:57.51	 & +16:44:08.07	 &
G131.72+09.7 & 02:39:57.52	 & +70:42:11.57	 \\
G132.07+08.8 & 02:39:18.18	 & +69:44:01.08	 &
G142.49+07.4 & 03:59:13.57	 & +62:58:52.39	 \\
G142.62+07.2 & 03:59:00.65	 & +62:45:12.55	 &
G144.44+08.5 & 04:17:58.86	 & +62:26:51.41	 \\
G144.66+06.1 & 04:05:12.80	 & +60:31:21.52	 &
G144.84+00.7 & 03:40:20.81	 & +56:16:28.14	 \\
G146.71+02.0 & 03:56:37.15	 & +56:07:23.12	 &
G147.96-08.0 & 03:24:02.12	 & +47:17:29.48	 \\
G150.35-38.0 & 02:22:11.93	 & +19:55:36.57	 &
G151.58-38.5 & 02:24:53.06	 & +19:01:47.06	 \\
G111.33+19.9 & 20:57:51.09	 & +77:10:53.64	 &
G058.07+03.2 & 19:27:32.78	 & +23:46:10.07	 \\
G111.04+14.0 & 22:00:56.89	 & +72:50:36.21	 &
G111.42+19.8 & 21:00:27.72	 & +77:10:37.45	 \\
G118.08-09.6 & 00:19:40.63	 & +52:53:01.30	 &
G121.50-20.9 & 00:44:17.38	 & +41:55:48.07	 \\
G121.88-08.7 & 00:44:21.66	 & +54:05:40.39	 &
G121.90-01.5 & 00:42:52.65	 & +61:18:23.22	 \\
G121.92-07.6 & 00:44:27.19	 & +55:11:17.99	 &
G122.01-07.4 & 00:45:01.81	 & +55:22:43.33	 \\
G122.18-10.2 & 00:46:37.80	 & +52:35:20.75	 &
G122.62-12.1 & 00:49:34.09	 & +50:43:53.45	 \\
G125.66-00.5 & 01:14:52.21	 & +62:11:16.63	 &
G125.77+04.1 & 01:20:18.22	 & +66:50:16.42	 \\
G126.27+05.1 & 01:26:45.93	 & +67:45:33.07	 &
G126.49-01.3 & 01:21:14.55	 & +61:21:34.60	 \\
G126.73+06.1 & 01:33:23.87	 & +68:44:47.63	 &
G127.79+13.6 & 02:09:57.71	 & +75:48:51.99	 \\
G128.21+13.7 & 02:16:18.66	 & +75:42:10.64	 &
G129.22+15.1 & 02:40:47.35	 & +76:41:22.13	 \\
G132.07+13.6 & 03:08:01.25	 & +74:01:34.48	 &
G132.27+13.2 & 03:07:42.40	 & +73:36:53.35	 \\
G132.29+09.3 & 02:44:21.76	 & +70:09:28.58	 &
G133.28+08.8 & 02:51:42.23	 & +69:14:13.43	 \\
G133.72-45.3 & 01:23:04.84	 & +16:53:32.68	 &
G135.08+12.9 & 03:35:45.37	 & +71:47:59.17	 \\
G135.28+12.9 & 03:37:34.32	 & +71:40:05.98	 &
G143.85+11.4 & 04:33:38.01	 & +64:53:16.90	 \\
G144.49+08.7 & 04:19:10.27	 & +62:31:24.89	 &
G145.81+10.9 & 04:41:31.46	 & +63:06:01.71	 \\
G146.18+08.8 & 04:29:36.20	 & +61:22:56.86	 &
G148.00+00.0 & 03:54:48.04	 & +53:47:19.87	 \\
G148.24+00.4 & 03:57:26.19	 & +53:52:36.34	 &
G149.52-01.2 & 03:56:52.62	 & +51:48:01.74	 \\
G149.58+03.4 & 04:18:23.92	 & +55:13:30.60	 &
G154.48-15.1 & 03:30:19.60	 & +37:46:39.05	 \\
G154.68-15.3 & 03:30:35.18	 & +37:31:28.11	 &
G154.95-15.1 & 03:32:04.91	 & +37:29:53.83	 \\
G155.45-14.5 & 03:35:51.95	 & +37:41:02.89	 &
G359.97+21.9 & 16:28:06.76	 & -16:12:17.13	 \\
G159.52+03.2 & 04:59:55.06	 & +47:40:52.24	 &
G127.22-02.2 & 01:26:10.18	 & +60:19:29.25	 \\

\hline 
\end{longtable}	

\begin{longtable}{crr|crr}

	\caption{180 sources without  DCO$^+$, DCN, H$^{13}$CO$^+$ and H$^{13}$CN 1-0 detection with the high velocity resolution mode (AROWS mode 13).}
	\label{180sources}\\
	\hline 
	\hline
	Source & R.A. & Decl. & Source & R.A. & Decl. \\
	Name     & (hh:mm:ss)  & (dd:mm:ss) & 	Name     & (hh:mm:ss)  & (dd:mm:ss) \\
\hline

\endfirsthead
\caption{continue.}\\
	         
	\hline 
	\hline
	Source & R.A. & Decl. & Source & R.A. & Decl. \\
	Name     & (hh:mm:ss)  & (dd:mm:ss) & 	Name     & (hh:mm:ss)  & (dd:mm:ss) \\
	         \hline
  
    \endhead
\hline
\endfoot
G121.92-01.7 & 00:43:06.34	 & +61:08:21.60	 &
G140.97+05.7 & 03:39:23.31	 & +62:33:59.78	 \\
G149.23+03.0 & 04:14:48.52	 & +55:12:03.26	 &
G149.41+03.3 & 04:17:09.05	 & +55:17:39.36	 \\
G149.65+03.5 & 04:19:11.24	 & +55:14:44.38	 &
G150.22+03.9 & 04:23:51.69	 & +55:06:22.47	 \\
G151.45+03.9 & 04:29:56.24	 & +54:14:51.71	 &
G153.34-08.0 & 03:48:42.73	 & +44:08:46.62	 \\
G153.34+11.2 & 05:19:01.04	 & +57:19:44.78	 &
G153.74+35.9 & 08:36:34.59	 & +62:26:27.09	 \\
G154.07+05.0 & 04:47:23.37	 & +53:03:31.39	 &
G154.07+05.2 & 04:47:57.79	 & +53:07:51.21	 \\
G154.55-01.2 & 04:19:58.35	 & +48:23:32.14	 &
G154.90+04.6 & 04:48:27.03	 & +52:06:30.37	 \\
G155.52-08.9 & 03:54:34.60	 & +42:03:48.96	 &
G155.67+05.1 & 04:54:08.58	 & +51:50:10.93	 \\
G156.04+06.0 & 05:00:19.24	 & +52:06:45.62	 &
G156.42+32.5 & 08:06:27.98	 & +60:34:20.05	 \\
G156.53-08.6 & 03:59:41.99	 & +41:38:26.63	 &
G156.90-08.4 & 04:01:38.87	 & +41:29:43.41	 \\
G156.92-09.7 & 03:57:26.44	 & +40:33:29.15	 &
G157.10-08.7 & 04:01:41.53	 & +41:12:37.45	 \\
G157.12-11.5 & 03:51:59.15	 & +39:01:55.10	 &
G157.19-08.8 & 04:01:38.24	 & +41:04:05.05	 \\
G157.25-01.0 & 04:32:09.45	 & +46:37:25.02	 &
G157.58-08.8 & 04:02:54.80	 & +40:45:04.96	 \\
G157.60-12.1 & 03:51:49.54	 & +38:15:38.76	 &
G157.91-08.2 & 04:06:32.28	 & +41:01:15.73	 \\
G158.24-21.8 & 03:25:13.67	 & +30:19:30.39	 &
G158.40-21.8 & 03:25:35.50	 & +30:11:29.55	 \\
G158.77-33.3 & 02:57:33.29	 & +20:38:31.01	 &
G158.86-21.6 & 03:27:54.15	 & +30:08:33.87	 \\
G158.88-34.1 & 02:55:48.23	 & +19:51:50.97	 &
G158.97-33.0 & 02:58:50.03	 & +20:47:23.13	 \\
G159.03-08.3 & 04:10:28.53	 & +40:11:46.78	 &
G159.14-08.7 & 04:09:20.14	 & +39:48:21.73	 \\
G159.16-05.5 & 04:21:08.38	 & +42:04:14.47	 &
G159.34-24.3 & 03:21:51.73	 & +27:38:19.30	 \\
G159.36+03.7 & 05:01:43.99	 & +48:06:42.77	 &
G159.41-34.3 & 02:56:54.67	 & +19:28:15.70	 \\
G159.58-32.8 & 03:01:05.17	 & +20:38:43.23	 &
G159.65+11.3 & 05:42:16.93	 & +52:07:48.14	 \\
G159.67-05.7 & 04:22:32.37	 & +41:37:15.79	 &
G159.67-34.3 & 02:57:46.94	 & +19:23:09.02	 \\
G159.76-19.6 & 03:36:37.75	 & +31:10:57.96	 &
G159.78-24.8 & 03:22:11.99	 & +27:04:15.40	 \\
G159.82-10.4 & 04:05:49.63	 & +38:05:19.15	 &
G160.33+03.2 & 05:02:45.93	 & +47:01:41.07	 \\
G160.35-06.3 & 04:22:34.71	 & +40:40:40.18	 &
G160.51-16.8 & 03:47:34.31	 & +32:53:12.17	 \\
G160.51-17.0 & 03:46:51.18	 & +32:42:28.95	 &
G160.53-19.7 & 03:38:58.11	 & +30:39:04.72	 \\
G160.62-16.7 & 03:48:22.46	 & +32:55:21.25	 &
G160.83-09.4 & 04:13:04.61	 & +38:08:51.19	 \\
G161.08-21.7 & 03:34:51.38	 & +28:43:51.87	 &
G161.21-08.7 & 04:17:02.75	 & +38:24:57.50	 \\
G161.34-09.3 & 04:15:21.52	 & +37:53:39.42	 &
G161.43-35.5 & 02:59:41.08	 & +17:30:58.04	 \\
G161.60-08.3 & 04:19:49.16	 & +38:24:23.05	 &
G162.24-09.0 & 04:19:33.27	 & +37:28:03.29	 \\
G162.46-08.7 & 04:21:32.60	 & +37:33:04.74	 &
G162.79+01.3 & 05:02:42.86	 & +43:55:05.65	 \\
G162.83-18.4 & 03:50:32.04	 & +30:13:17.24	 &
G162.86-05.2 & 04:35:48.53	 & +39:38:08.62	 \\
G162.90-08.6 & 04:23:20.51	 & +37:17:33.45	 &
G163.21-08.4 & 04:25:13.29	 & +37:13:48.93	 \\
G163.82-08.3 & 04:27:35.71	 & +36:50:26.29	 &
G164.13-08.8 & 04:26:43.25	 & +36:15:25.18	 \\
G164.57-24.4 & 03:38:12.61	 & +24:36:13.41	 &
G164.81-05.6 & 04:40:47.51	 & +37:53:40.02	 \\
G164.92-12.6 & 04:16:04.53	 & +33:03:27.18	 &
G164.94-05.7 & 04:41:05.14	 & +37:46:15.50	 \\
G165.34-07.5 & 04:35:37.43	 & +36:16:57.13	 &
G165.91-44.0 & 02:50:21.82	 & +08:41:43.64	 \\
G166.70-15.0 & 04:13:45.64	 & +30:08:25.67	 &
G166.99-15.3 & 04:13:42.01	 & +29:44:26.31	 \\
G168.46-17.9 & 04:09:48.87	 & +26:55:07.96	 &
G168.85-14.7 & 04:21:32.29	 & +28:52:48.43	 \\
G169.14-01.1 & 05:12:20.07	 & +37:20:57.12	 &
G169.82-19.3 & 04:09:12.03	 & +24:58:39.00	 \\
G169.84-07.6 & 04:49:23.31	 & +32:49:47.26	 &
G169.98-04.2 & 05:02:22.42	 & +34:48:06.81	 \\
G170.57-07.8 & 04:50:38.52	 & +32:07:05.91	 &
G170.77-08.5 & 04:48:49.00	 & +31:33:00.09	 \\
G170.81-18.3 & 04:15:23.21	 & +25:01:23.84	 &
G170.83-15.9 & 04:23:24.42	 & +26:39:55.87	 \\
G170.88-10.9 & 04:40:32.71	 & +29:55:42.61	 &
G171.03+02.6 & 05:33:35.42	 & +37:56:42.68	 \\
G171.32+04.4 & 05:41:51.47	 & +38:37:50.74	 &
G171.34+02.5 & 05:34:06.95	 & +37:38:47.28	 \\
G171.43-17.3 & 04:20:19.19	 & +25:15:38.94	 &
G171.78-08.4 & 04:52:01.90	 & +30:49:13.60	 \\
G171.80-09.7 & 04:47:15.73	 & +29:57:33.86	 &
G172.15-09.9 & 04:47:43.89	 & +29:35:43.25	 \\
G172.30-09.9 & 04:48:14.33	 & +29:29:23.26	 &
G172.85+02.2 & 05:36:51.80	 & +36:11:58.25	 \\
G172.94-05.4 & 05:06:15.78	 & +31:42:39.02	 &
G173.10+07.2 & 05:58:56.30	 & +38:31:58.92	 \\
G173.18-09.1 & 04:53:34.39	 & +29:18:35.60	 &
G173.45-05.4 & 05:07:52.92	 & +31:20:25.45	 \\
G173.78-05.2 & 05:09:25.64	 & +31:10:33.56	 &
G174.81+05.9 & 05:57:31.74	 & +36:25:24.39	 \\
G174.99-04.8 & 05:14:20.24	 & +30:27:09.32	 &
G175.20+01.2 & 05:38:55.10	 & +33:41:05.90	 \\
G175.53+01.3 & 05:39:59.24	 & +33:26:08.78	 &
G176.17-02.1 & 05:27:55.18	 & +31:01:35.04	 \\
G176.35+01.9 & 05:44:23.17	 & +33:02:58.95	 &
G176.37-02.0 & 05:28:38.86	 & +30:53:34.32	 \\
G176.94+04.6 & 05:57:00.76	 & +33:55:16.25	 &
G177.09+02.8 & 05:50:02.13	 & +32:53:35.85	 \\
G178.28-00.6 & 05:39:03.83	 & +30:04:05.92	 &
G178.72-07.0 & 05:15:59.71	 & +26:09:46.55	 \\
G179.14-06.2 & 05:19:43.91	 & +26:14:18.94	 &
G179.29+04.2 & 06:00:44.85	 & +31:40:15.04	 \\
G179.40+04.0 & 06:00:27.95	 & +31:30:39.71	 &
G180.92+04.5 & 06:05:48.75	 & +30:25:11.40	 \\
G181.16+04.3 & 06:05:31.15	 & +30:06:33.65	 &
G181.42-03.7 & 05:34:46.40	 & +25:44:48.16	 \\
G181.71+04.1 & 06:06:03.83	 & +29:32:54.89	 &
G181.88+04.4 & 06:07:48.53	 & +29:33:27.39	 \\
G182.02-00.1 & 05:49:42.48	 & +27:07:10.01	 &
G182.04+00.4 & 05:51:59.62	 & +27:23:47.47	 \\
G182.54-25.3 & 04:23:04.85	 & +12:13:03.09	 &
G185.29-02.2 & 05:49:08.92	 & +23:13:41.99	 \\
G188.04-03.7 & 05:49:52.71	 & +20:08:09.75	 &
G190.08-13.5 & 05:19:28.71	 & +13:15:43.32	 \\
G190.15-14.3 & 05:16:45.83	 & +12:45:44.57	 &
G190.17-13.7 & 05:18:44.25	 & +13:02:43.23	 \\
G190.50-02.2 & 06:00:19.46	 & +18:43:31.95	 &
G191.03-16.7 & 05:10:23.30	 & +10:45:03.71	 \\
G191.51-00.7 & 06:08:02.75	 & +18:35:43.91	 &
G192.12-10.9 & 05:32:55.83	 & +12:57:08.60	 \\
G192.28-11.3 & 05:31:43.36	 & +12:35:44.79	 &
G192.54-11.5 & 05:31:28.55	 & +12:15:22.39	 \\
G192.94-03.7 & 06:00:08.12	 & +15:53:50.85	 &
G194.69-16.8 & 05:17:37.50	 & +07:43:26.85	 \\
G194.80-03.4 & 06:05:04.24	 & +14:25:43.04	 &
G194.94-16.7 & 05:18:26.83	 & +07:34:38.62	 \\
G196.21-15.5 & 05:25:17.29	 & +07:10:11.87	 &
G196.65+13.9 & 07:13:49.08	 & +20:37:45.73	 \\
G198.03-15.2 & 05:29:45.07	 & +05:46:55.24	 &
G198.96+01.4 & 06:30:41.93	 & +13:03:55.51	 \\
G201.26+00.4 & 06:31:36.70	 & +10:34:48.62	 &
G201.84+02.8 & 06:41:11.23	 & +11:09:07.32	 \\
G202.21-09.1 & 05:59:02.91	 & +05:12:05.39	 &
G202.58-08.7 & 06:01:25.31	 & +05:06:09.99	 \\
G203.00-03.6 & 06:19:58.41	 & +07:06:23.11	 &
G203.11-08.7 & 06:02:25.08	 & +04:38:39.70	 \\
G203.57-30.0 & 04:47:58.28	 & -05:56:06.84	 &
G203.75-08.4 & 06:04:20.98	 & +04:11:17.07	 \\
G204.82-13.8 & 05:47:22.03	 & +00:42:34.83	 &
G205.29-06.0 & 06:16:00.65	 & +04:00:29.07	 \\
G205.42-08.1 & 06:08:40.02	 & +02:53:24.34	 &
G205.57-08.2 & 06:08:37.14	 & +02:42:41.26	 \\
G206.05-08.3 & 06:09:06.85	 & +02:14:08.74	 &
G206.21-08.4 & 06:09:11.90	 & +02:04:29.29	 \\
G206.34-08.5 & 06:08:54.46	 & +01:53:20.73	 &
G206.54-14.4 & 05:48:27.03	 & -01:01:42.28	 \\
G206.60-14.4 & 05:48:46.44	 & -01:03:28.63	 &
G206.87-04.3 & 06:24:47.70	 & +03:22:44.35	 \\
G208.56+02.3 & 06:51:55.20	 & +04:57:58.38	 &
G209.20+02.1 & 06:52:12.86	 & +04:17:19.67	 \\
G210.01-20.1 & 05:34:10.53	 & -06:35:04.69	 &
G210.30-00.0 & 06:46:30.11	 & +02:19:29.76	 \\
G210.67-36.7 & 04:34:00.52	 & -14:10:28.57	 &
G210.89-36.5 & 04:35:10.69	 & -14:14:37.73	 \\
G211.70-12.1 & 06:05:47.39	 & -04:26:58.01	 &
G213.92-13.6 & 06:04:31.04	 & -07:01:33.60	 \\
G214.21-13.4 & 06:05:46.53	 & -07:10:53.95	 &
G214.69-19.9 & 05:42:43.77	 & -10:26:42.40	 \\
G215.41-16.3 & 05:56:57.94	 & -09:32:27.84	 &
G215.70-15.0 & 06:02:20.00	 & -09:12:32.57	 \\
G215.88-17.5 & 05:53:22.51	 & -10:27:04.58	 &
G215.92-15.3 & 06:01:43.10	 & -09:31:03.11	 \\
G216.01-15.9 & 05:59:35.96	 & -09:51:42.95	 &
G216.32-15.7 & 06:00:58.21	 & -10:01:42.75	 \\
G216.51-13.8 & 06:08:00.58	 & -09:24:24.66	 &
G216.69-13.8 & 06:08:18.64	 & -09:33:37.84	 \\
G216.76-16.0 & 06:00:25.70	 & -10:33:36.89	 &
G217.13-12.5 & 06:13:57.69	 & -09:21:40.44	 \\

\hline 
\end{longtable}	
\section{The spectral lines of DCO$^+$, DCN, H$^{13}$CO$^+$ and H$^{13}$CN 1-0}

The spectra of deuterated molecules (black lines) and the spectra of $^{13}$C--isotopologue counterparts (red lines) are presented in the following figures, grouped by HCO$^+$ and HCN respectively.

Figures \ref{DCO+mode3_1}$\sim$\ref{DCO+mode13_2} show the spectral lines of DCO$^+$ 1-0. Figure \ref{DCO+mode3_1} displays the spectra for 8 sources where DCO$^+$ 1-0 was detected using the low velocity resolution mode (AROWS mode 3), but H$^{13}$CO$^+$ 1-0 was not observed. Figure \ref{DCO+mode3_2} presents spectra for 5 sources with DCO$^+$ 1-0 detections and no H$^{13}$CO$^+$ 1-0 detections using the low velocity resolution mode (AROWS mode 3). Figure \ref{DCO+mode13_2} shows spectra for 12 sources with DCO$^+$ 1-0 detections and no H$^{13}$CO$^+$ 1-0 detections using the high velocity resolution mode (AROWS mode 13).

Figures \ref{DCNmode3_1}$\sim$\ref{DCNmode13_2} show the spectral lines of DCN 1-0. Figure \ref{DCNmode3_1} shows spectra for 2 sources with DCN 1-0 detection and no H$^{13}$CN 1-0 observation using the low velocity resolution mode (AROWS mode 3). Figure \ref{DCNmode3_2} shows  spectra for 5 sources with DCN 1-0 detections but without H$^{13}$CN 1-0 detections using the low velocity resolution mode (AROWS mode 3). Figure \ref{DCNmode13_2} presents spectra for 2 sources with DCN 1-0 detections and no H$^{13}$CN 1-0 detections using the high velocity resolution mode (AROWS mode 13).

Figure \ref{H13CO+mode13_2} displays the spectral lines of H$^{13}$CO$^+$ 1-0, with 11 sources showing H$^{13}$CO$^+$ 1-0 detections but without DCO$^+$ 1-0 detections with the high velocity resolution mode (AROWS mode 13).

Figures \ref{H13CNmode3_2} and \ref{H13CNmode13_2} show the spectral lines of H$^{13}$CN 1-0. Figure \ref{H13CNmode3_2} shows spectra for 16 sources with H$^{13}$CN 1-0 detections but no DCO$^+$ 1-0 detections using the low velocity resolution mode (AROWS mode 3). Figure \ref{H13CNmode13_2} shows spectra for 6 sources with H$^{13}$CN 1-0 detections and no DCN 1-0 detections using the high velocity resolution mode (AROWS mode 13).

Figures \ref{DCO+H13CO+mode3_2} and \ref{DCO+H13CO+mode13_2} show the spectral lines of both DCO$^+$ and H$^{13}$CO$^+$ 1-0. Figure \ref{DCO+H13CO+mode3_2} shows spectra for 66 sources with detections of both DCO$^+$ and H$^{13}$CO$^+$ 1-0 using the low velocity resolution mode (AROWS mode 3). Figure \ref{DCO+H13CO+mode13_2} displays spectra for 46 sources with detections of both species using the high velocity resolution mode (AROWS mode 13). Due to errors in the Doppler tracking with mode 13, H$^{13}$CO$+$ 1-0 lines were manually aligned with DCO$+$ 1-0 lines in Figure \ref{DCO+H13CO+mode13_2}.

Figures \ref{DCNH13CNmode3_2} and \ref{DCNH13CNmode13_2} show the spectral lines of DCN and H$^{13}$CN 1-0. Figure \ref{DCNH13CNmode3_2} includes spectra for 4 sources with both DCN and H$^{13}$CN 1-0 detections using the low velocity resolution mode (AROWS mode 3). Figure \ref{DCNH13CNmode13_2} presents spectra for 7 sources with both detections using the high velocity resolution mode (AROWS mode 13). Due to errors in the Doppler tracking with mode 13, H$^{13}$CN 1-0 lines were manually aligned with DCN 1-0 lines in  Figure \ref{DCNH13CNmode13_2}.

\setcounter{figure}{0}
\renewcommand{\thefigure}{D\arabic{figure}}
\begin{figure}
\centering
\includegraphics[width=0.3\columnwidth]{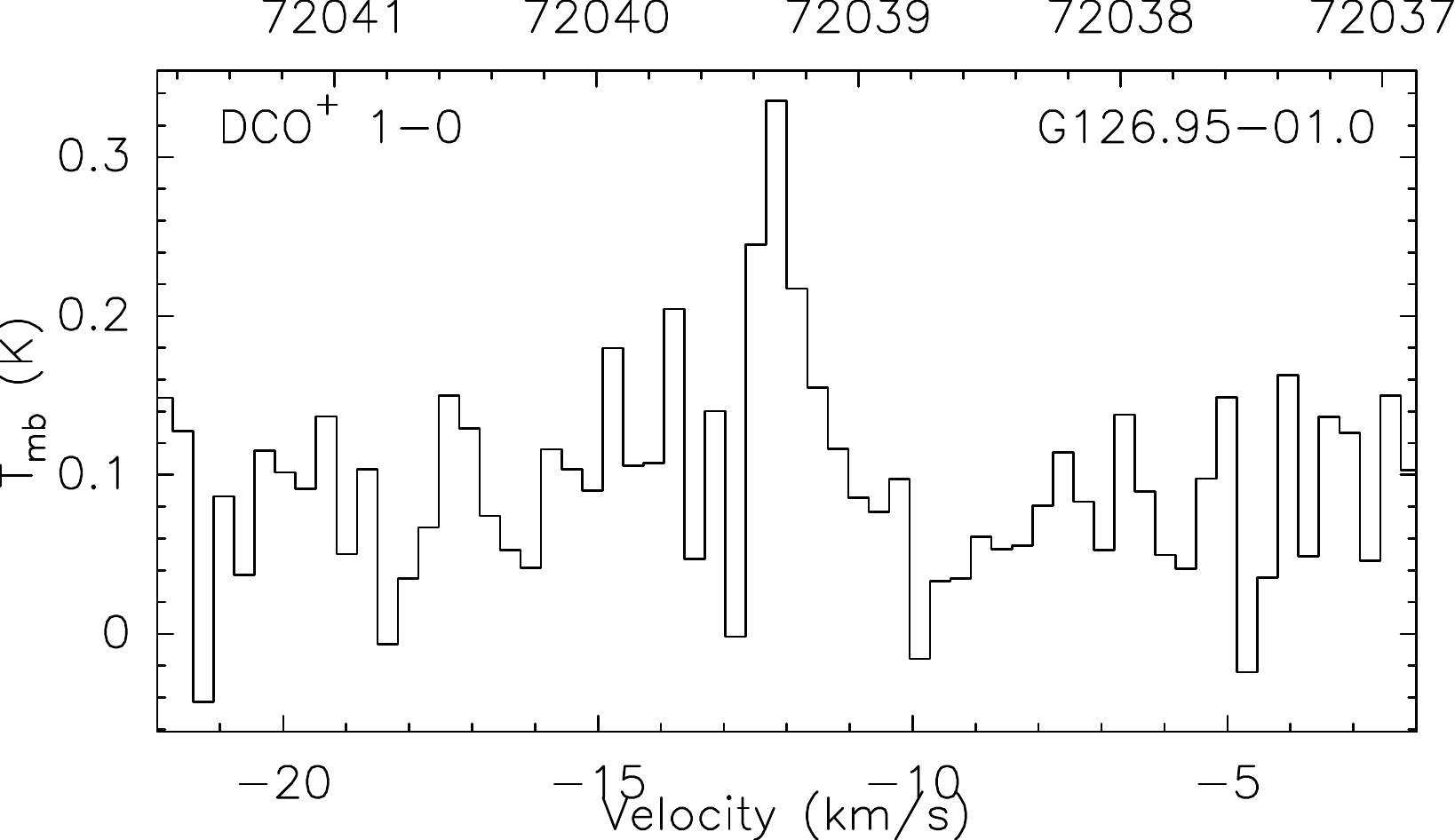}
\includegraphics[width=0.3\columnwidth]{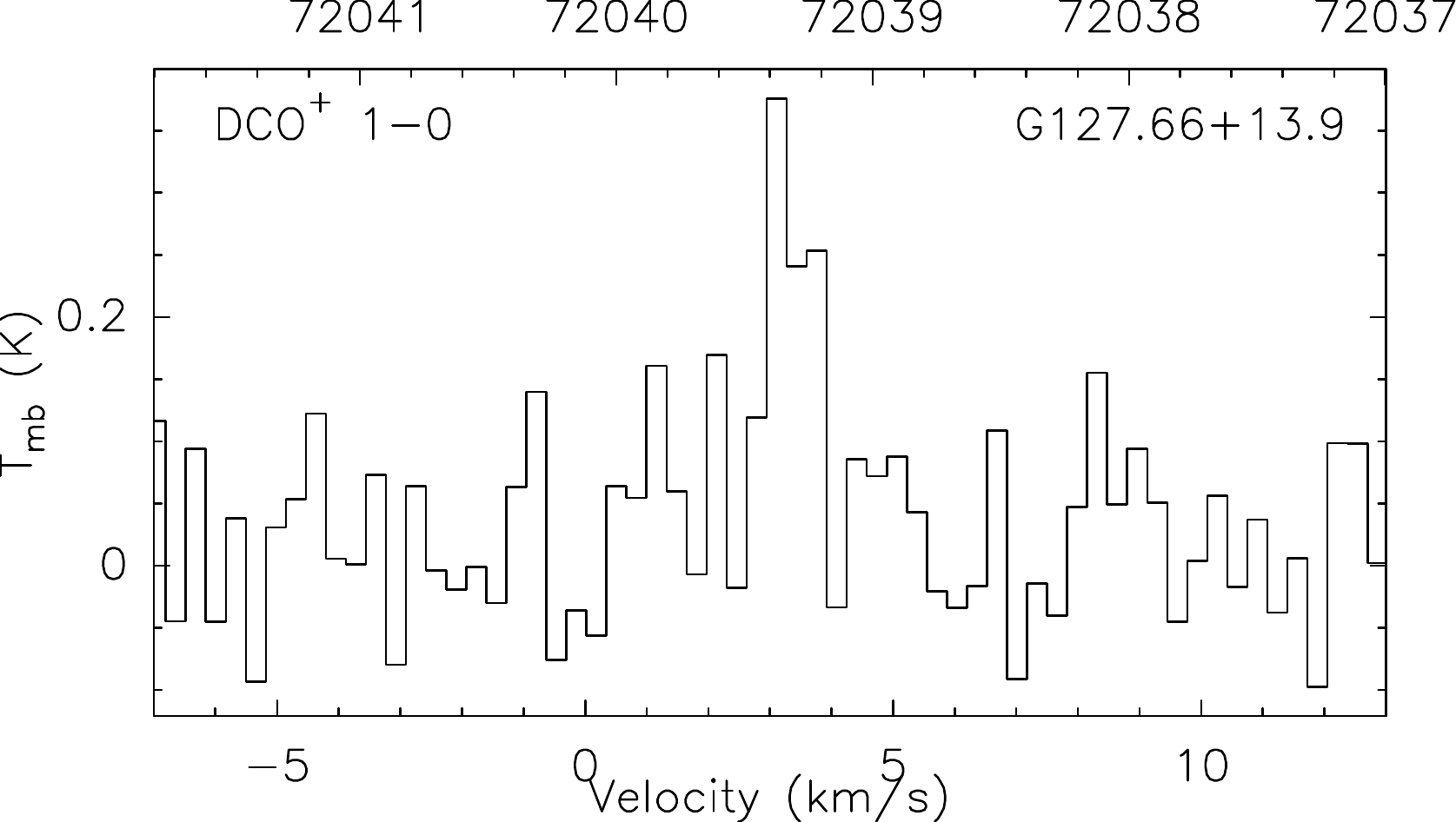}
\includegraphics[width=0.3\columnwidth]{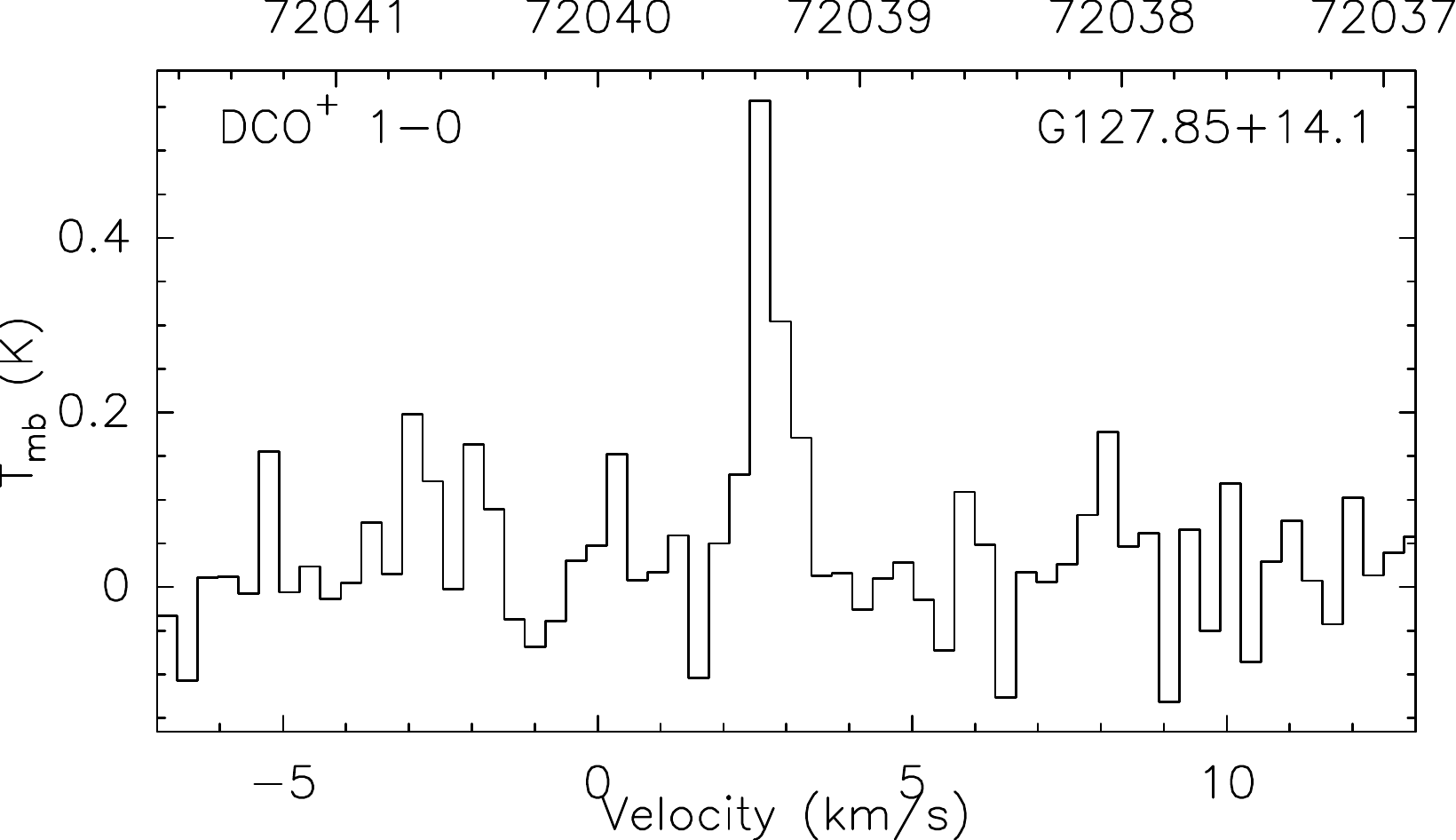}
\includegraphics[width=0.3\columnwidth]{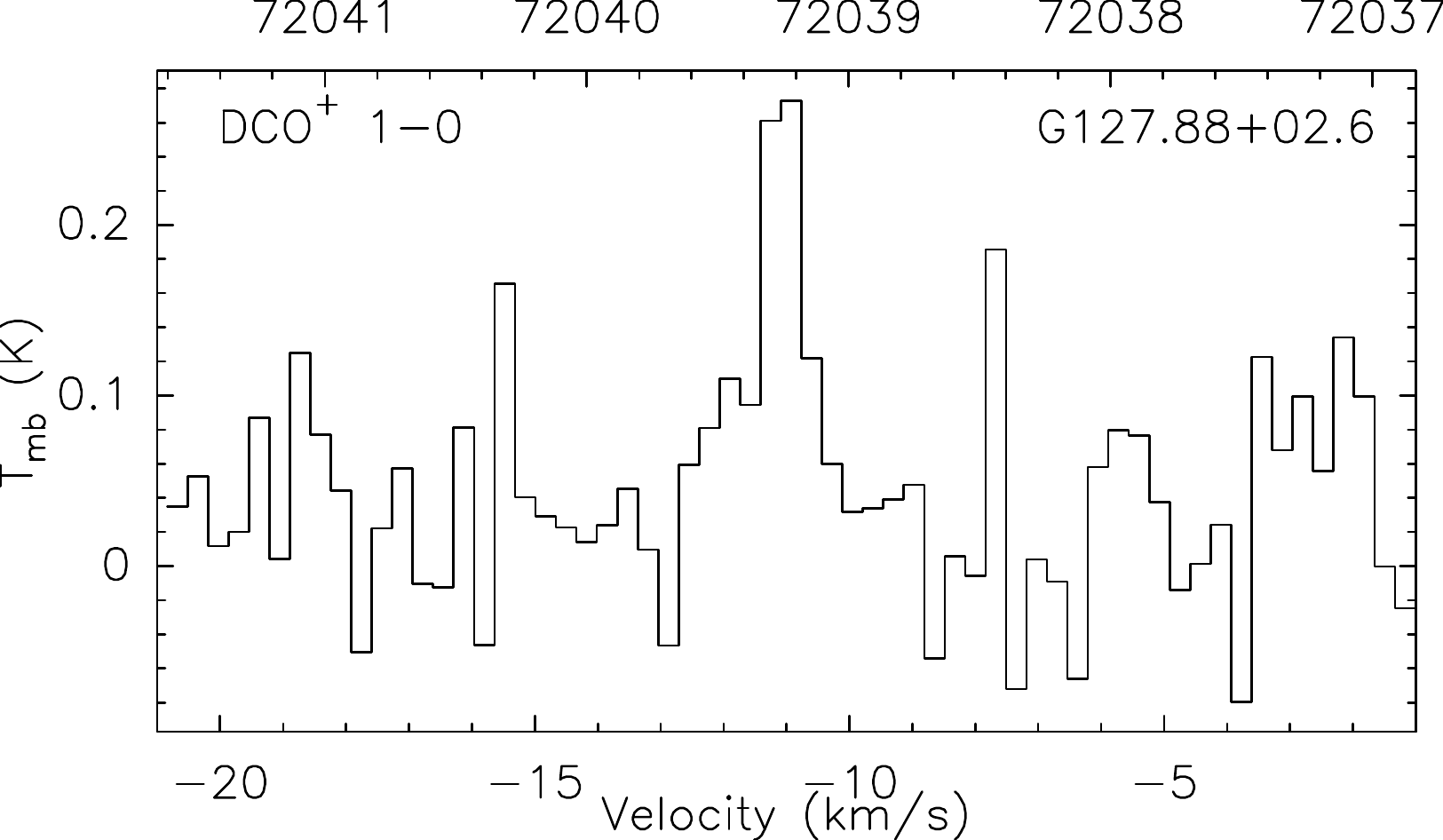}
\includegraphics[width=0.3\columnwidth]{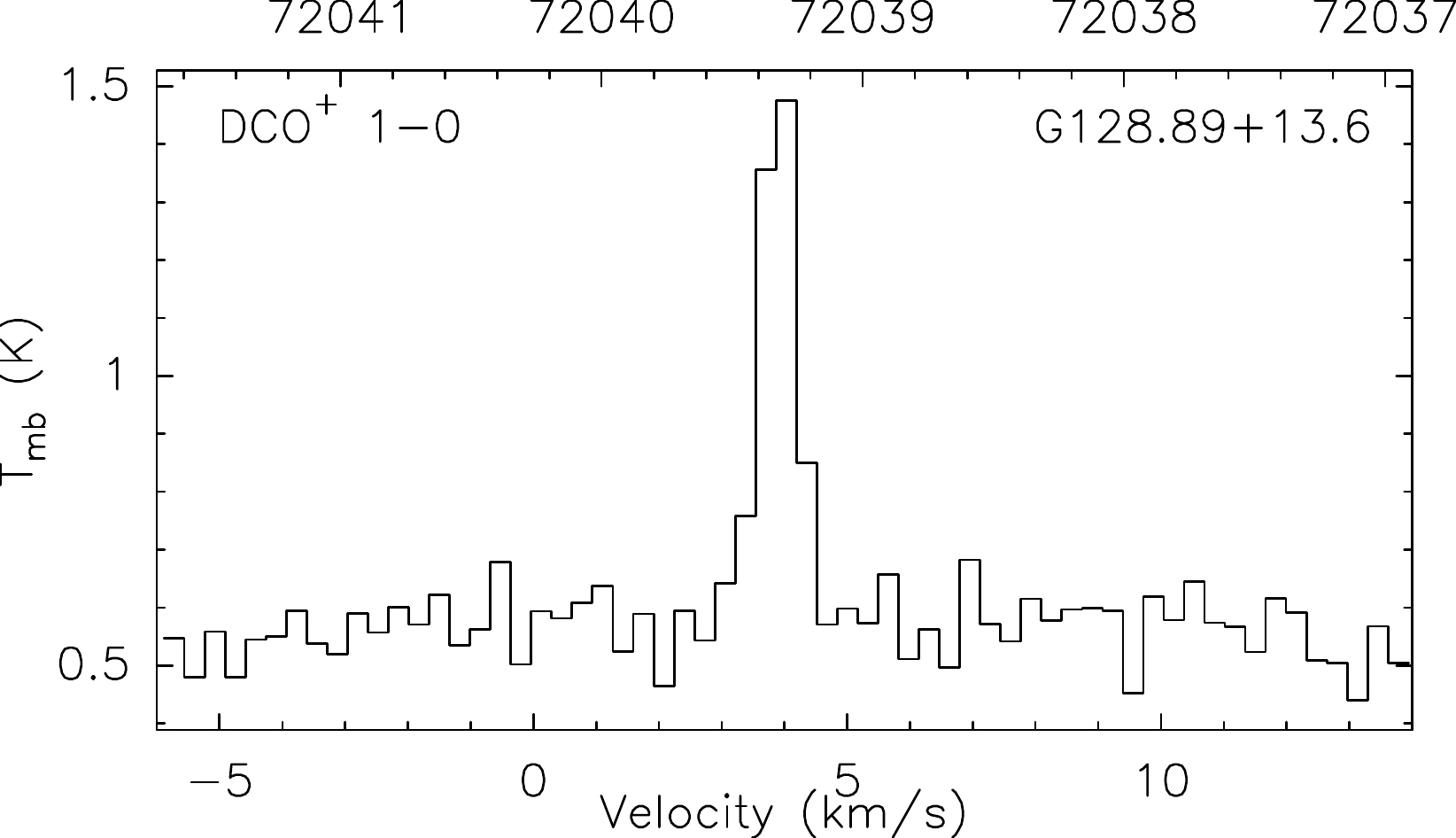}
\includegraphics[width=0.3\columnwidth]{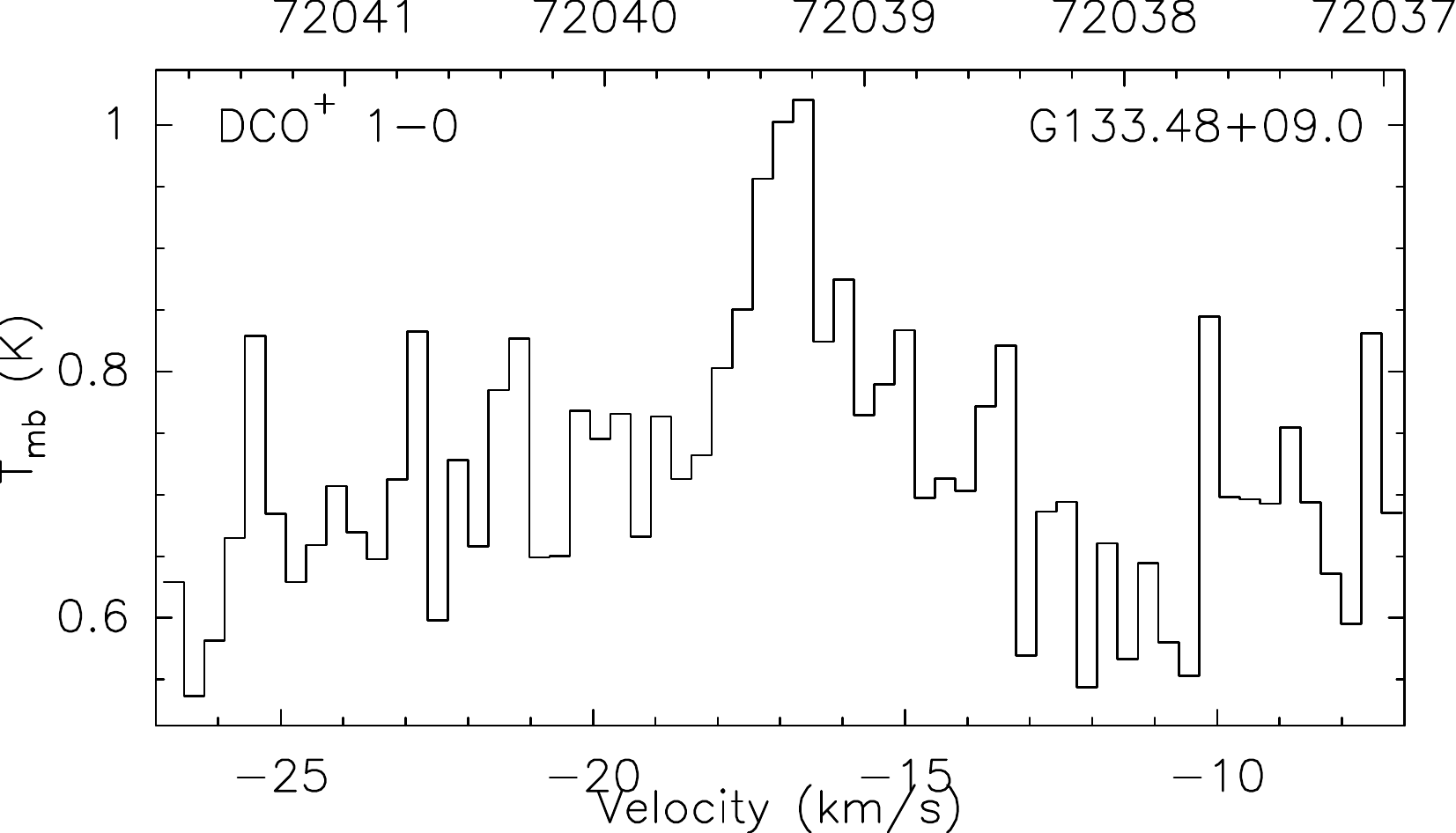}
\includegraphics[width=0.3\columnwidth]{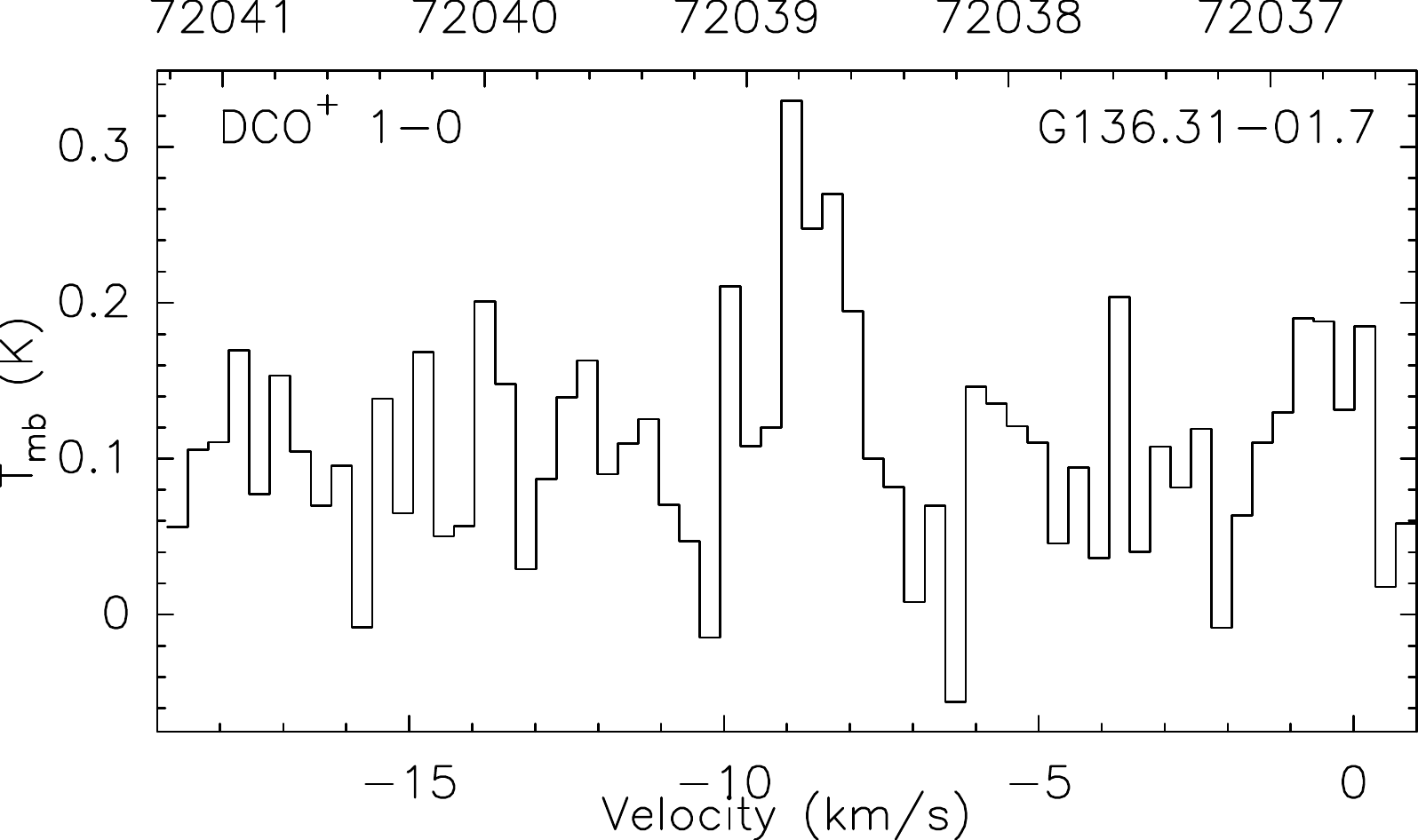}
\includegraphics[width=0.3\columnwidth]{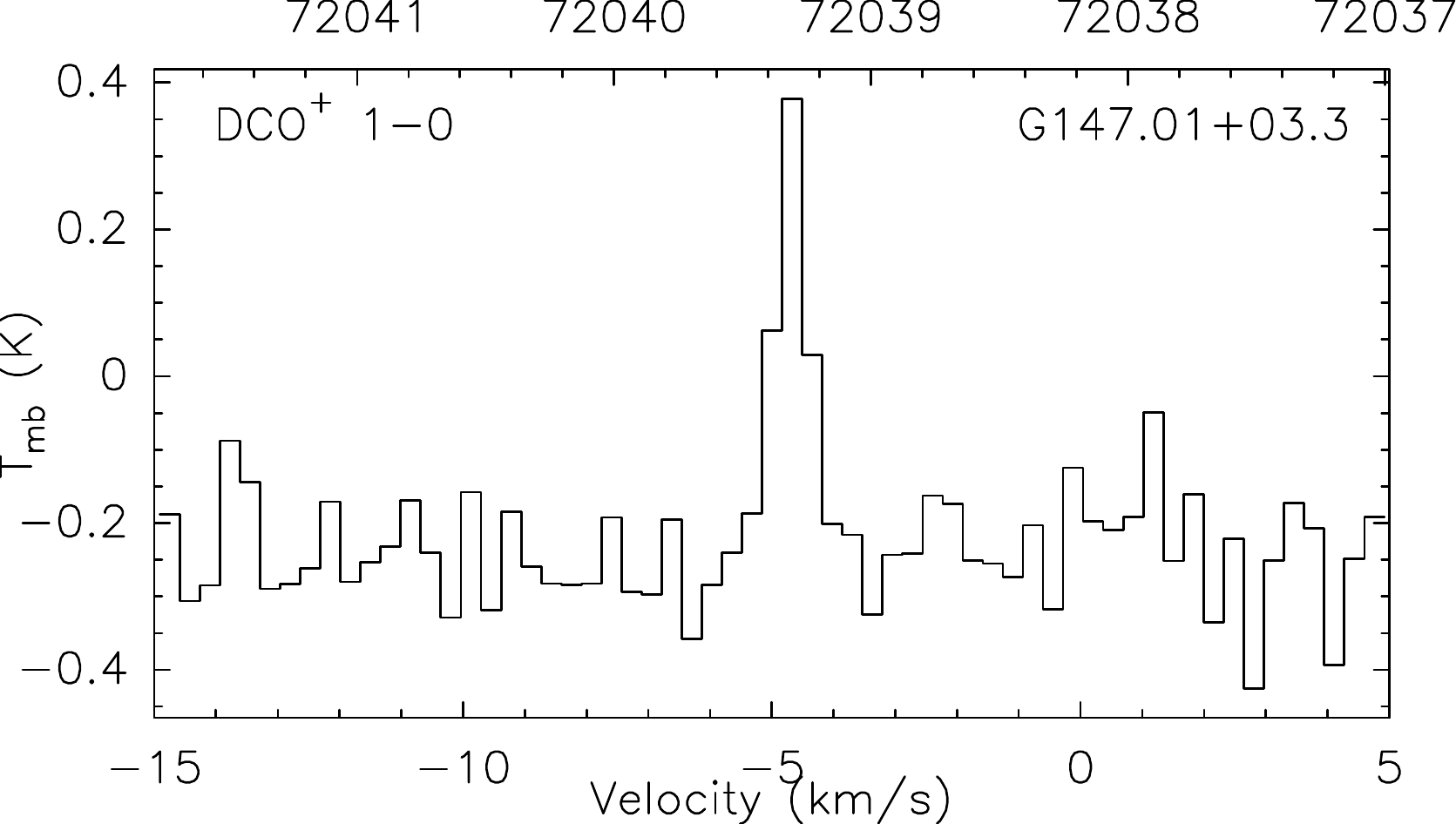}
\caption{Line profiles of DCO$^+$ 1-0 with the low velocity resolution mode (AROWS mode 3). The transitions of H$^{13}$CO$^+$ 1-0 have not been observed in these sources.\centering}
\label{DCO+mode3_1}
\end{figure}
\begin{figure}
\centering
\includegraphics[width=0.3\columnwidth]{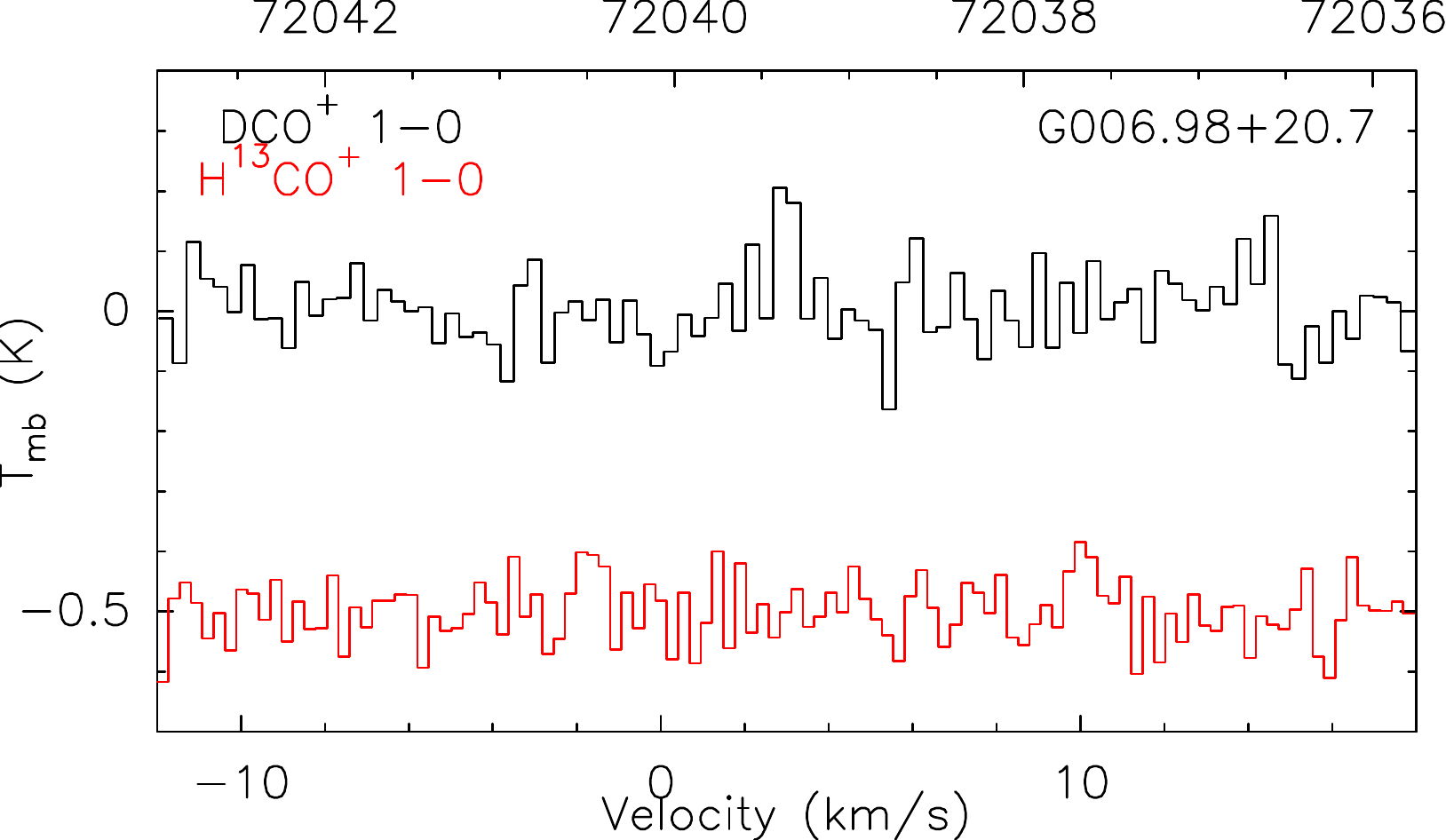}
\includegraphics[width=0.3\columnwidth]{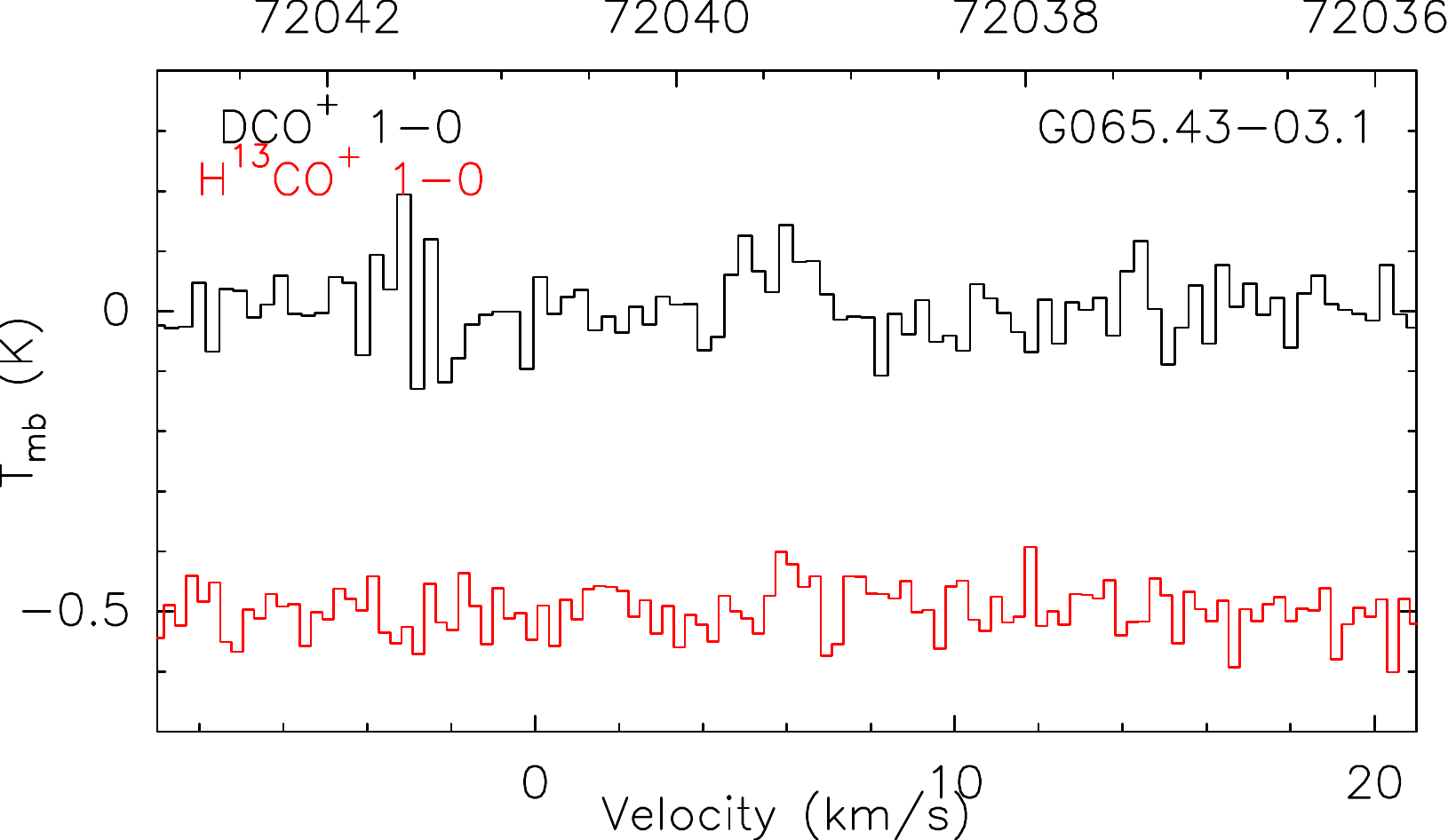}
\includegraphics[width=0.3\columnwidth]{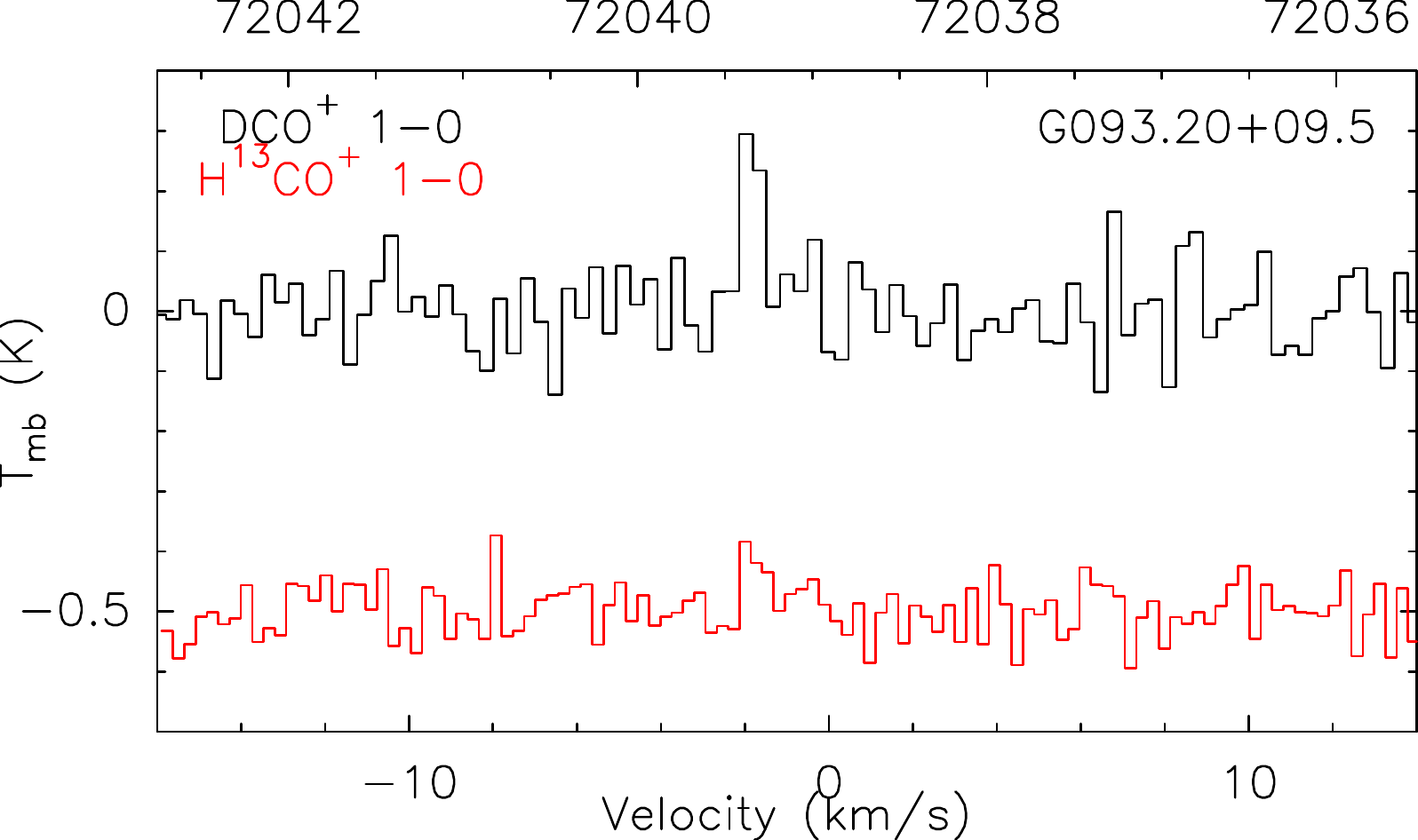}
\includegraphics[width=0.3\columnwidth]{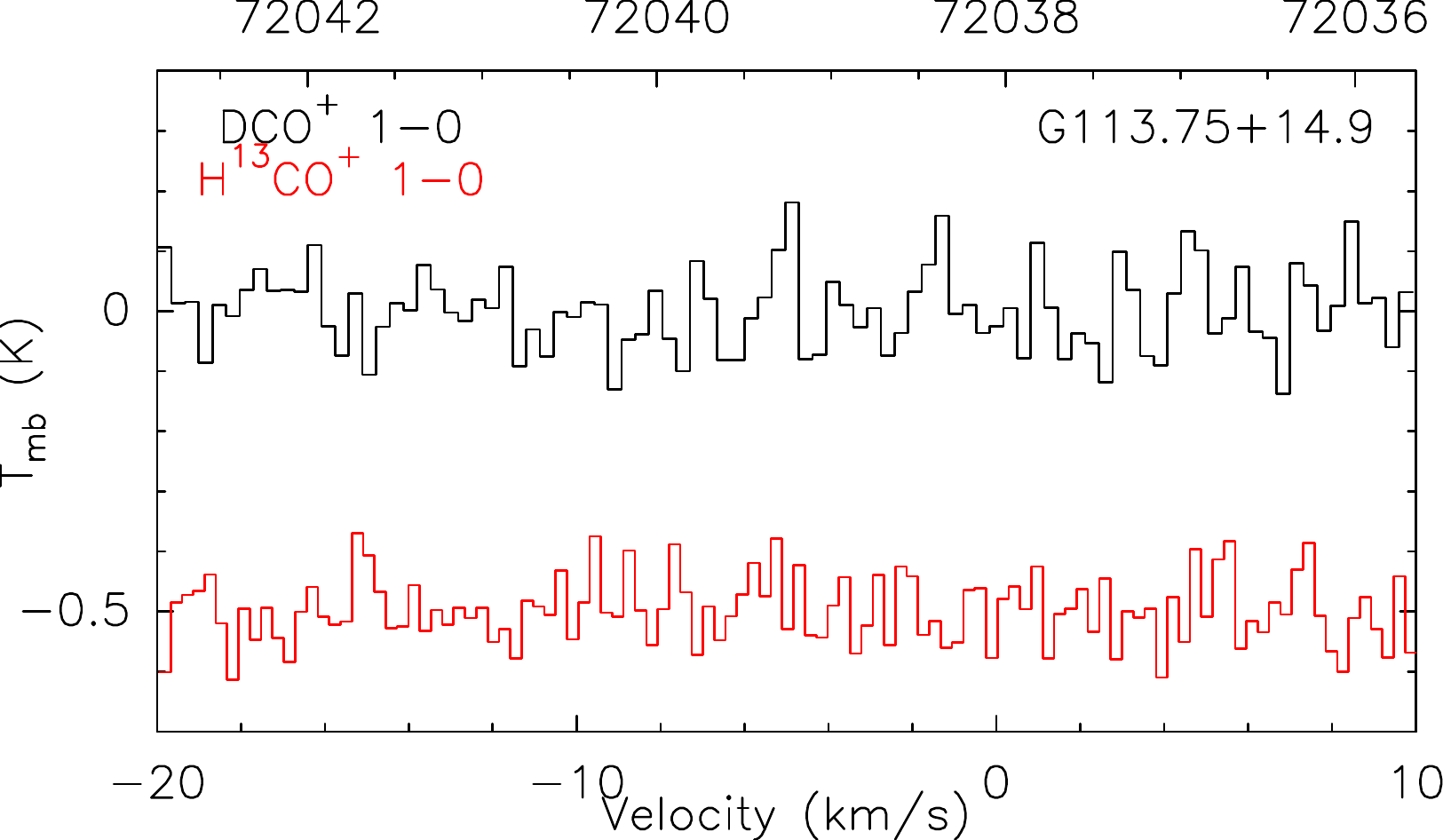}
\includegraphics[width=0.3\columnwidth]{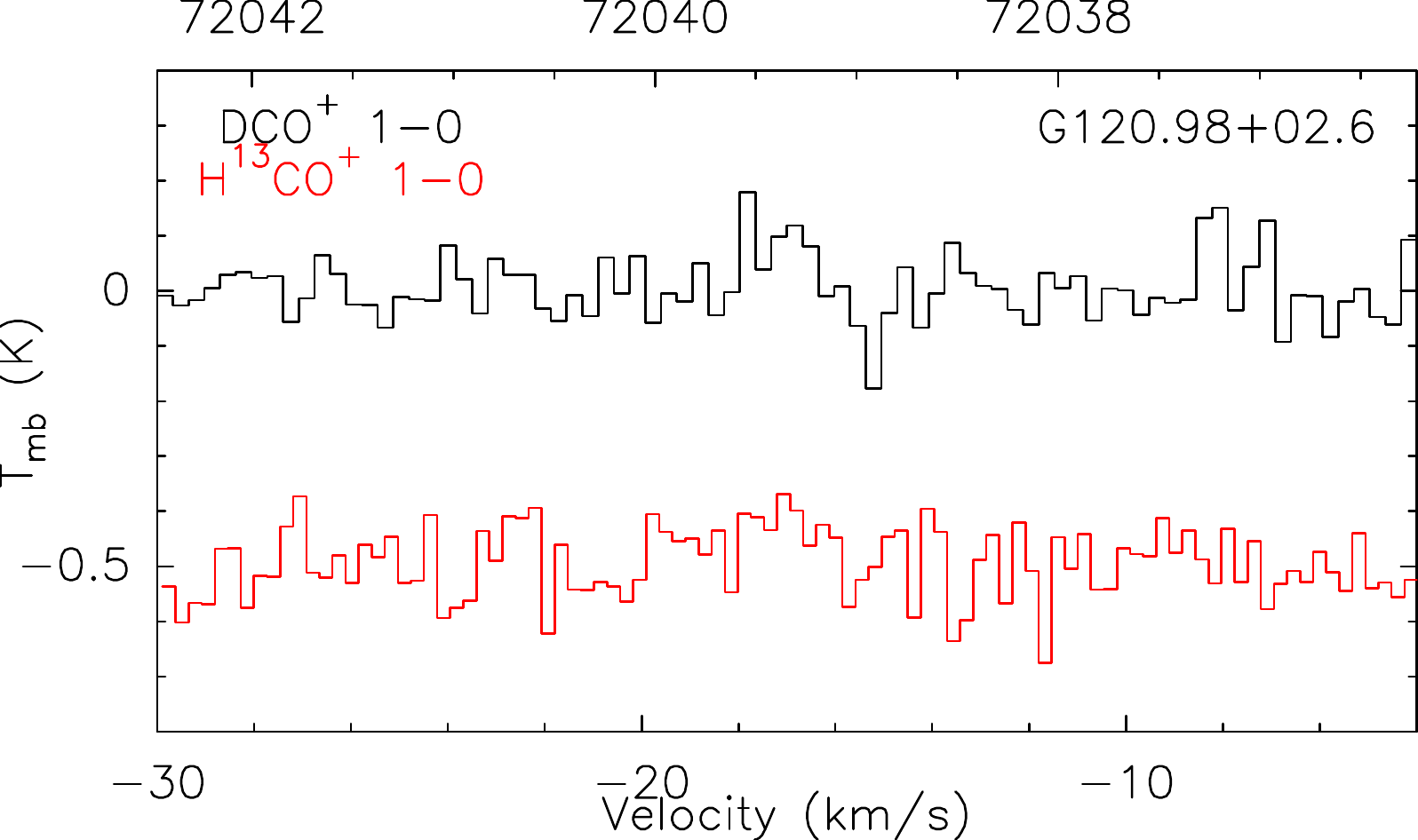}
\caption{Line profiles of DCO$^+$ 1-0 with the low velocity resolution mode (AROWS mode 3). The transitions of H$^{13}$CO$^+$ 1-0 have been observed but not detected in these sources.\centering}
\label{DCO+mode3_2}
\end{figure}
\begin{figure}
\centering
\includegraphics[width=0.3\columnwidth]{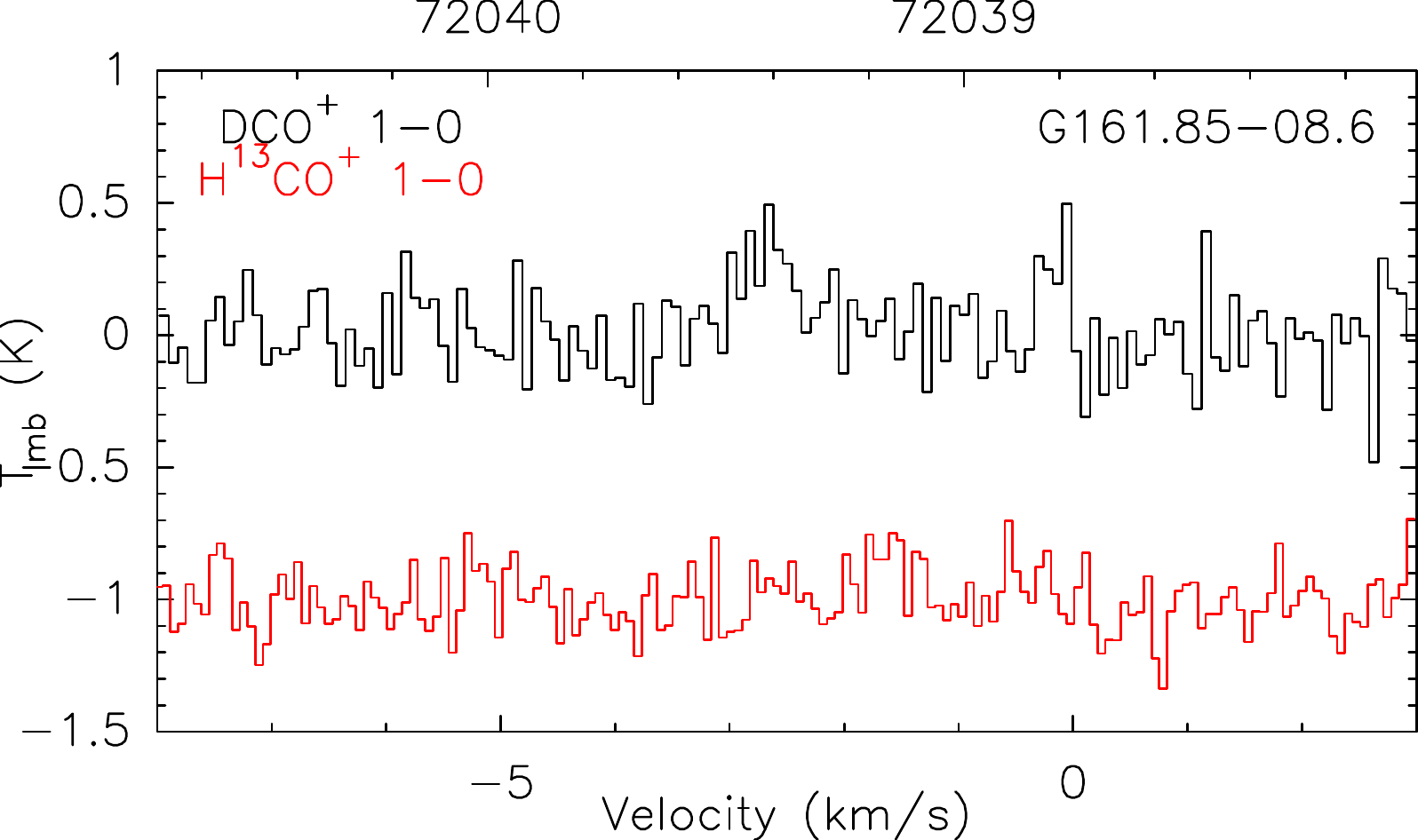}
\includegraphics[width=0.3\columnwidth]{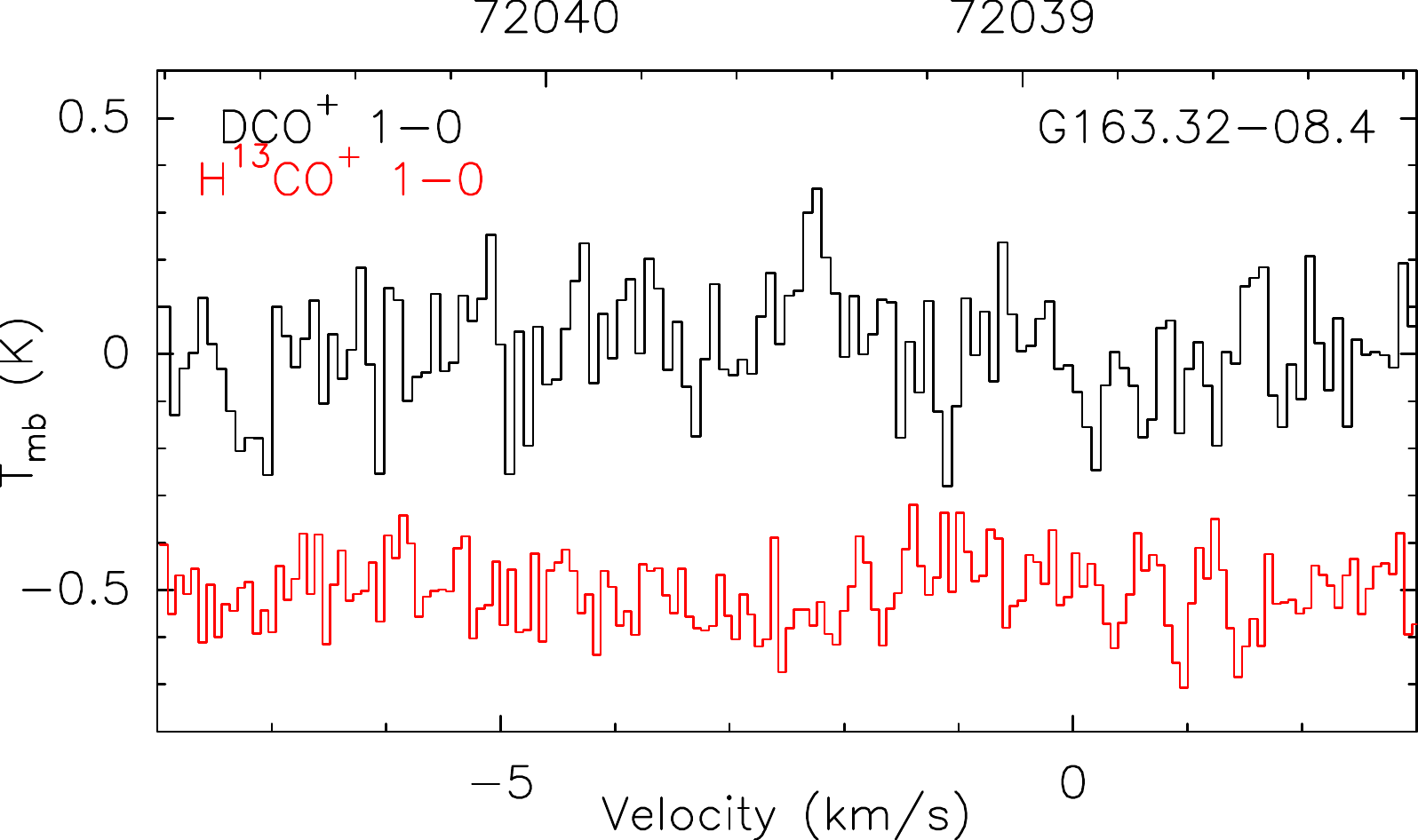}
\includegraphics[width=0.3\columnwidth]{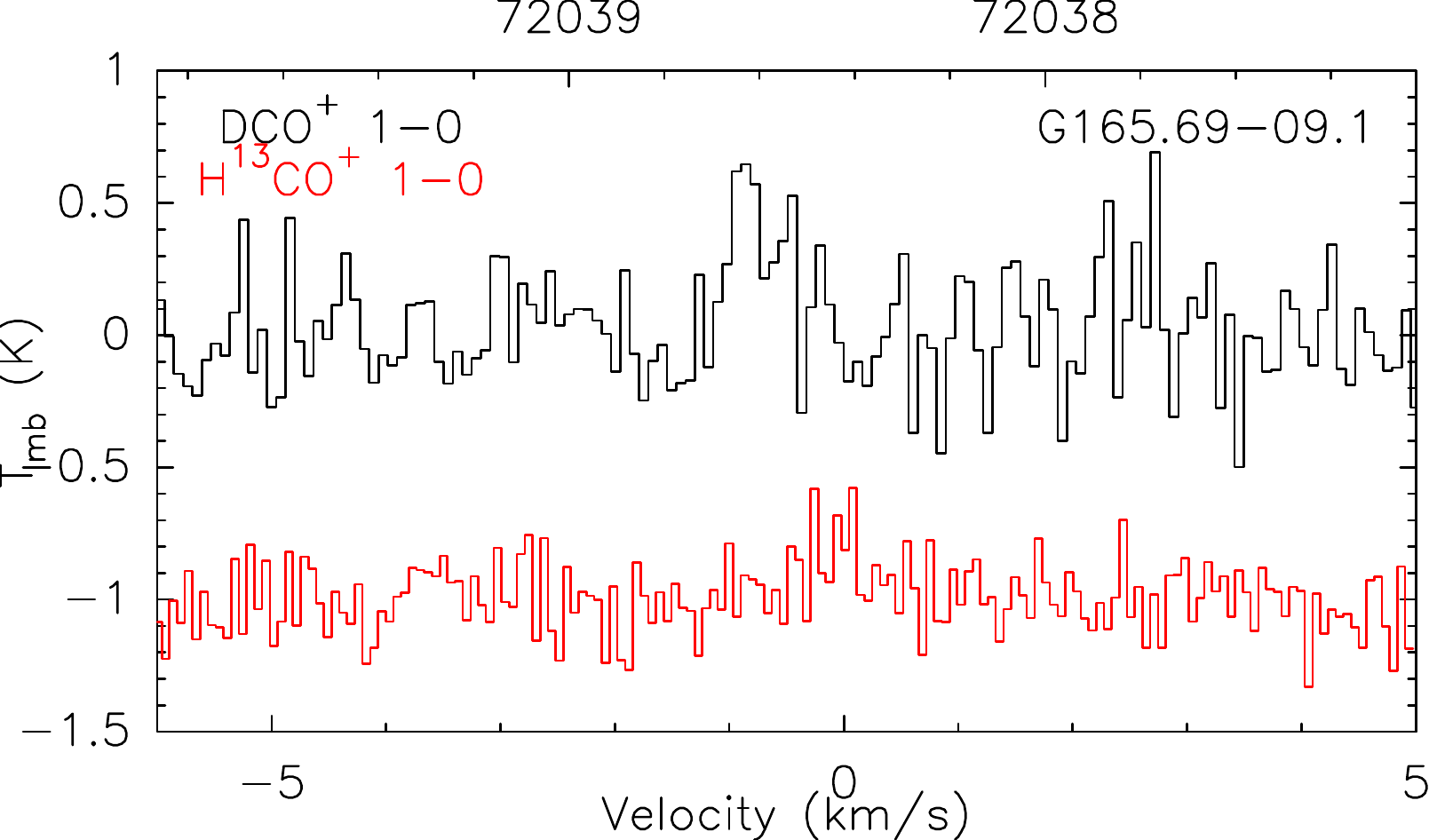}
\caption{Line profiles of DCO$^+$ 1-0 with the high velocity resolution mode (AROWS mode 13). The transitions of H$^{13}$CO$^+$ 1-0 have been observed but not detected in these sources.\centering}
\label{DCO+mode13_2}
\end{figure}
\addtocounter{figure}{-1}
\begin{figure}
\centering
\includegraphics[width=0.3\columnwidth]{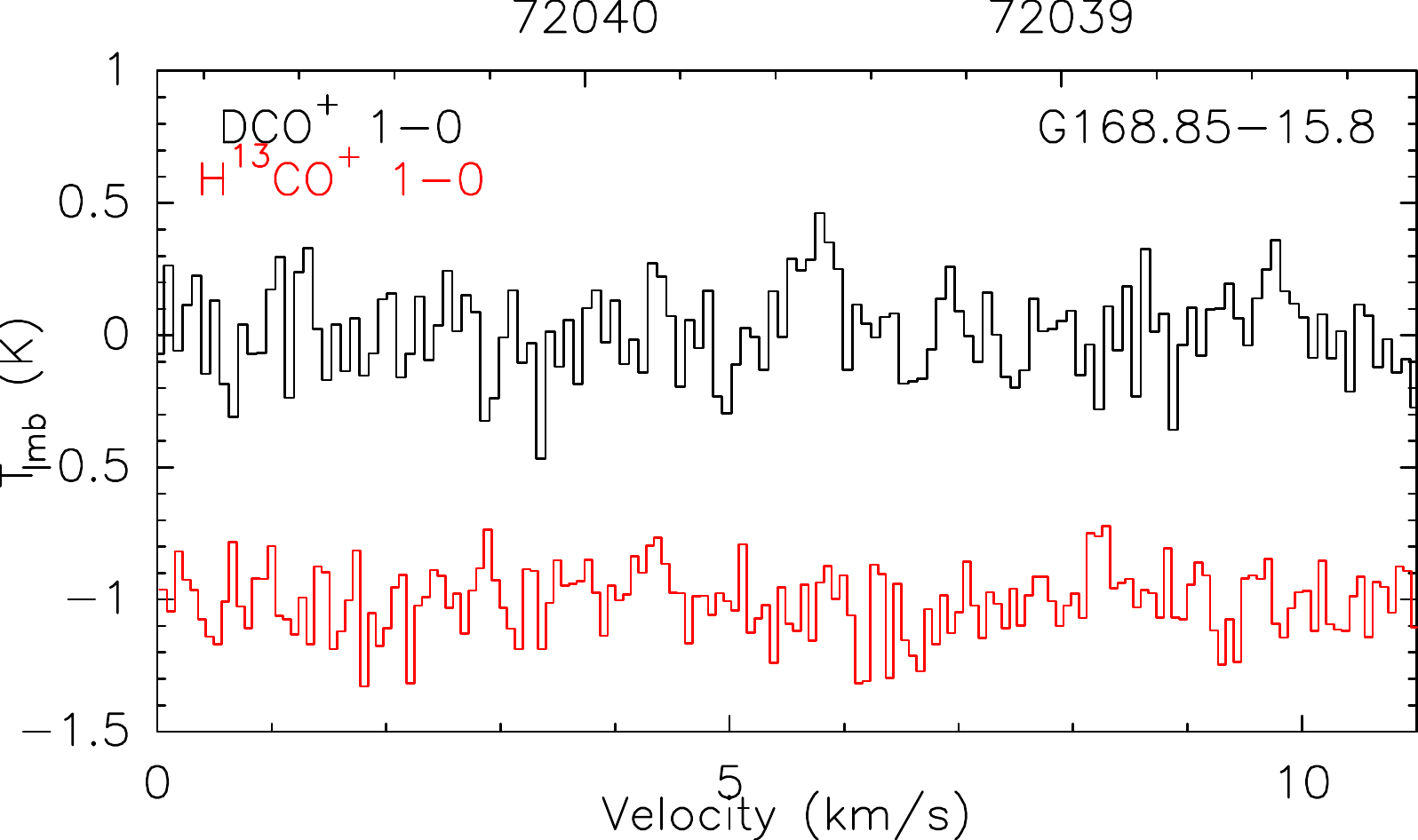}
\includegraphics[width=0.3\columnwidth]{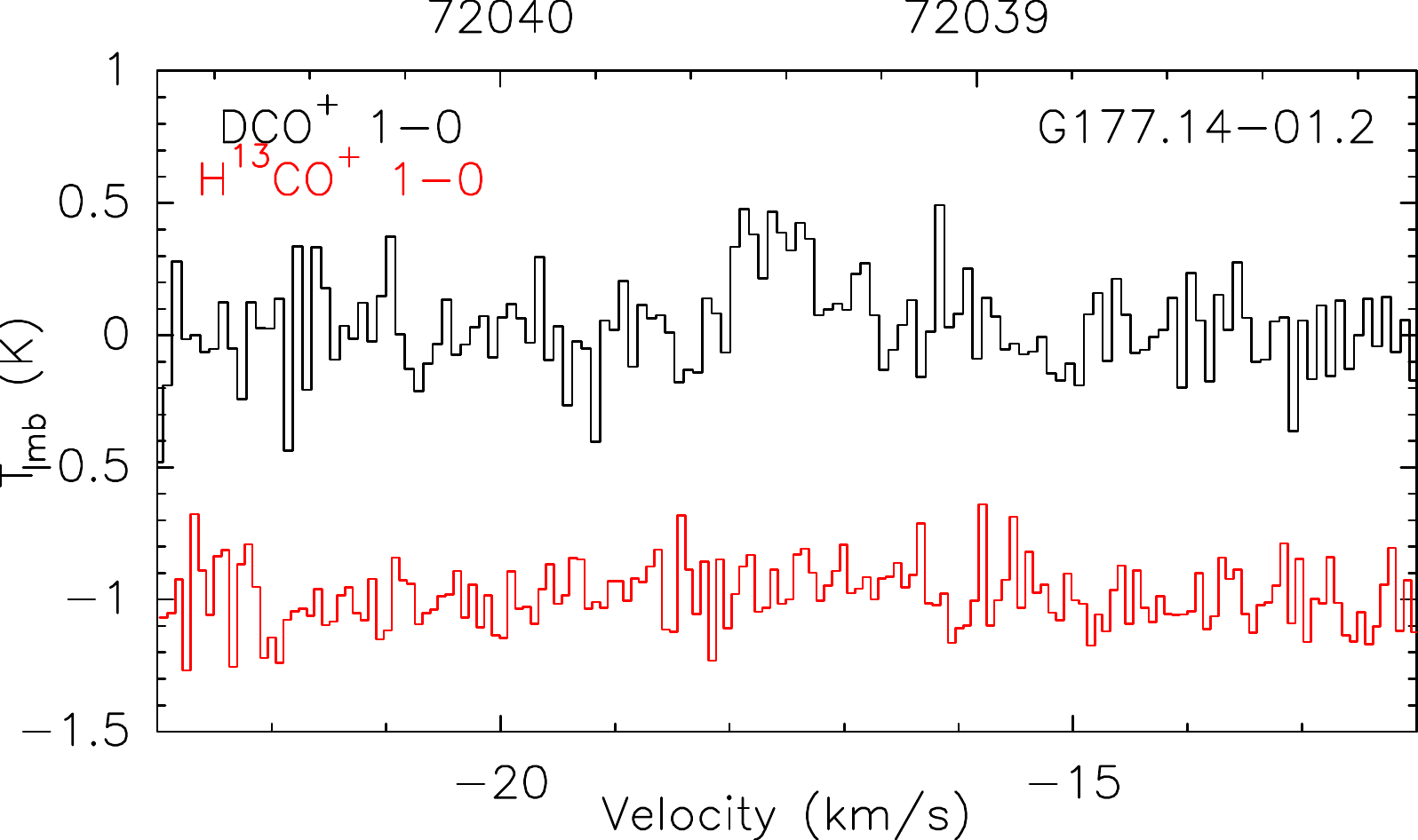}
\includegraphics[width=0.3\columnwidth]{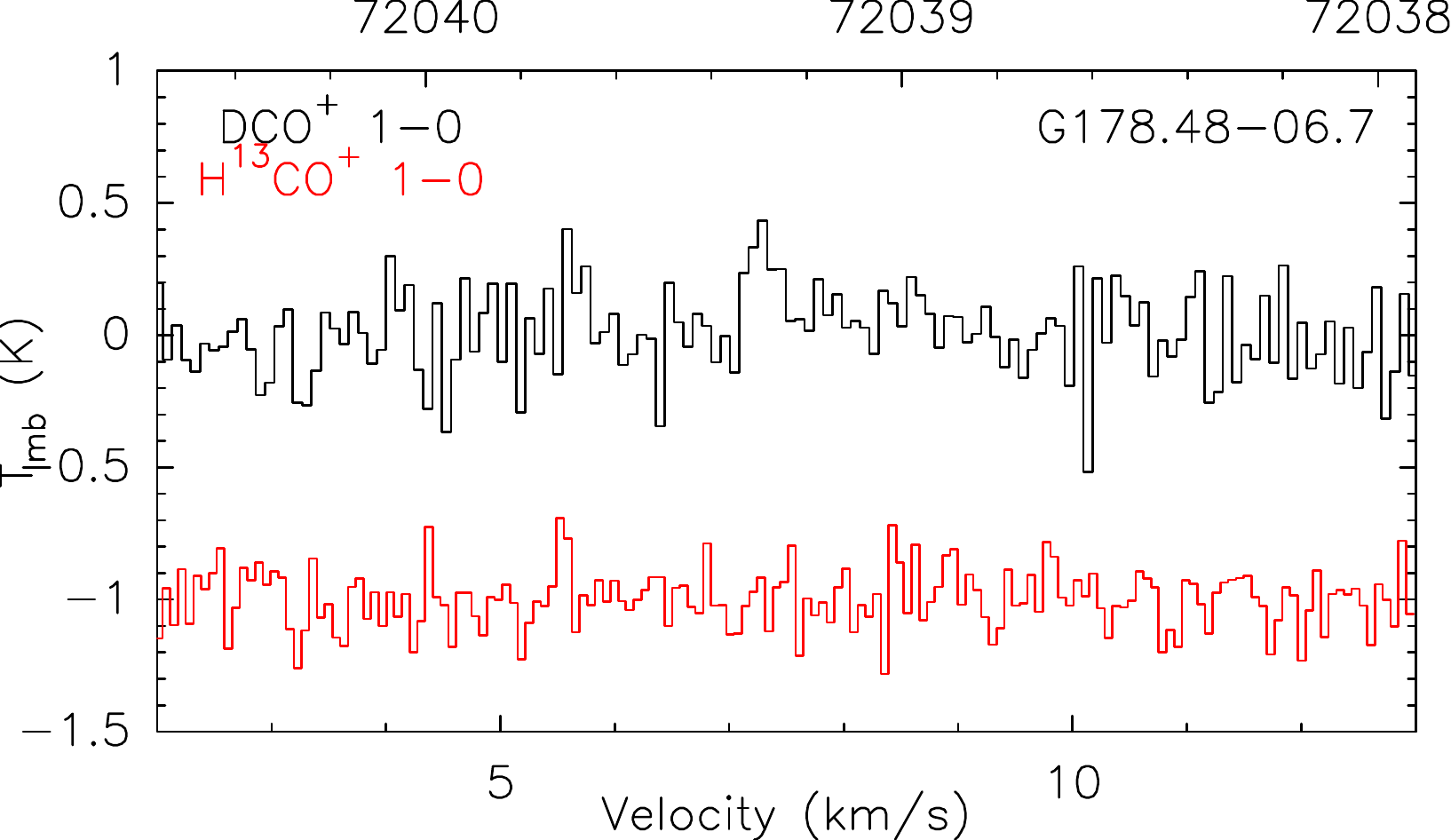}
\includegraphics[width=0.3\columnwidth]{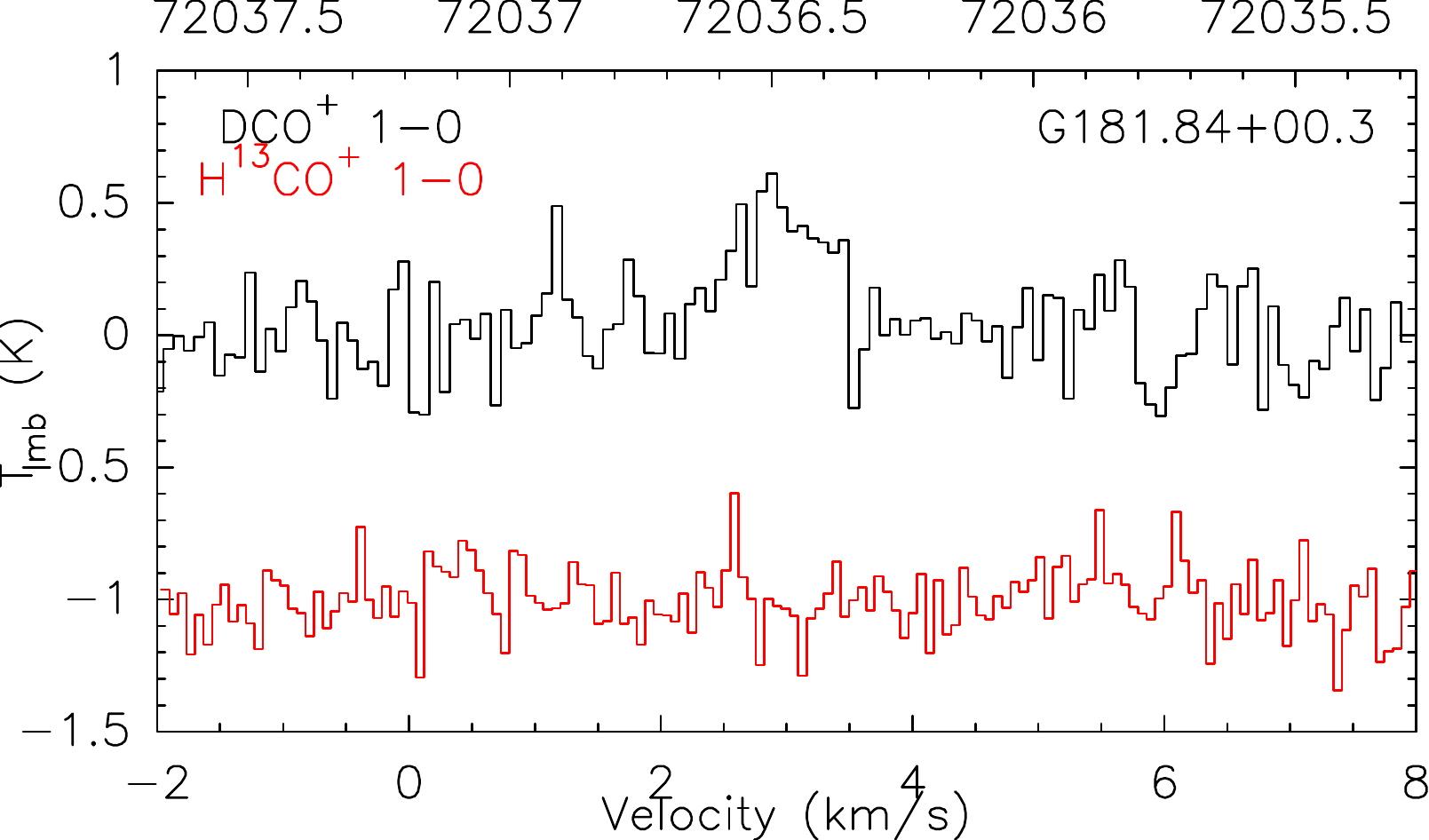}
\includegraphics[width=0.3\columnwidth]{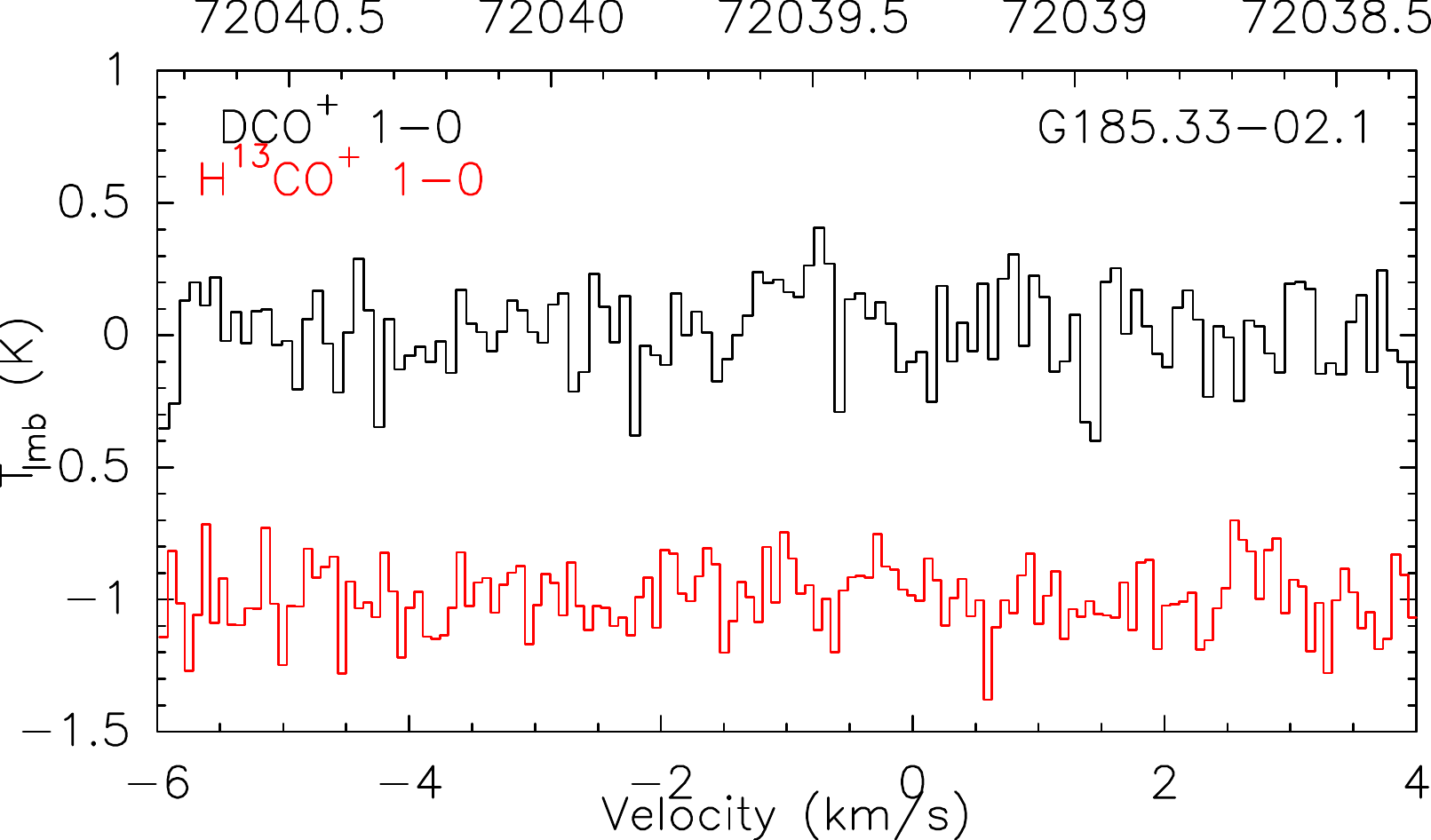}
\includegraphics[width=0.3\columnwidth]{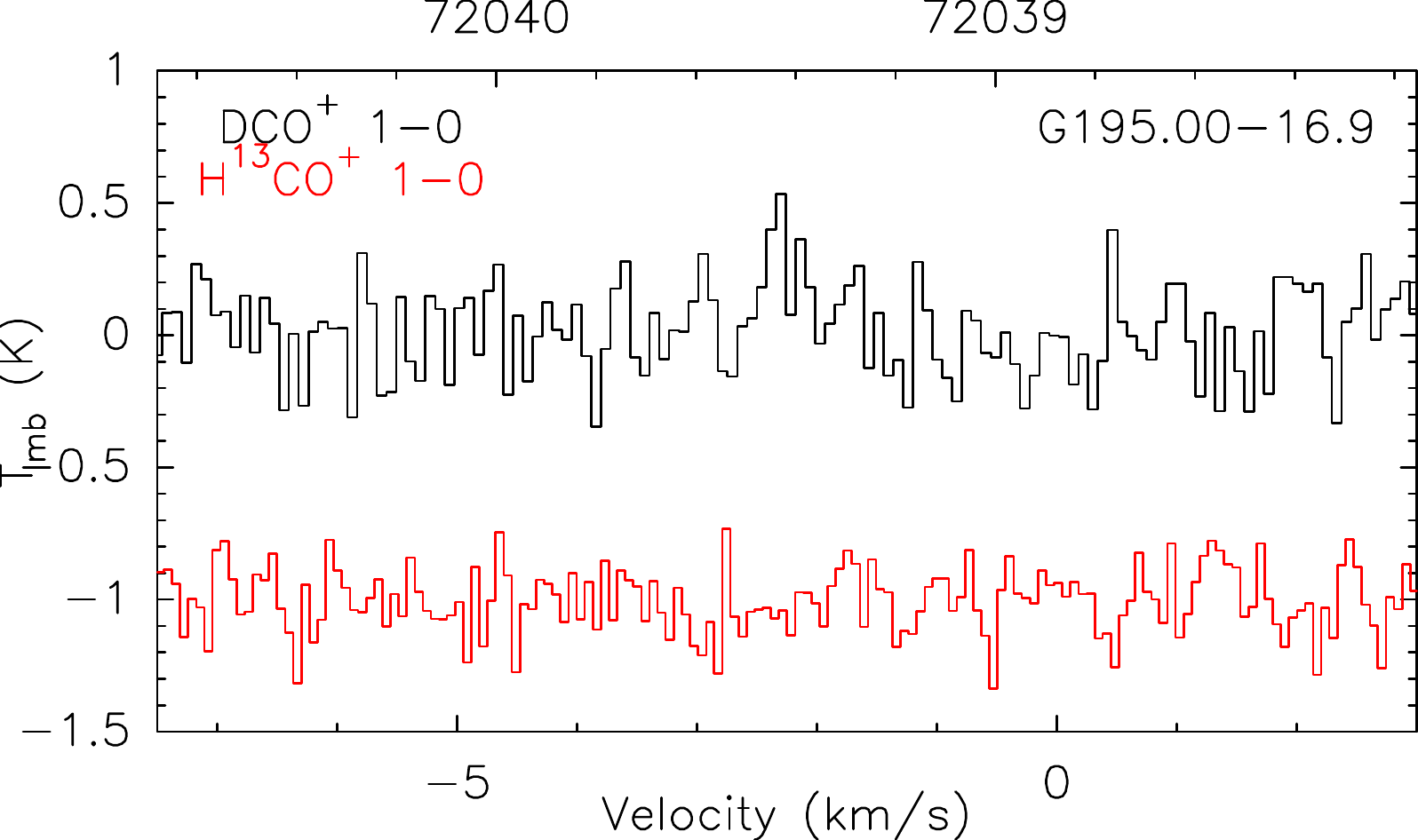}
\includegraphics[width=0.3\columnwidth]{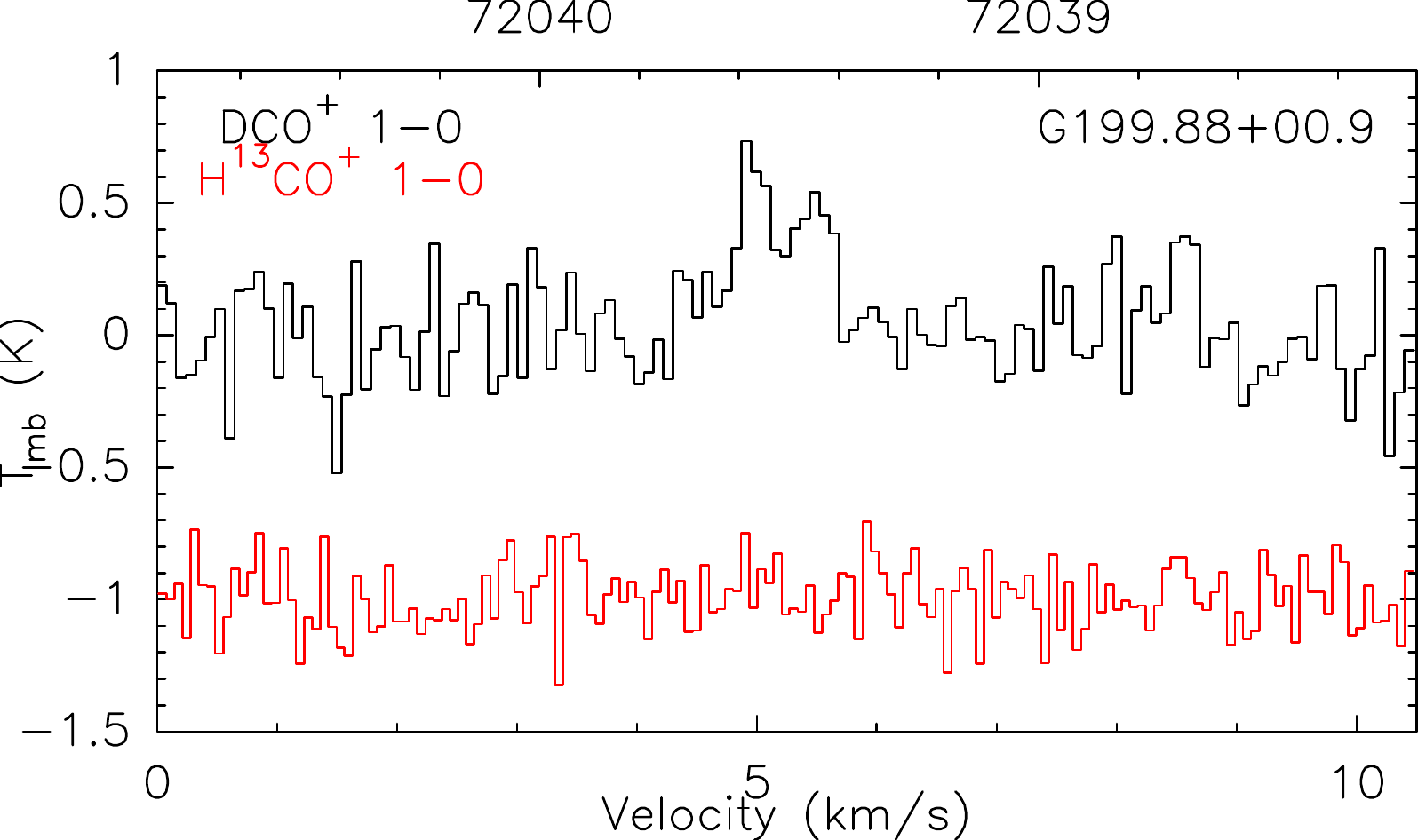}
\includegraphics[width=0.3\columnwidth]{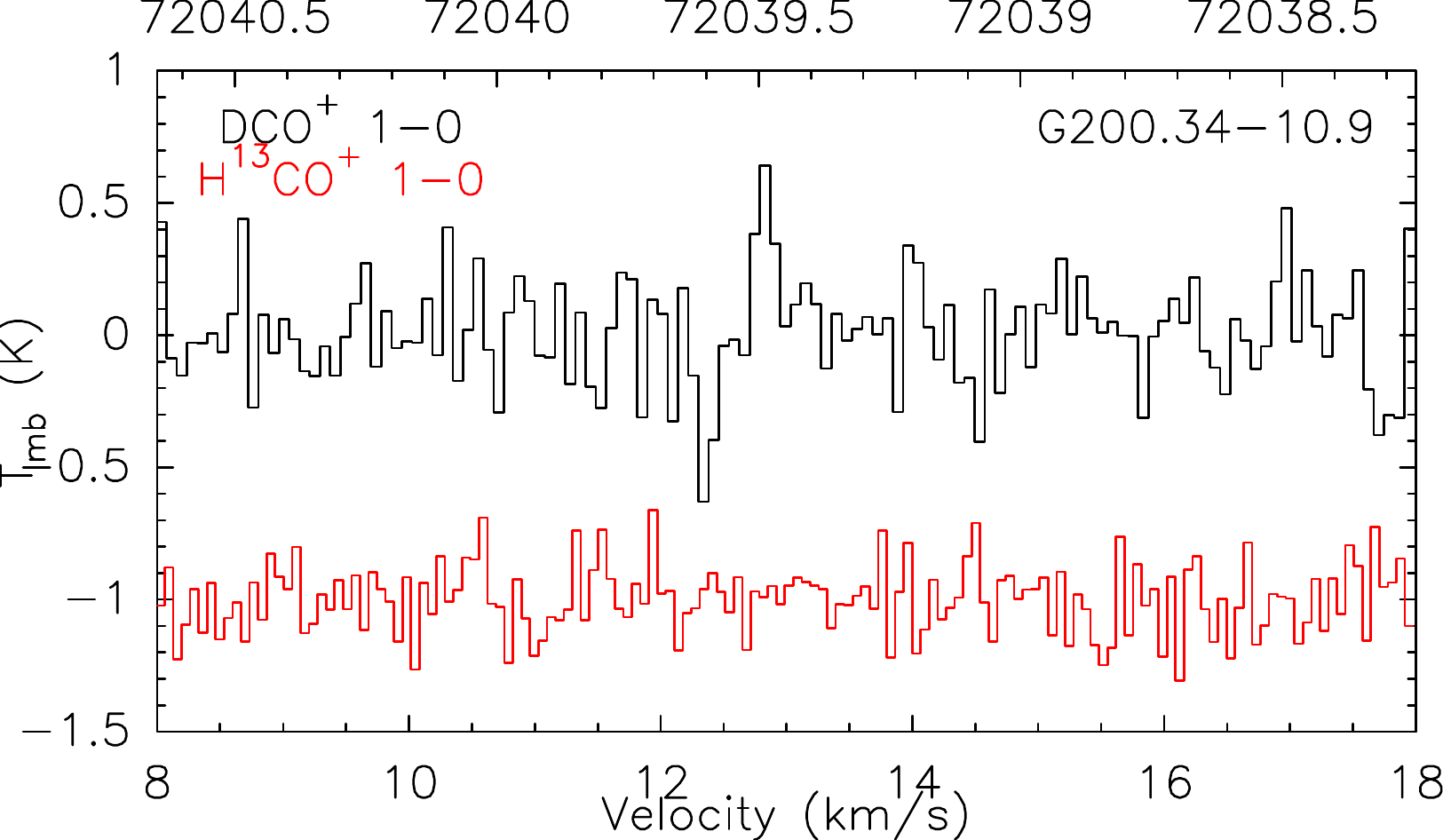}
\includegraphics[width=0.3\columnwidth]{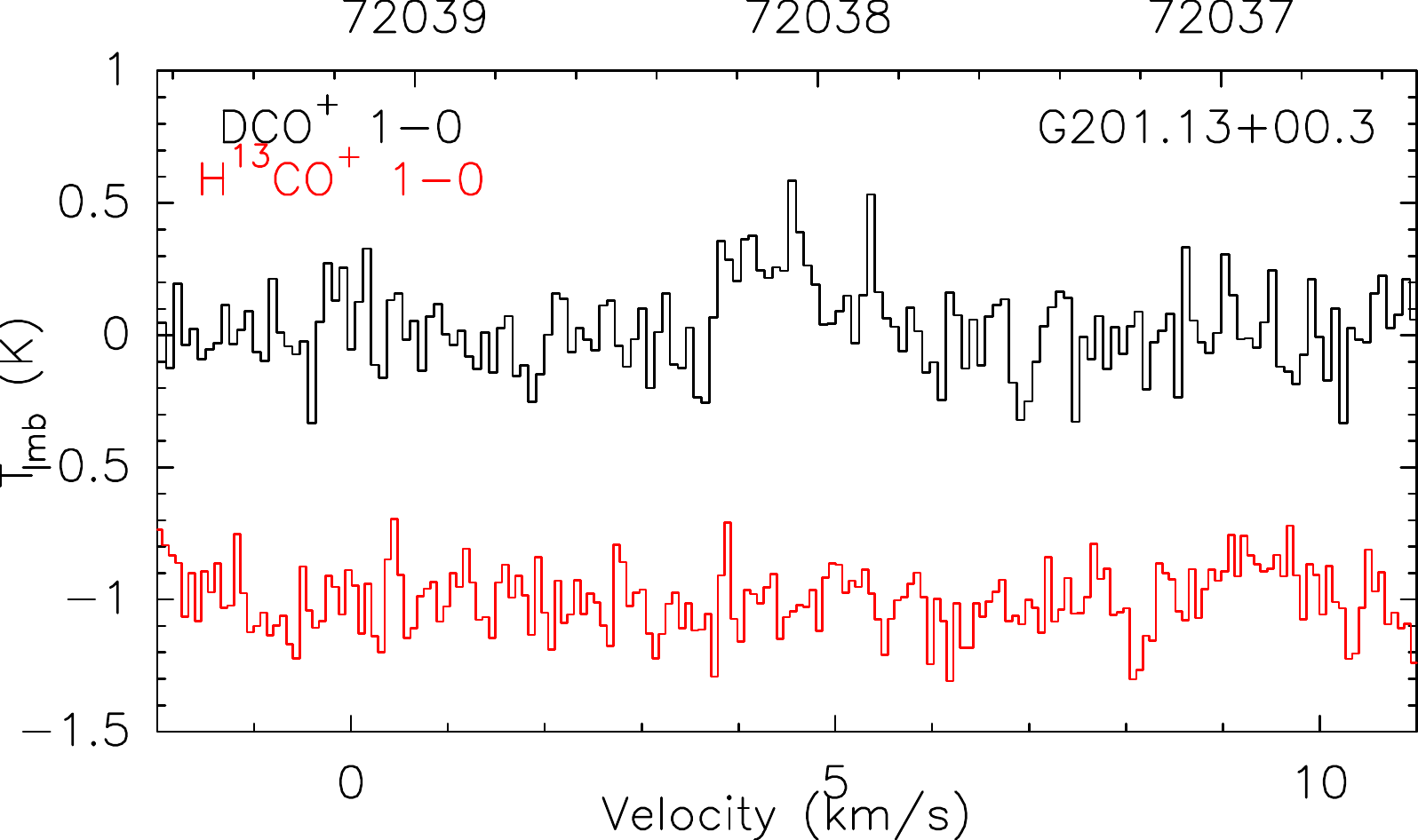}
\caption{Continued.\centering}
\label{DCO+mode13_2}
\end{figure}
\begin{figure}
\centering
\includegraphics[width=0.3\columnwidth]{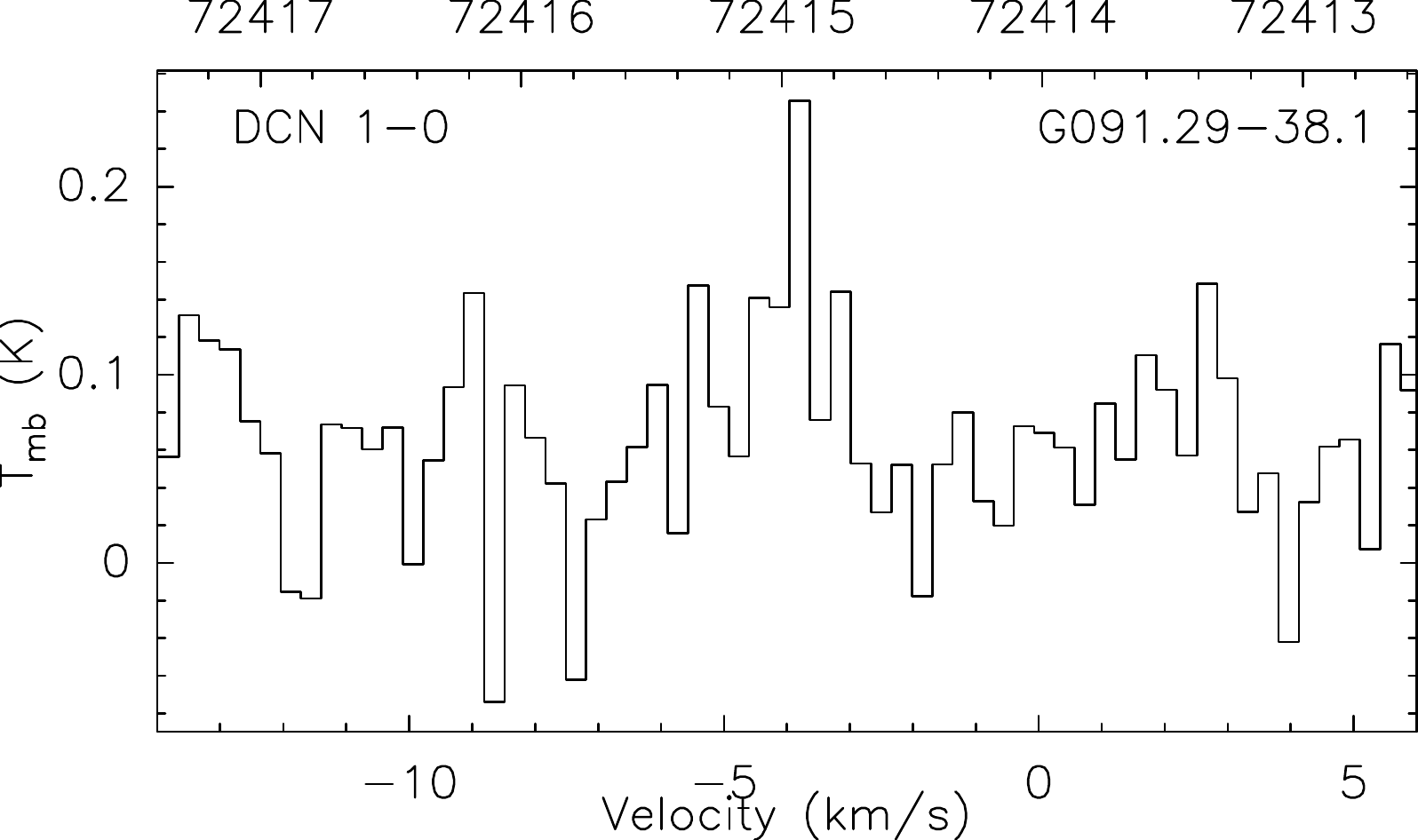}
\includegraphics[width=0.3\columnwidth]{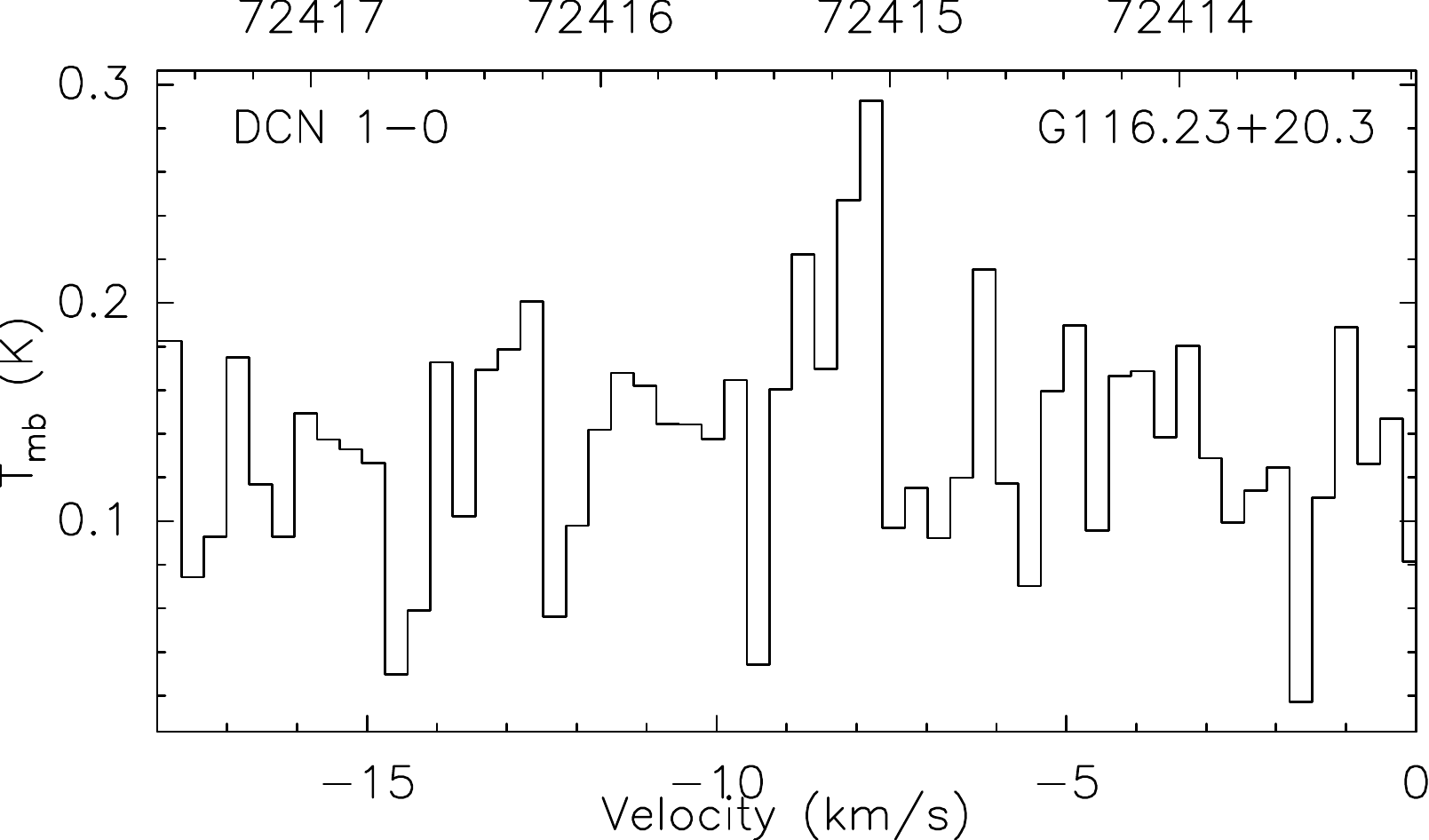}
\caption{Line profiles of DCN 1-0 with the low velocity resolution mode (AROWS mode 3). The transitions of H$^{13}$CN 1-0 have not been observed in these sources.\centering}
\label{DCNmode3_1}
\end{figure}
\begin{figure}
\centering
\includegraphics[width=0.3\columnwidth]{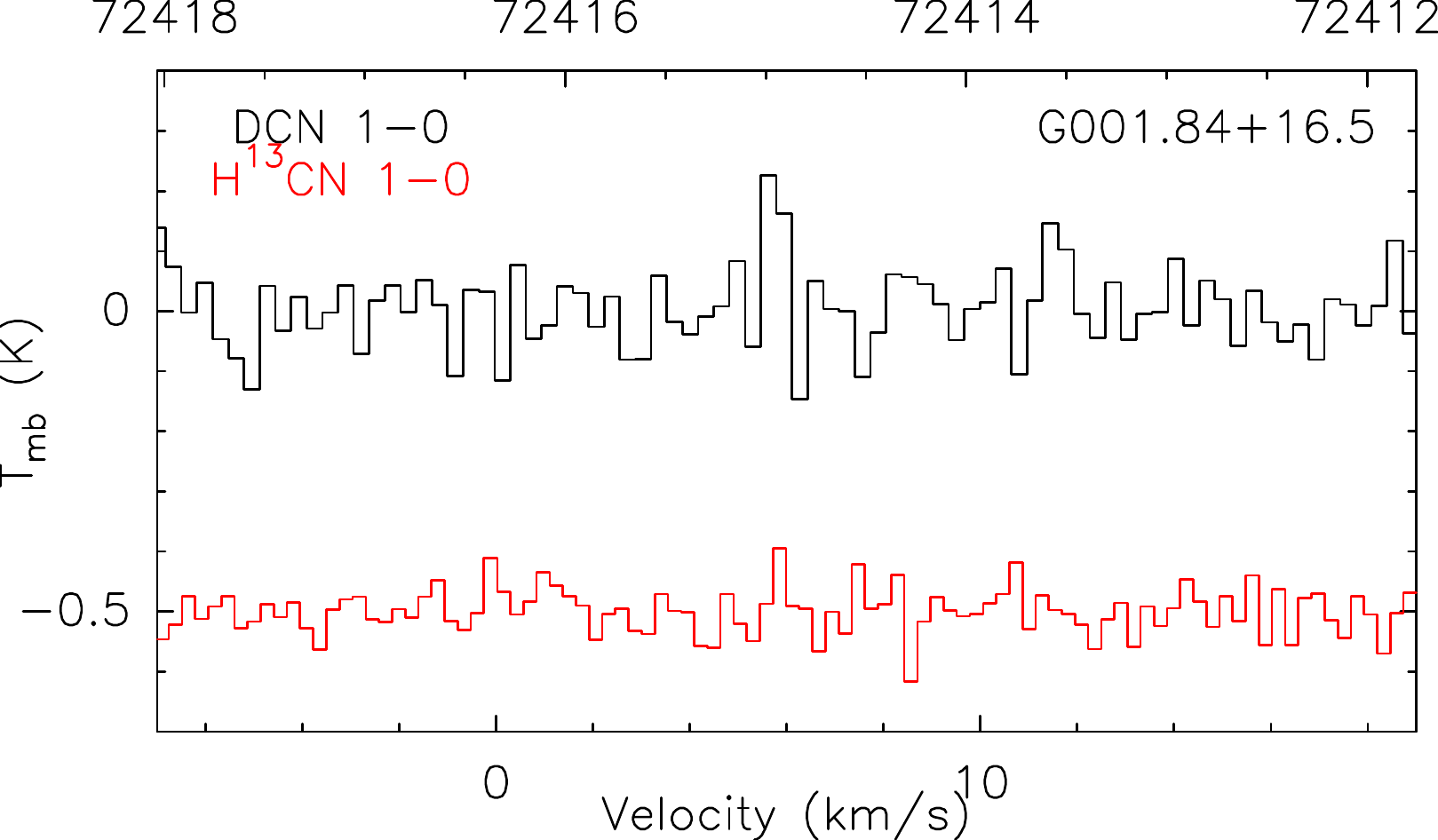}
\includegraphics[width=0.3\columnwidth]{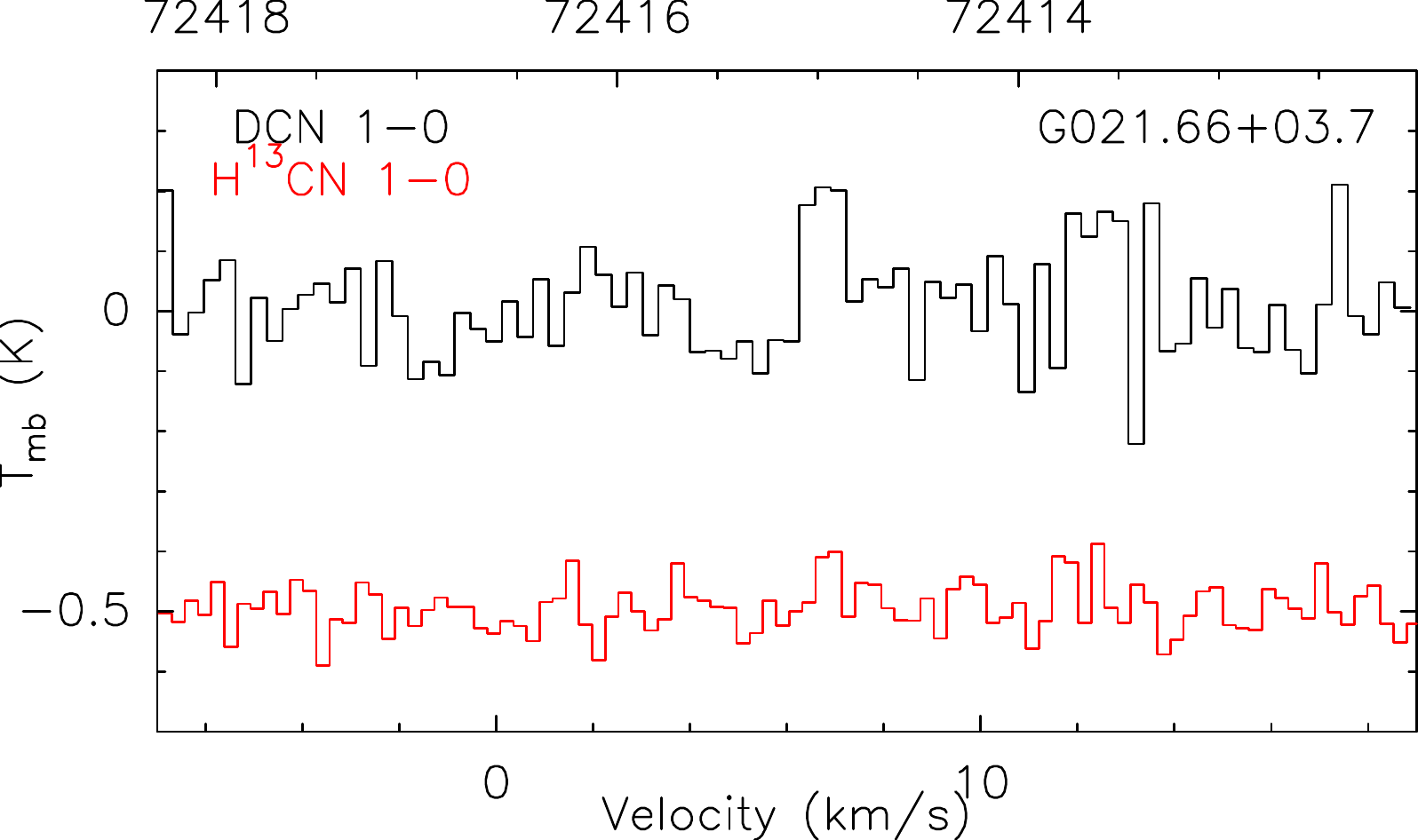}
\includegraphics[width=0.3\columnwidth]{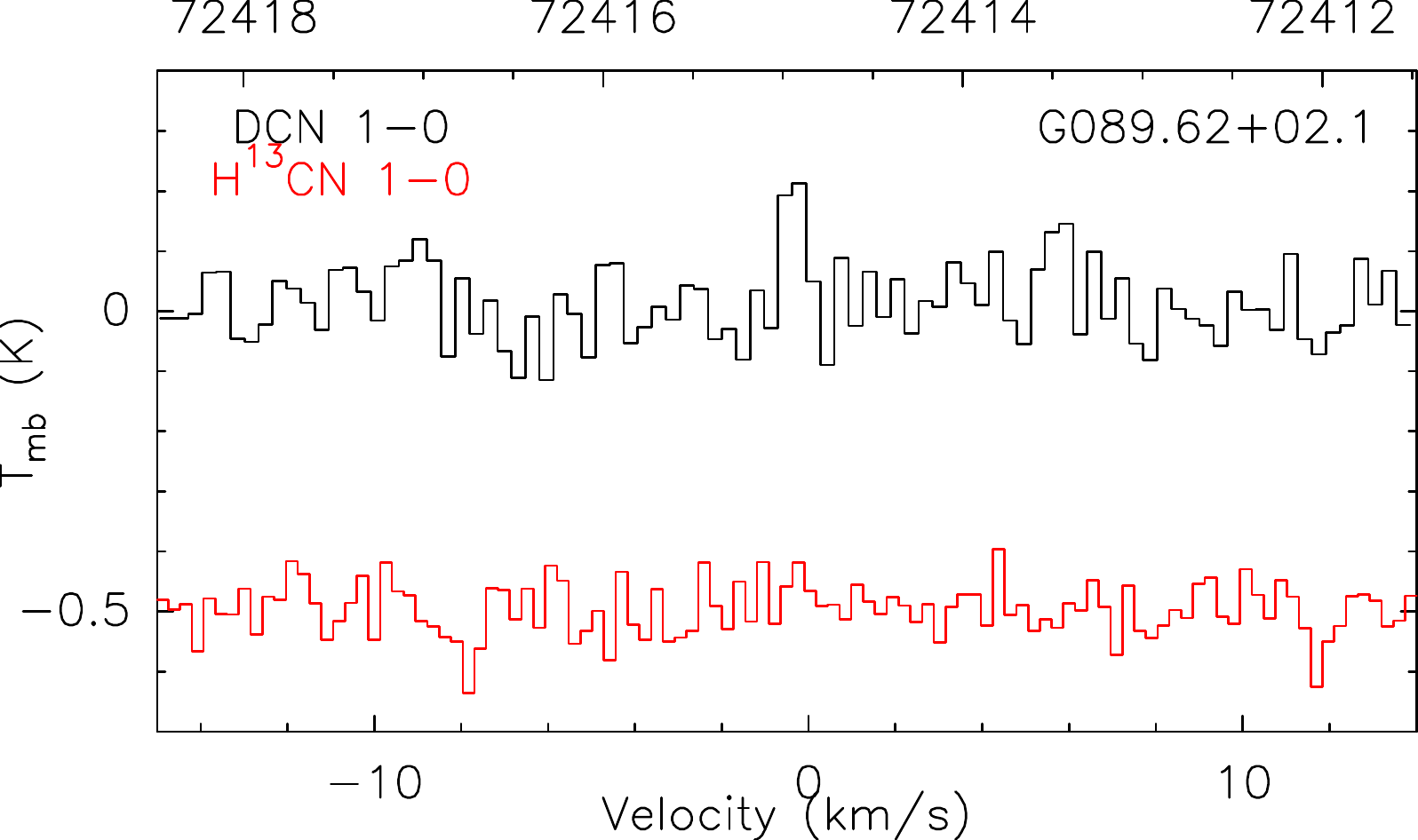}
\includegraphics[width=0.3\columnwidth]{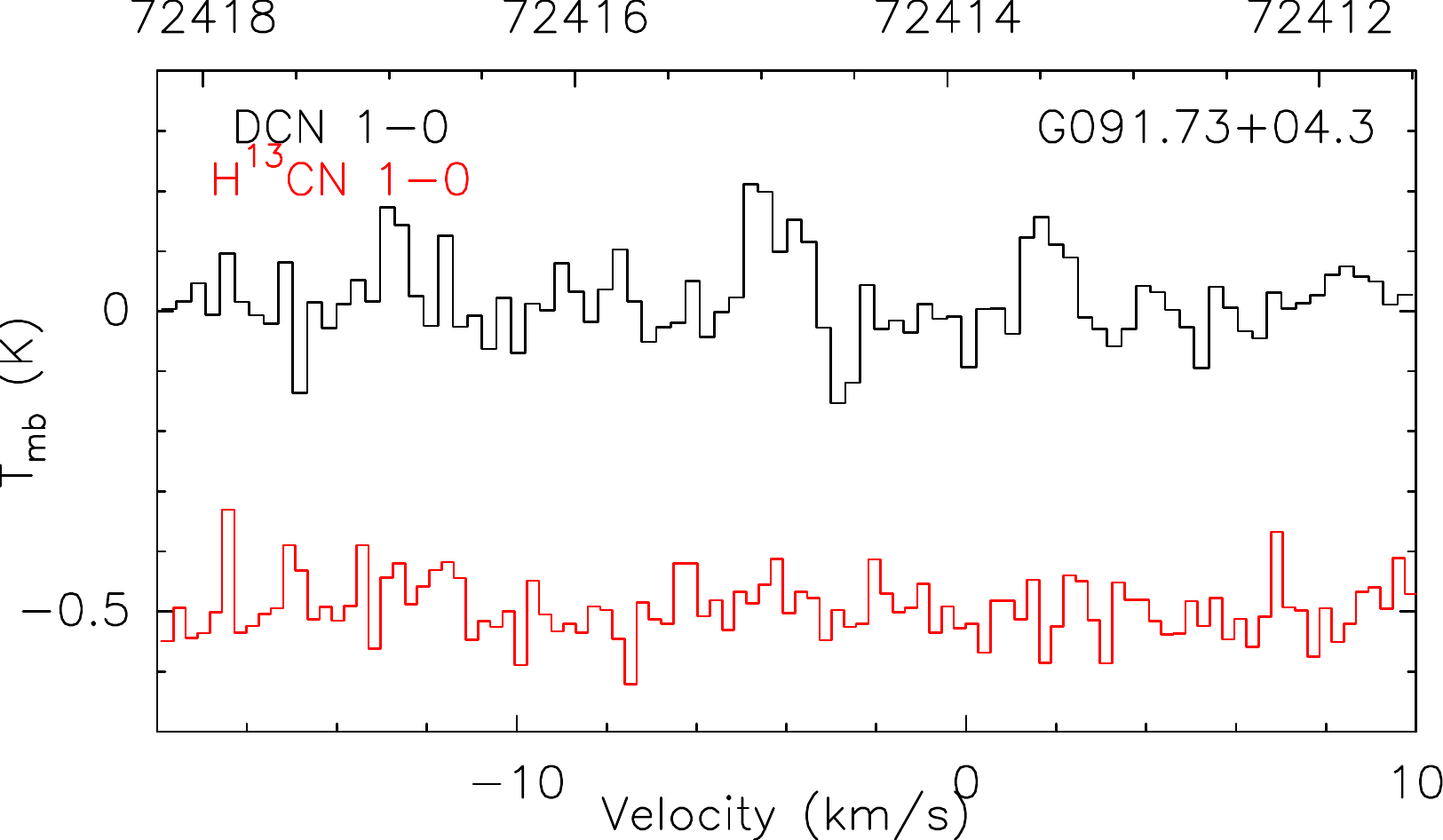}
\includegraphics[width=0.3\columnwidth]{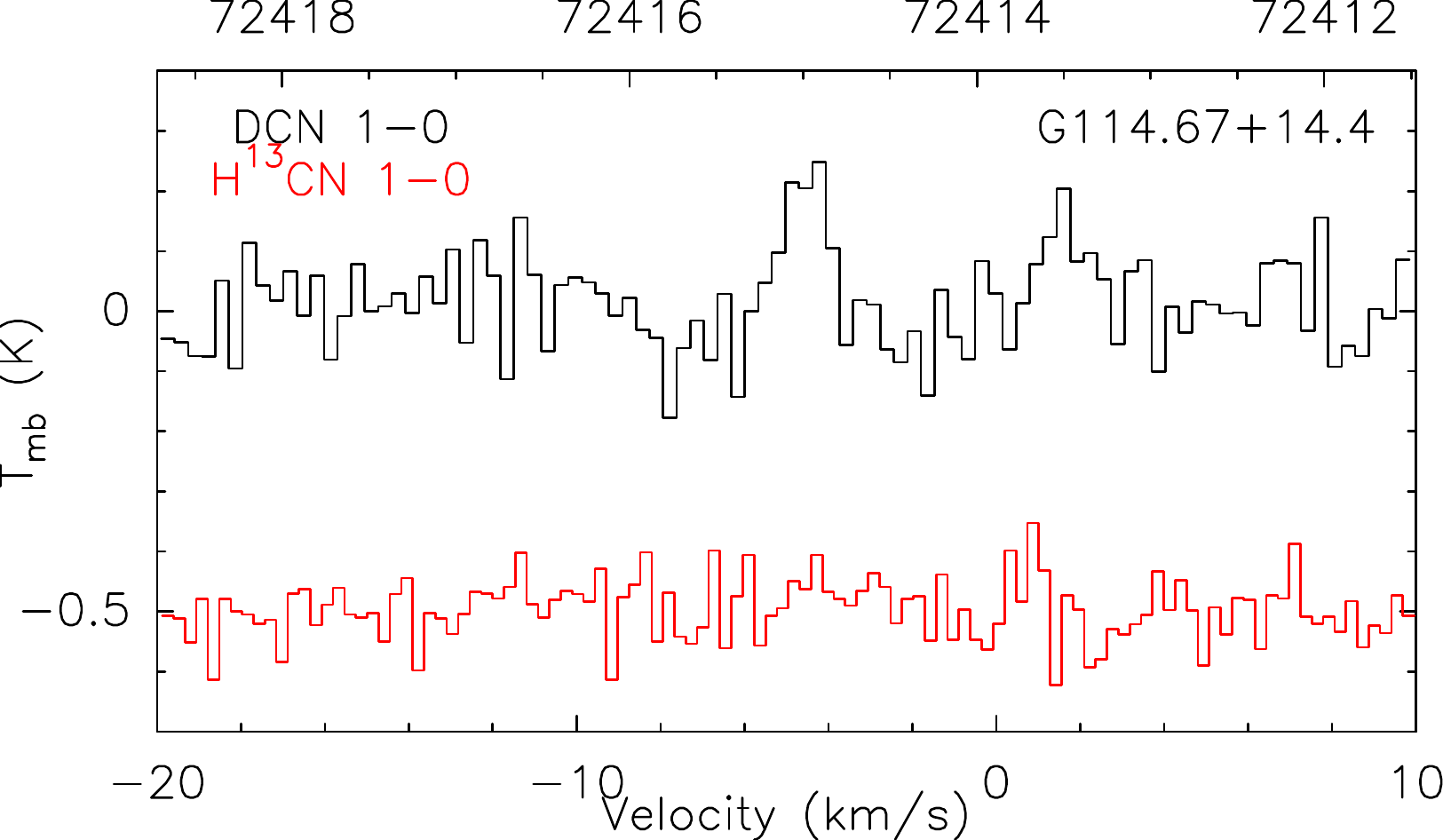}
\caption{Line profiles of DCN 1-0 with the low velocity resolution mode (AROWS mode 3). The transitions of H$^{13}$CN 1-0 have been observed but not detected in these sources.\centering}
\label{DCNmode3_2}
\end{figure}
\begin{figure}
\centering
\includegraphics[width=0.3\columnwidth]{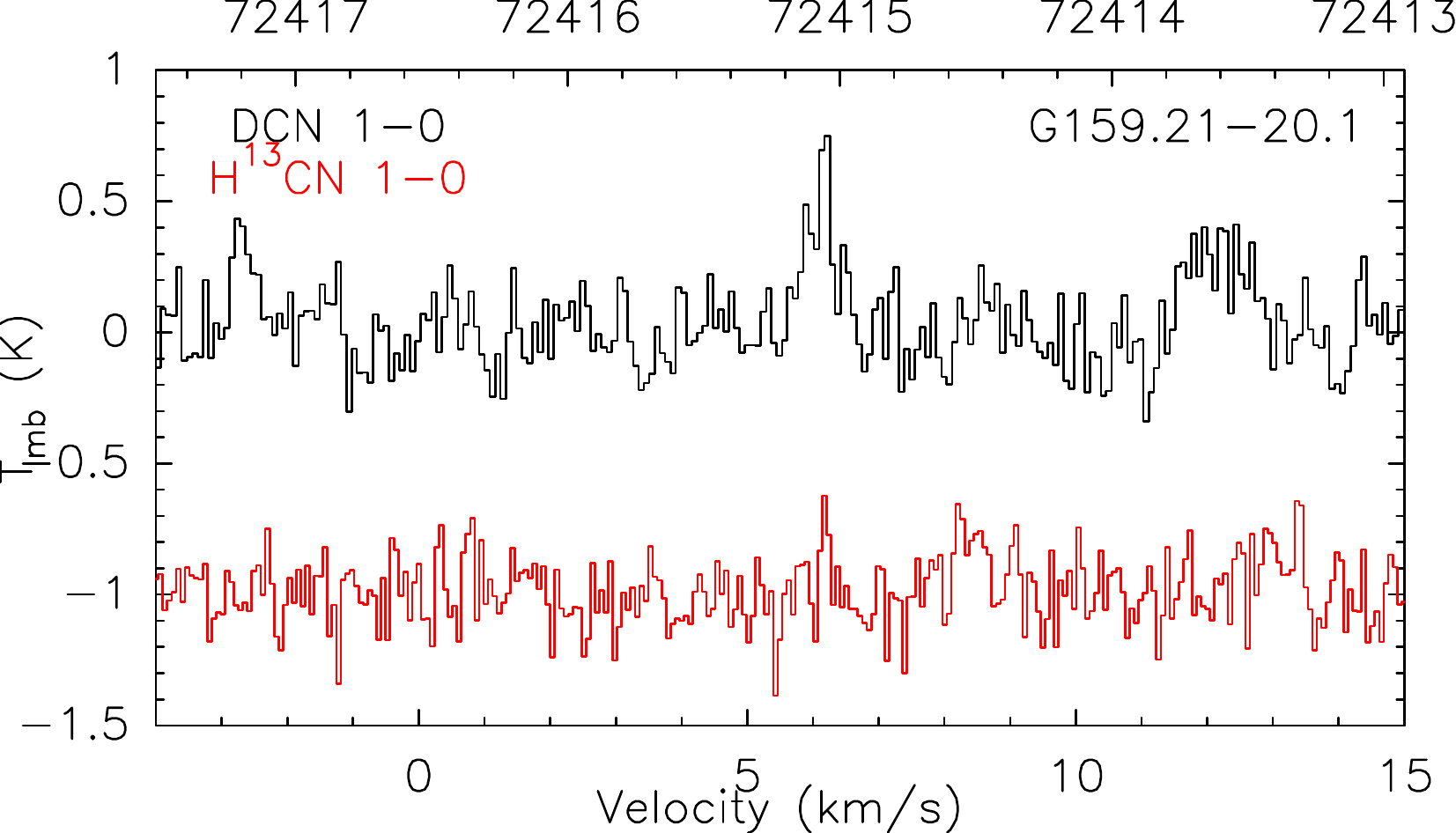}
\includegraphics[width=0.3\columnwidth]{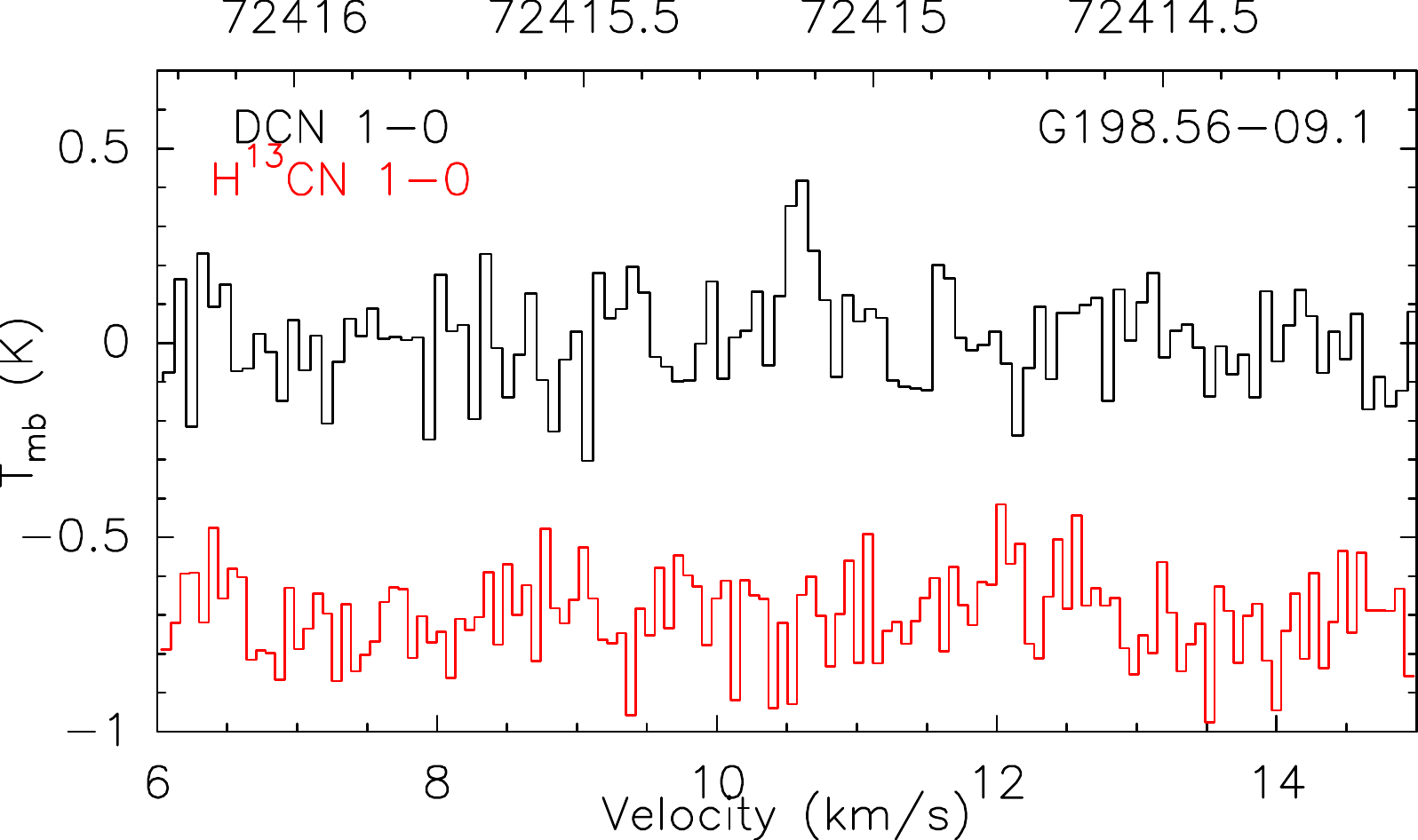}
\caption{Line profiles of DCN 1-0 with the high velocity resolution mode (AROWS mode 13). The transitions of H$^{13}$CN 1-0 have been observed but not detected in these sources.\centering}
\label{DCNmode13_2}
\end{figure}
\begin{figure}
\centering
\includegraphics[width=0.3\columnwidth]{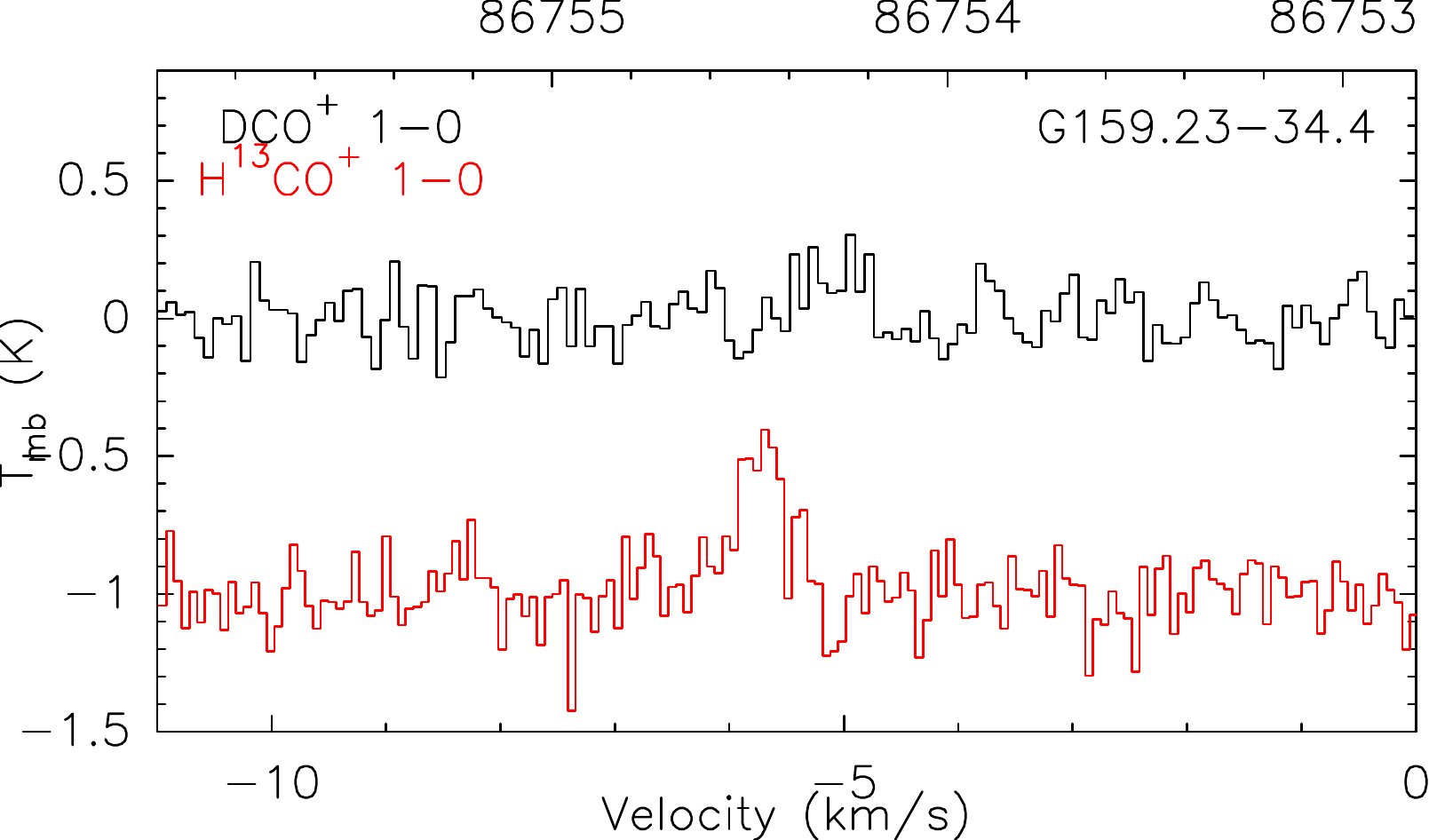}
\includegraphics[width=0.3\columnwidth]{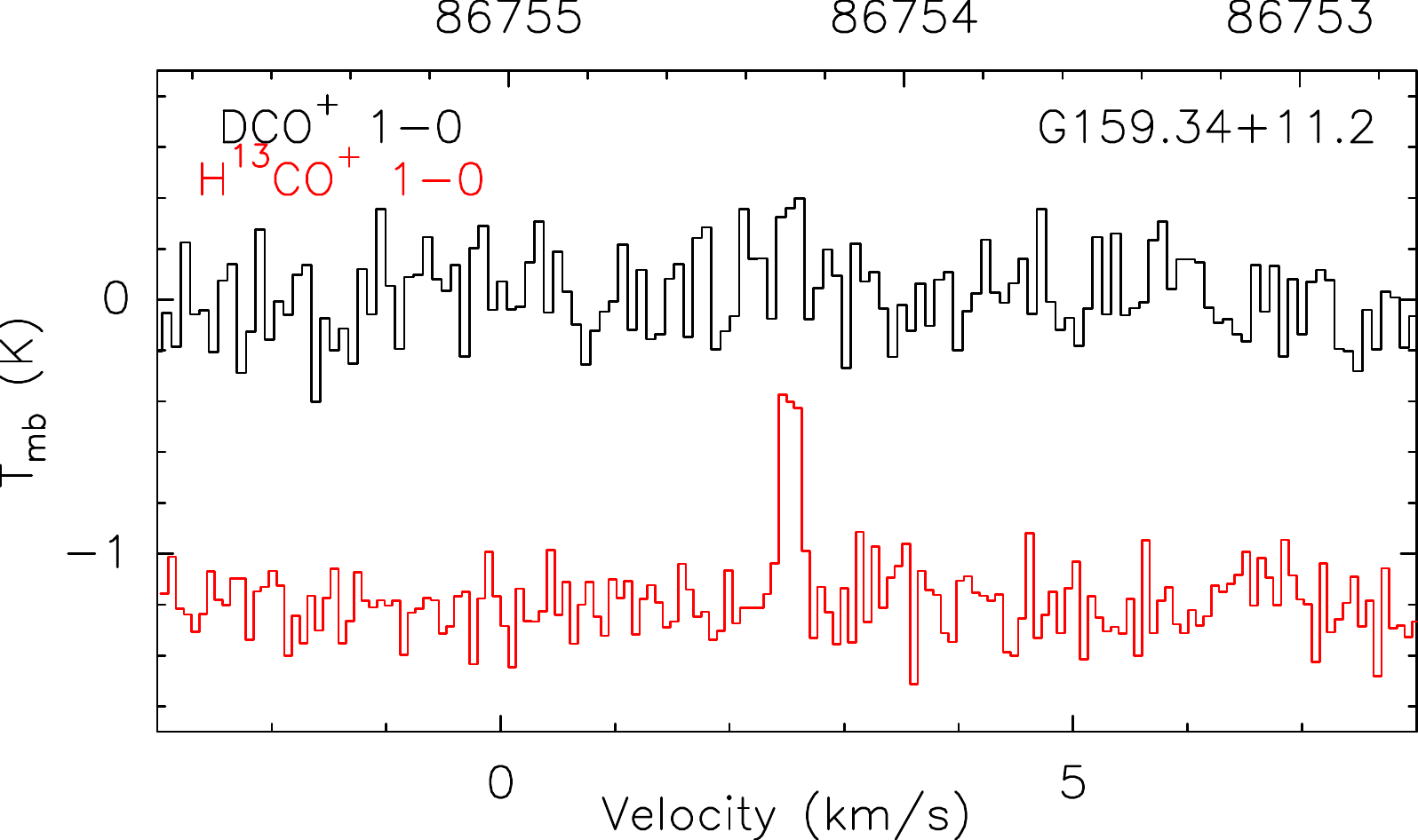}
\includegraphics[width=0.3\columnwidth]{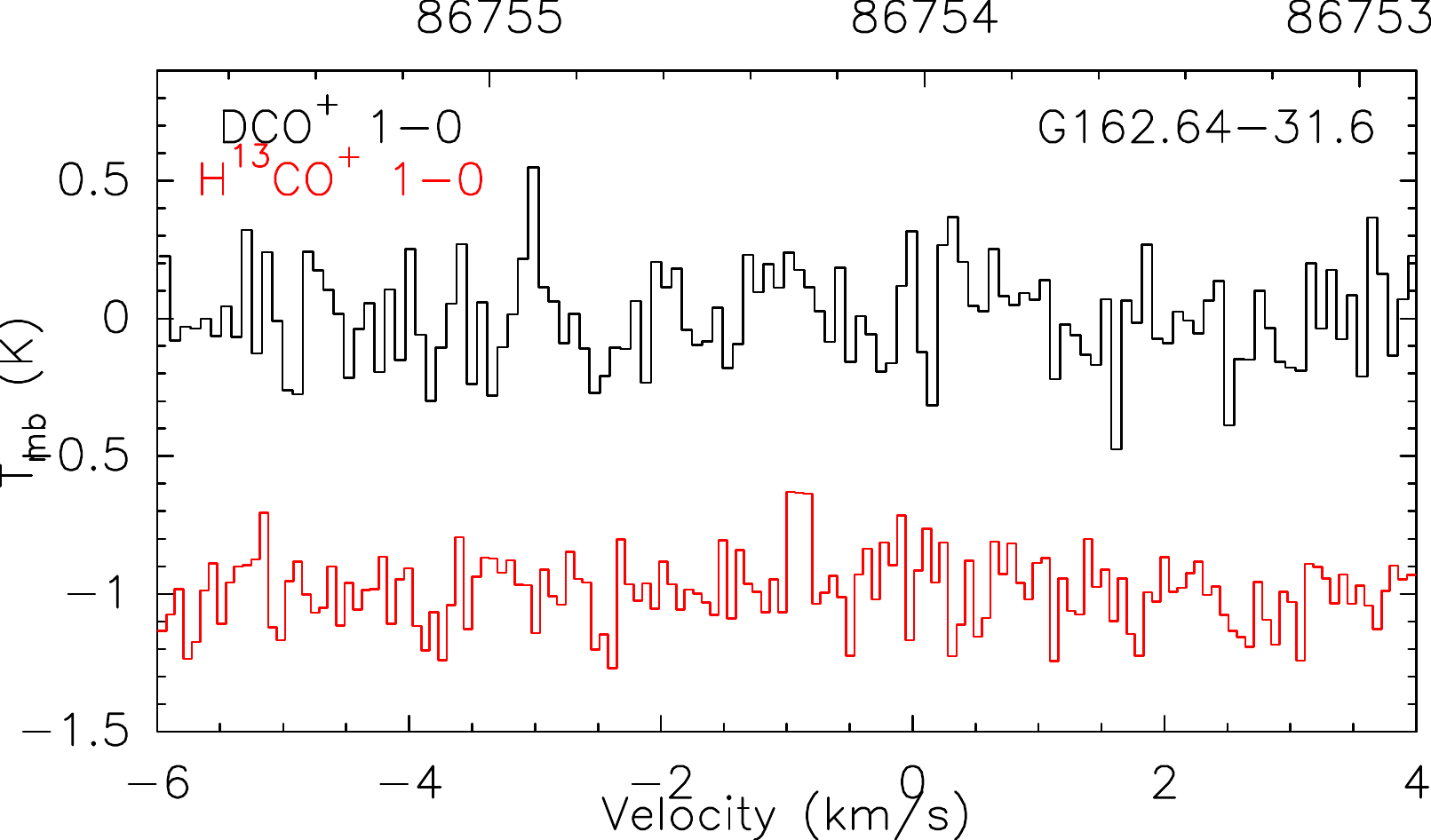}
\includegraphics[width=0.3\columnwidth]{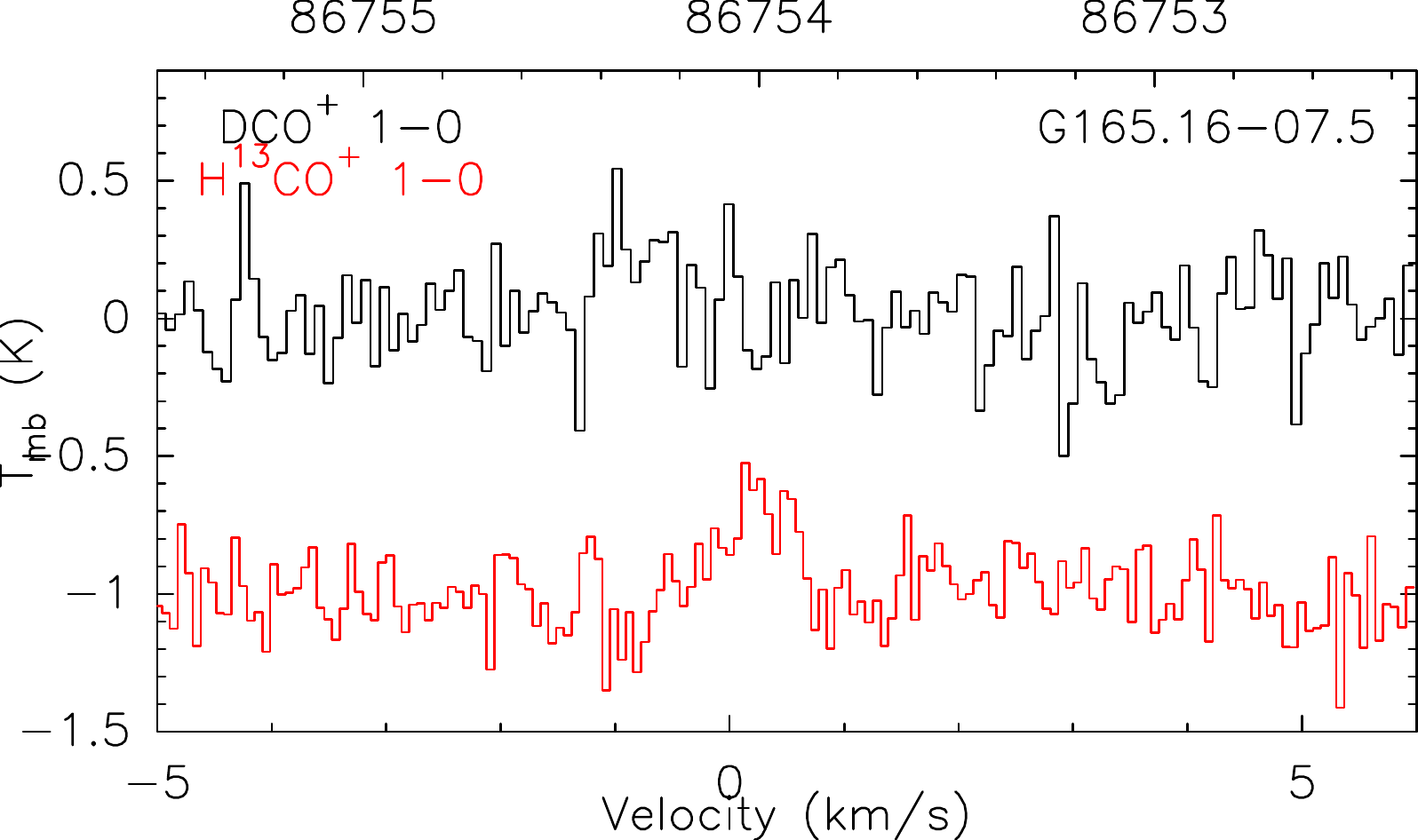}
\includegraphics[width=0.3\columnwidth]{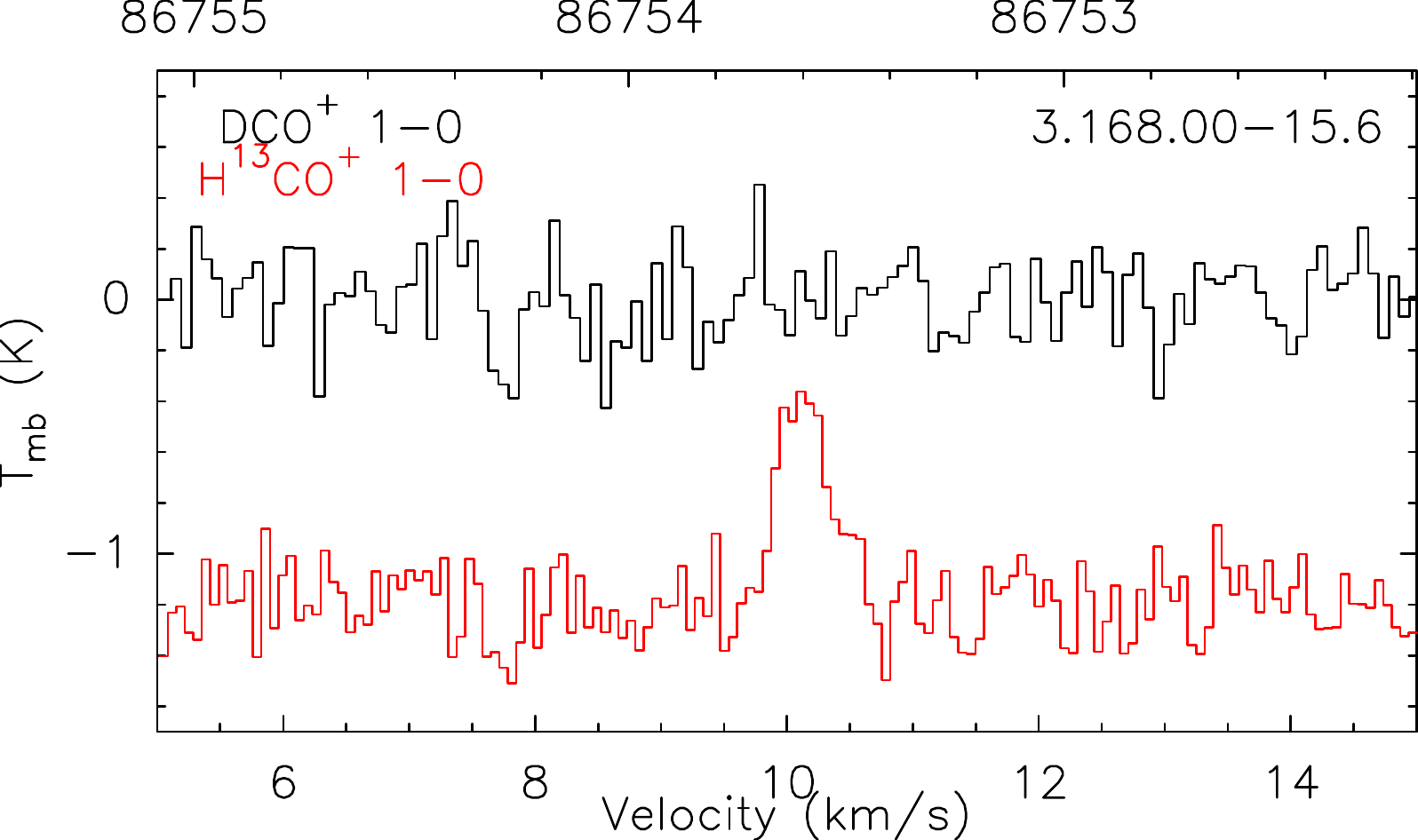}
\includegraphics[width=0.3\columnwidth]{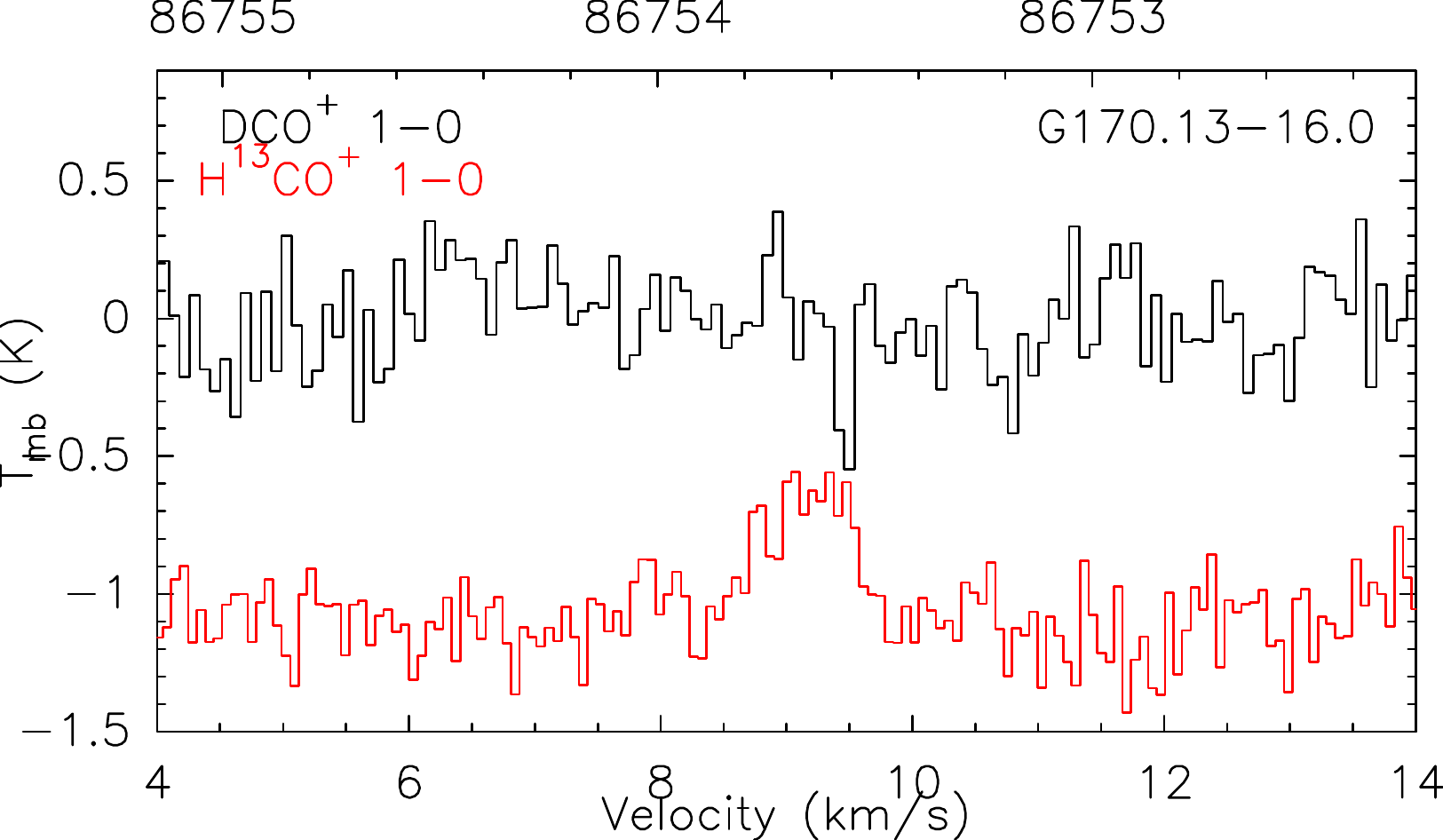}
\includegraphics[width=0.3\columnwidth]{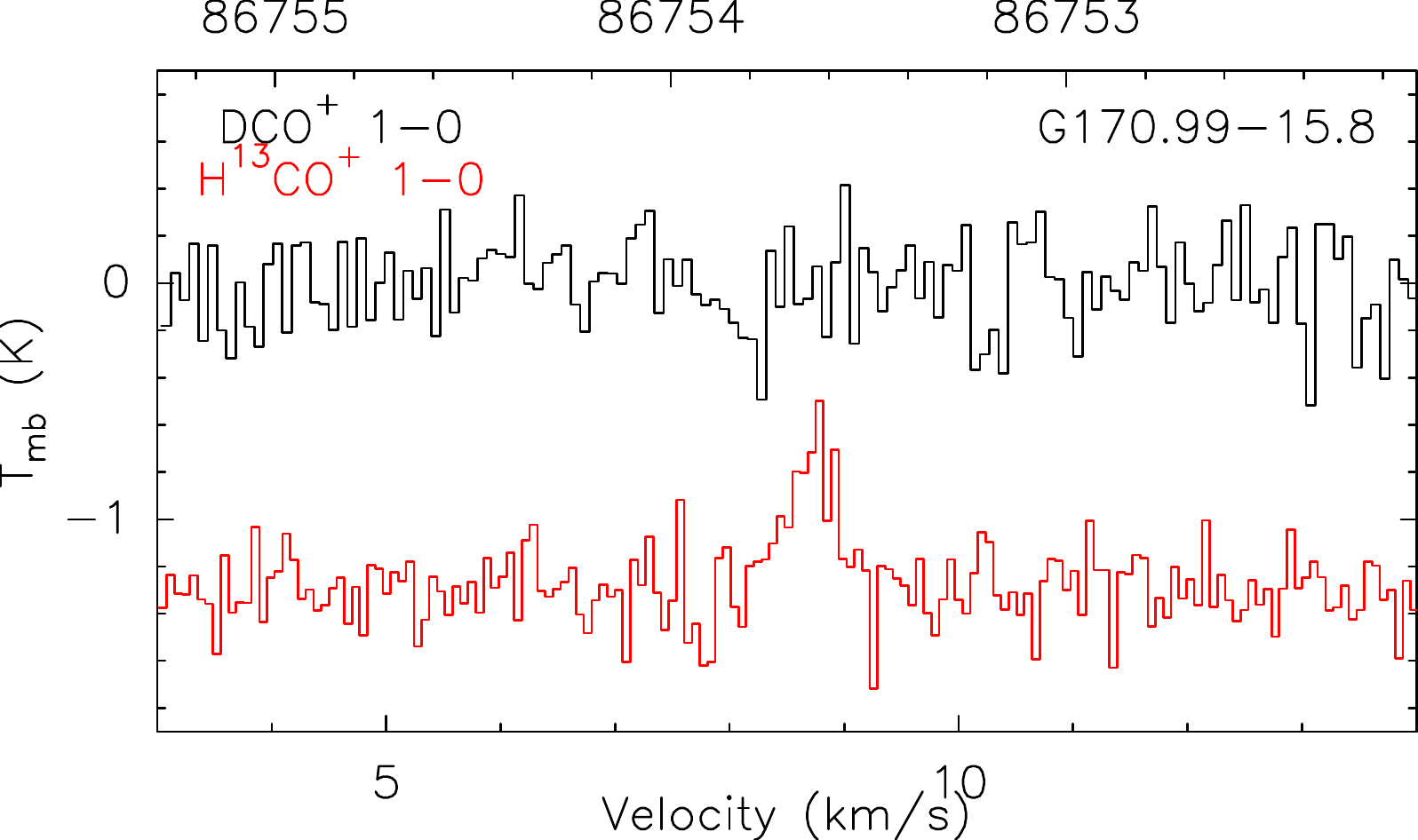}
\includegraphics[width=0.3\columnwidth]{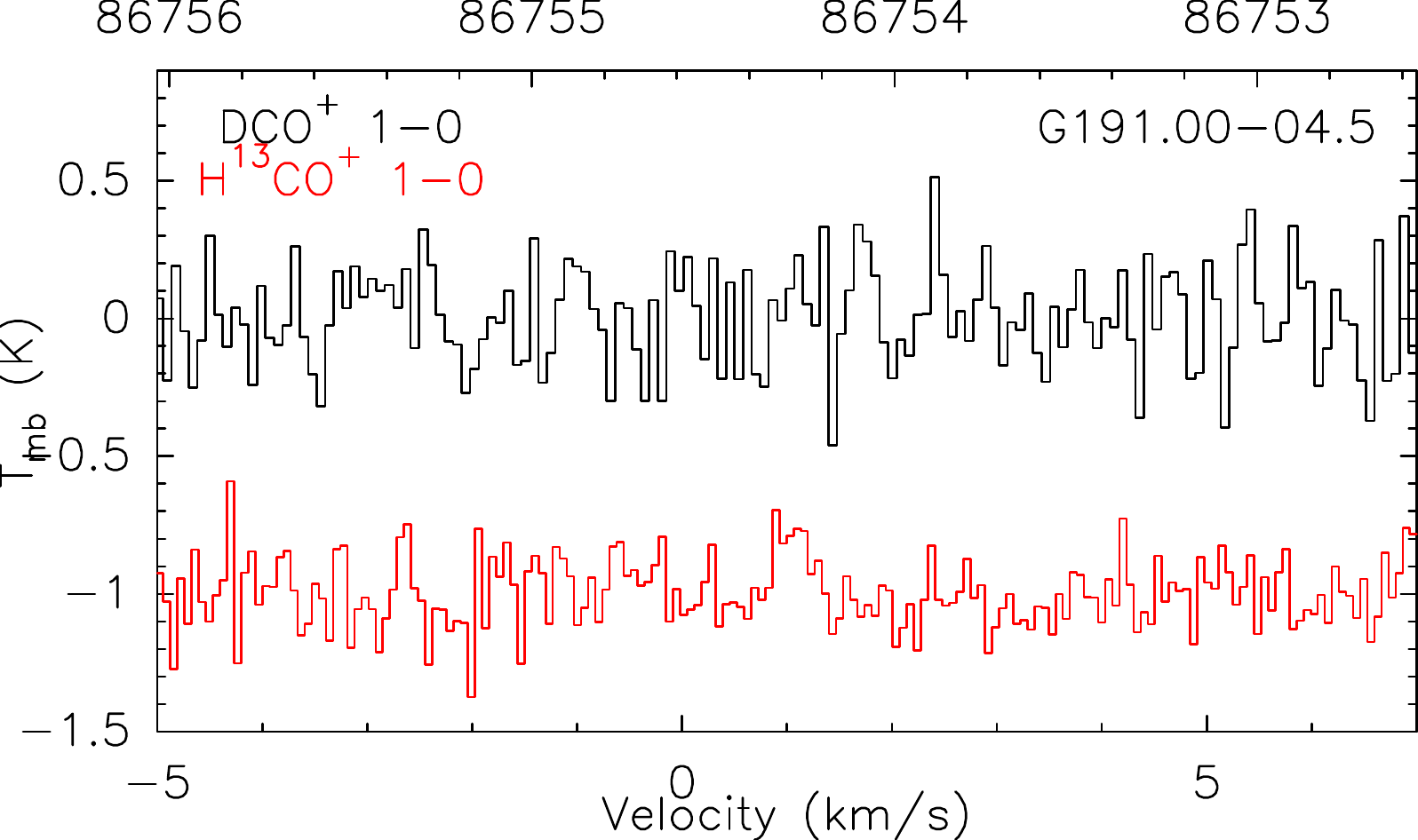}
\includegraphics[width=0.3\columnwidth]{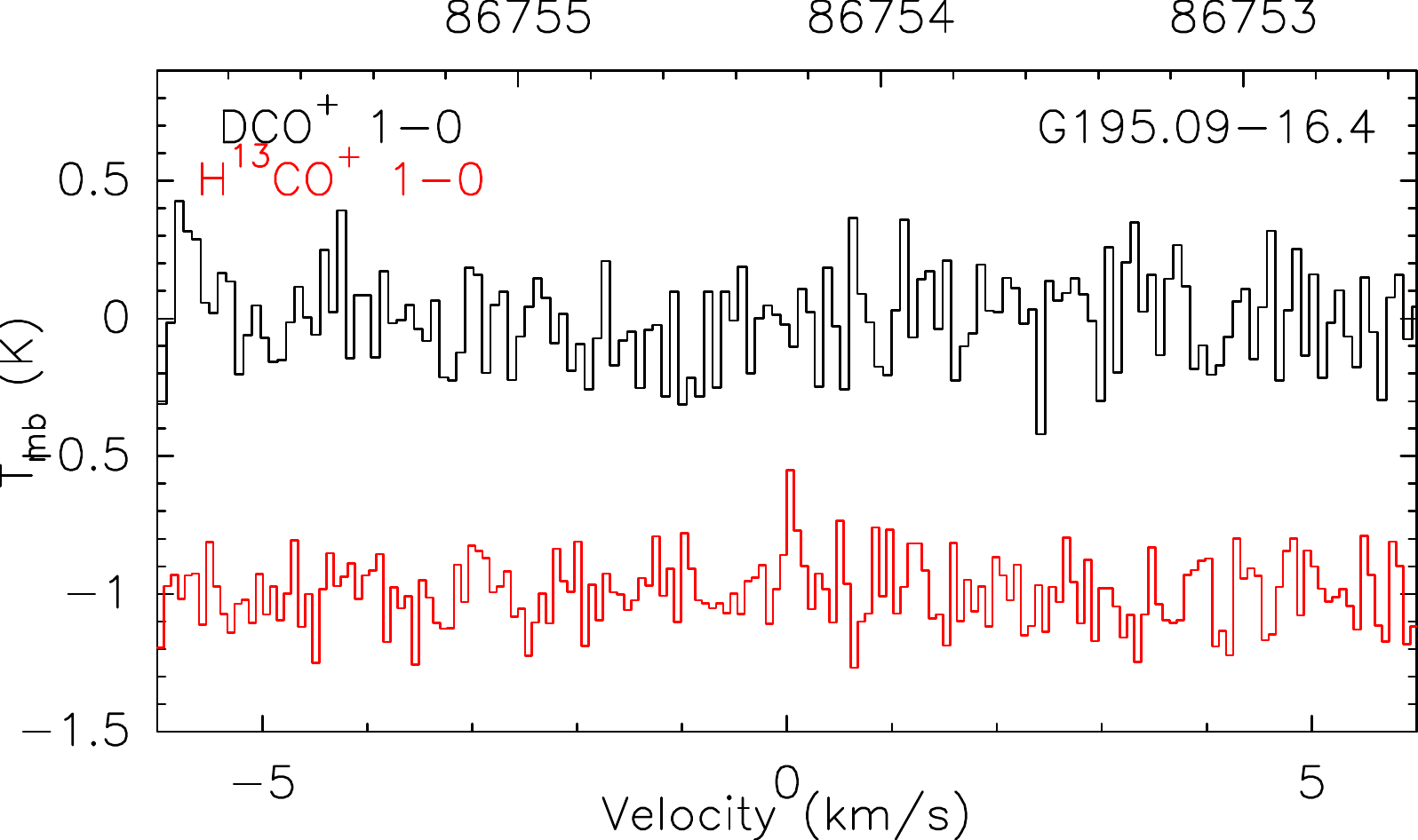}
\includegraphics[width=0.3\columnwidth]{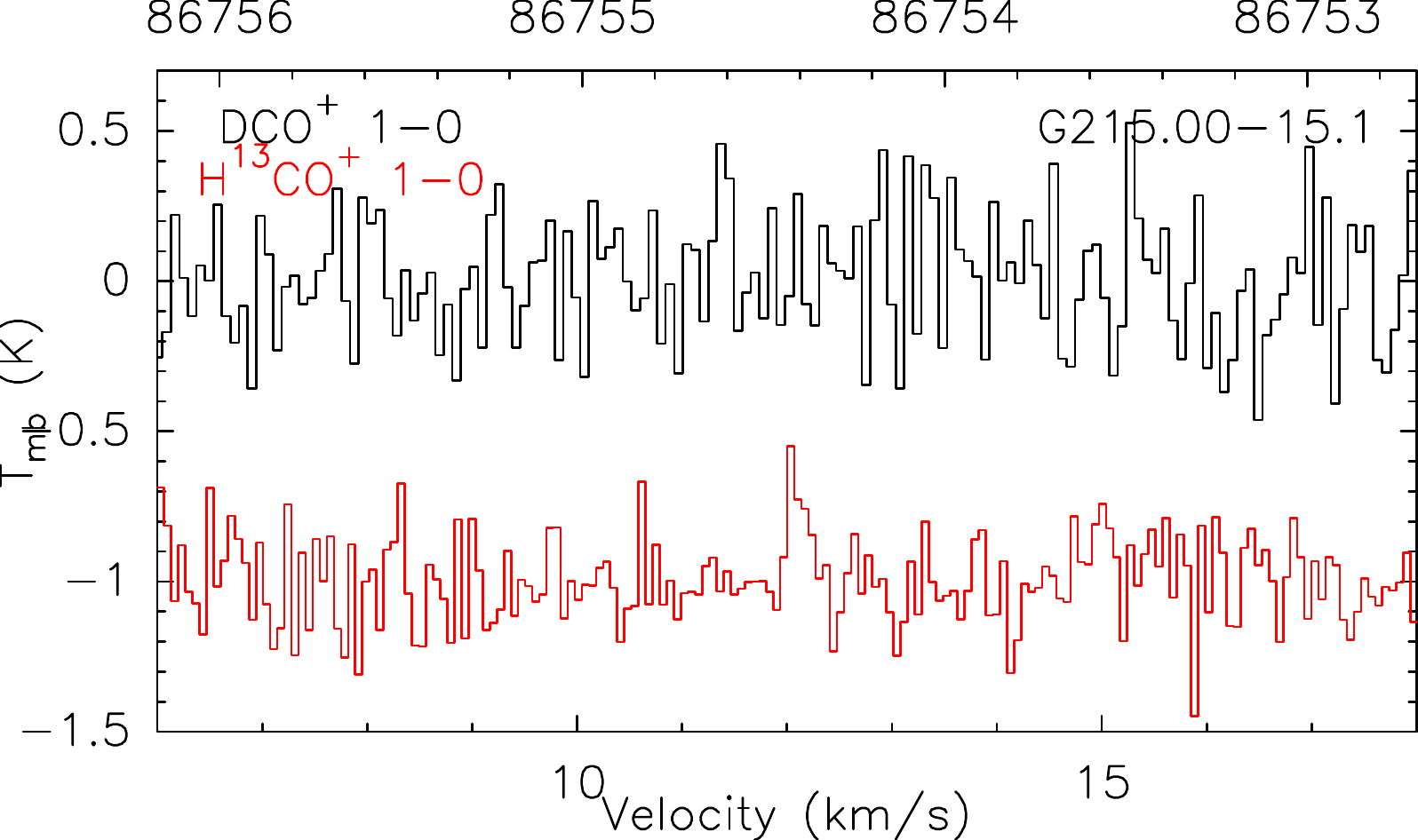}
\includegraphics[width=0.3\columnwidth]{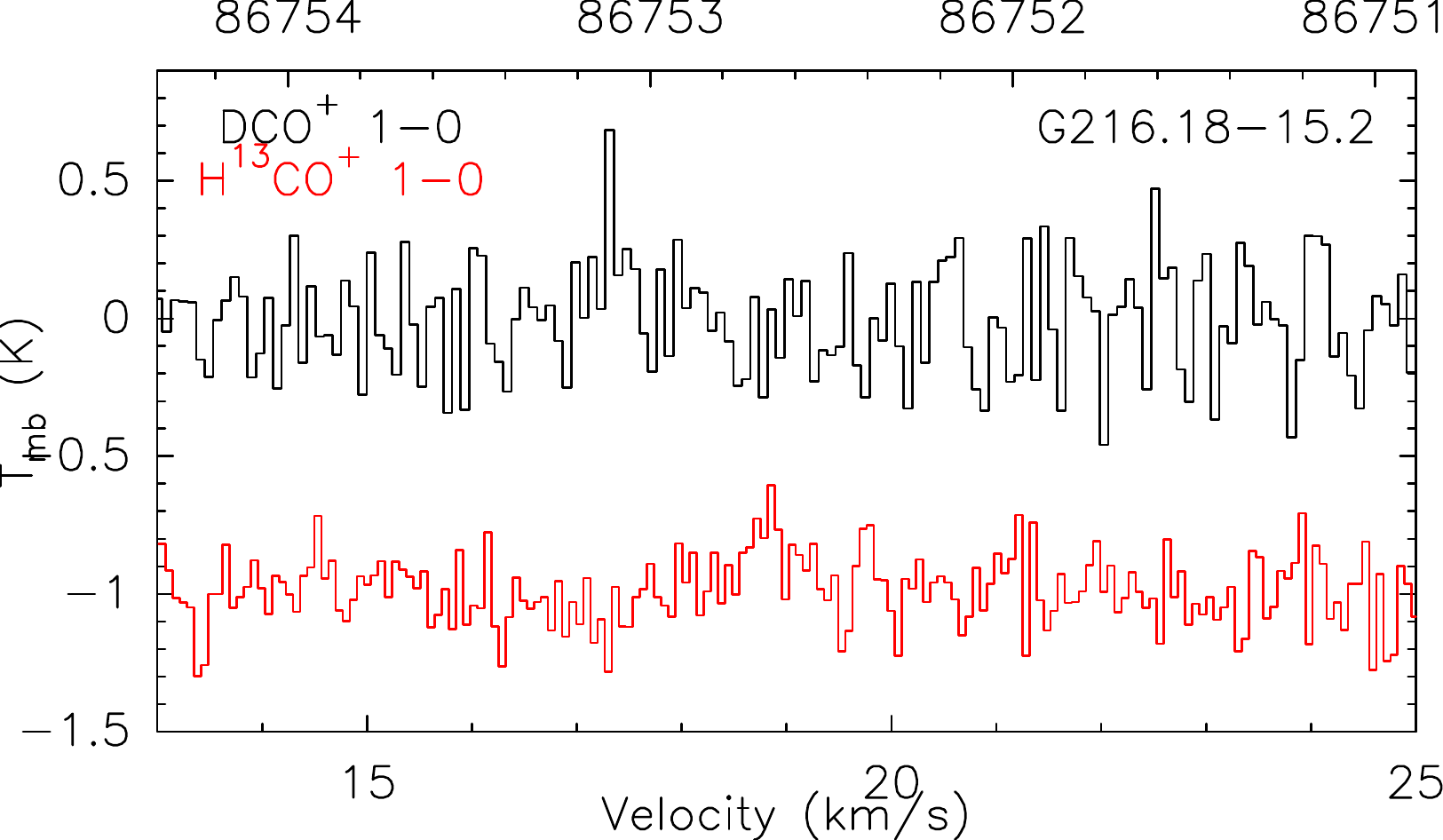}
\caption{Line profiles of H$^{13}$CO$^+$ 1-0 with the high velocity resolution mode (AROWS mode 13). The transitions of DCO$^+$ 1-0 have been observed but not detected in these sources.\centering}
\label{H13CO+mode13_2}
\end{figure}
\begin{figure}
\centering
\includegraphics[width=0.3\columnwidth]{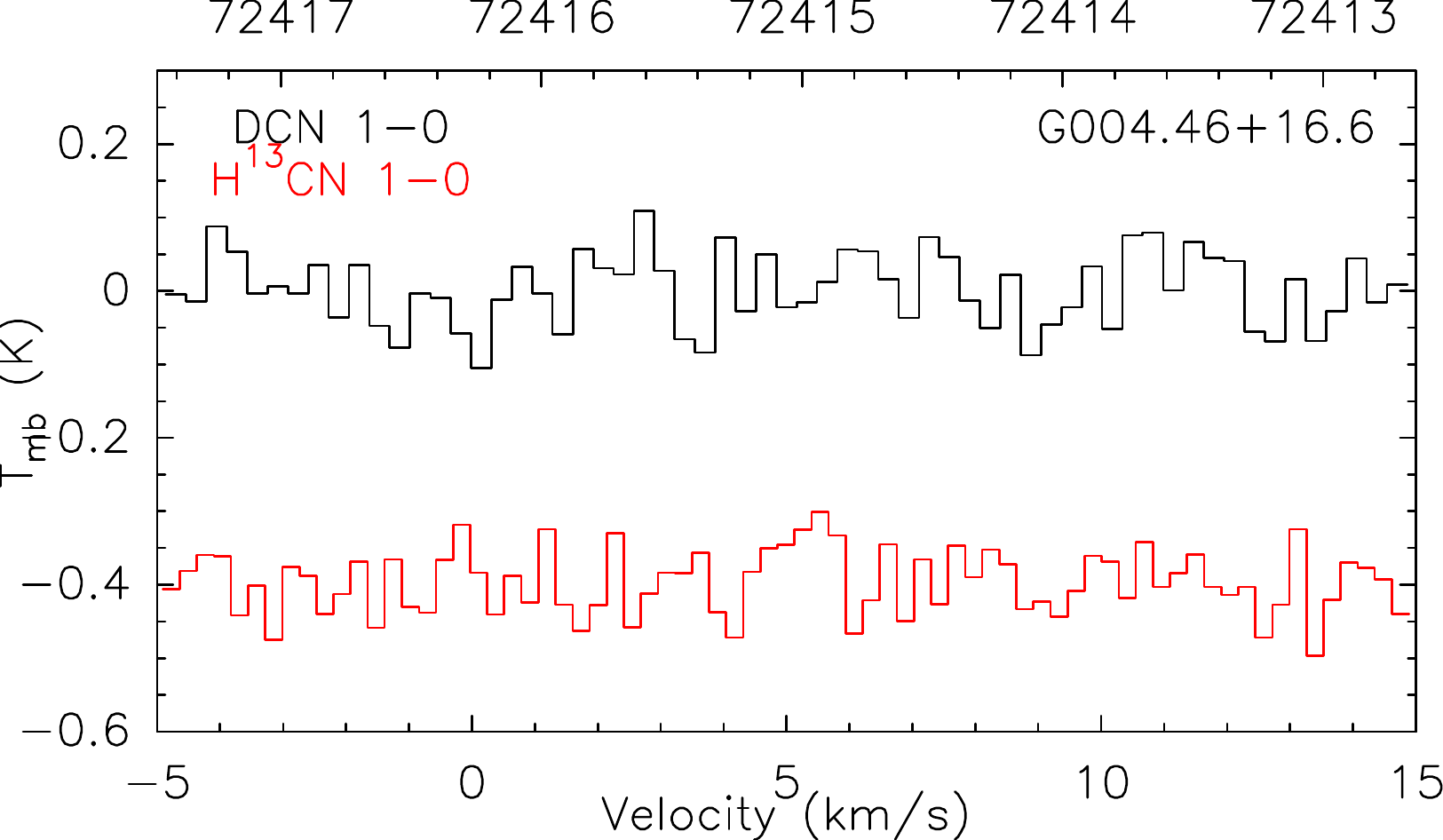}
\includegraphics[width=0.3\columnwidth]{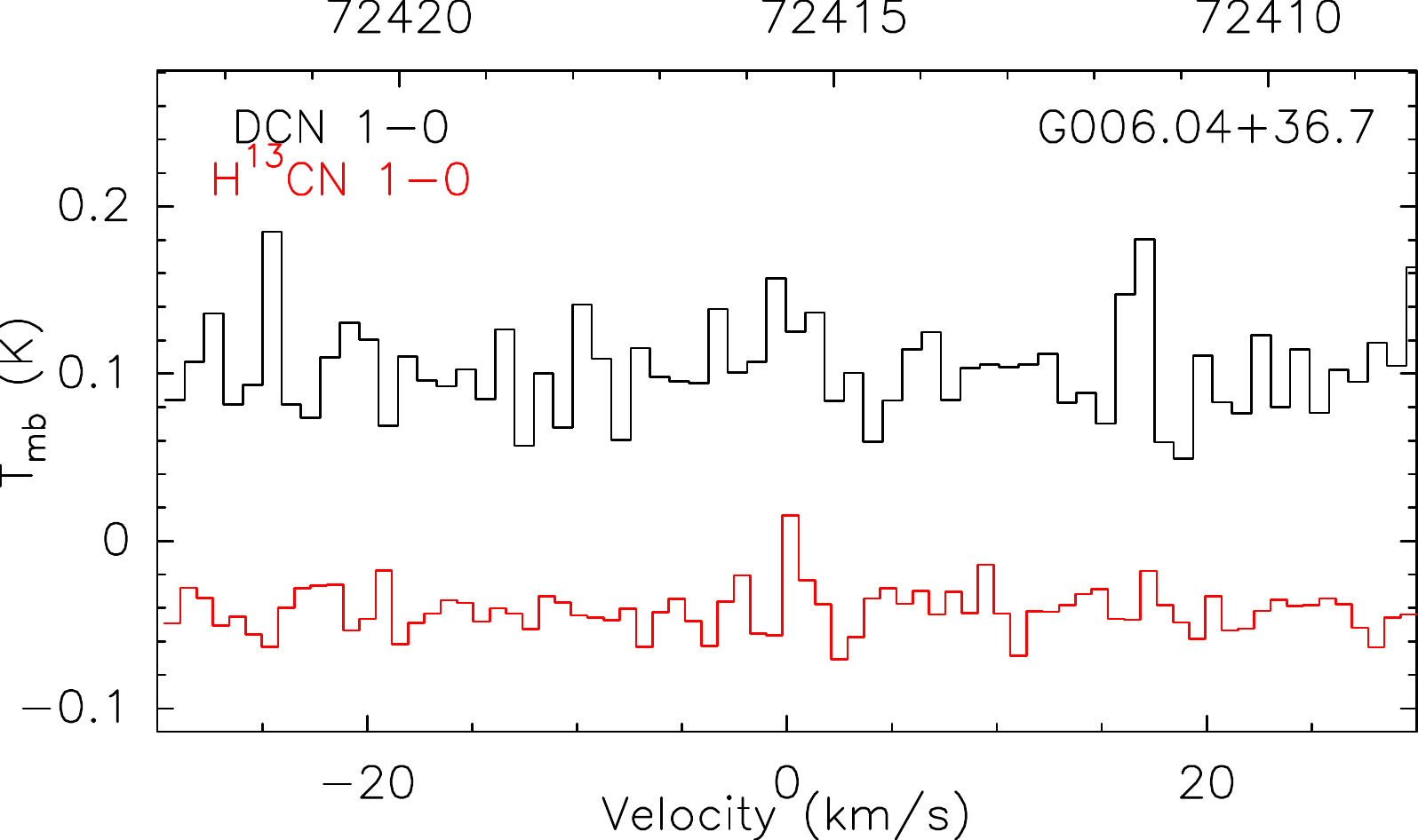}
\includegraphics[width=0.3\columnwidth]{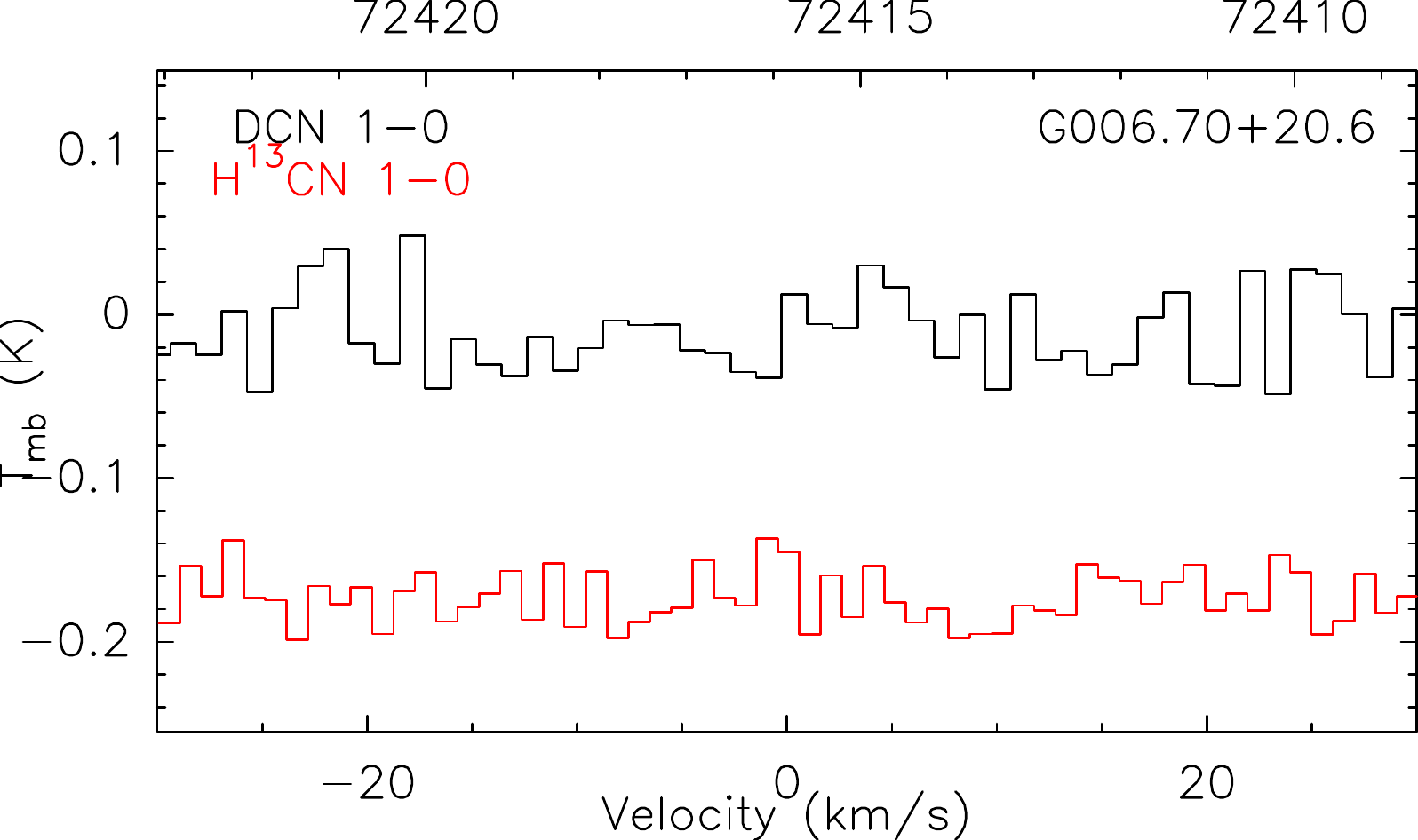}
\caption{Line profiles of H$^{13}$CN 1-0 with the low velocity resolution mode (AROWS mode 3). The transitions of DCN 1-0 have been observed but not detected in these sources.\centering}
\label{H13CNmode3_2}
\end{figure}
\addtocounter{figure}{-1}
\begin{figure}
\centering
\includegraphics[width=0.3\columnwidth]{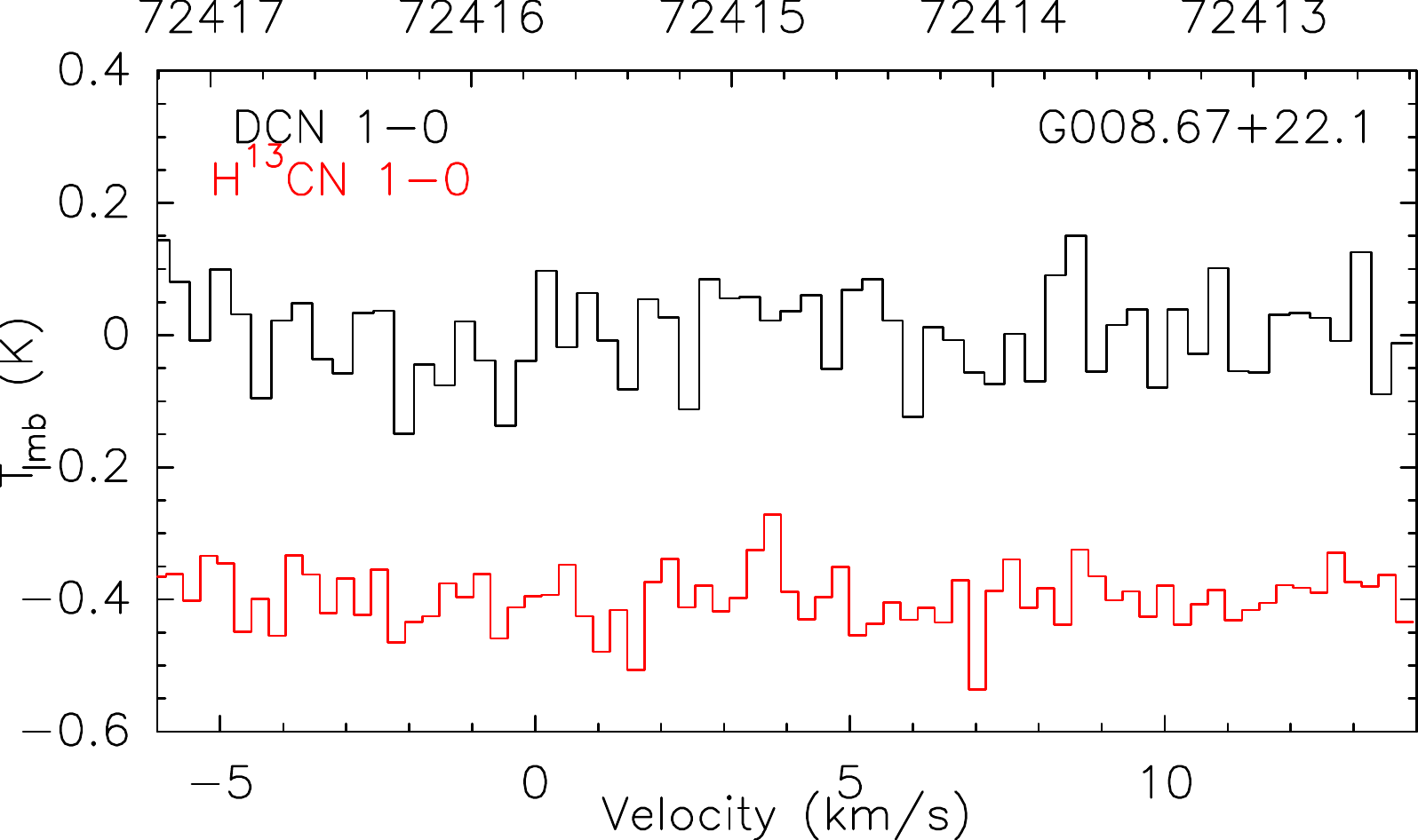}
\includegraphics[width=0.3\columnwidth]{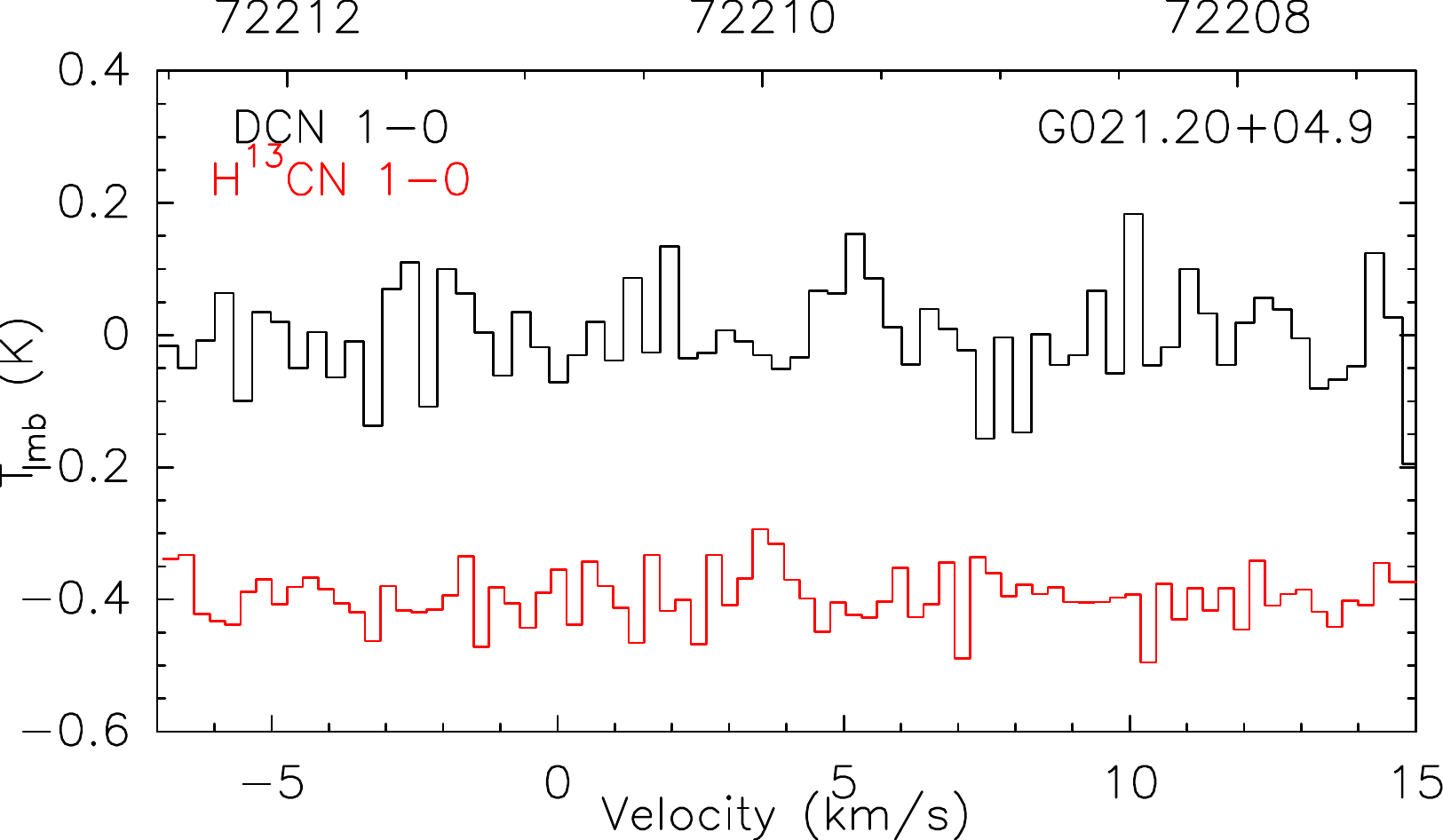}
\includegraphics[width=0.3\columnwidth]{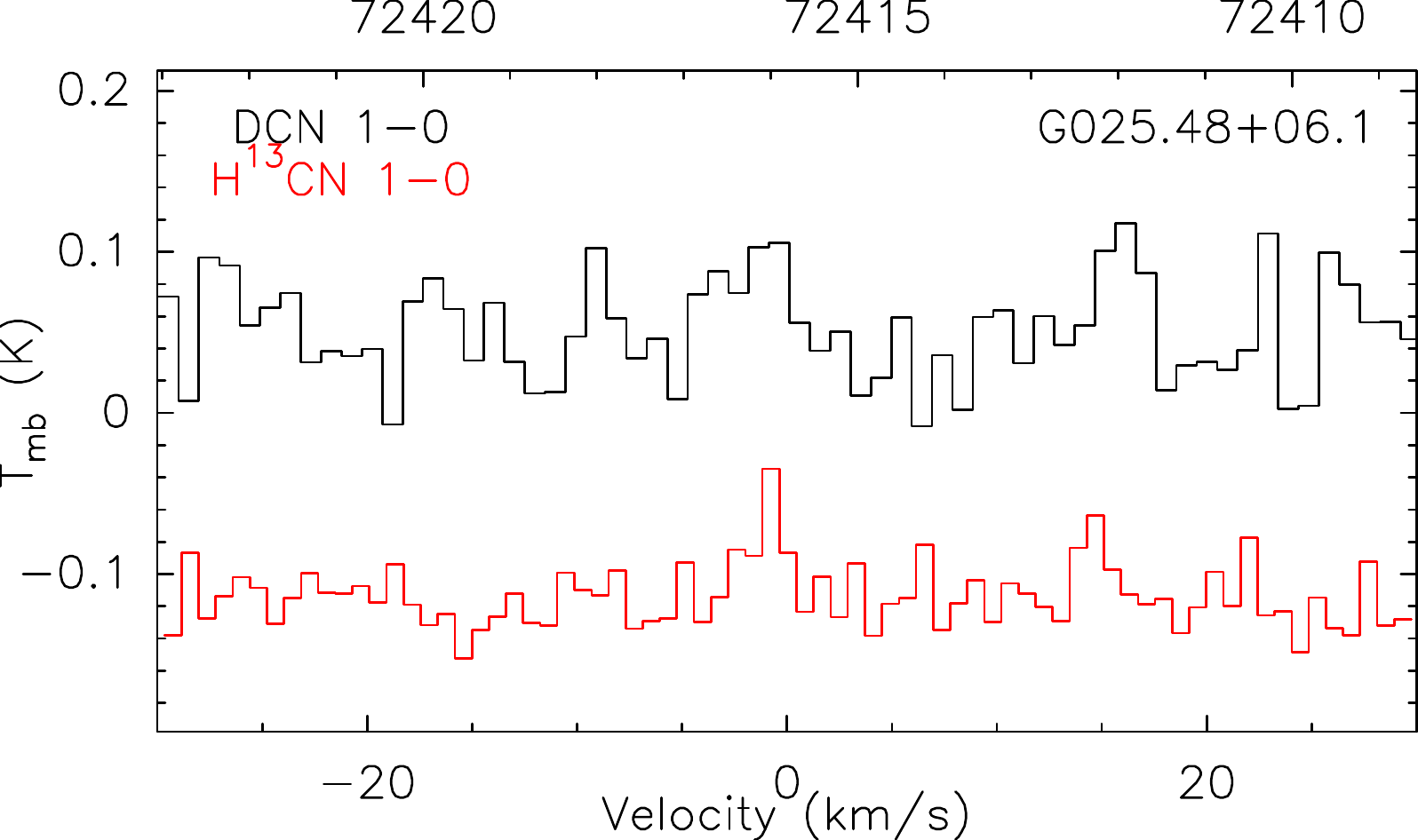}
\includegraphics[width=0.3\columnwidth]{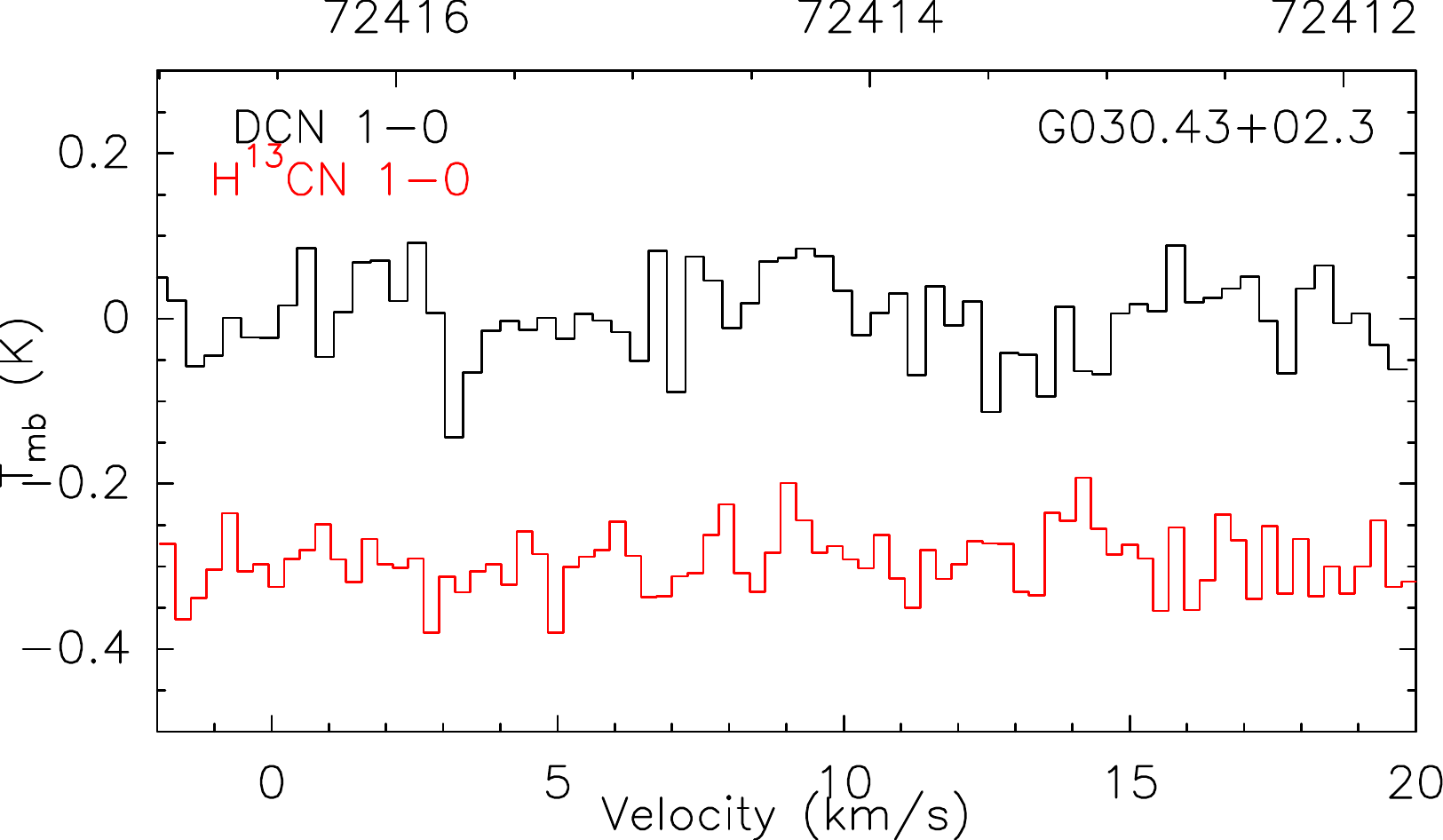}
\includegraphics[width=0.3\columnwidth]{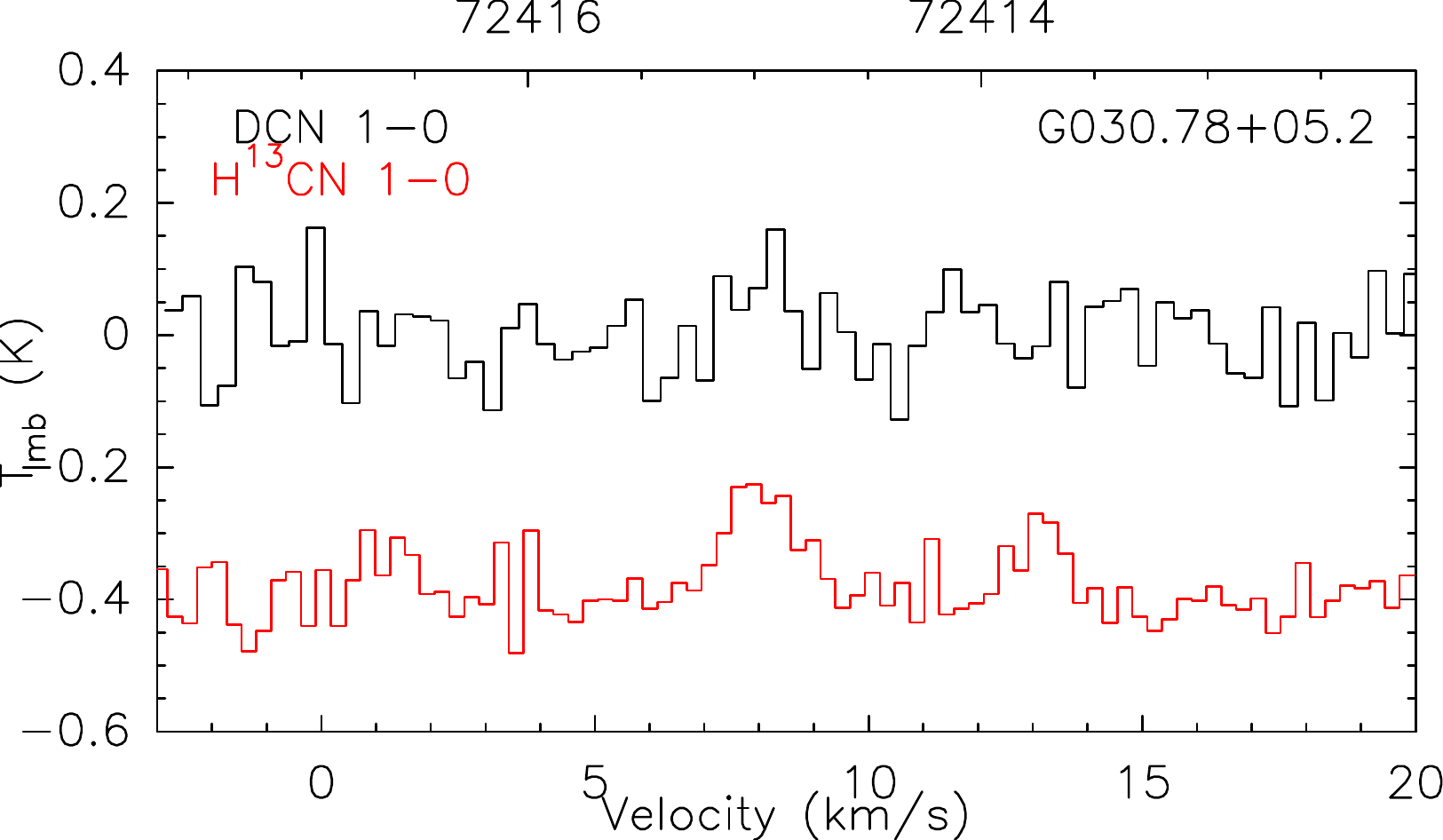}
\includegraphics[width=0.3\columnwidth]{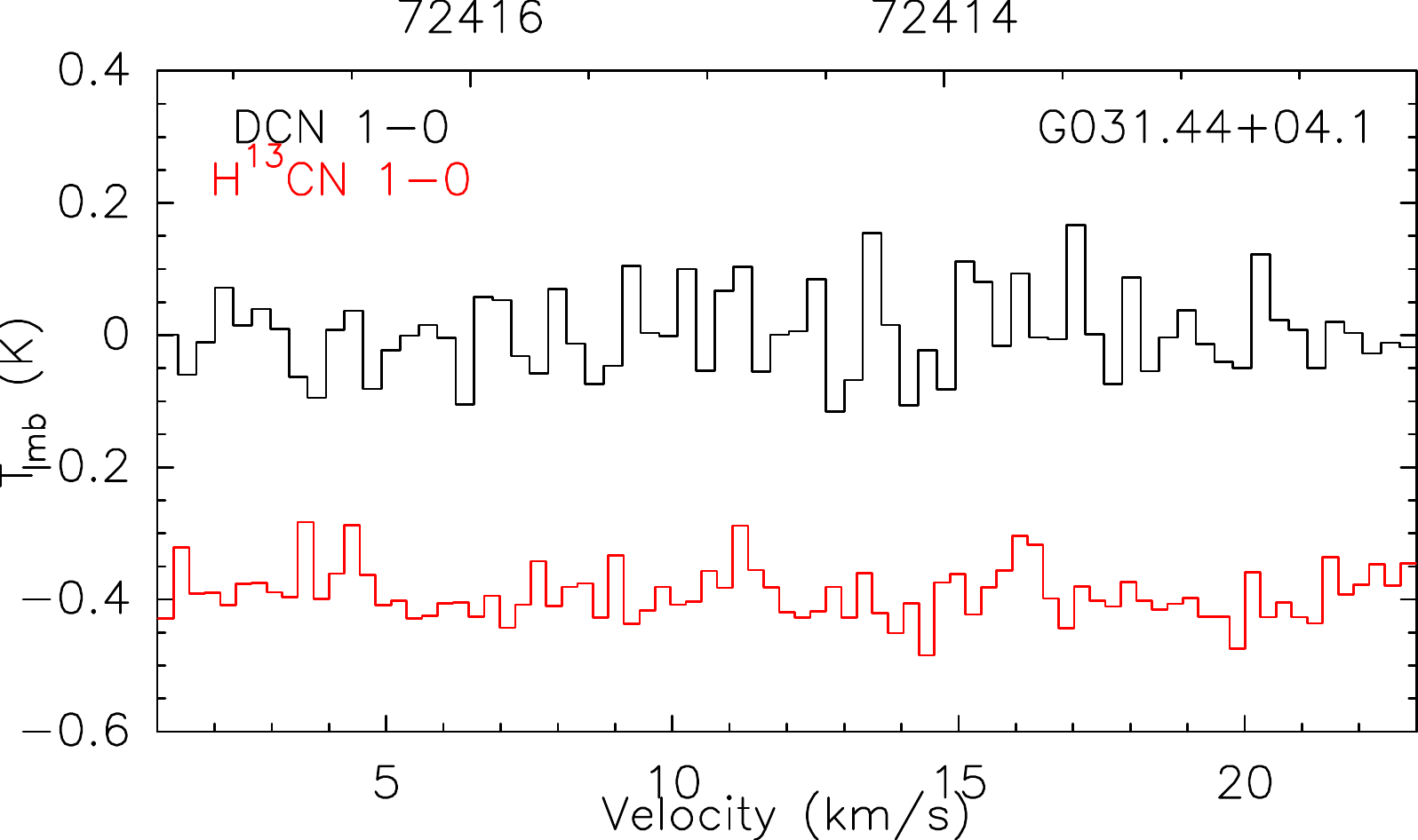}
\includegraphics[width=0.3\columnwidth]{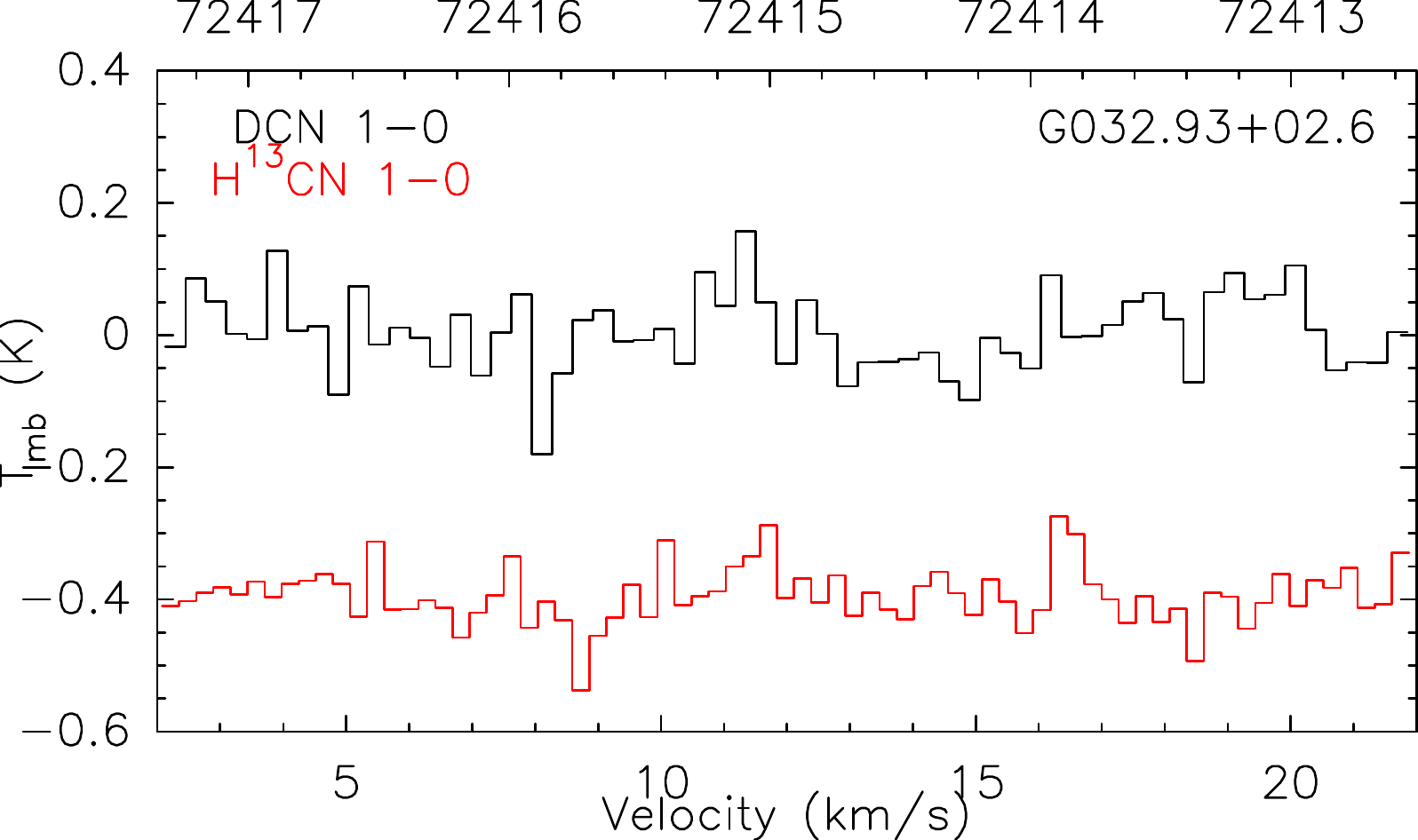}
\includegraphics[width=0.3\columnwidth]{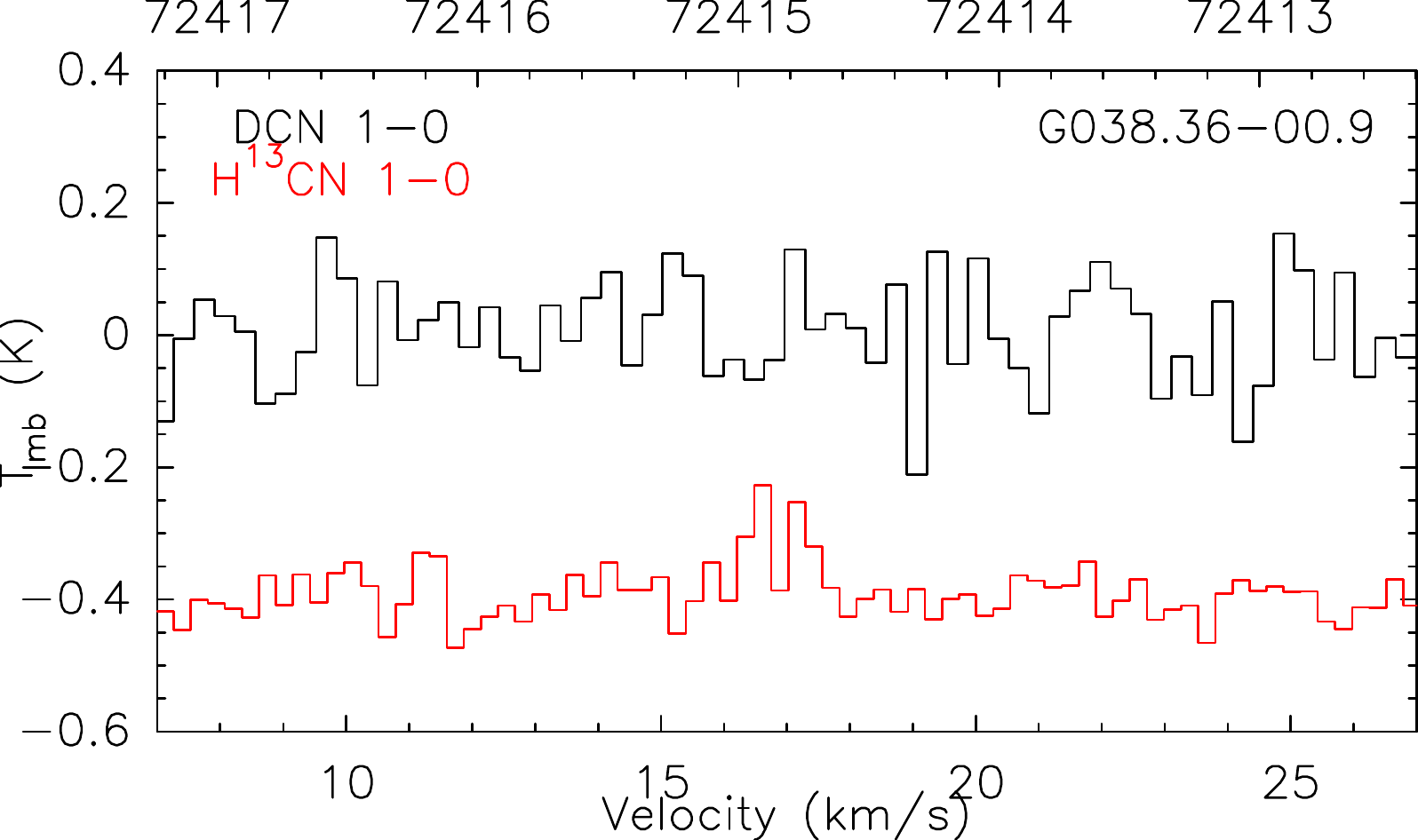}
\includegraphics[width=0.3\columnwidth]{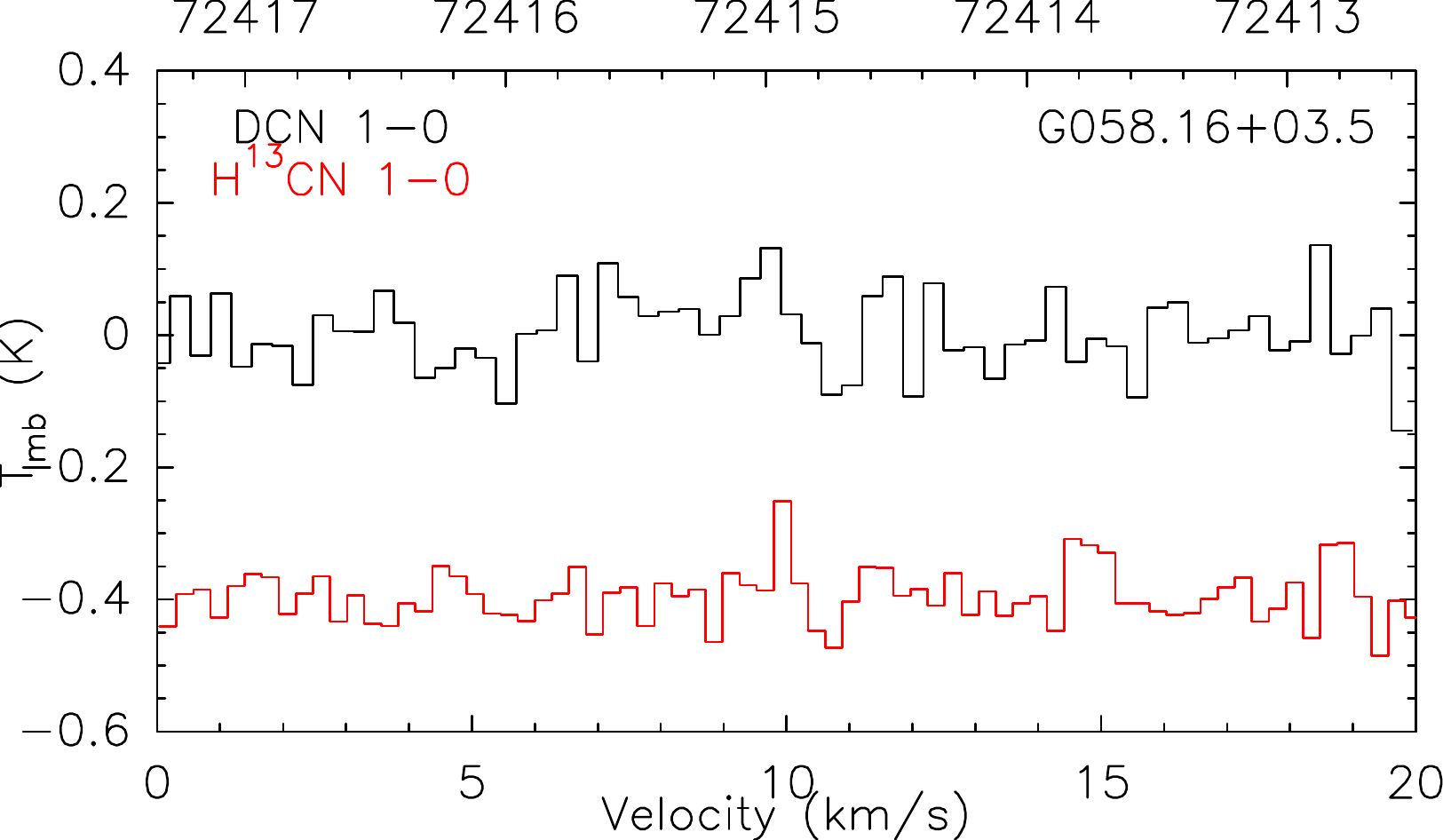}
\includegraphics[width=0.3\columnwidth]{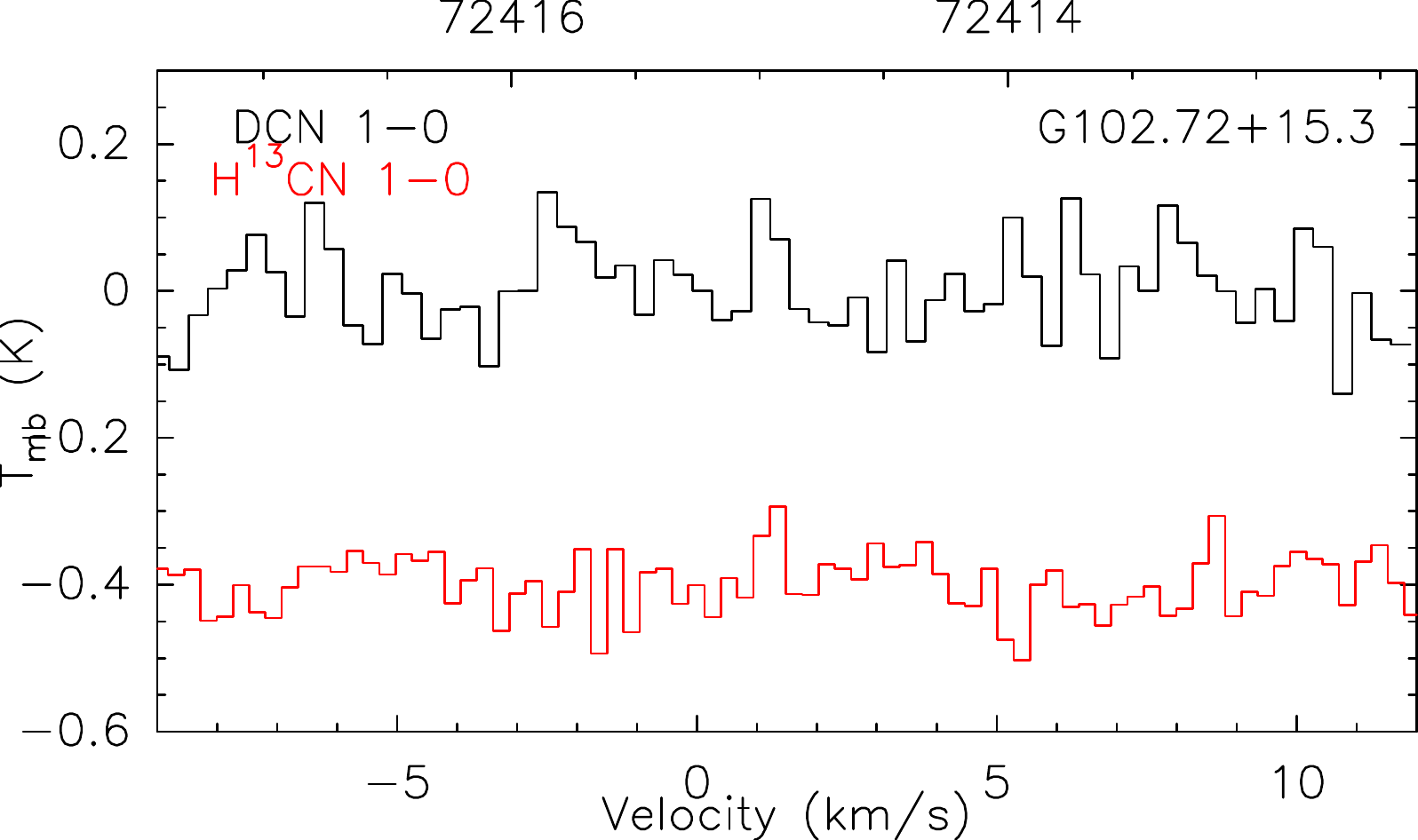}
\includegraphics[width=0.3\columnwidth]{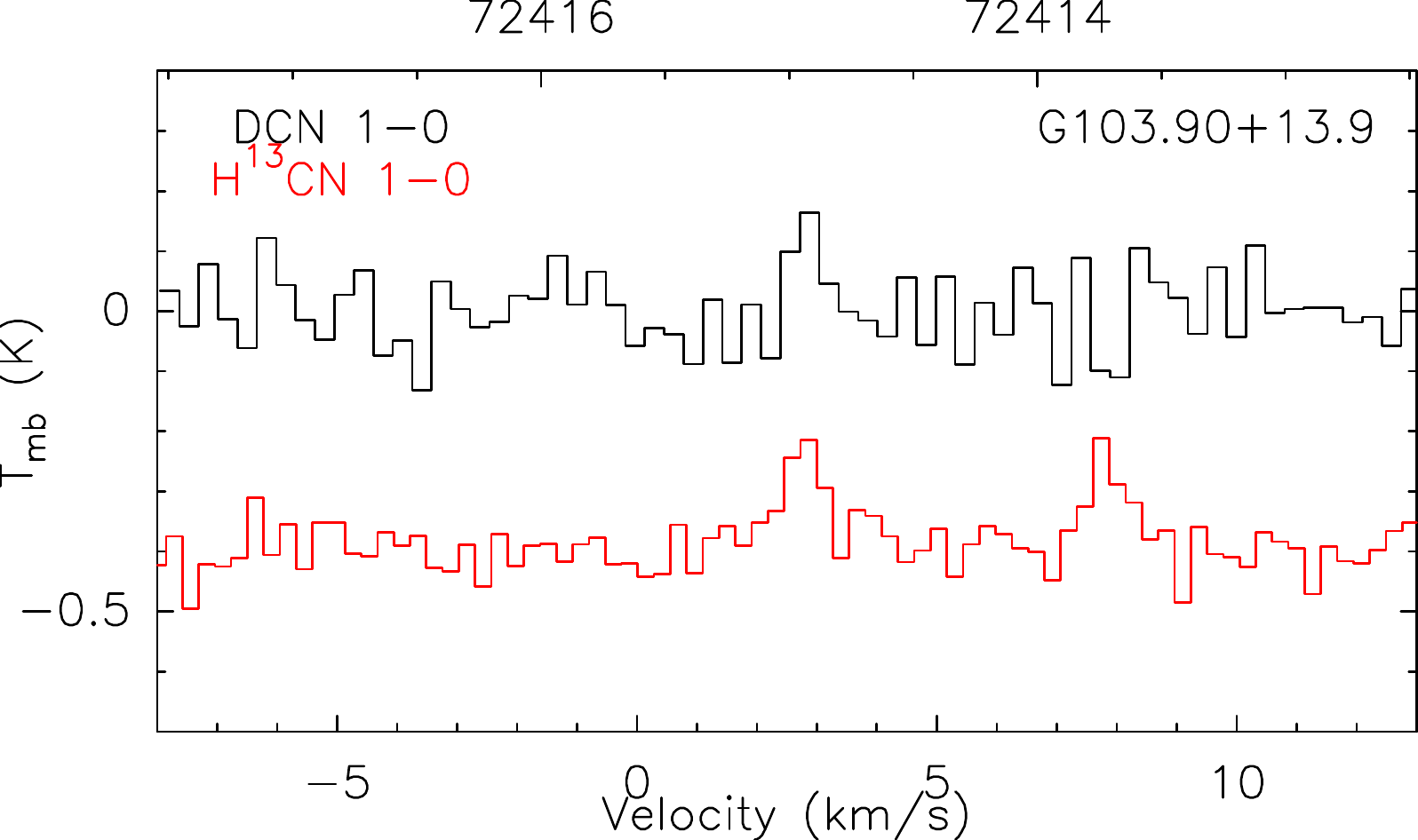}
\includegraphics[width=0.3\columnwidth]{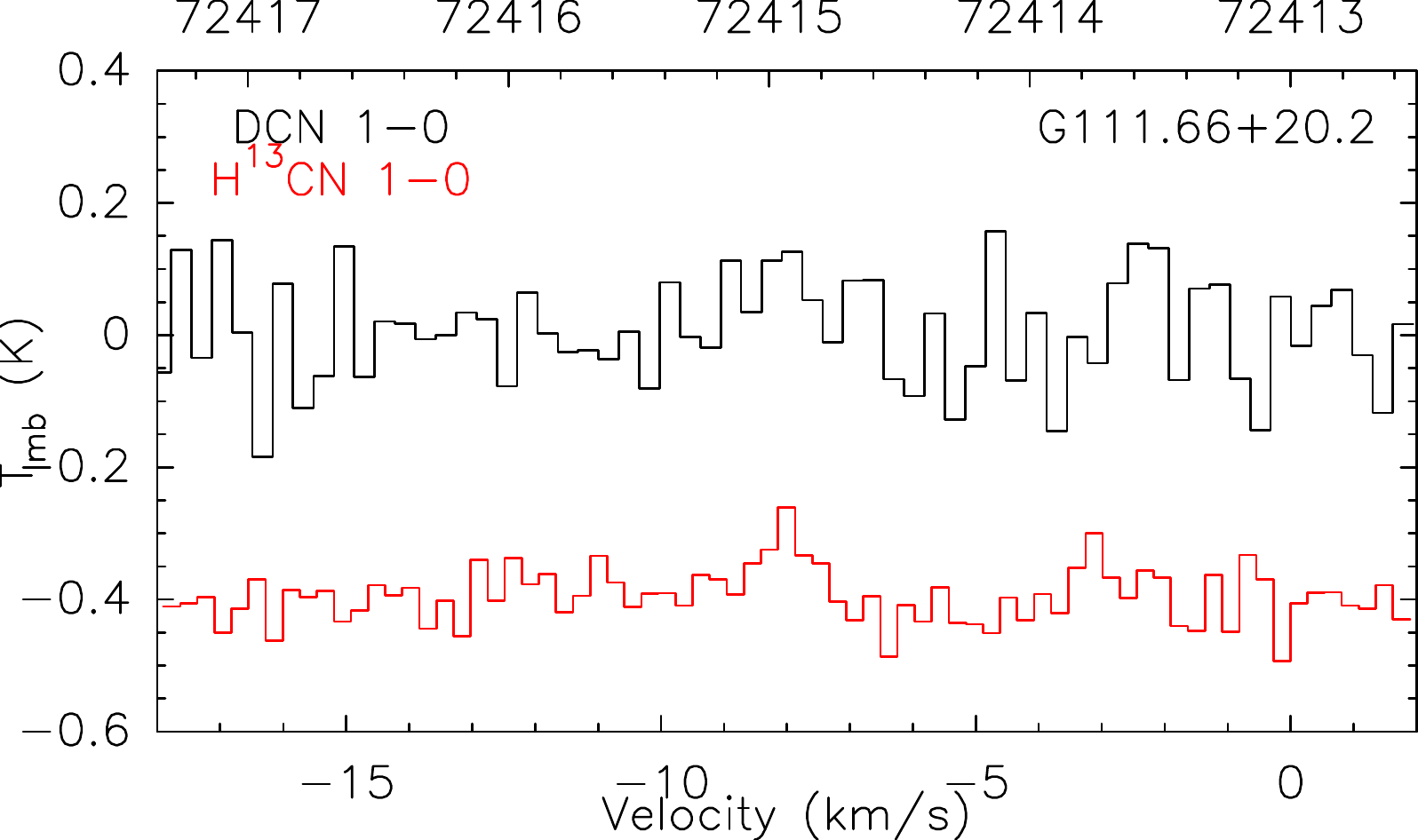}
\includegraphics[width=0.3\columnwidth]{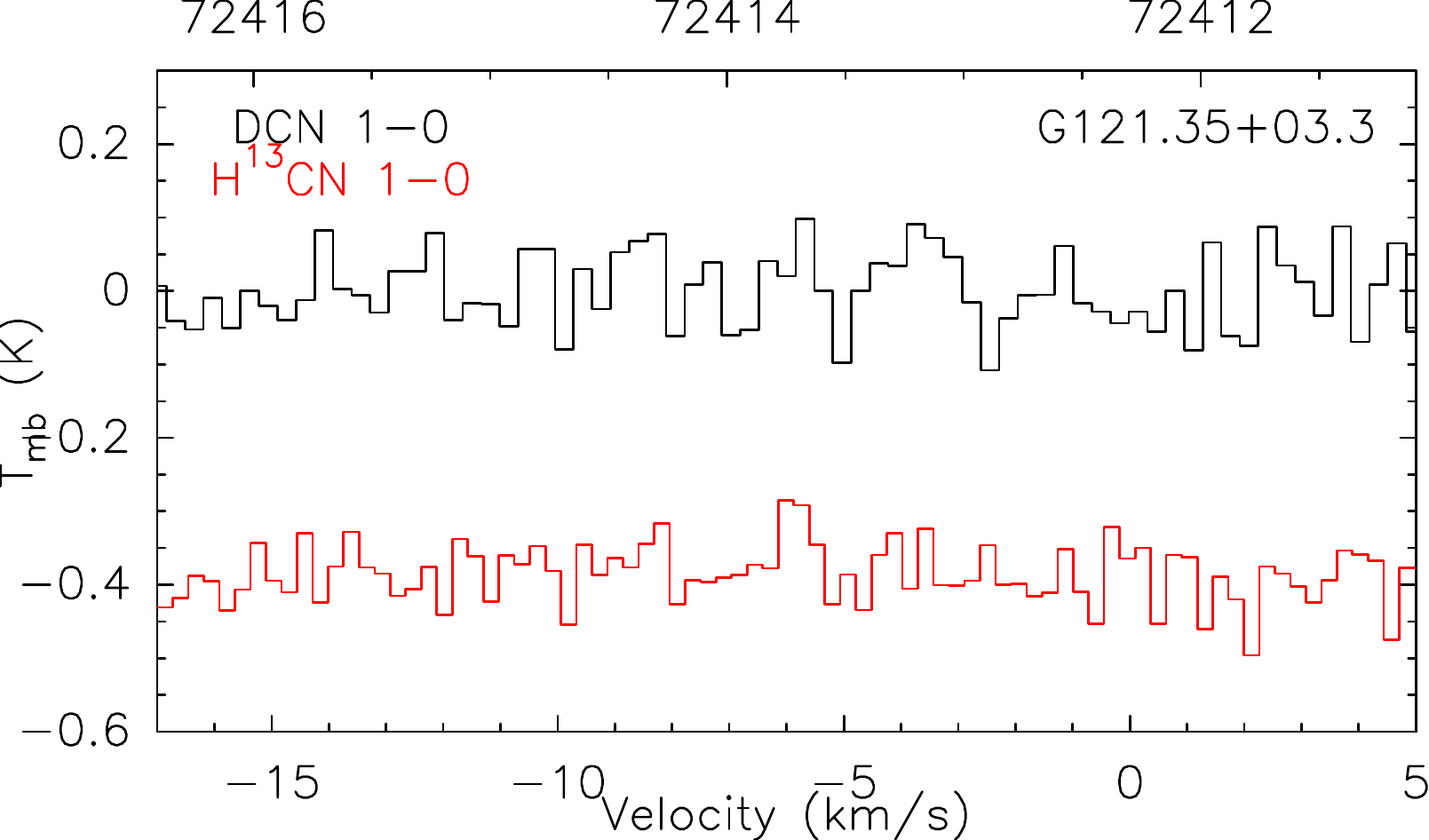}
\caption{Continued.\centering}
\label{H13CNmode3_2}
\end{figure}
\begin{figure}
\centering
\includegraphics[width=0.3\columnwidth]{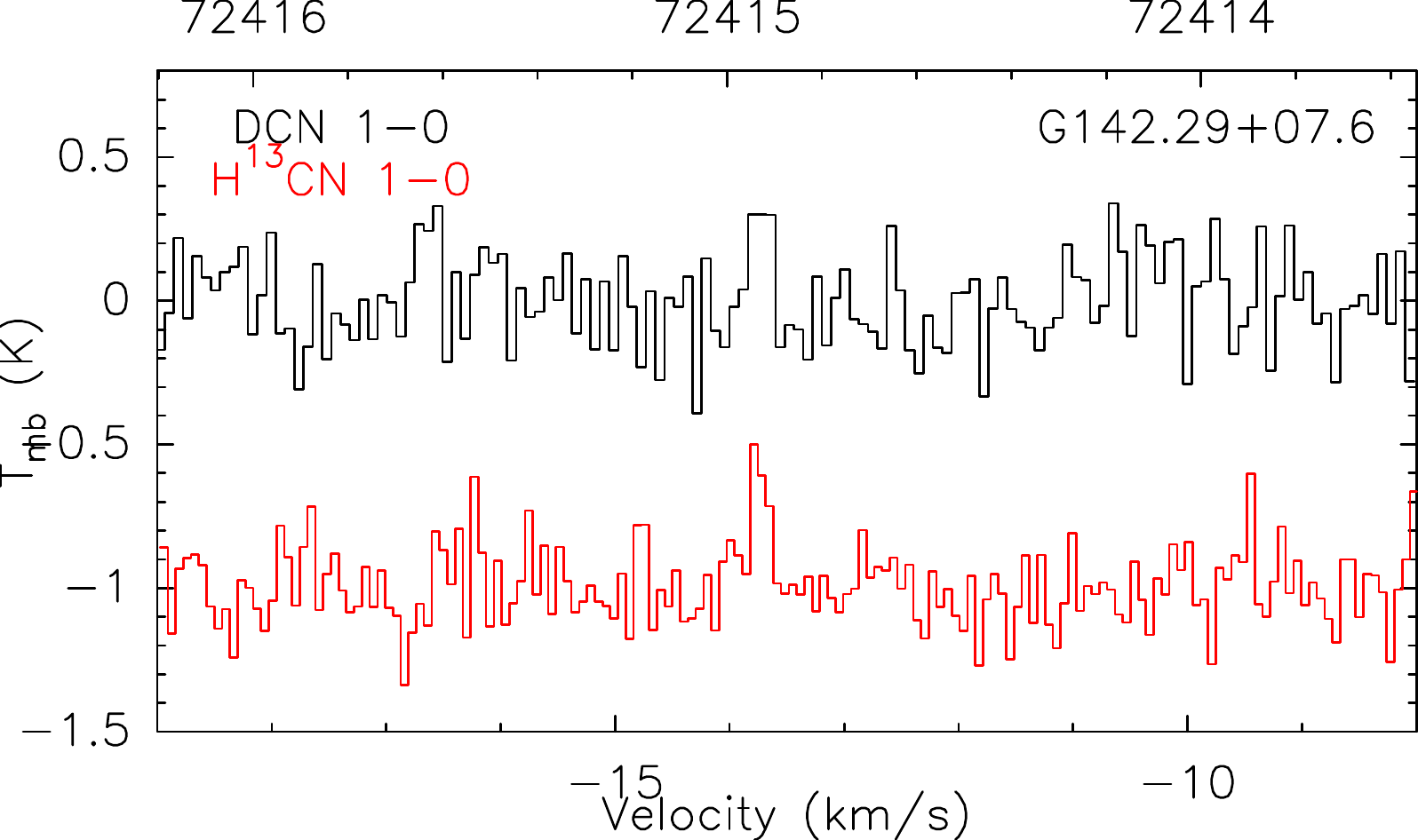}
\includegraphics[width=0.3\columnwidth]{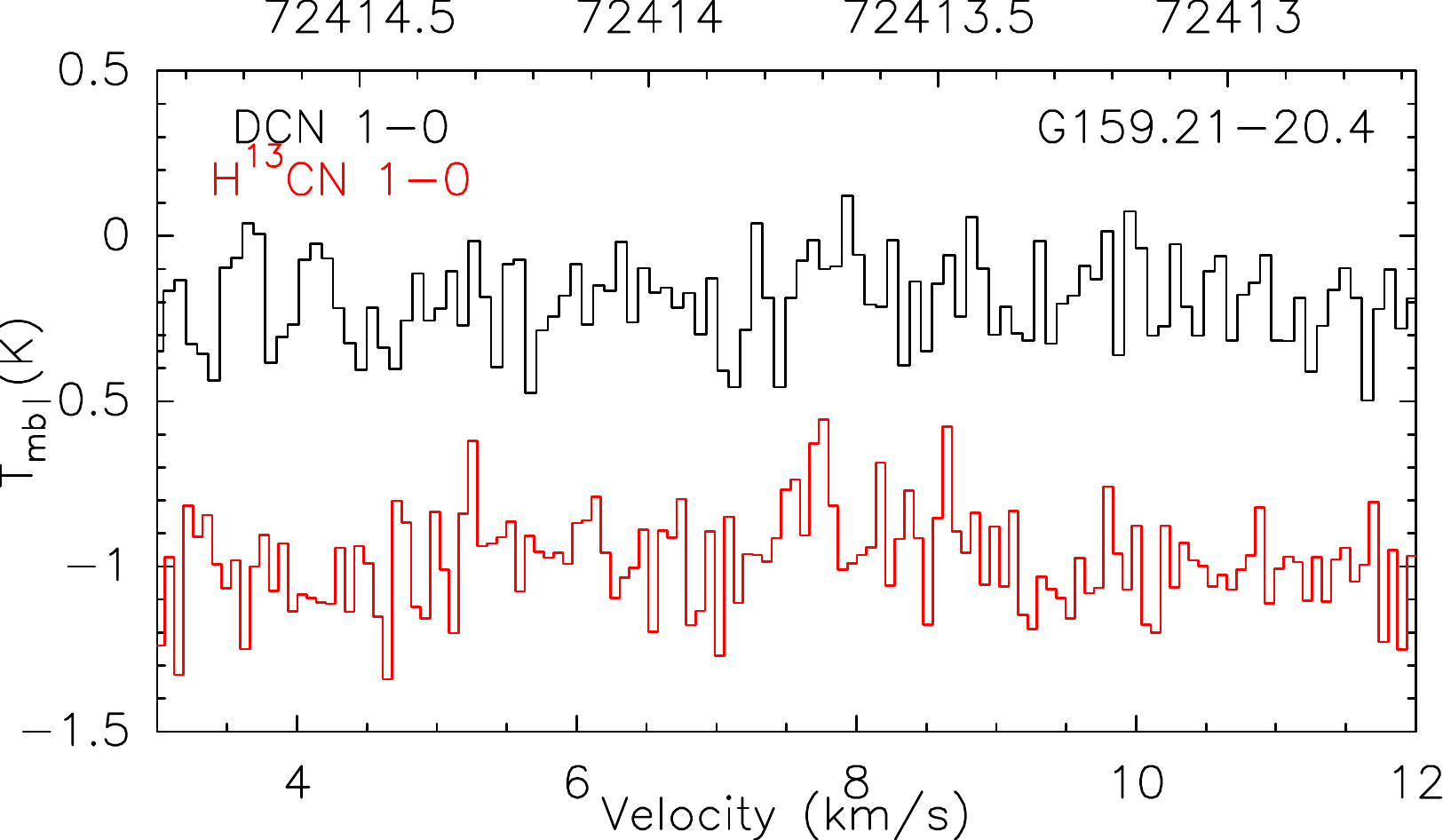}
\includegraphics[width=0.3\columnwidth]{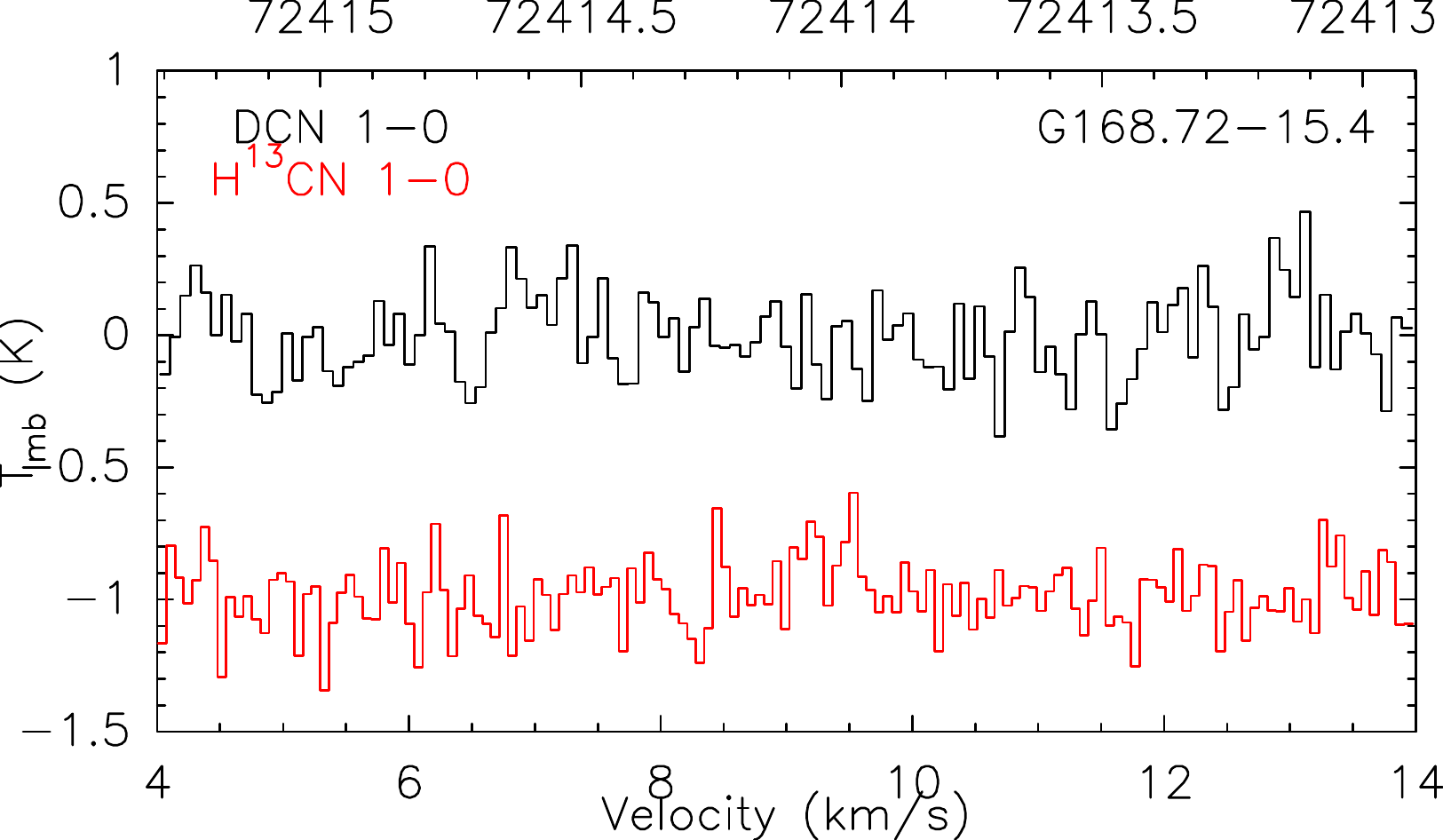}
\caption{Line profiles of H$^{13}$CN 1-0 with the high velocity resolution mode (AROWS mode 13). The transitions of DCN 1-0 have been observed but not detected in these sources.\centering}
\label{H13CNmode13_2}
\end{figure}
\addtocounter{figure}{-1}
\begin{figure}
\centering
\includegraphics[width=0.3\columnwidth]{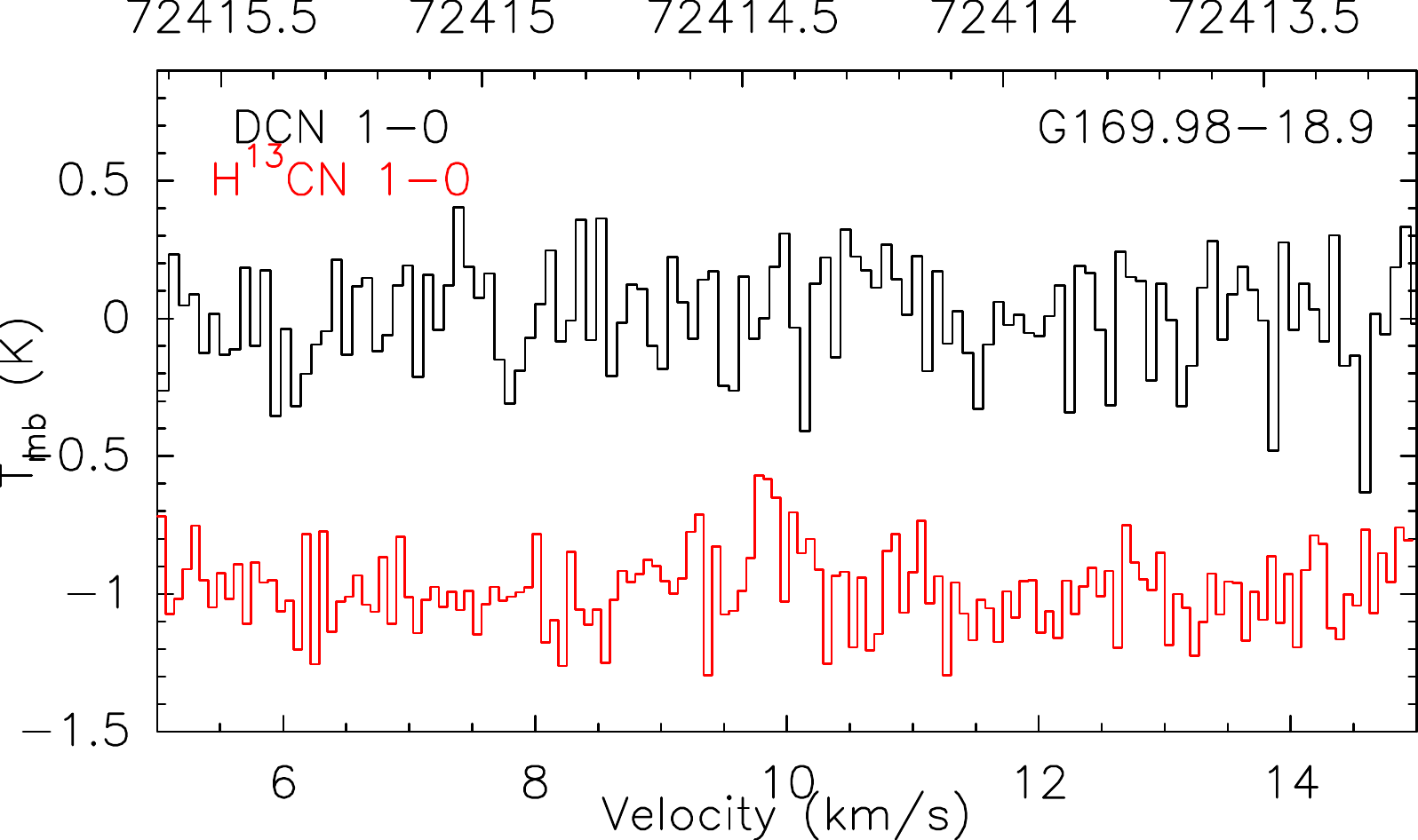}
\includegraphics[width=0.3\columnwidth]{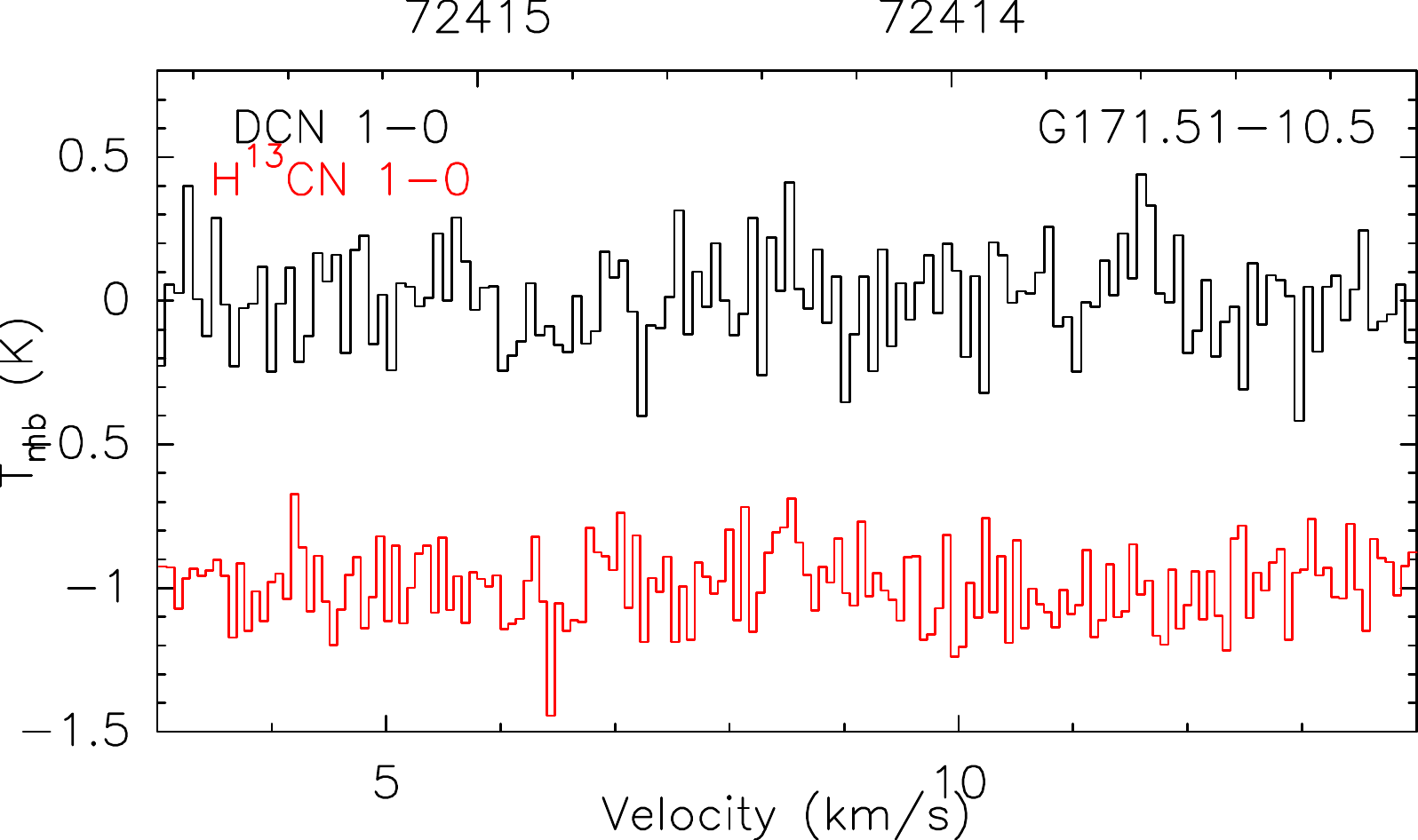}
\includegraphics[width=0.3\columnwidth]{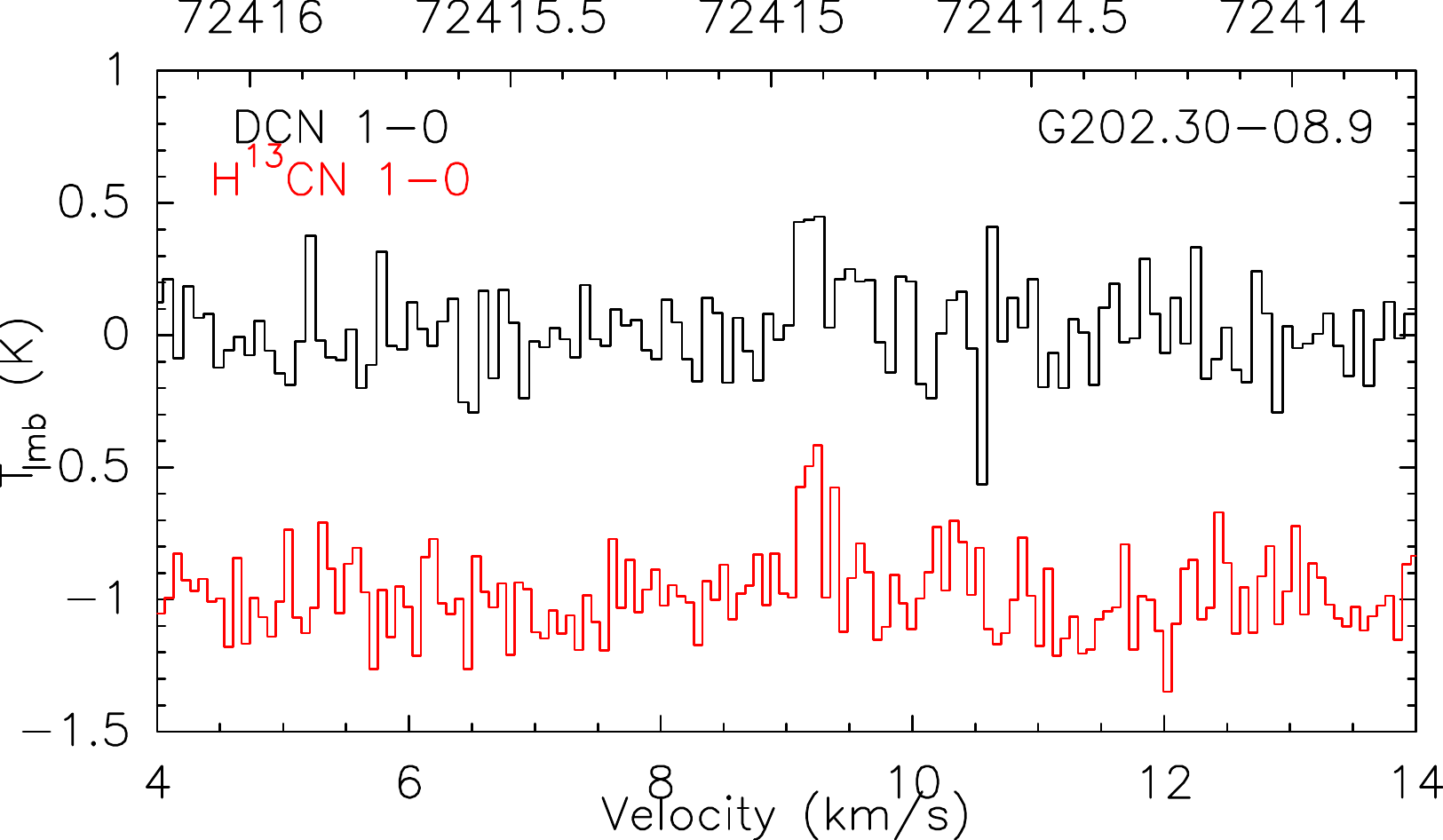}
\caption{Continued.\centering}
\label{H13CNmode13_2}
\end{figure}
\begin{figure}
\centering
\includegraphics[width=0.3\columnwidth]{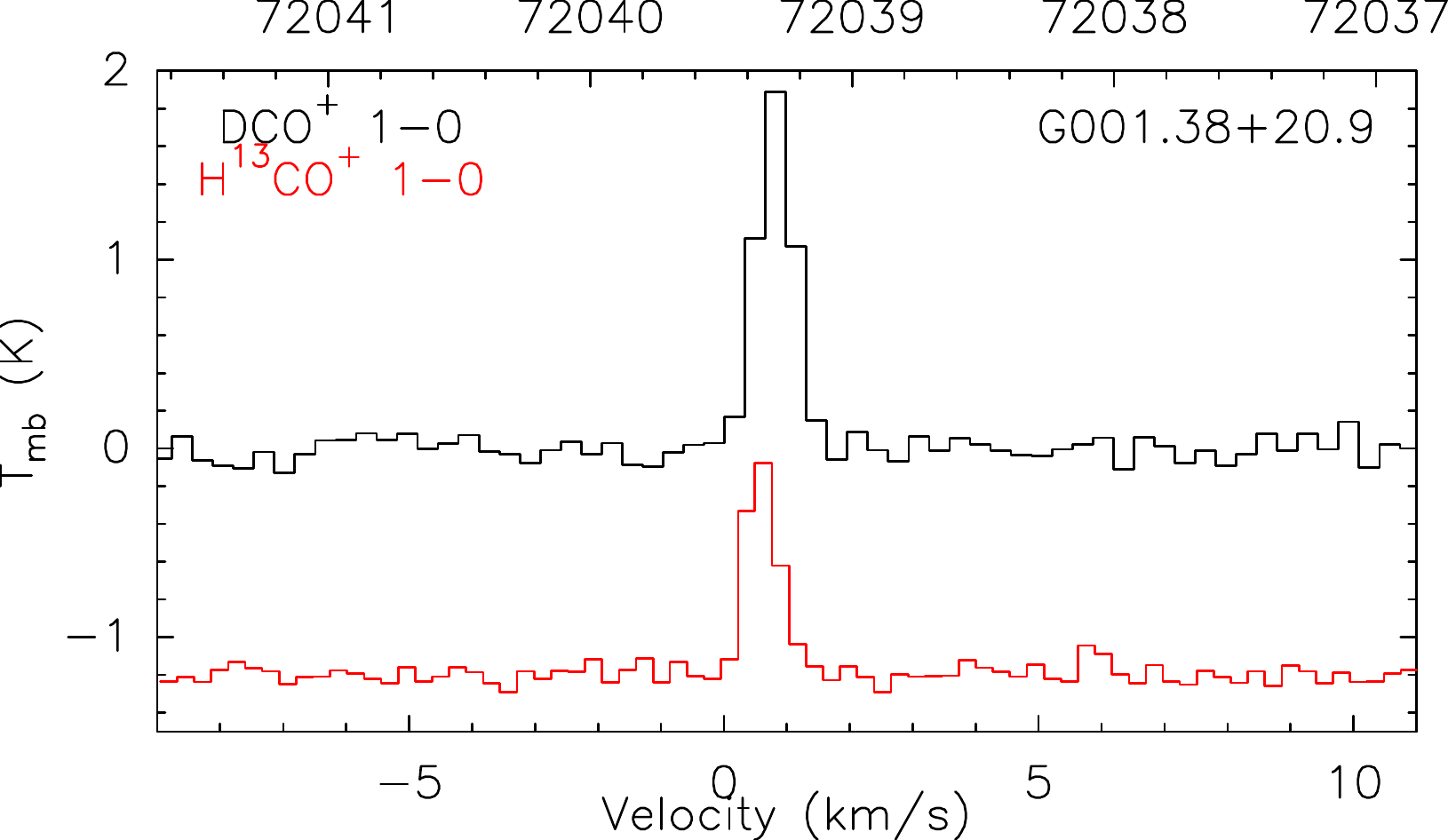}
\includegraphics[width=0.3\columnwidth]{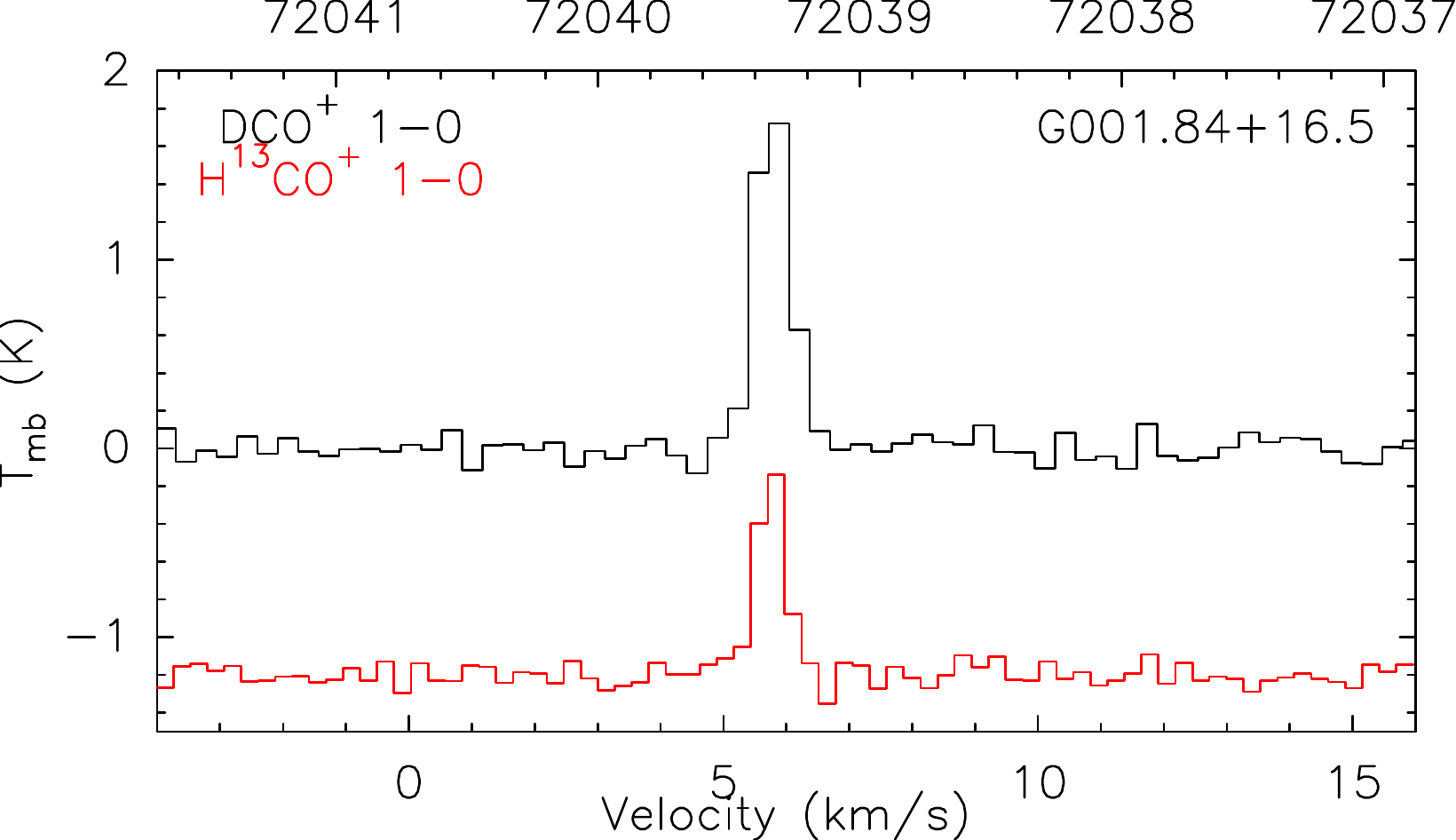}
\includegraphics[width=0.3\columnwidth]{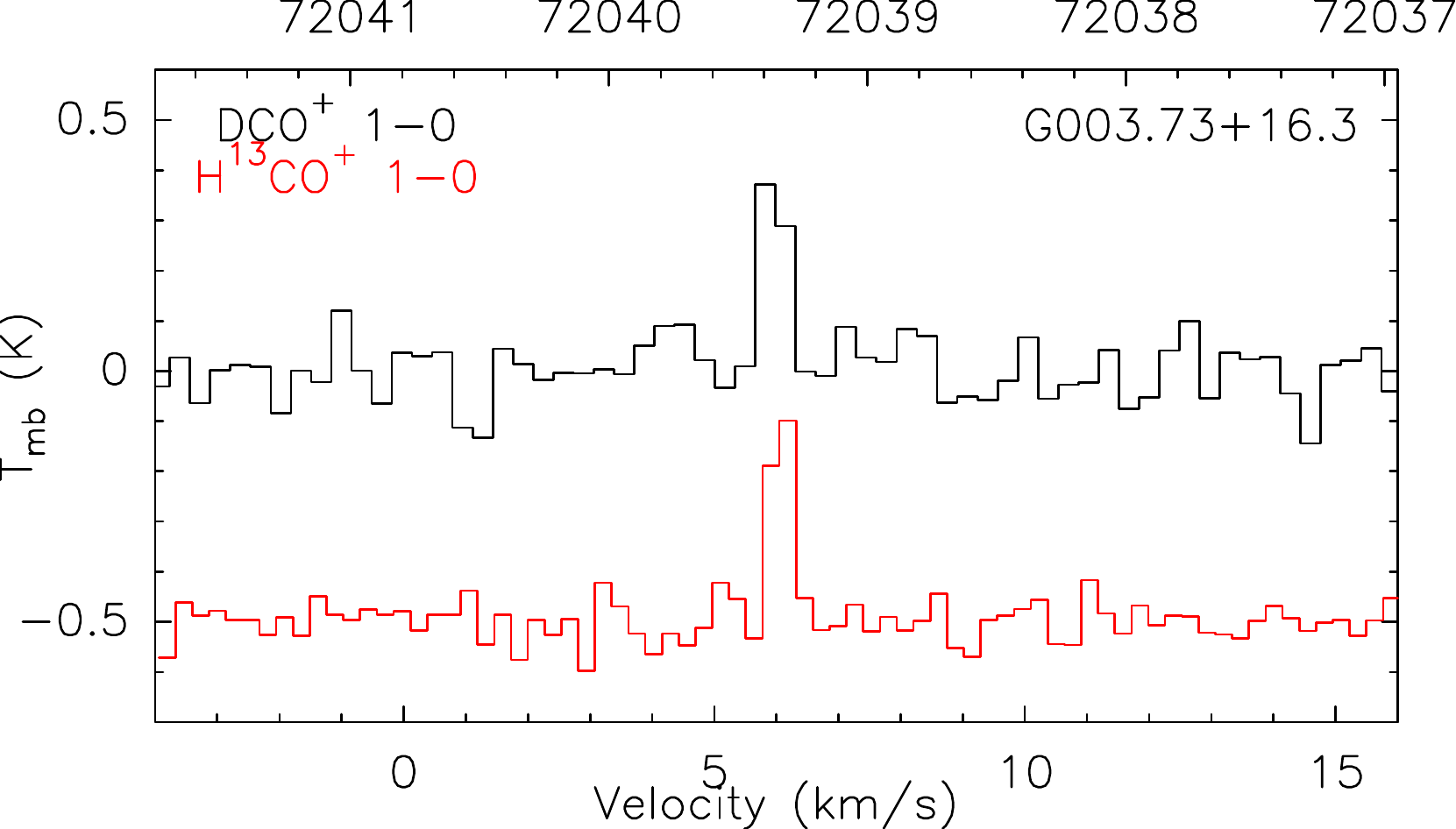}
\includegraphics[width=0.3\columnwidth]{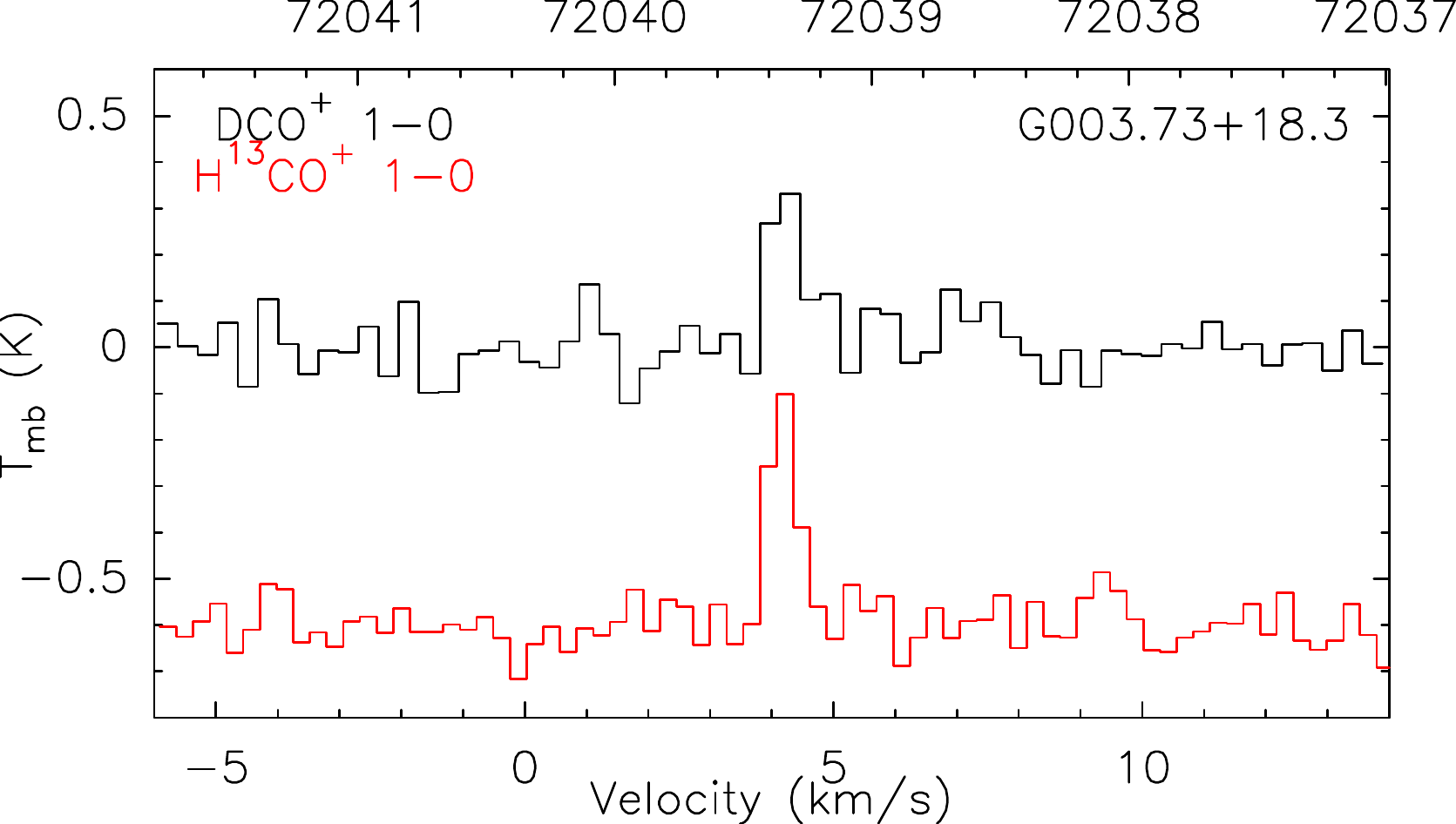}
\includegraphics[width=0.3\columnwidth]{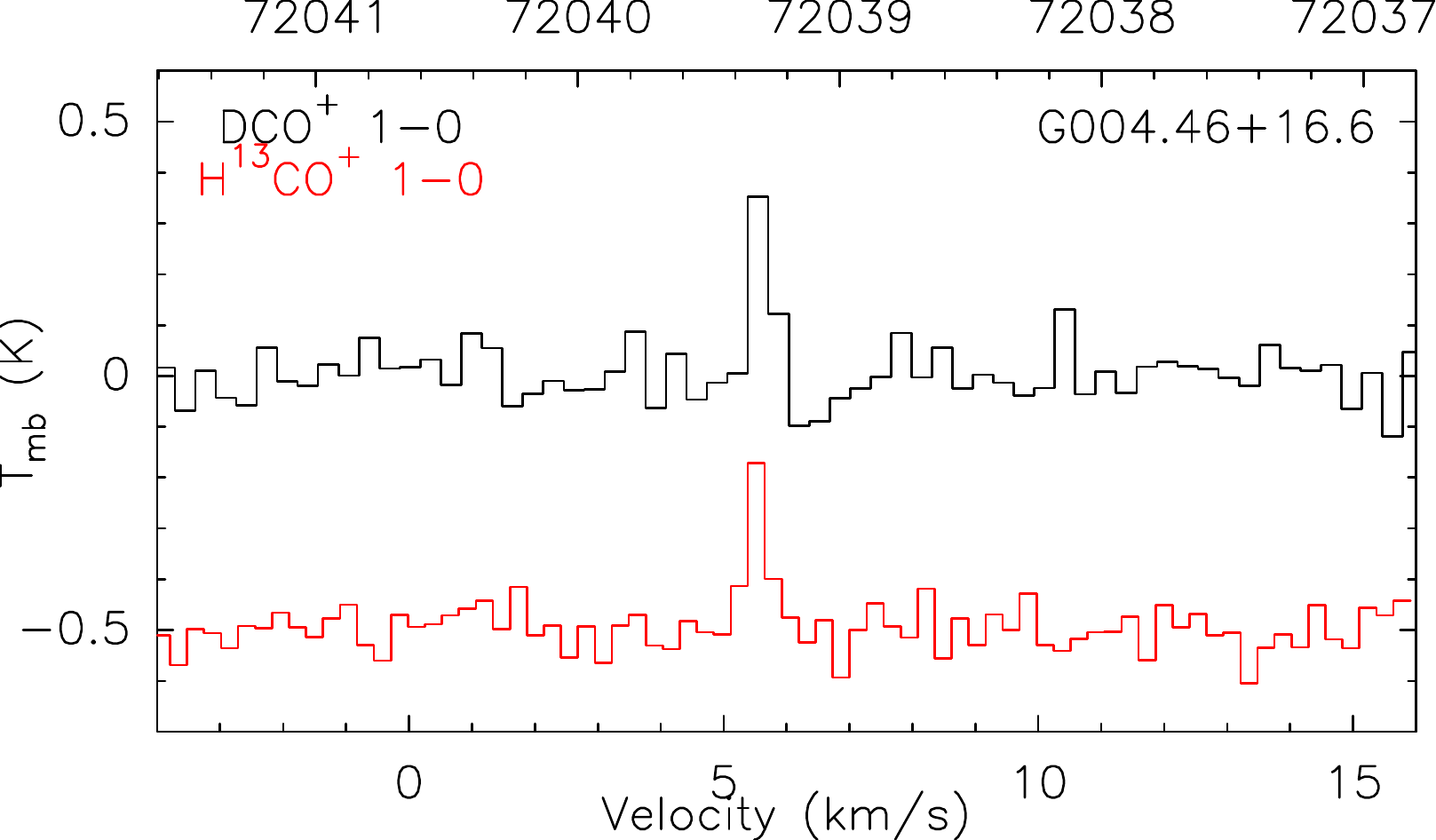}
\includegraphics[width=0.3\columnwidth]{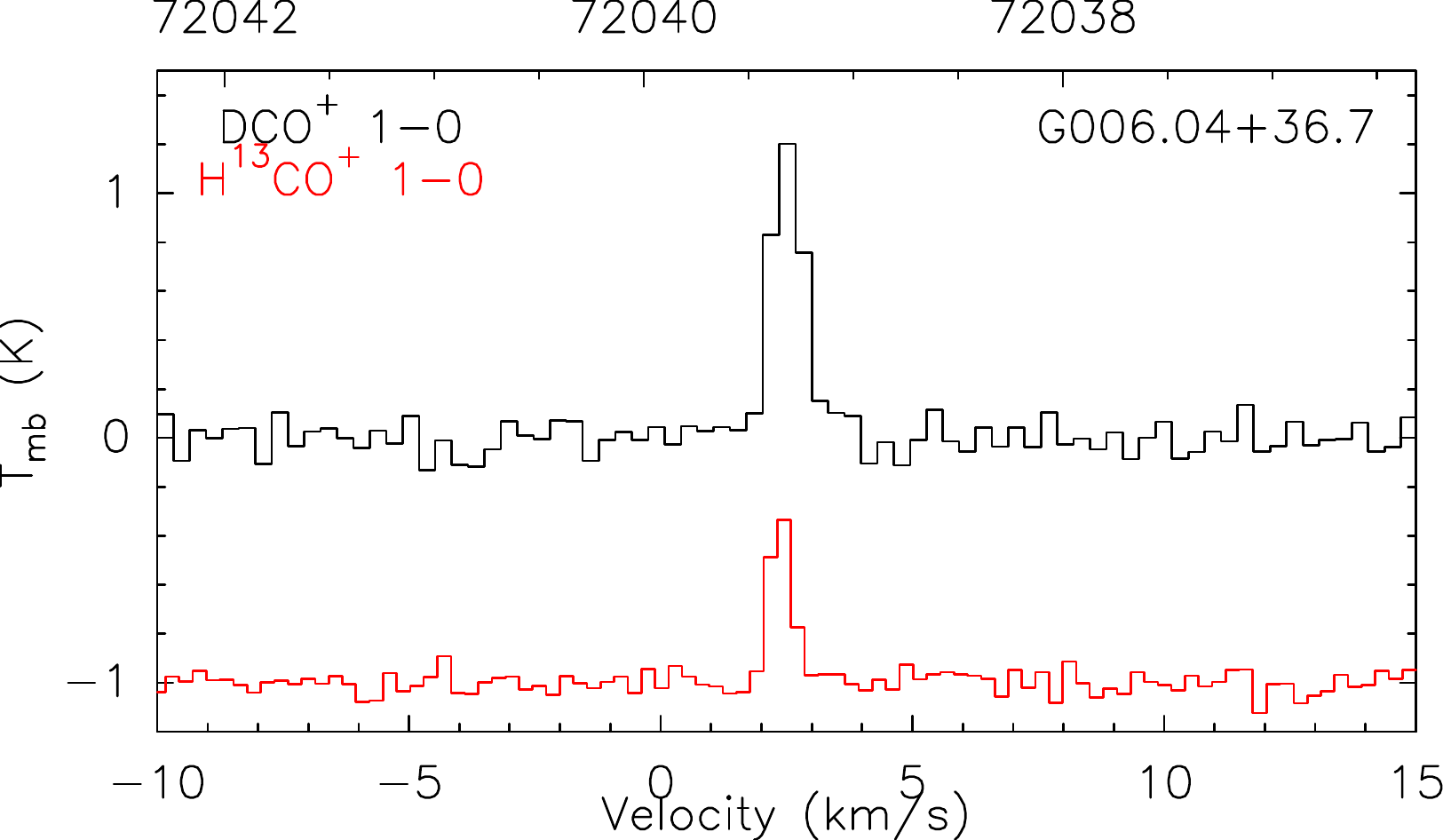}
\includegraphics[width=0.3\columnwidth]{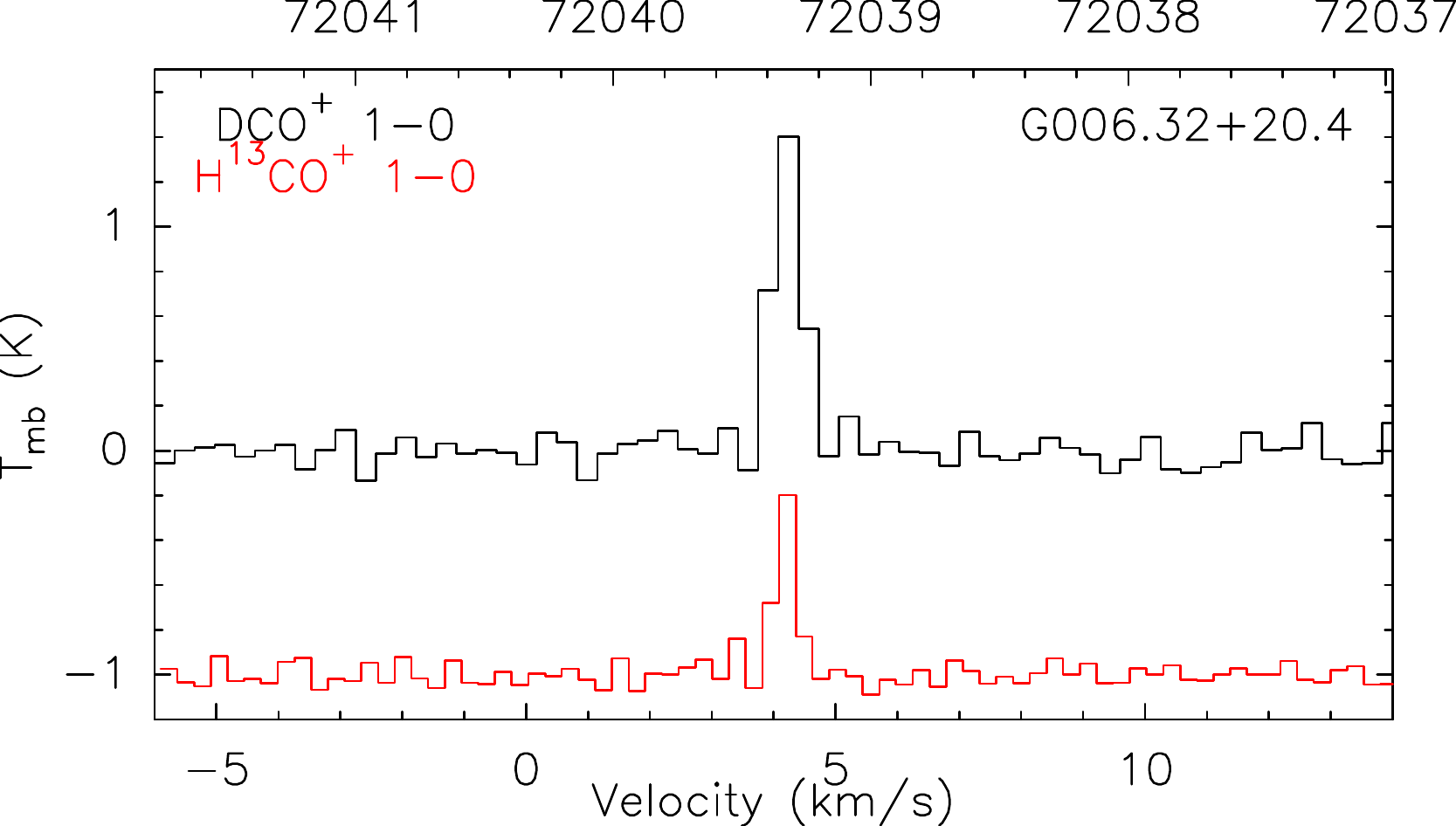}
\includegraphics[width=0.3\columnwidth]{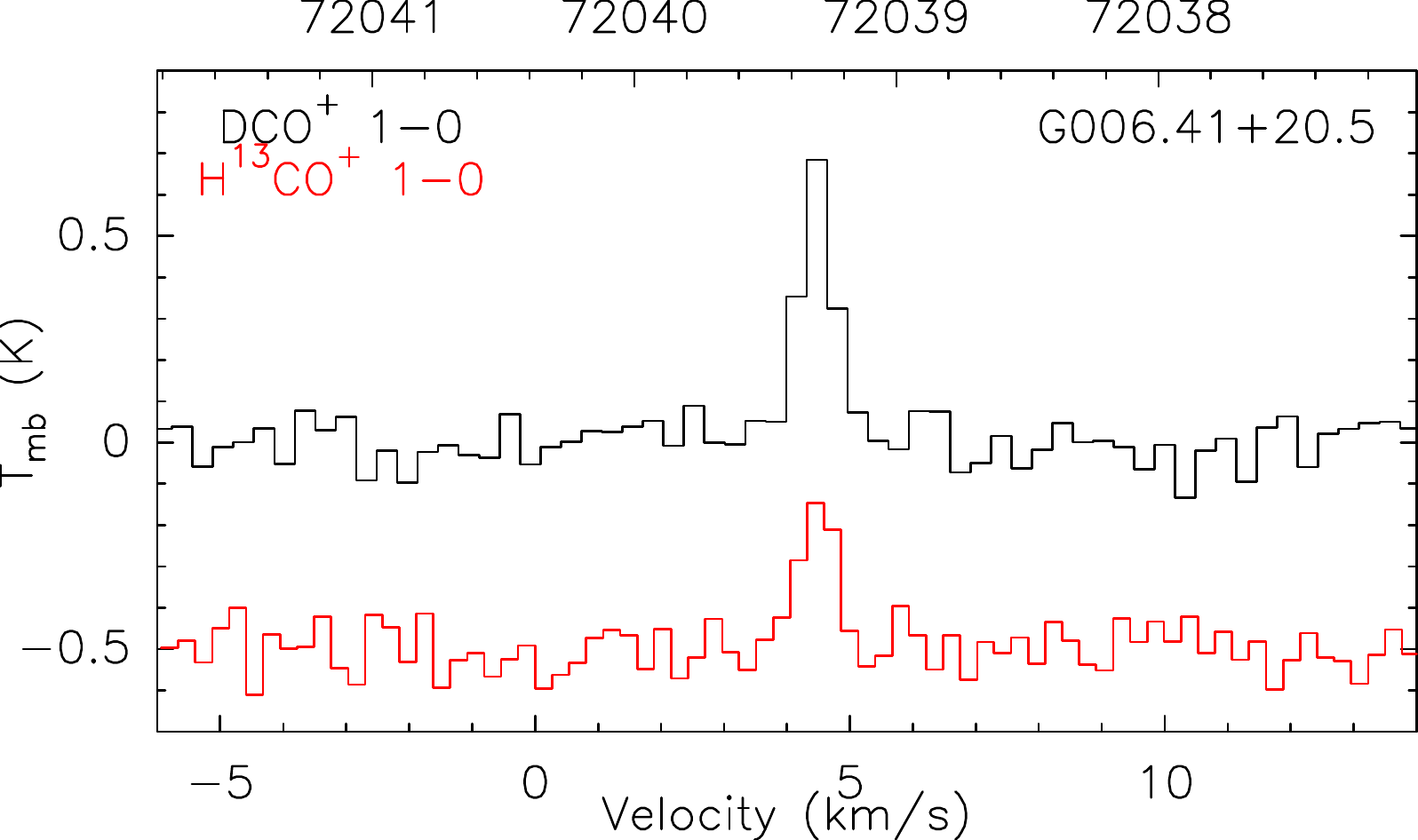}
\includegraphics[width=0.3\columnwidth]{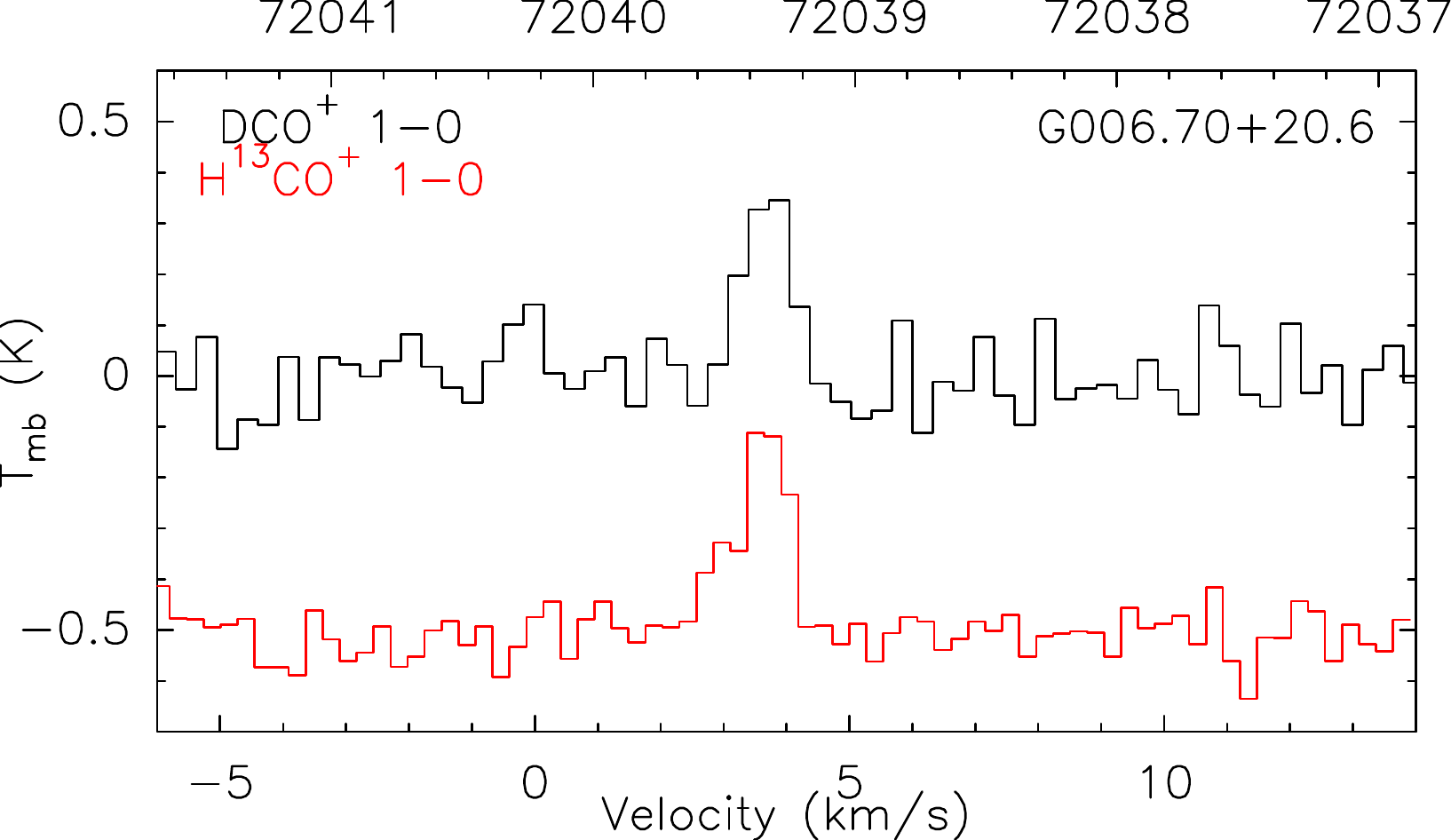}
\includegraphics[width=0.3\columnwidth]{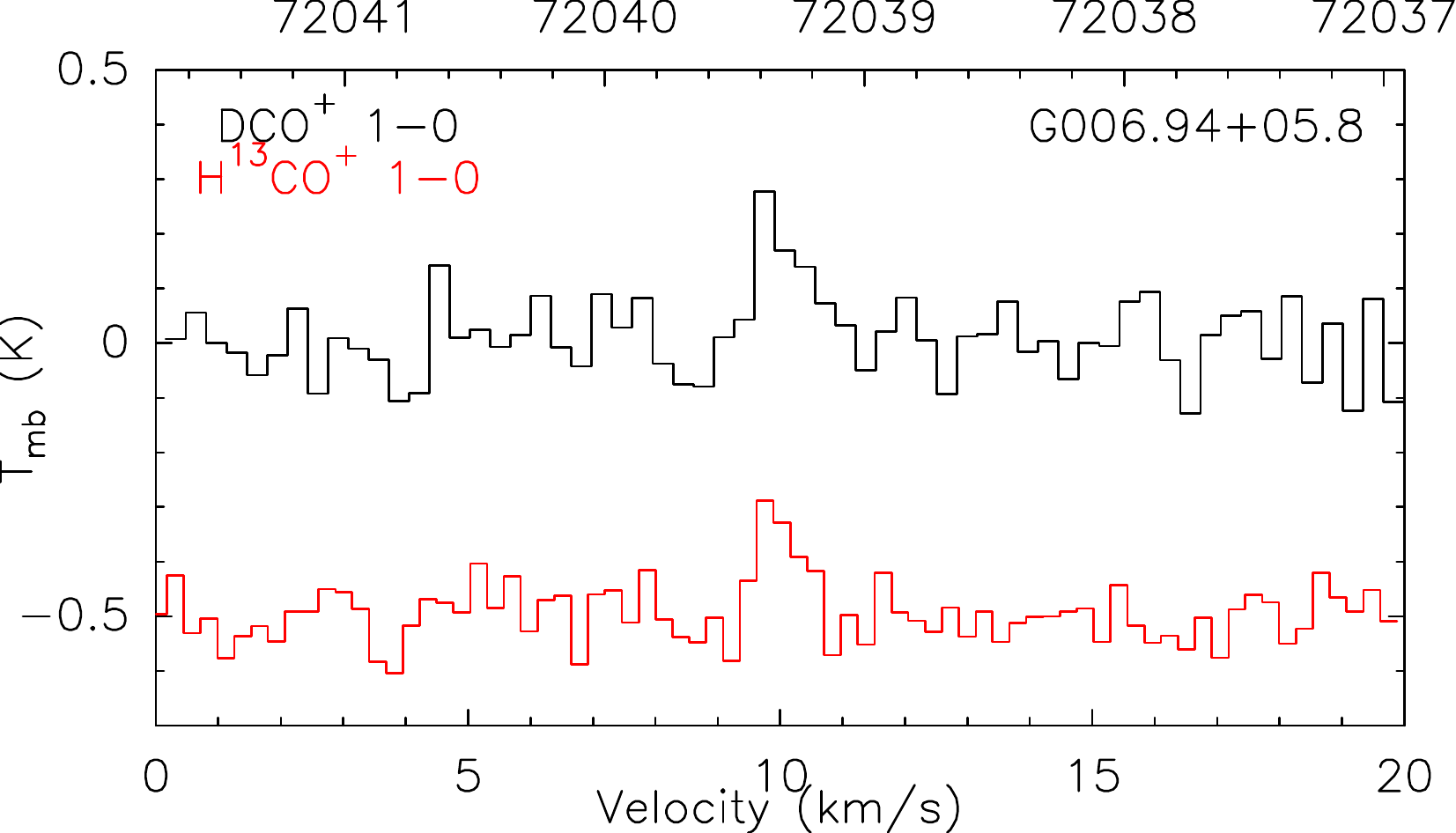}
\includegraphics[width=0.3\columnwidth]{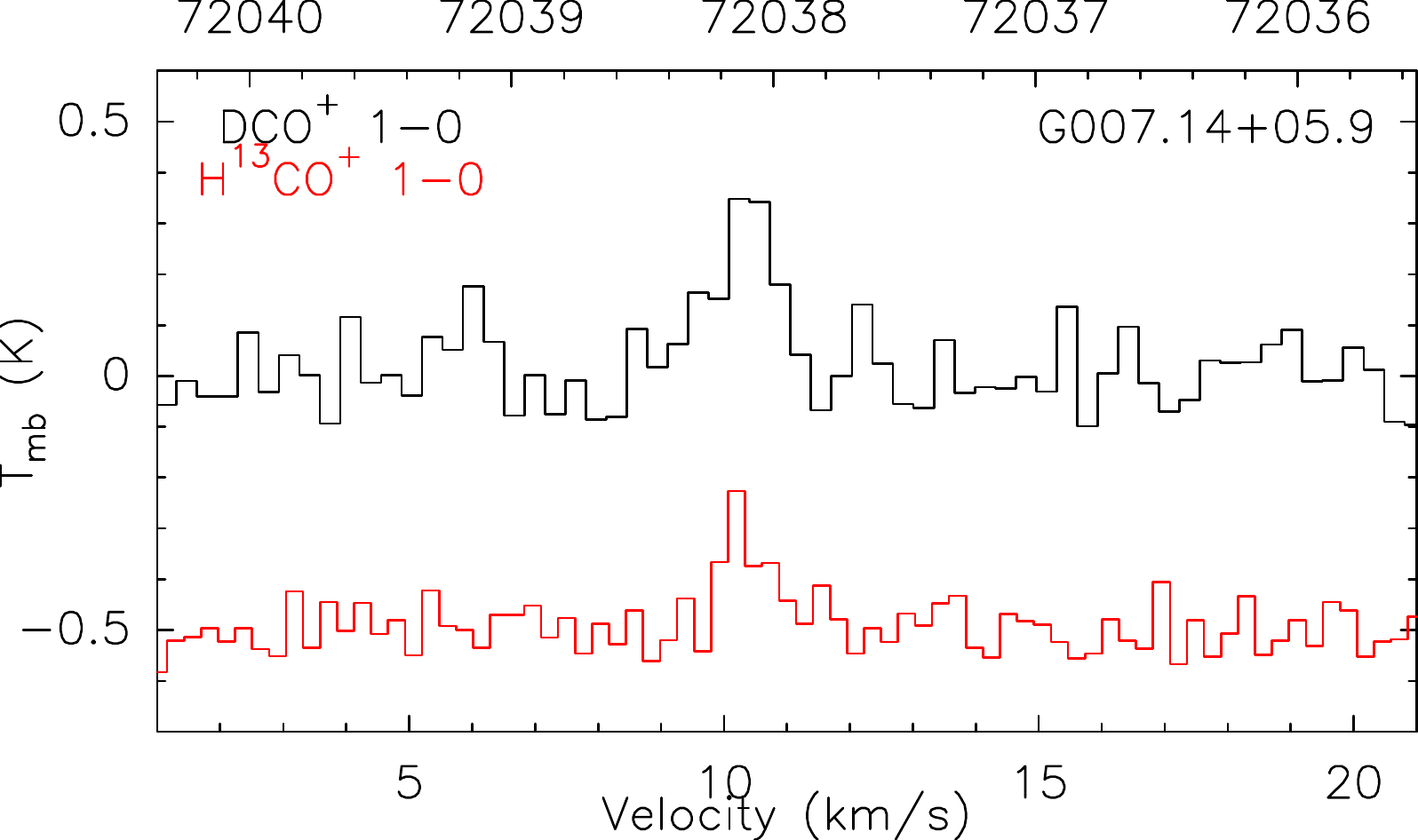}
\includegraphics[width=0.3\columnwidth]{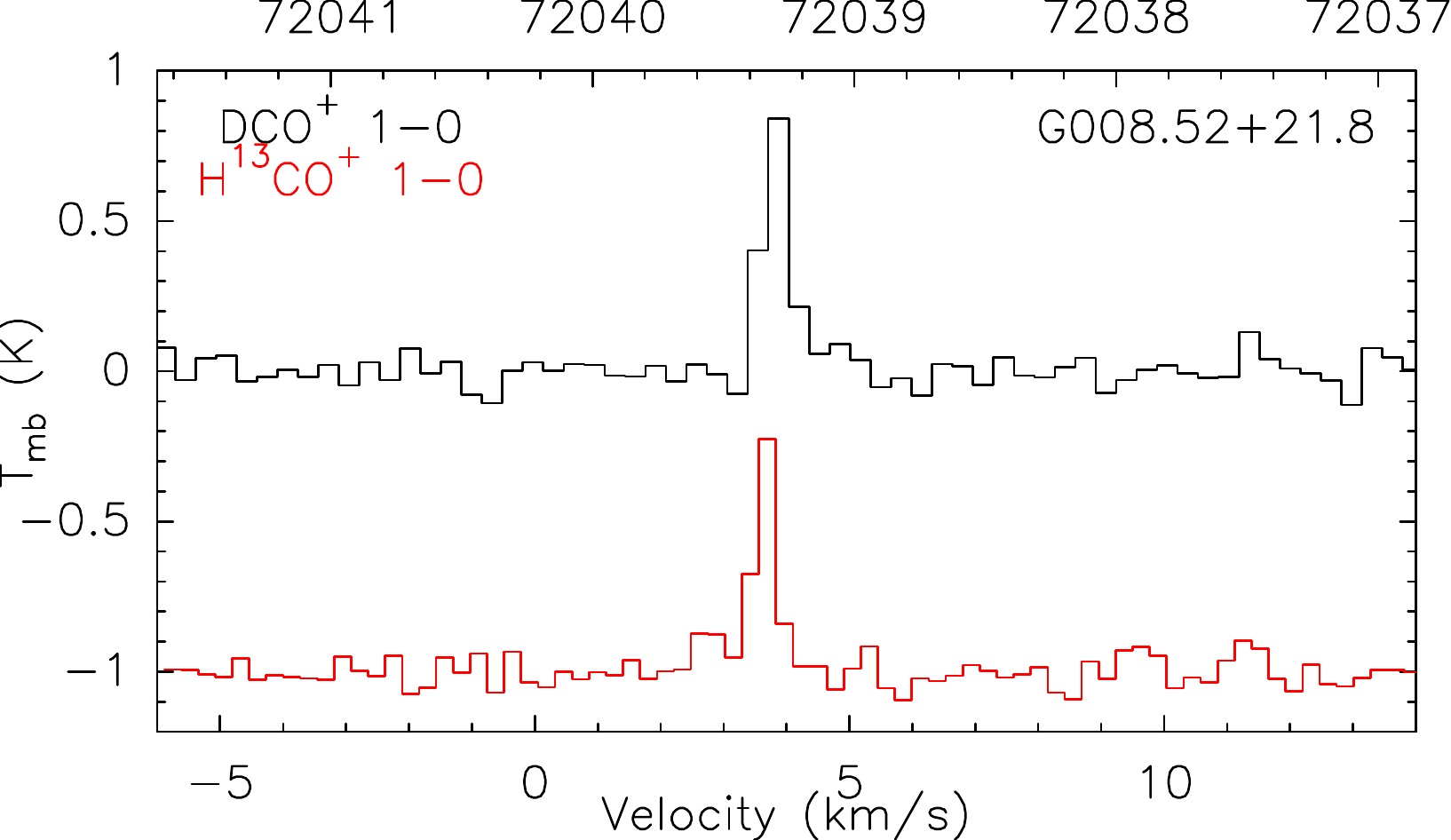}
\includegraphics[width=0.3\columnwidth]{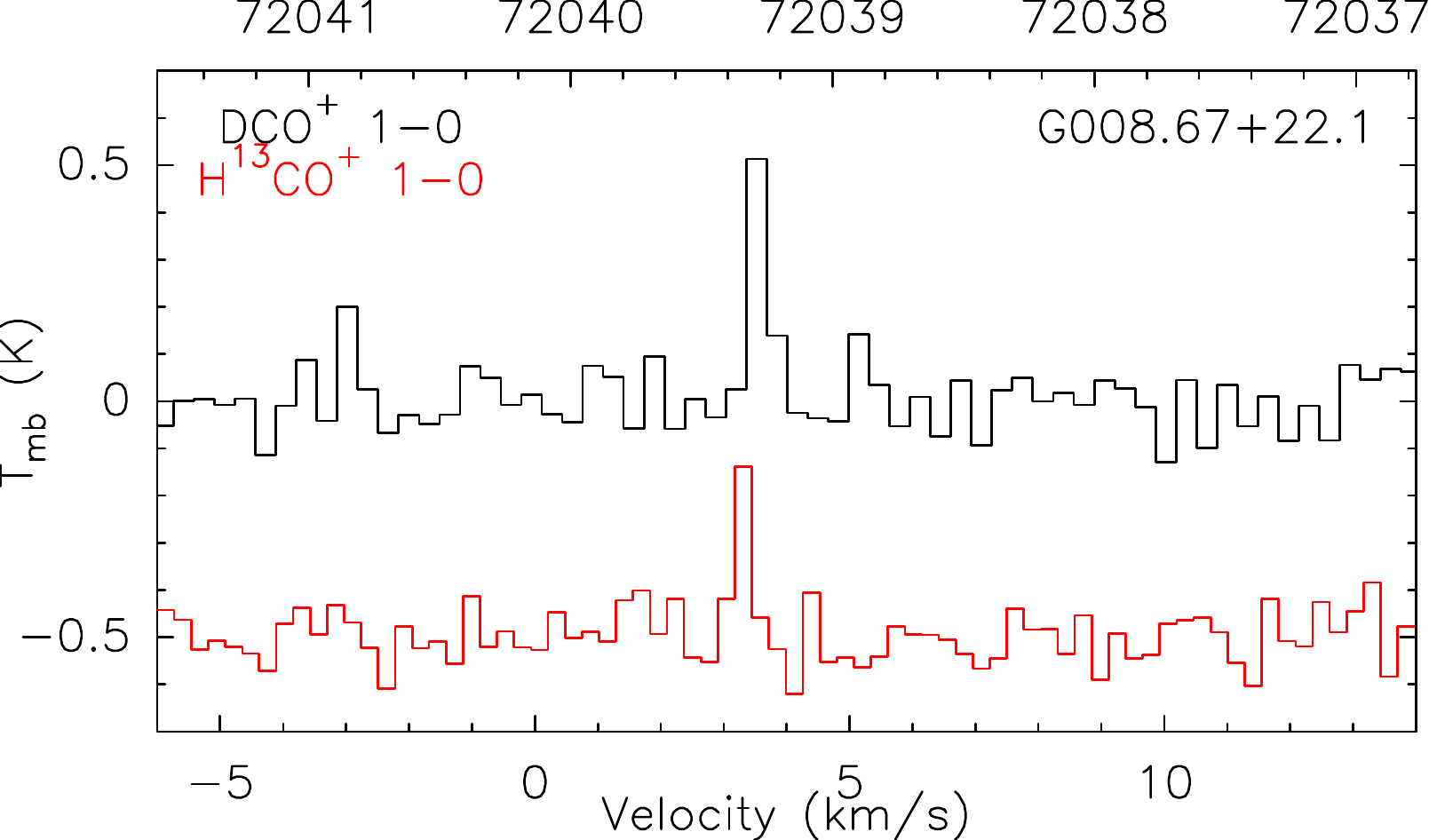}
\includegraphics[width=0.3\columnwidth]{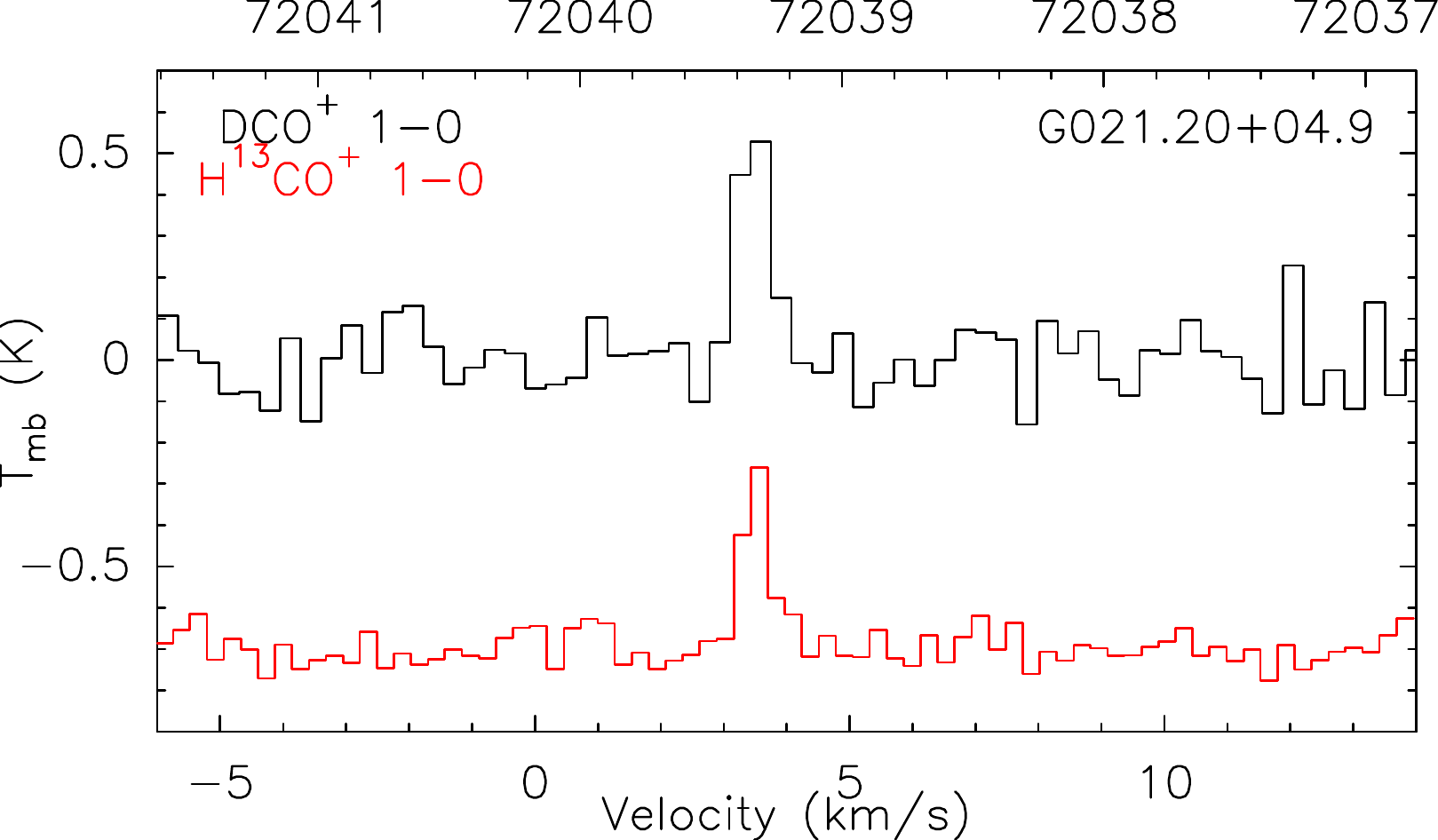}
\includegraphics[width=0.3\columnwidth]{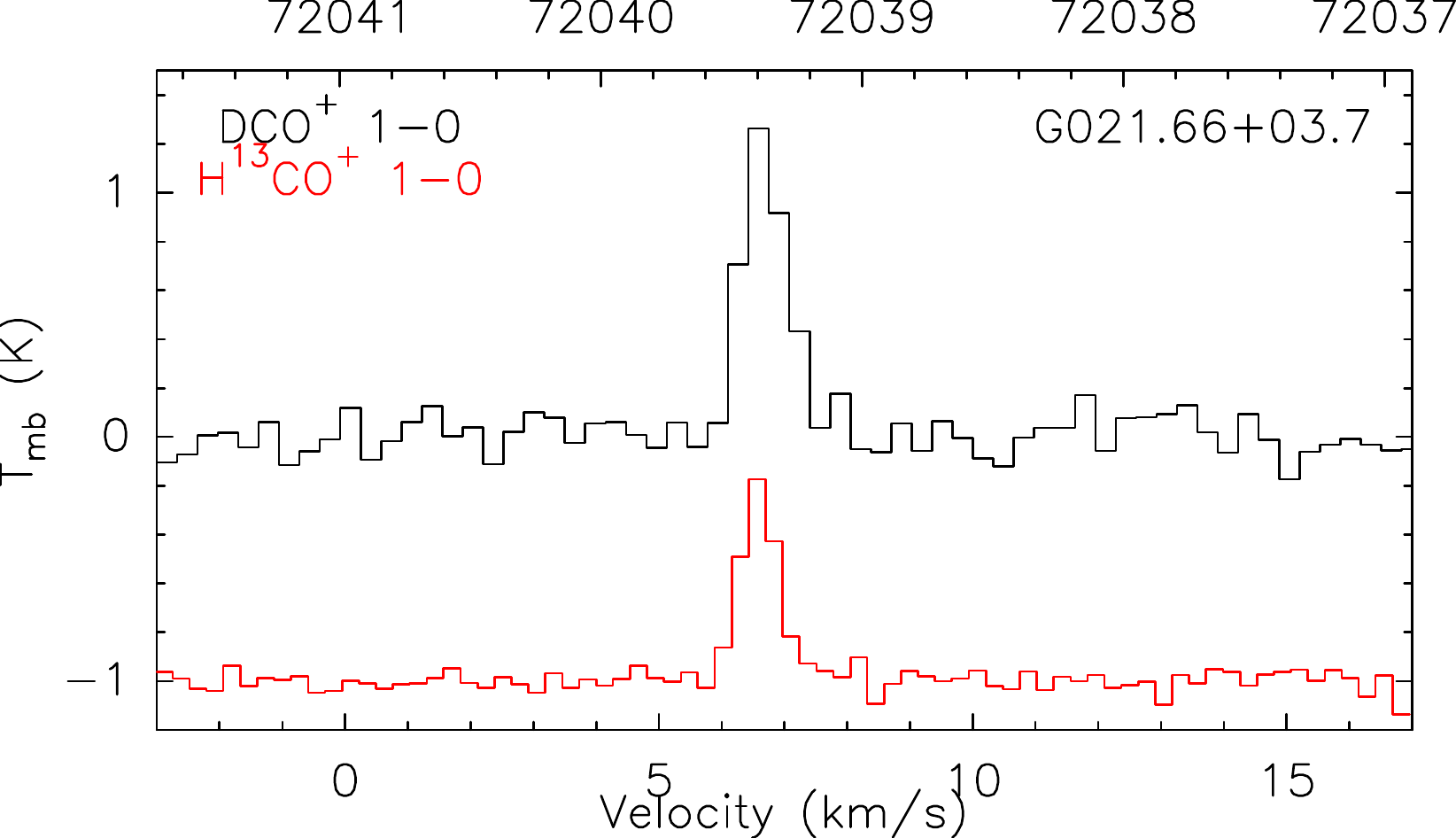}
\caption{Line profiles of DCO$^+$ and H$^{13}$CO$^+$ 1-0 with the low velocity resolution mode (AROWS mode 3).\centering}
\label{DCO+H13CO+mode3_2}
\end{figure}
\addtocounter{figure}{-1}
\begin{figure}
\centering
\includegraphics[width=0.3\columnwidth]{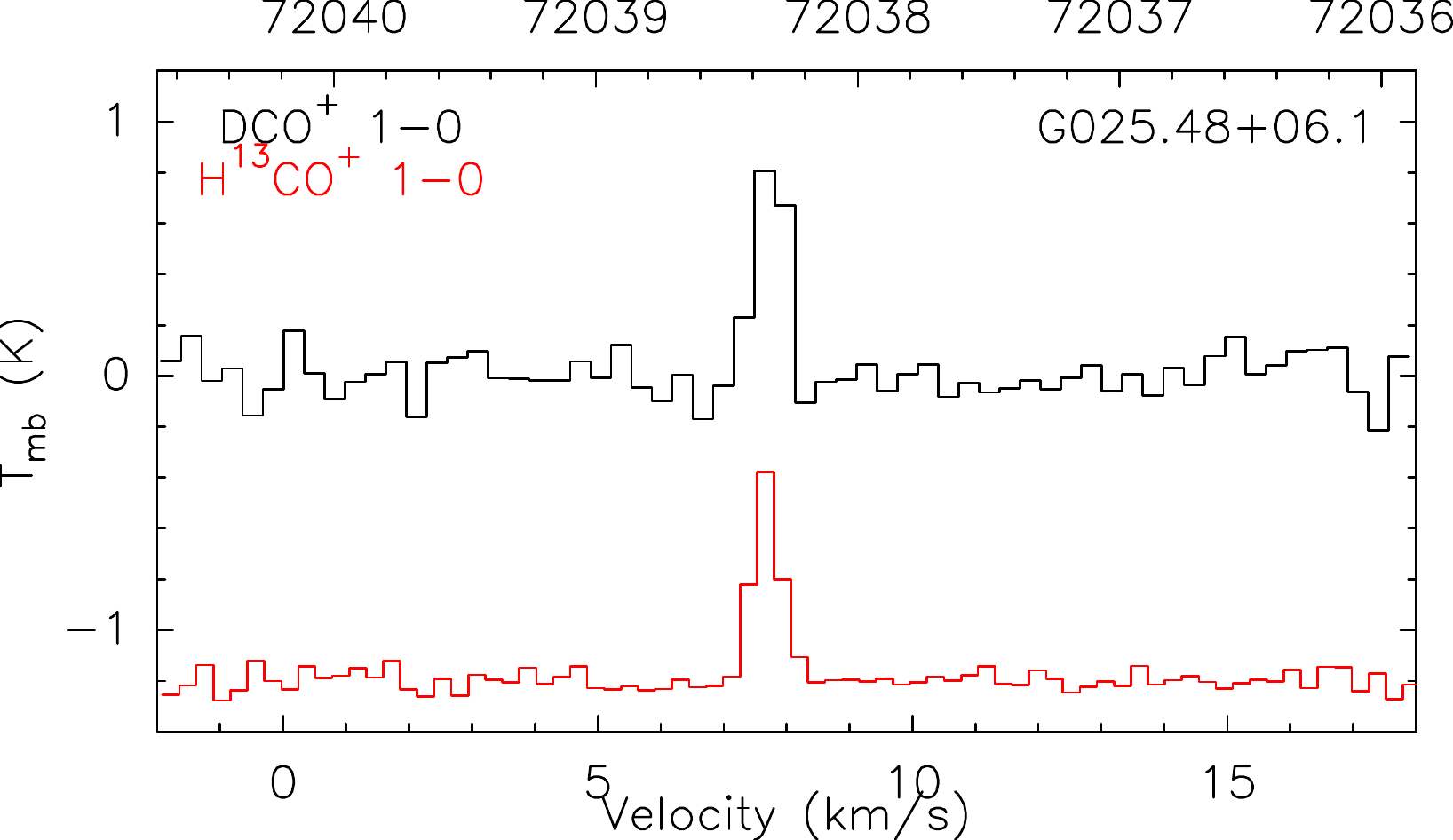}
\includegraphics[width=0.3\columnwidth]{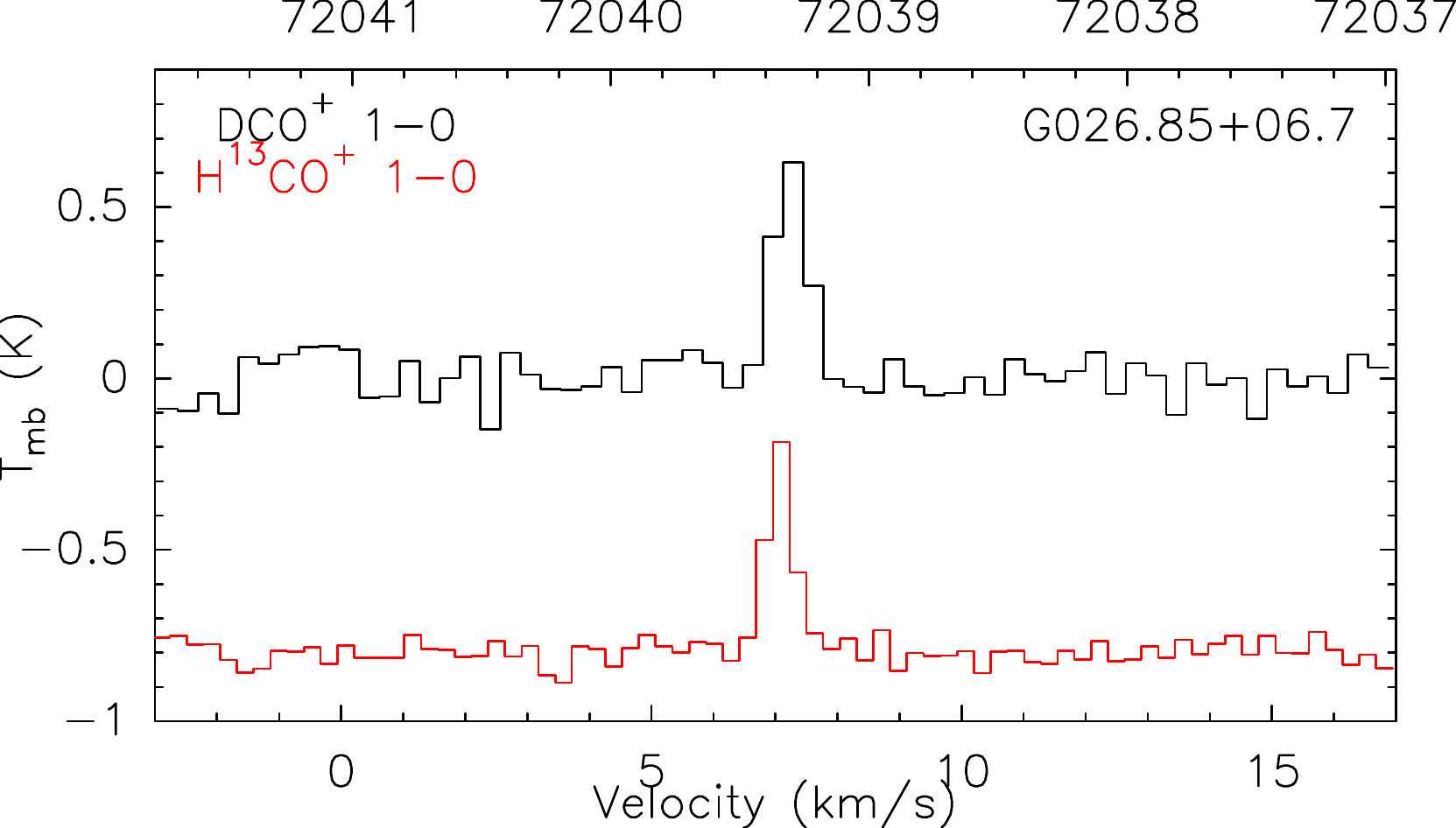}
\includegraphics[width=0.3\columnwidth]{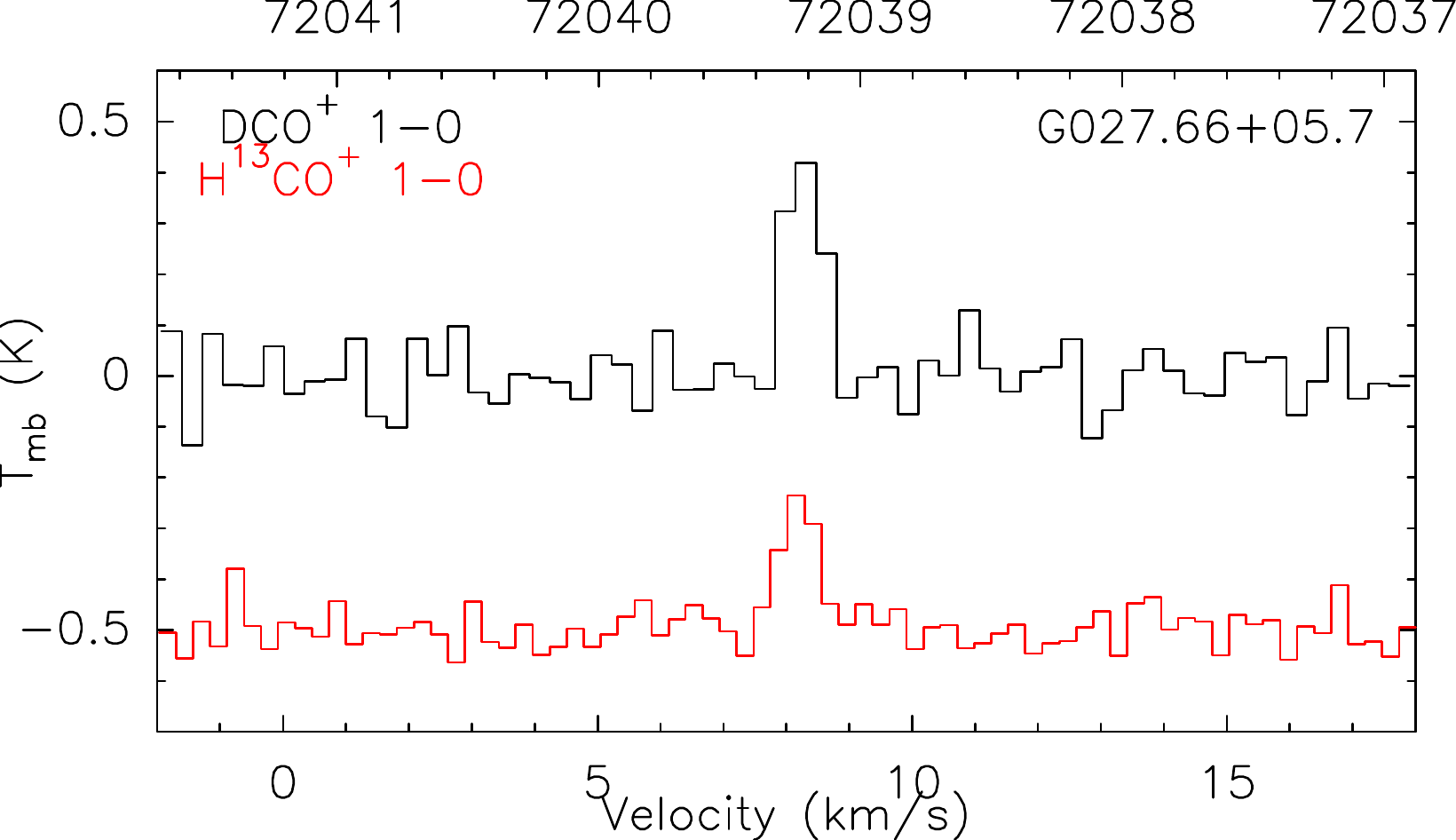}
\includegraphics[width=0.3\columnwidth]{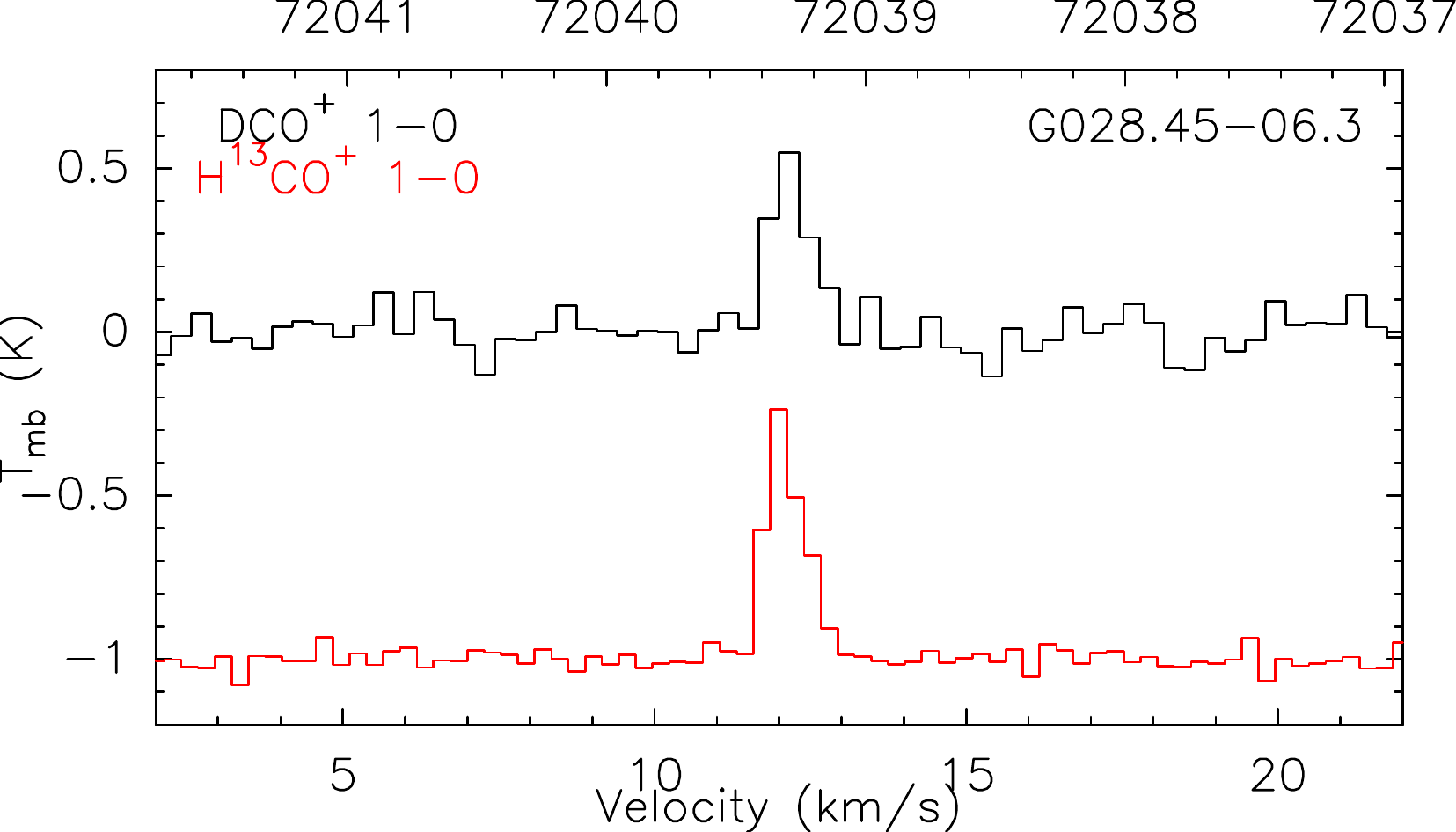}
\includegraphics[width=0.3\columnwidth]{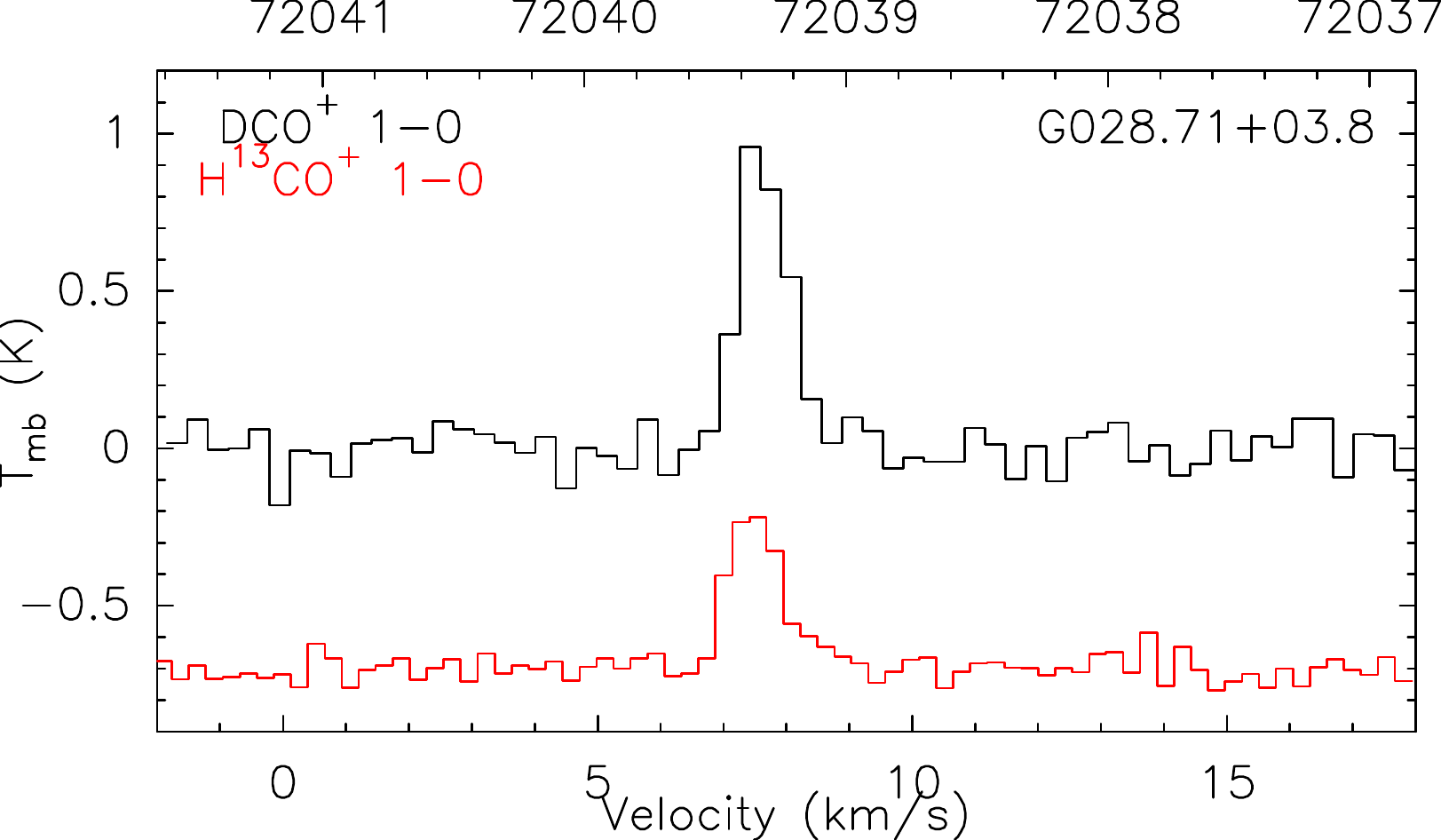}
\includegraphics[width=0.3\columnwidth]{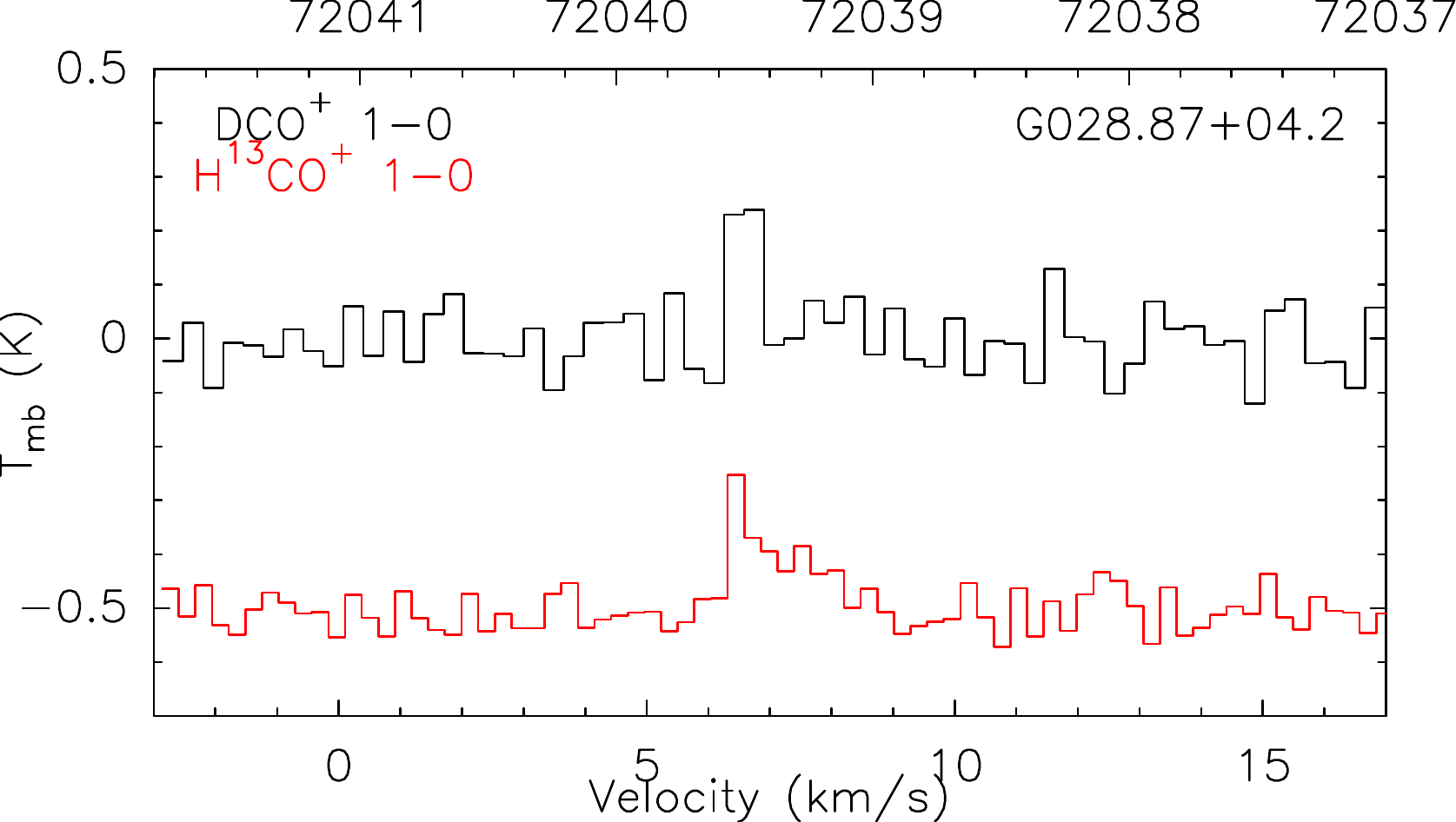}
\includegraphics[width=0.3\columnwidth]{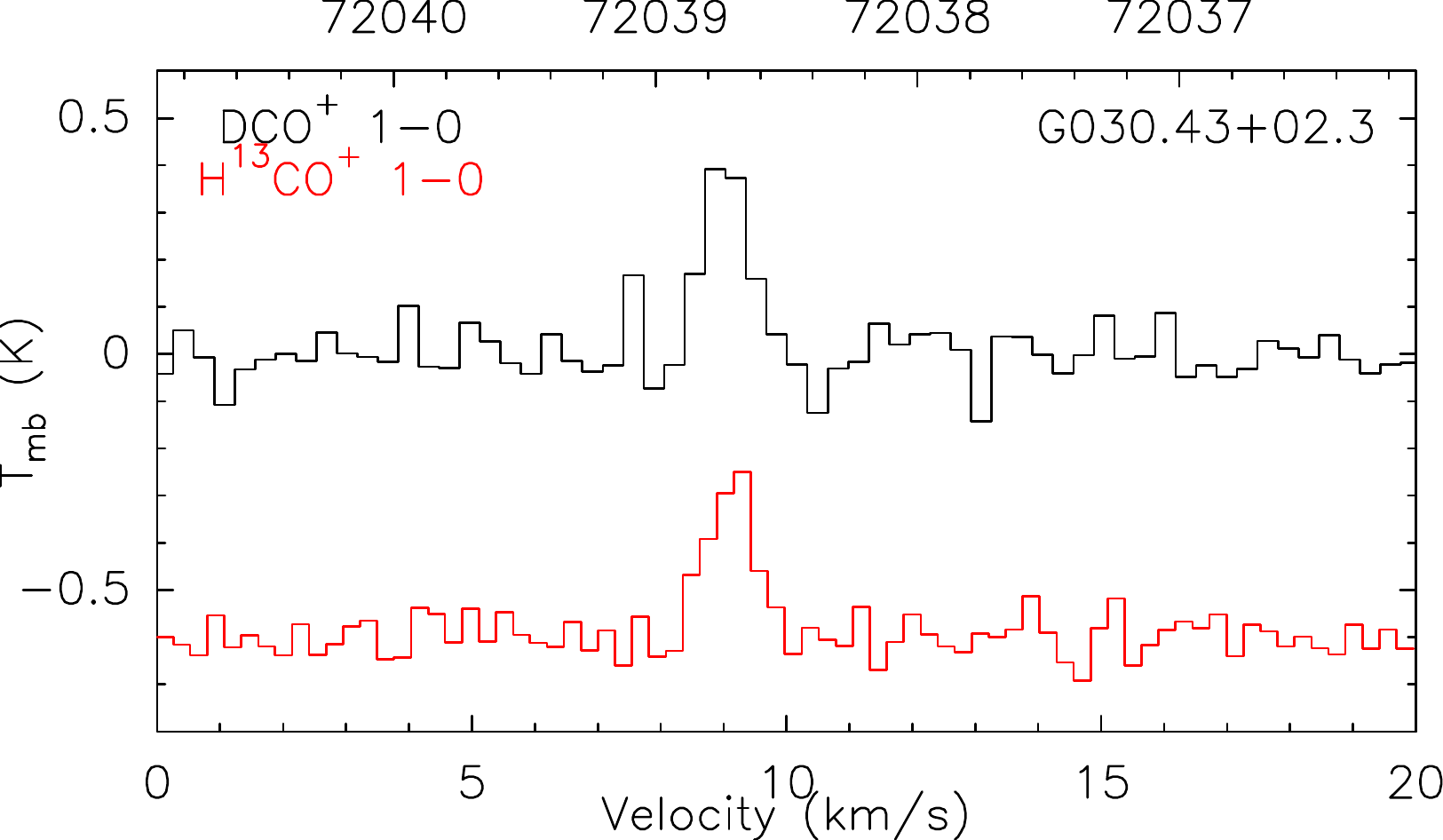}
\includegraphics[width=0.3\columnwidth]{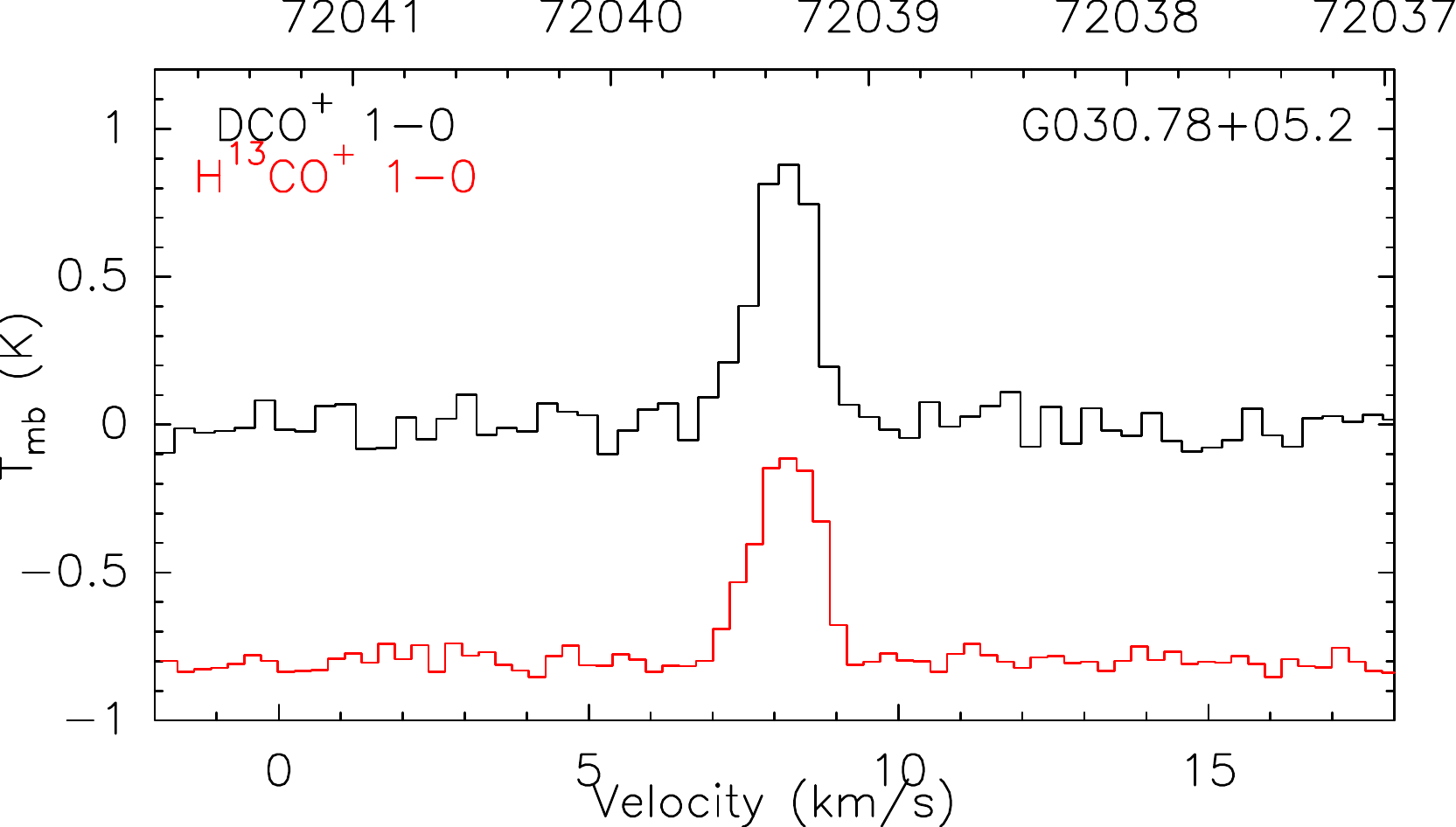}
\includegraphics[width=0.3\columnwidth]{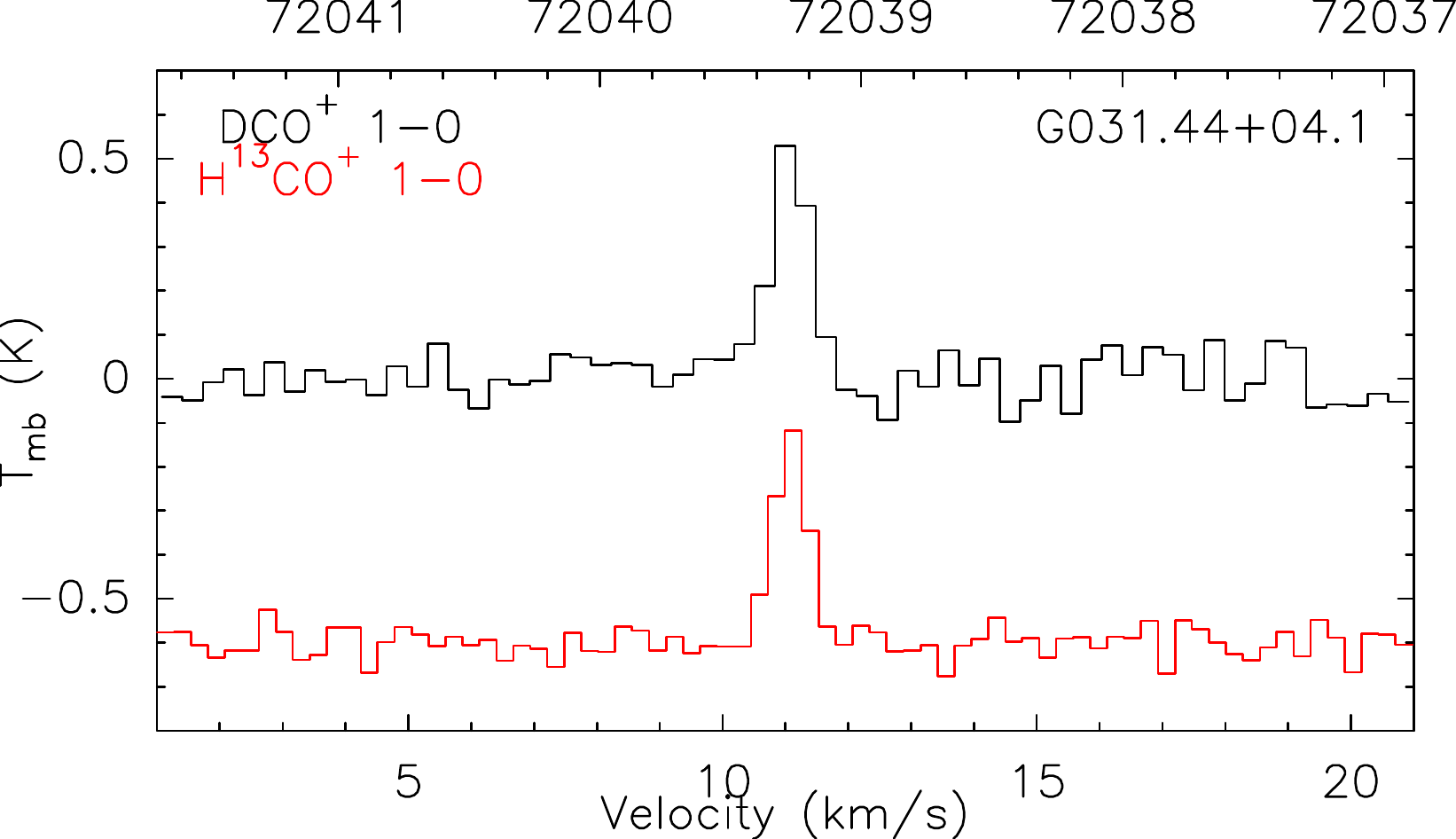}
\includegraphics[width=0.3\columnwidth]{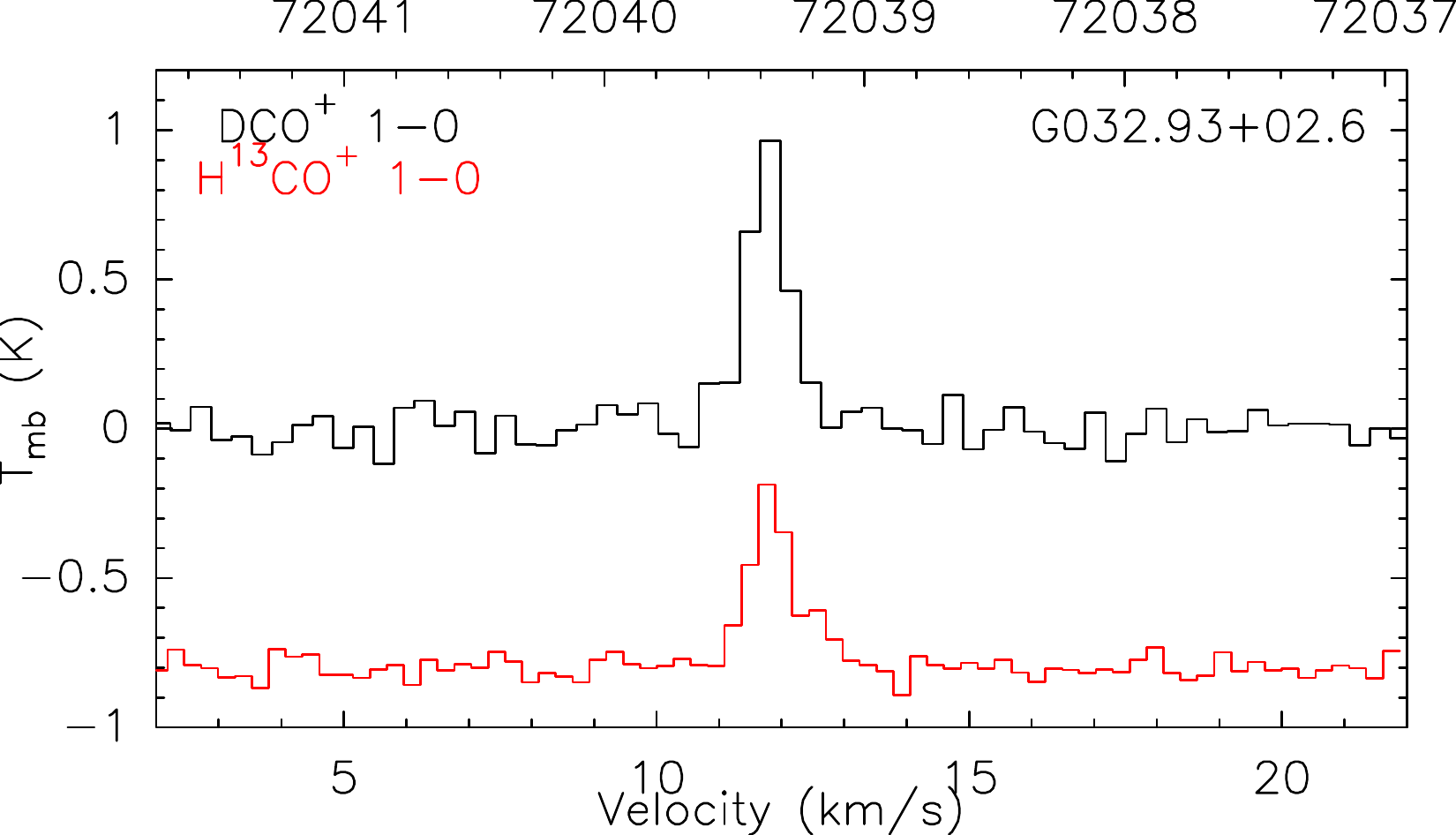}
\includegraphics[width=0.3\columnwidth]{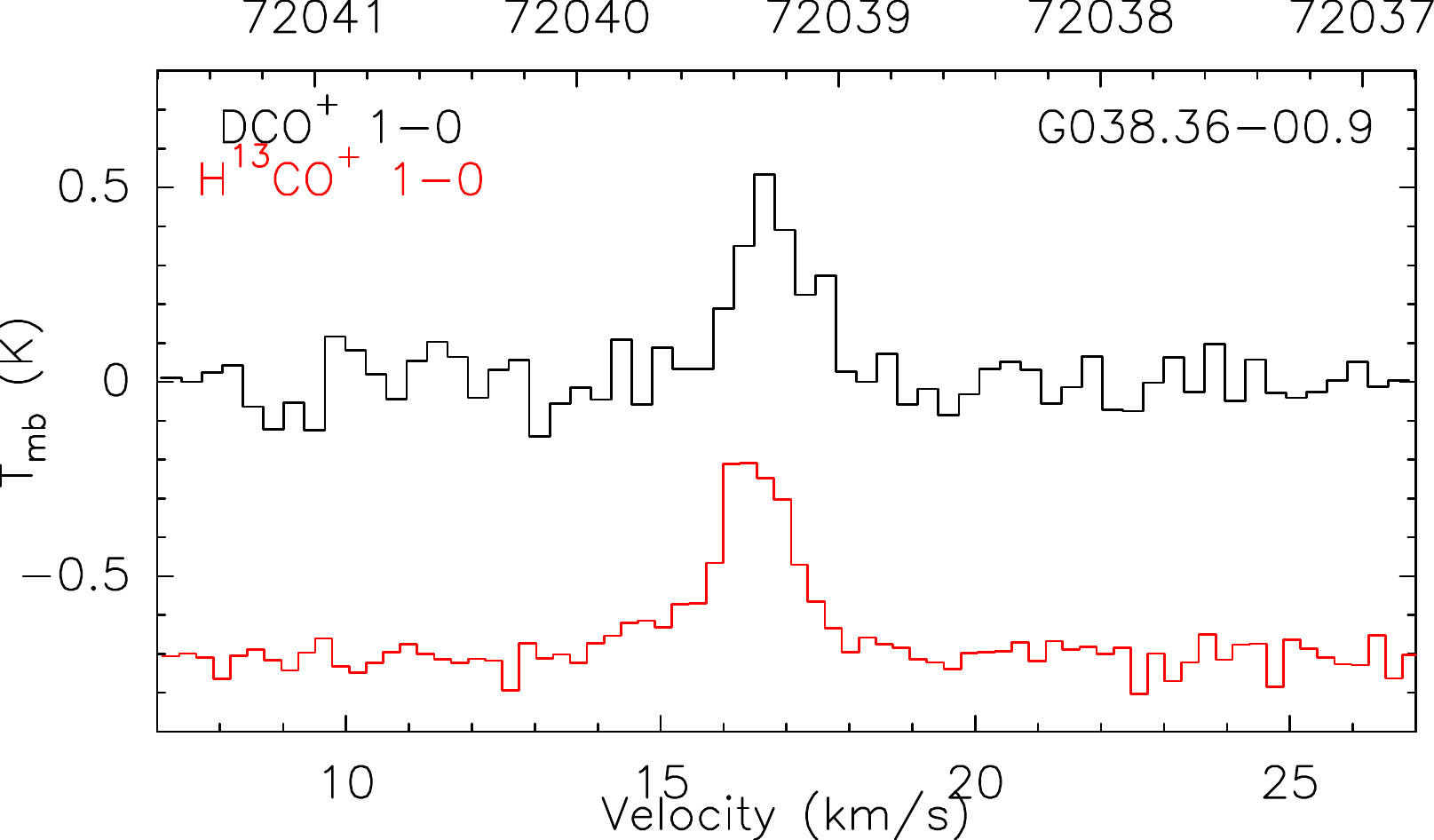}
\includegraphics[width=0.3\columnwidth]{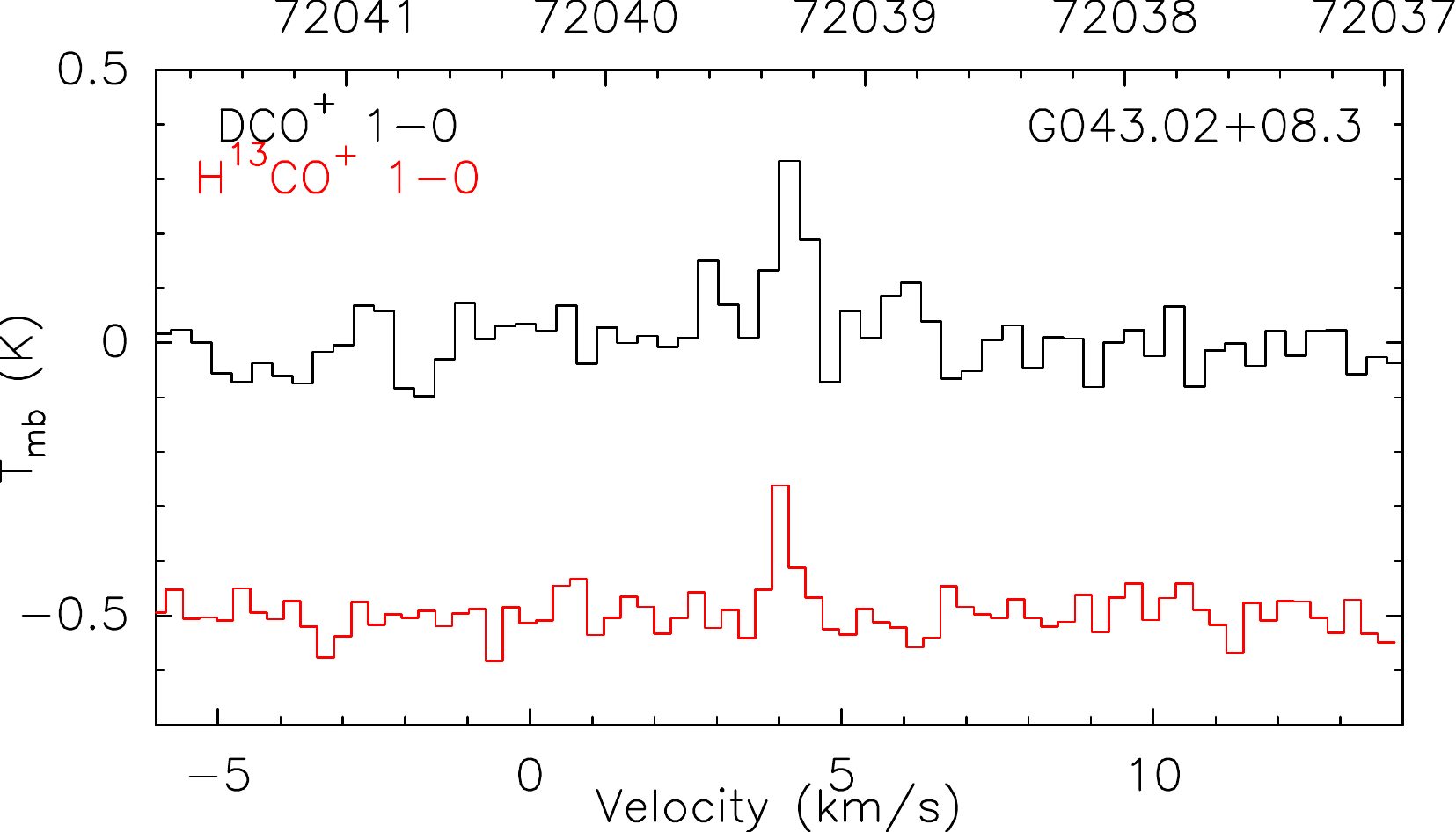}
\includegraphics[width=0.3\columnwidth]{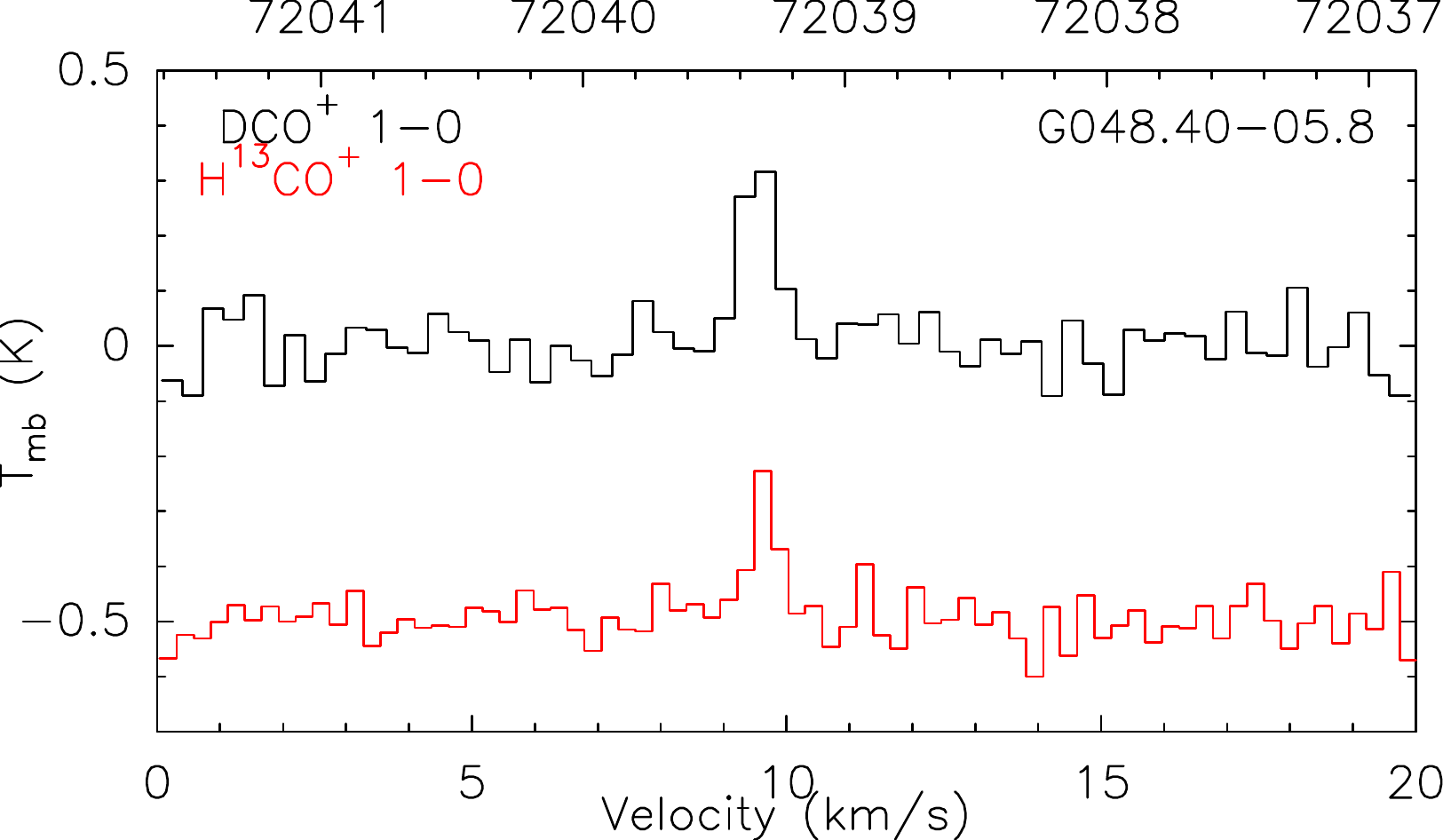}
\includegraphics[width=0.3\columnwidth]{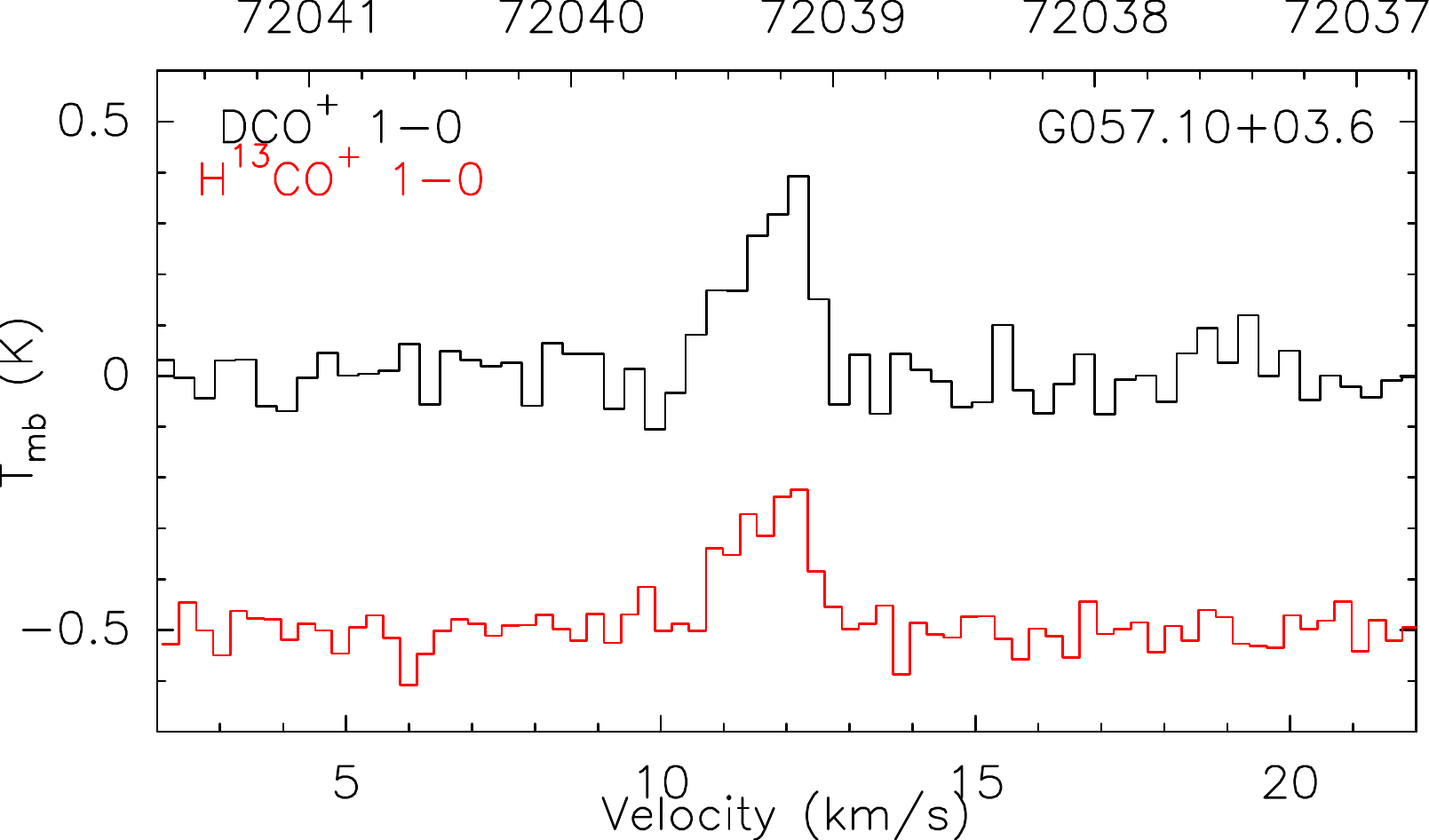}
\includegraphics[width=0.3\columnwidth]{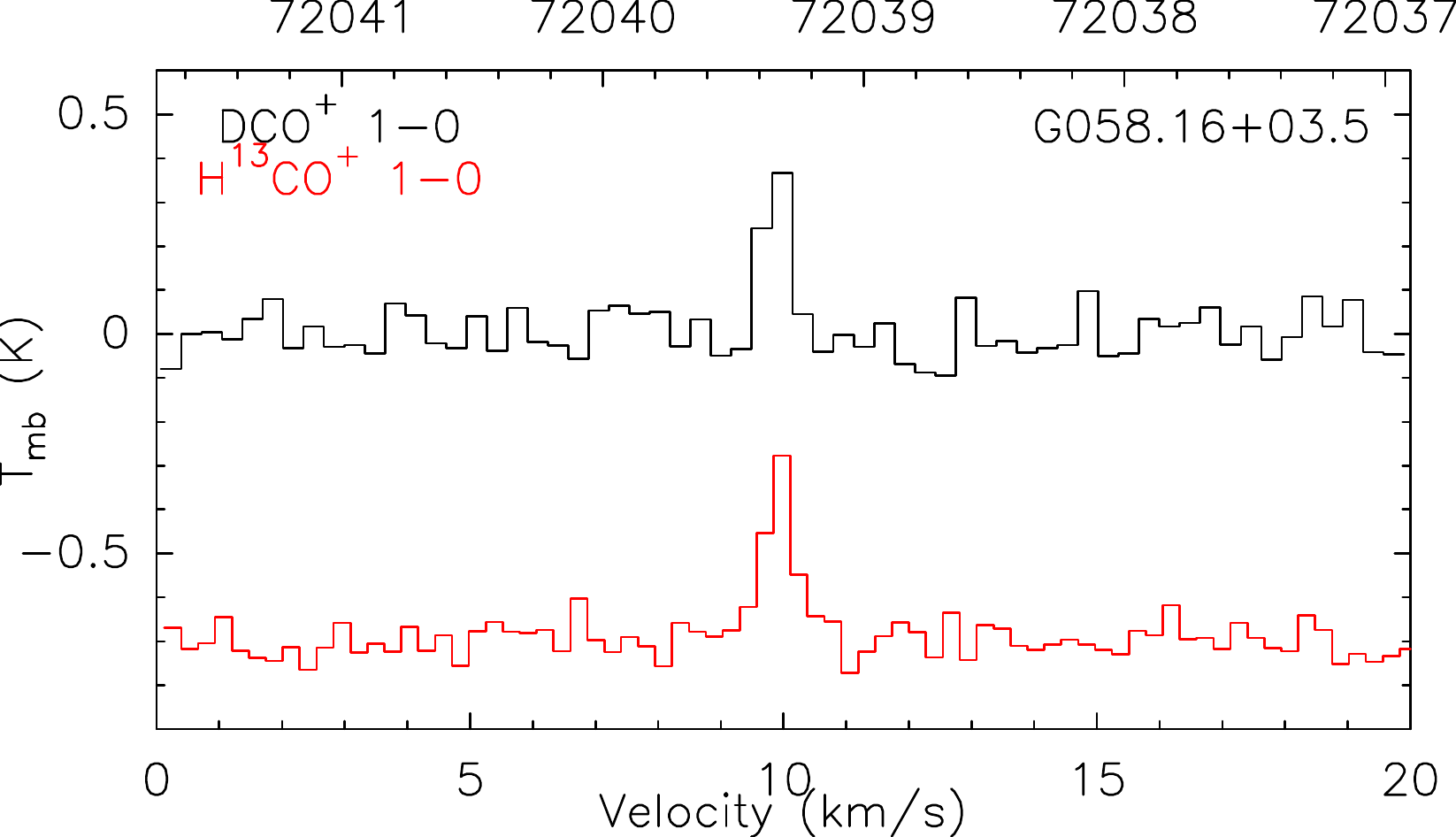}
\includegraphics[width=0.3\columnwidth]{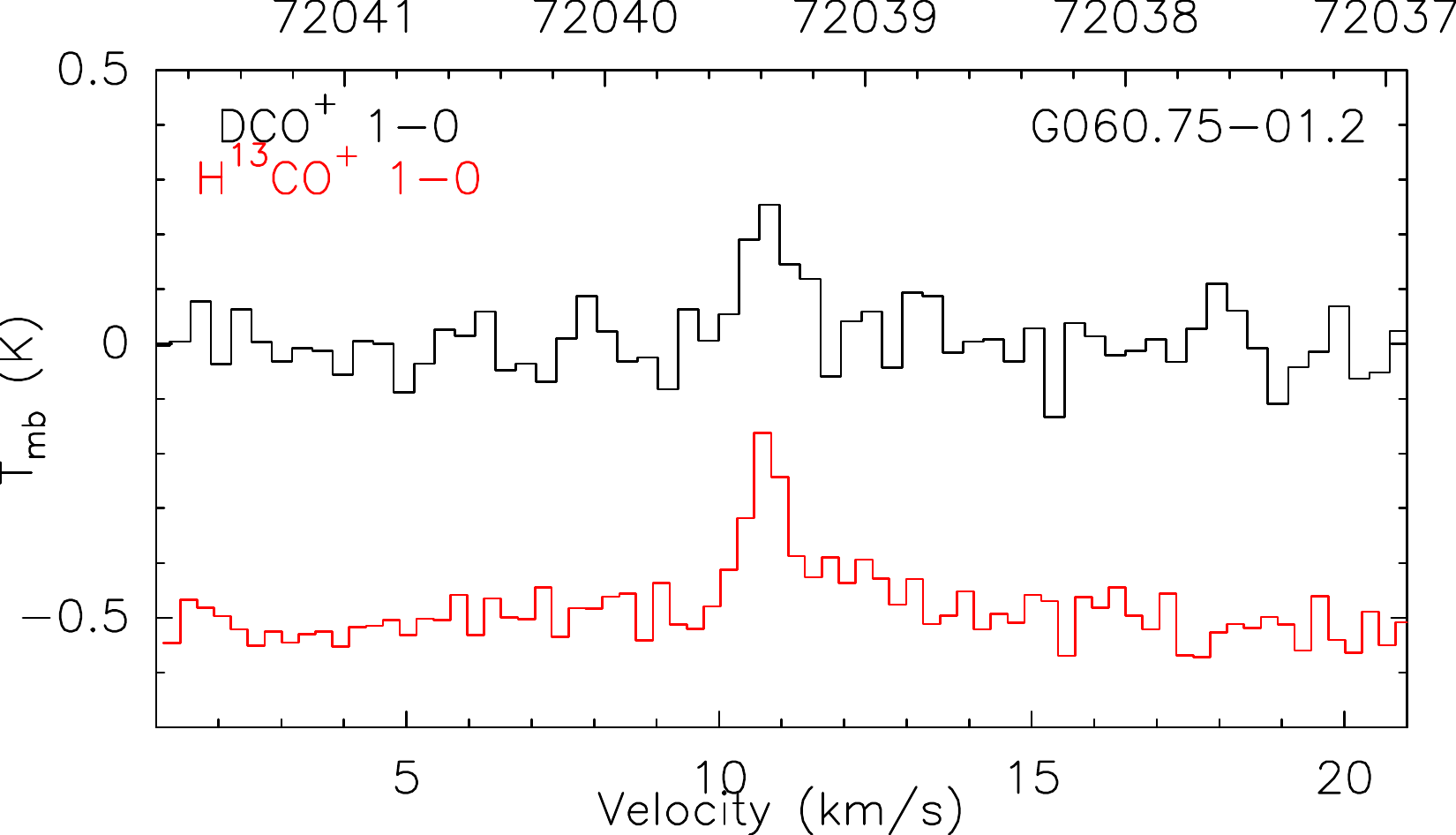}
\includegraphics[width=0.3\columnwidth]{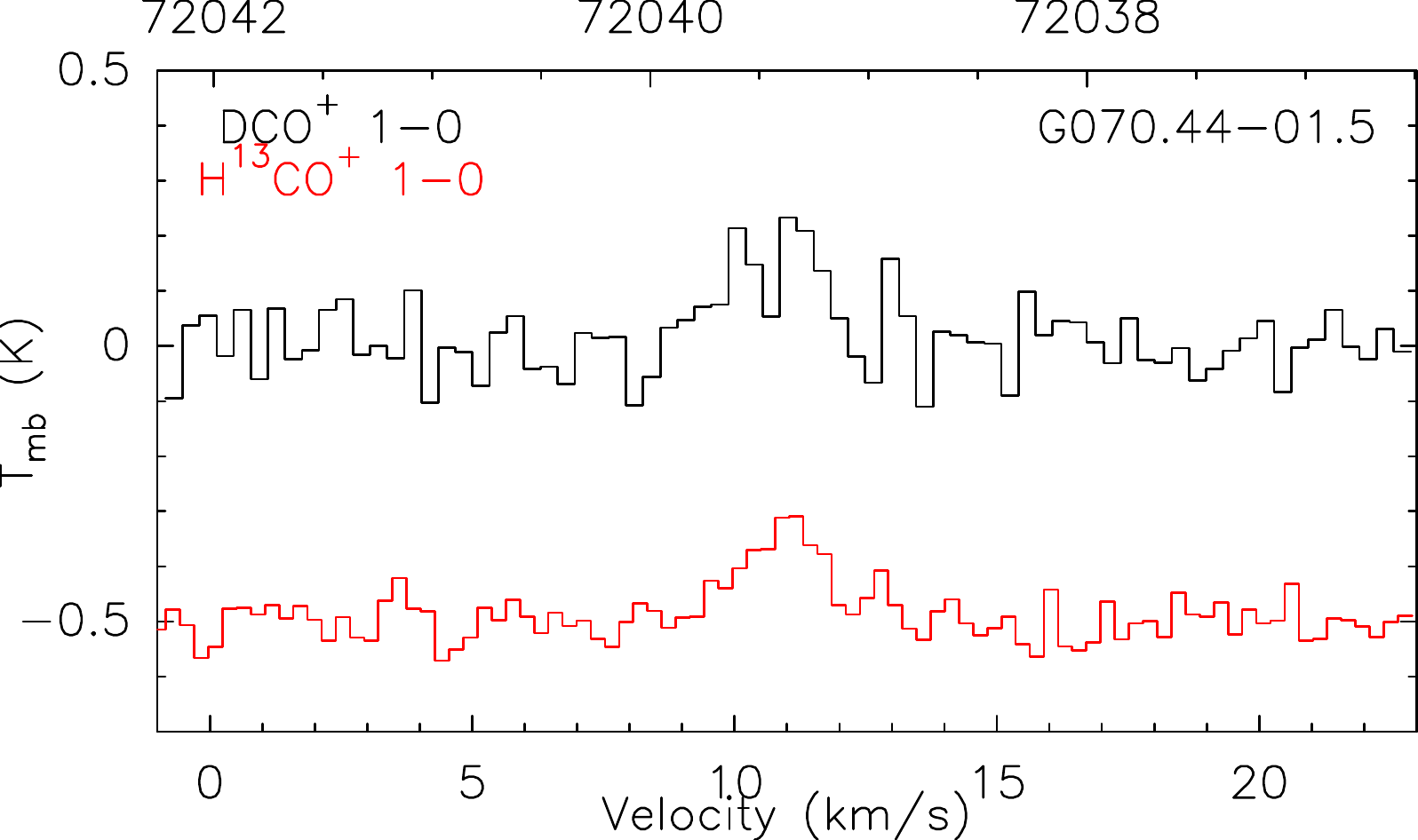}
\includegraphics[width=0.3\columnwidth]{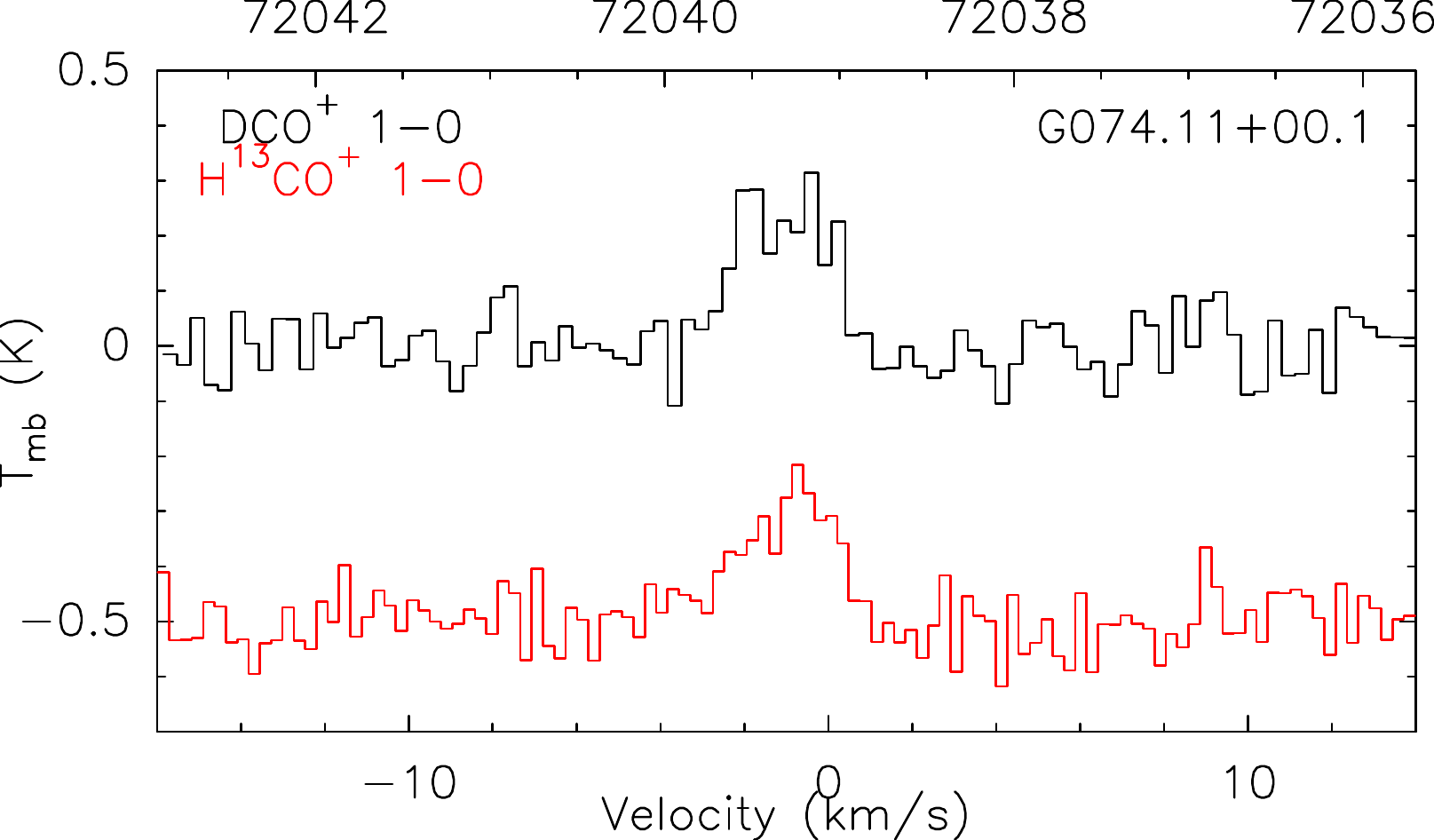}
\includegraphics[width=0.3\columnwidth]{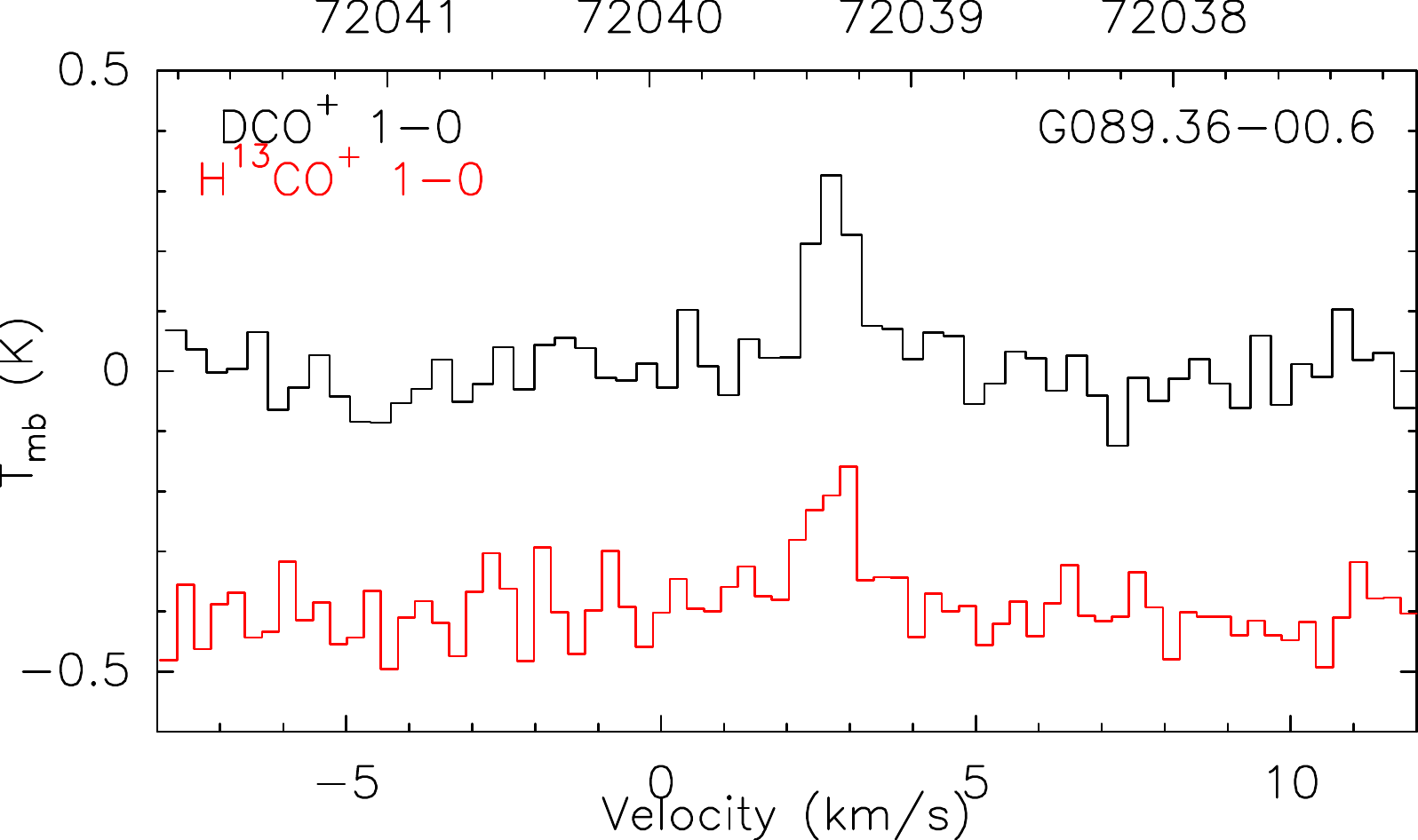}
\includegraphics[width=0.3\columnwidth]{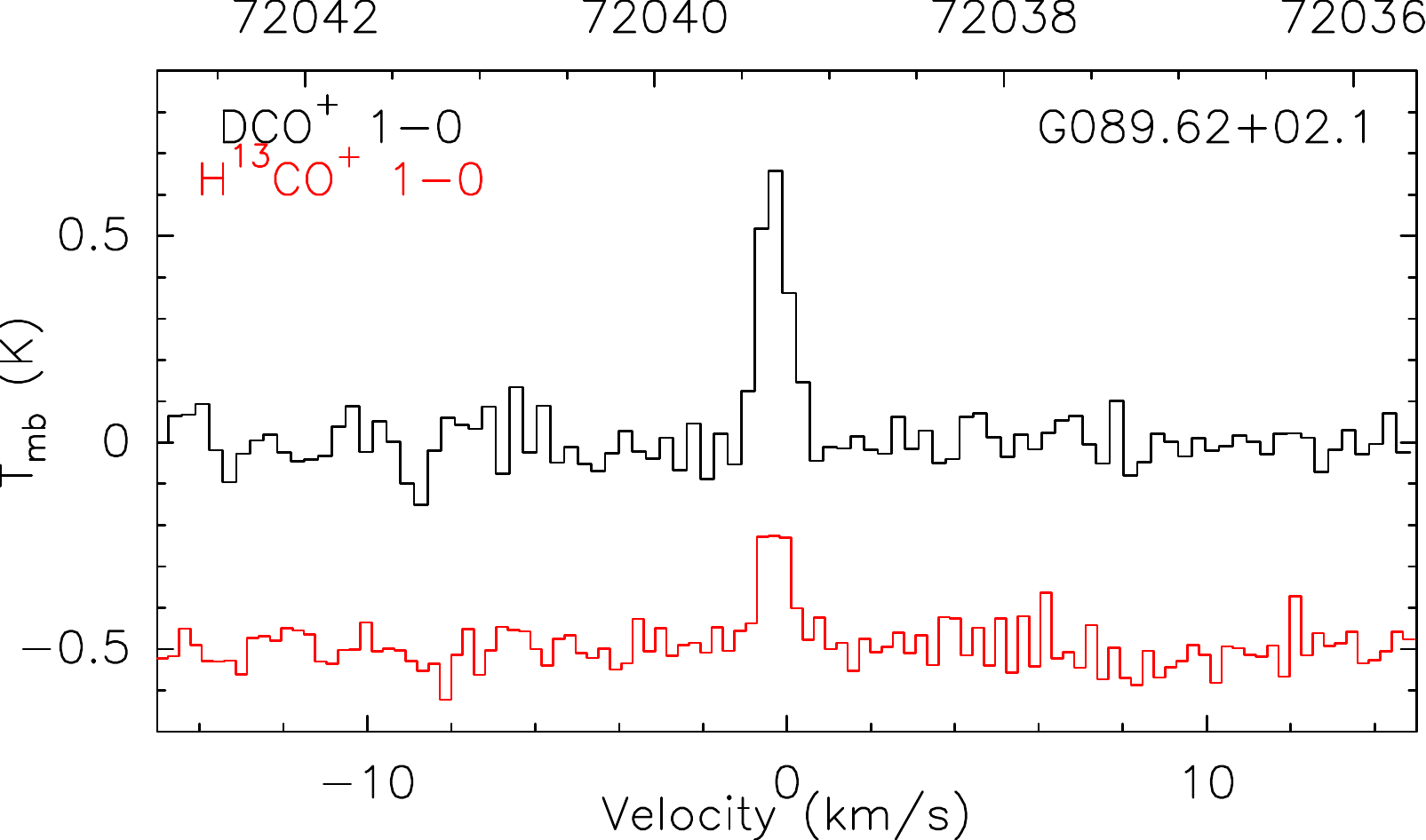}
\includegraphics[width=0.3\columnwidth]{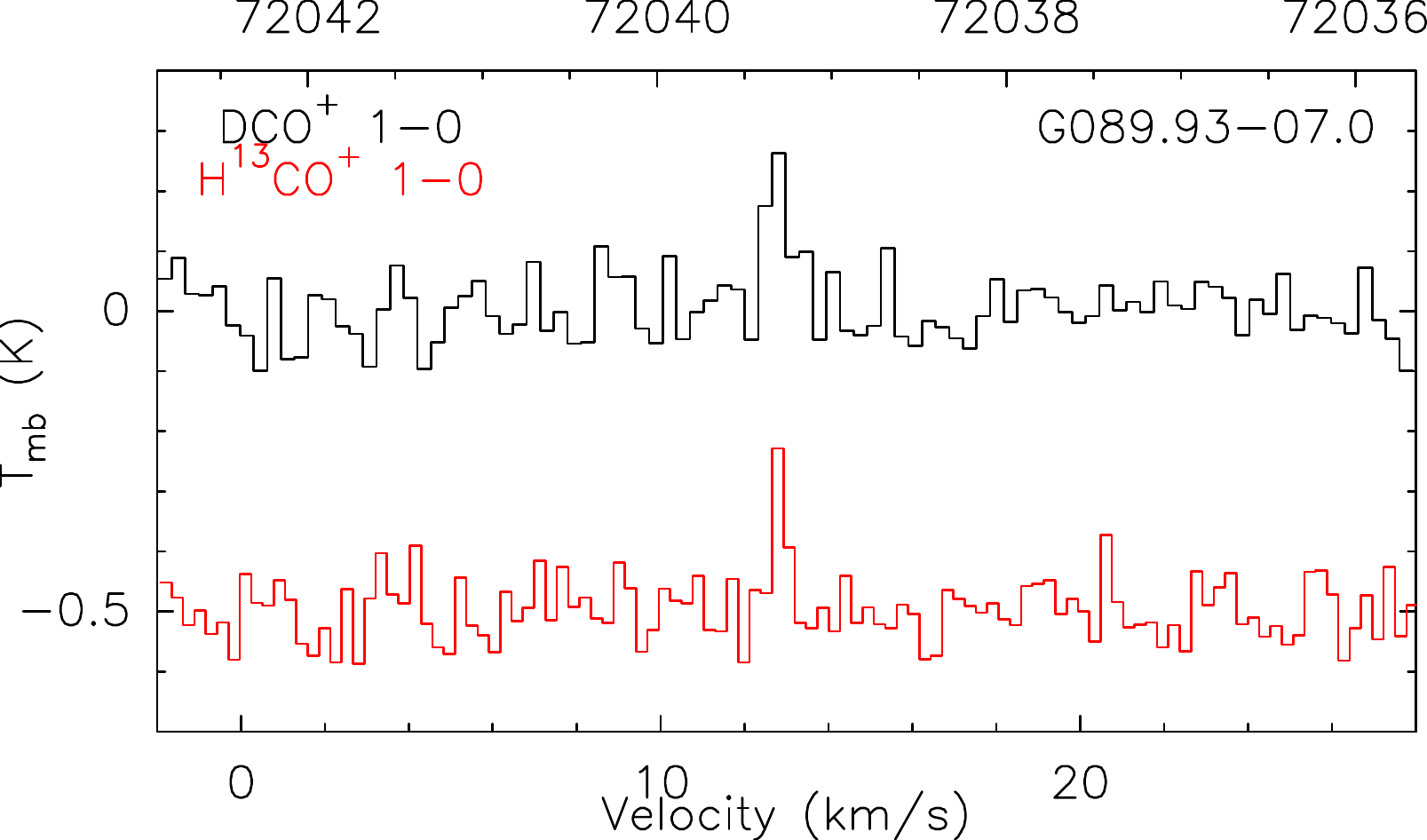}

\caption{Continued.\centering}
\end{figure}
\addtocounter{figure}{-1}
\begin{figure}
\centering
\includegraphics[width=0.3\columnwidth]{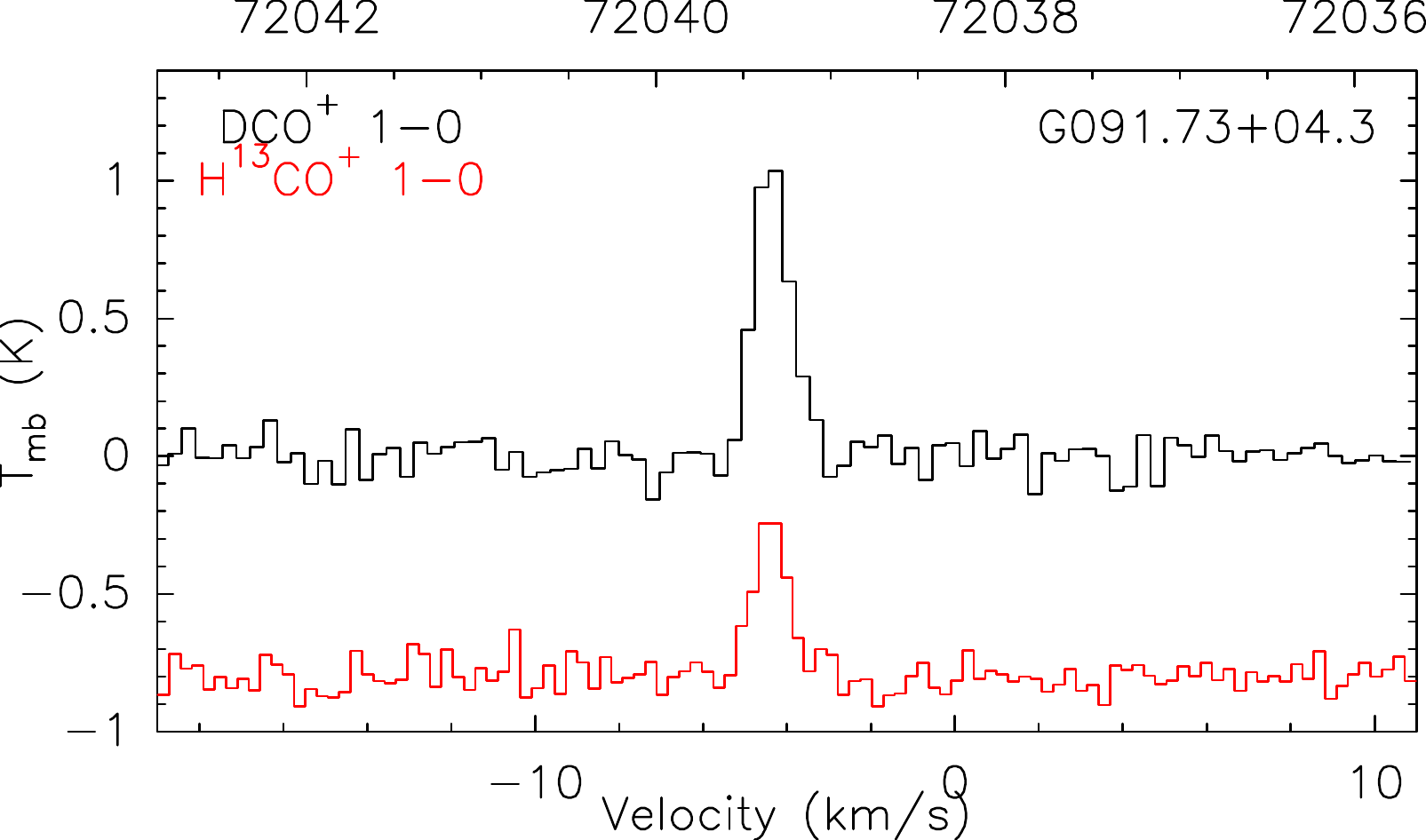}
\includegraphics[width=0.3\columnwidth]{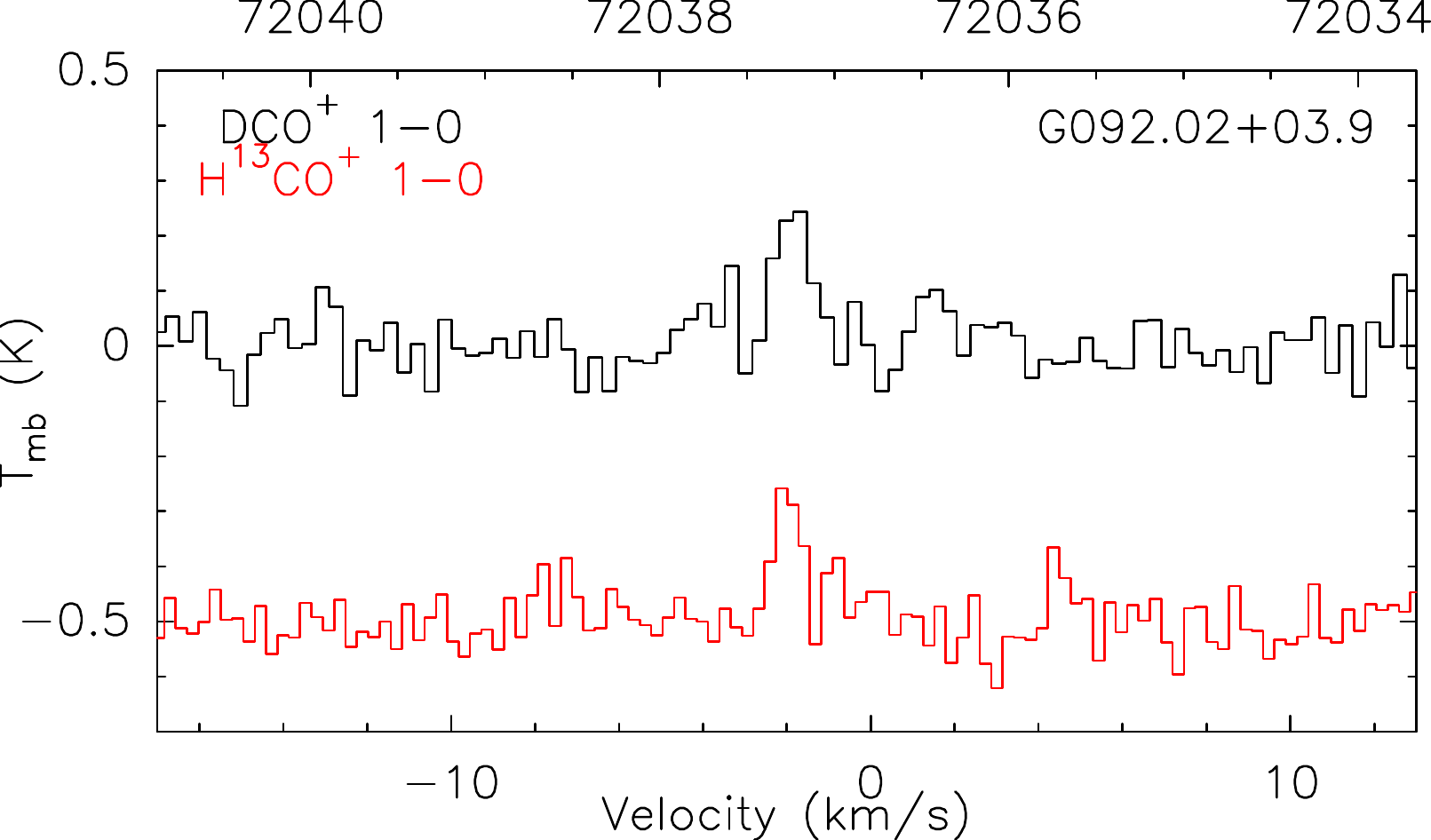}
\includegraphics[width=0.3\columnwidth]{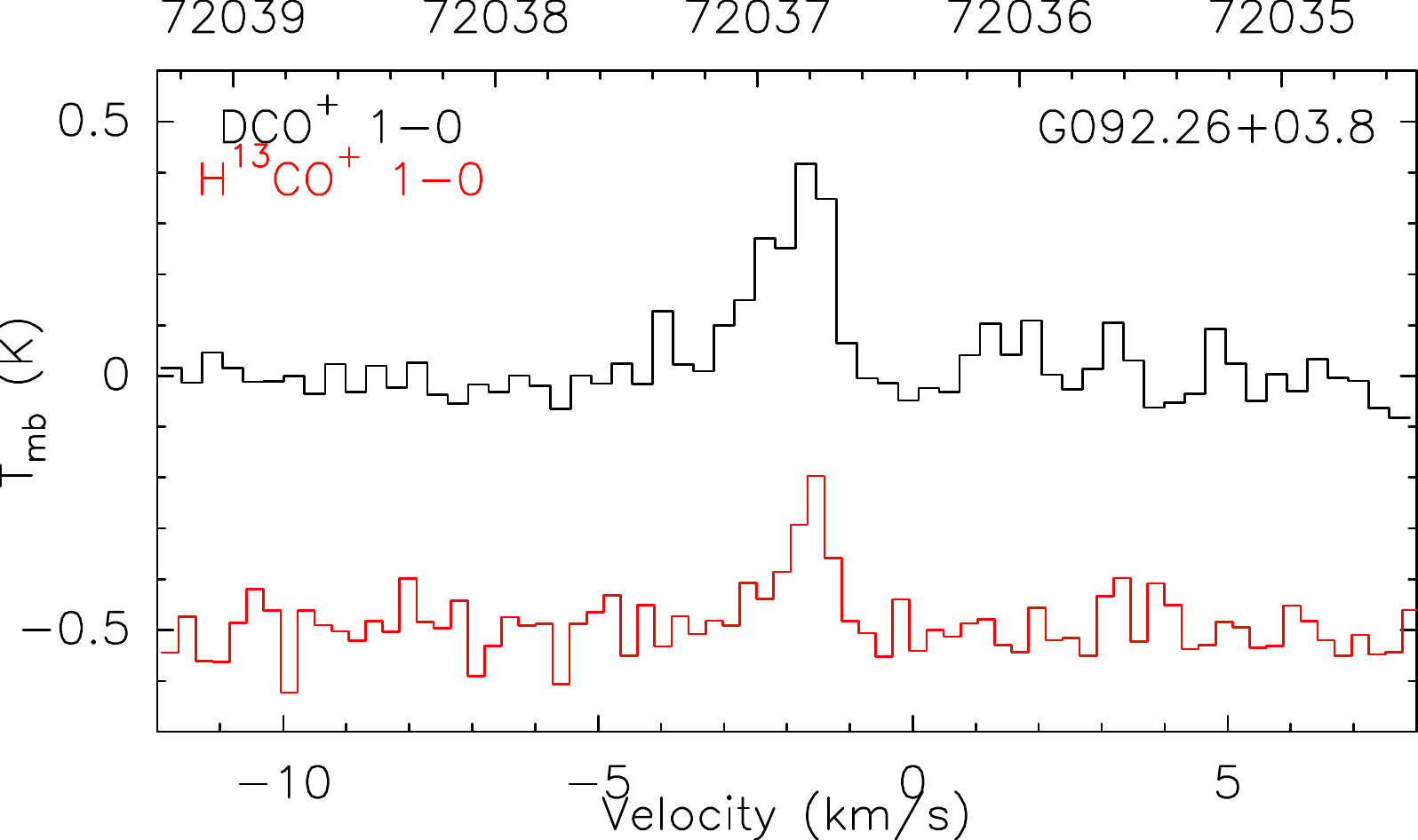}
\includegraphics[width=0.3\columnwidth]{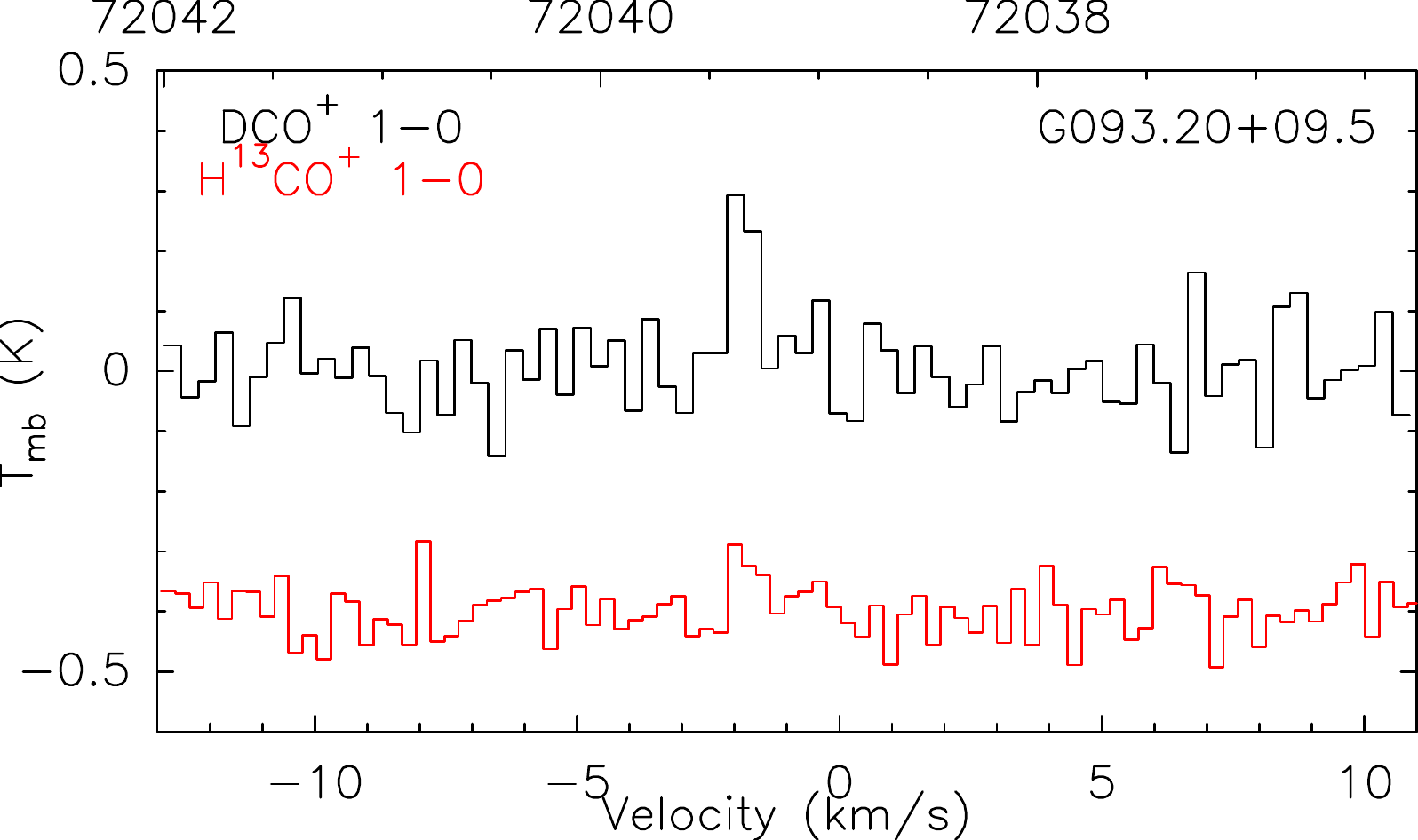}
\includegraphics[width=0.3\columnwidth]{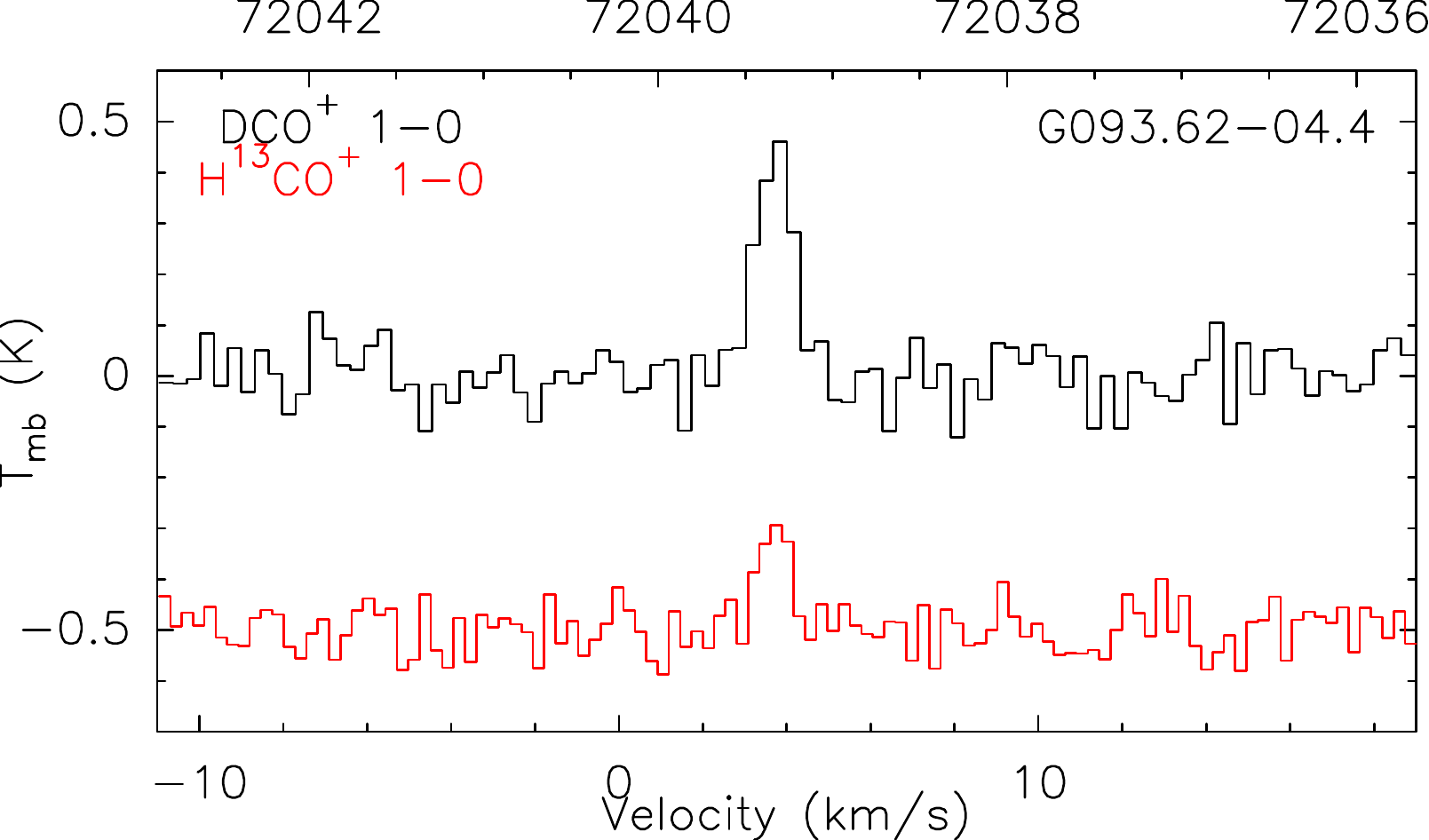}
\includegraphics[width=0.3\columnwidth]{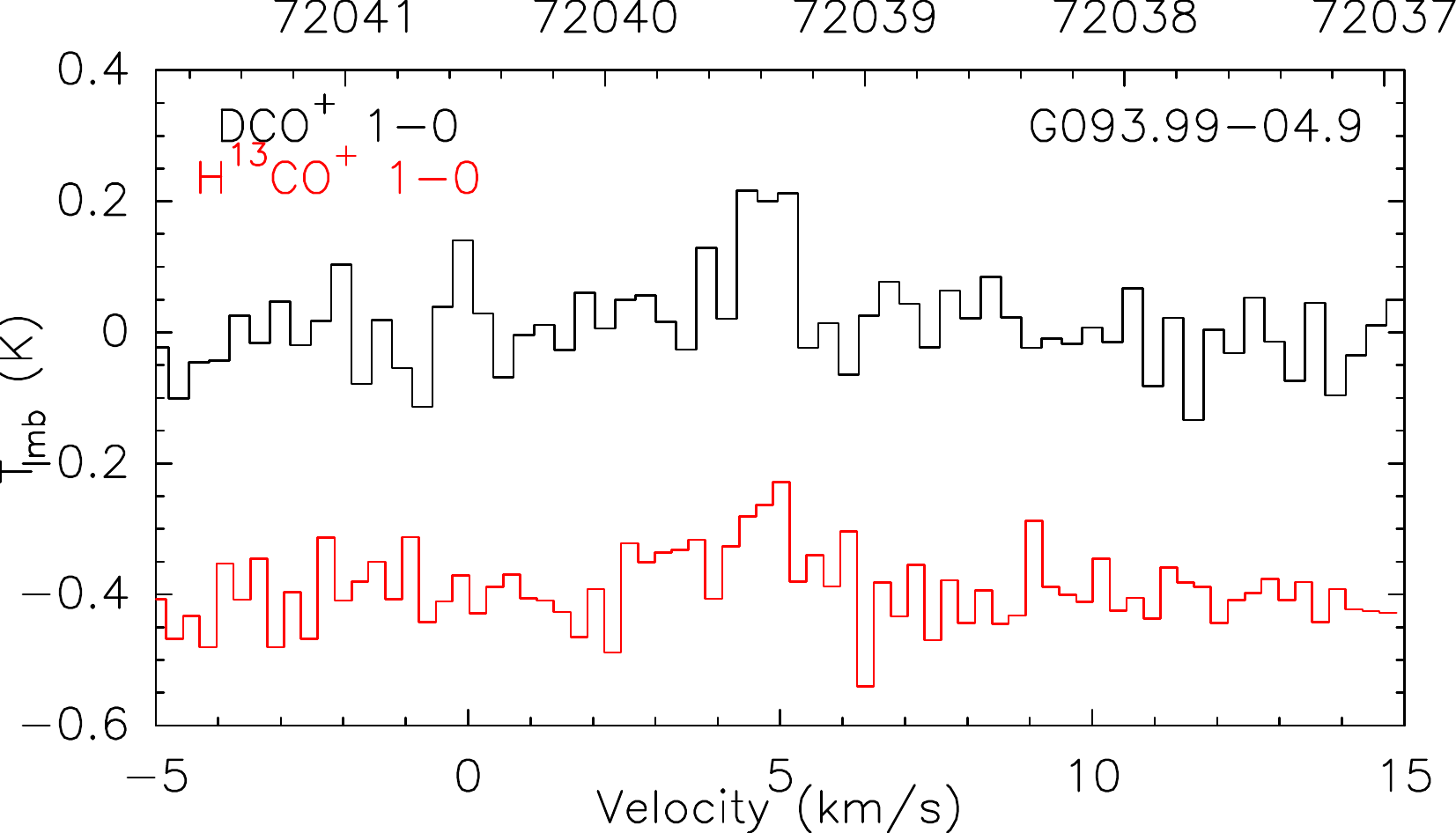}
\includegraphics[width=0.3\columnwidth]{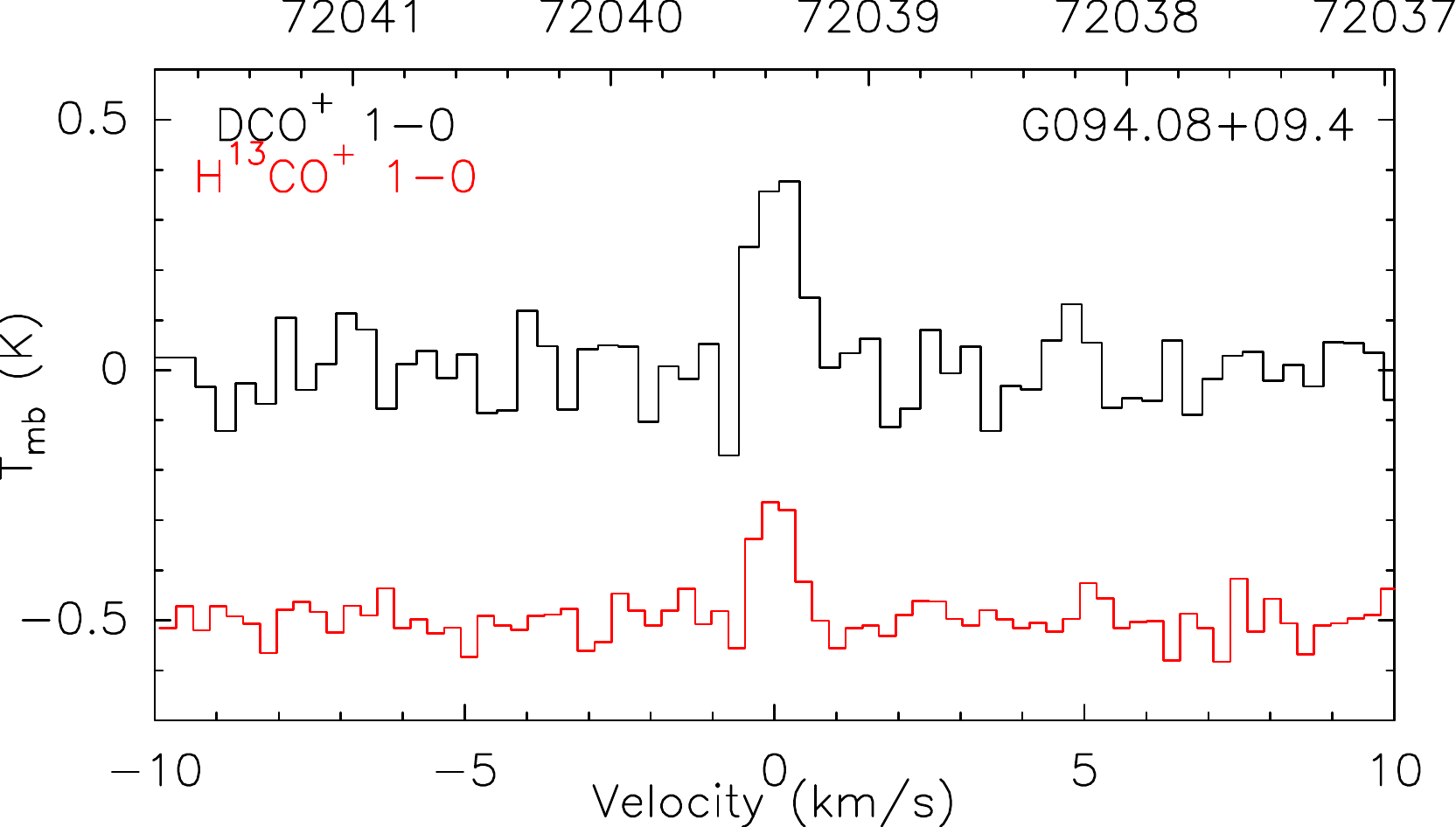}
\includegraphics[width=0.3\columnwidth]{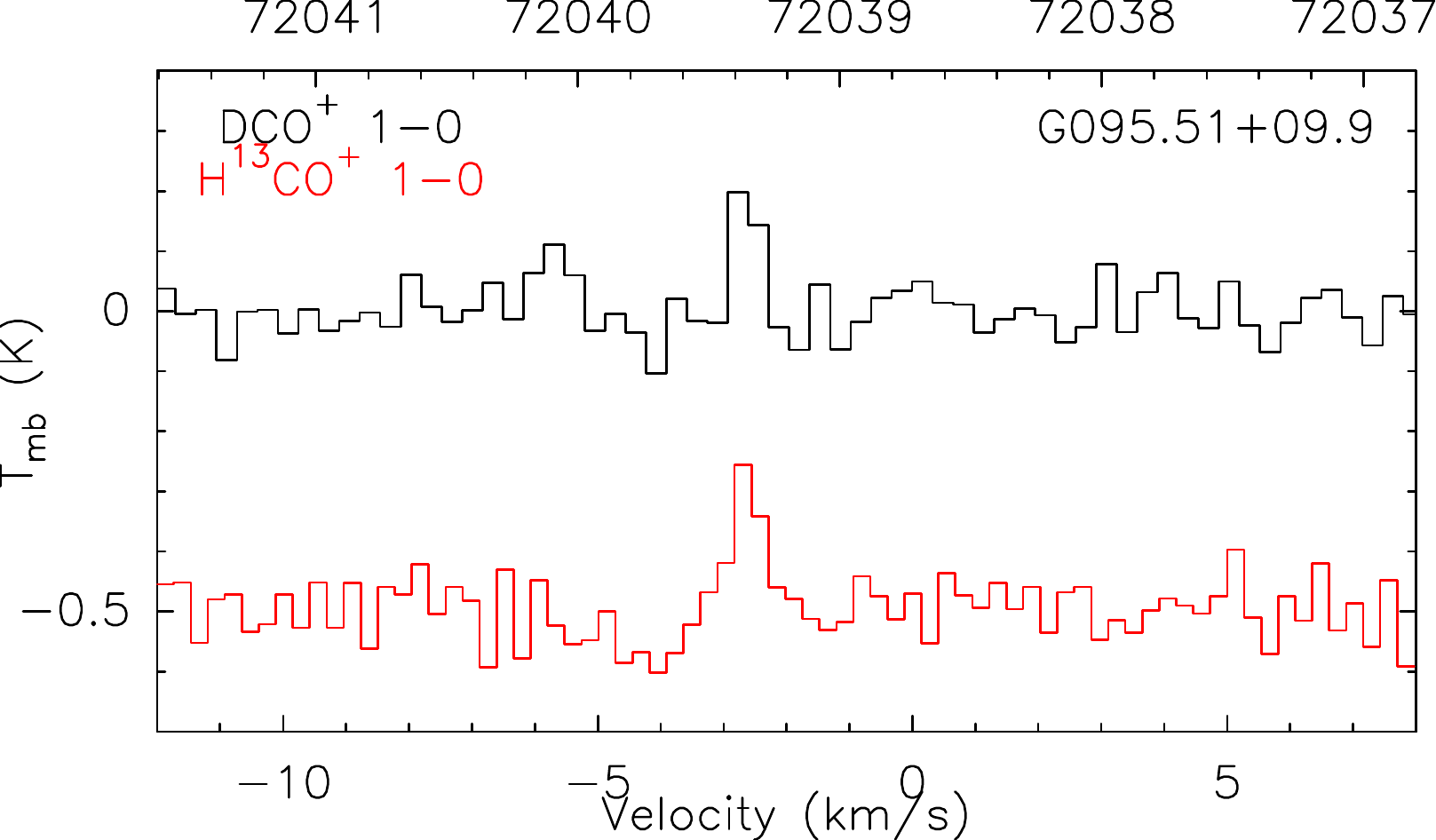}
\includegraphics[width=0.3\columnwidth]{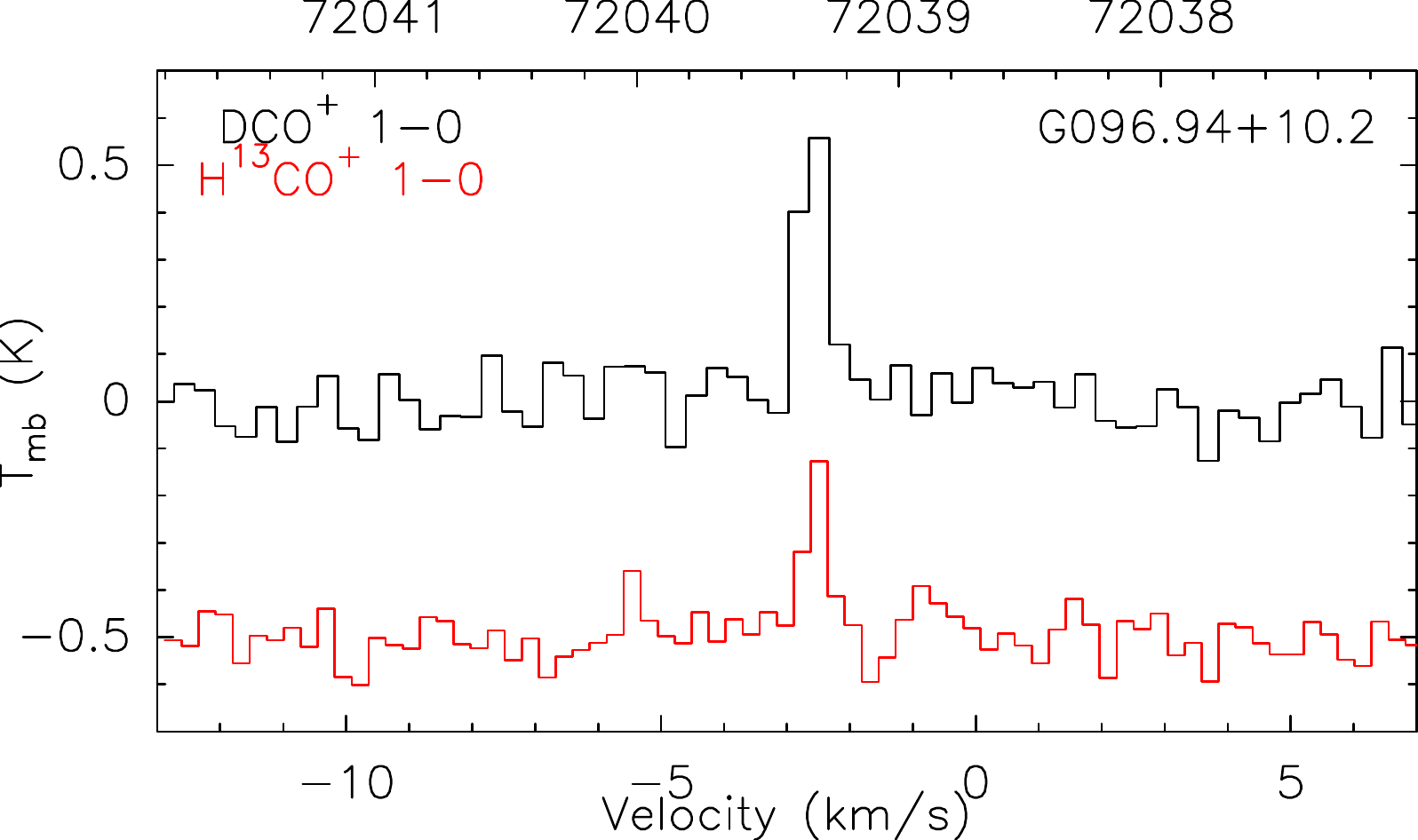}
\includegraphics[width=0.3\columnwidth]{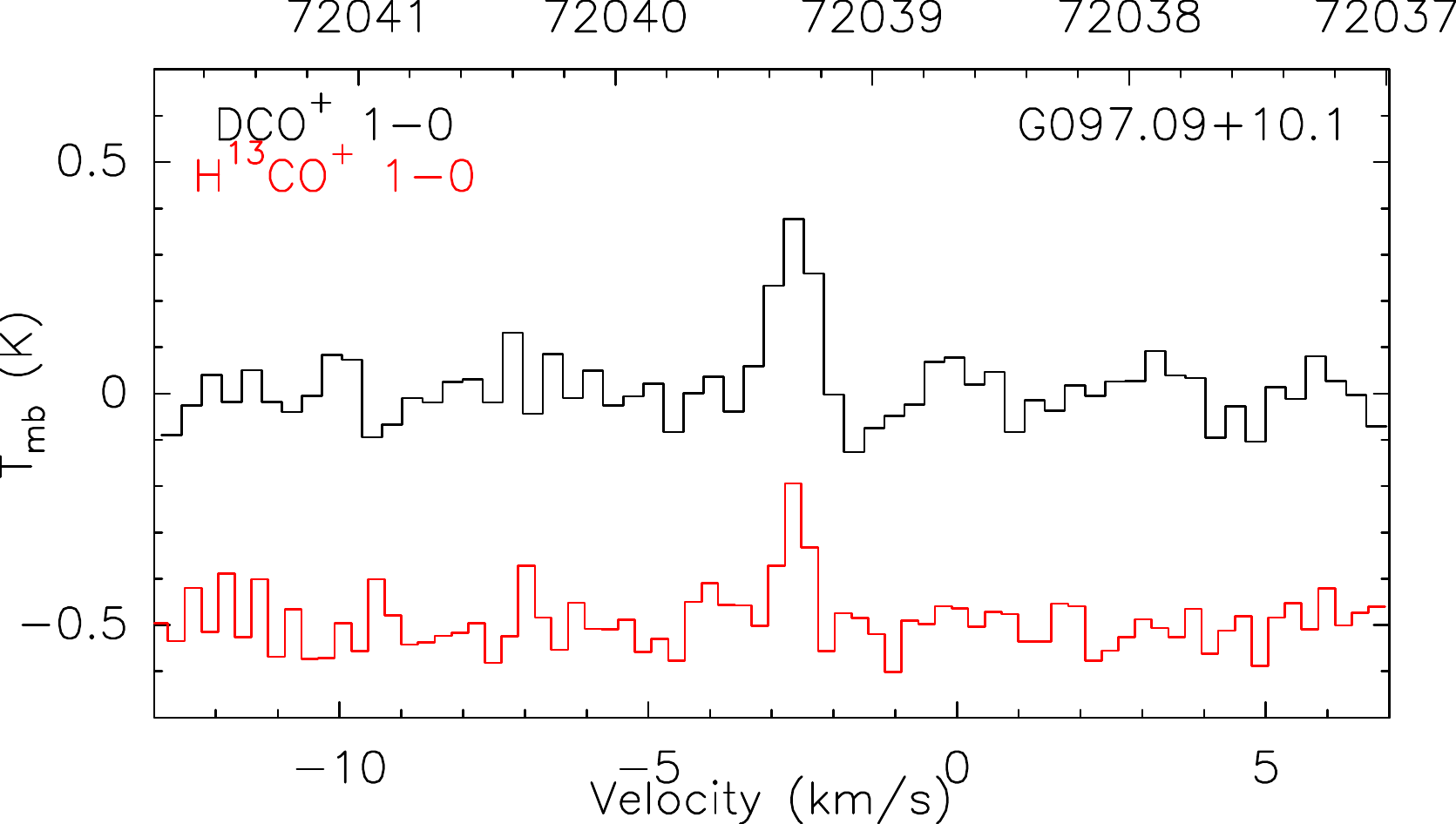}
\includegraphics[width=0.3\columnwidth]{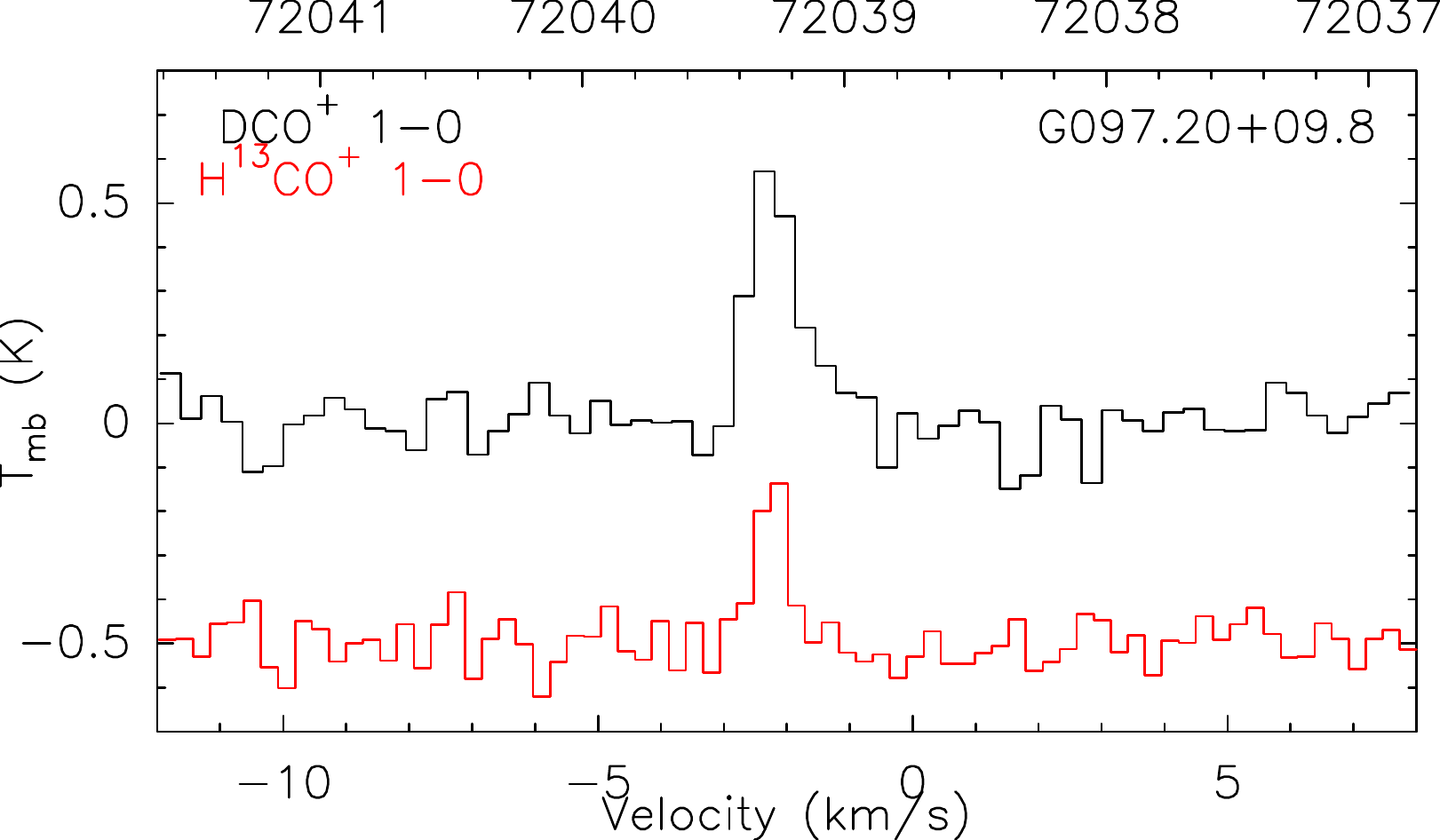}
\includegraphics[width=0.3\columnwidth]{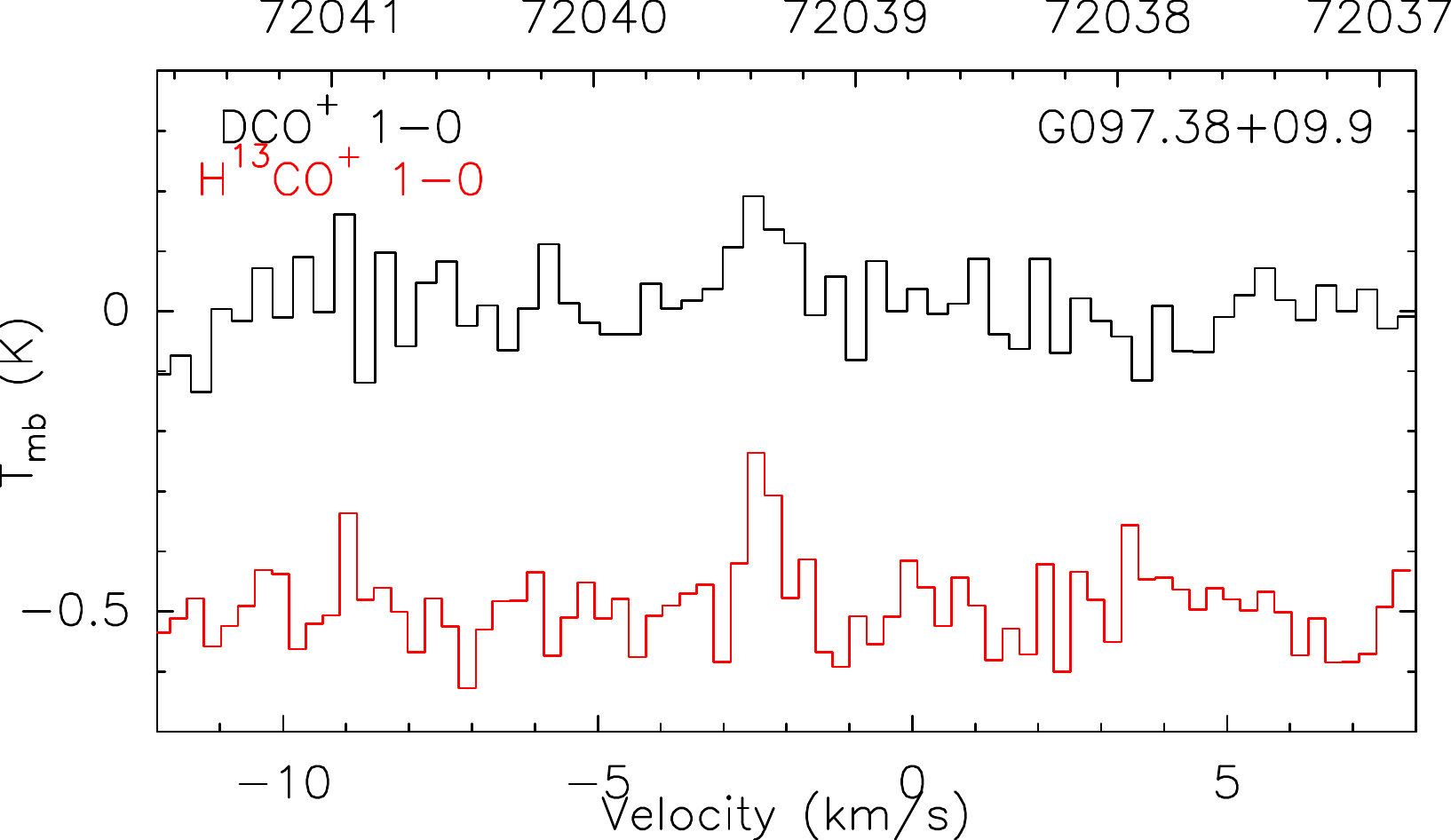}
\includegraphics[width=0.3\columnwidth]{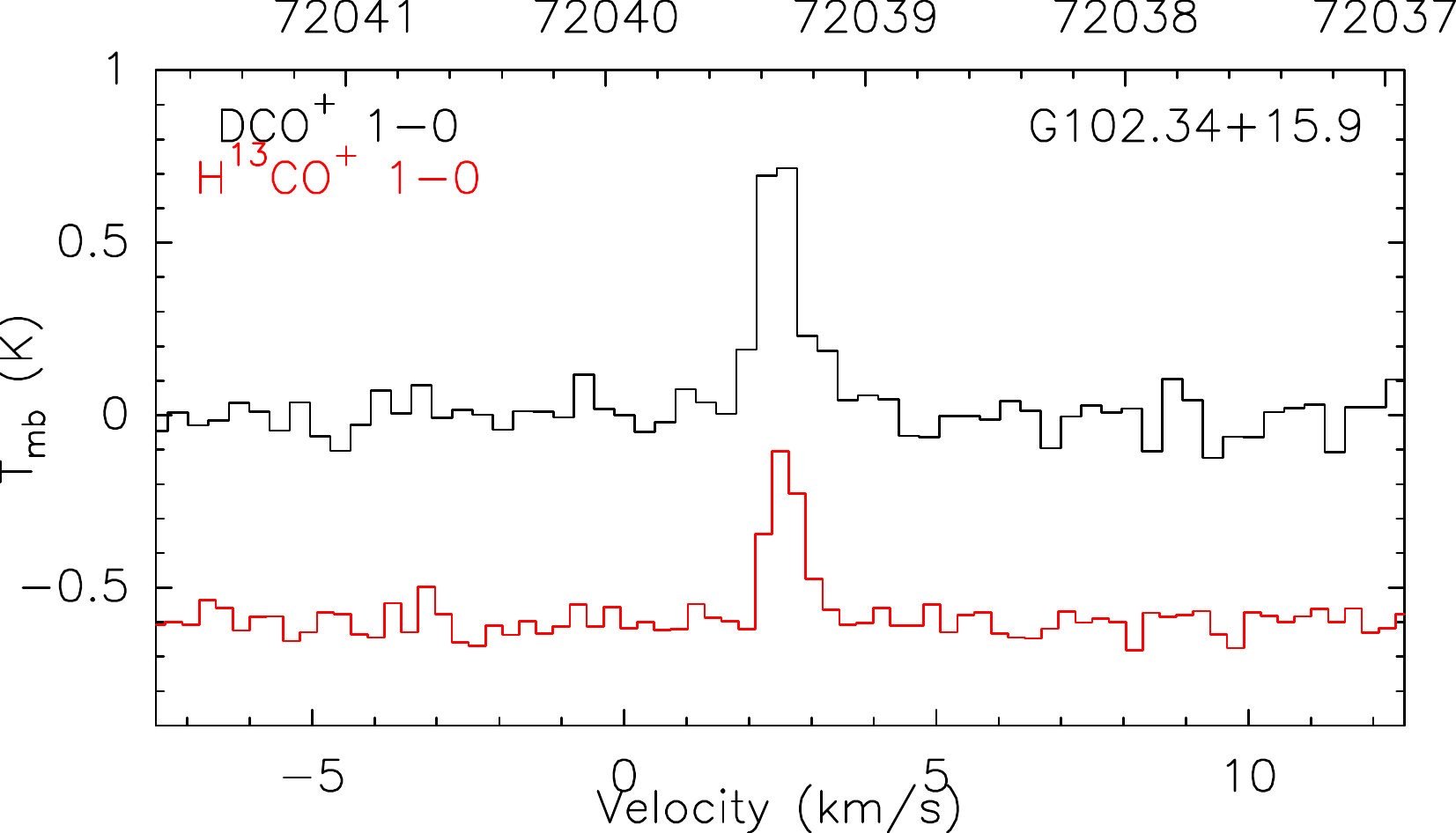}
\includegraphics[width=0.3\columnwidth]{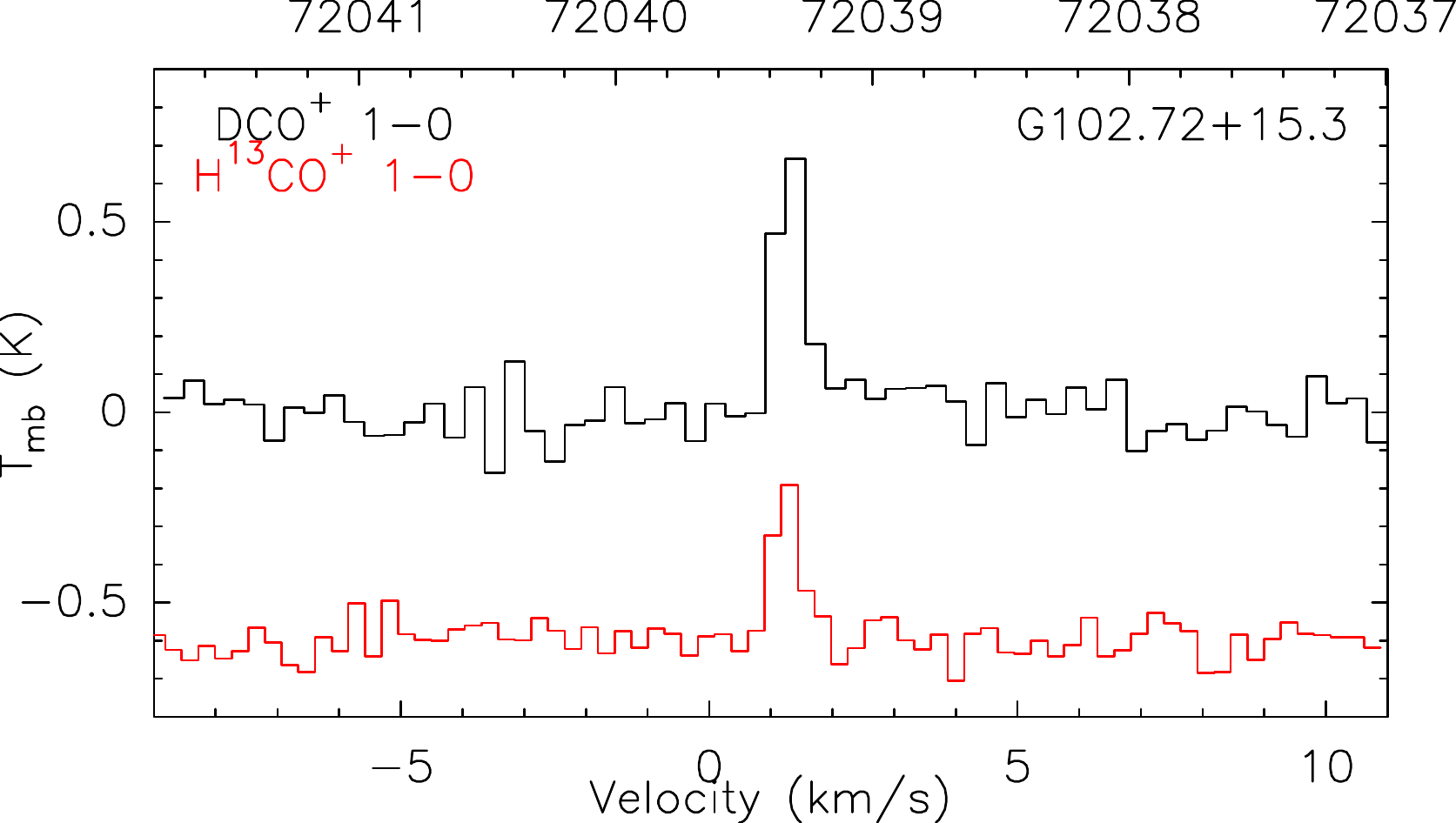}
\includegraphics[width=0.3\columnwidth]{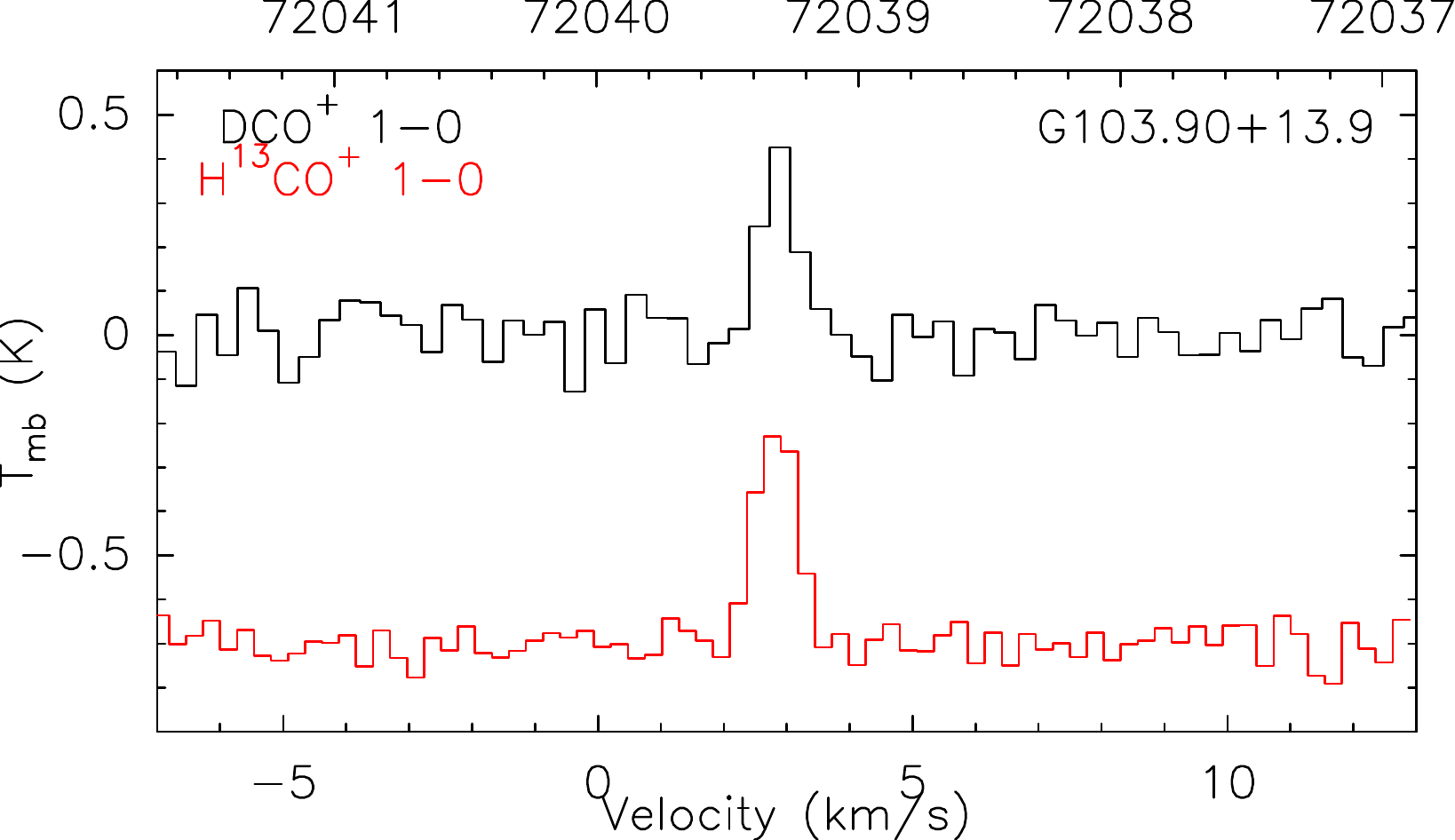}
\includegraphics[width=0.3\columnwidth]{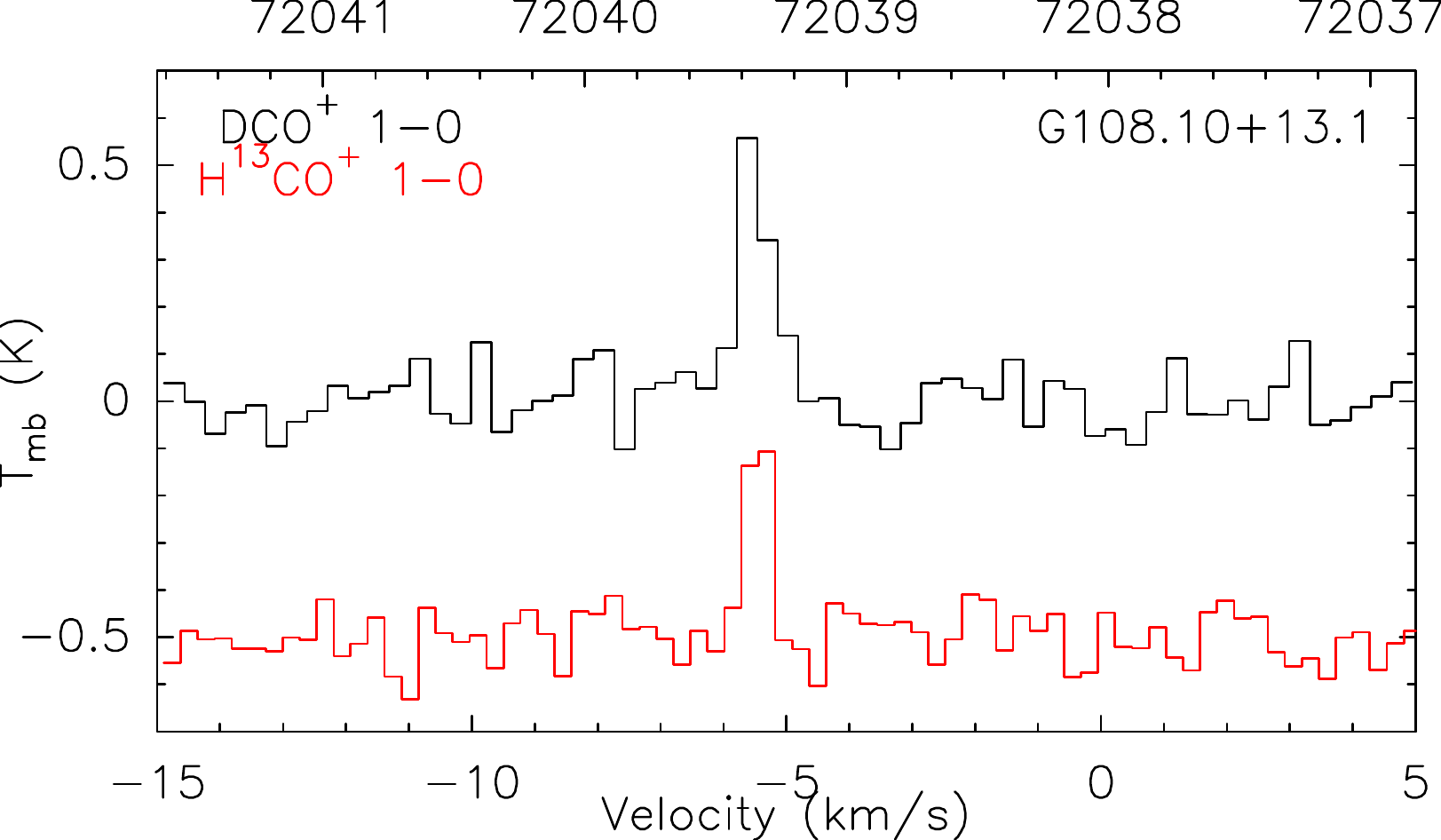}
\includegraphics[width=0.3\columnwidth]{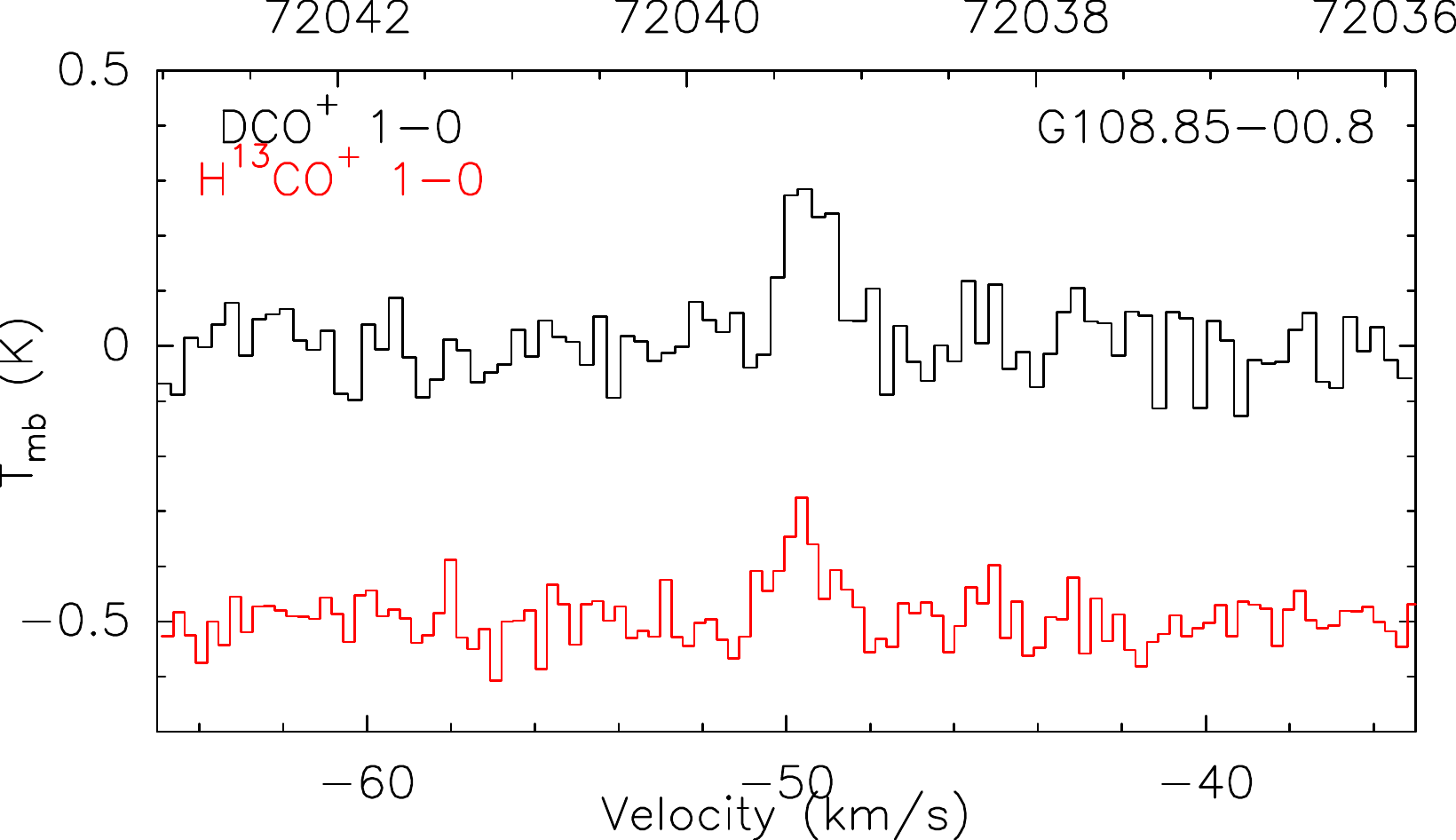}
\includegraphics[width=0.3\columnwidth]{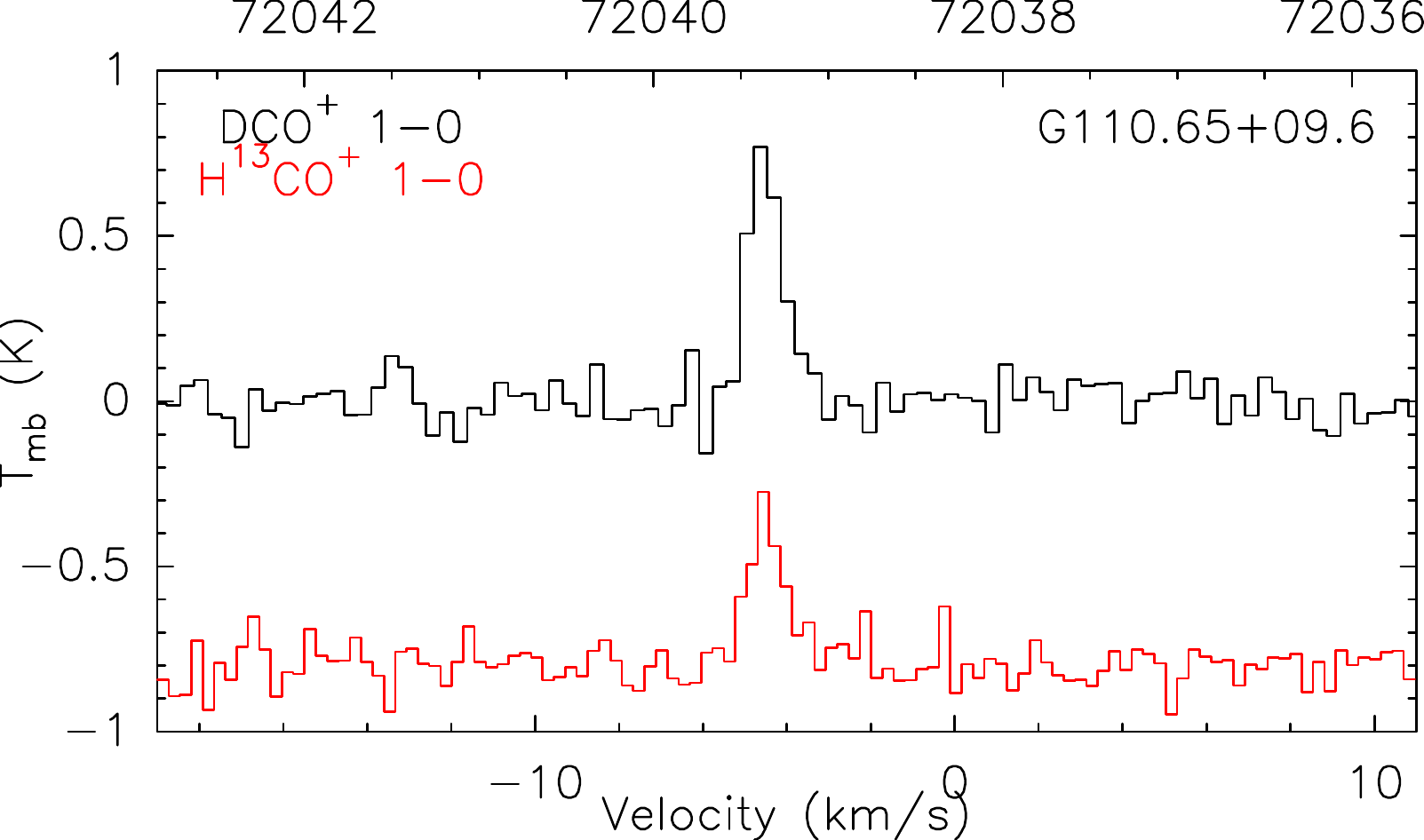}
\includegraphics[width=0.3\columnwidth]{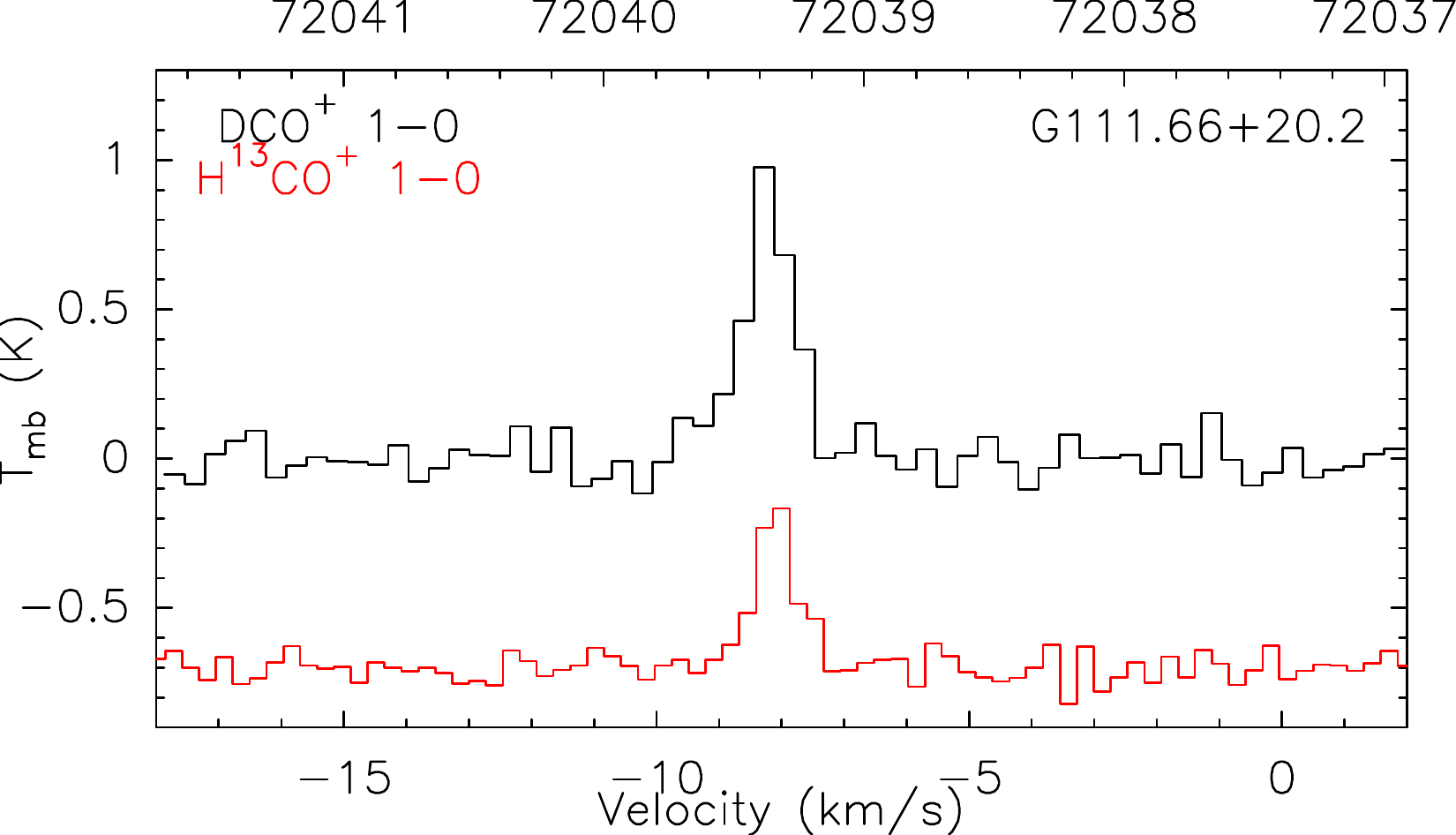}
\includegraphics[width=0.3\columnwidth]{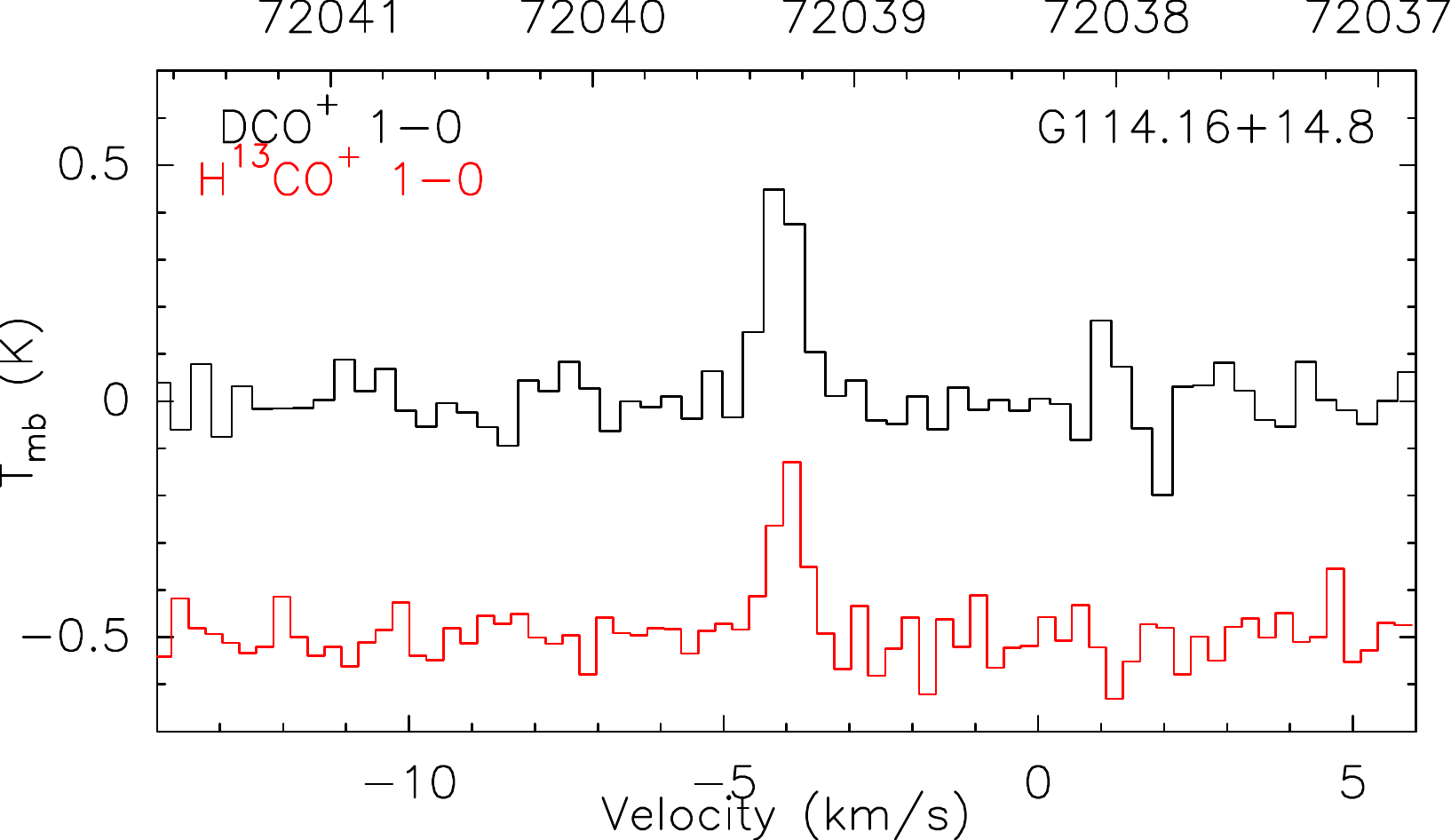}
\includegraphics[width=0.3\columnwidth]{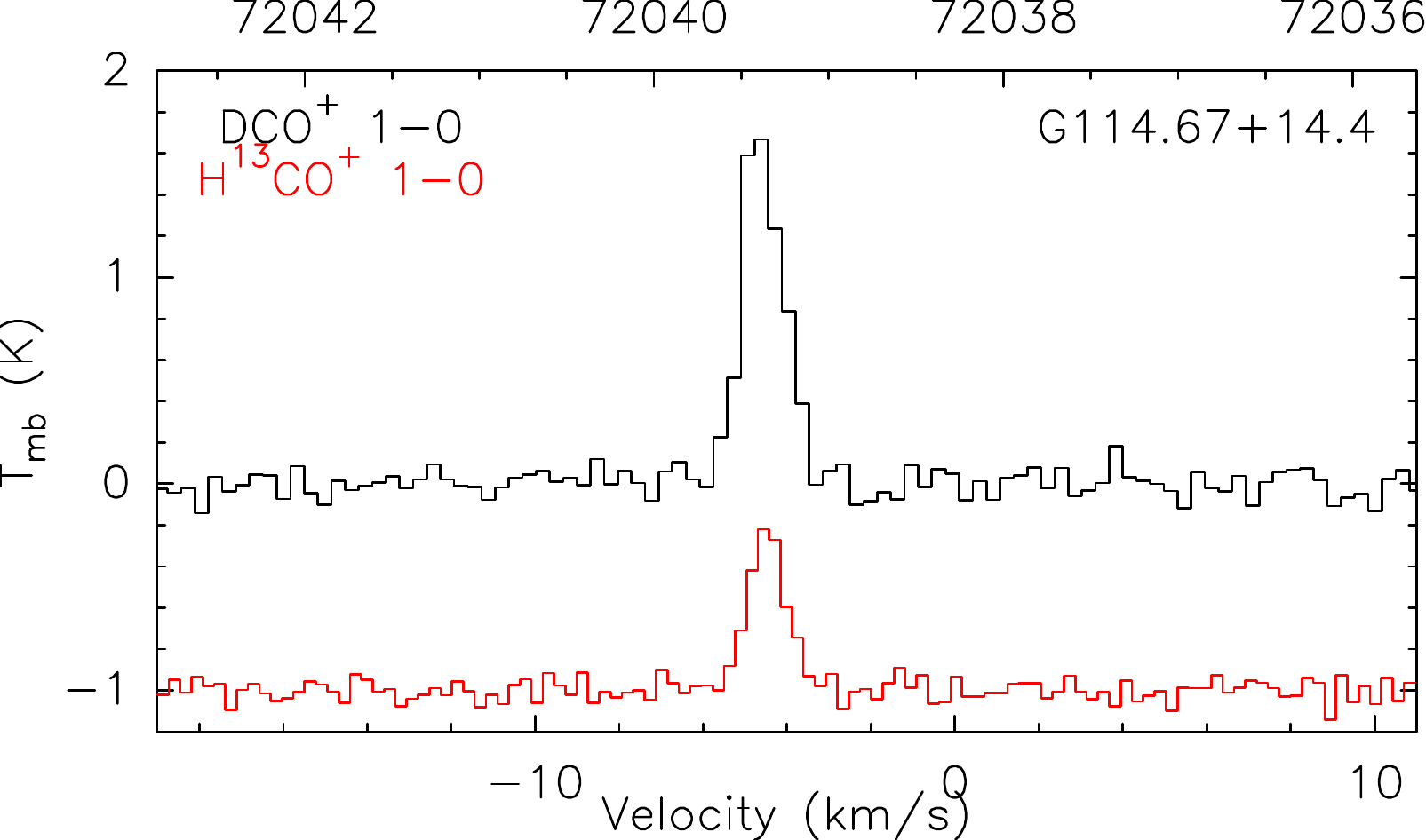}
\caption{Continued.\centering}
\end{figure}
\addtocounter{figure}{-1}
\begin{figure}
\centering
\includegraphics[width=0.3\columnwidth]{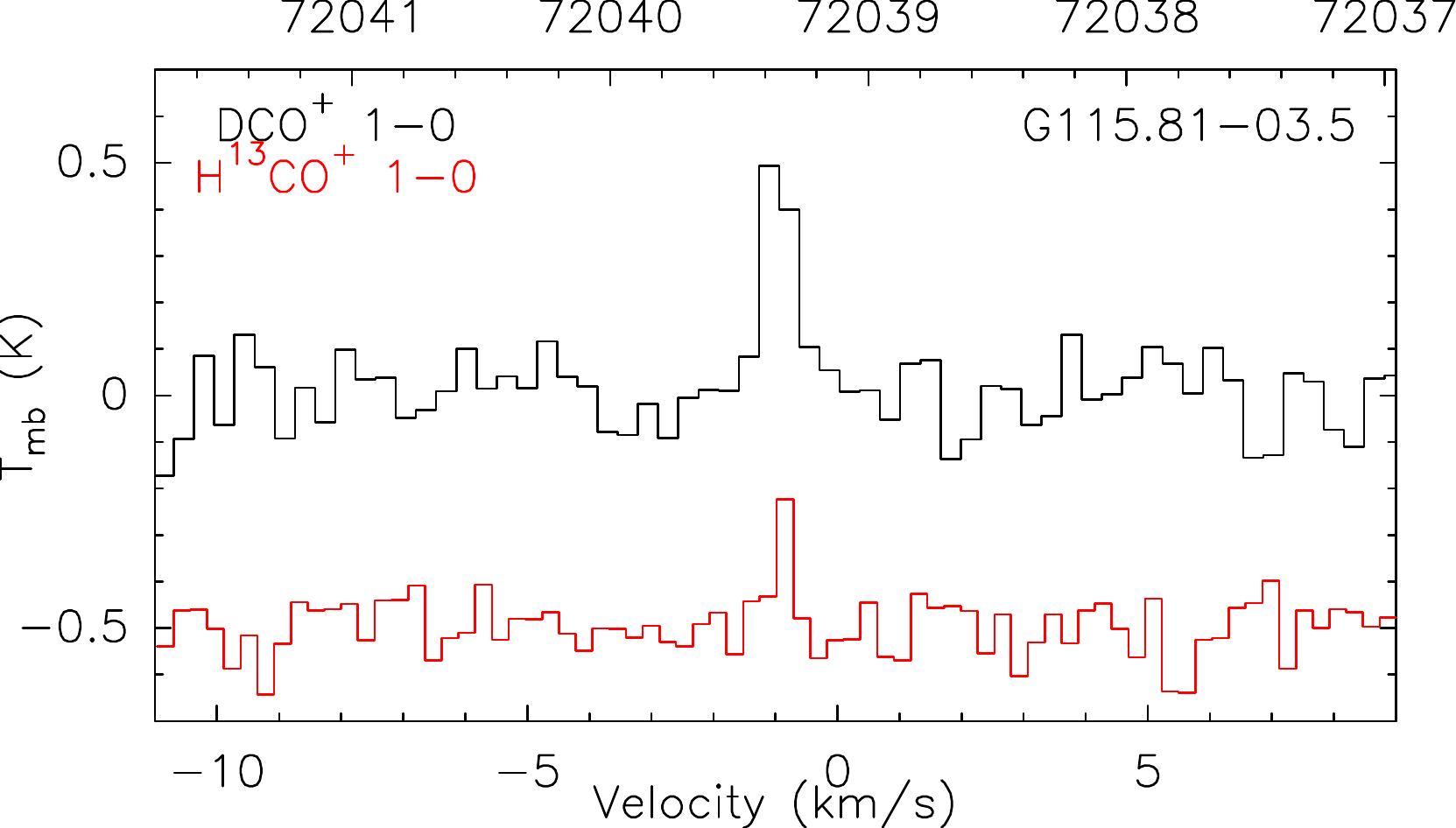}
\includegraphics[width=0.3\columnwidth]{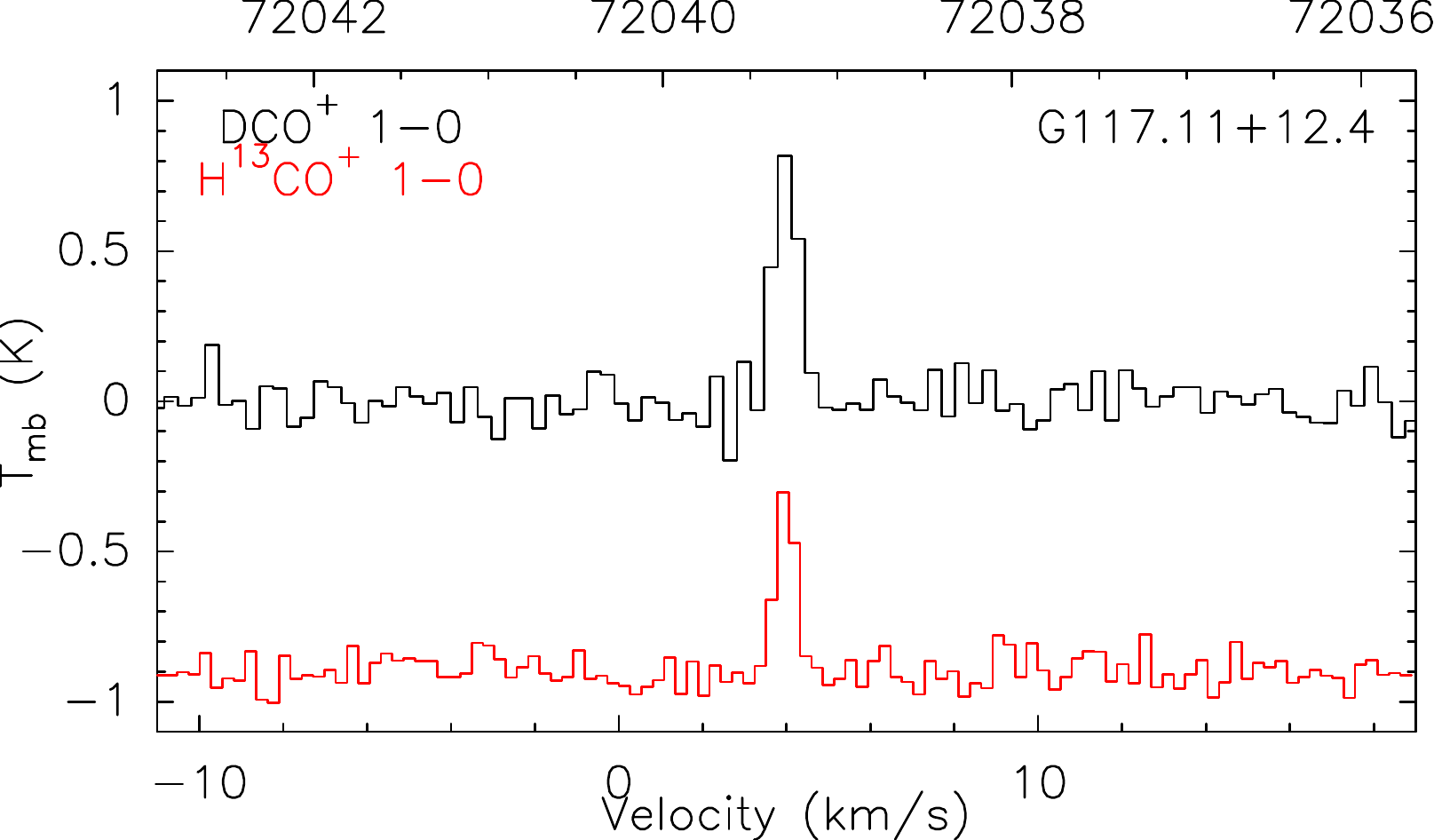}
\includegraphics[width=0.3\columnwidth]{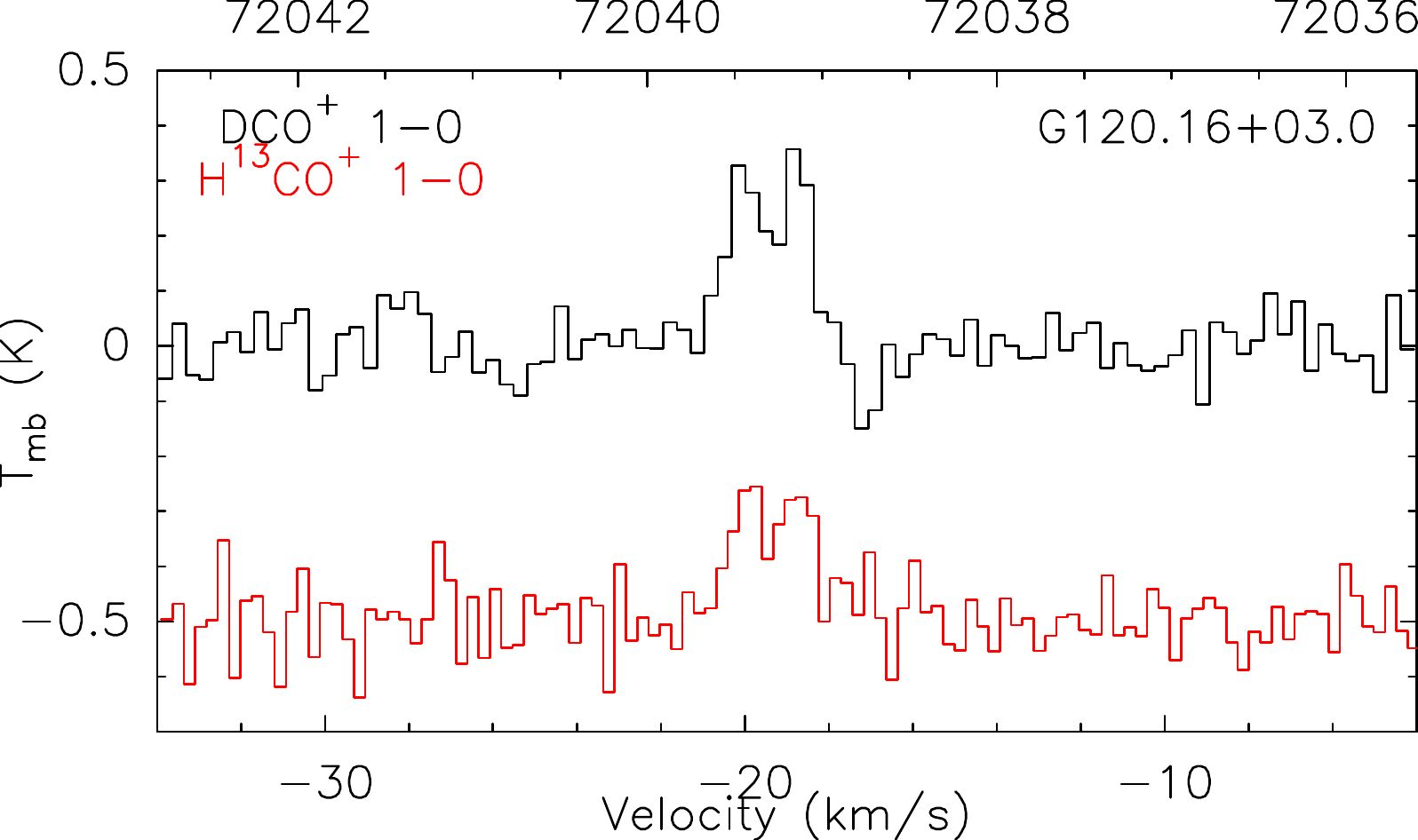}
\includegraphics[width=0.3\columnwidth]{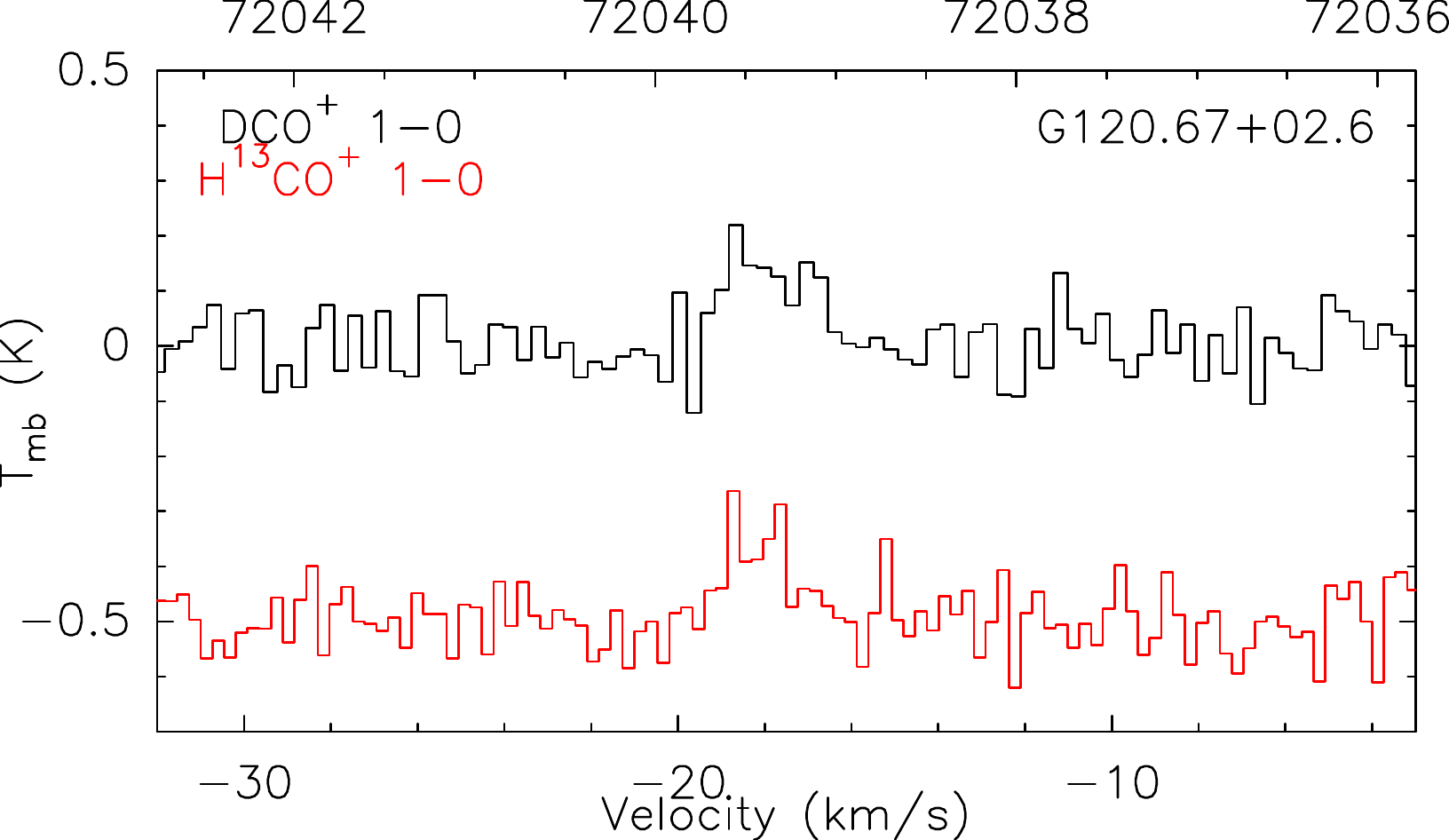}
\includegraphics[width=0.3\columnwidth]{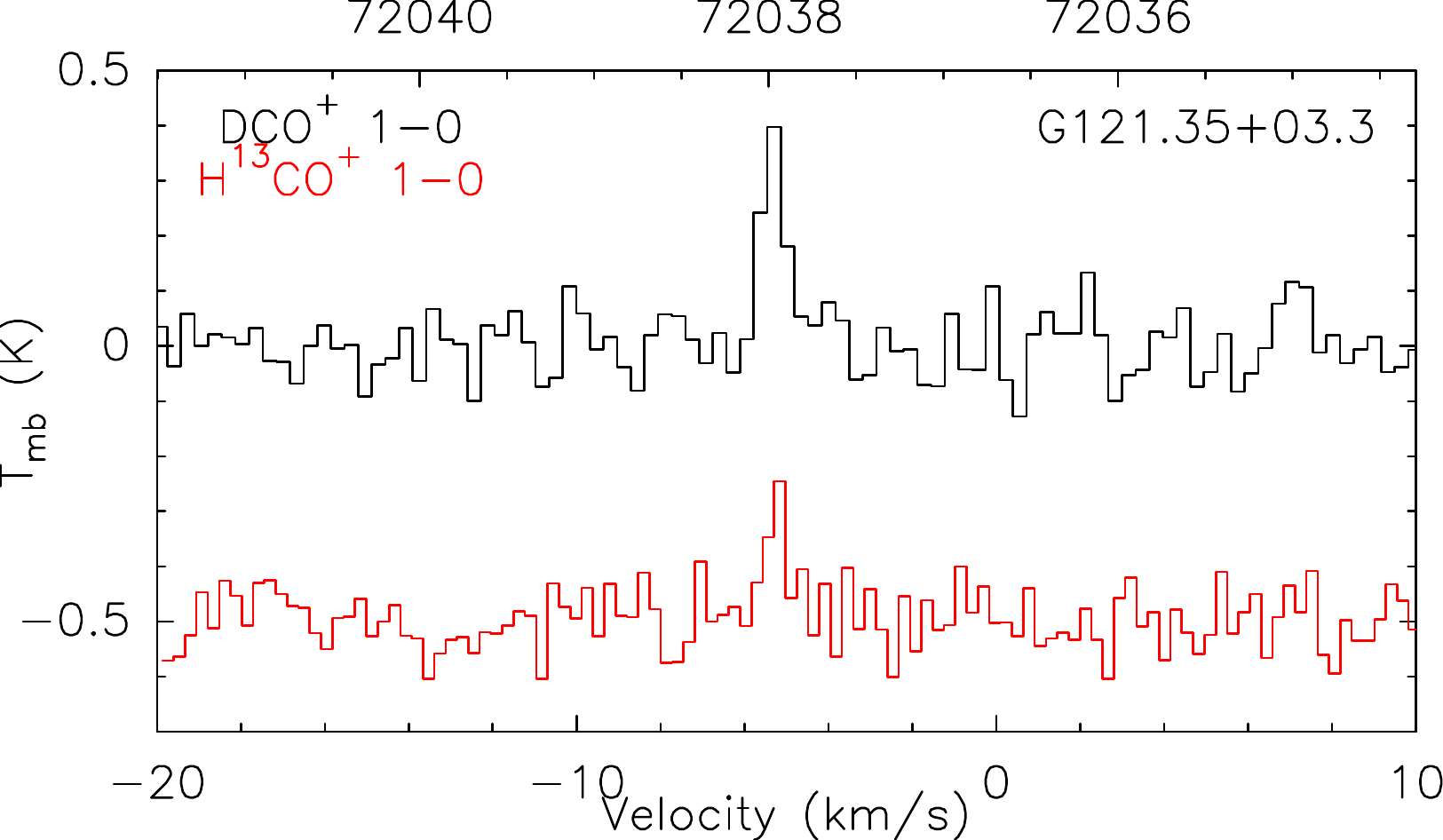}
\includegraphics[width=0.3\columnwidth]{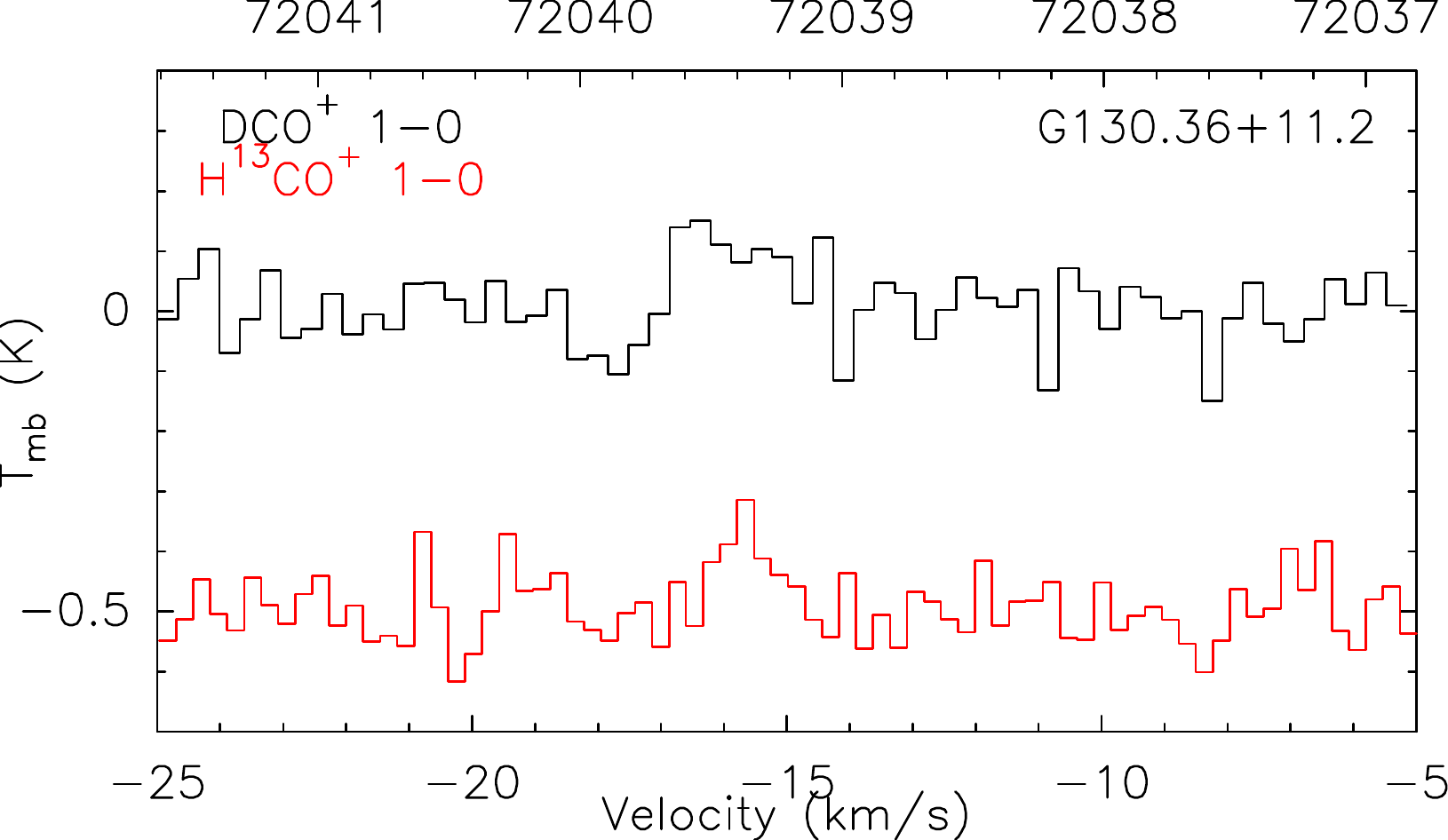}
\includegraphics[width=0.3\columnwidth]{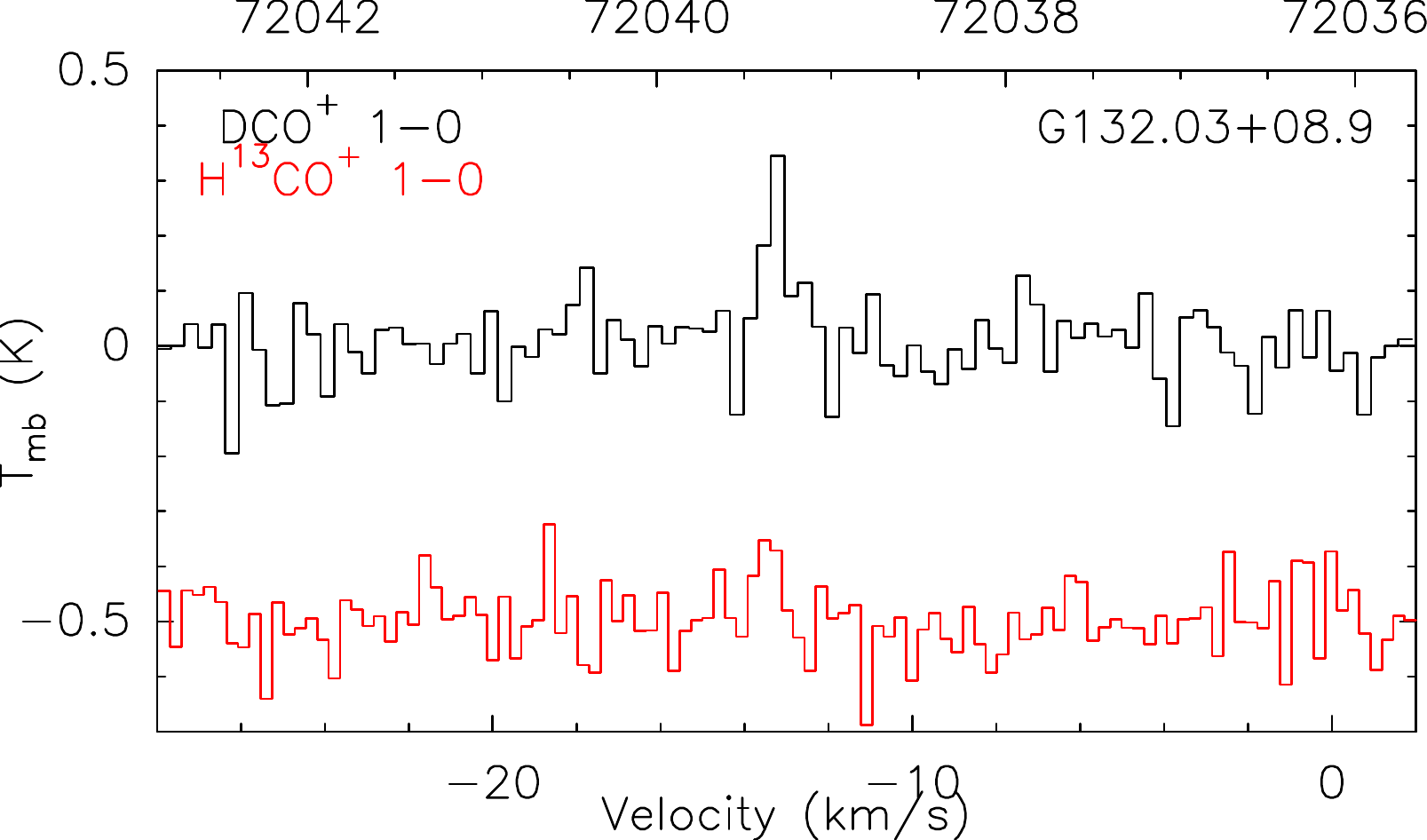}
\includegraphics[width=0.3\columnwidth]{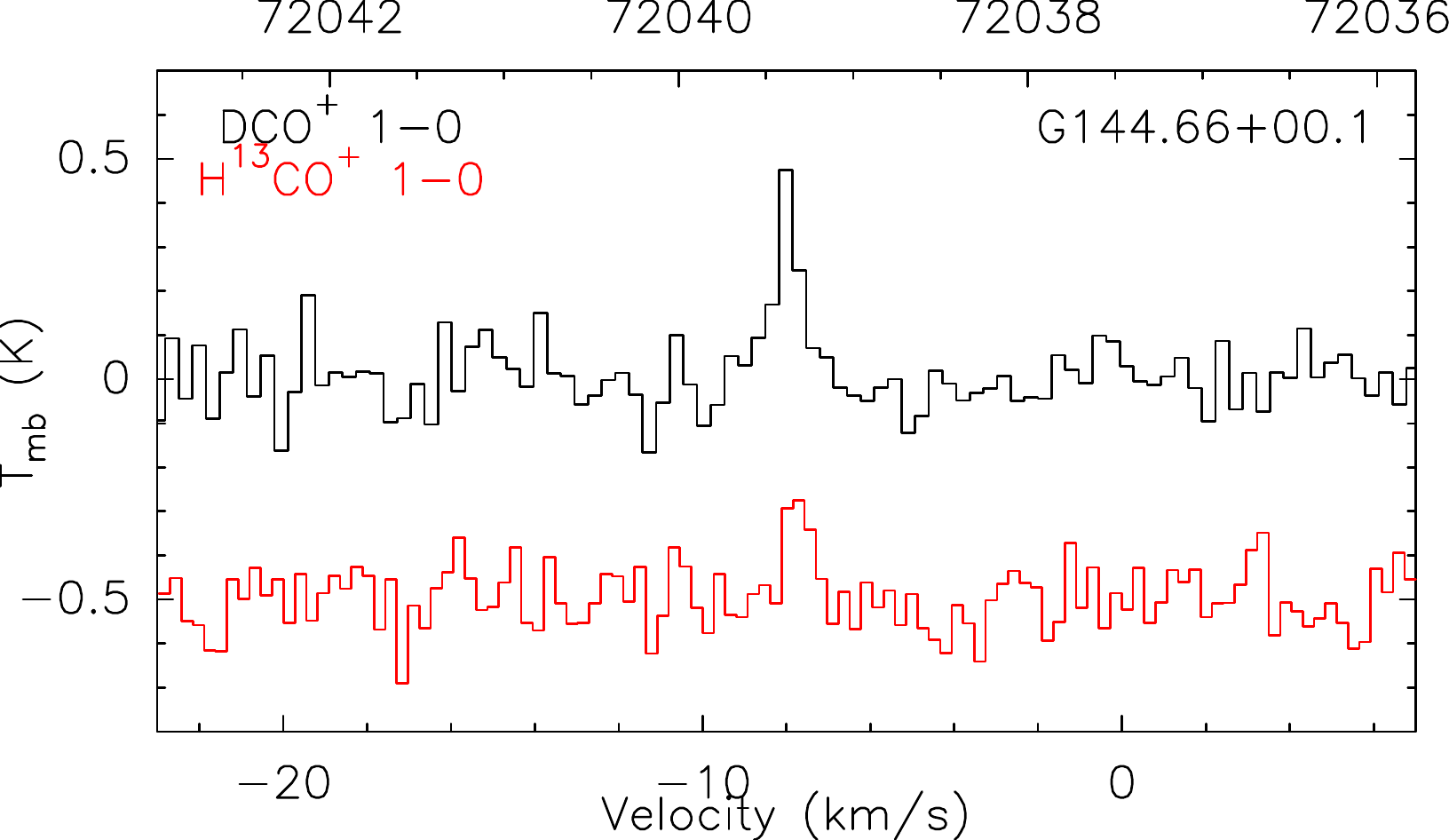}
\includegraphics[width=0.3\columnwidth]{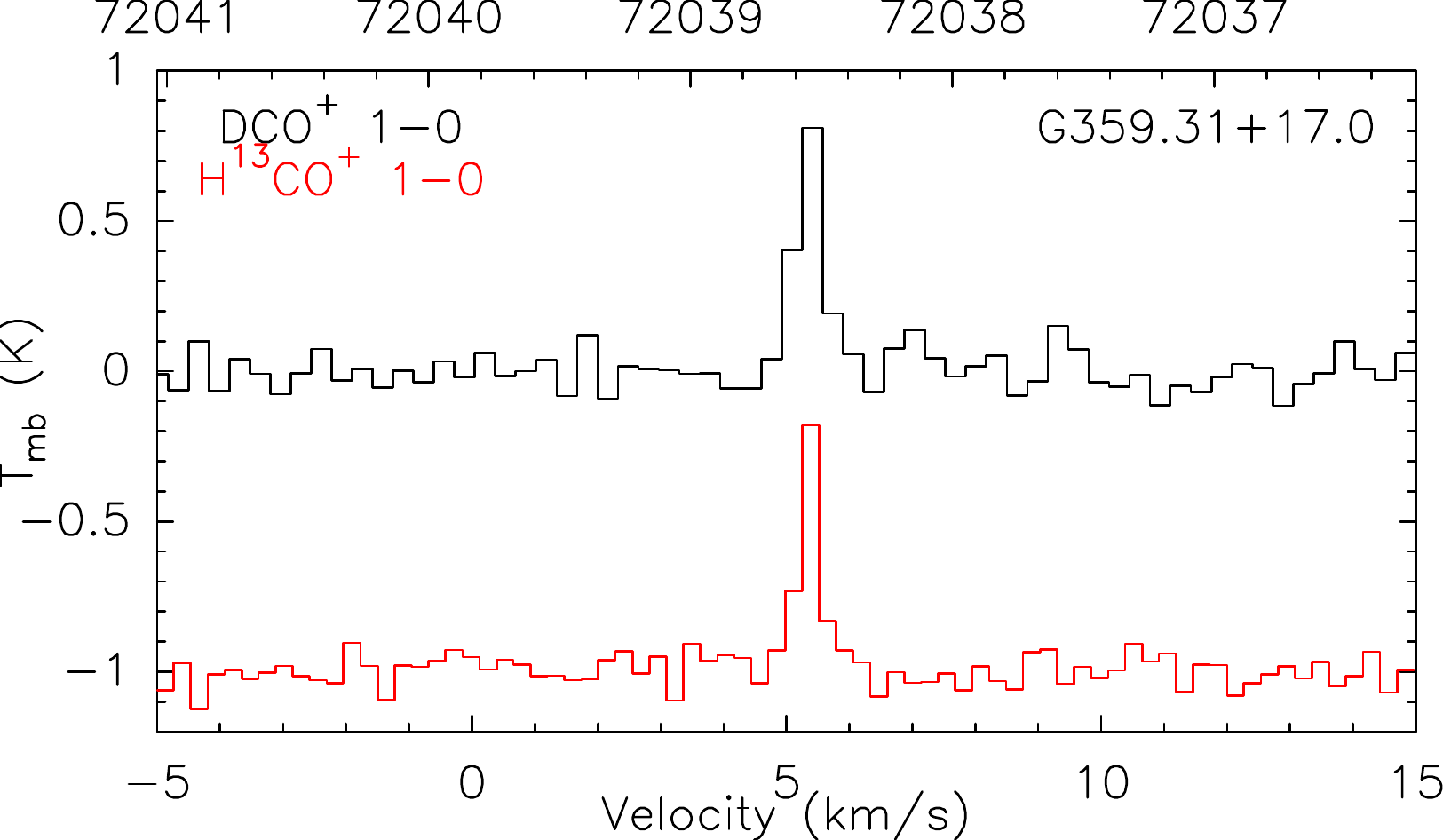}
\caption{Continued.\centering}
\end{figure}
\begin{figure}
\centering
\includegraphics[width=0.3\columnwidth]{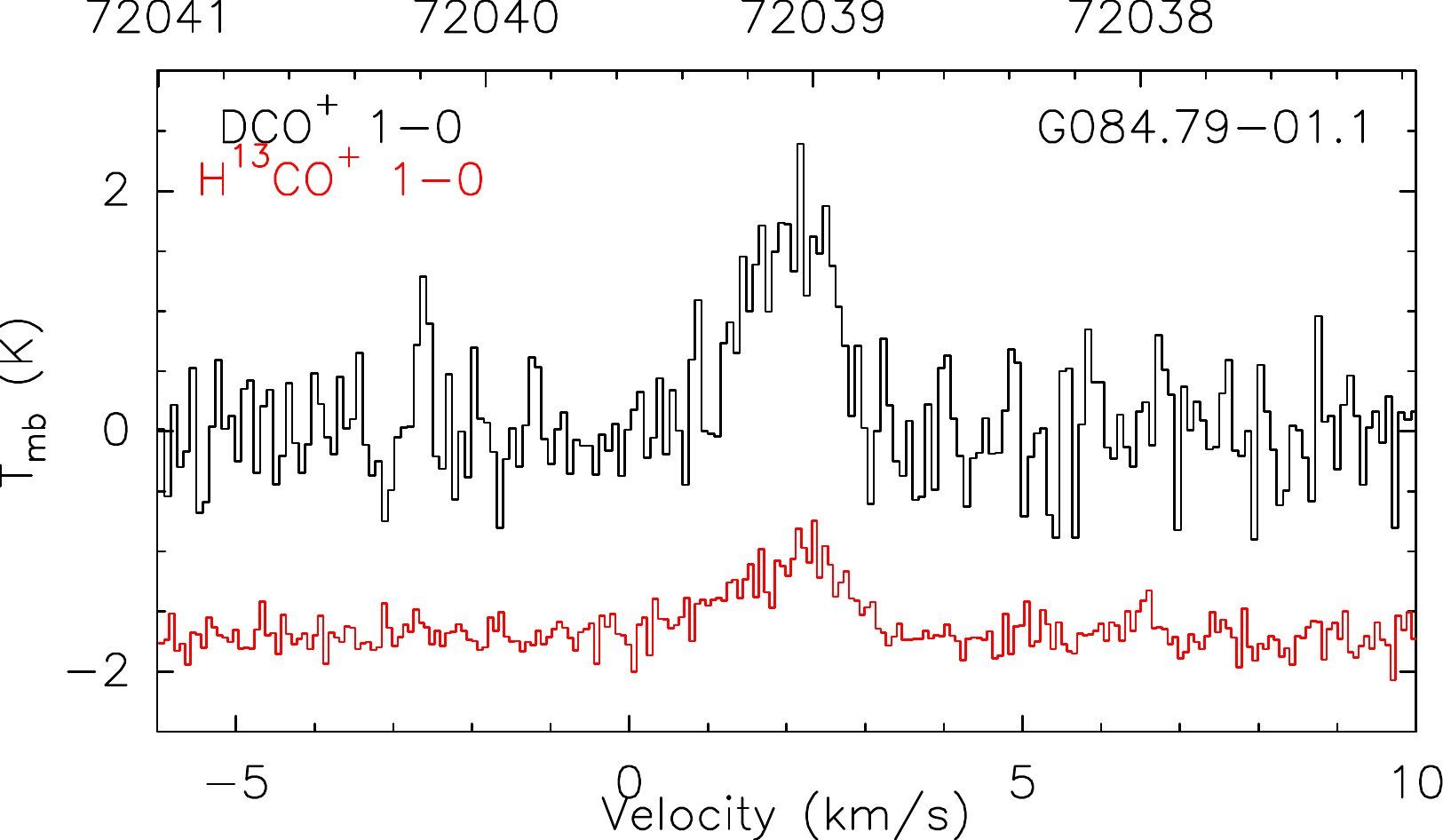}
\includegraphics[width=0.3\columnwidth]{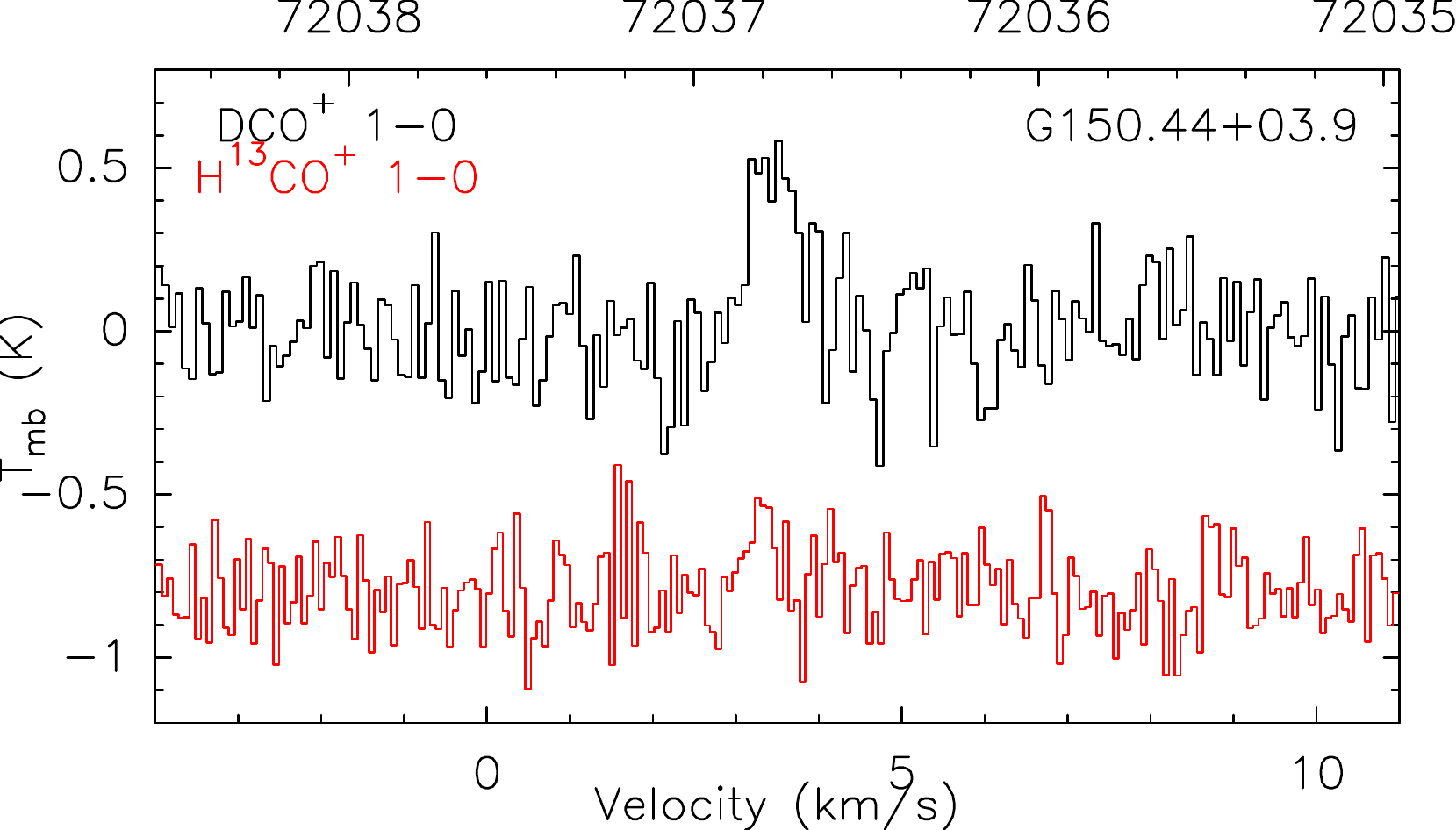}
\includegraphics[width=0.3\columnwidth]{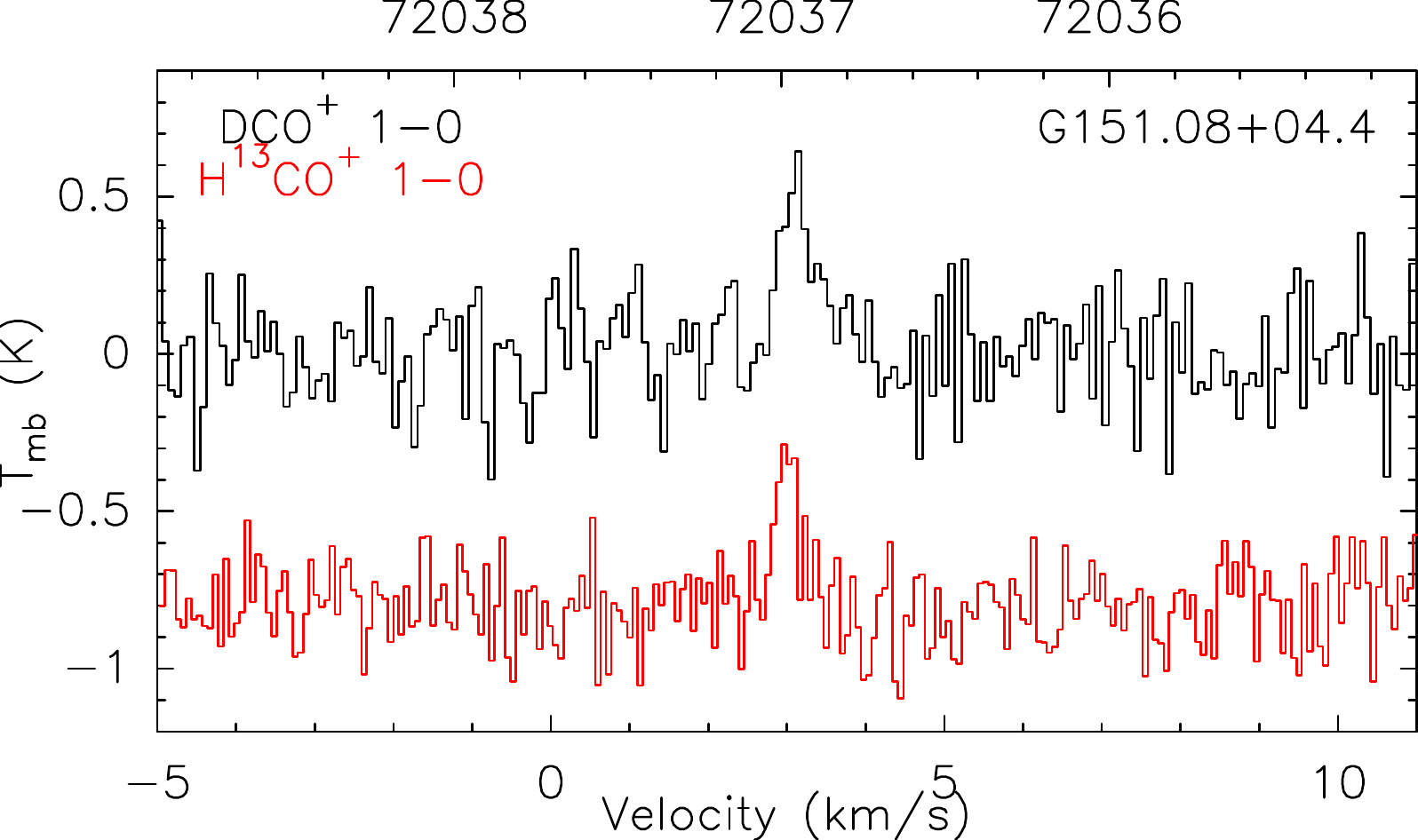}
\includegraphics[width=0.3\columnwidth]{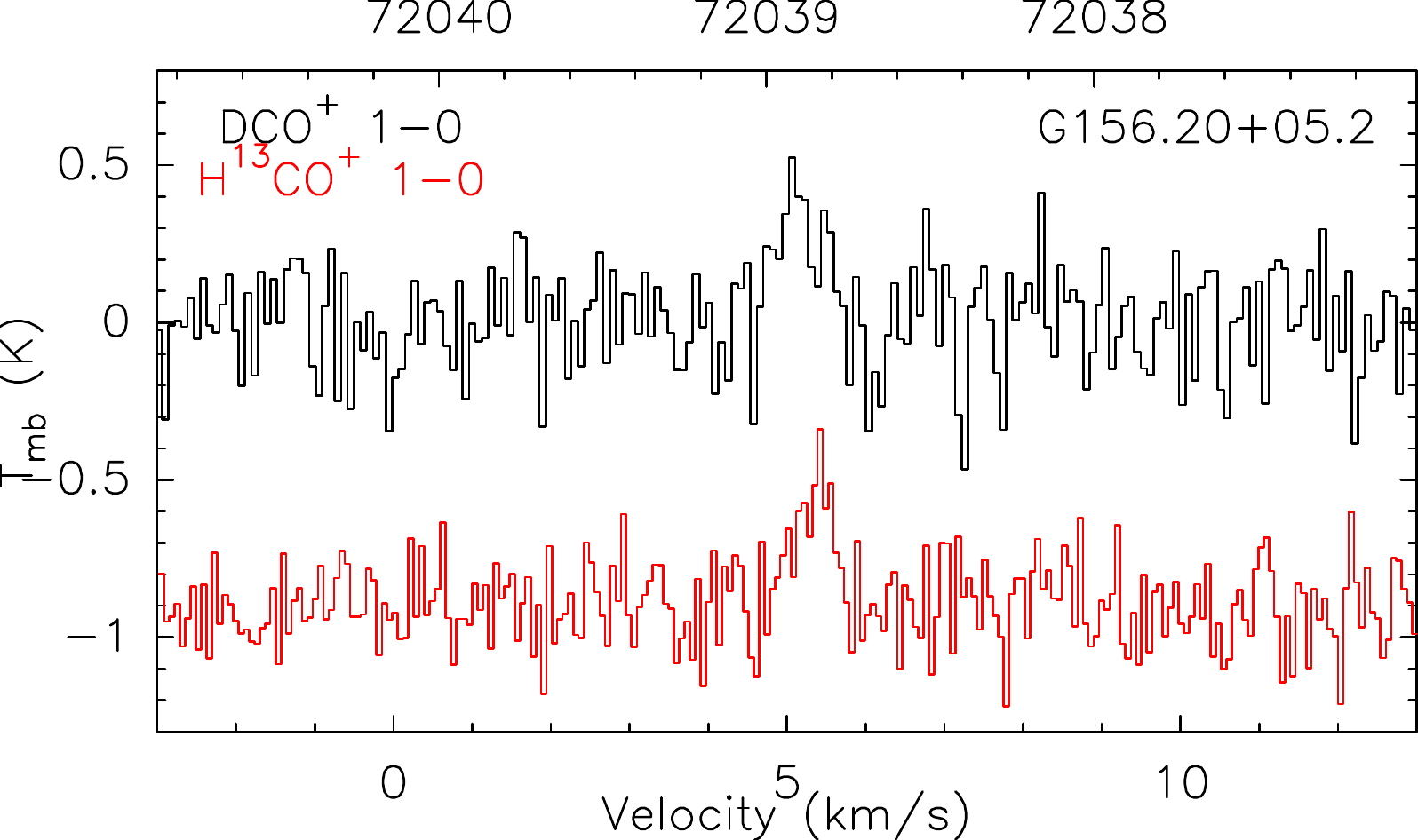}
\includegraphics[width=0.3\columnwidth]{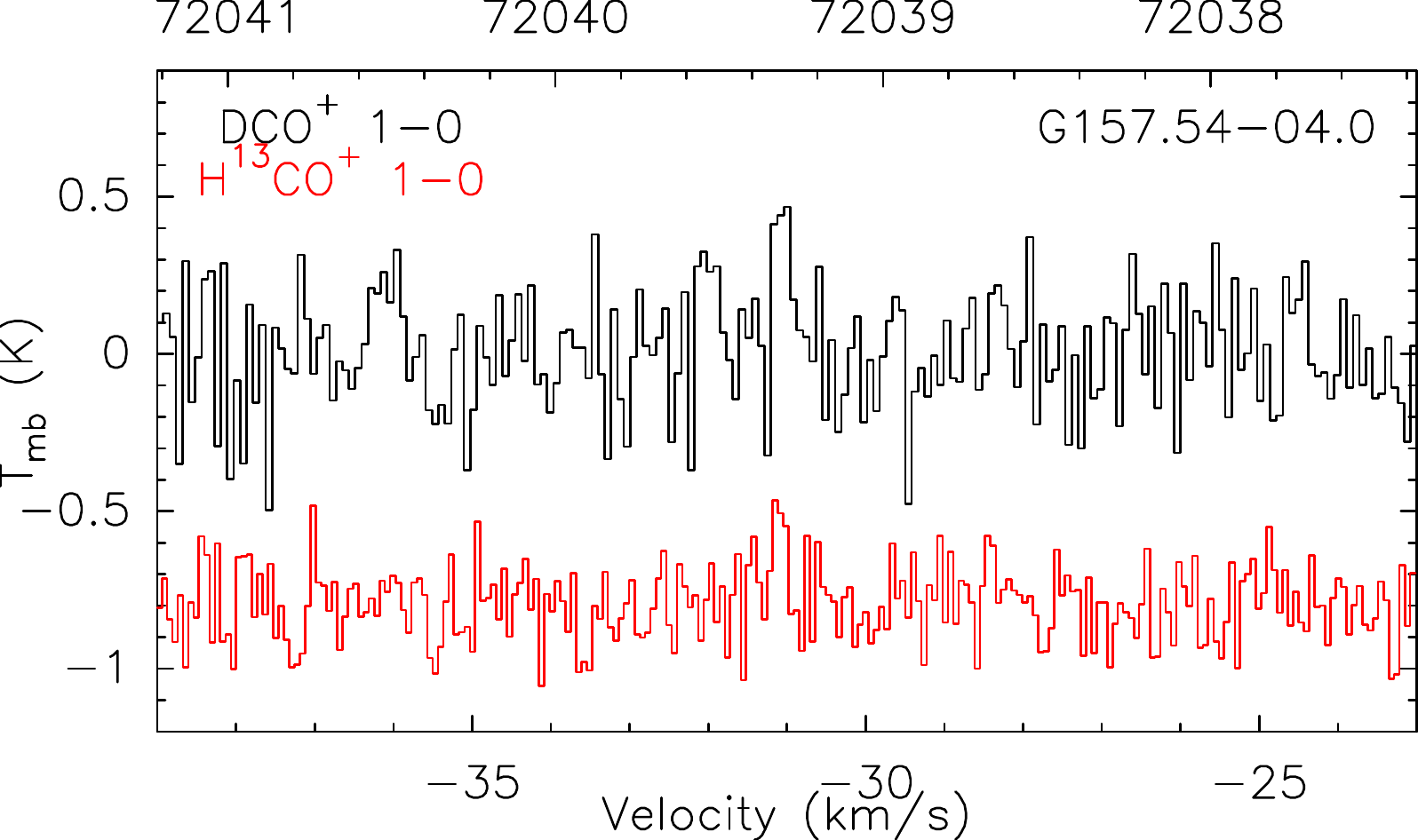}
\includegraphics[width=0.3\columnwidth]{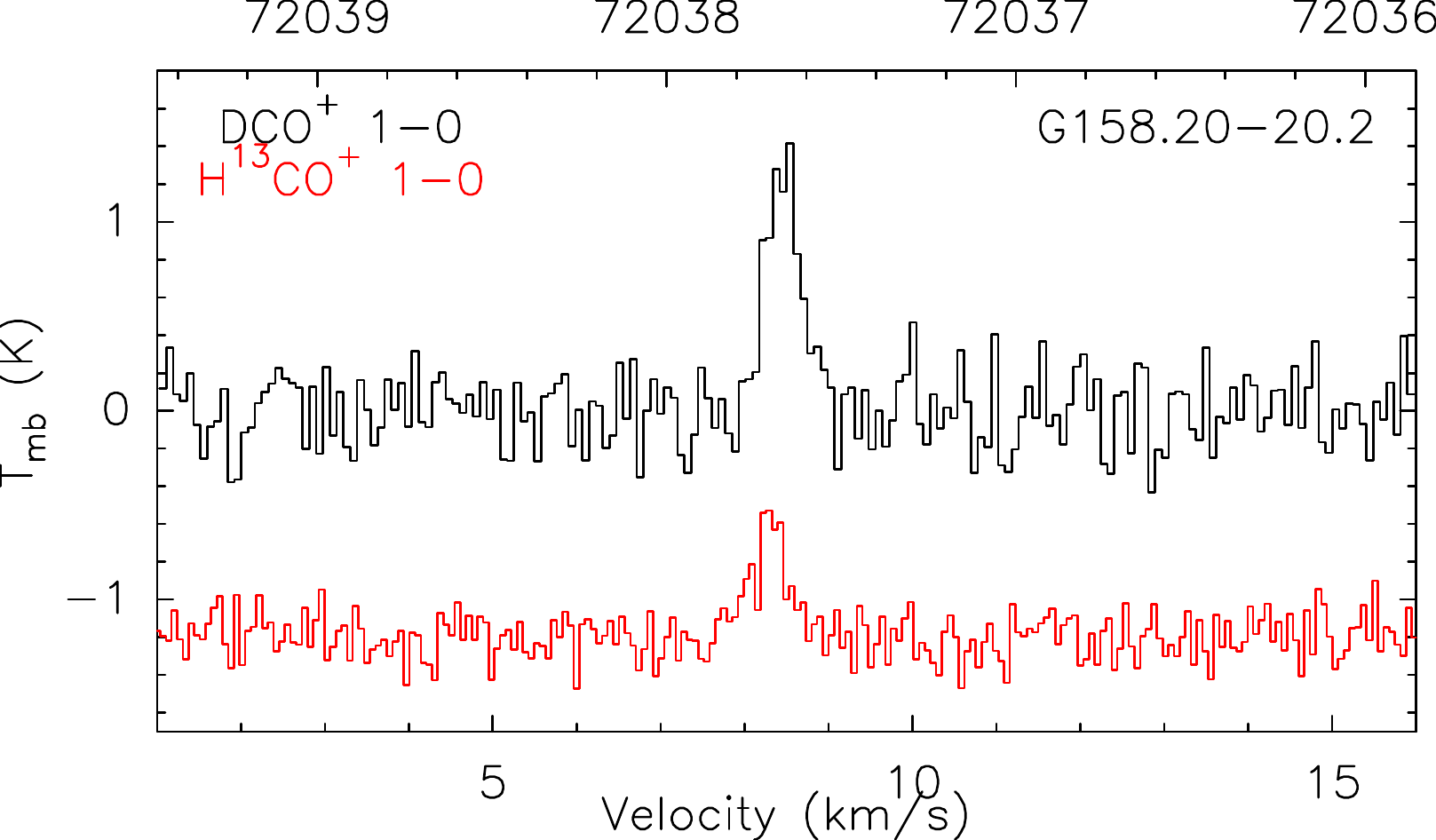}
\includegraphics[width=0.3\columnwidth]{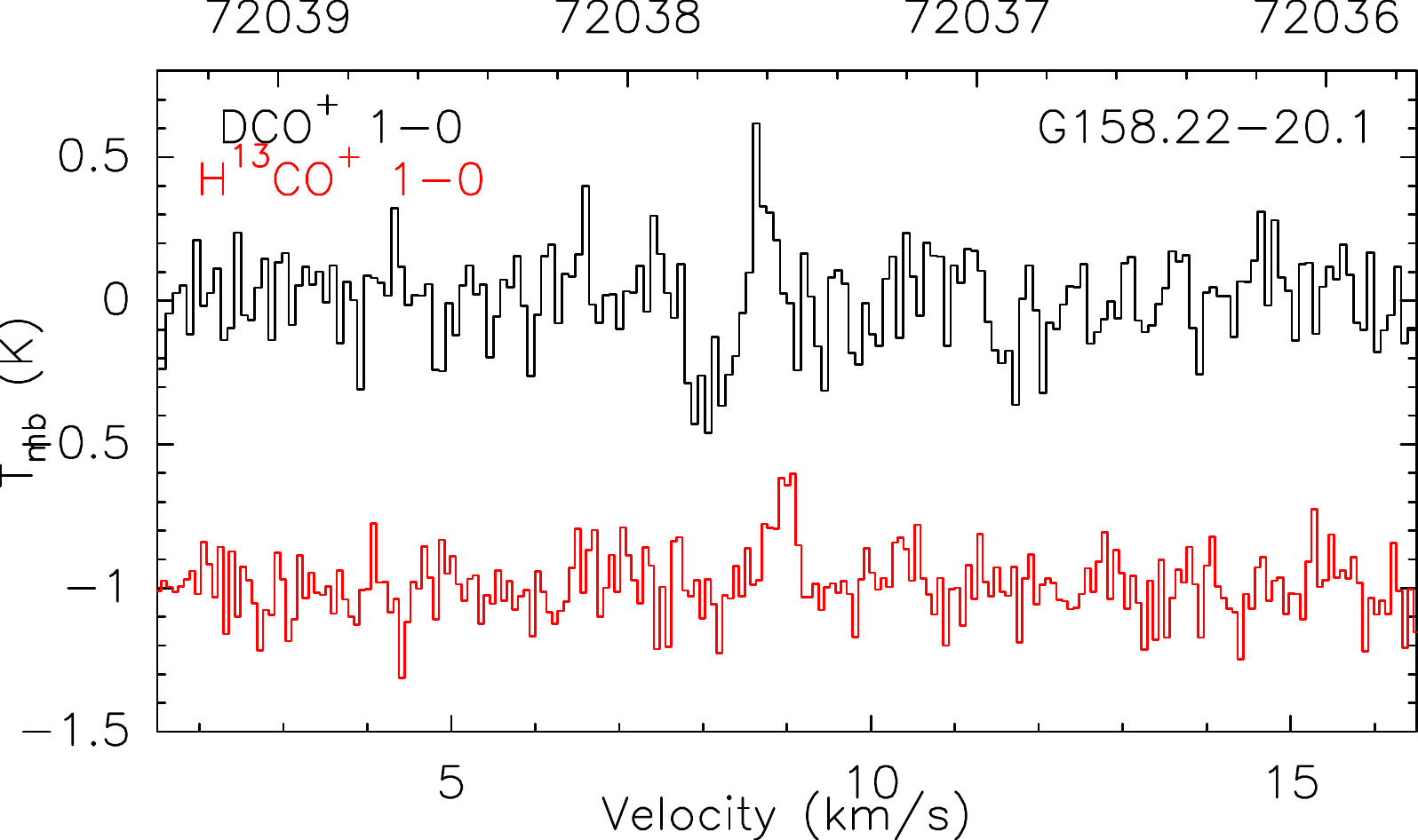}
\includegraphics[width=0.3\columnwidth]{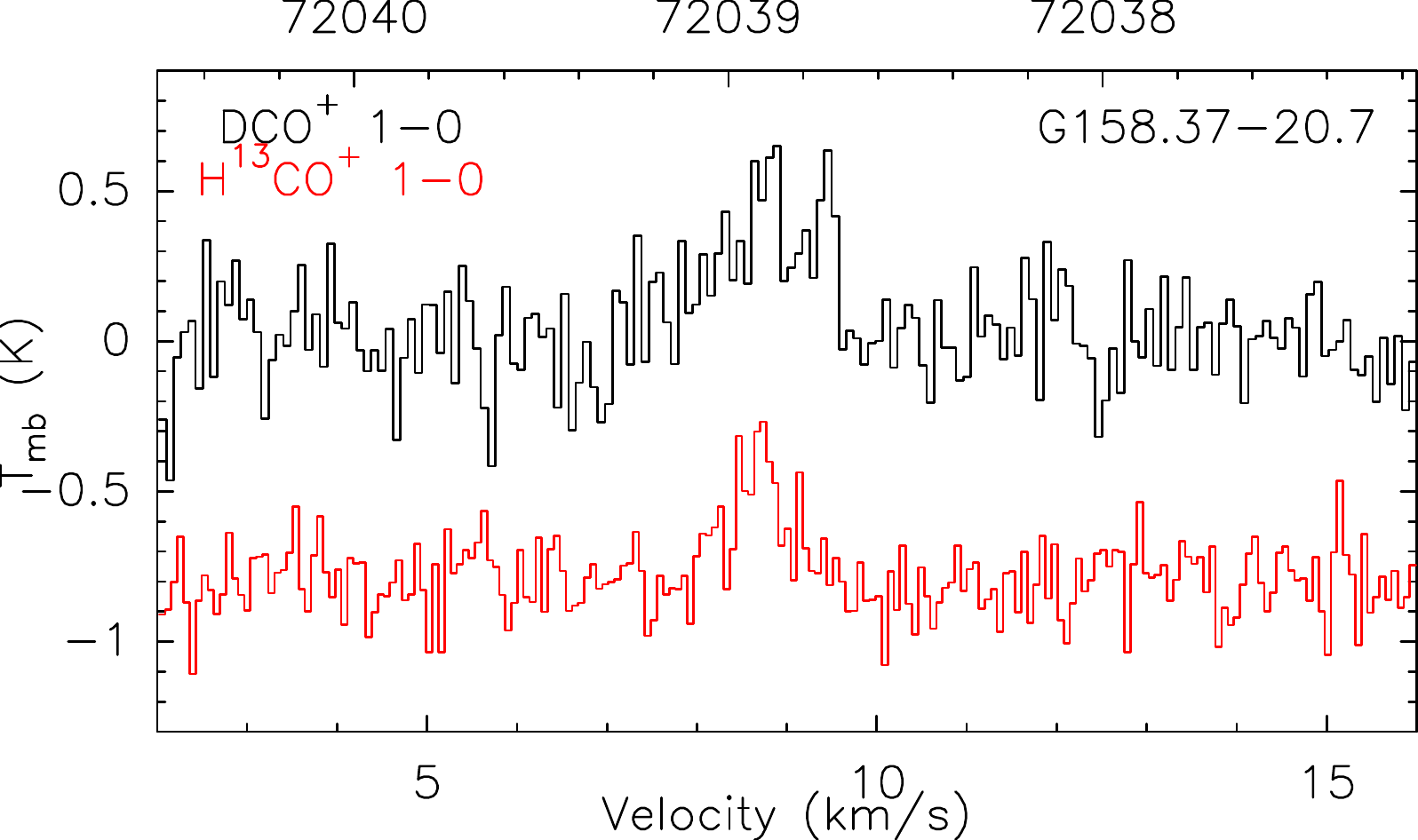}
\includegraphics[width=0.3\columnwidth]{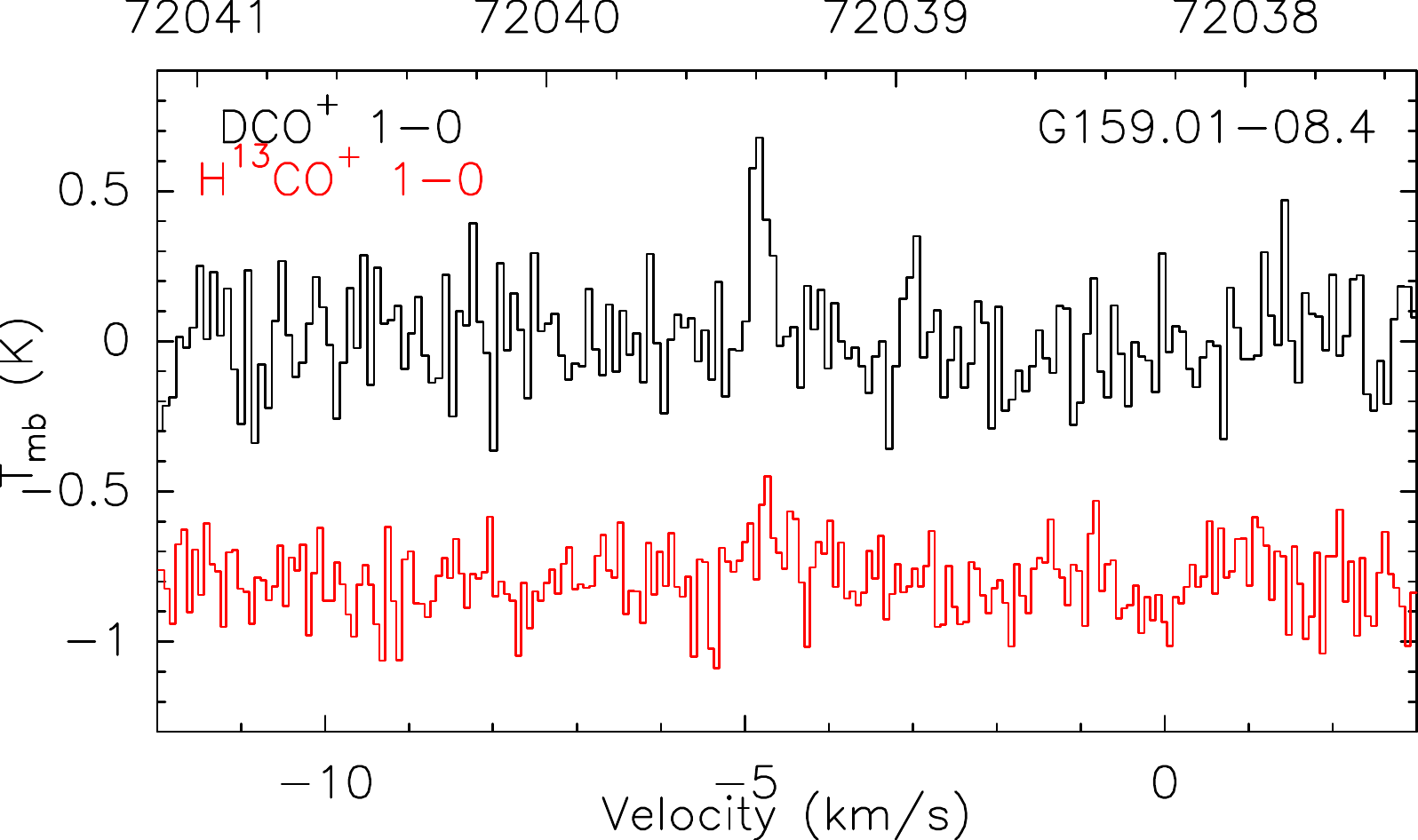}
\caption{Line profiles of DCO$^+$ and H$^{13}$CO$^+$ 1-0 with the high velocity resolution mode (AROWS mode 13). H$^{13}$CO$+$ 1-0 lines were manually aligned with DCO$+$ 1-0 lines, due to errors in the Doppler tracking.\centering}
\label{DCO+H13CO+mode13_2}
\end{figure}

\addtocounter{figure}{-1}
\begin{figure}
\centering
\includegraphics[width=0.3\columnwidth]{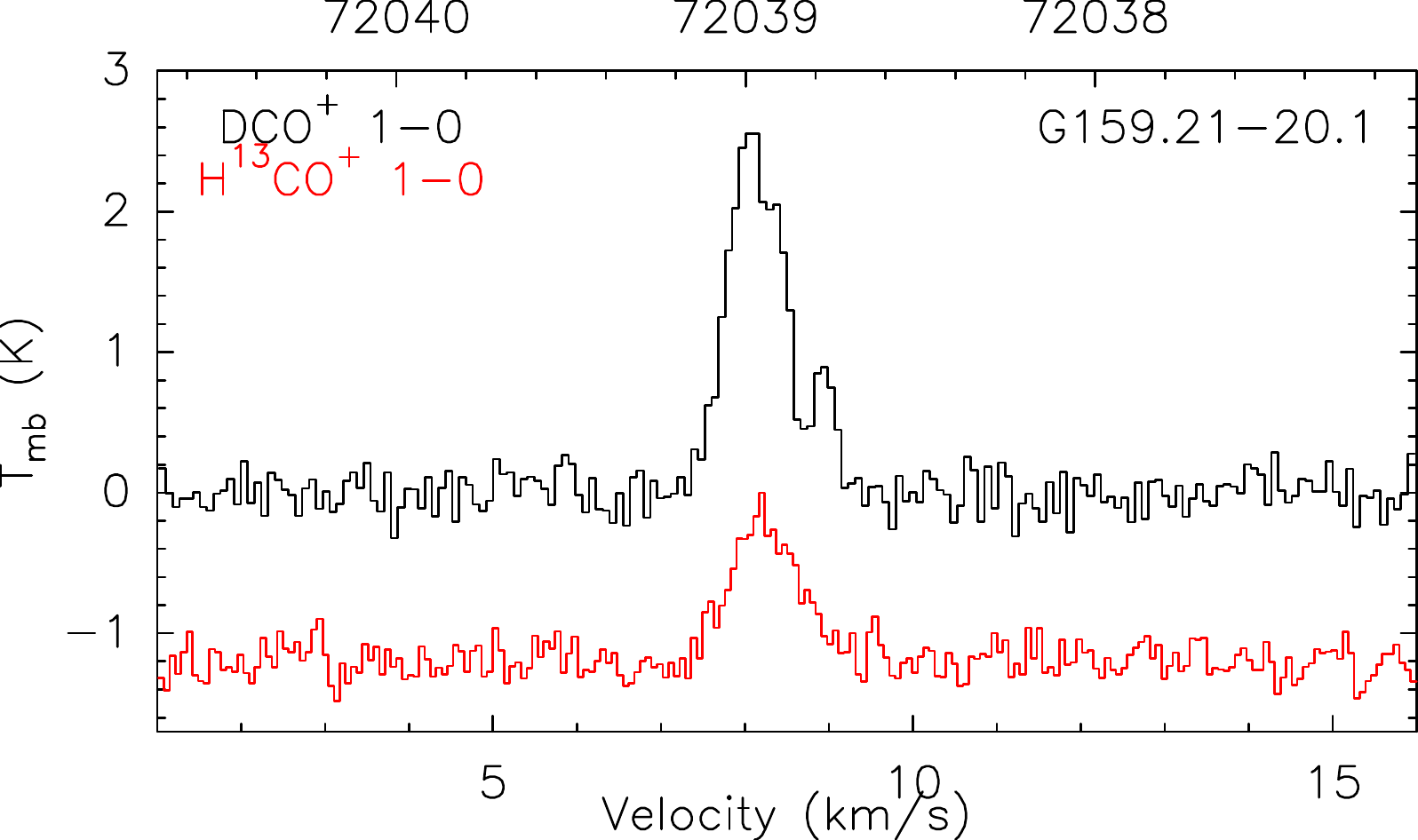}
\includegraphics[width=0.3\columnwidth]{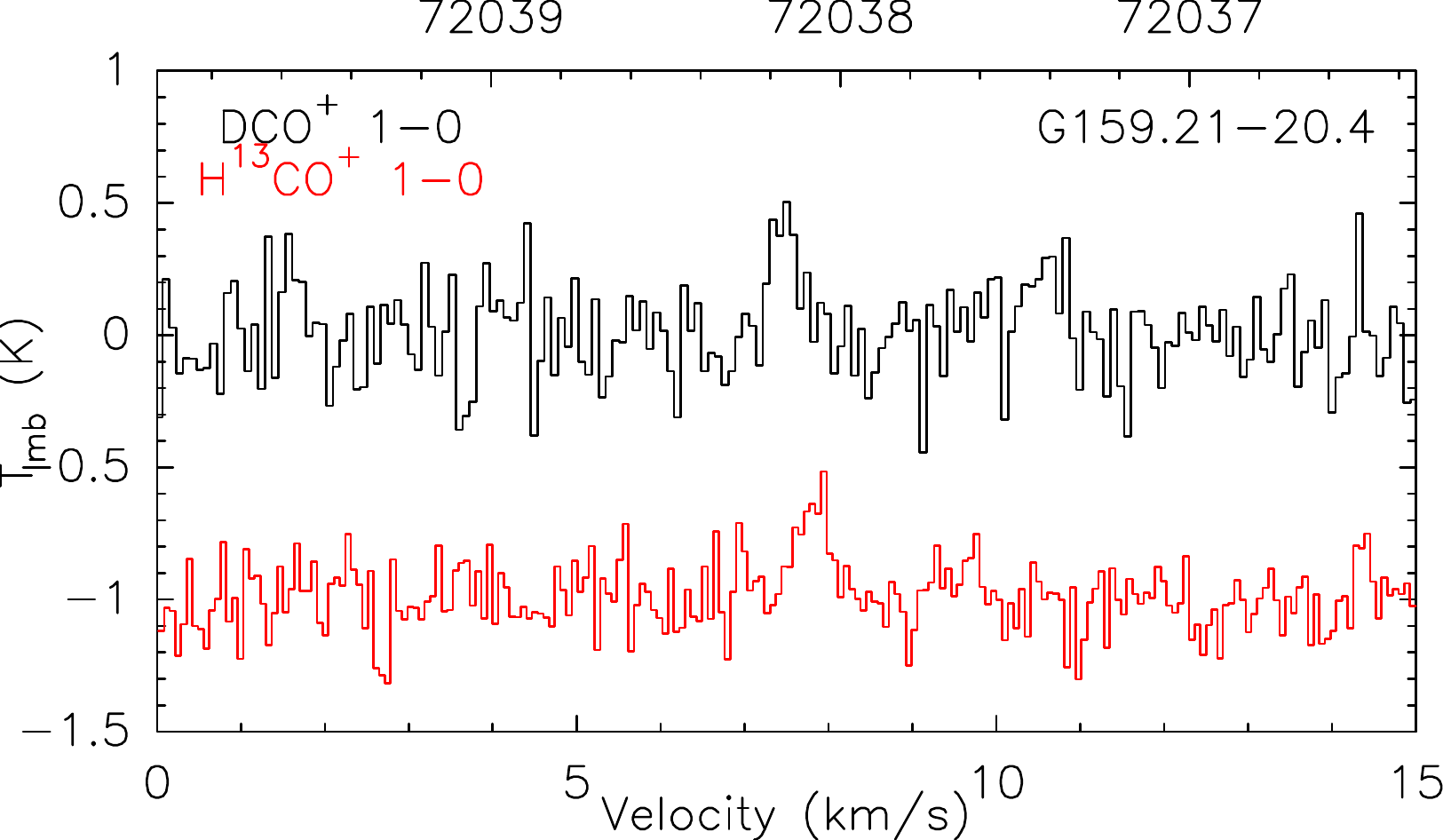}
\includegraphics[width=0.3\columnwidth]{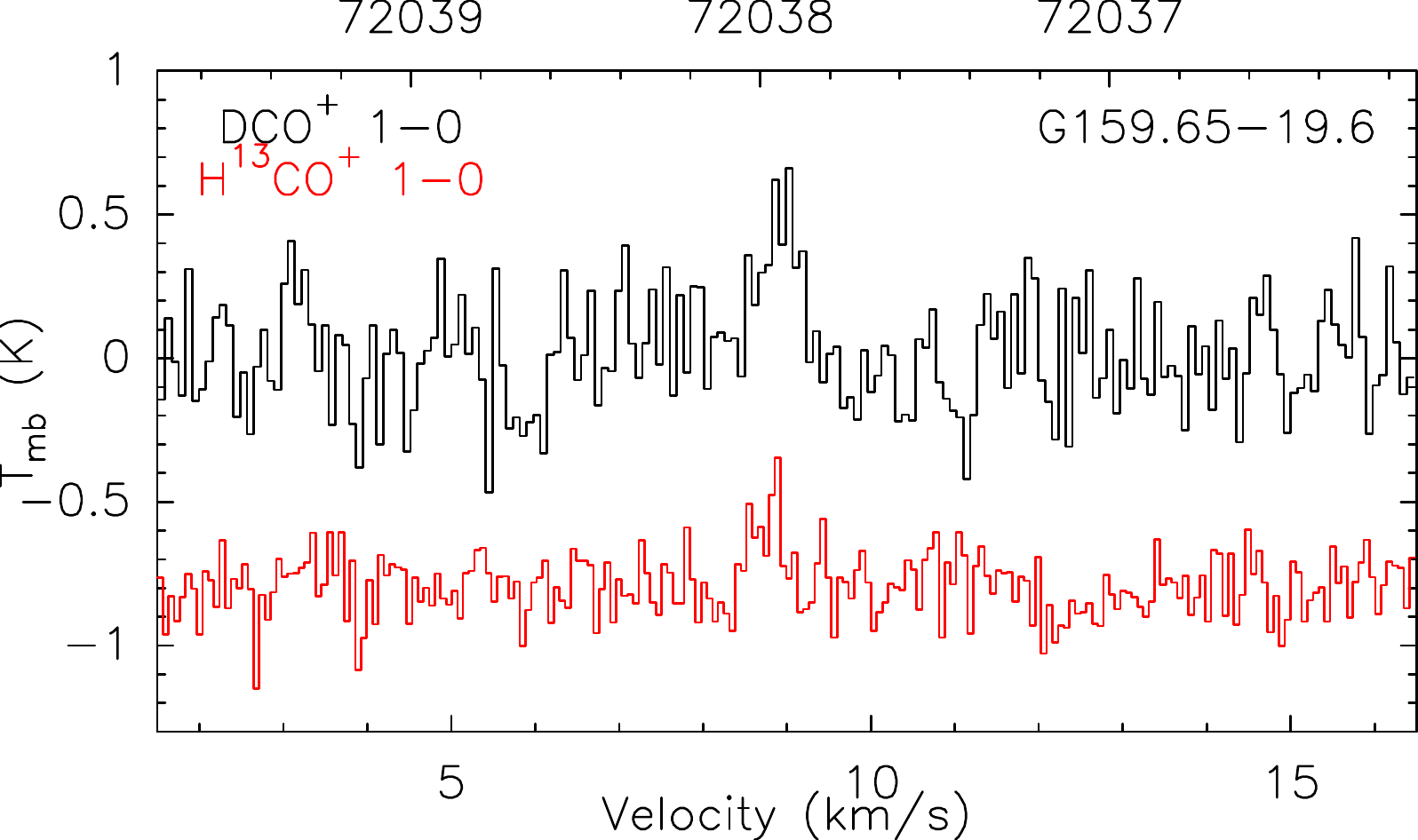}
\includegraphics[width=0.3\columnwidth]{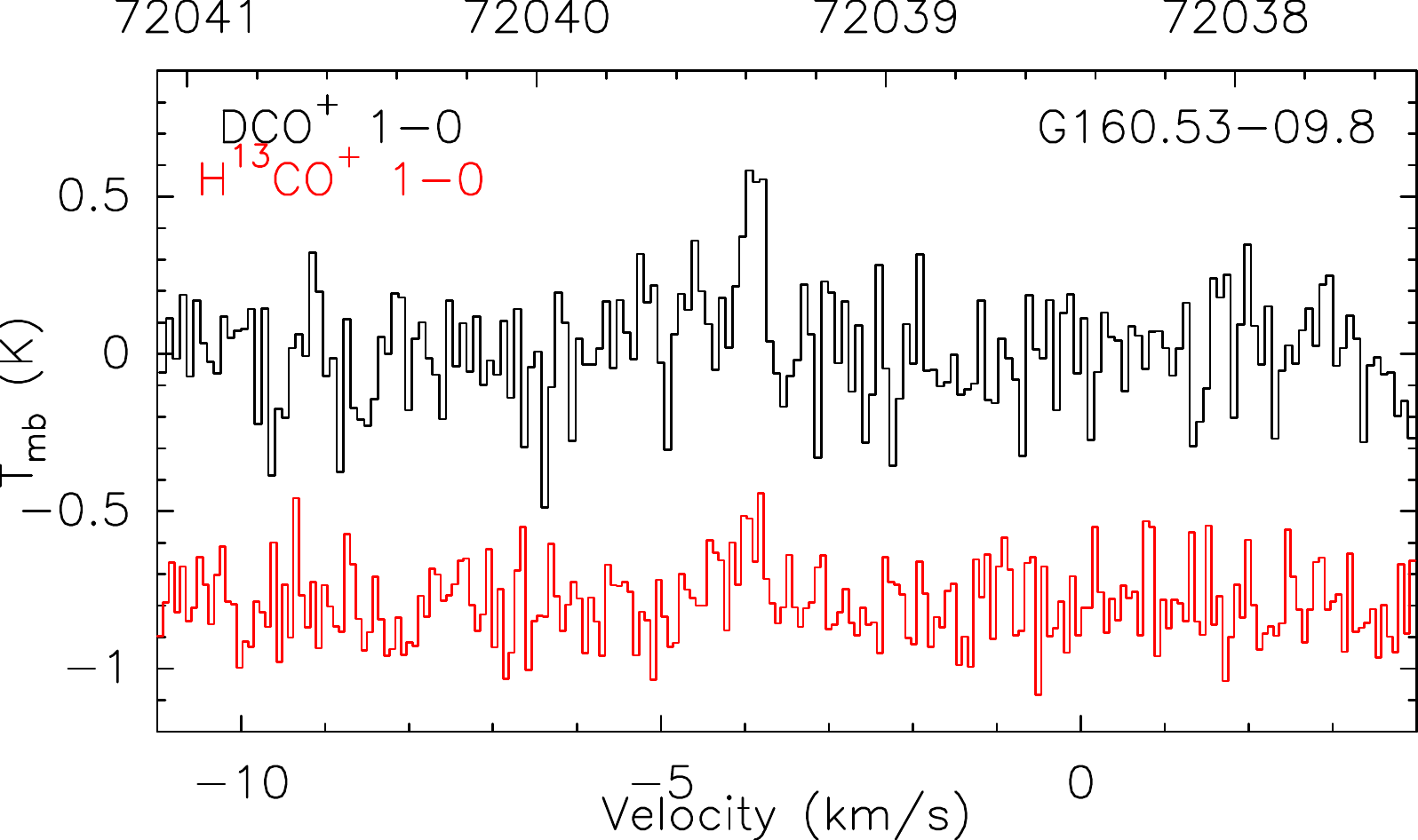}
\includegraphics[width=0.3\columnwidth]{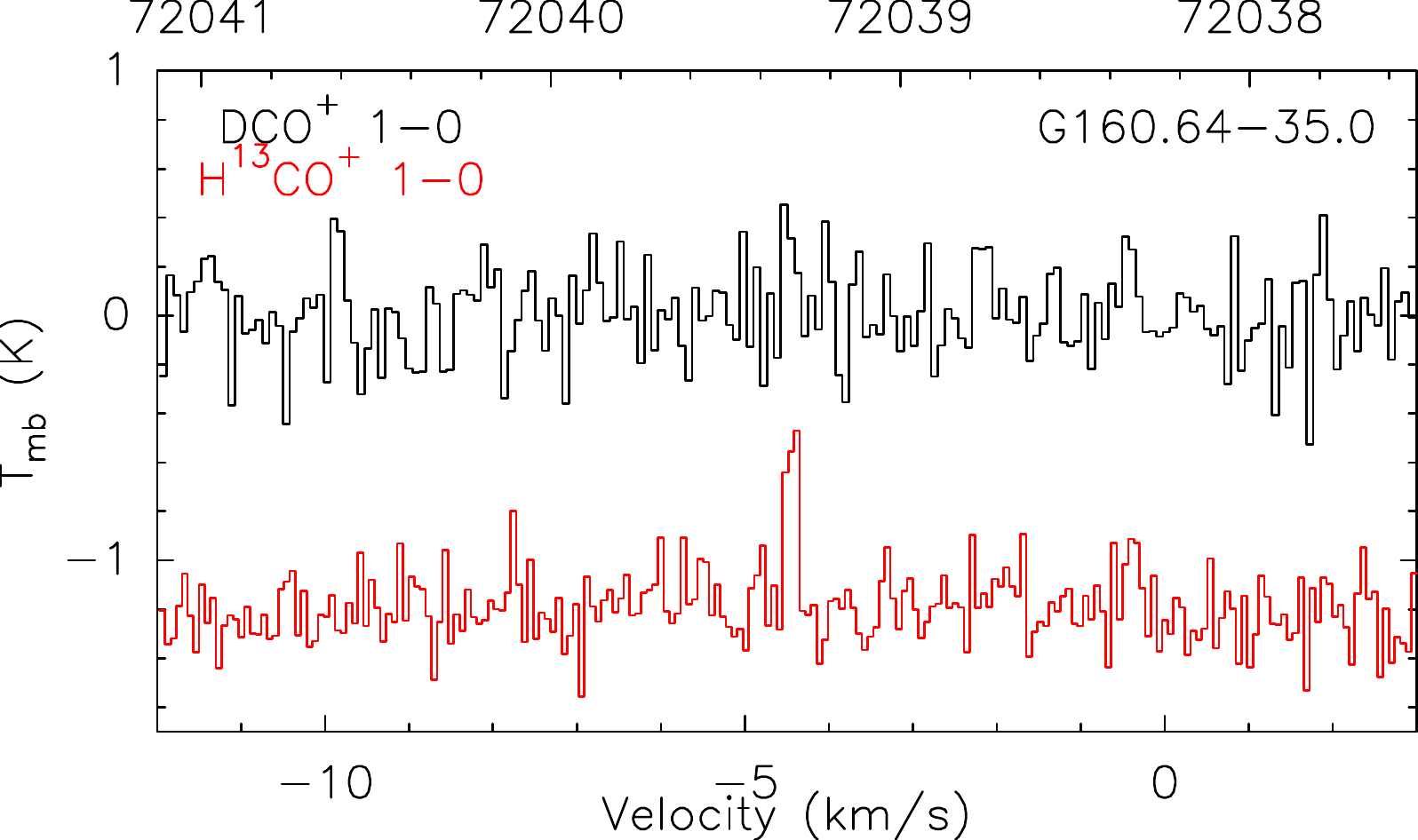}
\includegraphics[width=0.3\columnwidth]{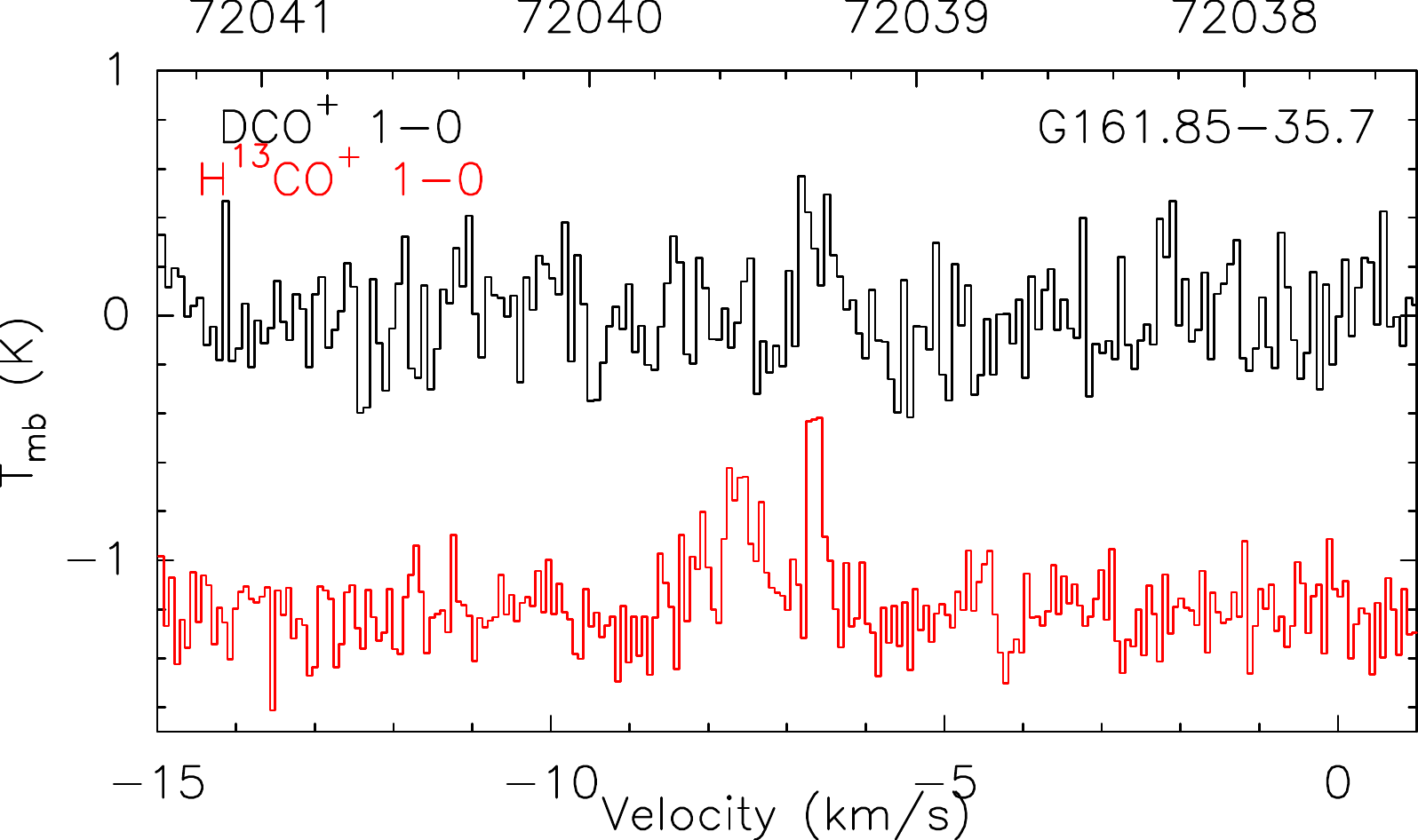}
\includegraphics[width=0.3\columnwidth]{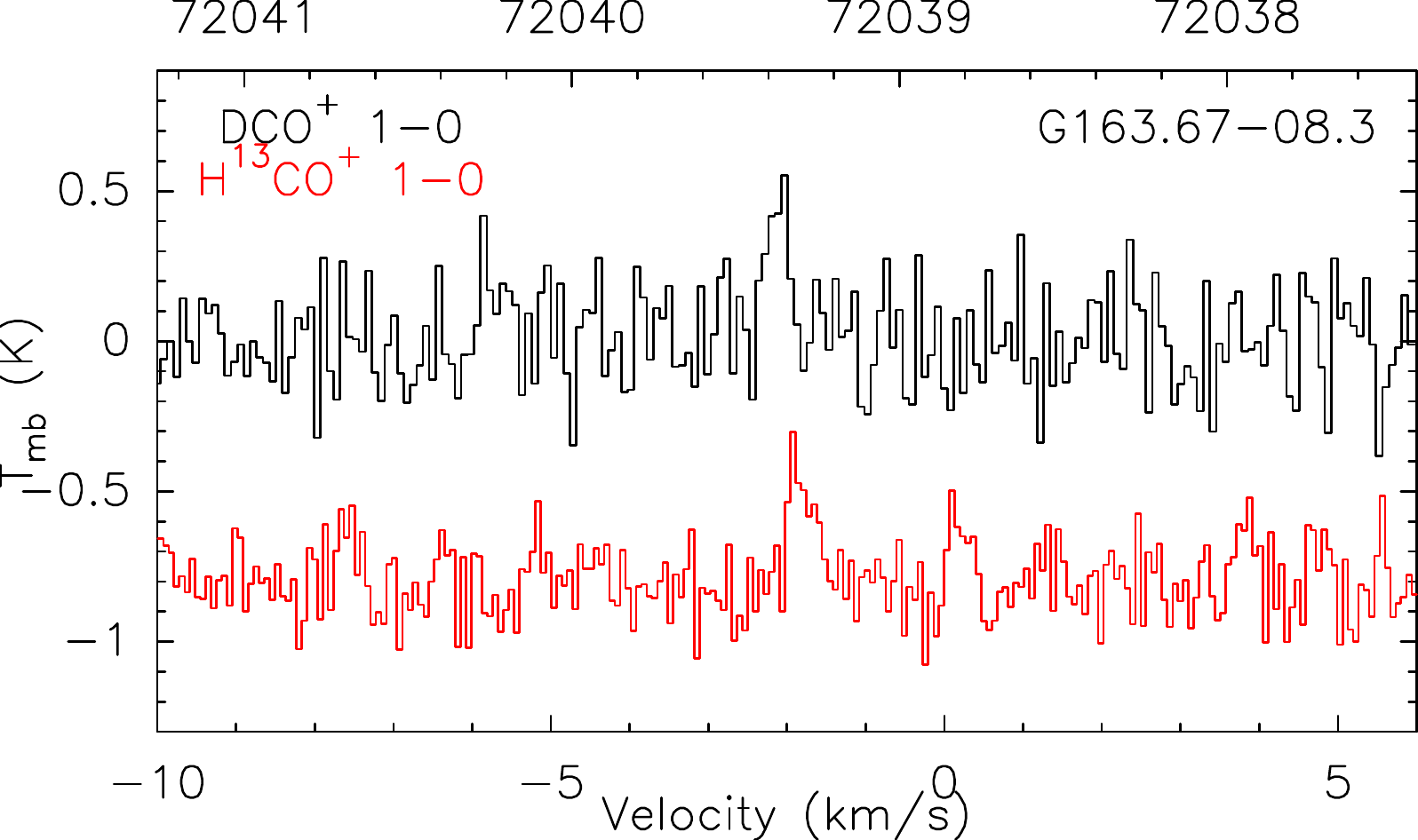}
\includegraphics[width=0.3\columnwidth]{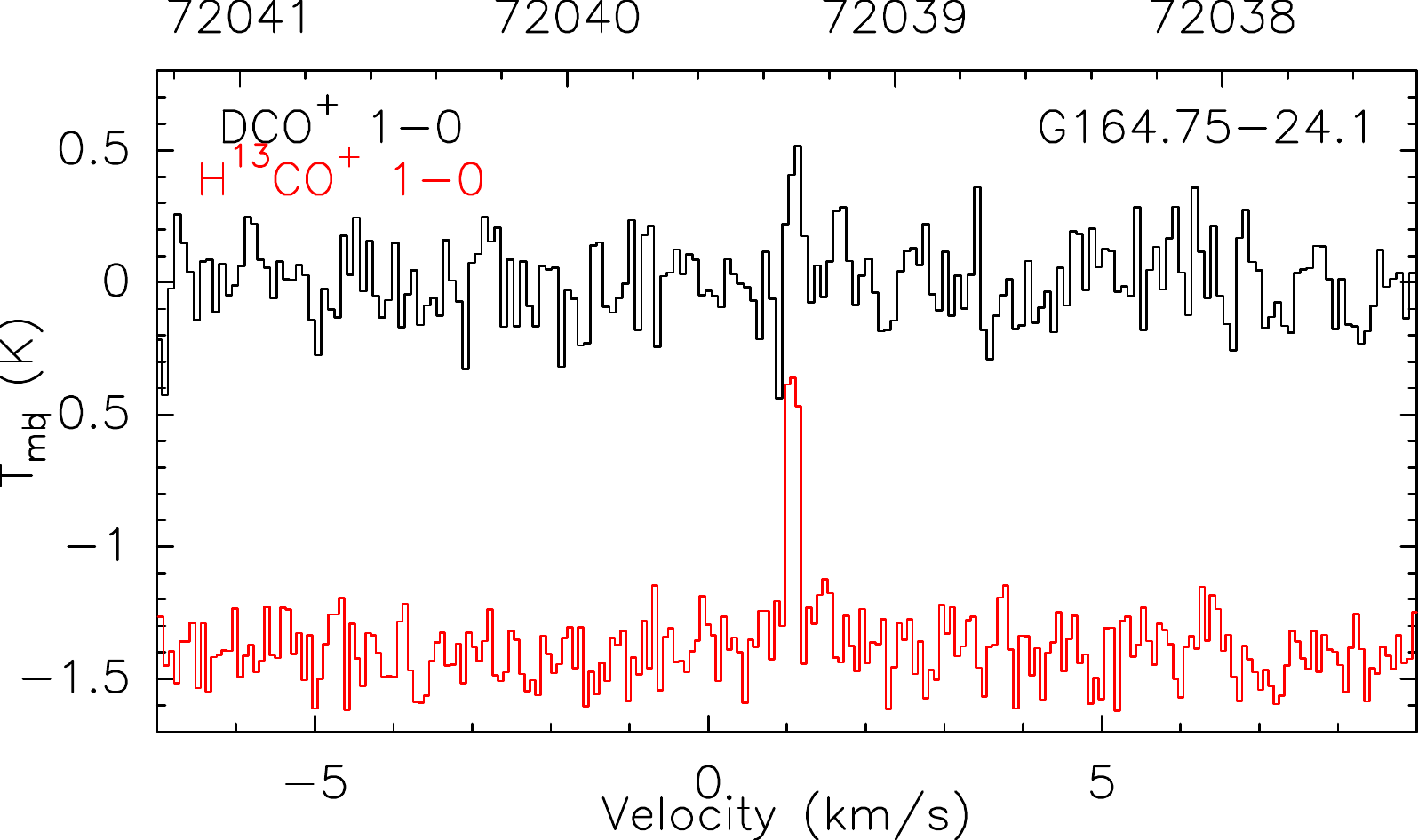}
\includegraphics[width=0.3\columnwidth]{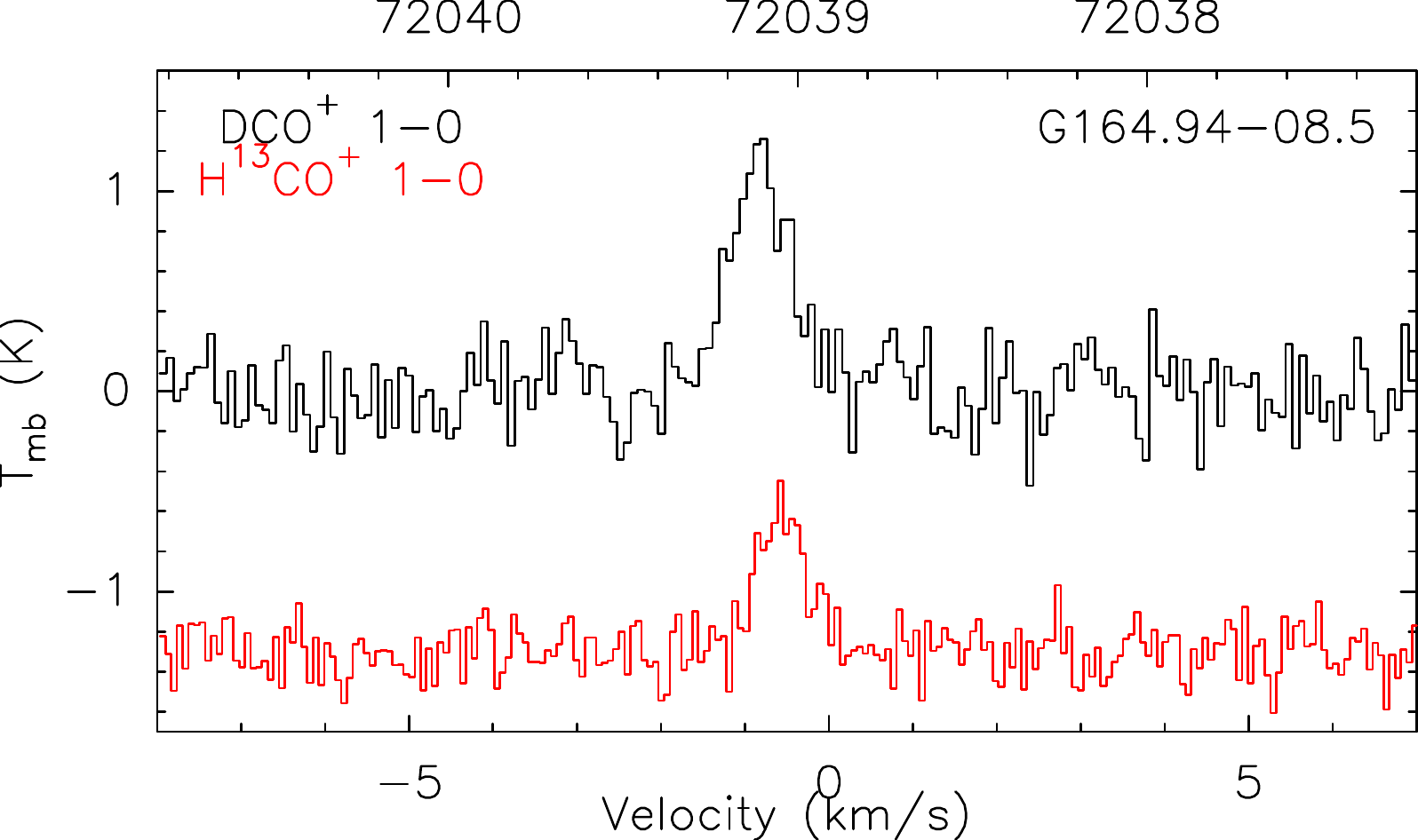}
\includegraphics[width=0.3\columnwidth]{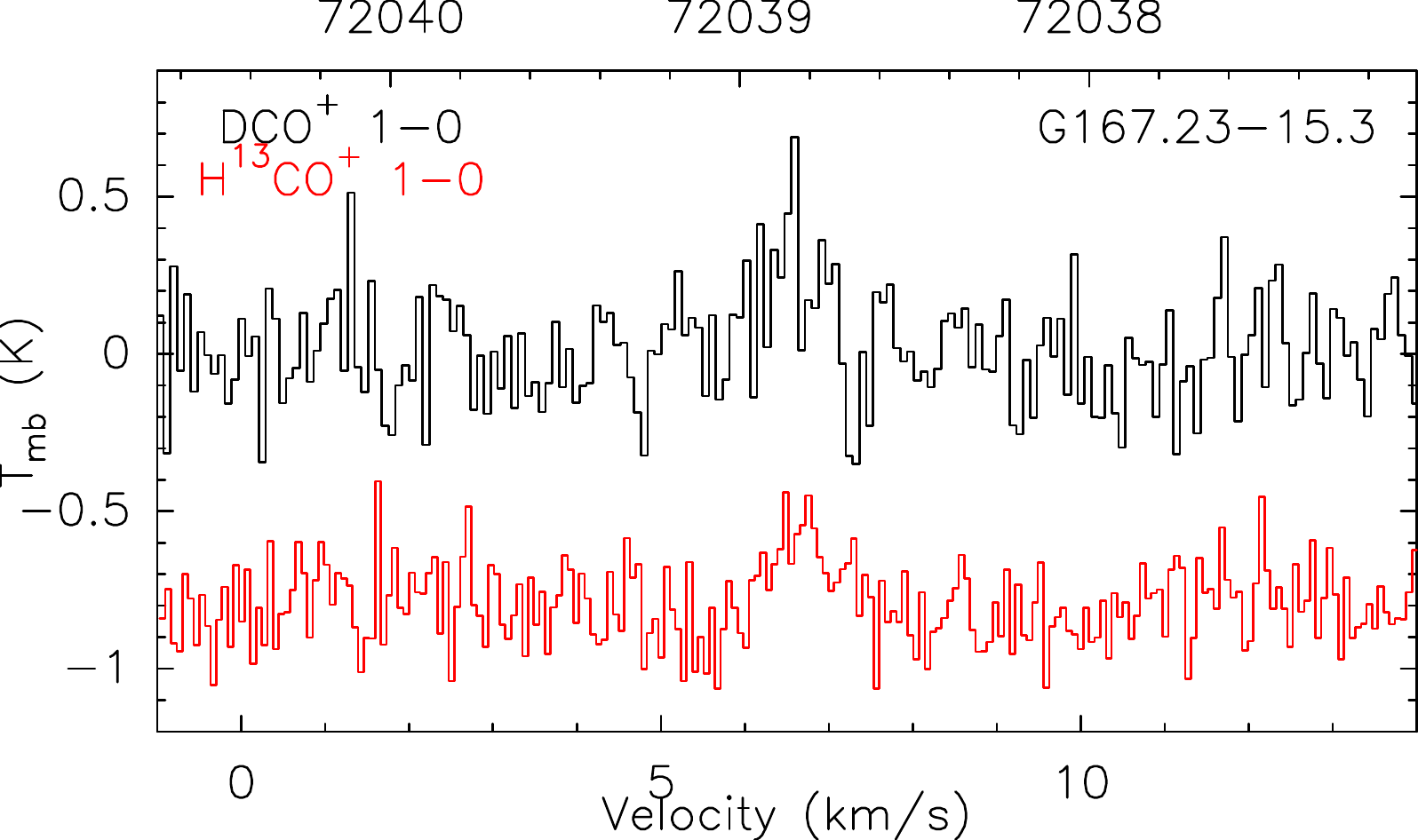}
\includegraphics[width=0.3\columnwidth]{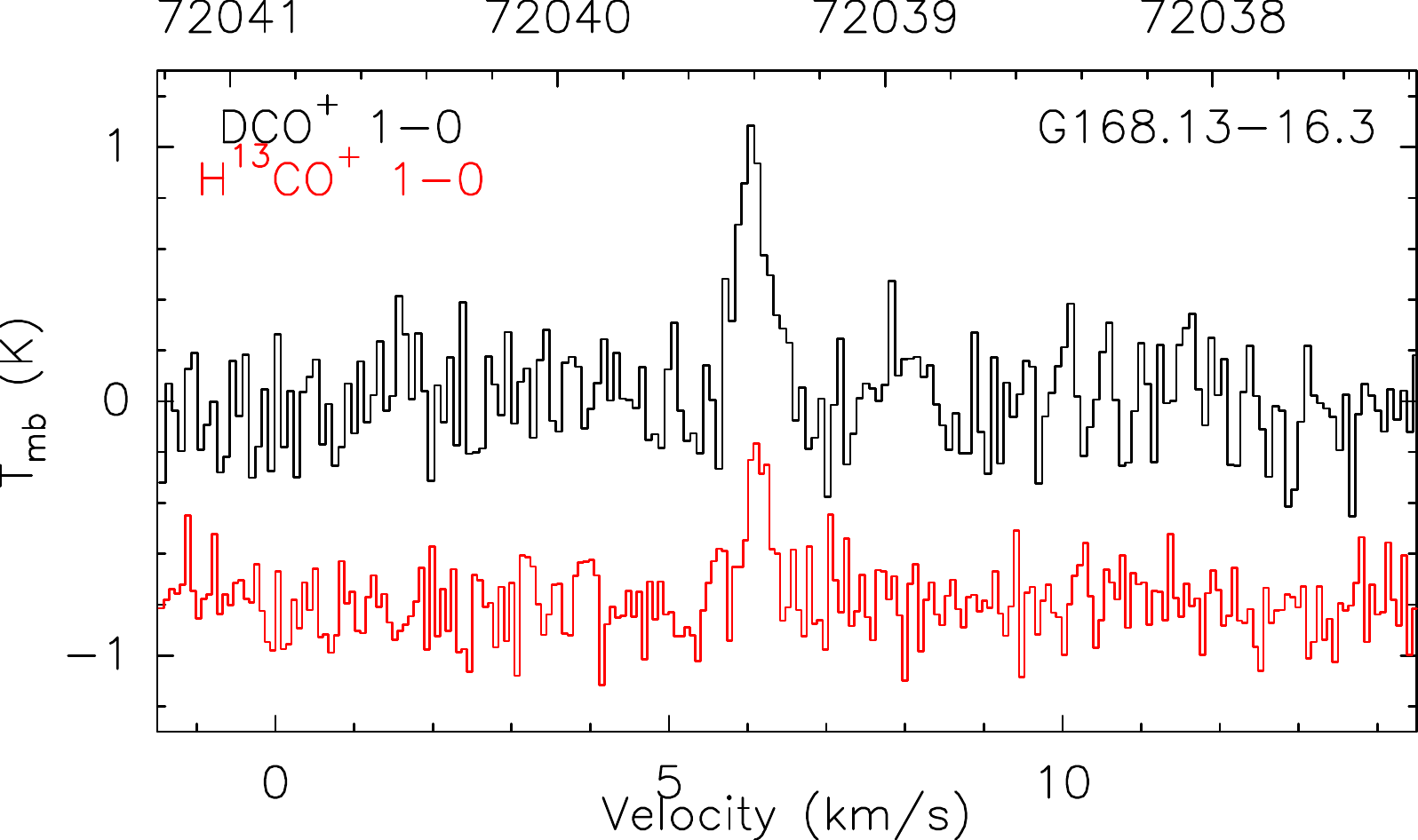}
\includegraphics[width=0.3\columnwidth]{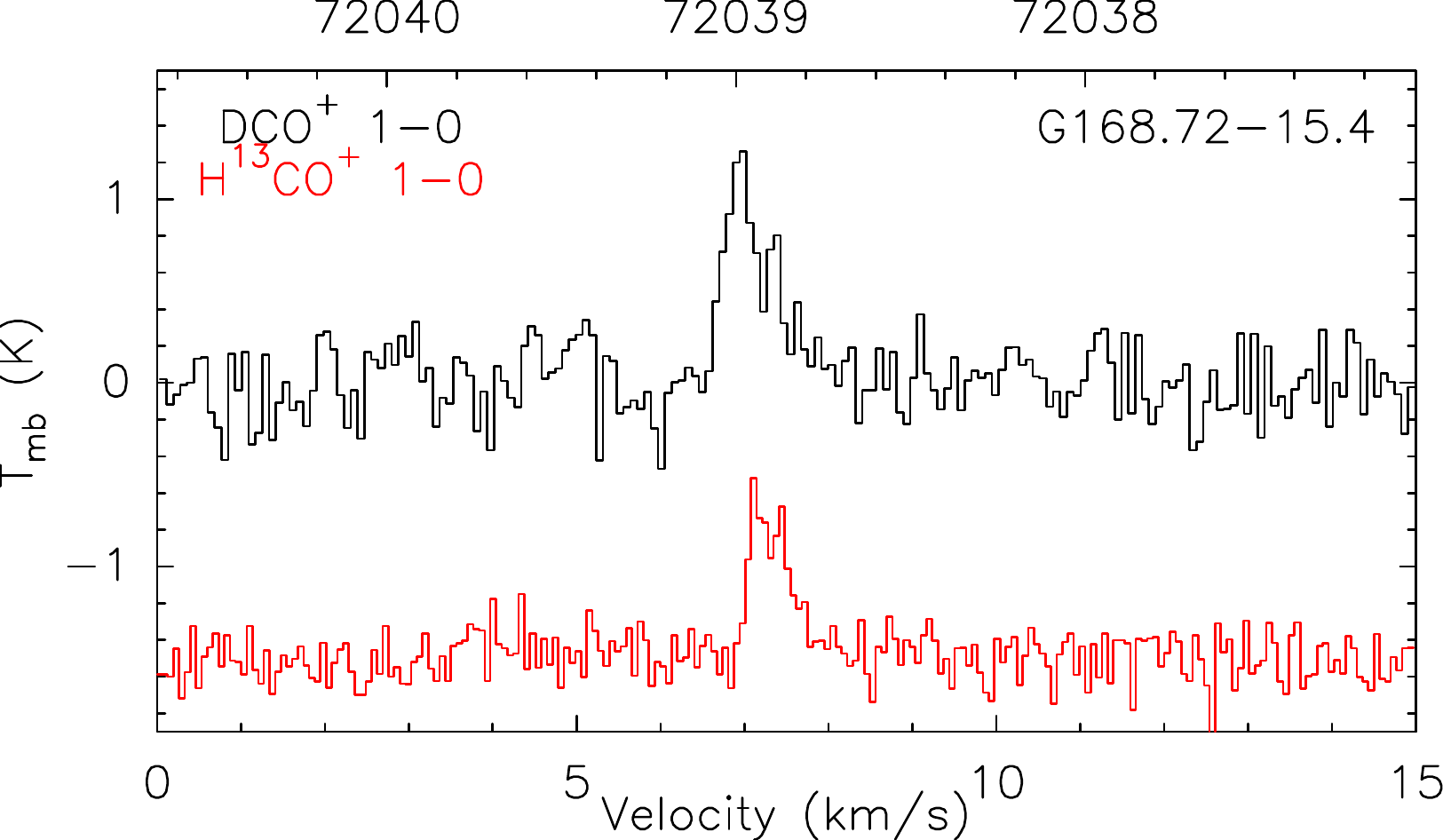}
\includegraphics[width=0.3\columnwidth]{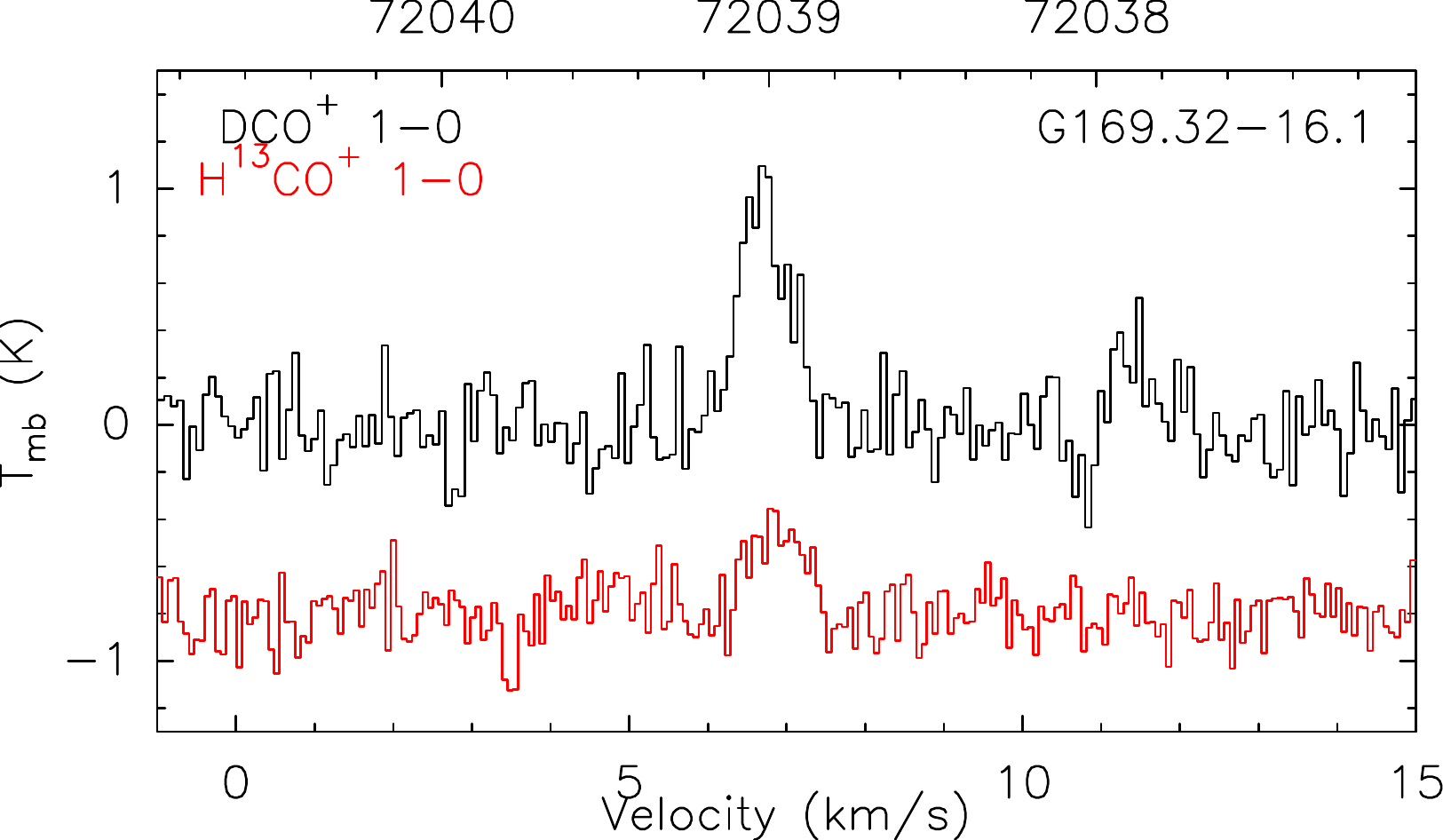}
\includegraphics[width=0.3\columnwidth]{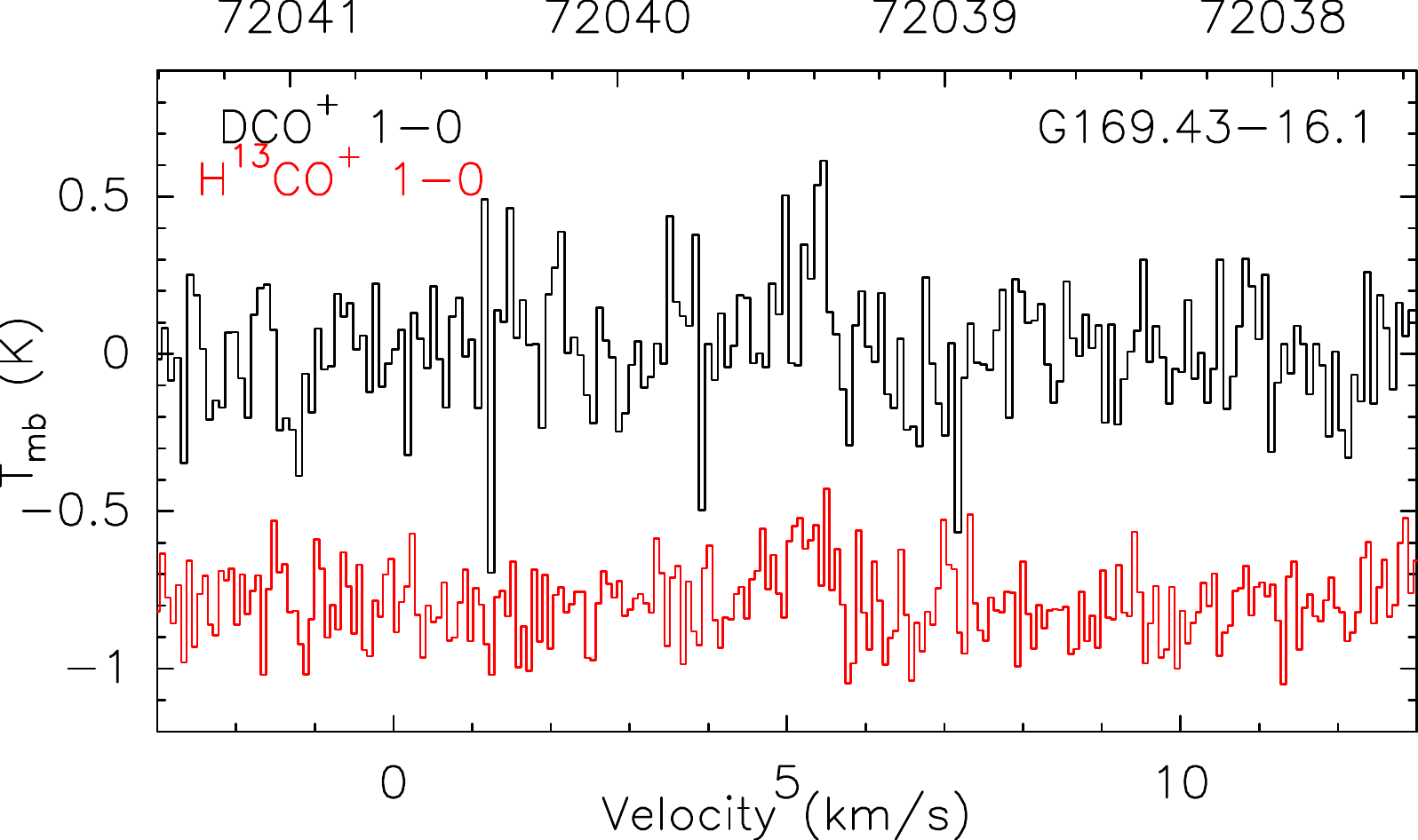}
\includegraphics[width=0.3\columnwidth]{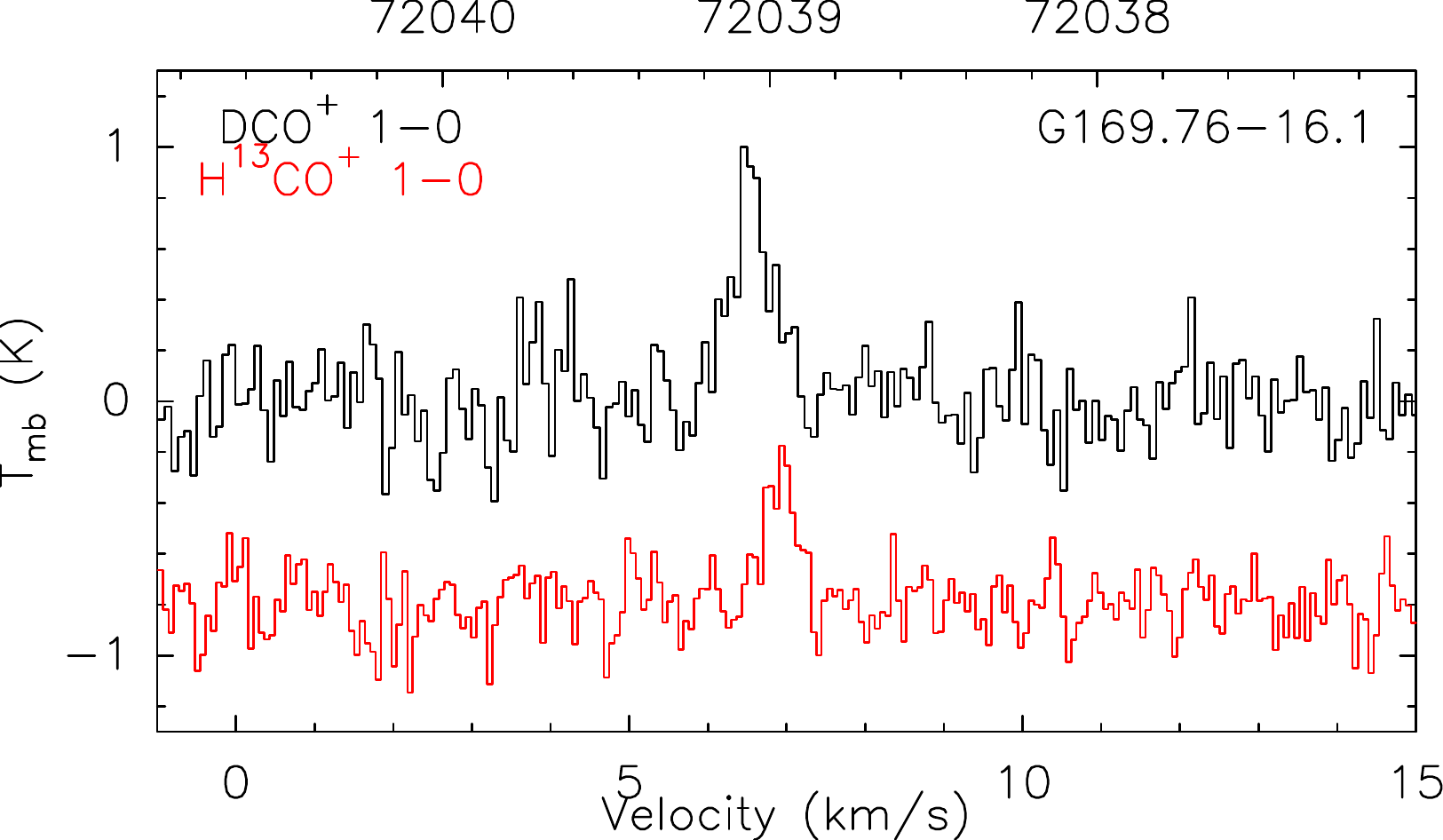}
\includegraphics[width=0.3\columnwidth]{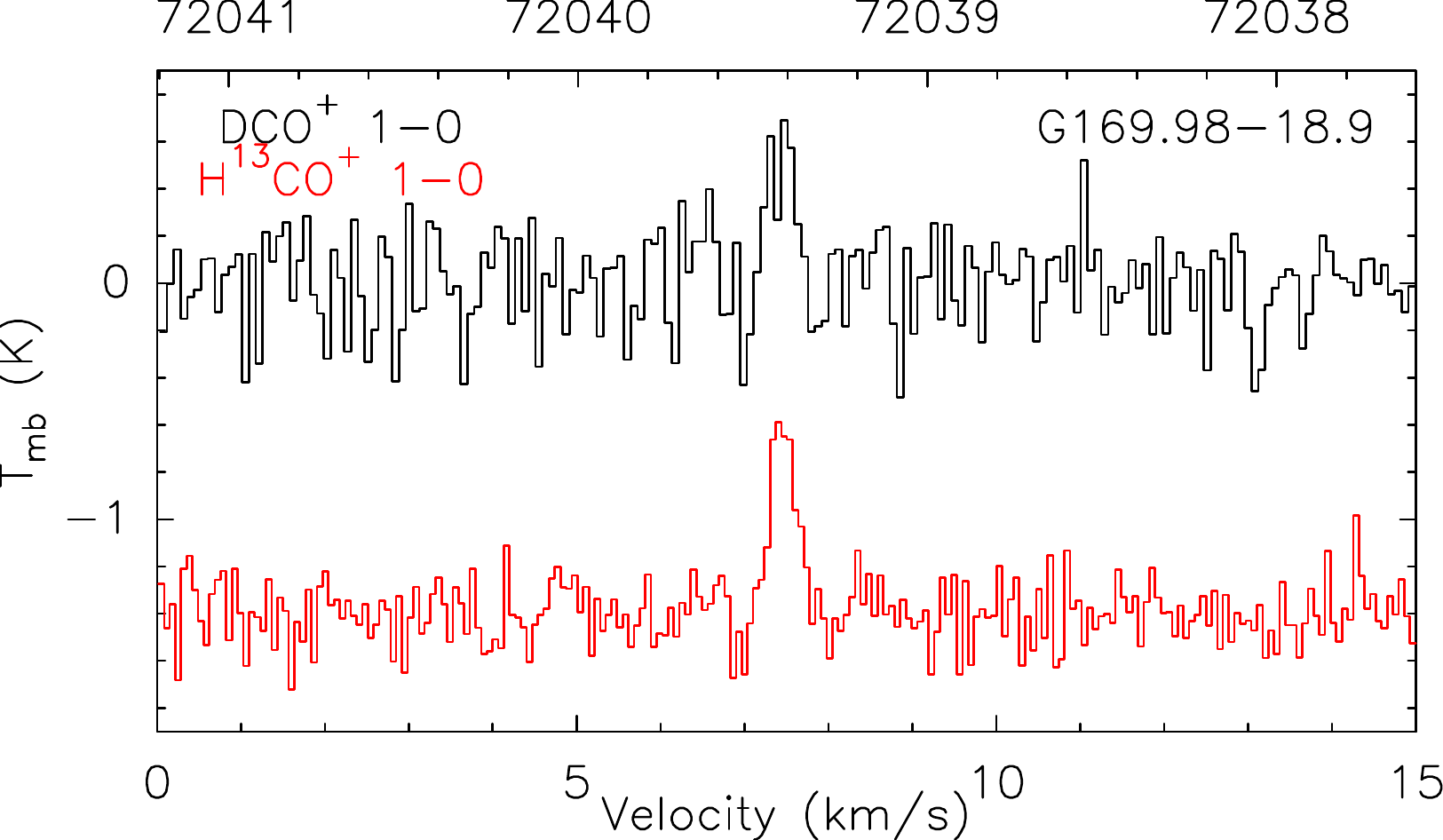}
\includegraphics[width=0.3\columnwidth]{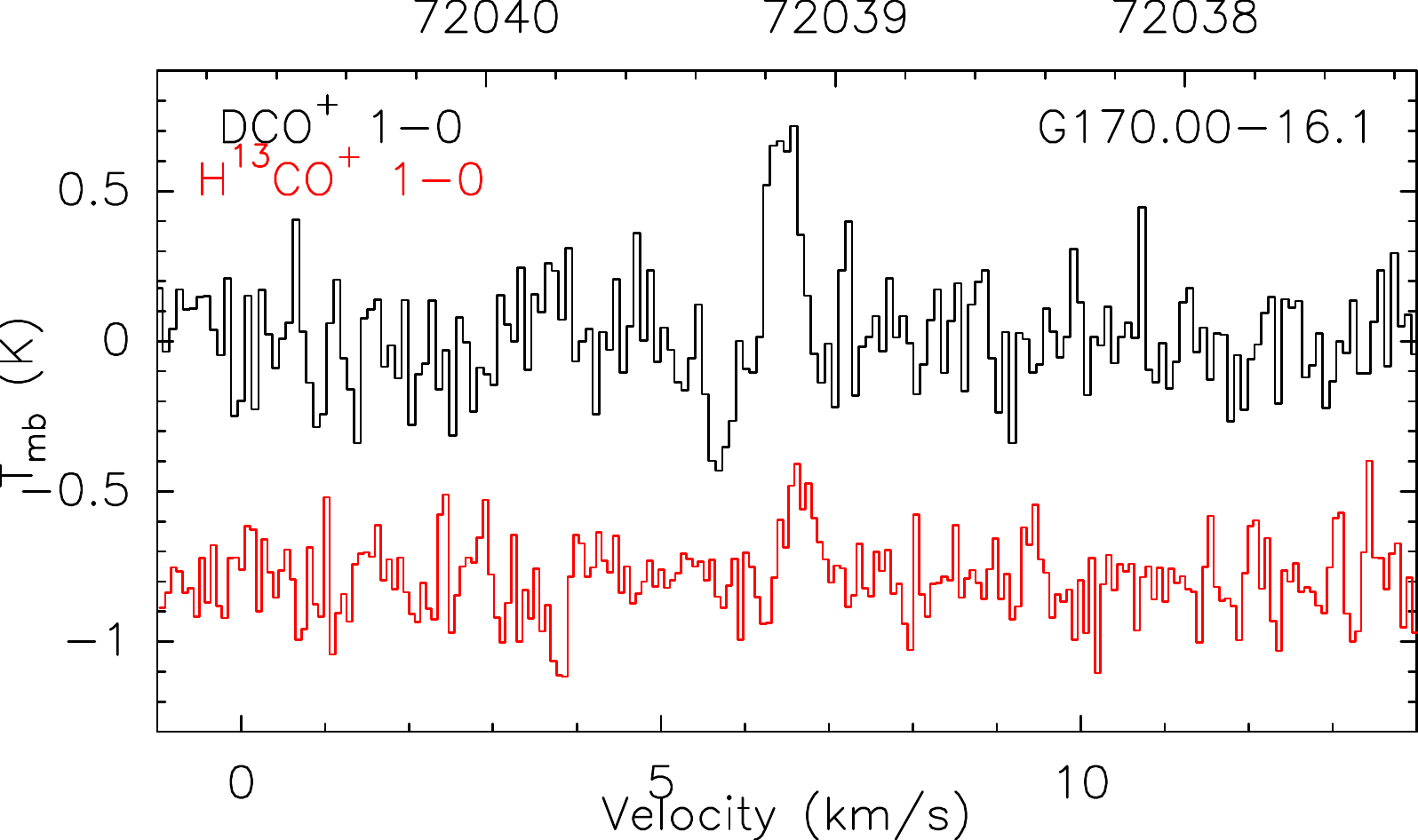}
\includegraphics[width=0.3\columnwidth]{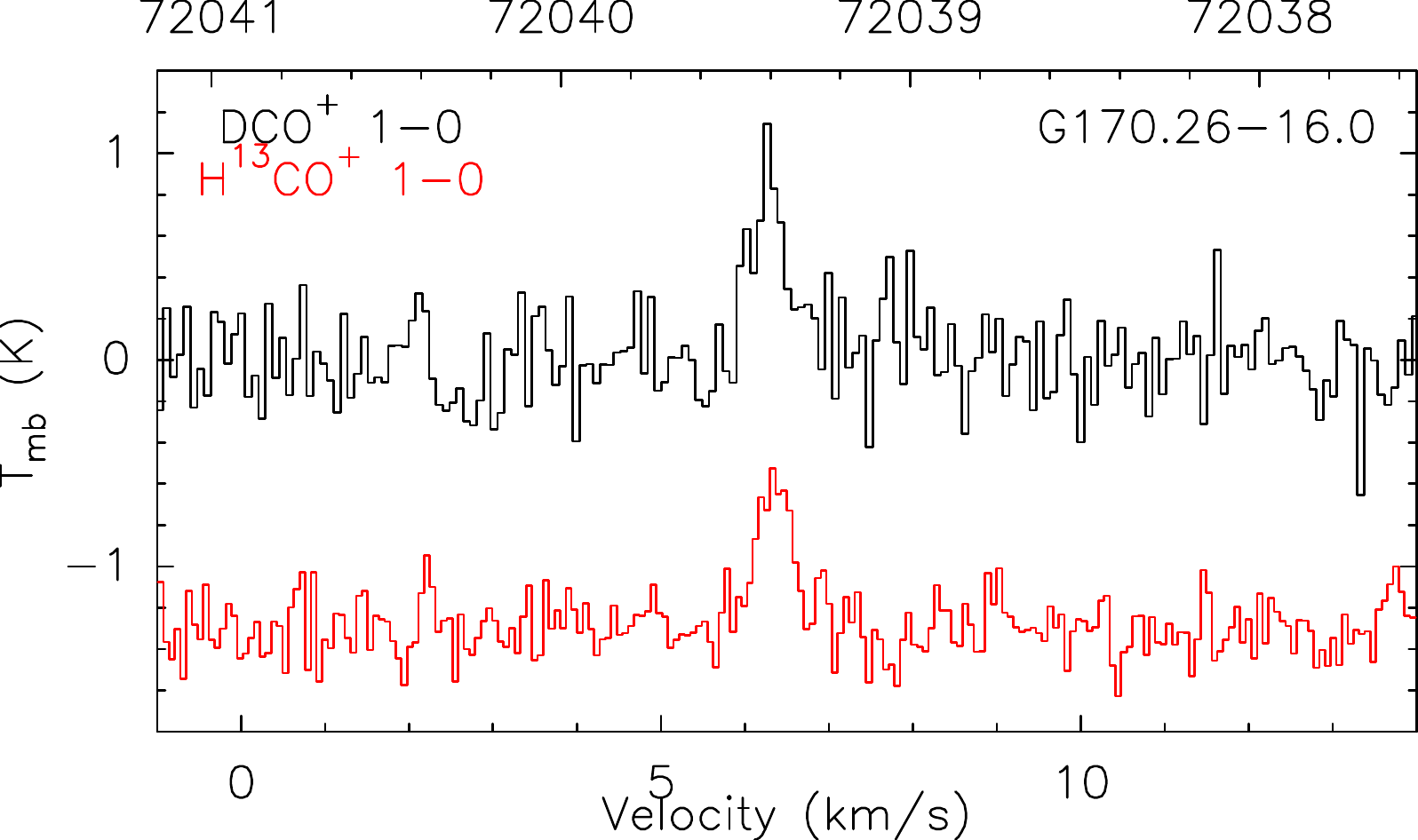}
\includegraphics[width=0.3\columnwidth]{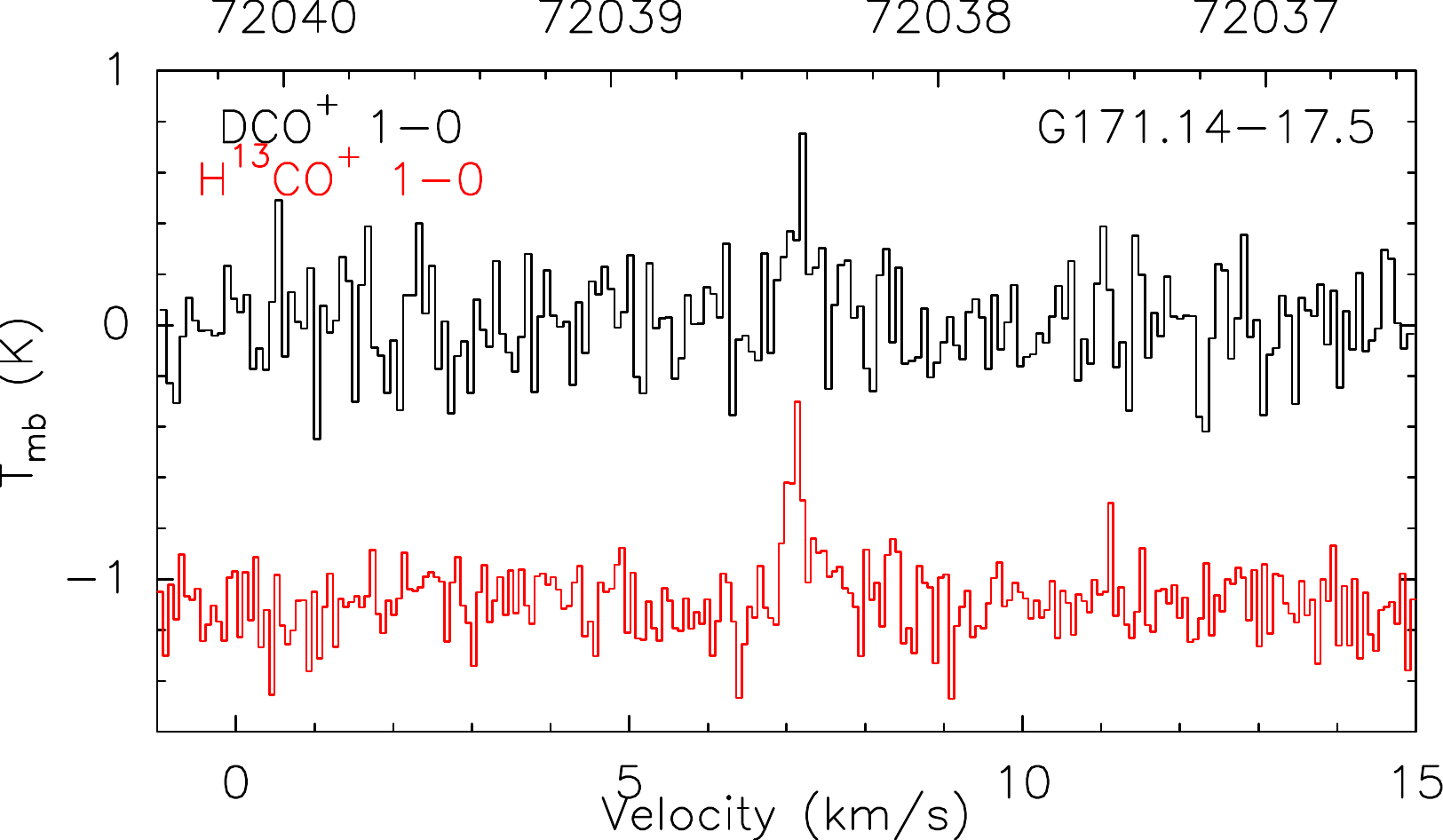}
\includegraphics[width=0.3\columnwidth]{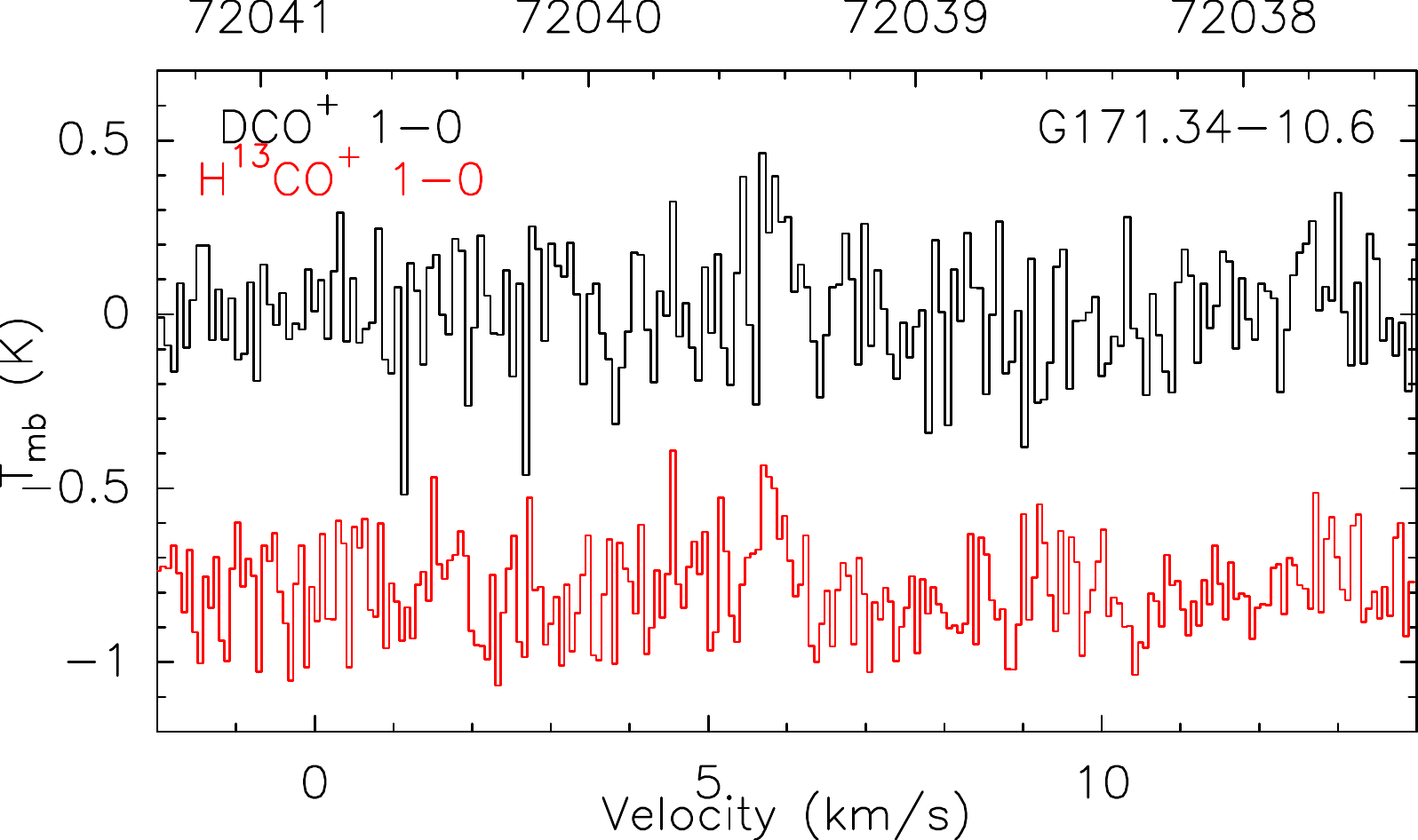}
\includegraphics[width=0.3\columnwidth]{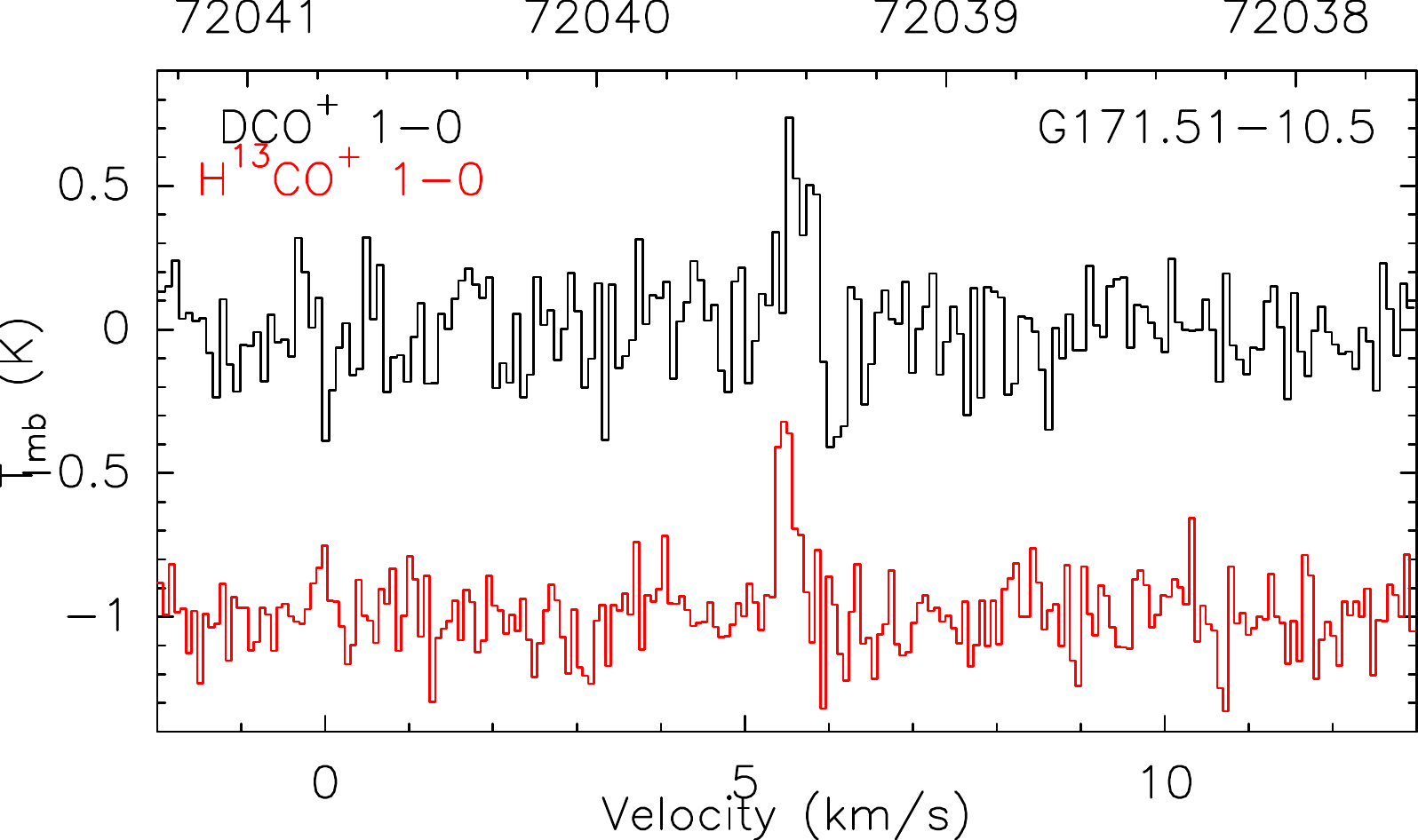}
\caption{Continued.\centering}
\end{figure}
\addtocounter{figure}{-1}
\begin{figure}
\centering
\includegraphics[width=0.3\columnwidth]{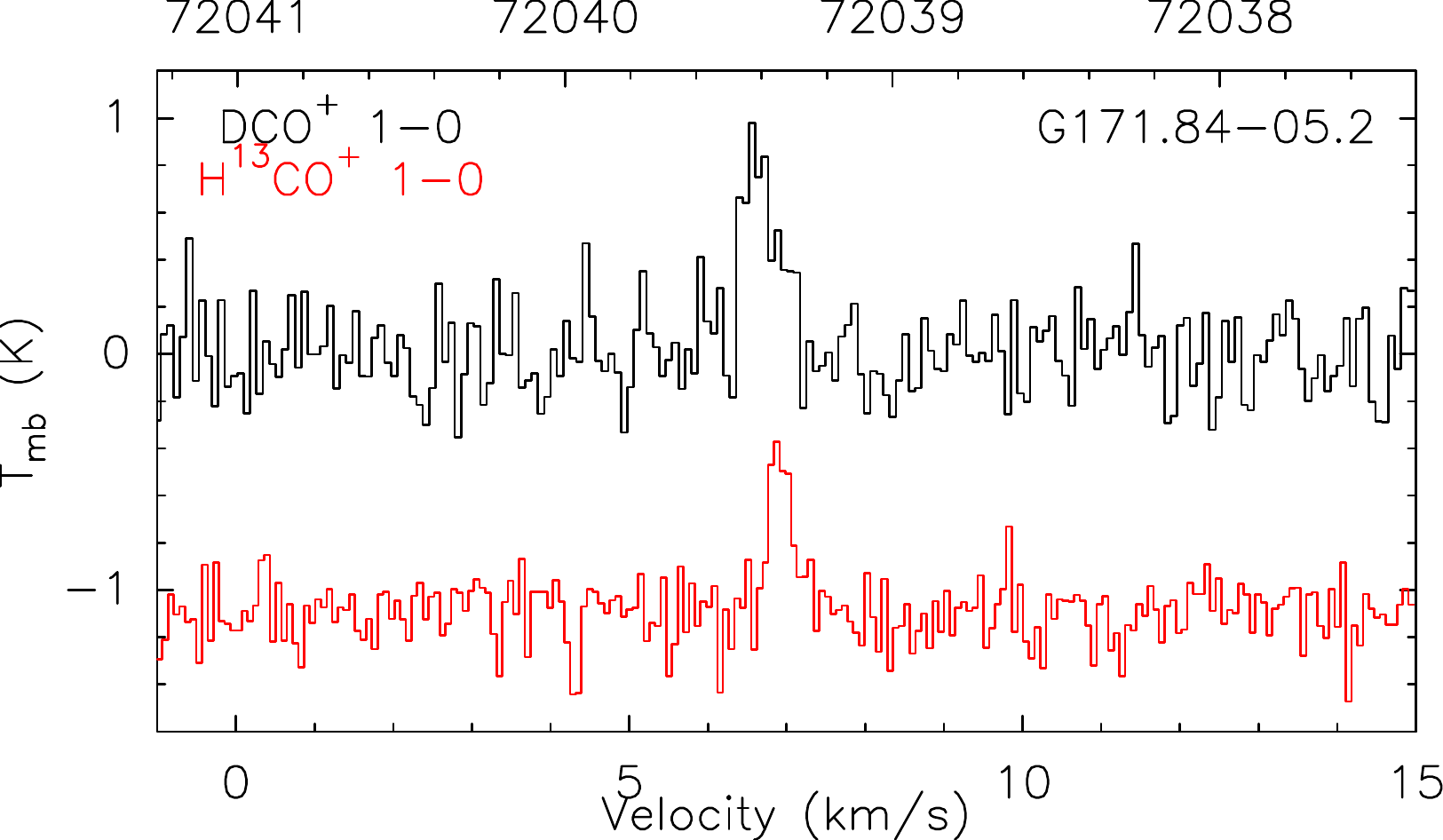}
\includegraphics[width=0.3\columnwidth]{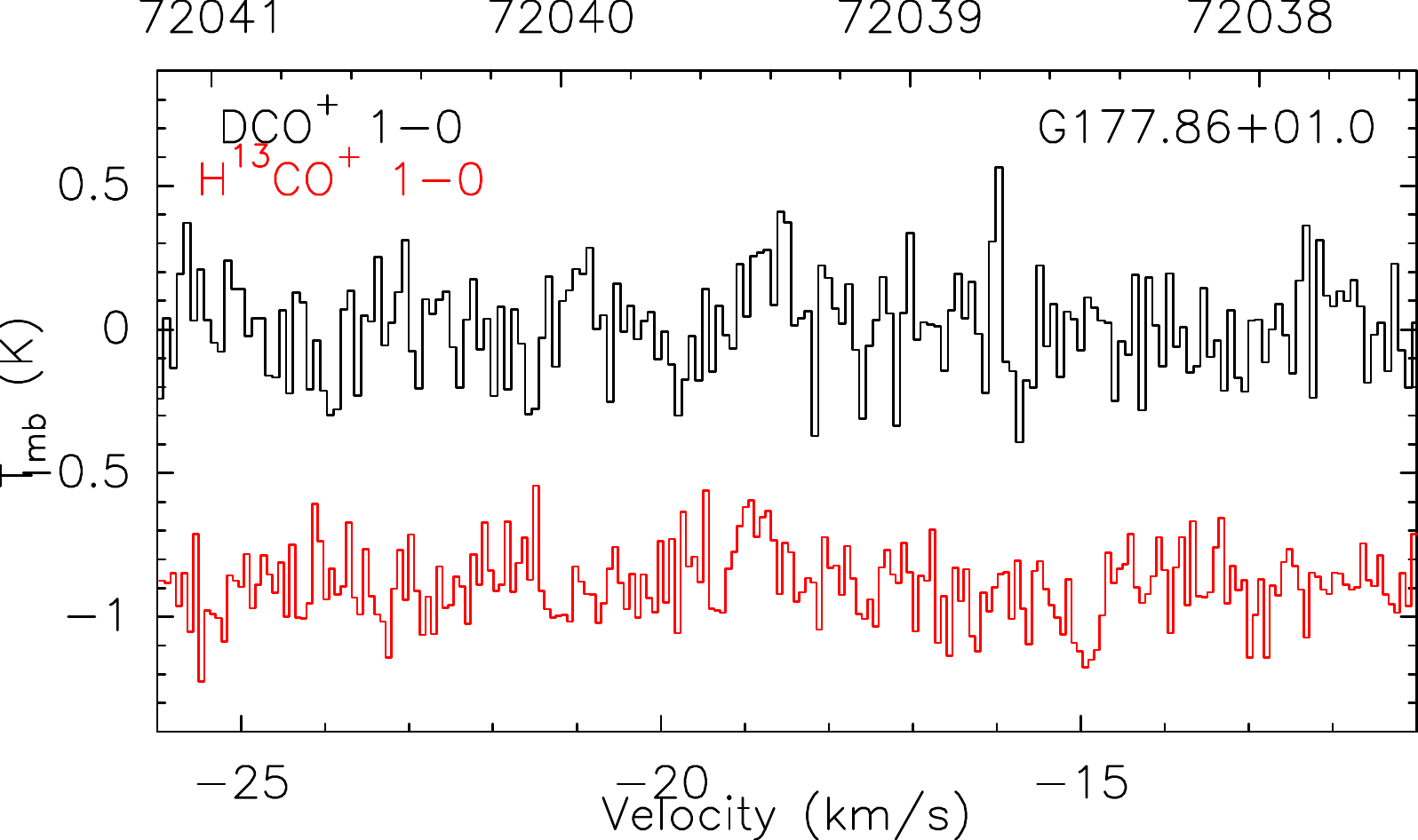}
\includegraphics[width=0.3\columnwidth]{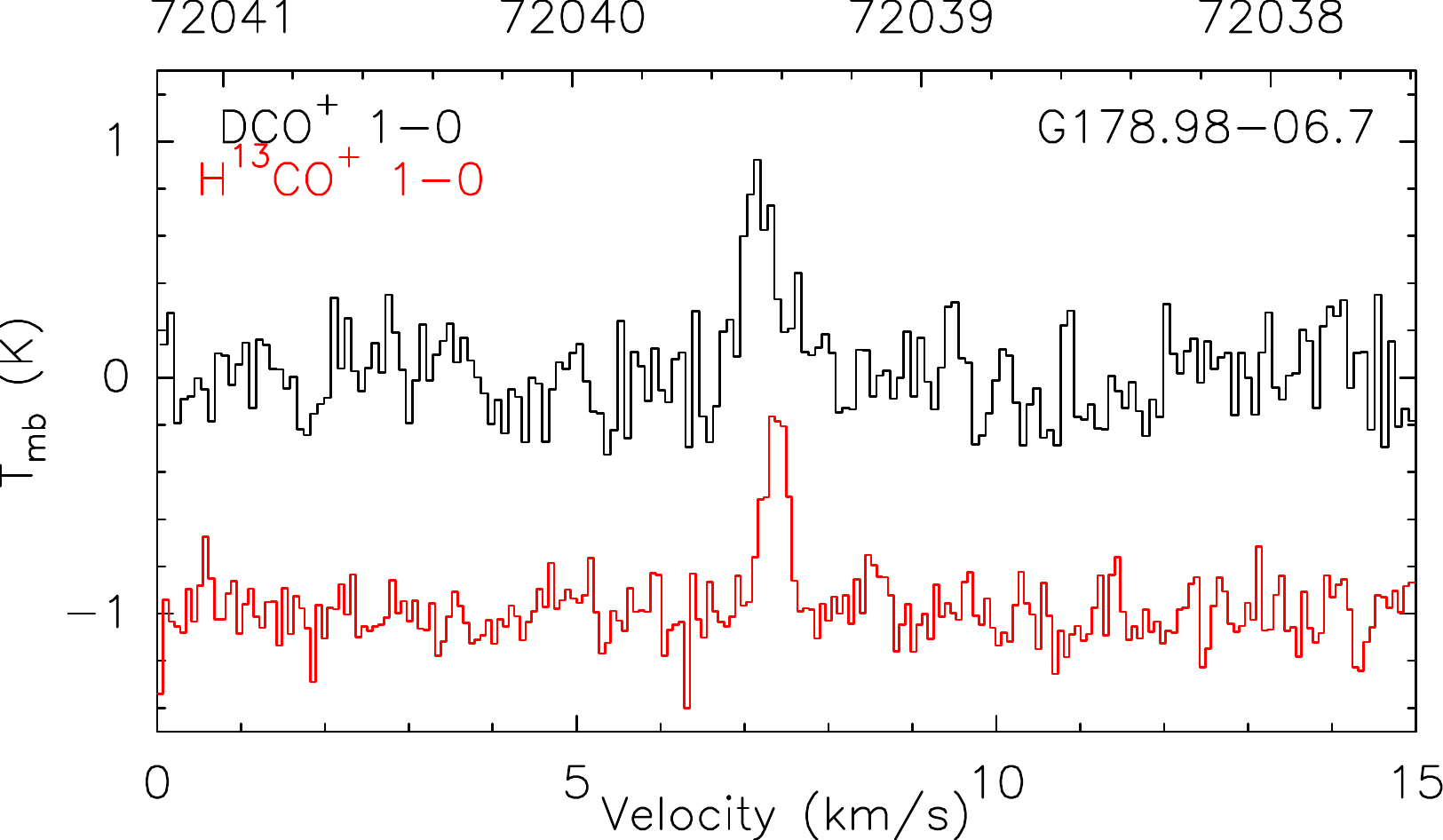}
\includegraphics[width=0.3\columnwidth]{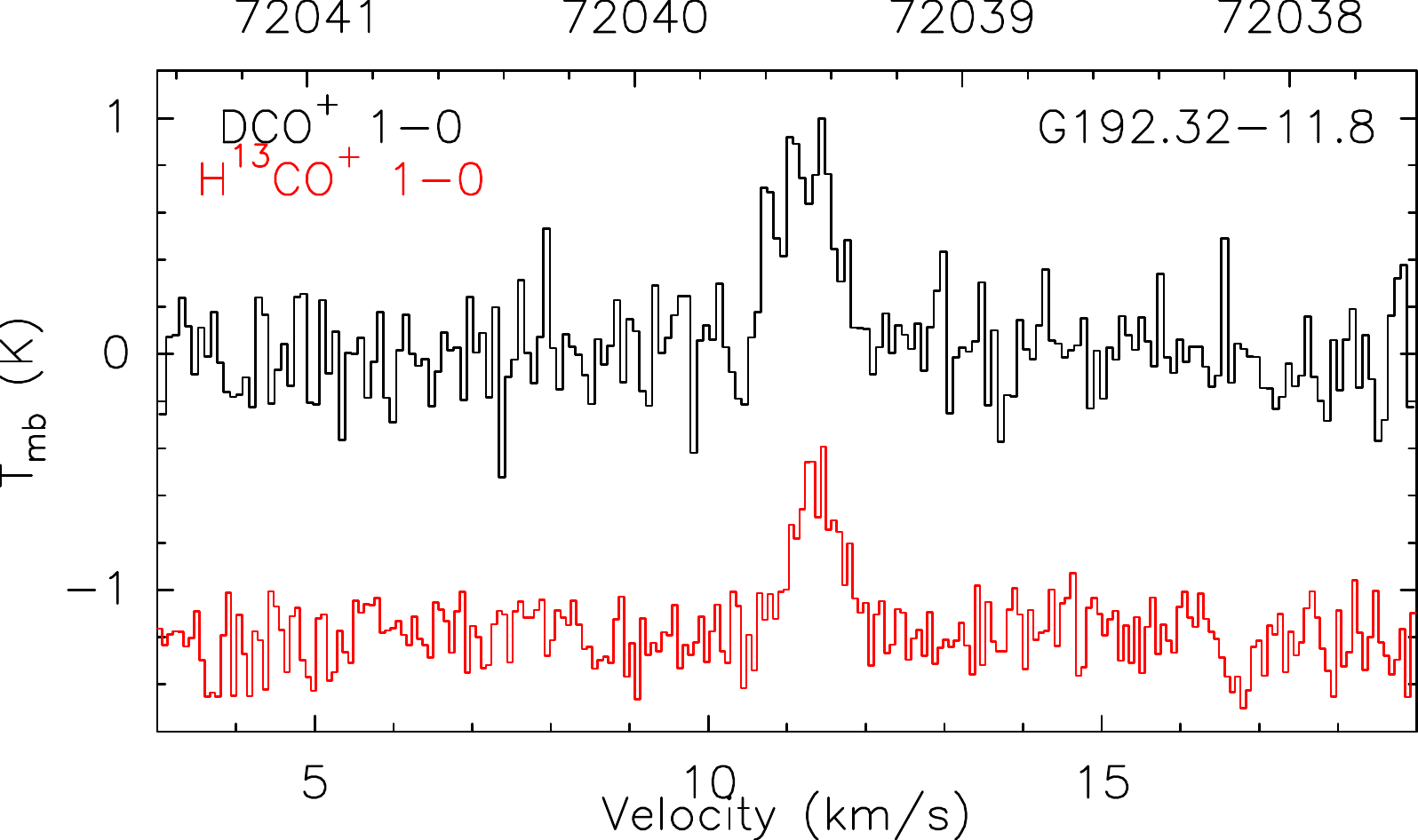}
\includegraphics[width=0.3\columnwidth]{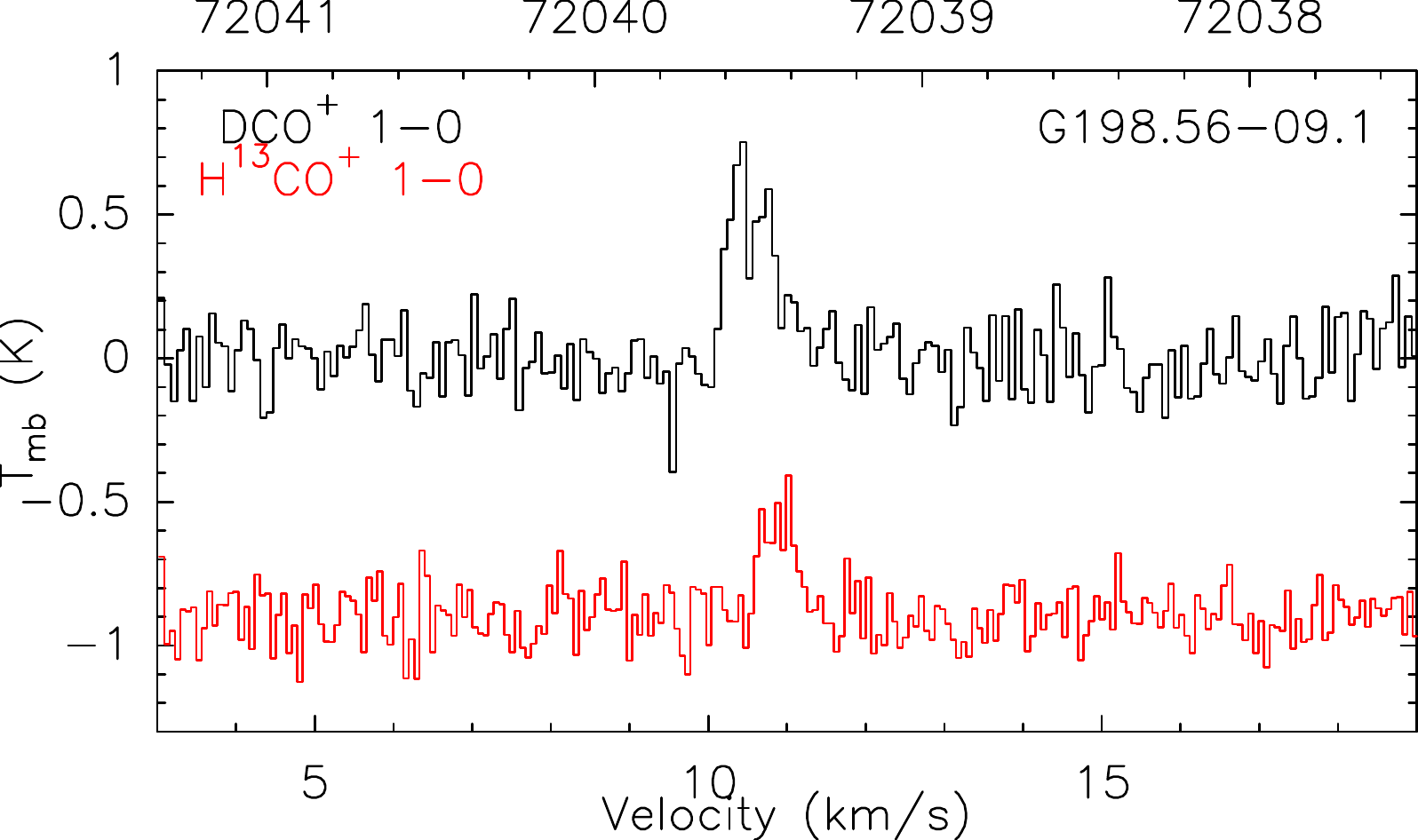}
\includegraphics[width=0.3\columnwidth]{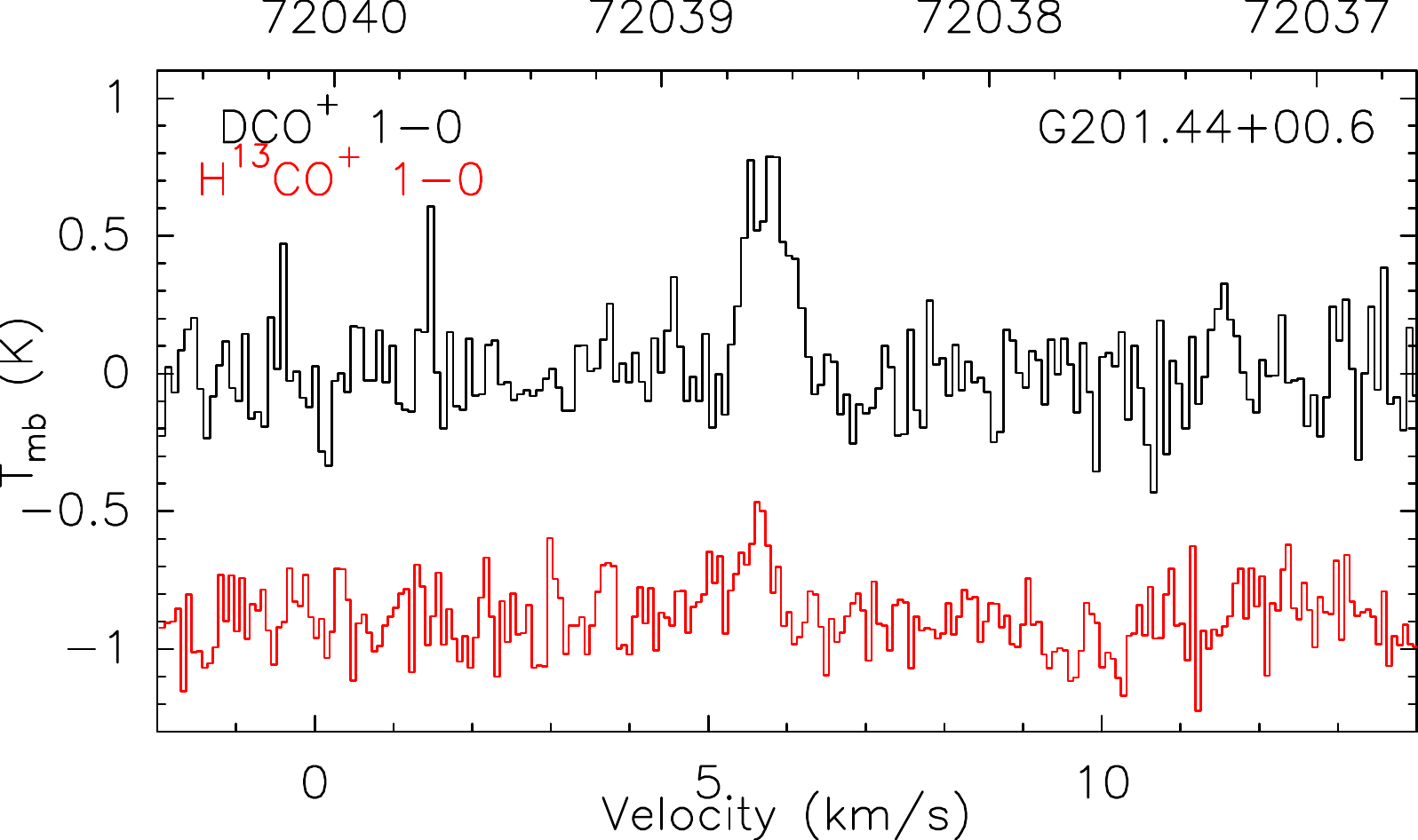}
\includegraphics[width=0.3\columnwidth]{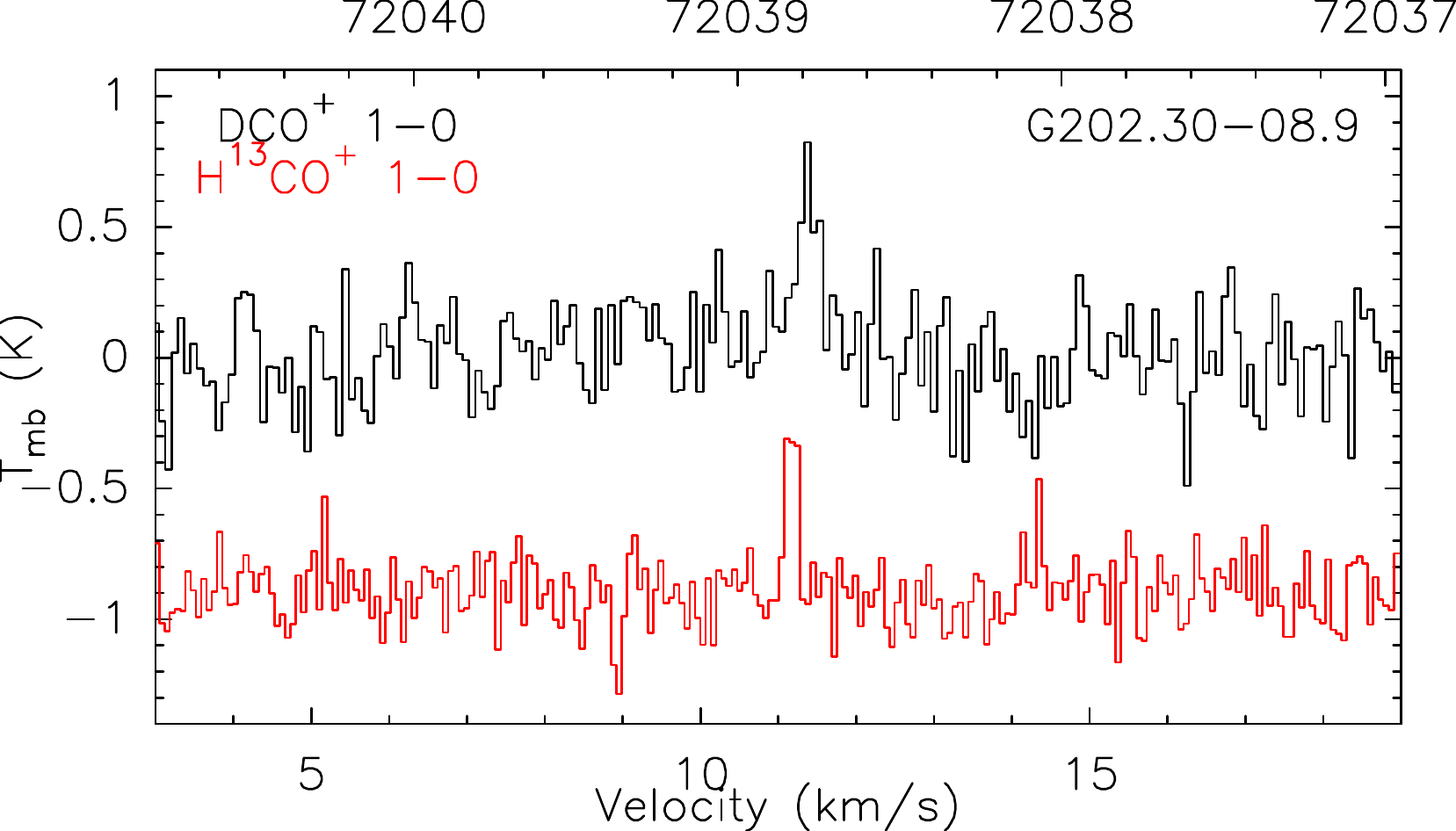}
\includegraphics[width=0.3\columnwidth]{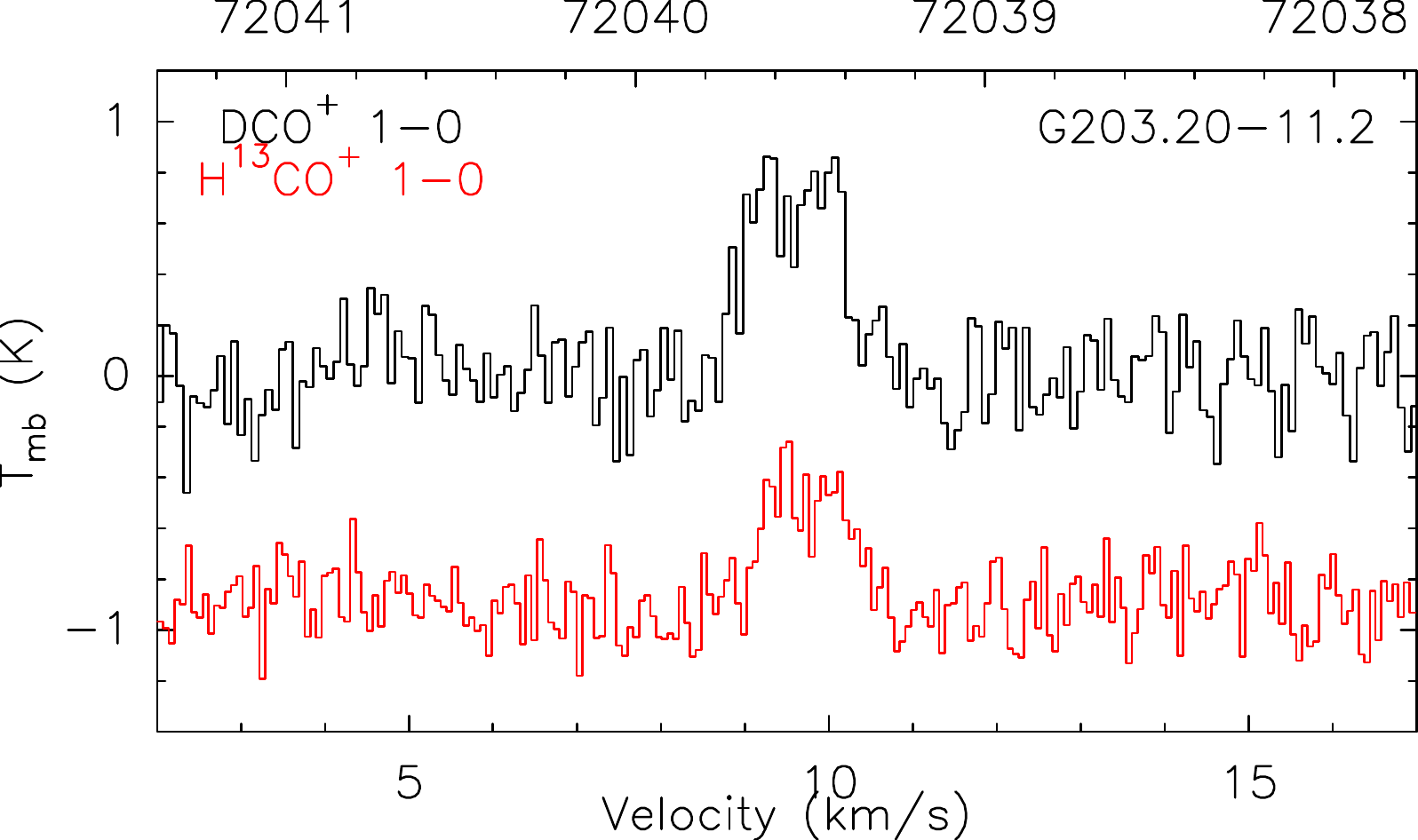}
\includegraphics[width=0.3\columnwidth]{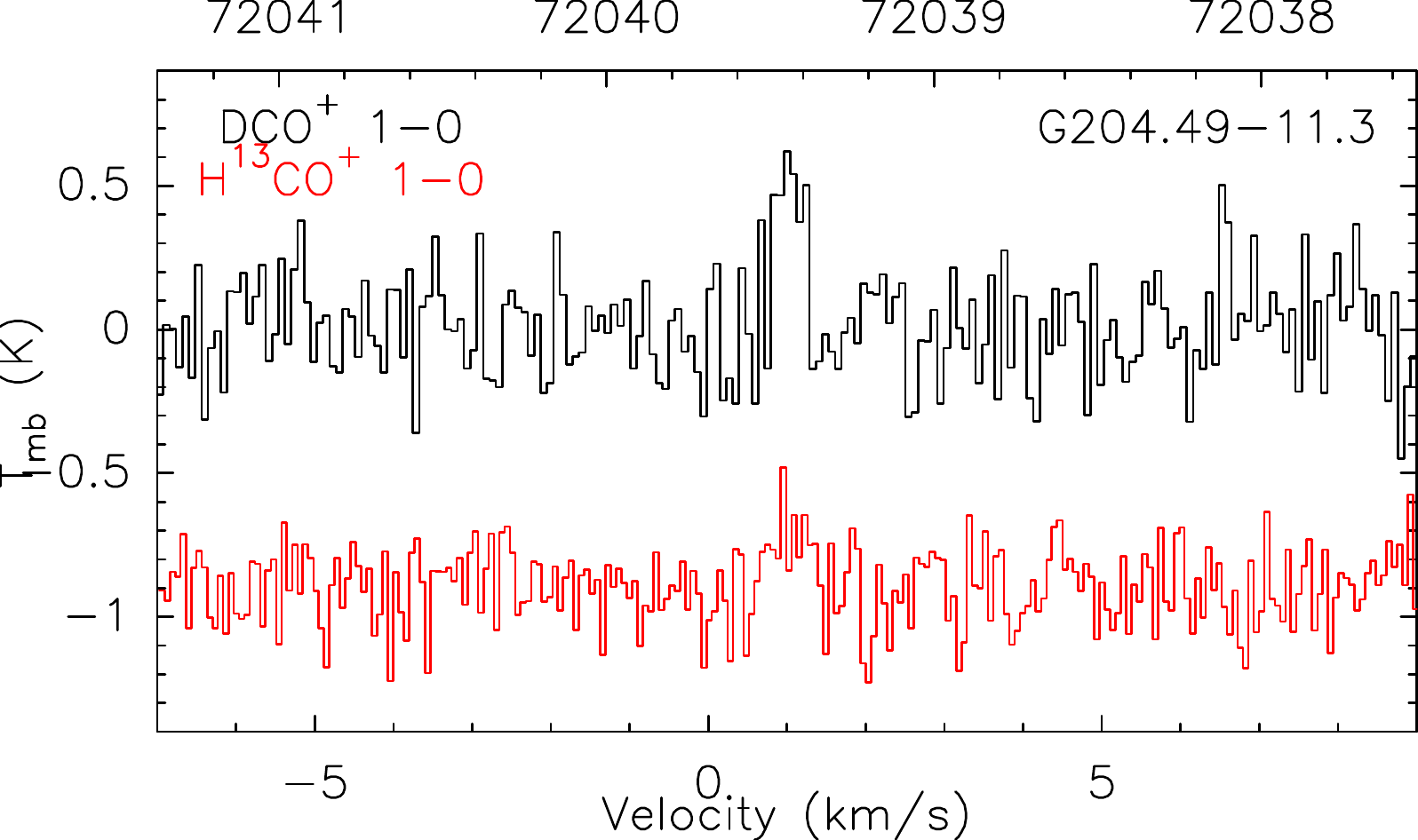}
\includegraphics[width=0.3\columnwidth]{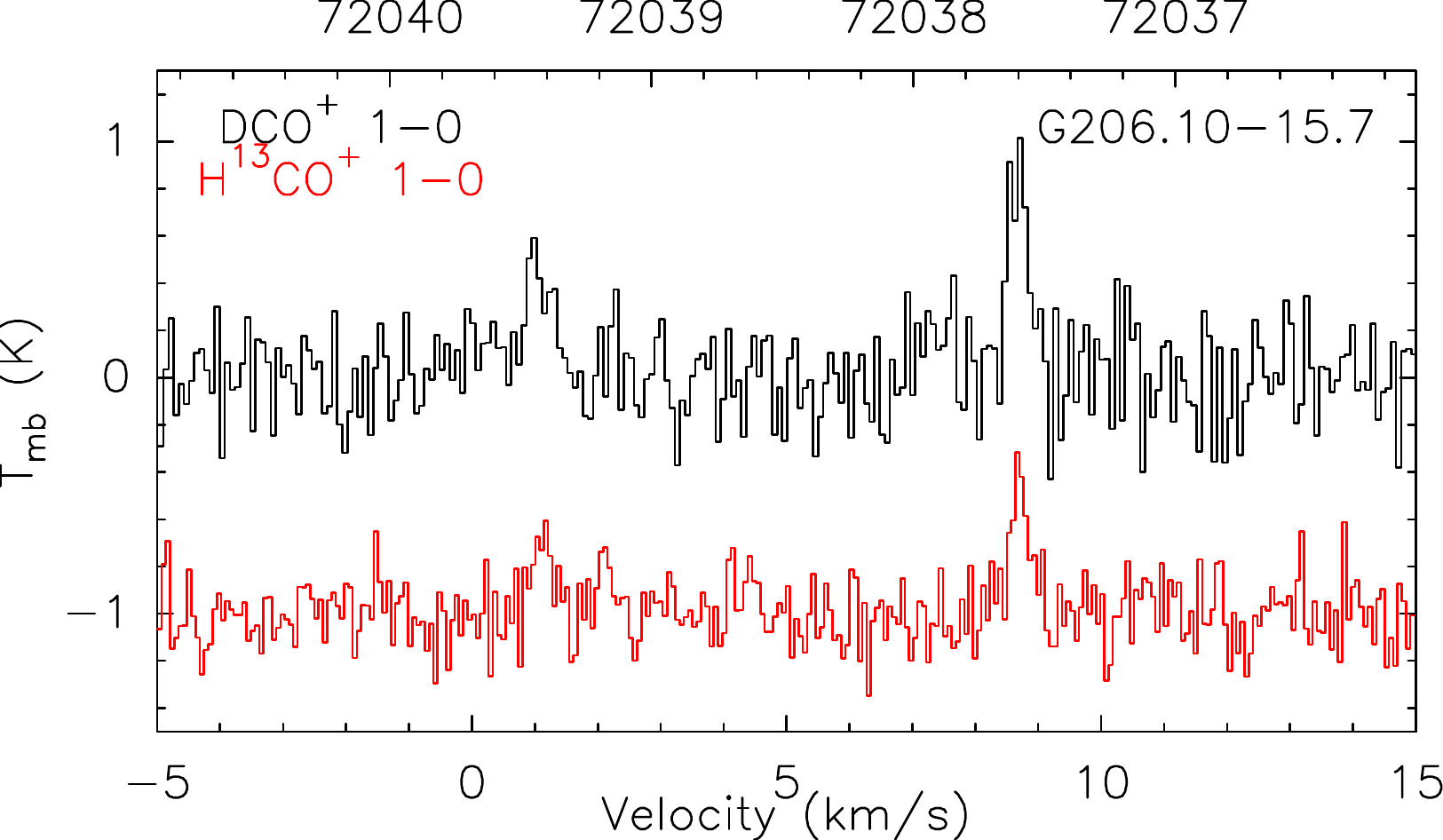}
\includegraphics[width=0.3\columnwidth]{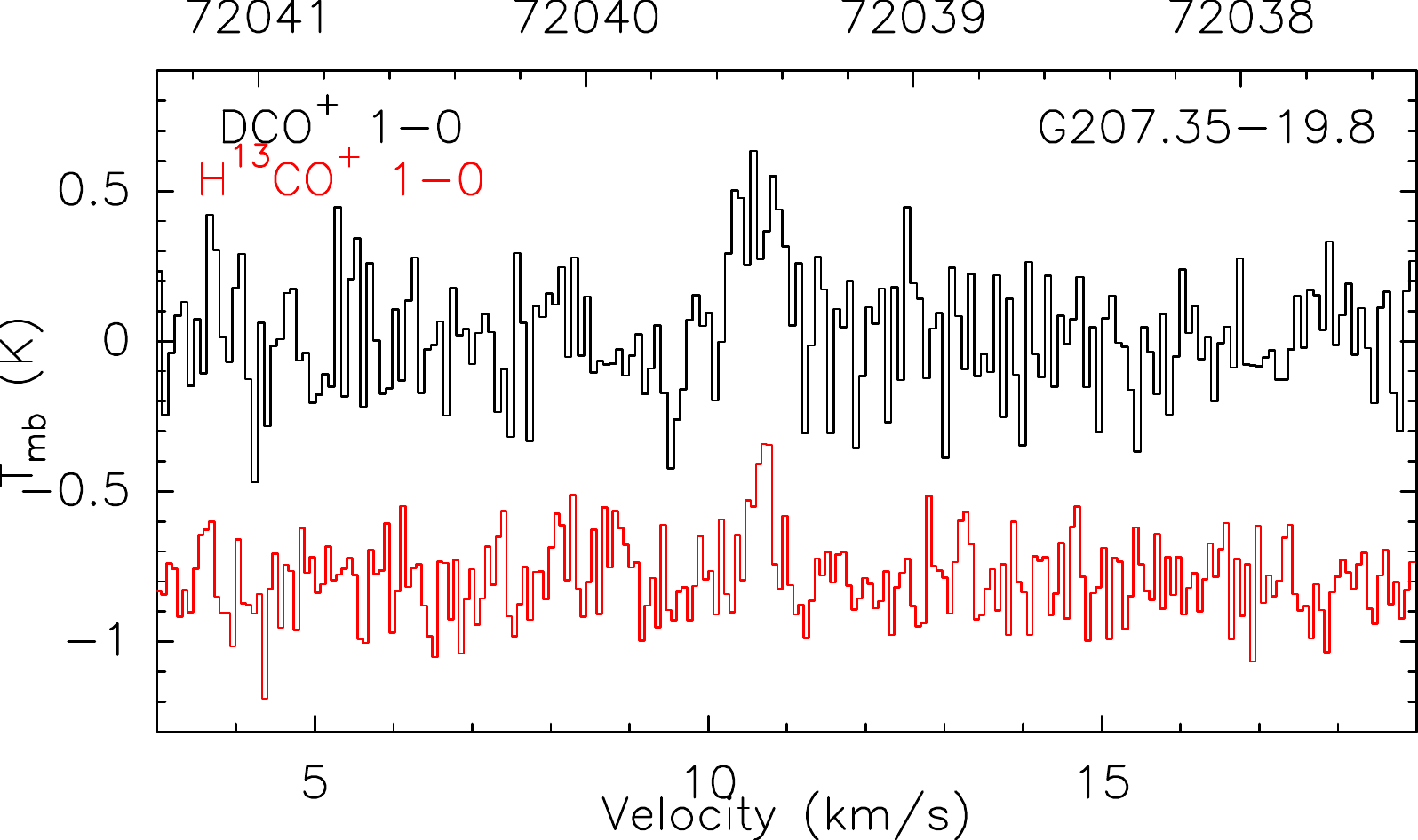}
\includegraphics[width=0.3\columnwidth]{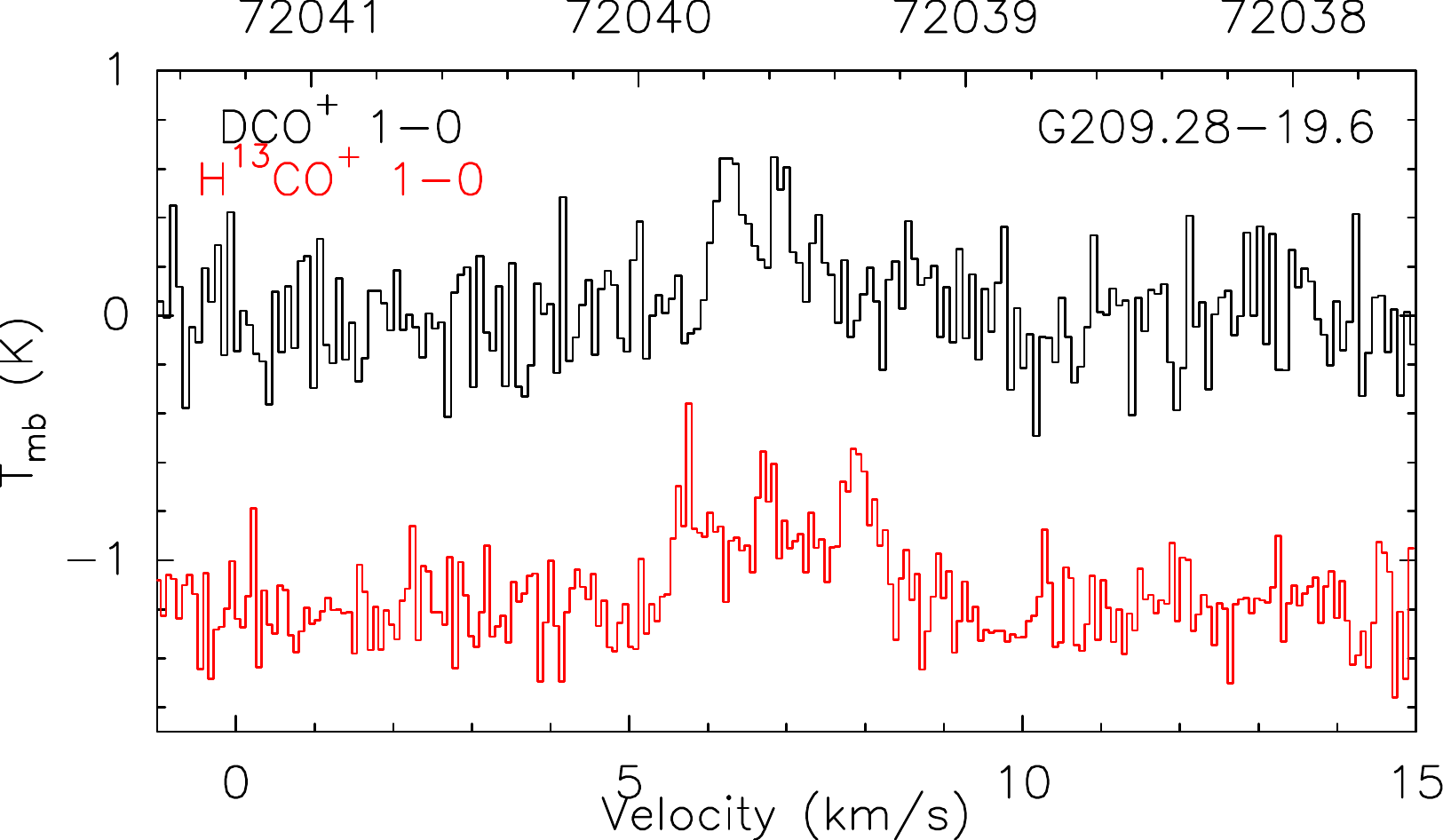}
\includegraphics[width=0.3\columnwidth]{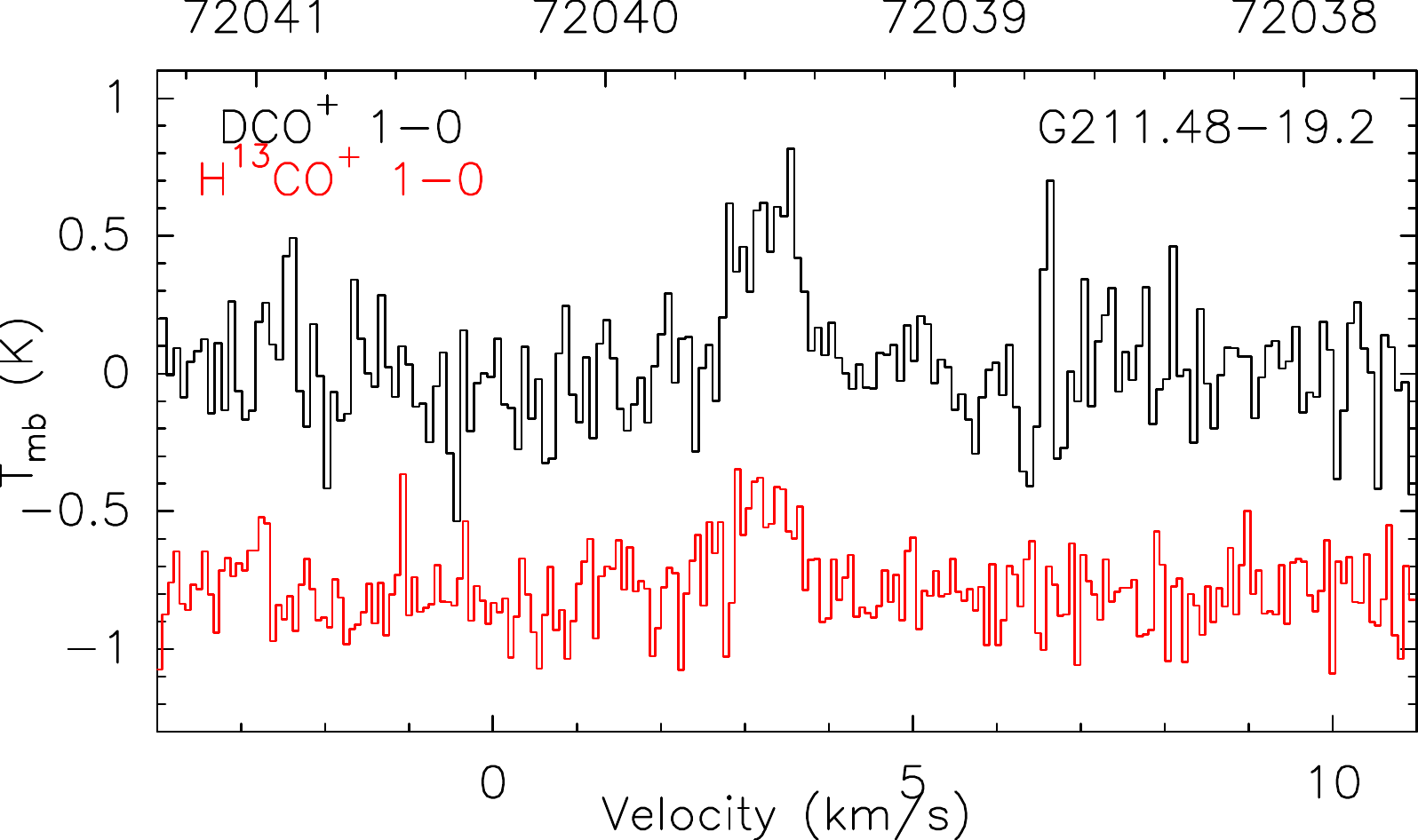}
\includegraphics[width=0.3\columnwidth]{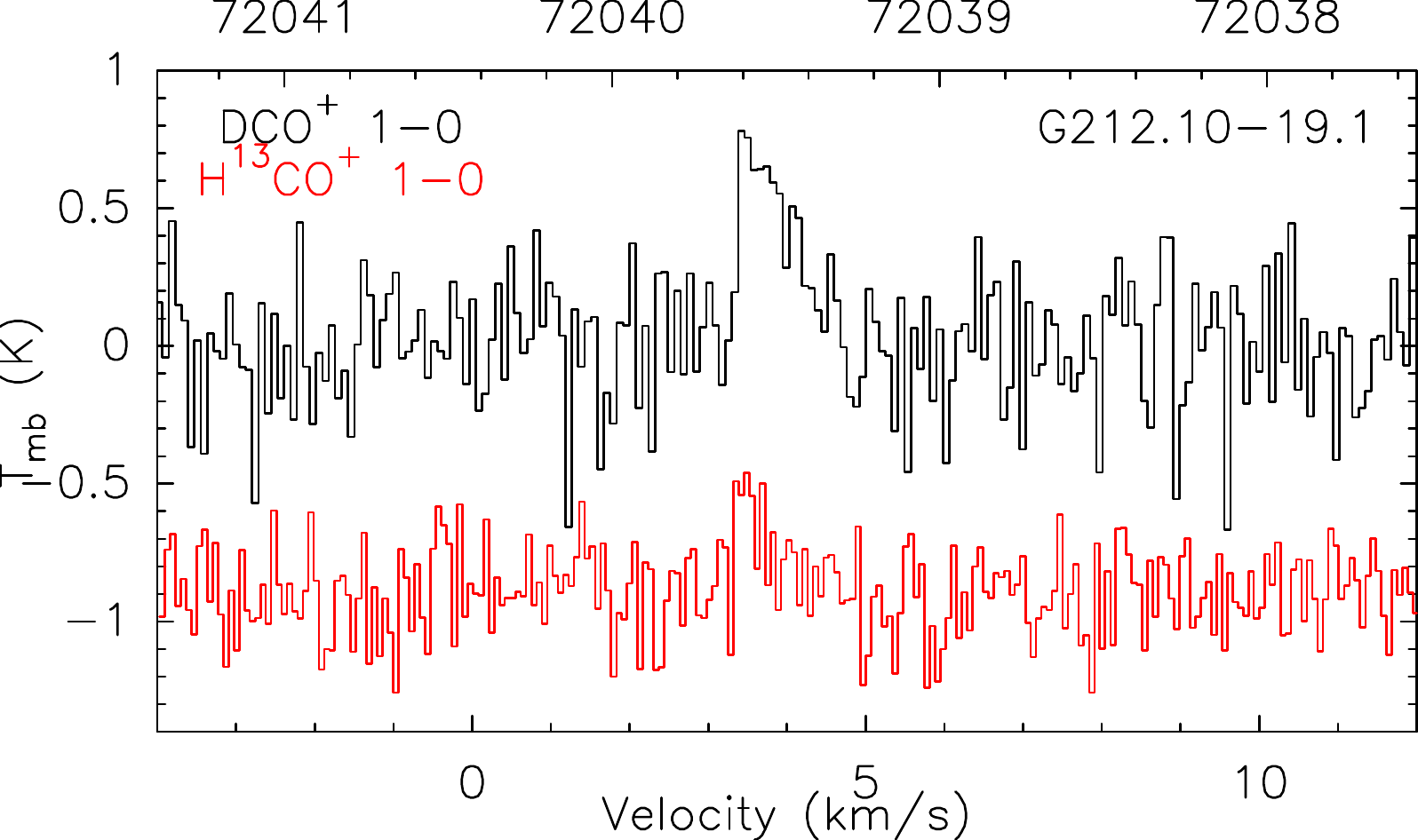}
\includegraphics[width=0.3\columnwidth]{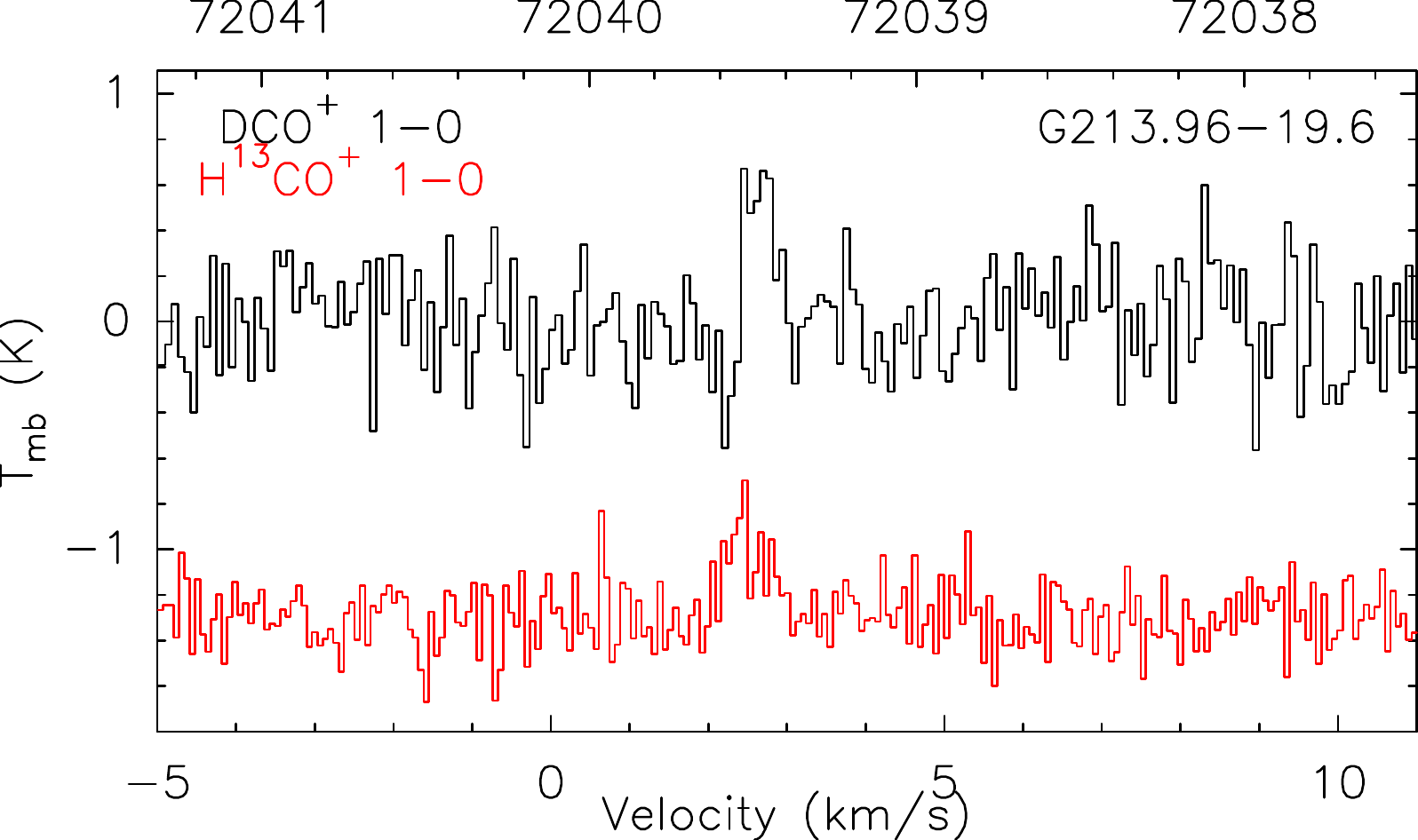}
\includegraphics[width=0.3\columnwidth]{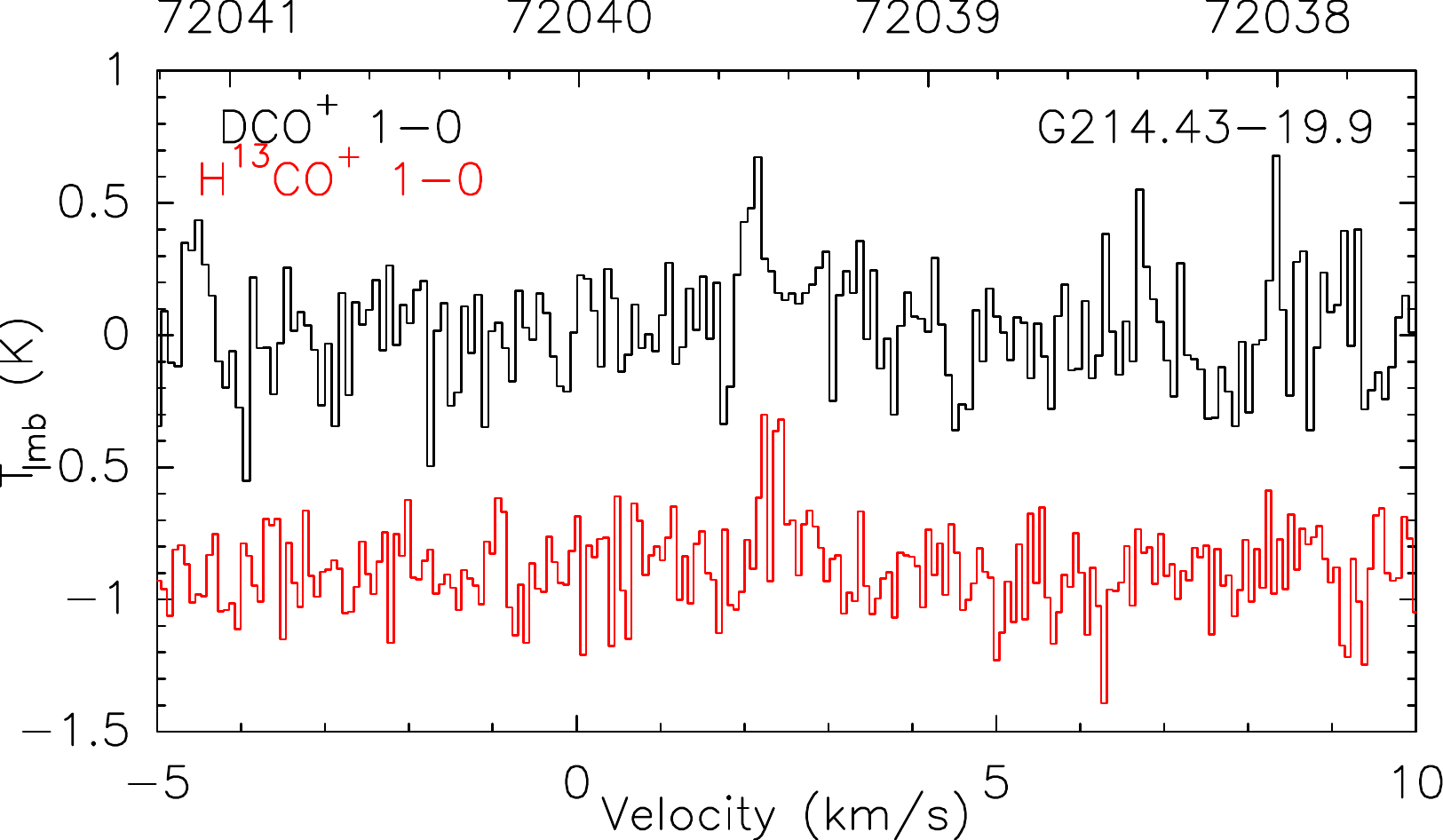}
\caption{Continued.\centering}
\end{figure}
\begin{figure}
\centering
\includegraphics[width=0.3\columnwidth]{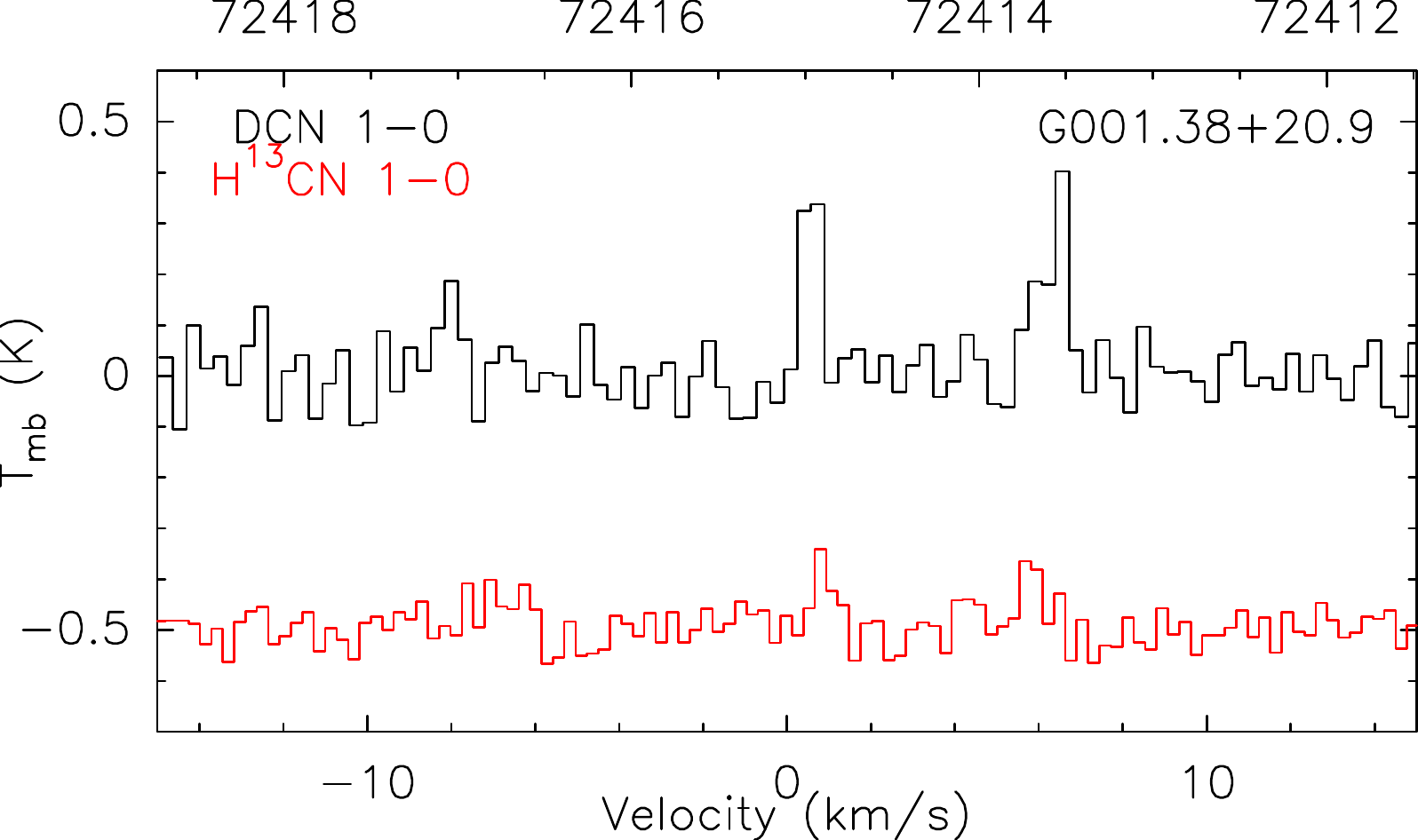}
\includegraphics[width=0.3\columnwidth]{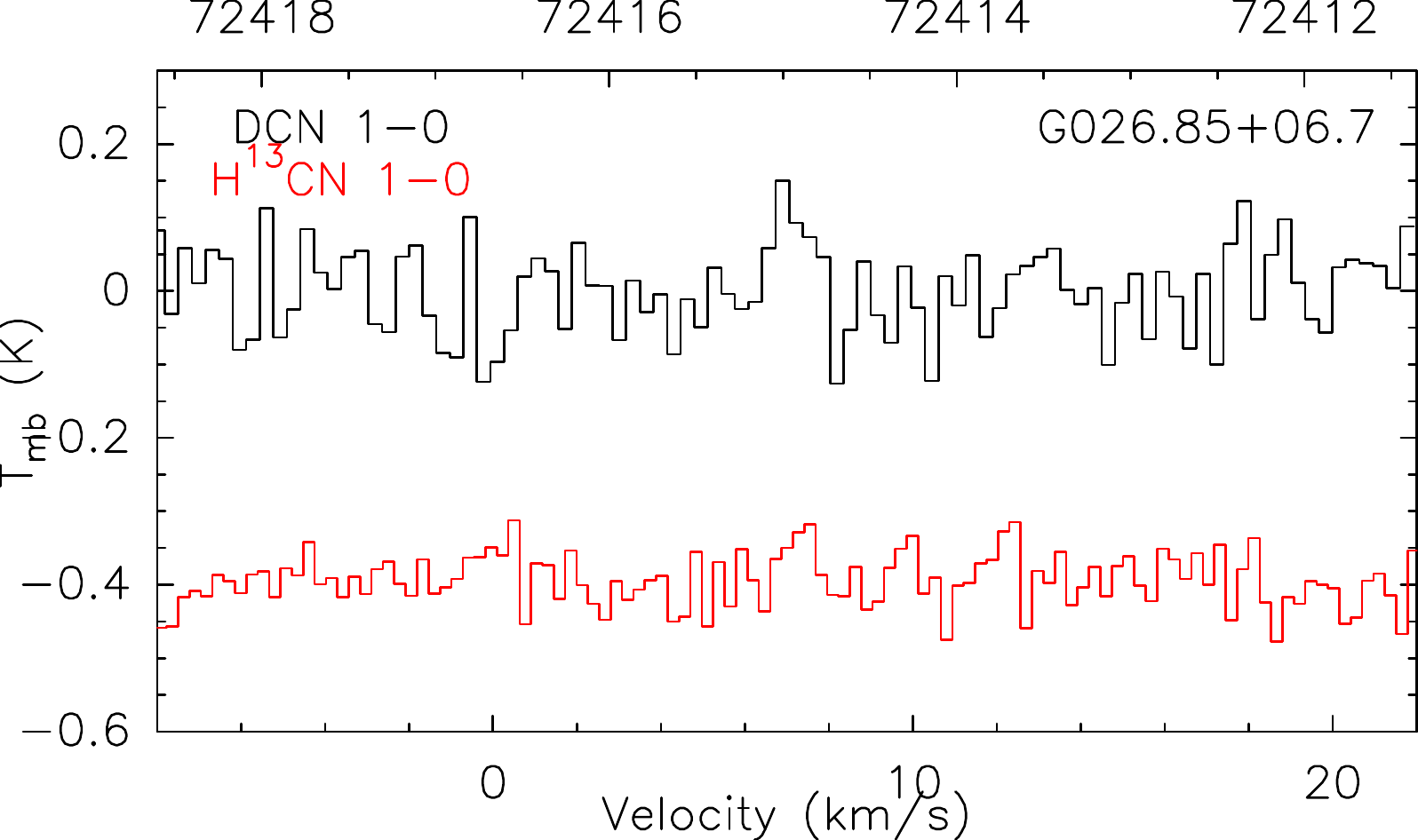}
\includegraphics[width=0.3\columnwidth]{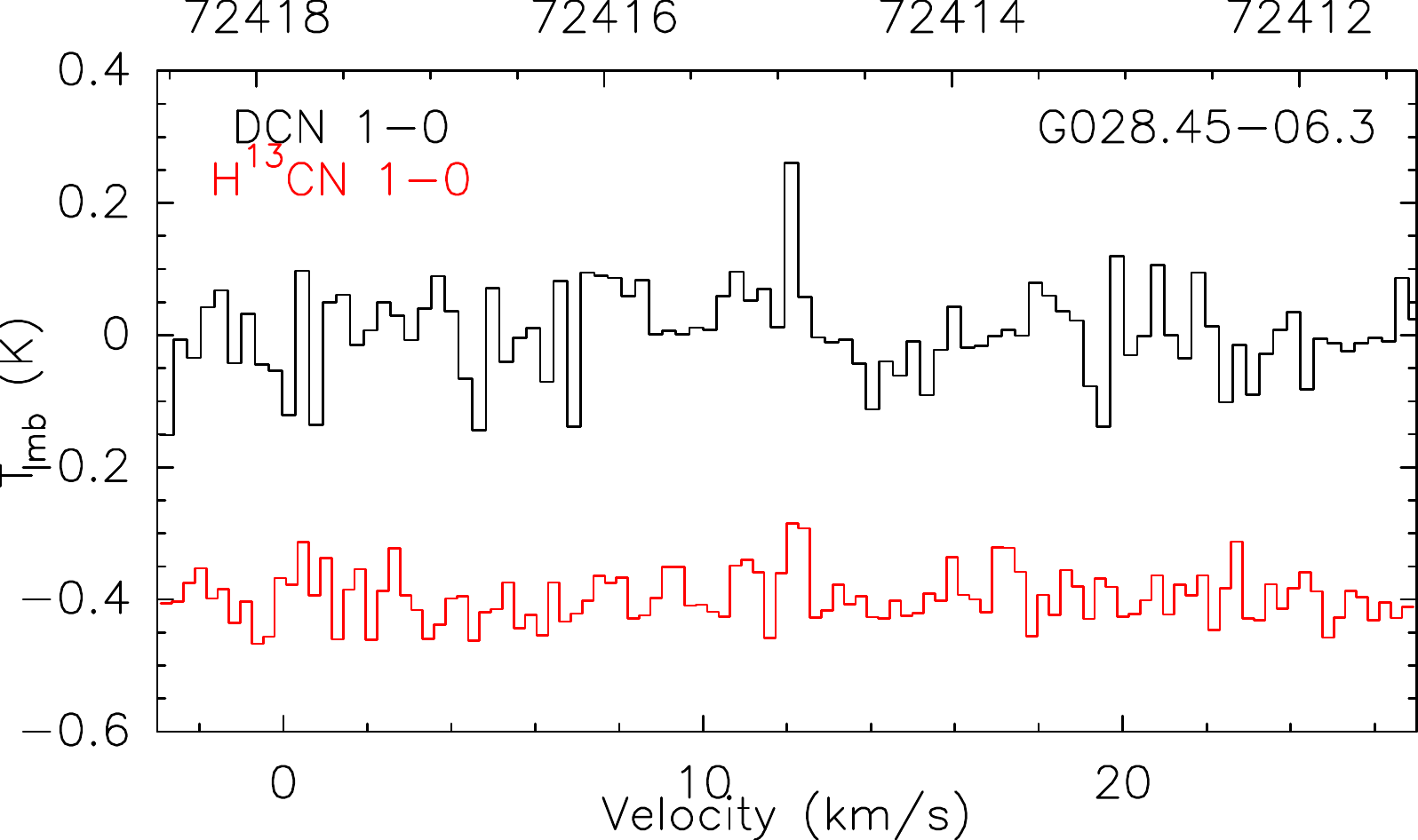}
\includegraphics[width=0.3\columnwidth]{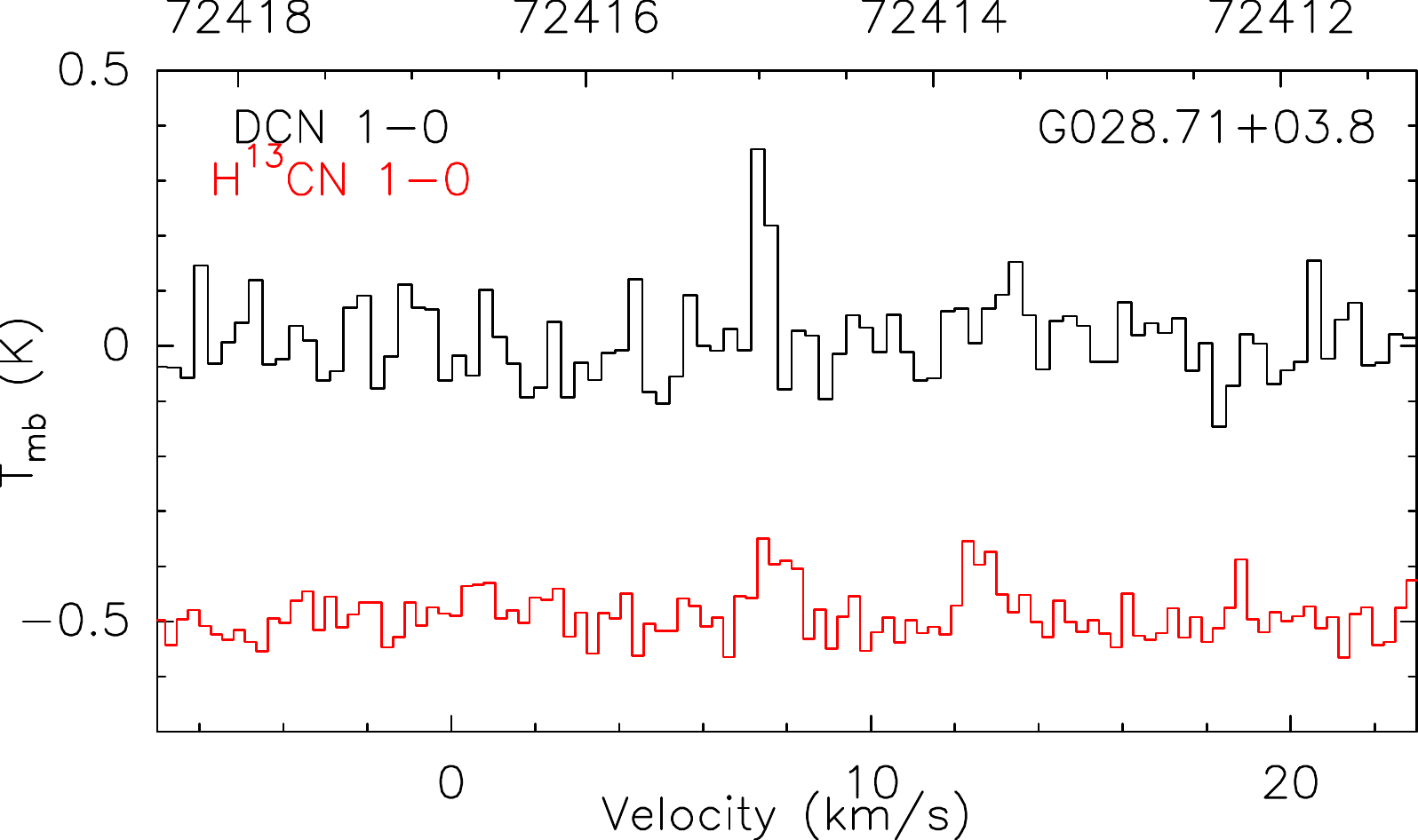}
\caption{Line profiles of DCN and H$^{13}$CN 1-0 with the low velocity resolution mode (AROWS mode 3).\centering}
\label{DCNH13CNmode3_2}
\end{figure}
\begin{figure}
\centering
\includegraphics[width=0.3\columnwidth]{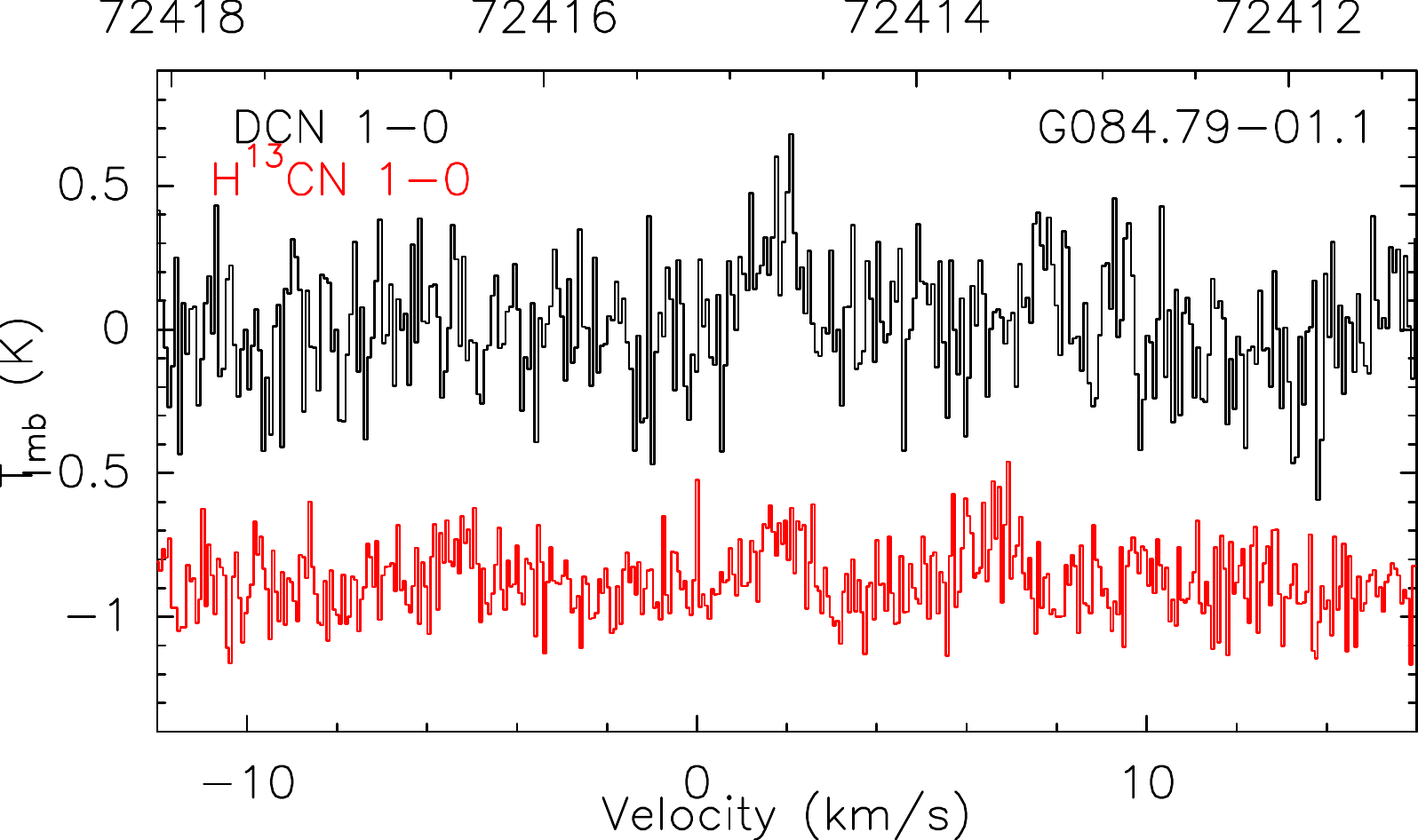}
\includegraphics[width=0.3\columnwidth]{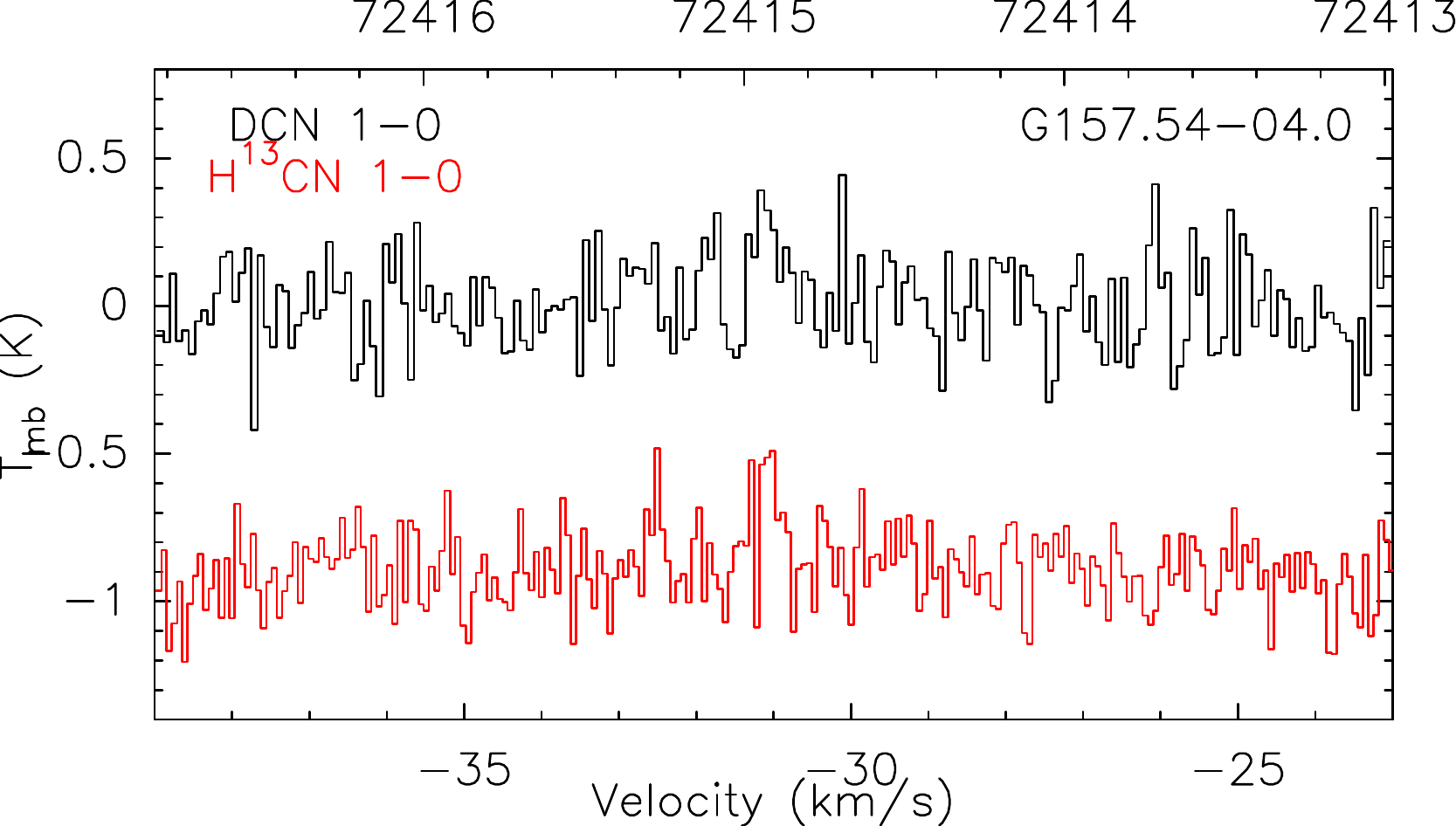}
\includegraphics[width=0.3\columnwidth]{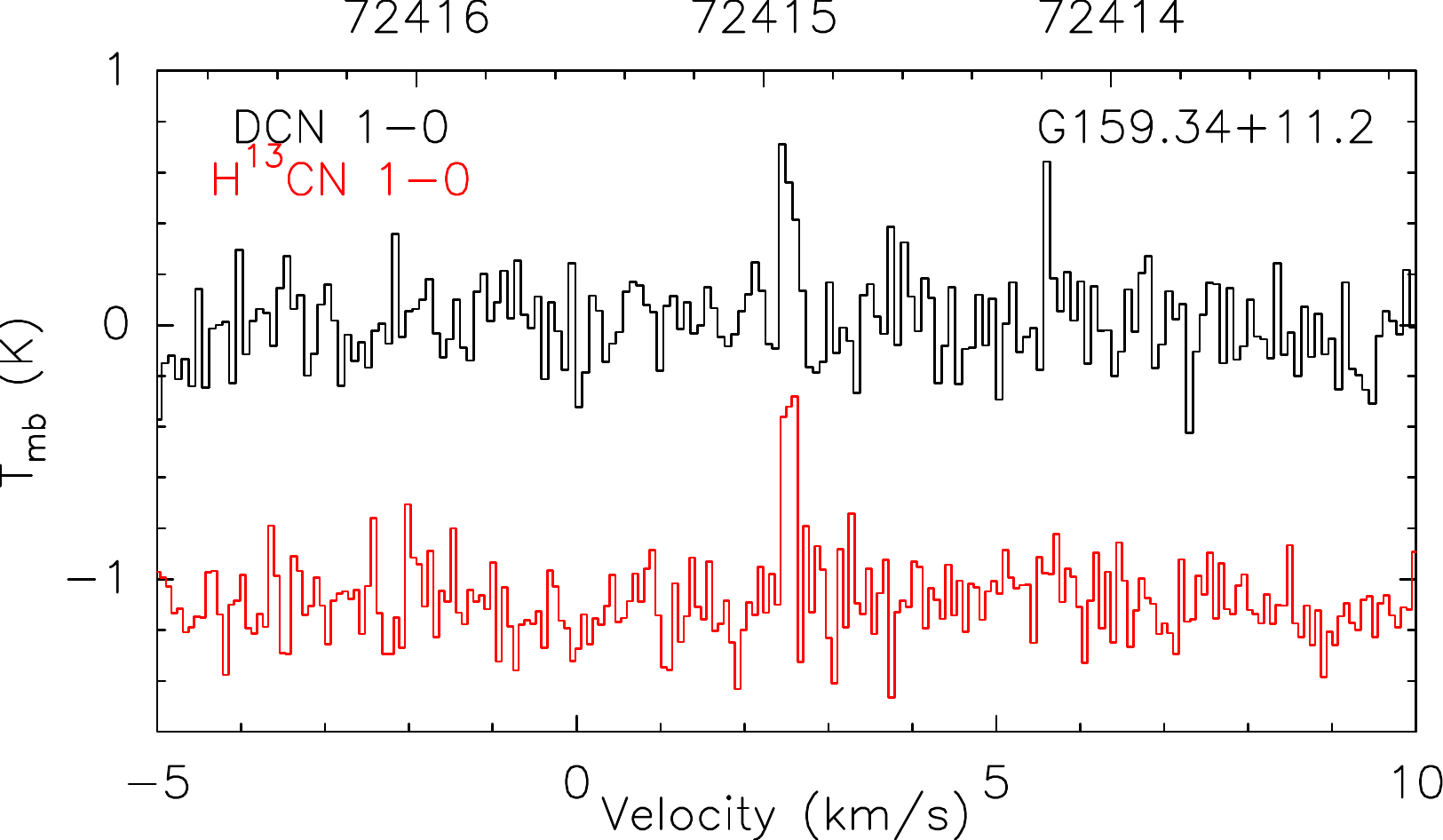}
\includegraphics[width=0.3\columnwidth]{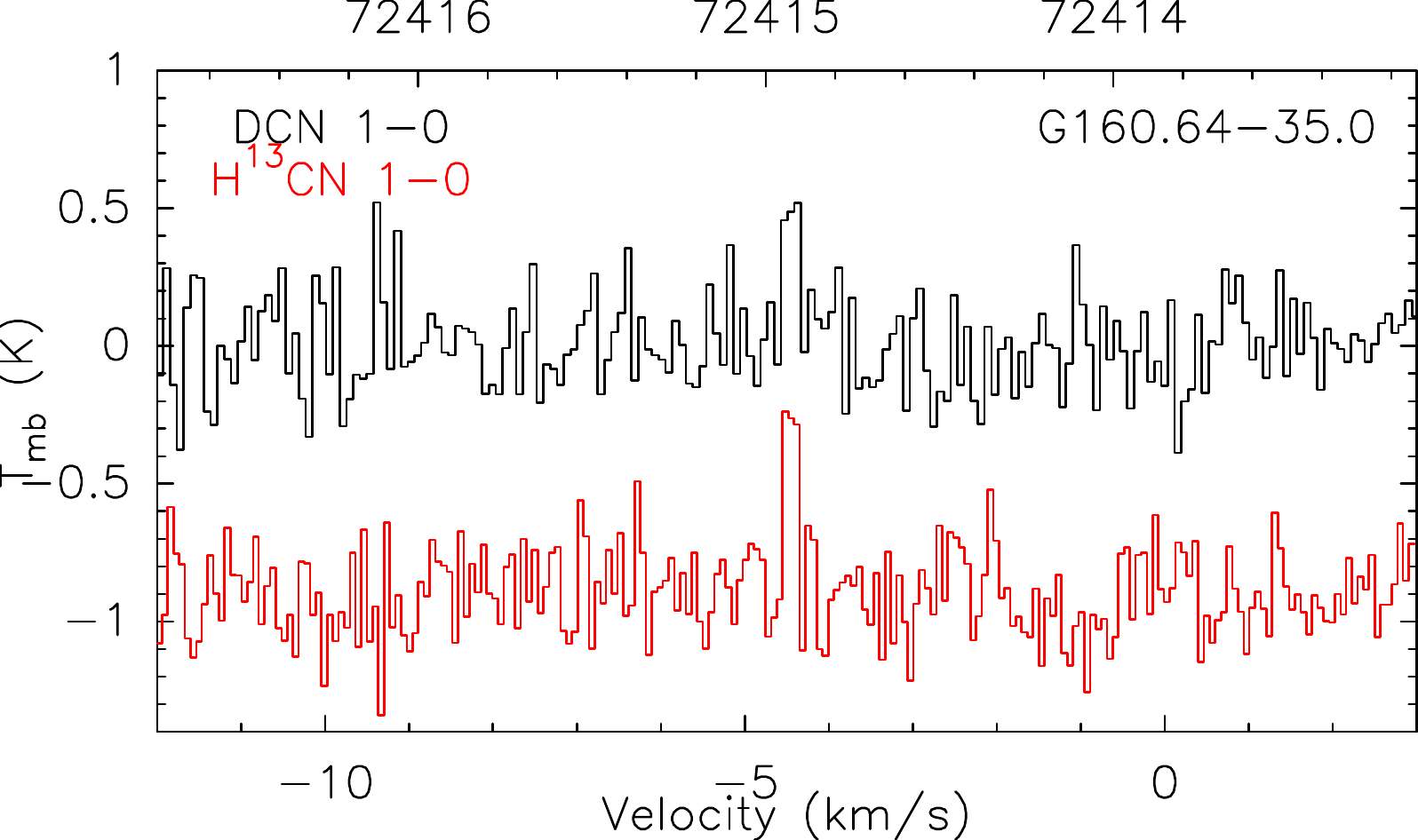}
\includegraphics[width=0.3\columnwidth]{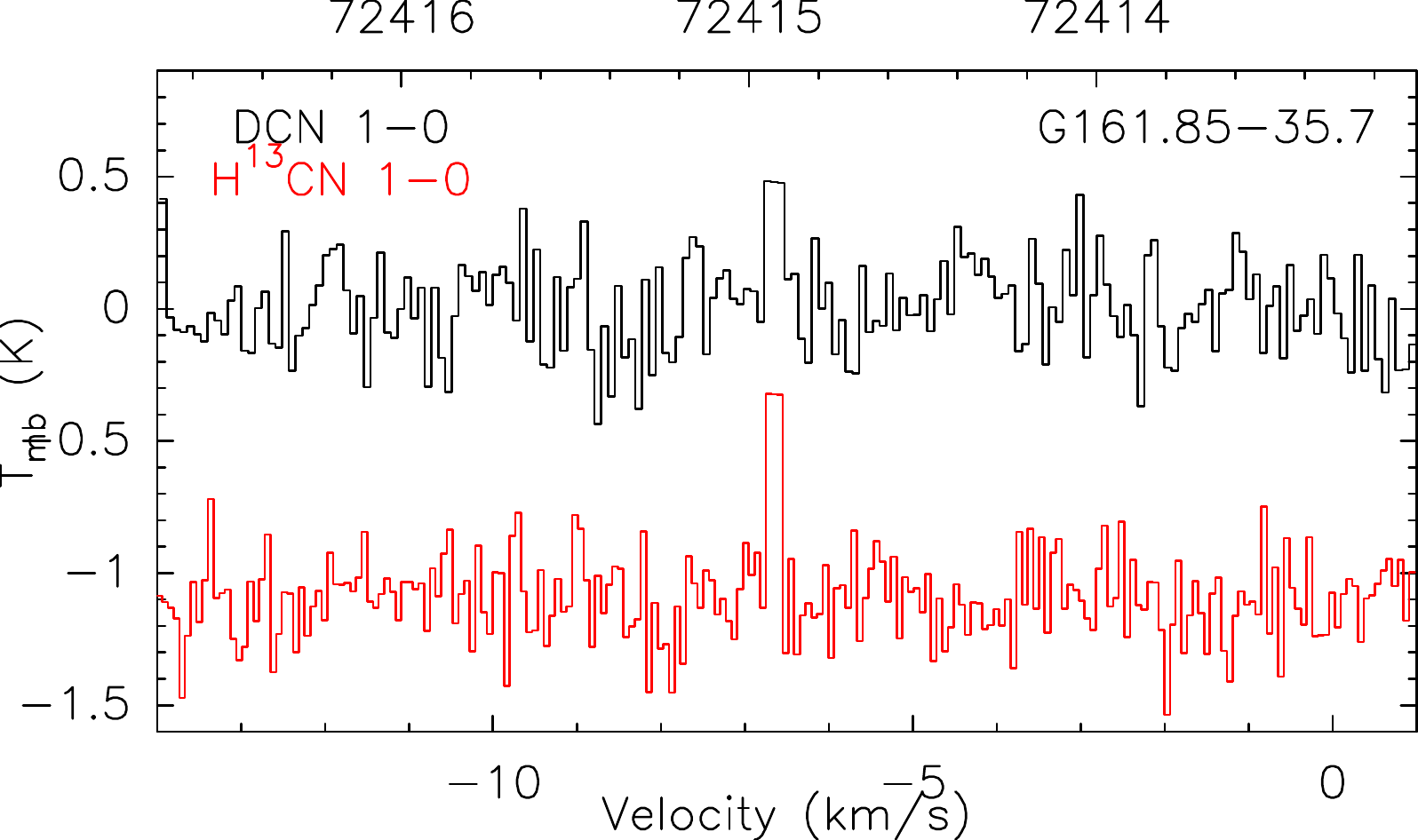}
\includegraphics[width=0.3\columnwidth]{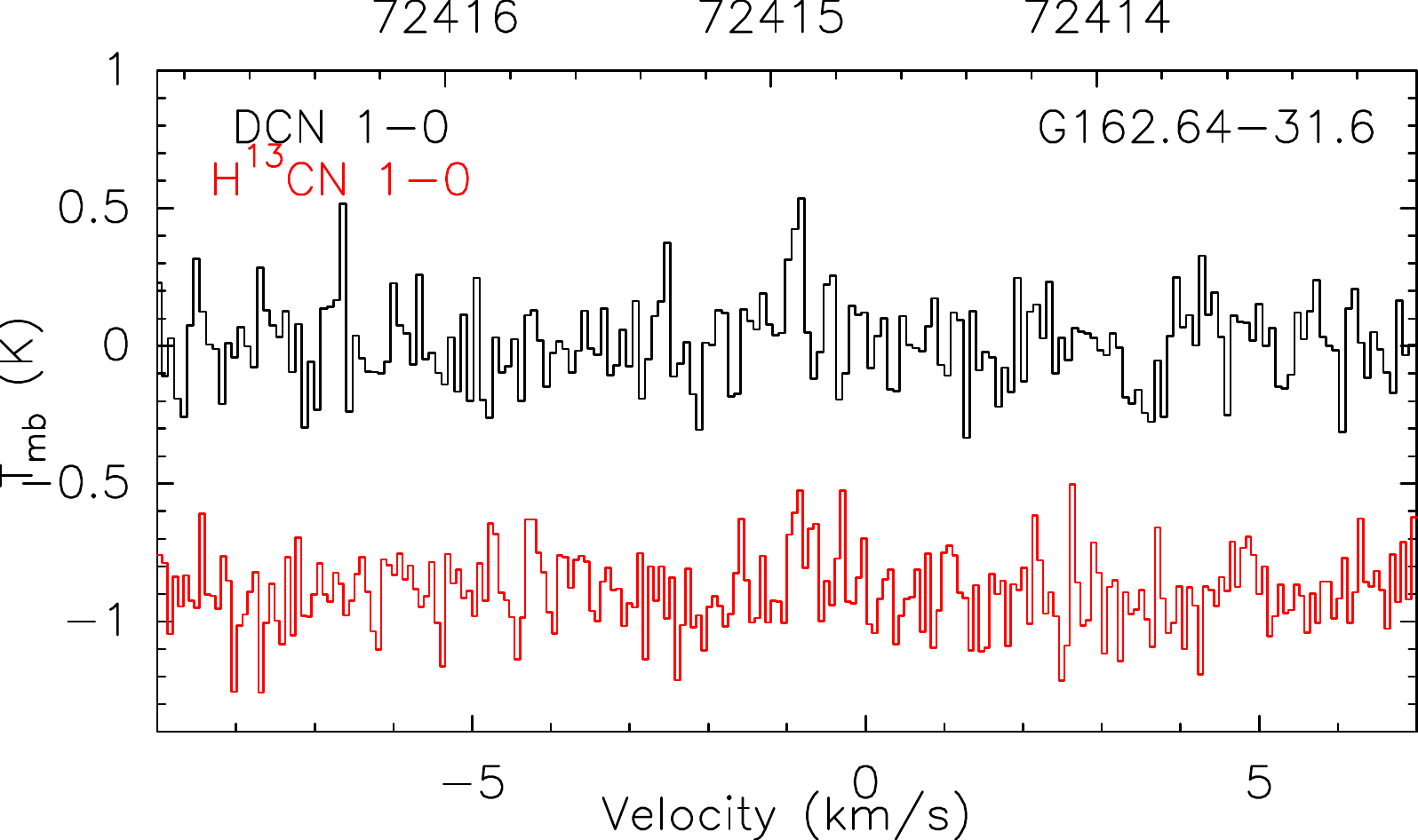}
\includegraphics[width=0.3\columnwidth]{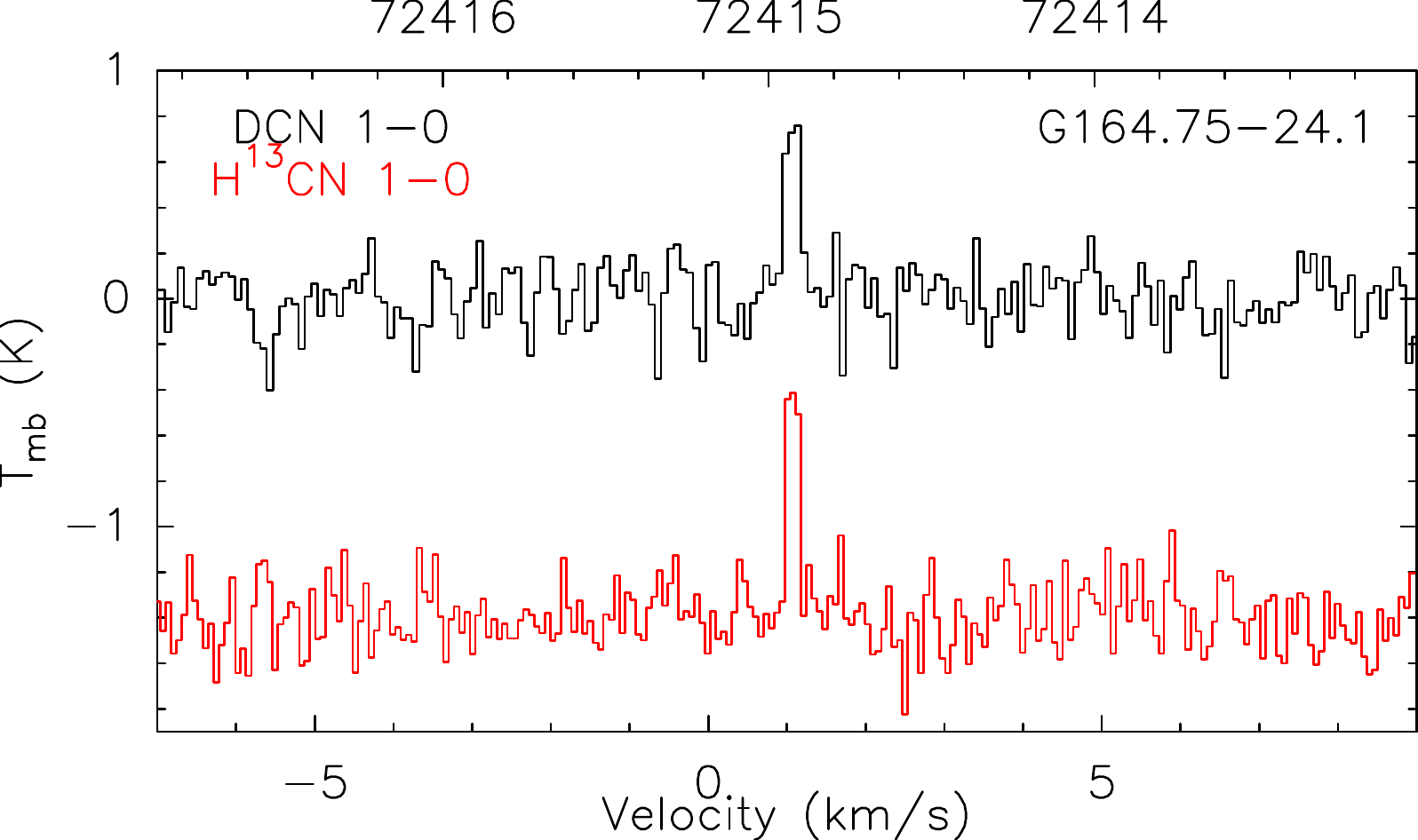}
\caption{Line profiles of DCN and H$^{13}$CN 1-0 with the high velocity resolution mode (AROWS mode 13). H$^{13}$CN 1-0 lines were manually aligned with DCN 1-0 lines, due to errors in the Doppler tracking.\centering}
\label{DCNH13CNmode13_2}
\end{figure}

\end{document}